\documentclass[11pt,a4paper]{book}
\pdfoutput=1
\usepackage[utf8]{inputenc}

\usepackage{stmaryrd,upgreek,mathrsfs}
\usepackage{amstext,amsmath,amssymb,amsfonts,amsthm, bbm}
\usepackage{extarrows}
\usepackage{hyperref}
\usepackage{authblk}
\usepackage{caption} 
\usepackage{epsfig} 
\usepackage{color} 
\usepackage{graphicx} 
\usepackage{braket} 
\usepackage{slashed} 
\usepackage{dsfont} 
\usepackage{young}
\usepackage{soul}
\usepackage{multirow}

\usepackage{pdfpages}

\usepackage{cite}
\usepackage{import}

\usepackage{appendix} 

\allowdisplaybreaks[4]

\usepackage{psfrag}
\usepackage{tikz} 
\usetikzlibrary{arrows}
\usetikzlibrary{plotmarks}
\usetikzlibrary{patterns}
\usetikzlibrary{external}
\usepackage{pgfplots}
\pgfplotsset{compat=1.9}
\usetikzlibrary{patterns}
\usetikzlibrary{calc}
\usetikzlibrary{automata,positioning}

\usepackage{float}  
\usepackage{subfig}
\usepackage[lmargin=55pt,rmargin=55pt,tmargin=60pt,bmargin=65pt]{geometry}
\usepackage[resetlabels]{multibib}
\newcites{publi}{Publications}
\usepackage{etoolbox}

\newcommand{\be}{\begin{equation}}
\newcommand{\ee}{\end{equation}} 
\newcommand{\bga}{\begin{gather}}
\newcommand{\ega}{\end{gather}} 
\newcommand{\nn}{\nonumber}
\newcommand{\Tr}{{\rm Tr}}
\newcommand{\tr}{{\rm tr}} 
\newcommand{\f}{\frac}
\newcommand{\p}{\partial}

\newcommand{\la}{\langle}
\newcommand{\ra}{\rangle}
\newcommand{\dd}{{\rm d}^d}

\let\a=\alpha \let\b=\beta  \let\g=\gamma  \let\d=\delta
\let\z=\zeta       \let\k=\kappa \let\l=\lambda
\let\m=\mu    \let\n=\nu          \let\r=\rho \let\om=\omega
\let\s=\sigma     \let\ph=\phi 

\let\G=\Gamma \let\D=\Delta   \let\L=\Lambda \let\X=F
         
\let\Om=\Omega  \let\eps=\epsilon

\newcommand{\phib}{\bar{\phi}}

\newcommand{\cB}{\mathcal{B}}

\newcommand{\cD}{\mathcal{D}}
\newcommand{\cE}{\mathcal{E}}
\newcommand{\cF}{\mathcal{F}}
\newcommand{\cG}{\mathcal{G}}

\newcommand{\cI}{\mathcal{I}}
\newcommand{\cJ}{\mathcal{J}}

\newcommand{\cN}{\mathcal{N}}
\newcommand{\cO}{\mathcal{O}}
\newcommand{\cP}{\mathcal{P}}

\newcommand{\cS}{\mathcal{S}}
\newcommand{\cT}{\mathcal{T}}

\newcommand{\cV}{\mathcal{V}}

\newcommand{\cZ}{\mathcal{Z}}

\newcommand{\gt}{\tilde{g}}

\newcommand{\rt}{\tilde{r}}
\newcommand{\htilde}{\tilde{h}}
\newcommand{\wtD}{\widetilde{\Delta}}

\DeclareMathOperator{\im}{\mathrm{i}} 

\newcommand{\sgn}{{\rm sgn}}
\newcommand{\mba}{\mathbf{a}}
\newcommand{\mbb}{\mathbf{b}}
\newcommand{\mbc}{\mathbf{c}}
\newcommand{\mbd}{\mathbf{d}}
\newcommand{\mbe}{\mathbf{e}}
\newcommand{\mbf}{\mathbf{f}}
\newcommand{\mbg}{\mathbf{g}}
\newcommand{\mbh}{\mathbf{h}}
\newcommand{\mbj}{\mathbf{j}}
\newcommand{\mbk}{\mathbf{k}}
\newcommand{\mbm}{\mathbf{m}}
\newcommand{\mbn}{\mathbf{n}}
\newcommand{\mbp}{\mathbf{p}}
\newcommand{\mbq}{\mathbf{q}}

\newcommand{\mbl}{\pmb{\lambda}}
\newcommand{\mbS}{\pmb{\Sigma}}
\newcommand{\mbG}{\pmb{G}}

\newcommand{\mbC}{\pmb{C}}
\newcommand{\mbK}{\pmb{K}}

\theoremstyle{plain}
\theoremstyle{definition}

\newtheorem{remark}{Remark}
\newtheorem{proposition}{Proposition}[section]
\newtheorem{definition}{Definition}
\newtheorem{lemma}{Lemma}[section]
\newtheorem{theorem}{Theorem}[section]

\numberwithin{remark}{chapter}
\numberwithin{theorem}{chapter}
\numberwithin{lemma}{chapter}
\numberwithin{proposition}{chapter}
\numberwithin{definition}{chapter}
\numberwithin{cor}{chapter}

\definecolor{orange}{rgb}{0.88,0.39,0.12} 
\definecolor{rouge}{rgb}{0.8, 0.0, 0.0}
\definecolor{vert}{rgb}{0.4, 0.69, 0.2}
\definecolor{bleu}{rgb}{0.19, 0.55, 0.91}
\definecolor{lavenderpurple}{rgb}{0.59, 0.48, 0.71}

\begin{document}
\frontmatter
\begin{titlepage}
\includepdf[pages=-,pagecommand={},width=\paperwidth]{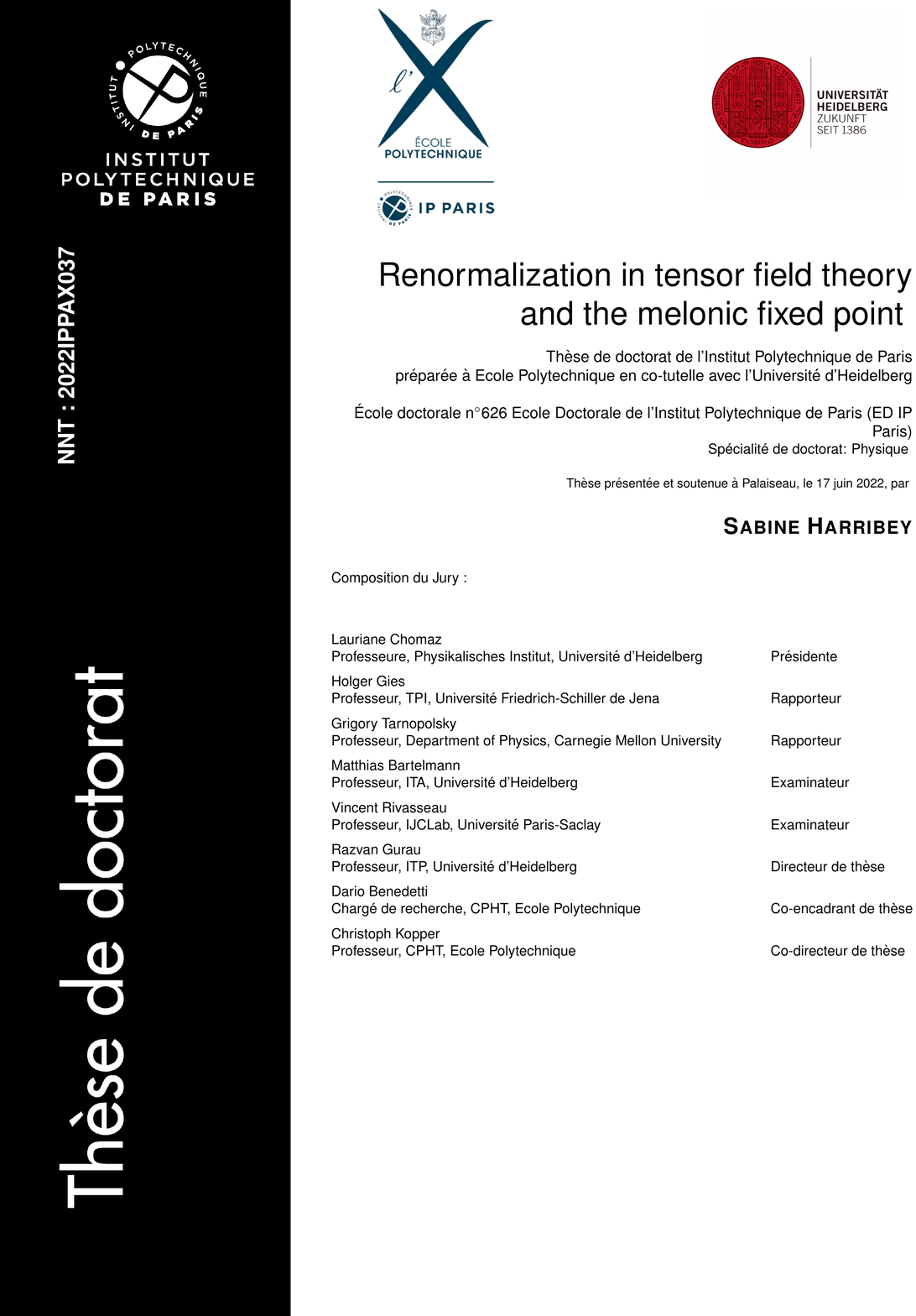}
\end{titlepage}

\chapter*{Acknowledgments}
This section is meant to express my gratitude towards the people that helped me during these four years in Paris and Heidelberg. 

I would first like to warmly thank my supervisors Razvan and Dario. With their great mentorship and personal involvement, they led me through the PhD journey. I was never left alone with questions or problems as they were always willing to spare some time and provide me with great advice. I am forever grateful for their excellent supervision of my PhD. I would also like to thank Christoph for accepting to be my third supervisor and for taking part in my PhD. 

Second, I would like to thank the members of the jury, Matthias Bartelmann, Lauriane Chomaz, Holger Gies, Grigory Tarnopolsky and Vincent Rivasseau for accepting to be part of it and for their participation in the defense. I thank in particular the two referees, Holger and Grigory, for their careful reading of my manuscript and Lauriane for presiding the jury. 

During my PhD, I also had great mentors outside my supervising team. In particular, I would like to thank Vincent and Marios for their support and encouragement. The director of the CPHT in Polytechnique, Jean-René, was also very welcoming and helpful when I first arrived. 

I am furthermore very grateful for getting to know and collaborating with the members of the \textit{tensor family}: Adrian, Ariane, Astrid, Carlos, Dario, Davide, Fabien, Hannes, Joseph, Kenta, Luca, Matteo, Nicolas, Razvan, Reiko, Romain, Sylvain, Thomas, Valentin, Vasily, Victor, Vincent. Particularly, thank you Sylvain for your great supervision during my master internship that led me to this PhD, thank you Nicolas for the warm welcome into the team and for all the cups of tea and thank you Davide for making the move to Heidelberg easier. The schools, workshops and conferences were a very rewarding part of my PhD, so I would like to thank Reiko, Adrian and Fabien for inviting me to Okinawa, Bordeaux and Lyon and all those that organized these conferences and others.

Exchanging with members of the CPHT (Andrea, Balt, Balthazar, Blaise, Charles, Debtosh, Emilian, Eric, Guillaume, Marios, Luca, Thibault, ...) over lunch or coffee was a gratifying experience during my PhD. I also have to thank all the administrative and IT staff of both my labs, in Polytechnique and Heidelberg University, who helped me through French and German bureaucratic formalities. So thank you Anja, Besma, Christine, Fadila, Florence, Jean-René, Malika, Melanie, Nadja, Sonja. 

Before starting my PhD, a few teachers helped me find my passion and achieve my goals, such as Mrs Matignon in middle school or Mr Baume in classes préparatoires. During my time at ENS Lyon, I would like to particularly thank François Gieres for his support and Dimitrios Tsimpis for his great supervision during my master thesis. 

Of course, the four years of my PhD would not have been the same without my friends. First, thanks to Eléa and Florence whom I met on my first day at the ENS in Lyon and are still by my side almost seven years later. I would also like to thank my Heidelberg friends, Marie and Maureen, who made my time in Germany wonderful.

I will conclude with a few words in French for my family that I am particularly glad to always have by my side. 
Je souhaite remercier vivement mes parents, Sylvie et Jean-Christophe, pour leur soutien tout au long de ma thèse et de mes études ainsi que pour m'avoir toujours poussée à suivre mes rêves. Je remercie également mes grand-parents, Jacqueline, André et Nicole, pour leurs encouragements. Mes pensées se tournent aussi vers ma tante, Cathy, qui aurait été si heureuse de me voir soutenir ma thèse. Je remercie finalement les membres de ma famille présents lors de ma soutenance: Gilberte, Jean-Luc, Laurence et Xavier ainsi que tous mes oncles, tantes, cousins, cousines qui m'accompagnent depuis toujours.

\tableofcontents

\bibliographystylepubli{JHEP-3}
\nocitepubli{*}
\bibliographypubli{publications}

\bigskip
\medskip 

Chapter \ref{chap:3loops} is based on \citepubli{Benedetti:2020rrq}. Chapter \ref{chap:CTKT} is an edited version of \citepubli{Benedetti:2019eyl} and \citepubli{Benedetti:2019ikb}. Chapter \ref{chap:trif} is based on \citepubli{Benedetti:2020sye} while chapter \ref{chap:Ftheorem} follows \citepubli{Benedetti:2021wzt}. Chapter \ref{chap:sextic} is based on \citepubli{Benedetti:2019rja} and \citepubli{Harribey:2021xgh}. Finally, chapter \ref{chap:rank5} follows \citepubli{Carrozza:2021qos}.

\mainmatter
\chapter*{Introduction}
\addcontentsline{toc}{chapter}{Introduction}
\markboth{INTRODUCTION}{INTRODUCTION}
The laws of our universe are encompassed by four fundamental forces: weak force, strong force, electromagnetism and gravity. The first three are described by the Standard Model of Physics while gravity is described by General Relativity. On the microscopic side, the Standard Model describes all elementary particles, the interactions between them as well as the Higgs boson. It was tested experimentally (in particular at the Large Hadron Collider) with great precision. On the other hand, General Relativity describes gravity and space-time at large scales. It is governed by Einstein equation:
\begin{equation}
R_{\mu \nu}-\frac{1}{2}(R-2\Lambda)g_{\mu \nu}=\frac{8\pi G}{c^4}T_{\mu \nu} \; ,
\end{equation}
where the left-hand side describes the geometry of space-time through the metric $g_{\mu \nu}$, with $R_{\mu \nu}$ and $R$ the Ricci tensor and scalar curvature respectively and the right-hand side describes the matter energy content through the stress-energy tensor $T_{\mu \nu}$. $\Lambda$ is the cosmological constant. The idea behind this equation is that space-time is deformed by matter. General Relativity explains with great precision phenomena that are not understandable using Newtonian gravity, such as the perihelion precession of Mercury or the existence of black holes. Predictions made by General Relativity were also observed, such as gravitational waves (that were recently detected by the LIGO experiment).
However, problems arise and these two theories are not complete. For example, the high number of parameters in the Standard Model is puzzling. Much more problematic are the existence of Landau poles in quantum electrodynamics (QED) and the appearance of many space-time singularities in General Relativity. This indicates that the Standard Model and General Relativity could be only an approximation of a more fundamental law. One of the main questions of theoretical physics is then to find this more fundamental framework in order to describe in a unified way both the Standard Model and gravity. However, these two theories live at very different energy scales. Indeed, gravity starts to compete with other forces at the Planck energy, $10^{16} \textrm{ TeV}$, which is far beyond what can be currently reached experimentally. The main issue is then to find a suitable way to describe gravity on a quantum level without experimental support. Many approaches have been tried to solve this problem among which we can mention string theory, asymptotic safety, random geometry or loop quantum gravity for example. 
Quantum gravity can also be approached through holography thanks to the AdS/CFT correspondence \cite{Maldacena:1997re,Aharony:1999ti}. This correspondence states that any theory of gravity in $d+1$ dimension living in the bulk of an anti-de Sitter (AdS) space is dual to a $d$-dimensional conformal field theory (CFT) without gravity living on the boundary of this AdS space. This is particularly useful as computations are often easier on the CFT side. This is a very successful program which has prompted numerous studies (see \cite{Ammon:2015wua,Erdmenger:2018xqz,DeWolfe:2018dkl} for more recent reviews). However, in this thesis, we will not focus on such approaches but rather on the tools and frameworks used to develop them.

\begin{figure}[htbp]
\centering
\captionsetup[subfigure]{labelformat=empty}
\subfloat[]{\includegraphics[scale=0.5]{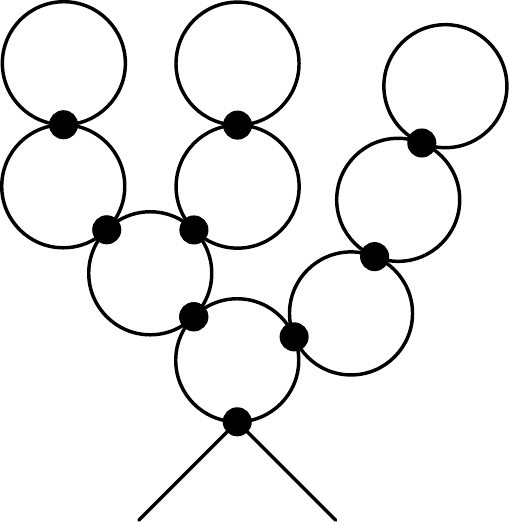}}
\hspace{1cm}
\subfloat[]{\includegraphics[scale=0.5]{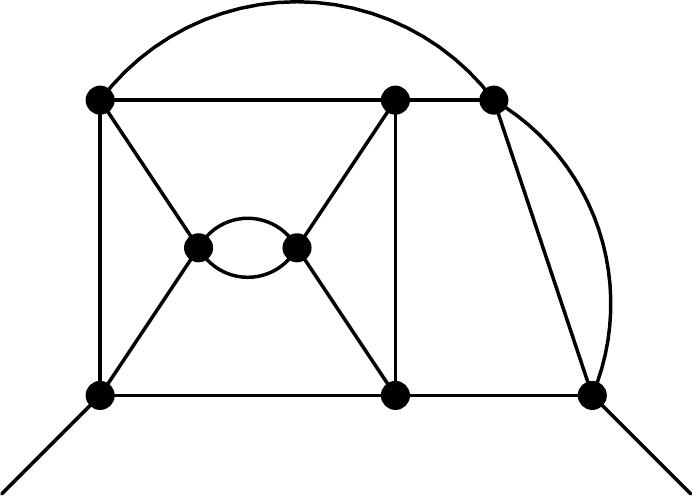}}
\hspace{1cm}
\subfloat[]{\includegraphics[scale=0.5]{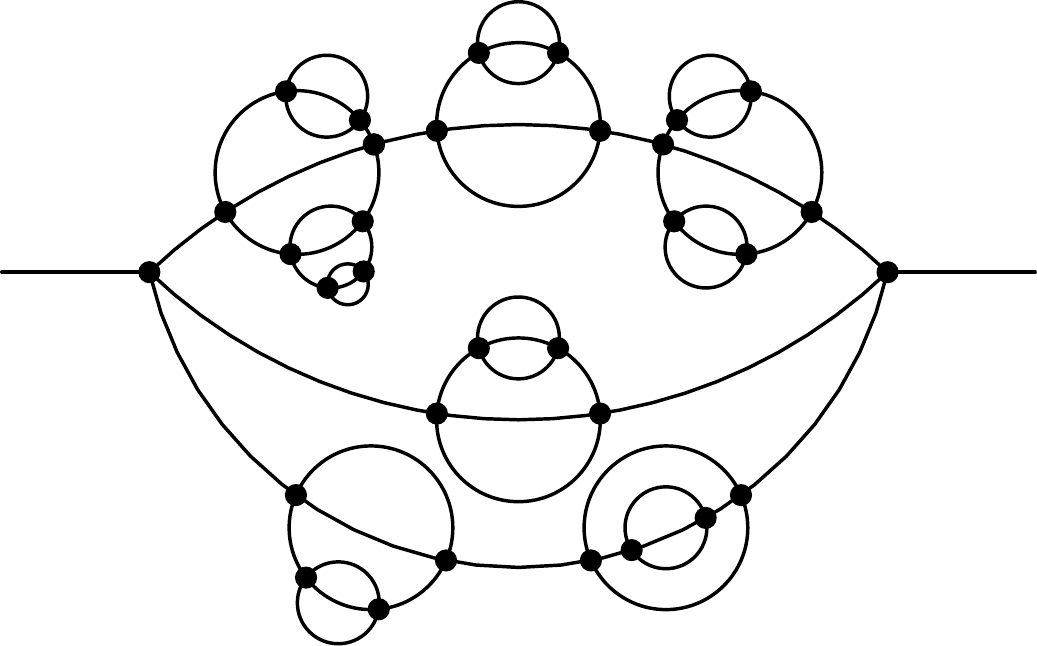}}
\caption{The large-$N$ limit of vector, matrix and tensor models from left to right.}
\label{fig:VMT}
\end{figure}

In all these approaches to quantum gravity, as well as in the Standard Model, one is confronted with a high number of degrees of freedom. The appropriate framework for describing both the Standard Model and interactions in String theory is Quantum Field Theory (QFT).\footnote{Many reviews and books were written on the subject. We refer for example to \cite{Peskin:1995ev,Weinberg:1995mt}.} In QFT, the fundamental objects are fields that describe particles and their interactions at the quantum level. However, QFT is ill-defined mathematically and many divergences arise. 
Renormalization was thus introduced in order to deal with those divergences. There exists many different renormalization methods but the main idea is that physics changes with the energy scale. Any field theory flows between different theories from the ultraviolet to the infrared: this is the renormalization group (RG) flow. One is then interested in fixed points of this flow: they correspond to field theories independent of the energy scale. However, the main methods rely on a perturbative expansion in the interaction coupling: one assumes that the interaction, or force, is weak and expands in the constant characterizing this interaction. An important question is then: how to perform renormalization analysis when the interaction is strong? 
One possible method, that we will study in this thesis, is to use large-$N$ theories: one has an extra parameter $N$ that can be taken large so that an expansion in $1/N$ is possible. $N$ can be the number of different fields or the size of a vector field for example. Such methods are also used in quantum chromodynamics (QCD) where one expands in the number of colors $N=3$. In the case of vector fields, the large-$N$ limit is dominated by bubble or cactus diagrams: they describe only local interactions \cite{Guida:1998bx,Moshe:2003xn}. Even though vector models have been linked to higher spin theories \cite{Vasiliev:1999ba}, their large-$N$ limit is not very rich from a QFT perspective, as for example the RG flow is one-loop exact in the strict large-$N$ limit. One can then go one step further and consider matrices of size $N$ with $N$ large. The leading order is in this case dominated by planar graphs: this is a much richer limit but computations are hard and resummation is not always possible \cite{'tHooft:1973jz,Brezin:1977sv,DiFrancesco:1993nw}. Finally, one can consider tensor fields. It could be expected that their large-$N$ limit is more complicated than the one of matrices as they are more complicated algebraic objects. However, the surprise was that tensor models admit a large-$N$ limit dominated by melon graphs \cite{Bonzom:2011zz,RTM}. This is simpler than the planar limit of matrices as melons are a subset of planar graphs but richer than the large-$N$ limit of vector models as melons describe bi-local interactions. We pictured in figure \ref{fig:VMT} a graphical representation of the leading order Feynman graphs of vector, matrix and tensor models. 

We have just presented tensor models as large-$N$ field theories that provide interesting toy models for strongly-coupled QFTs. However, they were introduced in a different context. They were studied in zero dimension as a way to describe random geometry and quantum gravity \cite{Ambjorn:1990ge,Sasakura:1990fs,RTM}. In quantum theories, path integrals are fundamental objects: they sum over all possible configurations available to the system. The idea is then to consider the space-time as discrete and sum over all possible geometries. In order to obtain a suitable theory of quantum gravity one should recover General Relativity at large distances, compared to the Planck scale. More precisely, one has to sum over piecewise linear $D$-dimensional manifolds up to diffeomorphism. These geometries are obtained by gluing $D$-dimensional simplices and appear as the Feynman graphs of the perturbative expansion of colored tensor models of rank $D$. When $D=2$, this reduces to matrix models which have led to a better understanding of two-dimensional quantum gravity \cite{DiFrancesco:1993nw,Distler:1988jt,David:1988hj} and string theory. In order to go to higher dimensions, one needs to increase the dimension of the simplices and hence the rank of the tensors. Tensor models could thus be used to describe quantum gravity in dimension higher than two. However, they only generate branched polymers in the continuous limit \cite{Gurau:2013cbh}. These are tree-like phases of Hausdorff dimension $2$ and spectral dimension $4/3$. They are therefore insufficient to build higher-dimensional geometries needed to describe quantum gravity in dimensions higher than two.

The study of tensor models was then halted until some large-$N$ melonic limit appeared in the Sachdev-Ye-Kitaev (SYK) model \cite{kitaev,Sachdev:1992fk}, a one-dimensional quantum mechanical model of strongly interacting fermions. This model is of particular interest as it could provide a holographic description of the near-horizon limit of near-extremal black holes. However, this model requires a quenched average over disorder meaning that it is not a fundamental theory. A tensor model in one dimension was then used by Witten to generalize the SYK model without disorder \cite{Witten:2016iux}. This prompted extensive studies of tensor models in one dimension \cite{Gurau:2016lzk,Peng:2016mxj,Krishnan:2016bvg,Krishnan:2017lra,Bulycheva:2017ilt,Choudhury:2017tax,Halmagyi:2017leq}. Tensor models were then generalized to higher dimensions to proper quantum field theories \cite{Giombi:2017dtl,Prakash:2017hwq,Benedetti:2017fmp,Giombi:2018qgp,Benedetti:2018ghn}. In this context, they were shown to give rise to a new kind of conformal field theories, called melonic CFTs. This is of particular interest as CFTs are quantum field theories with a high number of symmetries allowing to do analytical computations. Indeed, finding interesting models that are solvable, at least in some limits, is of great theoretical appeal. We can thus hope, by studying exactly solvable toy models built on tensors, to gain insight on a vast array of applications ranging from statistical physics and strongly-coupled matter to quantum gravity. 

The goal of this thesis is to investigate the properties of melonic CFTs. How do they differ from vector and matrix models? Can we obtain truly tensorial fixed points with a new critical behavior? What class of tensor models leads to melonic CFTs?

\vspace{5mm}

Before tackling these questions, we give in chapter \ref{ch:background} an extended overview of the notions we have just introduced. As it is the main subject of the thesis, we start in section \ref{sec:renormalization} with a short review of the Wilsonian renormalization group for quantum field theories. We present specific renormalization schemes through the standard example of the Wilson-Fisher fixed point. We conclude this section with a few comments on non-perturbative methods. In section \ref{sec:CFT}, we give some basics of conformal field theories. We detail the conformal group and the form of the two and three-point functions. We then present the operator product expansion and explain how it is used to determine the four-point functions and the spectrum of primary operators. Finally, in section \ref{sec:RTM}, after introducing colored graphs, we present the two main types of tensor models as well as their large-$N$ expansions. We finish this section by giving a few example of other melonic limits in one-dimensional quantum models and conclude the chapter with a detailed summary of the rest of the thesis. In chapter \ref{chap:3loops}, we study the long-range multi-scalar model with quartic interactions and compute its beta functions up to three loops. In chapter \ref{chap:CTKT}, we consider a long-range bosonic tensor model and study its renormalization group flow at large $N$ as well as the properties of the CFT at the fixed point. In chapter \ref{chap:trif}, we compute the $1/N$ corrections for this model by first considering a generic trifundamental model both in short- and long-range. In chapter \ref{chap:Ftheorem}, we investigate further the properties of the melonic CFT at the fixed point of the $O(N)^3$ model and in particular test the $F$-theorem. In chapter \ref{chap:sextic}, we now consider a tensor field theory with sextic interactions. We compute the beta functions and fixed points for two models, in rank $3$ and $5$ at large $N$ as well as the $1/N$ corrections for the rank-$3$ model. Finally, in chapter \ref{chap:rank5}, we extend the class of models exhibiting a melonic large-$N$ limit to irreducible tensor models in rank $5$. We end the thesis by giving some concluding remarks and future perspectives.

\chapter{Background and motivations}
\label{ch:background}
\section{Renormalization}
\label{sec:renormalization}
In Quantum Field Theory, the fundamental degrees of freedom are represented by quantum fields $\phi(x)$ at the space-time point $x$. The interesting objects to consider are then correlation functions between those fields:

\begin{equation}
\langle \mathcal{O}_1(x_1)\dots \mathcal{O}_n(x_n)\rangle \; ,
\end{equation}
where $\mathcal{O}_i$ are operators depending on $\phi$.
These correlations are the analogs of the moments of a probability distribution. We will compute them using a perturbative approach. 

We start with an action of the following form in $d$ dimensions:

\begin{equation}
S[\phi]=\int d^d x \left[ \phi(x) \left( -\partial_{\mu} \partial^{\mu} + m^2 \right)\phi(x) +\lambda \phi(x)^q\right] \; . 
\label{eq:action_generic}
\end{equation}
The first term is the free part of the action, with covariance $C(x,y)^{-1}=\left( -\partial_{\mu} \partial^{\mu} + m^2 \right)\delta(x-y)$, while the second term is the interacting part. It is useful to represent the covariance in momentum space:\footnote{We denote $x,y$ and so on positions, $\int_x \equiv \int d^dx$ and
$p,q$ and so on momenta and $\int_p \equiv \int \frac{d^dp}{(2\pi)^d}$. The Fourier transform is 
$  f(p) = \int_x  e^{\im p x} f(x)$ with inverse $f(x) = \int_p  e^{-\im px} f(p)$; we denote them by the same symbol, but context and argument of the function should lift any ambiguity.
}
\begin{equation}
C(p)=\int d^d x e^{i p (x-y)}C(x,y)=\frac{1}{p^2+m^2} \, ,
\end{equation}
as well as using Schwinger parameters:
\begin{equation}
C(x-y) = \frac{1}{(4\pi)^{d/2} } \int_0^{\infty} d\alpha \;\alpha^{-d/2 } e^{- \frac{(x-y)^2}{4 \alpha}-m^2\alpha}\;.
\end{equation}

The perturbative expansion of the correlations will come by first performing a Taylor expansion on the interaction part. Then, for each order $p$ in the interaction coupling,  we perform a Wick contraction of the fields with respect to the Gaussian term. This leads to the usual expansion of the correlations in terms of Feynman graphs. The free energy of the model is then:
\begin{equation}
\begin{split}
 {\mathcal F} & =-  \ln \bigg\{  \int [d\phi]\; e^{-S[\phi]}\bigg\} = -\sum_{\cG}  \frac{\left(-\lambda\right)^p}{ p!} A(\cG)   \;, \crcr
 A(\cG)& =  
 \int \prod_{ 1\leq k \leq p } dy_{k} \prod_{(y_i,y_j) \in E(\cG) } C(y_i,y_j) \;,
\end{split}
\end{equation}
where $\cG$ runs over connected vacuum Feynman graphs with $p$ labeled vertices $y_i \, , \, i=1,\dots,p$.

The $n$-point correlation functions are written similarly but the sum will run over connected Feynman graphs with $p$ internal vertices and $n$ external points. In this case, the $p$ internal vertices are $q$-valent (connected to $q$ half-edges) and the $n$ external points are one-valent vertices identified with the external points $(x_i)_{1\leq i \leq n}$. The half-edges are then connected pair-wise by propagators to form the edges $E(\mathcal{G})$ of the graph. 
In direct space, the integration is done over all internal vertices. In momentum space, due to conservation of momenta at the vertices, one integrated momentum is associated to every loop. Finally, the propagators attached to the external vertices are often removed. We call the corresponding graphs \textit{amputated}.

Perturbative approaches in quantum field theory give rise to a number of problems through divergences. First, at fixed order $p$, some Feynman graphs can have divergent amplitudes. Second, divergences also arise when resumming the perturbative expansion, due to the factorial growth of Wick contractions to sum over or to some correlation functions behaving as $p!$. We will see in the following how those divergences can be cured.

\subsection{Wilsonian renormalization}
\label{subsec:WR}

\paragraph{Power counting}
Let us first look at UV divergences arising in individual Feynman graphs. 
Let $\mathcal{G}$ be a connected amputated Feynman graph with $V(\mathcal{G})$ vertices, $E(\mathcal{G})$ edges and $n(\mathcal{G})$ external points. In momentum space one counts an independent integral $d^dp$ for every loop and a propagator $p^{-2}$ for every edge. Under a global rescaling by a factor $t$ of all the momenta, the amplitude is rescaled as:

\begin{equation}
t^{d(E(\mathcal{G})-V(\mathcal{G})+1)-2E(\mathcal{G})}=t^{d+n(\mathcal{G})\left(1-\tfrac{d}{2}\right)+V(\mathcal{G})\left(\tfrac{d(q-2)}{2}-q\right)} \; ,
\end{equation}
where $E(\mathcal{G})-V(\mathcal{G})+1$ is the number of loops and we used $2E(\mathcal{G})=qV(\mathcal{G})-n(\mathcal{G})$ as vertices are $q$-valent.
We can then define a UV degree of divergence $\text{deg}(\mathcal{G})$:
\begin{equation}
\text{deg}(\mathcal{G})=d+n(\mathcal{G})\left(1-\tfrac{d}{2}\right)+V(\mathcal{G})\left(\tfrac{d(q-2)}{2}-q\right) \; .
\label{eq:degUV}
\end{equation}

If the degree is positive, the graph is said to be power divergent. If it is negative, the graph is UV convergent. If it is exactly zero, the graph is logarithmically divergent. Choosing $d=\frac{2q}{q-2}$, the power counting does not depend on the number of internal vertices: the theory is then marginal. 

\paragraph{Wilsonian renormalization}
One of the main renormalization schemes is Wilsonian renormalization \cite{RevModPhys.55.583}. 
%
The idea behind Wilsonian renormalization is the fact that the best way to study a subset of degrees of freedom of a system is to build an effective theory by integrating out the others \cite{Delamotte:2007pf}.  

If we consider a field theory with an UV cutoff $\Lambda$, this would consist in defining an IR scale $k$ and integrating the ``high" modes of the field $\phi(p)$ (corresponding to $p$ between $k$ and $\Lambda$). We would be left with an effective Hamiltonian for the low energy modes, corresponding to $p<k$. 

More precisely, we introduce an explicit UV cutoff $\Lambda$:
\be 
C^{\Lambda}(p)  = \frac{1}{p^{2}} \, \Theta\left(\frac{p^2}{\Lambda^2}\right)  \; , \qquad C^{\Lambda}(x) =  \int_p \; \frac{e^{-\imath p x}}{ p^{2} } \,\Theta\left(\frac{p^2}{\Lambda^2}\right)  \; .
\ee
The partition function is then:
\begin{equation}
Z=\int d\mu_{ C^{\Lambda}}  (\phi) \; e^{-S_{\Lambda}[\phi]} \;,
\end{equation}
where $d\mu_{ C^{\Lambda}  }$ denotes the normalized Gaussian measure with covariance $C^{\Lambda}$. We denote by convention $S_{\Lambda}[\phi] \equiv S^{\rm int}[\phi]$ the bare potential of our model \eqref{eq:action_generic}.

The ultraviolet divergences are regularized by the multiplicative 
cutoff function $\Theta(p^2/\Lambda^2)$. $\Theta(u)$ is monotonic and takes values between 0 and 1,  such that $\Theta(u)\simeq 1$ for $u<1$, and $\Theta(u)\simeq 0$ for $u>1$.
Typical choices are the exponential cutoff $\Theta(u)= e^{-u}$, the sharp cutoff $\Theta(u)=\theta(1-u)$ or a smooth approximation of it for example.
While the specific choice of the cutoff function should not affect the main results, we will choose once and for all to use the standard exponential cutoff.


Let $k\le \Lambda$ be an infrared scale.
To implement the Wilsonian renormalization idea, we want to separate the modes with $p\in [0,k]$ and $p\in [k,\Lambda]$. In order to do so, we split the covariance:

\begin{equation}
C^{\Lambda}=C^{k}+C_k^{\Lambda} \; .
\end{equation}
This would be equivalent to introducing the slice cutoff function $\chi^{\Lambda}_k(p) = \Theta\left( p^2/\Lambda^2 \right)- \Theta\left( p^2/k^2 \right) $
and considering $C^{\Lambda}_k$, the covariance of the fluctuations (the modes with momenta between $\Lambda$ and $k$):
\be
 C^{\Lambda}_k (p) = \frac{1}{p^{2}} \,\chi^{\Lambda}_k(p) =  \int_{\Lambda^{-2}}^{k^{-2}}  d\alpha \; e^{-\alpha p^2 } \; .
\ee
Then, the partition function becomes (see for example \cite{Gurau:2014vwa}): 
\begin{equation}
Z=\int d\mu_{ C^{k}}  (\phi_{<}) \int d\mu_{C_k^{\Lambda}}(\phi_{>}) \; e^{-S_{\Lambda}[\phi_{<}+\phi_{>}]} \;.
\end{equation}
Now, we integrate the ``high" modes $\phi_{>}$ and we obtain the effective action for the ``low" modes $\phi_{<}$:
\begin{equation}
Z=\int d\mu_{ C^{k}}  (\phi_{<})e^{-S_{k}[\phi_{<}]} \; ,
\label{eq:effaction}
\end{equation}
with
\begin{equation}
e^{-S_{k}[\phi_{<}]}=\int d\mu_{C_k^{\Lambda}}(\phi_{>}) \; e^{-S_{\Lambda}[\phi_{<}+\phi_{>}]} \;,
\end{equation}
where $S_{k}[\phi_{<}]$ is the interaction part of the effective action of the theory. We can then expand it into powers of the field and use translation invariance to write:
\be
\begin{split}
S_{k}[\phi] & = \sum_{n\geq 0} \frac{1}{n!} \int_{x_1,\ldots, x_n} S_{k}^{(n)}(x_1,\ldots,x_n) \phi(x_1) \cdots \phi(x_n)\\
&= \sum_{n\geq 0} \frac{(2\pi)^d}{n!} \int_{p_1,\ldots, p_n} S_{k}^{(n)}(p_1,\ldots,p_n) \phi(p_1) \cdots \phi(p_n) \delta(p_1+\ldots+p_n) \;.\end{split}
\ee
Assuming $\mathbb{Z}_2$ invariance, the $n$-point functions $S_{k}^{(n)}$ will vanish for odd $n$. 

\begin{figure}[htbp]
\centering
\includegraphics[scale=1]{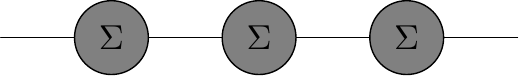}
\caption{Example of an amputated chain with two propagators and three 1PI vertices (self-energy $\Sigma$).}
\label{fig:chain}
\end{figure}

Now, let us look at $S_{k}^{(2)}$. It is minus the amputated connected two-point function for the theory with the UV covariance. Therefore it can be written as a sum of chains of self-energy vertices (sum of non-trivial one-particle irreducible graphs) and propagators (see figure \ref{fig:chain}). We can thus write:
\begin{equation}
S_{k}^{(2)}=-\Sigma-\Sigma C_k^{\Lambda}\Sigma - \dots \, .
\end{equation}
This effective two-point vertex must be combined with the infrared covariance $C^k$ from the measure of the effective action. 
The effective covariance is then $\left((C^k)^{-1}-\Sigma-\Sigma C_k^{\Lambda}\Sigma - \dots\right)^{-1}$. 
This truncates to the first term, $\left((C^k)^{-1}-\Sigma\right)^{-1}$: all the corrections involve the term $C^k\Sigma  C_k^{\Lambda}\Sigma$ and because of the conditions on the cutoff function, $C^k C_k^{\Lambda}$ is small (even zero for a sharp cutoff). 

We now define the wave function by $\Sigma=-(Z_k^{-1}-1)p^{2\zeta}+\dots $. The effective covariance is then $Z_kC^k$. All the remaining two-point terms (coming from $\Sigma  C_k^{\Lambda}\Sigma$ and from the other terms of the self-energy) are left in a two-point rest term. This rest term is usually ignored and we will neglect it here. To retrieve the original covariance $C^k$ from the effective one $Z_k C^k$, we can rescale the fields $\phi \rightarrow \sqrt{Z_k}\phi$. 

The last step of the Wilsonian renormalization is to rescale the momenta by $\Lambda/k$ in order to recover the bare covariance $C^{\Lambda}$ of the leftover modes $p<k$ in \eqref{eq:effaction}. 
The effective action for the modes $p<k$ is then:
\begin{equation}
S_{k}[\phi] = \sum_{n\geq 0} Z_k^{n/2}\frac{(2\pi)^d}{n!} \int_{p_1,\ldots, p_n} S_{k}^{(n)}(p_1,\ldots,p_n) \phi(p_1) \cdots \phi(p_n) \delta(p_1+\ldots+p_n) \; .
\end{equation}

Now that we have obtained an effective action for our theory, we can express the corresponding effective couplings. In order to do so, we have to choose a scale $\mu$ to do the subtraction, such that the effective coupling is defined as $Z_k^{n/2}S_{k}^{(n)}(\mu,\mu,\mu,\mu)$.
Usually, in massless theories the subtraction is done at a scale $\mu \neq 0$.  This is particularly relevant in dimensional regularization: the scale $\mu$ is required in order to tame the infrared divergences. As it is the only scale in the problem, the renormalization flow is then studied with respect to this subtraction scale. 

However, in the Wilsonian picture, the infrared divergences are tamed by the IR cutoff $k$ and integrating the modes down to the scale $k$ has provided us an effective theory for the modes $p<k$. Therefore, it is natural to do the subtraction at zero momentum. Indeed, imposing a subtraction scale $\mu$ does not make sense if the cutoff scale $k$ is smaller than the subtraction scale $\mu$: the remaining modes $p<k$ of the effective action cannot reach the scale $\mu$. 

Doing the subtraction at zero momentum means that the effective coupling for an interaction of order $n$ is $Z_k^{n/2}S_{k}^{(n)}(0,0,0,0)$. 
These couplings are dimensionful: the fields have dimension $[k^{(d-2)/2)}]$ so that the coupling constants of interactions of order $n$ have dimension $[k^{d-n\frac{d-2}{2}}]$. This choice of scheme is often called the BPHZ subtraction scheme at zero momentum.

Finally we obtain the dimensionless coupling constants at scale $k$, $g^{(n)}$:
\begin{equation}
g^{(n)}=k^{\frac{n-2}{2}d-n}Z_k^{n/2}S_{k}^{(n)}(0,0,0,0) \; .
\end{equation}
Later, we will be interested in the effective coupling constants of quartic interactions:
\begin{equation}
g^{(4)}=k^{d-4}Z_k^2S_k^{(4)}(0,0,0,0) \; .
\label{eq:def_ren_g}
\end{equation}

The change of the values of the coupling constants with the scale is called the renormalization group flow. It is characterized by the beta functions:
\begin{equation}
\beta_{g^{(n)}}=k\frac{\partial g^{(n)}}{\partial k} \; .
\end{equation}

Usually, we are interested in computing the points $g^{\star}$ such that $\beta_g(g^{\star})=0$ because they describe a theory that does not depend on the scale. We thus call those points, fixed points. It is then interesting to compute the critical exponents of the fixed points: $\beta_g^{'}(g^{\star})=\frac{\partial \beta_{g}}{\partial g}\mid _{g^{\star}}$. Their sign determines if the fixed points are attractive or repulsive. To see this, we can write the linearized flow equation:
\begin{equation}
k\frac{\partial (g-g^{\star})}{\partial k}=\beta_g^{'}(g^{\star})(g-g^{\star})\; .
\end{equation}
Solving this equation for $g(k)$, we find:
\begin{equation}
g(k)=g^{\star}+k^{\beta_g^{'}(g^{\star})}\; .
\end{equation}
Therefore, if $\beta_g^{'}(g^{\star})$ is positive the theory will flow towards $g^{\star}$ when $k$ is sent to $0$: $g^{\star}$ is an attractive IR fixed point. On the contrary, if $\beta_g^{'}(g^{\star})$ is negative the theory will flow away from $g^{\star}$ when $k$ is sent to $0$: $g^{\star}$ is a repulsive IR fixed point. 


\paragraph{The renormalized series}

In \eqref{eq:def_ren_g}, we defined the renormalized coupling $g^{(n)}$ in terms of the effective $n$-point function $S_k^{(n)}(0,0,0,0)$. The latter is a series in the bare coupling $\lambda$ and is thus called the \textit{bare series}. To express the beta function in terms of the renormalized coupling, one thus need to invert this series. 
The BPHZ theorem states that it is indeed possible to invert the bare series to obtain a renormalized series with finite coefficients when lifting the ultraviolet cutoff. 

Moreover, this inversion is immediate using the Bogoliubov-Parasuk recursion \cite{Rivasseau:1991ub}. The renormalized series is identical to the bare one, up to exchanging the roles of the renormalized and bare constants and replacing the bare amplitudes by counterterms:
\be
\begin{split}
 g
 & = \mu^{-\epsilon} \lambda - \sum_G
 s(G)\left(- \mu^{-\epsilon}\lambda\right)^V \; {\cal \hat A}(\mathcal{G}) \,, \crcr
  \mu^{-\epsilon} \lambda
 & = g - \sum_G
 s(\mathcal{G})\left(-g\right)^V\; K_{\mathcal{G}} \,,
\end{split}
\label{eq:BPrecursion}
\ee
where the sums run over one particle irreducible $n$-point graphs $\mathcal{G}$ and
\begin{itemize}
 \item  $V$ is the number of vertices of $\mathcal{G}$ and 
 $s(\mathcal{G})$ is the symmetry factor of $\mathcal{G}$,
 \item $ {\cal \hat A}(\mathcal{G})$ is the bare amplitude and $K_{\mathcal{G}}$ is the counterterm of $\mathcal{G}$ which we define below.
\end{itemize}
The point is that $s(\mathcal{G})$ is the \emph{same} in the two formulas.
The counterterms $K_{\mathcal{G}}$ are defined recursively:
\be
\qquad K_{\mathcal{G}} = - \sum_{\mathcal{G}_1,\dots \mathcal{G}_k} 
   {\cal \hat A} ( \mathcal{G}/{\cup \mathcal{G}_i} )  \prod_{i} K_{\mathcal{G}_i} \,,
   \label{eq:rec_counter}
\ee
where the sum runs over all the families of (vertex) disjoint primitively divergent subgraphs, $\mathcal{G}_i$ of $\mathcal{G}$, including the empty family ($\mathcal{G}$ itself is not its own subgraph), and $ \mathcal{G}/\cup \mathcal{G}_i$ is the graph $\mathcal{G}$ where all the $\mathcal{G}_i$ have been contracted to vertices. This definition is recursive: the counterterms $K_{\mathcal{G}_i}$ have already been defined at this stage.

\subsection{Dimensional regularization and minimal subtraction scheme}

The textbook example of renormalization is the Wilson-fisher fixed point \cite{Wilson:1971dc,Wilson:1973jj}. It is obtained for a quartic scalar field in $d$ dimensions and describes the critical points of models in the Ising universality class. We will use this model to present the dimensional regularization scheme. 

We start with the following bare action in the UV:
\begin{equation}
S=\int d^d x\left[ \frac{1}{2}\left(\partial \phi_0\right)^2+\frac{1}{2}m_0^2\phi_0^2+ \frac{\lambda_0}{4!}\phi_0^4\right] \, .
\end{equation}

We then introduce a renormalized field via the rescaling $\phi_0=\sqrt{Z}\phi$ with $Z$ the wave function renormalization. We also introduce a renormalized mass $m$ and dimensionless renormalized quartic coupling $\lambda$ such that the renormalized action writes:

\begin{equation}
S=\int d^d x\left[ \frac{Z}{2}\left(\partial \phi\right)^2+\frac{Z_m}{2}m^2\phi^2+ \frac{m^{4-d}Z_{\lambda}\lambda}{4!}\phi^4\right] \, ,
\end{equation}
with 
\begin{equation}
m_0^2=m^2\frac{Z_m}{Z} \, , \quad 
\lambda_0=\lambda m^{4-d}\frac{Z_{\lambda}}{Z^2} \, . 
\end{equation}

With this parametrization, the renormalized constants $Z$, $Z_m$ and $Z_{\lambda}$ are dimensionless and thus depend only on the renormalized coupling $\lambda$. They will be defined through renormalization group conditions on the diverging amplitudes of one-particle irreducible diagrams at some energy scale $\mu$. 
Different schemes can be used to fix those renormalization conditions and a convenient one is minimal subtraction \cite{ZinnJustin:2002ru}. In this case, the renormalized constants are defined to cancel exactly the divergent part of the amplitudes. Setting $d=4-\epsilon$, they will then have the following form:
\begin{equation}
Z_i=1+\sum_{n\geq 1}\frac{f(\lambda)}{\epsilon^n}+ \text{ (regular terms in }\epsilon) \, ,
\end{equation} 
where $f$ is a polynomial in $\lambda$. 

\begin{figure}[htbp]
\centering
\includegraphics[scale=1]{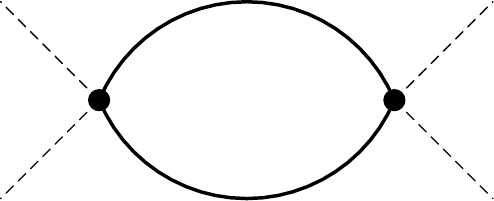}
\caption{Four-point Feynman graphs contributing to the beta function at one-loop.}
\label{fig:WF}
\end{figure}

At one-loop, there is only one four-point graph to consider, depicted in figure~\ref{fig:WF}. Following the minimal subtraction scheme, the beta function writes:
\begin{equation}
\beta(\lambda)=-\epsilon\lambda+ \alpha \lambda^2 \, ,
\end{equation}
with $\alpha$ some numerical constant. 
Besides a trivial fixed point $\lambda=0$, this beta function gives the Wilson-Fisher fixed point $\lambda=\epsilon/\alpha$. At this fixed point, the quartic operator is irrelevant and has critical exponent $-\epsilon$. 

This procedure allowed to determine the critical exponents of many physical systems in dimension three by setting $\epsilon=1$ and using Borel summation. 

\paragraph{Massless theories}
Up to now, we discussed only UV divergences as the mass played the role of an IR cutoff. In order to study renormalization of massless theories we thus need to introduce an IR regulator. This can be done by putting an IR cutoff by hand as in section \ref{subsec:WR} or a mass parameter as in chapter \ref{chap:3loops}. This can also be implemented by defining the renormalization conditions at non-zero external momenta. This choice is called the Gell-Mann and Low subtraction at non-zero momentum. For four-point functions in four-dimension, the standard subtraction point is the symmetric one \cite{Brezin:1974eb,Kleinert:2001ax}:
\begin{equation}
k_ik_j=\mu^2\left(\delta_{ij}-\frac{1}{4}\right)\; .
\end{equation}

\subsection{Non-perturbative methods}

Perturbative renormalization is a very strong tool to describe a vast array of systems. One of its first big successes was the proof by Dyson that the theory of quantum electrodynamics is renormalizable at all orders \cite{Dyson:1949bp}. Moreover, it is also possible to compute the anomalous magnetic dipole moment of the electron for example. It was first done up to order two by Schwinger \cite{schwinger1948quantum} and has been performed nowadays up to the 10\textsuperscript{th} order \cite{Aoyama:2014sxa}. These computations match the experimental value up to ten significant digits, constituting one of the most precise agreements between theory and experiment.

Nevertheless, studying QFTs non-perturbatively is also of interest. It is particularly necessary when considering strongly-coupled interactions where one cannot do a Taylor expansion in the coupling constant. We only presented so far methods curing divergences in individual Feynman graphs but divergences can still occur when resumming the perturbative series. The goal is then to show the Borel summability of the perturbative series. This is the basis of constructive quantum field theory \cite{Summers:2012vr,Rivasseau:2014zpm}. This is a very challenging domain, nevertheless, results were recently obtained with the proof of the triviality of the dimension four $\phi^4$ and Ising models \cite{Aizenman:2019yuo}.

One of the main tools of constructive QFT is the multiscale analysis. It is not far from the idea of Wilsonian renormalization and consists in slicing the free propagator in energy scales. The maximal scale will then correspond to the UV cutoff and the amplitudes will decompose as sums over the scales. It is then possible to show the BPHZ theorem \cite{bogoliubow1957multiplikation} that we introduced above. The proof of this theorem can be found in\cite{Hepp:1966eg} but we will not present it here (see also \cite{Rivasseau:2014zpm,Vignes-Tourneret:2006rav} for reviews). It however relies on the notion of high subgraphs and the use of subtraction operators which we will introduce in chapter \ref{chap:CTKT} in order to cure divergences in section \ref{sec:divergences}.

\section{Aspects of conformal field theory}
\label{sec:CFT}

\subsection{Conformal symmetry}

A conformal field theory is a particular case of a quantum field theory with conformal symmetry. It is a stronger symmetry extending the scale invariance of the theories at fixed points of the RG flow.\footnote{Many reviews were made on this subject, we refer for example to \cite{DiFrancesco:1997nk} and references therein.} Conformal symmetry includes invariance under translations, rotations, dilatations and special conformal transformations: 

\begin{equation}
x'^{\mu}=\frac{x^{\mu}+x^2b^{\mu}}{1+2b\cdot x+b^2x^2} \; . 
\end{equation}
This means that in Euclidean $\mathbb{R}^d$ with $dx^2=\delta_{\mu\nu}dx^{\mu}dx^{\nu}$, a conformal transformation will preserve the line element up to a local scale factor: $dx'^2=\Omega(x)^2dx^2$. The corresponding infinitesimal transformations are $x'^{\mu}=x^{\mu}+v^{\mu}$, $\Omega=1+\sigma$ with:
\begin{equation}
v_{\mu}(x)=a_{\mu}+R_{\mu\nu}x^{\nu}+\Lambda x_{\mu}+b_{\mu}x^2-2x_{\mu}b\cdot x\; , \quad R_{\mu\nu}=-R_{\nu\mu} \; , \quad \sigma=\Lambda-2b\cdot x \; . 
\end{equation}

The conformal group has $\binom{d+2}{2}$ generators and is locally isomorphic to $SO(d+1,1)$. 
The primary operators are then $\mathcal{O}^{\bar{\mu}}_{h,J}$ characterized by their spin $J$ and dimension $h$. In the rest of this section we will consider only operators of integer spin $J$ in the symmetric traceless representation of $SO(d)$.\footnote{It is also possible to consider other irreducible representations of $SO(d)$, see for example \cite{Costa:2016hju}.} We denote their multi-indices in the compact way $\bar{\mu}=\mu_1,\dots ,\mu_{J}$. Descendants are obtained by applying derivatives on primary operators.

\subsection{Two- and three-point functions}

The two- and three-point functions are fixed by conformal invariance. The two-point function between two primary operators is non-zero only if they have same dimension and spin:\footnote{We use the notation:
\begin{equation*}
V_{(\nu_1\dots \nu_n)}=\frac{1}{n!}\sum_{p \in \mathcal{S}_n}V_{\nu_{p(1)}\dots \nu_{p(n)}}\, ,
\end{equation*}
for $V$ any object with $n$ indices. \label{fn:sym} }
\begin{equation}
\begin{split}
\left\langle \mathcal{O}^{\bar{\mu}}_{h,J}(x_1)\mathcal{O}_{h,J,\bar{\nu}}(x_2)\right\rangle&=\frac{I^{\mu_1}_{(\nu_1}(x_{12})\dots I^{\mu_J}_{\nu_J)}(x_{12})-\textrm{traces}}{\left\vert x_{12}\right\vert ^{2h}} \; , \crcr
I^{\mu}_{\nu}(x)&=\delta^{\mu}_{\nu}-2\frac{x^{\mu}x_{\nu}}{\vert x\vert^2} \; ,
\end{split}
\end{equation}
where we use the shorthand notation $x_{ij}=x_i-x_j$.
For scalar fields of dimension $\Delta_{i}$ this simplifies to:
\begin{equation}
\left\langle \phi_{\Delta_{i}}(x_1)\phi_{\Delta_{i}}(x_2)\right\rangle= \frac{1}{\left\vert x_{12}\right\vert ^{2\Delta_{i}}} \; .
\end{equation}

The three-point functions are also strongly constrained by conformal invariance. For example, the three-point function between two fields and a primary operator is given by:
\begin{equation}
\begin{split}
\left\langle \phi_1(x_1)\phi_2(x_2)\mathcal{O}_{h,J,\bar{\mu}}(x_3)\right\rangle &= C^{\Delta_1,\Delta_2}_{h,J}\frac{Z_{\mu_1}\dots Z_{\mu_J}-\text{traces}}{\left\vert x_{12}\right\vert^{\Delta_1+\Delta_2-h+J}\left\vert x_{13}\right\vert^{h+\Delta_1-\Delta_2-J}\left\vert x_{23}\right\vert^{h+\Delta_2-\Delta_1-J}} \; , \crcr
Z_{\mu}&=\frac{\left(x_{13}\right)_{\mu}}{\left\vert x_{13}\right\vert^{2}}-\frac{\left(x_{23}\right)_{\mu}}{\left\vert x_{23}\right\vert^{2}} \; ,
\end{split}
\label{eq:3ptCFT}
\end{equation}
where the coefficients $C^{\Delta_1,\Delta_2}_{h,J}$ are numbers. 
Again for scalar fields this simplifies to:
\begin{equation}
\left\langle \phi_{1}(x_1)\phi_{2}(x_2)\phi_{3}(x_3)\right\rangle= \frac{C^{\Delta_1,\Delta_2}_{\Delta_3}}{\left\vert x_{12}\right\vert ^{\Delta_1+\Delta_2-\Delta_3}\left\vert x_{23}\right\vert ^{\Delta_2+\Delta_3-\Delta_1}\left\vert x_{31}\right\vert ^{\Delta_3+\Delta_1-\Delta_2}} \; .
\end{equation}

\subsection{Four-point function and operator product expansion}
\label{sec:ope}

The coefficients of the three-point function also appear in the operator product expansion (OPE) which is the product of two operators at nearby points. In particular, for a CFT, this product is strongly constrained by conformal invariance and the sum restricts to primary operators:

\begin{equation}
\phi_1(x_1)\phi_2(x_2)=\sum_{h,J} C_{h,J}^{\Delta_1,\Delta_2} P_{h,J}^{\Delta_1,\Delta_2,\bar{\mu}}(x_{12},\partial_{x_2})\mathcal{O}^{\bar{\mu}}_{h,J}(x_2) \; ,
\end{equation}
where $P_{h,J}^{\Delta_1,\Delta_2,\bar{\mu}}$ is a universal differential operator that can be computed explicitly. This equality should be understood in the weak sense and is valid when inserted in correlations. 

This is a very important feature of CFT as it allows to compute arbitrary correlation functions. More precisely, it means that a CFT is entirely defined by the dimensions of its primaries and the set of OPE coefficients. 

In particular, expanding the four-point function in the $s$-channel one obtains:\footnote{Equating this with the four-point function expanded in the $t$-channel gives rise to the so-called crossing equations used in the bootstrap program (see for example \cite{Poland:2018epd} and references therein).}
\begin{equation}
\left\langle \phi_{1}(x_1)\phi_{2}(x_2)\phi_{3}(x_3)\phi_4(x_4)\right\rangle=\sum_{h,J} C_{h,J}^{\Delta_1,\Delta_2}C_{h,J}^{\Delta_3,\Delta_4}G_{h,J}^{\Delta_i}(x_i) \; , 
\end{equation}
where the functions $G_{h,J}^{\Delta_i}$, the conformal blocks, are known explicitly \cite{Simmons-Duffin:2017nub,Liu:2018jhs}. 

\paragraph{Conformal Partial Wave expansion}
We are now going to present a method allowing to compute the OPE coefficients and dimensions of primaries. This is based on the conformal partial wave expansion and we follow the presentation of \cite{Gurau:2019qag} (see also \cite{Dolan:2000ut,Pappadopulo:2012jk,Costa:2011dw} and \cite{Simmons-Duffin:2016gjk} for a detailed review).

The first step is, for a primary $\mathcal{O}_{\Delta,J}$, to introduce the shadow operator $\mathcal{O}_{\tilde{h},J}$ with dimension $\tilde{h}=d-h$. The conformal partial waves (CPW) are then defined as:
\begin{equation}
\Psi^{\Delta_i}_{h,J}(x_i)=\int dx_0\Big\langle\phi_1(x_1)\phi_2(x_2)\mathcal{O}_{h,J}(x_0)\Big\rangle_{cs}\Big\langle\tilde{\mathcal{O}}_{\tilde{h},J}(x_0)\phi_3(x_3)\phi_4(x_4)\Big\rangle_{cs} \; ,
\end{equation}
where the subscript $cs$ indicates a conformal structure with OPE coefficients set to one. 

The CPW form an orthogonal set. Moreover, if we include all integer spins $J$ and dimensions $h=\tfrac{d}{2}+ir$ (principal series) for $d>1$, they form a complete set. In $d=1$, one needs to add the extra discrete dimensions $h=2n$ for $n\geq 1$. The CPW are then a basis for four-point functions of scalars. We give more detail on the CPW expansion of generic functions in appendix \ref{app:CPW} of chapter \ref{chap:Ftheorem}. 
Here we will focus on the four-point function. 

The shadow coefficient of three operators is defined by:
\begin{equation}
\int d^d y \Big\langle \tilde{\mathcal{O}}_{\tilde{h},J}(x_1)\tilde{\mathcal{O}}_{\tilde{h},J}(y)\Big\rangle_{cs}\Big\langle \mathcal{O}_{h,J}(y)\mathcal{O}_2(x_2)\mathcal{O}_3(x_3)\Big\rangle=S^{\mathcal{O}_2\mathcal{O}_3}_{\mathcal{O}}\Big\langle \tilde{\mathcal{O}}_{\tilde{h},J}(x_1)\mathcal{O}_2(x_2)\mathcal{O}_3(x_3)\Big\rangle \; ,
\end{equation}
and explicitly by:
\begin{equation}
S^{\Delta_1,\Delta_2}_{h,J}=\pi^{\tfrac{d}{2}}\frac{\Gamma(h-\tfrac{d}{2})\Gamma(h+J-1)\Gamma(\tfrac{\tilde{h}+\Delta_1-\Delta-2+J}{2})\Gamma(\tfrac{\tilde{h}-\Delta_1+\Delta-2+J}{2})}{\Gamma(h-1)\Gamma(d-h+J)\Gamma(\tfrac{h+\Delta_1-\Delta-2+J}{2})\Gamma(\tfrac{h-\Delta_1+\Delta-2+J}{2})} \; . 
\end{equation}
Using this shadow formalism, it is possible to show that the CPW can be expressed in terms of a conformal block and its shadow:
\begin{equation}
\Psi^{\Delta_i}_{h,J}(x_i)=\left(-\frac{1}{2}\right)^J S^{\Delta_3,\Delta_4}_{\tilde{h},J}G^{\Delta_i}_{h,J}(x_i)+\left(-\frac{1}{2}\right)^J S^{\Delta_3,\Delta_4}_{h,J}G^{\Delta_i}_{\tilde{h},J}(x_i) \; .
\end{equation}

Let us now consider a field theory for a scalar field $\phi$ with zero one-point function and four-point kernel in the channel $12\rightarrow 34$ given by:
\begin{equation}
K(x_1,x_2,x_3,x_4)=\int d^dx_a d^d x_b G(x_{1a})G(x_{1b})\frac{\delta \Sigma(x_{34})}{\delta G(x_{ab})} \; ,
\end{equation}
where $\Sigma$ and $G$ are respectively the self-energy and the two-point function.

The connected four-point function between $12$ and $34$ is then given by:
\begin{equation}
\left\langle \phi(x_1)\phi(x_2)\phi(x_3)\phi(x_4)\right\rangle=\int d^dx_a d^d x_b \left(\frac{1}{1-K}\right)(x_1,x_2,x_a,x_b)\left(G(x_{a3})G(x_{b4})+a\leftrightarrow b\right) \; .
\label{eq:4ptfctkernel}
\end{equation}
Note that this formula is valid for any field theory and not only conformal field theories. 
For a conformal field theory, this four-point function can also be expressed in terms of conformal partial waves, as the field is, in this case, a primary operator of dimension $\Delta_{\phi}$:
\begin{equation}
\left\langle \phi(x_1)\phi(x_2)\phi(x_3)\phi(x_4)\right\rangle=\frac{1}{\left\vert x_{12}\right\vert^{2\Delta_{\phi}}\left\vert x_{34}\right\vert^{2\Delta_{\phi}}}+\sum_{J}\int_{d/2}^{d/2+i\infty}\frac{dh}{2\pi i}\rho(h,J)\Psi^{\Delta_{\phi}}_{h,J}(x_i) \; ,
\end{equation}
where the first disconnected term comes from the contribution of the identity operator with dimension and spin $0$ in the CPW expansion and we assumed that there was no other physical operators of dimension smaller than $d/2$.

The density $\rho(h,J)$ captures the properties of all other operators and OPE coefficients. One can compute it by expanding \eqref{eq:4ptfctkernel} on CPW. The first two terms in the channel $12\rightarrow 34$ are then:

\begin{equation}
\left\langle \phi(x_1)\phi(x_3)\right\rangle \left\langle \phi(x_2)\phi(x_4)\right\rangle + 1 \leftrightarrow 2= \sum_{J}\int_{d/2}^{d/2+i\infty}\frac{dh}{2\pi i}\rho^0(h,J)\Psi^{\Delta_{\phi}}_{h,J}(x_i) \; ,
\end{equation}
with 

\begin{equation}
\rho^0(h,J)=\frac{1+(-1)^J}{n_{h,J}}t_0S^{\tilde{h}_{\phi}(h,J)}_{\tilde{h}_{\phi}}S^{\Delta_{\phi}(h,J)}_{\Delta_{\phi}} \; ,
\end{equation}
and 
\begin{equation}
\begin{split}
t_0&=\frac{1}{\text{Vol}(SO(d-1))}\frac{\Gamma(\tfrac{d-2}{2})\Gamma(J+d-2)}{2^J\Gamma(d-2)\Gamma(J+\tfrac{d-2}{2})} \; , \crcr
n_{h,J}&=\pi\frac{S^{\Delta_3,\Delta_4}_{\tilde{h},J}S^{\tilde{\Delta_3},\tilde{\Delta_4}}_{h,J}\text{Vol}(S^{d-2})}{\text{Vol}(SO(d-1))}\frac{(2J+d-2)\Gamma(J+1)\Gamma(J+d-2)}{2^{2J+d-2}\Gamma(J+\tfrac{d}{2})^2} \; .
\end{split}
\end{equation}
For the other terms in \eqref{eq:4ptfctkernel}, we use the fact that the three-point functions are eigenvalues of the kernel due to conformal invariance:
\begin{equation}
\int d^d x_3 d^d x_4 K(x_1,x_2,x_3,x_4)\left\langle \phi(x_3) \phi(x_4)\mathcal{O}_{h,J}(x_0)\right\rangle =k(h,J)\left\langle \phi(x_1)\phi(x_2)\mathcal{O}(x_0)\right\rangle \; .
\end{equation}
This implies that:
\begin{equation}
\rho(h,J)=\frac{1}{1-k(h,J)}\rho^0(h,J)\, .
\end{equation}

Putting everything together, the four-point function in a CFT with a real field of dimension $\Delta_{\phi}$ can be written as :
\begin{equation} \label{eq:4ptCFT}
\begin{split}
 \langle{\phi(x_1) \phi(x_2) \phi(x_3) \phi(x_4)} \rangle
 = & \langle{\phi(x_1) \phi(x_2) \rangle \; \langle \phi(x_3) \phi(x_4)} \rangle  \crcr
 & \qquad  +  \sum_J 
  \int_{\frac{d}{2}-\imath \infty}^{\frac{d}{2}+\imath\infty} \frac{dh}{2\pi \imath}
  \;\frac{1}{1-k(h,J)} \; \mu_{\Delta_{\phi}}^d(h,J)
     G^{\Delta_{\phi}}_{h,J}(x_i)\;,
\end{split}
\end{equation}
with $\mu_{\Delta_{\phi}}^d(h,J)$ the measure:
\begin{equation}\label{eq:measure}
\begin{split}
\mu_{\Delta_{\phi}}^d(h, J)  \, = &  \,  \left( \frac{ 1 + (-1)^J }{2} \right)
   \frac{  \Gamma(J+\frac{d}{2}) 
  }
  {  \Gamma(J+1)} 
   \crcr
&  \ \times    \frac{
 \Gamma( \frac{d}{2} - \Delta_{\phi})^2 
    \Gamma(\frac{ 2\Delta_{\phi} -d +h+J}{2}) 
    \Gamma(\frac{2\Delta_{\phi}-h+J}{2})
    \Gamma(h-1)\Gamma(d-h+J)\Gamma(\frac{h +J}{2})^2
 }{ \Gamma( \Delta_{\phi})^2 
  \Gamma(\frac{2d-2\Delta_{\phi}-h+J}{2})\Gamma(\frac{d-2\Delta_{\phi} +h+J}{2}) 
  \Gamma(h-\frac{d}{2})\Gamma(h+J-1)\Gamma(\frac{d- h  +J}{2})^2
 }  \; .
\end{split}
\end{equation}
As we explained above, the case $d=1$ is special \cite{Maldacena:2016hyu}: in that case one gets an extra contour for $h$ around the even integers.

The OPE expansion is obtained by deforming the integration contour to the right and picking up the poles in the integrand. All the poles coming from the measure and the conformal block are spurious \cite{Simmons-Duffin:2017nub}. Only the poles from $\frac{1}{1-k(h,J)}$ are physical. We denote $h_{m,J}$ the solutions of $k(h,J)=1$: these are the dimensions of the physical primary operators contributing in the OPE of $\phi\phi$. We then obtain:
\begin{equation}
  \langle{\phi(x_1) \phi(x_2) \phi(x_3) \phi(x_4)} \rangle
 =  \langle{\phi(x_1) \phi(x_2) \rangle \; \langle \phi(x_3) \phi(x_4)} \rangle 
+  \sum_{m,J} c_{m,J}^2  \; G^{\Delta_{\phi}}_{h_m,J}(x_i) \;,
\end{equation}
and the squares of the OPE coefficients are the residues at the poles:
\begin{equation}\label{eq:OPE coef}
c_{m;J}^2=-\mu_{\Delta_{\phi}}^d(h_{m,J} ,J) \text{Res}\left[\frac{1}{1-k(h,J)}\right]_{h=h_{m,J} } 
=\frac{\mu_{\Delta_{\phi}}^d(h_{m,J},J)}{ k'( h_{m,J},J) }  \;,
\end{equation}
where the prime denotes a derivative with respect to $h$. As we deform the contour to the right, we only pick up the poles with ${\rm Re}(h_{m,J}) \ge d/2$. If some poles lie on the original integration contour, that is ${\rm Re}(h_{m,J}) = d/2$, or on the left of the contour, extra care must be taken. We give more details on this situation in appendix~\ref{app:CPW}.

\subsection{Unitarity}

A Lorentzian CFT is unitary if all states have positive norm. For an Euclidean CFT, unitarity is ensured by the reflection positivity of correlators. This condition imposes that the central charge $c$ must be positive. Moreover, the conformal dimensions $h$ of primary operators of spin $J$ must obey:
\begin{equation}
\begin{split}
h&\geq \frac{d-2}{2} \; , \; \text{if } J=0 \; ,\crcr
h& \geq J+d-2 \; , \; \text{if } J>0 \; .
\end{split}
\end{equation}
Unitary CFTs are of particular interest but statistical systems are not necessarily described by unitary CFTs. Non-unitary CFTs can for example occur in non-integer dimensions, for interactions with imaginary couplings or in the case of logarithmic CFTs. In the following chapters, we will also encounter non-unitary CFTs for tensor field theories where an imaginary spectrum of conformal dimensions is obtained.

\section{Random tensors}
\label{sec:RTM}

The idea to consider models built on tensors of rank $r\geq  3$ started in the 1990s as a generalization of matrix models \cite{Ambjorn:1990ge,Sasakura:1990fs,Boulatov:1992vp}. However, considering models with indistinguishable indices led to singular geometries and no proper large-$N$ limit \cite{Gurau:2010nd}. This was solved twenty years later by introducing colored tensor models where indices are distinguishable. Such models were first formulated in the language of Group Field Theory \cite{Gurau:2009tw} before being written as proper tensor models. We start this section by giving combinatorial tools on colored graphs in subsection \ref{sec:colored_graphs}. We then present the two main types of tensor models: colored and uncolored in subsections \ref{sec:colored_model} and \ref{sec:uncolored} respectively. Finally, in subsection \ref{sec:other_melonic}, we present a few other models exhibiting melonic limits.

\subsection{Colored graphs and combinatorial maps}
\label{sec:colored_graphs}

%
\paragraph{Edge colored graphs}

\begin{definition}
A bipartite, closed, edge $(D+1)$-colored graph is a graph $\mathcal{G}=(V(\mathcal{G}),E(\mathcal{G}))$ with vertex set $V(\mathcal{G})$ and edge set $E(\mathcal{G})$ such that:
\begin{itemize}
\item $V(\mathcal{G})$ is bipartite, i.e there are only edges between black and white vertices. (The numbers of black and white vertices are equal.)
\item There are $D+1$ different types of edges with color $0,1,\dots ,D$.
\item All vertices are $(D+1)$-valent and have exactly one incident edge of each color.
\end{itemize}
\end{definition}

\begin{figure}[htbp]
\centering
\includegraphics[scale=0.75]{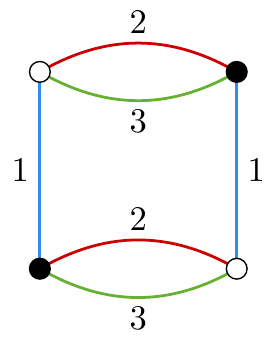}
\caption{Example of a 3-colored graph with four vertices.}
\label{fig:graph}
\end{figure}

An example of a $(D+1)$-colored graph for $D=2$ is given in figure~\ref{fig:graph}. There are $\left(\frac{V(\mathcal{G})}{2}\right)^{D+1}$ bipartite edge $(D+1)$-colored graphs with $V(\mathcal{G})$ labeled vertices.

\begin{definition}
The faces of a $(D+1)$-colored graph $\mathcal{G}$ are its maximal connected subgraphs made of edges of two fixed (distinct) colors.
\end{definition}

\begin{figure}[htbp]
\centering
\includegraphics[scale=0.75]{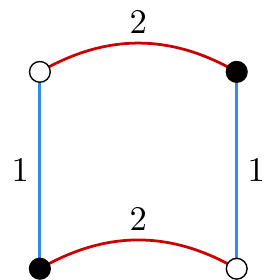}
\caption{Face of color (1,2) of the graph represented in figure~\ref{fig:graph}.}
\label{fig:face}
\end{figure}

As shown in figure \ref{fig:face}, this definition means that a face is a cycle made of edges of two alternating colors.

\paragraph{Combinatorial maps and $(D+1)$-colored graphs}

We are going to define the degree, an important characteristic of colored graphs that will allow us to classify them. In order to do so, we first need to introduce combinatorial maps.

\begin{definition}
A combinatorial map is:
\begin{itemize}
\item A finite set $\mathcal{S}$ of half-edges of even cardinality
\item A permutation $\sigma$ on $\mathcal{S}$
\item An involution $\alpha$ on $\mathcal{S}$ with no fixed points (i.e a pairing of half-edges)
\end{itemize}
\end{definition}

To build the graph corresponding to $\mathcal{S}$, a vertex is associated to each cycle of $\sigma$ and an edge to each cycle of $\alpha$. The faces of the graph are then given by the cycles of $\sigma \circ \alpha$.
More intuitively, it can be shown that this definition is equivalent to saying that a combinatorial map yields a graph embedded on a two-dimensional surface with an orientation \cite{ZVONKIN1997281}. Indeed, there is a local embedding of each vertex defined by the cycles of $\sigma$. They are then glued together, according to the cycles of $\alpha$, to form an orientable surface canonically associated to the map. Let us now define the Euler characteristic of a combinatorial map.

\begin{definition}
The Euler characteristic $\chi$ of a connected combinatorial map (equivalently its genus g) with $V$ vertices, $E$ edges and $F$ faces is:
\begin{equation}
\chi=V-E+F=2-2g \; .
\label{eq:euler_charac}
\end{equation}
\end{definition}

The genus of a map is a topological invariant: it depends only on the topology and not on the geometry of the surface. It is a positive number, and it is an integer if the embedding surface is orientable (half-integer if it is non-orientable). It corresponds to the genus of the surface on which the combinatorial map embeds a graph. 

Now, we want to associate a combinatorial map to a $(D+1)$-colored graph. A $(D+1)$-colored graph $\mathcal{G}$ yields a well-defined set of half-edges but nothing fixes the order in which the colors should be listed in the cycle of $\sigma$. Indeed, each vertex has one half-edge for each color but the order of the colors in the cycle is not fixed. The involution $\alpha$ is however completely fixed by the pairing of the half-edges. The only constraint for the cycles of $\sigma$ is that there are no fixed points. Therefore, there are $D!$ choices for the ordering of the cycles of $\sigma$: there are $D!$ permutations of $D+1$ elements with no fixed points. This leads to $D!$ combinatorial maps canonically associated to $\mathcal{G}$. We call them the jackets of $\mathcal{G}$ and denote them $\mathcal{J}$.

We now define the degree of a $(D+1)$-colored graph as follows.
\begin{definition}
The degree of a connected $(D+1)$-colored graph $\mathcal{G}$ is the half-sum of the genera of its jackets (which are integers: as $\mathcal{G}$ is bipartite its jackets are orientable, we can for example orient edges from white to black vertices):
\begin{equation}
\omega(\mathcal{G})=\frac{1}{2}\sum_{\mathcal{J}}^{}g(\mathcal{J}) \; .
\end{equation}
\end{definition}
The degree is a positive integer: this is a very important characteristic for its future use in tensor models. However, it is not a topological invariant.
We can give a more practical formula to compute the degree of a $(D+1)$-colored graph \cite{RTM}, using the relation between genus and Euler characteristic:
\begin{equation}
\omega(\mathcal{G})=\frac{(D-1)!}{2}\left(D+\frac{D(D-1)}{4}V(\mathcal{G})-F(\mathcal{G})\right) \; ,
\label{eq:gurau_degree}
\end{equation}
with $V(\mathcal{G})$ the number of vertices and $F(\mathcal{G})$ the number of faces of the graph $\mathcal{G}$. 
For example, for the graph of figure \ref{fig:graph}, $D=2$, $k=2$ and $F=4$. Therefore, equation~\eqref{eq:gurau_degree} gives $\omega=0$, which is the minimal value.

\paragraph{Melonic graphs}

\begin{figure}[H]
\centering
\subfloat[Prime melonic graph]{\includegraphics[scale=0.75]{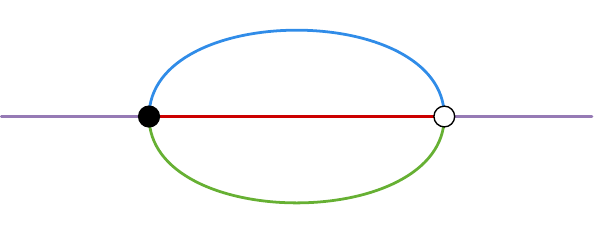}}
\hspace{1cm}
\subfloat[Melonic graph]{\includegraphics[scale=0.75]{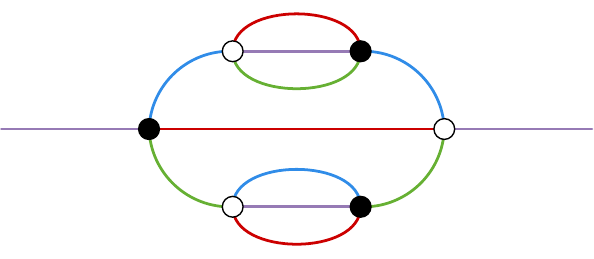}}
\caption{Prime melonic graph and example of melonic graph for $D=3$.}
\label{fig:melon_backgnd}
\end{figure}

\begin{definition}
A prime melonic graph is an open $(D+1)$-colored graph made of $D$ parallel edges, two vertices and two half-edges.
Melonic graphs are constructed recursively by iteratively inserting prime melonic graphs on the edges of a melonic graph (figure~\ref{fig:melon_backgnd}).
\end{definition}
The degree of a melonic graph will be the degree of the closed graph obtained by joining the two half-edges of the melonic graph.
We now give a fundamental theorem which proof can be found in \cite{Bonzom:2011zz,RTM}.

\begin{theorem}
A rooted closed colored graph is melonic if and only if it has degree 0.
\end{theorem}
This theorem will be very important in the next sections because it shows that melonic graphs will dominate the $1/N$ expansion of tensor models. It can also be extended for non-bipartite graphs.

%
%
%
%
%
%
%

\subsection{Colored tensor models}
\label{sec:colored_model}

A covariant tensor of rank $D$ is a multi-linear form on a tensor product of $D$ distinct Hilbert spaces $\mathcal{H}_c, ~~c=1,...D$:

\begin{equation}
\Psi: \mathcal{H}_1 \otimes \dots \otimes \mathcal{H}_D \rightarrow \mathbb{C} \; .
\end{equation}
We take the dimensions of all the Hilbert spaces  $\mathcal{H}_c$ equal to $N$.
A colored tensor model consists of $D+1$ tensors of rank $D$: $\Psi^c_{\textbf{a}^c}$ with $\textbf{a}^c=\lbrace a^{c}_0, \dots,a^{c}_{c-1},a^{c}_{c+1},\dots,a^{c}_D\rbrace$. Thus, each index $a^{c}_{i}$ takes value from $1$ to $N$.
We are going to study the independent identically distributed  (i.i.d) colored tensor model in rank $D$ \cite{Gurau:2011xp}.

\begin{definition}
A $(D+1)$-colored model is a probability measure $d\nu$: \\
\begin{equation*}
d\nu=\prod_{c}d\mu_{C^{c}}(\Psi^c,\overline{\Psi}^c)e^{-S},~~~~ S=\sum_{c=0}^D\sum_{\textbf{a}^c,\overline{\textbf{a}}^c}\Psi^c_{\textbf{a}^c}C^{c}_{\textbf{a}^c,\overline{\textbf{a}}^c}\overline{\Psi}^c_{\overline{\textbf{a}}^c}+\lambda\sum_{\textbf{a}^c}K_{\textbf{a}^0\dots \textbf{a}^D}\prod_{c=0}^D\Psi^c_{\textbf{a}^c}+\overline{\lambda}\sum_{\overline{\textbf{a}}^c}\overline{K}_{\overline{\textbf{a}}^0\dots \overline{\textbf{a}}^D}\prod_{c=0}^D\overline{\Psi}^c_{\overline{\textbf{a}}^c} \, ,
\end{equation*}
where $\Psi^c$ are complex random fields, $d\mu_{C^{c}}$ is the normalized Gaussian probability measure of covariance $C^{c}$ and $K_{\textbf{a}^0\dots \textbf{a}^D}$ is a vertex kernel.
\end{definition}
It is invariant under $U(N)^{\otimes (D+1)}$: each tensor transforms trivially under one $U(N)$ (the one associated to its color) and non-trivially under the $D$ other $U(N)$.
We make the following choice for the covariance and the vertex kernel:
\begin{equation}
C^c_{\textbf{a}^c,\overline{\textbf{a}}^c}=\prod_{c_1\neq c}\delta_{a^{c}_{c_1},\overline{a}^{c}_{c_1}},~~~~ K_{\textbf{a}^0\dots \textbf{a}^D}=\frac{1}{N^{D(D-1)/4}}\prod_{c_1<c_2}\delta_{a^{c_1}_{c_2},a^{c_2}_{c_1}} \; .
\label{eq:kernel_colored}
\end{equation}
The scaling in $N$ of the vertex kernel has been chosen so to have later a sensible $1/N$ expansion.

\begin{figure}[htbp]
\centering
\includegraphics[scale=0.75]{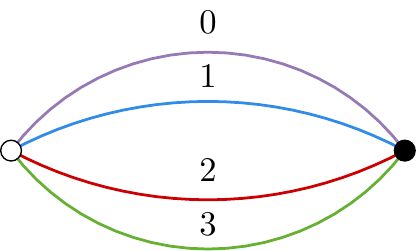}
\caption{Example of a simple graph with only two vertices for $D=3$.}
\label{fig:exemple2}
\end{figure}

Therefore, in the case $D=3$, we have a probability measure and an action of the form:\footnote{From now on we use Einstein's summation convention.}
\begin{align}
d\nu&=\frac{1}{Z_0}\prod_{c,\textbf{a}^c}\frac{d\Psi^c_{\textbf{a}^c}d\overline{\Psi}^c_{\overline{\textbf{a}}^c}}{2\pi}e^{-S} \, , \crcr
S&=S_0+S_I \crcr
&=\Psi^i_{abc}\overline{\Psi}^i_{abc}+\left(\frac{\lambda}{N^{3/2}}\Psi^0_{abc}\Psi^1_{ade}\Psi^2_{bdf}\Psi^3_{cef}+\frac{\overline{\lambda}}{N^{3/2}}\overline{\Psi}^0_{abc}\overline{\Psi}^1_{ade}\overline{\Psi}^2_{bdf}\overline{\Psi}^3_{cef}\right)\; .
\end{align}
The observables are then the $2n$-point correlation functions:
\begin{equation}
\langle\Psi^{c_1}_{\textbf{a}^{c_1}}\dots\Psi^{c_n}_{\textbf{a}^{c_n}}\overline{\Psi}^{c_1}_{\overline{\textbf{a}}^{c_1}}\dots\overline{\Psi}^{c_n}_{\overline{\textbf{a}}^{c_n}}\rangle=\int d\nu \Psi^{c_1}_{\textbf{a}^{c_1}}\dots\Psi^{c_n}_{\textbf{a}^{c_n}}\overline{\Psi}^{c_1}_{\overline{\textbf{a}}^{c_1}}\dots\overline{\Psi}^{c_n}_{\overline{\textbf{a}}^{c_n}}e^{-S} \; .
\end{equation}
The correlation functions are evaluated through Feynman graphs: each white vertex represents an interaction with four fields $\Psi$ while each black vertex represents an interaction term with four fields $\bar{\Psi}$. Each edge of color $c$ represents a propagator $C^c_{\textbf{a}^c,\overline{\textbf{a}}^c}$ identifying the indices of the fields $\Psi^c$ and $\bar{\Psi}^c$ of its end vertices. This means that the Feynman graphs of the model are bipartite $(D+1)$-colored graphs.
Moreover, each face of $\mathcal{G}$ will give a free sum and thus yield a factor $N$. This can be easily seen for example considering the face of color $(0,1)$ of the graph in figure~\ref{fig:exemple2}. The amplitude corresponding to this face is:
\begin{equation}
\sum_{a^0_{1},a^1_{0},\bar{a}^0_{1},\bar{a}^1_{0}} \delta_{a^0_{1}a^1_{0}} \delta_{a^0_{1}\bar{a}^0_{1}}\delta_{a^1_{0}\bar{a}^1_{0}} \delta_{\bar{a}^0_{1}\bar{a}^1_{0}}=\sum_{a^0_{1}} \delta_{a^0_{1}a^0_{1}}=N \,. 
\end{equation} 

The amplitude of a Feynman graph $\mathcal{G}$ can then be written as:

\begin{equation}
\mathcal{A}(\mathcal{G})=(\lambda\overline{\lambda})^pN^{F-pD(D-1)/2}=(\lambda\overline{\lambda})^pN^{D-\frac{2}{(D-1)!}\omega(\mathcal{G})}\; ,
\label{amp}
\end{equation}
using equation~\eqref{eq:gurau_degree} with $p$ the number of white vertices.

This shows that in a large-$N$ expansion, the dominant graphs will be the graphs of degree $0$, the melonic graphs.

For example, let us compute the amplitude of the graph $\mathcal{G}$ of figure \ref{fig:exemple2}. This graph has $6$ faces and $2$ vertices and we are in the case $D=3$. Thus: $\mathcal{A}(\mathcal{G})=(\lambda\overline{\lambda})N^3$.

\subsection{Uncolored tensor models}
\label{sec:uncolored}

The existence of a large-$N$ limit has also been proven for models with only one tensor field, called uncolored tensor models (see \cite{Rivasseau:2013uca, Rivasseau:2014ima,  Eichhorn:2018phj} for reviews). They have been extensively studied in the context of random geometry and quantum field theory but have also triggered rigorous developments in the broader context of tensorial group field theory \cite{BenGeloun:2011rc, Samary:2012bw, Carrozza:2012uv, Carrozza:2013wda,Krajewski:2016svb,Rivasseau:2017xbk}.

In these models, each tensor index transforms independently under a representation of a group. For example, a tensor model invariant under $O(N)^D$ is obtained by considering a rank-$D$ tensor $\Phi$ transforming with a distinct group $O(N)$ per index:

\begin{equation}
\Phi_{a_1\dots a_D}\rightarrow \Phi'_{b_1\dots b_D} =O^{(1)}_{b_1a_1}\dots O^{(D)}_{b_Da_D}\Phi_{a_1\dots a_D} \; .
\end{equation}

Observables of the theory are then trace invariants or bubbles which are built by contracting in all possible ways pairs of indices in a product of $k$ tensors. Contrary to the colored models we saw above, contractions are allowed only between indices in the same position. We can thus define a generic trace invariant between $k$ tensors by giving $D$ permutations of $k$ elements:
\begin{equation}
\mathcal{B}(\Phi)=\sum_{a}\left(\prod_{j=1}^k\Phi_{a_1^j \dots a_D^j}\right)\left(\prod_{c=1}^D\prod_{j=1}^k\delta_{a_c^ja_c^{\tau^{(c)}(j)}}\right)\; ,
\end{equation}
with $\tau^{(1)},\dots ,\tau^{(D)}$ permutations of $k$ elements. 
The trace invariants are a complete set: that means that any $O(N)^{D}$ invariant can be written as a sum over trace invariants.

\begin{figure}[ht]
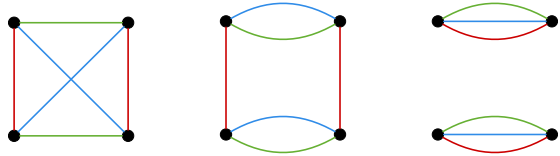

\captionsetup[subfigure]{labelformat=empty}
\begin{center}
\subfloat[]{\includegraphics[scale=0.75]{tetra.pdf}}
\hspace{1cm}
\subfloat[]{\includegraphics[scale=0.75]{pillow.pdf}}
\hspace{1cm}
\subfloat[]{\includegraphics[scale=0.75]{double_trace.pdf}}
 \caption{Graphical representation of the quartic $O(N)^3$ invariants. From left to right: the tetrahedron, the pillow, and the double-trace (there are three pillow contractions, distinguished by the color of the vertical edge).}
\label{fig:trace_invariants}
 \end{center}
\end{figure}

A trace invariant, specified by $D$ permutations $\tau^{(c)}$ of $k$ elements, is canonically associated to a $D$-colored graph by the following procedure:
\begin{itemize}
\item we represent $\Phi_{a_1^j \dots a_D^j}$ by a vertex labeled $v_j$.
\item we connect the vertex $v_j$ with the vertex $v_{\tau^{(c)}(j)}$ by an edge of color $c$.
\end{itemize}
Conversely, a trace invariant can also be defined from a $D$-colored graph $\mathcal{G}$ by associating a tensor to each vertex and an identification of the index number $c$ to each edge of color $c$.
For example, the trace invariant associated to the middle graph of figure~\ref{fig:trace_invariants} is:
\begin{equation}
\sum_{a_1,\dots, a_6}\Phi_{a_1a_2a_3}\Phi_{a_4a_2a_3}\Phi_{a_4a_5a_6}\Phi_{a_1a_5a_6} \; .
\end{equation}

We can now give the action of an uncolored tensor model with one interaction bubble:

\begin{equation}
S=\Phi_{a_1\dots a_D}\Phi_{a_1\dots a_D}+\lambda_{\mathcal{B}}N^{-s(\mathcal{B})}\mathcal{B}(\Phi) \; , 
\end{equation}
with
\begin{equation}
s(\mathcal{B})=\frac{D(D-1)}{4}+\frac{F(\mathcal{B})-\tfrac{D(D-1)}{2}}{D-1} \; ,
\end{equation}
and $F(\mathcal{B})$ the number of faces of the bubble $\mathcal{B}$. We defined the scaling $s(\mathcal{B})$ as a generalization of the optimal scaling of \cite{Carrozza:2015adg}.

For example, in the case $D=3$, there are three different quartic trace invariants, represented in figure~\ref{fig:trace_invariants}. They are called respectively tetrahedron, pillow and double-trace and will play an important role in the rest of this thesis.

The partition function can then be expressed in terms of Feynman graphs:
\begin{equation}
Z=\sum_{\mathcal{G}}(-\lambda_{\mathcal{B}})^{n(\mathcal{B})}\mathcal{A}(\mathcal{G})\; ,
\end{equation}
with $n(\mathcal{B})$ the number of bubbles $\mathcal{B}$ in $\mathcal{G}$ and $\mathcal{A}(\mathcal{G})$ the amplitude of the graph $\mathcal{G}$.

\begin{figure}[htbp]
\centering
\subfloat[Contribution with $F=5$]{\includegraphics[scale=0.75]{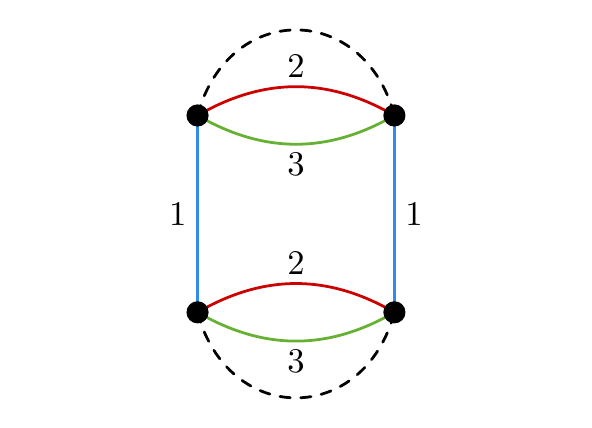}}
\hspace{1cm}
\subfloat[Contribution with $F=4$]{\includegraphics[scale=0.75]{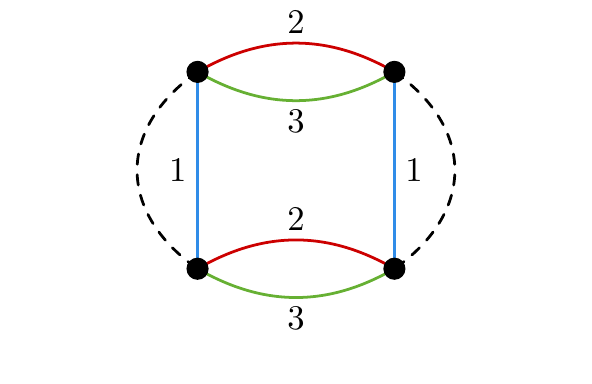}}
\caption{Graphs corresponding to the first order term in the computation of the partition function in the case $D=3$.}
\label{fig:feyn}
\end{figure}

As for colored models, $Z$ will generate $(D+1)$-colored graphs but their construction is different. For each bubble $\mathcal{B}$, we draw the corresponding $D$-colored graph. Then, the vertices of the bubbles are connected by propagators represented by dashed (or black) lines of color $0$. These propagators represent the different Wick pairings. For example, for the pillow bubble of figure~\ref{fig:trace_invariants}, the first order terms are represented in figure~\ref{fig:feyn}.

Therefore, to compute the amplitude of a Feynman graph, each vertex corresponds to a sum over the indices of the tensors associated and to a factor $N^{s(\mathcal{B})}$. Each edge of color $c$ yields a contraction of the index number $c$ of the corresponding tensors while each dashed edge (color $0$) yields a contraction of all the indices of the tensors it connects.

It is then easy to see that each face of color $(0,c)$ will yield a factor $N$ in the amplitude. Indeed, if we consider a face composed of $2n$ vertices, the amplitude corresponding to this face will be, by denoting $a_k$ the index number $c$ of the tensor $k$ in the face of color $(0,c)$:
\begin{equation}
\sum_{a_1,\dots,a_{2n}}\delta_{a_1a_2}\dots\delta_{a_{2n-1}a_{2n}}\delta_{a_{2n}a_1}=\sum_{a_1}\delta_{a_1a_1}=N \; .
\end{equation}

The amplitude of a graph $\mathcal{G}$ is then:
\begin{equation}
\mathcal{A}(\mathcal{G})=N^{\sum_{c=1}^D F_{0c}+ \sum_{\mathcal{B}\in \mathcal{G}}s(\mathcal{B})} \; ,
\end{equation}
where $F_{0c}$ is the number of faces of color $(0,c)$.

These models have a large-$N$ expansion and we will present here the proof in the special case $D=3$ with only tetrahedral bubbles which is referred to as the Carrozza-Tanasa model \cite{Carrozza:2015adg}. 
We make this choice because the subclass of uncolored tensor models of interest for large-$N$ quantum field theory applications are the ones that generate \emph{bilocal} melonic radiative corrections \cite{Dartois:2013he, Carrozza:2015adg, Ferrari:2017ryl, Ferrari:2017jgw, Benedetti:2020iyz,Prakash:2019zia}. Rank-$3$ tensors transforming under the trifundamental representation of $\mathrm{O}(N)$ is then a simple and popular choice. 

The scaling $s(\mathcal{B})$  for a tetrahedron bubble is $3/2$ and we have for any Feynman graph $\mathcal{G}$:

\begin{equation}
\mathcal{A}(\mathcal{G})=N^{\sum_{c=1}^3 F_{0c}-\tfrac{3}{2}n_t(\mathcal{G})} \; ,
\end{equation}
where we denoted $n_t({\mathcal{G}})$ the number of tetrahedral bubbles of $\mathcal{G}$.

Starting from $\cG$ one can build three \emph{jackets} \cite{Gurau:2010ba,Carrozza:2015adg} ${\cal J}^c$, that is ribbon graphs\footnote{The ribbon graphs are made evident in the stranded representation, where one replaces each black line and vertex by three parallel red, green, and blue lines: a jacket ${\cal J}^c$ is then obtained by simply deleting color $c$.} obtained by ignoring the faces of color $(0,c)$.
Each jacket has a non-orientable genus $k({\cal J}^i) \ge 0$ and the number of faces is:\footnote{We used here the relation between the genus and Euler characteristic of a non-orientable surface $\chi=2-g$.}
\begin{equation}
F(\cJ^c) =  n_t(\cG) + 2 - k({\cal J}^c) \; ,
\end{equation}
where we have used the fact that the jackets are connected.
As every face belongs to two jackets, the total number of faces of color $(0,c)$ of $\cG$ is:
\[
 F(\cG) = \frac{3}{2} n_t(\cG) + 3 - \frac{1}{2} \sum_{c} k({\cal J}^c) \; .
\]
Denoting $ \omega(\cG) = \frac{1}{2} \sum_{c} k({\cal J}^c) \ge 0$ the degree of the graph $\cG$, the scaling with $N$ of a connected vacuum graph is:
\[
 N^{3 - \omega(\cG)} \; .
\]

As we explained in section \ref{sec:colored_graphs}, a $(D+1)$-colored graph $\cG$ has degree zero if and only if it is melonic. 
We finally conclude that an uncolored tensor model with tetrahedron interaction has a well-defined large-$N$ limit dominated by melon graphs.

\subsection{Other types of melonic limits}
\label{sec:other_melonic}

\subsubsection{Irreducible tensor models}

When tensor models were first introduced, in zero dimension and in the context of random geometry and quantum gravity \cite{Ambjorn:1990ge,Sasakura:1990fs}, a large-$N$ expansion was initially lacking. A proper generalization of the genus expansion of matrix models was only discovered later for the colored tensor models presented in section \ref{sec:colored_model}. This then led to different realizations of the melonic limit such as uncolored tensor models presented in section \ref{sec:uncolored}. 

More recently, and in rank $3$, it was understood how to generalize the bilocal melonic limit to ordinary tensor representations of $O(N)$ and $Sp(N)$ \cite{Klebanov:2017nlk, Gurau:2017qya, Benedetti:2017qxl, Carrozza:2018ewt, Carrozza:2018psc}, thereby going beyond colored and uncolored models. It might at first appear that completely symmetric rank-$3$ tensor models (such as the ones initially introduced in the nineties \cite{Ambjorn:1990ge,Sasakura:1990fs}) cannot support a bilocal melonic limit. Indeed, they instead support a vector-like (and ultralocal) large-$N$ limit \cite{Benedetti:2017qxl}. However, removing the vector modes contained in the traces of the tensor is sufficient to reach a melonic regime, as was initially proposed in \cite{Klebanov:2017nlk} and proven in \cite{Benedetti:2017qxl}. In a similar way, one can conjecture that any \emph{irreducible} tensor representation (of $O(N)$ or $Sp(N)$) can support a melonic large-$N$ limit, as was proven rigorously in rank $3$ \cite{Benedetti:2017qxl, Carrozza:2018ewt, Carrozza:2018psc}. In chapter \ref{chap:rank5}, we will present a generalization in rank $5$ of this proof. Steps have also been taken to extend those results to Hermitian multi-matrix models \cite{Carrozza:2020eaz}, in the spirit of \cite{Ferrari:2017ryl}.

\subsubsection{SYK model}

The SYK model consists in $N$ Majorana fermions interacting randomly. It was proposed by Kitaev in a series of talks \cite{kitaev} and is a variant of a model introduced by Sachdev and Ye~\cite{Sachdev:1992fk}. We give here a quick introduction to this model (see \cite{Rosenhaus:2018dtp} for a more detailed review).

The SYK model was originally defined with the following quartic Hamiltonian:
\begin{equation}
H_{SYK}=j_{abcd}\psi_a\psi_b\psi_c\psi_d \; ,
\end{equation}
where the couplings $j_{abcd}$ are random variables following a Gaussian distribution with mean zero and width $J/N^{3/2}$.

We will consider here a generalization with an interaction of order $q$ even. The action $S$ at temperature $1/\beta$ is then:
\begin{equation}
S=\int_{-\beta/2}^{\beta/2} d\tau \left(\frac{1}{2}\sum_{i=1}^N \psi_i\frac{d}{d\tau}\psi_i+\frac{i^{\frac{q}{2}}}{q!}\sum_{i_1,\dots ,i_q=1}^Nj_{i_1\dots i_q}\psi_{i_1}\dots \psi_{i_q}\right) \; ,
\end{equation}
where the coupling $j_{i_1\dots i_q}$ is totally antisymmetric and is a random variable following a Gaussian distribution for each $i_1,\dots ,i_q$.

We take the normalization $\langle j_{i_1\dots i_q}j_{i_1\dots i_q}\rangle=(q-1)!\frac{J^2}{N^{q-1}}$ where the scaling in $N$ has been chosen in order to have a non-trivial large-$N$ limit. The other factors are just for convenience in the computations. 
In the rest of this section, we take the large-$N$ limit and then the zero temperature limit.

We now want to compute the connected two-point correlation function $G(\tau)=\langle\psi_a(\tau)\psi_a(0)\rangle$. We will follow the method developed in \cite{Maldacena:2016hyu}.
For a free Majorana fermion this is quite simple. Indeed, in this case, the connected two-point function is the fermionic propagator $C_{\psi}(\tau,\tau')$ which is translation invariant and antisymmetric:
\begin{equation}
C_{\psi}^{-1}(\tau, \tau ')=\delta(\tau - \tau ')\partial_{\tau '} ,~~~~ C_{\psi}(\tau, \tau ')=\frac{1}{2}\sgn(\tau - \tau ') \; .
\end{equation}
Its Fourier transform is:
\begin{equation}
\widehat{C}_{\psi}(\omega)=\int d\tau e^{i\omega\tau}C_{\psi}(\tau)=\frac{1}{-i\omega}\; .
\end{equation}

\begin{figure}[htbp]
\centering
\includegraphics[scale=1]{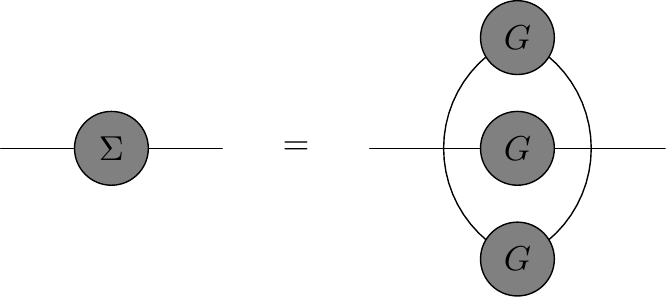}
\caption{Graphical representation of the one-particle irreducible two-point function $\Sigma$ when melonic graphs dominate for $q=4$. $G$ is the full two-point function.}
\label{fig:sigma}
\end{figure}

The full two-point function can then be computed at leading order in $1/N$ as it was shown that melon graphs dominate the large-$N$ expansion \cite{Maldacena:2016hyu}. The self-energy at large $N$ is (see figure \ref{fig:sigma}):
\begin{equation}
\Sigma(\tau)=J^2G^{q-1}(\tau)\; .
\label{eq:sig}
\end{equation}
We can show this by noticing that the most general one-particle irreducible ($1$PI) two-point melonic graph is a prime melon with edges decorated by two-point melonic graphs. Hence, the sum over all $1$PI graphs is obtained by summing over all melonic graphs. Yet, the sum over all melonic graphs gives the two-point function. Therefore, we obtain the equality of figure~\ref{fig:sigma}.

The Schwinger-Dyson equation then writes:
\begin{equation}
\begin{split}
G^{-1}(\omega)&=\widehat{C}_{\psi}(\omega)^{-1}-\Sigma(\omega)\\
&=-i\omega-\Sigma(\omega) \; .
\end{split}
\label{eq:genericsde}
\end{equation}
This, along with \eqref{eq:sig}, gives:
\begin{equation}
\begin{split}
G(\tau)=C(\tau)+(J^2C\circ G^{q-1}\circ G)(\tau) \; .
\end{split}
\end{equation}


We now consider the infrared limit $\omega \rightarrow 0$. In this limit, the term $-i\omega$ from the free propagator can be neglected. 
Taking the inverse Fourier transform of \eqref{eq:genericsde} and using the translation invariance of $G$, we obtain:
\begin{equation}
-\delta(t_1-t_2)=J^2\int dt G(t_1-t)G^{q-1}(t-t_2) \; .
\label{eq:cf}
\end{equation}
Written in this form, this equation is invariant under reparametrization~\cite{Maldacena:2016hyu}:
\begin{equation}
G(t,t')\rightarrow [f'(t)f'(t')]^{\Delta}G(f(t),f(t')) \; ,
\label{eq:rescale}
\end{equation}
with $f$ strictly increasing. 

By injecting this equation in \eqref{eq:cf}, we obtain $\Delta=\frac{1}{q}$. 
Then, if we choose $f(t)=at$ with $a$ a constant, \eqref{eq:rescale} gives:
\begin{equation}
G(t,t')=a^{2\Delta}G(at,at') \; .
\label{eq:inv}
\end{equation}

Using \eqref{eq:inv} and knowing that $G$ must be antisymmetric and invariant under translation, we obtain:
\begin{equation}
G(t)=b\frac{\sgn(t)}{|t|^{2\Delta}}=b\frac{\sgn(t)}{|t|^{2/q}} \; ,
\label{eq:ansatzSYK}
\end{equation}
with $b$ a constant defined by:


\begin{equation}
J^2b^q\pi=(\frac{1}{2}-\frac{1}{q})\tan(\frac{\pi}{q}) \; .
\label{eq:final}
\end{equation}

We thus obtain in the infrared a conformally invariant theory. This is of particular interest in the context of holographic duality. Indeed, further studies of this model \cite{Maldacena:2016hyu}, showed that it saturates the chaos bound at large $N$ and thus describes at low temperature the near-horizon geometry of near-extremal black holes. Higher point correlation functions were also computed as well as conformal data \cite{Polchinski:2016xgd,Gross:2016kjj,Gross:2017aos}. More recently the SYK model was linked to random matrices and Jackiw-Teitelboim gravity \cite{Saad:2019lba}.

\subsubsection{SYK-like models}

The melonic large-$N$ limit of the SYK model prompted the generalization of tensor models in one dimension. Such models reproduced the low energy behavior of the SYK model, dispensing of the quenched disorder. 
This was first done in the colored case by Witten \cite{Witten:2016iux}. We consider $D+1$ Majorana fields $\psi^{(0)}, \dots ,\psi^{(D)}$ defined as rank $D$ tensors. The Gurau-Witten model is then given by the action:
\begin{equation}
S=\int dt\left( \frac{i}{2}\sum_{k=0}^{D}\psi_{\mba^k}^{(k)}\partial_t \psi_{\mba^k}^{(k)}-i^{\tfrac{D+1}{2}}J K_{\mba^0 \dots \mba^D}\psi_{\mba^0}^{(0)}\dots \psi_{\mba^D}^{(D)}\right) \; ,
\end{equation}
where $\mba^k=(a_1^k \dots a_D^k)$ and $K_{\mba^0 \dots \mba^D}$ is defined in \eqref{eq:kernel_colored}.
The structure of the two- and four-point correlations functions were determined in \cite{Gurau:2016lzk} in terms of melons and ladders respectively.

An uncolored version was also introduced in \cite{Klebanov:2016xxf} for rank-$3$ tensors:
\begin{equation}
S=\int dt \left(\frac{i}{2}\psi_{abc}\partial_t\psi_{abc} +\frac{g}{4}\psi_{abc}\psi_{ade}\psi_{fbe}\psi_{fdc}\right) \; .
\label{eq:fermionCTKT}
\end{equation}
As the zero-dimensional version of this model was introduced by Carrozza-Tanasa in \cite{Carrozza:2015adg}, this model is referred as the Carrozza-Tanasa-Klebanov-Tarnopolsky (CTKT) model. It was extensively studied and the four-point ladder kernel was diagonalized. This reproduced the conformal spectrum of the bilinears for the CFT reached in the infrared. Moreover, this spectrum contains the mode dual to the stress-energy tensor: the dynamical gravitational dual and saturation of chaos characteristic of the SYK model are still present.

SYK-like models were extensively studied and many more models were defined, with tensors, in higher dimensions or with modified covariance for example \cite{Gurau:2016lzk,Peng:2016mxj,Krishnan:2016bvg,Krishnan:2017lra,Bulycheva:2017ilt,Choudhury:2017tax,Halmagyi:2017leq} (see also \cite{Klebanov:2018fzb,Delporte:2018iyf} for reviews).

\section{Summary of the thesis}
\label{sec:summary}

We saw in this chapter that tensor models present an interesting large-$N$ limit, the melonic limit. Moreover, because of the tensor structures, the number of interaction invariants is greater than for other large-$N$ theories such as vector or matrix models with bigger symmetry groups. Studying them as proper field theories is then of great interest as we can expect a new critical behavior. 
Indeed, in this context, they give rise at large $N$ to a new family of conformal field theories, called melonic CFTs, which can be studied analytically. For example, an explicit solution for the infrared two-point function and a list of the scaling dimensions of the bilinear operators were found. The first result is derived from the Schwinger-Dyson equation (SDE), while the second is derived from the Bethe-Salpeter equation (BSE). The treatment of the two equations is  similar \cite{Giombi:2017dtl,Bulycheva:2017ilt,Giombi:2018qgp,Klebanov:2018fzb}. First, at large $N$, both equations truncate to the first non-trivial term. Then, the infrared free term is neglected and the equation can be solved self-consistently. However, the results obtained by this method are somewhat formal, as both the SDE and the BSE have divergences. For fermionic models, some of the divergences are tamed by anticommutation. However, no such mechanism works for bosonic models. Dimensional regularization was used in \cite{Giombi:2017dtl} but the resulting fixed-point CFT was unstable.
The goal of this thesis is then to study rigorously the renormalization group flow of bosonic tensor field theories as well as to extend the class of models giving rise to melonic CFTs.

\subsection{Long-range models}

In the next chapter, we start by introducing a particular kind of quantum field theory that will be useful to study tensor models: long-range models. Long-range models have a vast array of applications \cite{Campa:2009rev}, and display several interesting features from a theoretical standpoint. 
They are defined as models having in the Lagrangian a kinetic term of the form $\phi (\p^2)^\z\phi$, with $0<\z<1$. The parameter $\zeta$ has to be positive to guarantee a well-defined thermodynamic limit and smaller than one to preserve reflection positivity.\footnote{Models with negative $\z$ are also of phenomenological interest. Models with positive and negative $\z$ are also known as ``weak'' and ``strong'' long-range models, respectively \cite{Mukamel:2009notes}. Notice that in most of the literature a different notation is used for the power of the Laplacian, with $\z\equiv\s/2$.}
One interesting feature of such models is the existence of phase transitions where they are forbidden in their short-range analogs ($\z=1$), for example in dimension $d=1$, as proved for the long-range Ising model by Dyson \cite{Dyson:1968up}. Moreover, their critical exponents depend on $\z$, thereby defining one-parameter families of universality classes. Tuning this parameter, one can study, at fixed dimension, interesting phenomena such as the transition at some $\z^\star<1$ from a long-range to a short-range universality class \cite{Sak:1973,Blanchard:2012xv,Angelini:2014,Brezin:2014,Defenu:2014,Behan:2017dwr,Behan:2017emf}, or construct rigorous renormalization group results in three dimensions \cite{Brydges:2002wq,Abdesselam:2006qg,Slade:2017,Lohmann:2017}.
However, the non-integer value of $\zeta$ gives rise to a number of challenges in the analytical treatment of long-range models. For example, the discussion of conformal invariance at the fixed point is complicated by the absence of a local stress-energy tensor. This issue has only been recently addressed in \cite{Paulos:2015jfa} for the long-range Ising model. Another, very practical, complication arises in the analytic evaluation of Feynman integrals; which will be the subject of chapter \ref{chap:3loops}.
For these reasons, long-range models have not been as thoroughly studied as their short-range counterparts, even though their interesting features have led to their investigation in many contexts (see for example \cite{Gawedzki:1985jn,Gross:2017vhb,Gubser:2019uyf,Heydeman:2020ijz,Giombi:2019enr,Defenu:2020umv}). In particular, short-range multi-scalar models with quartic interactions have been extensively studied by renormalization group methods both in their general version as well as with various symmetry restrictions (see for example \cite{Pelissetto:2000ek,Kleinert:2001ax,Vicari:2006xr,Osborn:2017ucf,Rychkov:2018vya} and references therein), and computations of critical exponents have reached the six-loop approximation \cite{Kompaniets:2017yct}, with the beta function and anomalous dimensions of the $O(N)$ model being available even at seven loops \cite{Schnetz:2016fhy}. 
On the contrary, their long-range versions have been analyzed much less, and the renormalization group analysis has been halted at the two-loop computations done in the 1970s \cite{Fisher:1972zz,Yamazaki:1977pt}. Other methods have also been underdeveloped as compared to the short-range case, with Monte Carlo simulations being mostly limited to the Ising model in one or two dimensions \cite{Glumac:1989,Luijten:1997-thesis,Angelini:2014,uzelac2001critical,tomita2009monte,Luijten:1997,Loscar:2018,rodriguez2009study,Xu:1993}, and with only occasional excursions from other methods, such as the functional renormalization group \cite{Defenu:2014} or the conformal bootstrap \cite{Behan:2018hfx}.

In chapter \ref{chap:3loops}, we then study the renormalization group of the long-range multi-scalar model with quartic interactions. We first compute the beta functions up to three loops, the computation of the Feynman amplitudes of the graphs contributing to the renormalization of the four-point function at three loops being the main result of the chapter. To do so we use dimensional regularization in the weak relevant case $\zeta=\frac{d+\epsilon}{4}$ with $d<4$ fixed. We then specialize to various symmetry restrictions: Ising model, $O(N)$ vector model, cubic model and $O(N_1)\times O(N_2)$ bifundamental model.

\subsection{Melonic CFTs}

\paragraph{The bosonic $O(N)^3$ model} The simplest model giving rise to a melonic CFT is a quartic bosonic tensor model in rank $3$ with $O(N)^3$ symmetry. In the short-range case (usual Laplacian propagator), it was proven in \cite{Giombi:2017dtl} using dimensional regularization that this model has a melonic fixed point. However, the fixed points are complex and the CFT is not unitary as there are operators with complex dimensions. The aim of chapter \ref{chap:CTKT} is then to treat melonic conformal field theories rigorously, and in order to deal with the divergences that appear in the perturbative expansion, we use the Wilsonian renormalization group picture.
More precisely, we consider a long-range version of the quartic bosonic tensor model of \cite{Klebanov:2016xxf,Giombi:2017dtl} and choose in particular a critical value of the power of the Laplacian, reproducing the infrared scaling and rendering the interactions marginal.
A similar idea has been applied to the SYK model by Gross and Rosenhaus in \cite{Gross:2017vhb}. One of the main differences of our model to that of Gross and Rosenhaus is that we have not just one marginal interaction but three (as in \cite{Giombi:2017dtl}): while we find that at large $N$ one of them remains exactly marginal, the other two have a non-trivial renormalization group flow, and in order to find a CFT we need to look for fixed points. We prove rigorously the existence of an infrared fixed point of the RG flow. More precisely, we show, non-perturbatively in the coupling constant, but at large $N$, that this model admits an infrared stable line of fixed points. The tetrahedron coupling is purely imaginary but all other critical couplings are real as well as the critical exponents. We then study the CFT at this IR fixed point. We find real dimensions of bilinears and real OPE coefficients in agreement with a unitary CFT at large $N$. 

In the next chapter we go one step further and study the next-to-leading order. We start with a generic trifundamental model with symmetry $O(N_1)\times O(N_2)\times O(N_3)$ both in short- and long-range. We compute fixed points up to next-to-leading order for different large-$N$ scaling limits looking for a stable fixed point with non-zero tetrahedral coupling at small $\epsilon$. In the short-range case, $\epsilon$ is the deviation from the critical dimension while in the long-range case it is the deviation from the critical scaling of the free propagator. The conclusion is that in order to find such fixed point, we need to consider complex couplings. In particular, in the homogeneous case $N_1=N_2=N_3=N$, corresponding to the $O(N)^3$ model, our analysis relied on identifying a suitable hierarchy between $N$ and $\epsilon$. In the long-range case, for $\epsilon N \ll 1$ and considering a purely imaginary tetrahedral coupling, we found a stable fixed point at large $N$ with real pillow and double-trace couplings. However, at next-to-leading order, the fixed point and critical exponents get complex corrections. Unitarity is broken at next-to-leading order. In the short-range case, the situation is somewhat reversed. For $\epsilon N^2 \gg 1$, at leading order, the fixed point is complex and unstable. However, critical exponents acquire a real part at next-to-leading order so that the fixed point of \cite{Giombi:2017dtl} becomes a genuine infrared fixed point at finite $N$.

In chapter \ref{chap:Ftheorem}, we investigate the properties of melonic CFTs. More precisely, we test the F-theorem for the long-range bosonic $O(N)^3$ tensor model. This theorem, in its weak form, states that, for a CFT in dimension $3$, the free energy on the sphere at the UV fixed point is greater than that at the IR fixed point \cite{Klebanov:2011gs}. It was not clear if this theorem also applied to long-range models, mainly because it was proven for unitary theories while the long-range $O(N)^3$ model can be unitary only at large $N$. However, we prove that it holds for the long-range bosonic $O(N)^3$ tensor model, thus providing a new highly non-trivial example. As a warm up we first studied a flow between two Gaussian CFTs. Then we tested our computation method for the free energy on the $O(N)$ model for which the result is known. We finally computed the free energy for the $O(N)^3$ model. This was challenging as it involved resumming an infinite series of ladder diagrams but doable thanks to the use of conformal partial wave expansion. Moreover, we showed that the free energy decreases between the UV and IR due to the inclusion of a non-normalizable state in the UV. This can also be seen in perturbation theory as the reversal of the sign of an infinite class of diagrams due to the flow of a coupling constant.

\paragraph{Sextic tensor models in rank $3$ and $5$}
We can wonder how the previous results depend on the rank of the tensors and on the order of the interaction. We start answering this question in chapter \ref{chap:sextic} by studying sextic tensor field theories. We considered two sextic models, one with bipartite interactions in rank $3$ and one with non-bipartite interactions in rank $5$. For both models, we considered the short- and long-range cases. In rank $3$, we found two infrared stable fixed points in short-range and a line of infrared stable fixed points in long-range. The situation is similar to the quartic case but the fixed points are here real in both cases and we find a window with real spectrum of bilinear operators. However, in the short-range case, the stability matrix is non-diagonalizable implying that the CFT at the fixed point is logarithmic. Therefore, as for the quartic case, the CFT at the large-$N$ fixed point in the short-range case is non-unitary. Surprisingly, the only fixed point in rank $5$ is the Gaussian one, in both short- and long-range cases. 

In the second part of the chapter, we study the $1/N$ corrections for the rank-$3$ model both in short- and long-range. In the short-range case, we still find two infrared stable fixed points at next-to-leading order. Moreover, the stability matrix is now diagonalizable. However, this does not mean that the theory can be unitary at finite $N$ as corrections in $1/N$ cannot cancel the logarithmic terms from the leading order. In the long-range case, the situation is very different from the quartic model. Indeed, the corrections to the fixed point are non-perturbative and hence unreliable. We found no precursor of the large-$N$ fixed point.

\subsection{Irreducible tensor model in rank $5$}

To use tensor models in the context of quantum mechanics and quantum field theory, a key feature is the existence of a melonic large-$N$ limit. As we saw in subsection \ref{sec:other_melonic}, the existence of a melonic large-$N$ limit for irreducible models was proven only recently in rank $3$ \cite{Benedetti:2017qxl,Carrozza:2018ewt}. In chapter \ref{chap:rank5}, we extend this proof in rank $5$.
This was a highly non-trivial generalization relying on recursive bounds derived from a detailed combinatorial analysis of Feynman graphs involved in the perturbative expansion of our model. We follow the combinatorial methods developed in \cite{Benedetti:2017qxl}. As in rank $3$, technical difficulties arise from the existence of Feynman graphs which violate the maximum scaling naively allowed by the large-$N$ limit. Such contributions cancel out when resumming because of the irreducibility condition but they still complicate greatly the recursive strategy used to bound Feynman amplitudes. Considering a rank-$5$ model simplifies some aspects of the combinatorial construction with respect to the rank-$3$ models (we do not need to consider triangle subgraphs for example). However, other aspects are much more involved in rank $5$ and we need, for example, to bound some eight-point functions in addition to two- and four-point functions. 

\chapter{Long-range multi-scalar models at three loops}
\label{chap:3loops}

In this chapter we study the long-range multi-scalar model with quartic interactions and compute its beta functions up to three loops. 
The general framework and the main results are presented in section \ref{sec:model3l}, while  some detailed computations are included in the appendices. 
We then specialize the beta functions to various symmetry restrictions. 
First, in section \ref{sec:ising model}, we study the long-range Ising model. We give the fixed points and critical exponents in the $\epsilon$ expansion up to order $\epsilon^3$ and compare them with numerical simulations at $d=1$ \cite{Glumac:1989,Luijten:1997-thesis,uzelac2001critical,tomita2009monte} and $d=2$ \cite{Angelini:2014}. 
Second, in section \ref{sec:vector model}, we study the long-range $O(N)$ vector model (the long-range Ising model being the special case $N=1$). We compute the beta functions up to three loops as well as the fixed points and critical exponents and give numerical values for the critical exponents for different dimensions and values of $N$. Our two-loop results agree with \cite{Fisher:1972zz} and \cite{Yamazaki:1977pt}, while the three-loop results are new. We also give the expression for the critical exponents in the large-$N$ expansion. Again our $1/N$ result agrees with \cite{Fisher:1972zz} while the $1/N^2$ contribution is new.
Next, in section \ref{sec:cubic}, we consider the long-range cubic model which is obtained by breaking explicitly the $O(N)$ symmetry with an interaction of the form $\sum_{\mba} \phi_{\mba}^4$. This results in the (hyper-)cubic symmetry group $(\mathbb{Z}_2)^N \rtimes S_N$. We again compute the beta functions, fixed points and critical exponents up to three loops. Our two-loop results agree with \cite{Yamazaki:1978-cubic,Yamazaki:1981-cubic,Chen:2001} while the three-loop results are new. This model admits three non-trivial fixed points: a $O(N)$, or Heisenberg, fixed point with the cubic coupling being zero, an Ising fixed point with the $O(N)$ coupling being zero, and a cubic fixed point when both couplings are non-zero. We then compute the critical value of $N$ at which the Heisenberg and cubic fixed points collapse and exchange stability. 
Finally, in section \ref{sec:bifundamental}, we consider the long-range $O(M)\times O(N)$, or  bifundamental, model. This model has two couplings that are associated to the single-trace and double-trace quartic invariants, in a matrix terminology. We again compute the beta functions and fixed points. However, we do not write explicitly the three-loop contributions as they are too lengthy. There are three non-trivial fixed points: a Heisenberg fixed point with the single-trace coupling being zero and two chiral fixed points with both couplings non-zero. We also compute the three critical values of $N$ delimiting four regimes of criticality at fixed $M$, depending on the stability of the Heisenberg and chiral fixed points. We give their expansions up to three loops as well as numerics in three dimensions for $M=2$. For this long-range model we are not aware of any previous results even at two loops.
We conclude in section \ref{sec:concl3} with a brief summary and some concluding remarks.

\section{The long-range multi-scalar model}
\label{sec:model3l}
The long-range multi-scalar model with quartic interactions in dimension $d$ is defined by the action:
\begin{equation} \label{eq:actionms}
		S[\phi]  \, = \,  \int d^dx \, \bigg[ \frac{1}{2} \phi_\mba(x) ( - \partial^2)^{\zeta}\phi_{\mba}(x) +\frac{1}{2}\, \kappa_{\mba \mbb}\phi_{\mba}(x) \phi_{\mbb}(x) + 
		\frac{1}{4!} \, \lambda_{\mba \mbb \mbc \mbd}
		\phi_{\mba}(x) \phi_{\mbb}(x) \phi_{\mbc}(x) \phi_{\mbd}(x) \bigg] \, ,
\end{equation}
where the indices take values from 1 to $\cN$, and a summation over repeated indices is implicit.
The  coupling $\lambda_{\mba\mbb\mbc\mbd}$ and the mass parameter $\kappa_{\mba\mbb}$ are symmetric tensors, thus corresponding in general to $\binom{\cN+3}{4}$ and $\tfrac{\cN(\cN+1)}{2}$ couplings, respectively.
As explained in section \ref{sec:summary}, the model is called long-range due to the non-integer power of the Laplacian, 
$0< \zeta < 1$.  The short-range model is defined analogously, but with $\z=1$.
From now on the dimension $d$ is fixed to be smaller than (and not necessarily close to) four. 

We treat the mass parameter $\kappa$ as a perturbation, hence the covariance (or propagator) of the free theory is 
$ C_{\mba \mbb}(x,y) =  \delta_{\mba \mbb} \;  C(x-y) $, with:
\be\label{eq:cov3} 
\begin{split}
 & C(x-y) = \int\frac{d^dp}{(2\pi)^d}\; e^{ - \im p (x-y) } C(p) = \f{\G\left(\f{d-2\z}{2}\right)}{2^{2\z}\pi^{d/2}\G(\z)} \; \f{1}{|x-y|^{d-2\z}}\,, \\
 & C(p) = \frac{1}{p^{2\zeta}}  = \frac{1}{\Gamma(\zeta)} \int_0^{\infty} da \;a^{\zeta -1 } e^{- a p^2} \,.
\end{split}
\ee 
In the last equality we have introduce also the Schwinger parametrization that we will use to evaluate the Feynman integrals.

The canonical dimension of the field is:
\be
\Delta_{\phi} = \frac{d-2\zeta}{2} \,.
\ee
Therefore, the quartic interaction is irrelevant for $\z<d/4$ leading to mean-field behavior (as rigorously proved in \cite{Aizenman:1988}), while for $\z>d/4$ it is relevant, and a non-trivial IR behavior is expected. The marginal case is $\z=d/4$. We will be interested in the weakly relevant case:
\be \label{eq:zeta-eps}
 \zeta = \frac{d+\epsilon}{4} \,,
\ee
with small $\epsilon$. The ultraviolet dimension of the field is thus fixed to $\Delta_{\phi}=\frac{d-\epsilon}{4}$.

\paragraph{Divergences and regularization.} In order to define an $\epsilon$ expansion, we need to make sense of the theory at $\epsilon=0$ first. However, at $\zeta = d/4$ the model exhibits logarithmic divergences. Let us first consider graphs with $\lambda$ vertices only (we will include the $\kappa$ vertices soon after). Let $\cG$ be a connected amputated Feynman graph with $V(\cG)$ vertices, $E(\cG)$ edges and $n(\cG)$ external points.
In momentum space one counts an independent integral $d^dp$ for every loop and a propagator $p^{-2\zeta}$ for every edge. Adapting \eqref{eq:degUV}, with $q=4$ and $\zeta=d/4$, we obtain the following UV degree of divergence:
\begin{equation}
\text{deg}(\mathcal{G})=d \left( 1 - \frac{ n(\cG)} {4} \right) \, .
\label{eq:power_counting}
\end{equation}
The theory is marginal, that is, the power counting does not depend on the number of internal vertices. The two-point graphs are superficially power divergent, the four-point graphs are superficially logarithmically divergent and the higher-point graphs are superficially convergent.

The power divergences can be ignored: one can either use  dimensional regularization or add a two-point power divergent local counterterm to cancel them. Once the local power divergence is subtracted, the two-point graphs are convergent, that is, \emph{there is no wave function renormalization}.\footnote{There is however a coefficient of the full two-point function $\mathcal{Z}$, such that $G(p)=\mathcal{Z}p^{-2\zeta}$. However, this coefficient is finite and does not contribute to an anomalous dimension. We will see this in more detail in chapter \ref{chap:CTKT}.} This is a key feature of long-range models (see for example \cite{Fisher:1972zz,suzuki1972wilson,brezin2014crossover,Behan:2017emf} and references therein or \cite{Brydges:2002wq,lohmann2017critical} for a study of long-range models at a rigorous level).

The $\kappa$ vertices represent the $\phi^2$ perturbation with respect to the critical theory. For power counting purposes they can be seen as quartic vertices with two external half-edges carrying zero momentum, and thus the previous power counting goes trough with little alteration. The only superficially divergent graphs are:
\begin{itemize}
 \item four-point graphs with only $\lambda$ vertices,
 \item two-point graphs with exactly one $\kappa$ vertex.
\end{itemize}
Both types of graphs are logarithmically divergent, hence we need to regularize both the ultraviolet and the infrared, and choose a renormalization prescription. 

The ultraviolet is naturally regularized by reintroducing $\epsilon>0$.\footnote{ A different approach, working directly at finite $\epsilon$ is discussed in \cite{Belim_2003}. The lack of a small-parameter however makes it less rigorous, and prone to the appearance of spurious solutions, as discussed for the short-range case in \cite{Delamotte_2010}.}
As a renormalization prescription, we  use the zero momentum BPHZ subtraction scheme \cite{Rivasseau:1991ub}, made explicit in \eqref{eq:bphz} below.
However, since we are working with a massless propagator, an infrared regulator is required. We introduce that by modifying the propagator as: 
\be\label{eq:paramcov}
 C_{\mu}(p) = \frac{1}{(p^2 + \mu^2)^{\zeta}} = \frac{1}{\Gamma(\zeta)} \int_0^{\infty} d a \;a^{\zeta -1 } e^{- a p^2 -a \mu^2} \,,
\ee
for some mass parameter $\mu>0$.
One could wonder why we choose this prescription and not the usual Gell-Mann and Low subtraction at non-zero momentum. The reason is that in such case we were not able to obtain analytic results for the amplitudes of graphs at three loops for $\z<1$.

Before continuing let us comment on the mass term we included in our action. Our regulator $\mu$ is  not a mass: the regulated quadratic part is $\phi(p^2 + \mu^2)^{\zeta}\phi$ while a massive propagator corresponds to $\phi(p^{2\zeta} + m^{2\zeta}) \phi $.  $\mu$ and $m$ are identified in the short-range case $\zeta=1$, but not in the long-range case $\zeta<1$. For the long-range model the mass is the coupling of the $\phi^2$ operator, while $\mu$ is relegated to an infrared cutoff.\footnote{In the counterterm picture we also tune to criticality by subtracting to zero the mass power divergence, that is we split $m^{2\zeta}$ into a power divergent counterterm and the logarithmic mass coupling $\kappa$. In full detail the quadratic part of the regulated theory is 
$\phi(p^{2} + \mu^2)^{\zeta} \phi + \kappa \phi^2 - m_b^{2\zeta} \phi^2$, with the bare mass  $m_b$ tuned such that $\Braket{\phi(x)\phi(x)}_{\kappa,\mu=0} =0$  
 } 
 
 The advantage of using the $\mu$ regulated propagator instead of the massive one is threefold. Contrary to the massive propagator ours is analytic at $p^2=0$, it admits a 
 K\"all\'en--Lehmann representation (hence it is Osterwalder Schrader positive):
 \be
  \frac{1}{(p^2+\mu^2)^{\zeta}} \sim \int_{\mu^2}^{\infty} dx \; \frac{ (x-\mu^2)^{-\zeta} }{p^2 + x} \;,
 \ee
 and finally (and most importantly) it allows us to compute the three loop integrals analytically.
 
Once this is settled we have two options. As we are interested in determining the mass critical exponent we can either treat $\kappa$ as a fully fledged coupling of the model (which is what we do below) or we can set it to zero, and look at the anomalous scaling of the renormalized mass operator $[\phi^2]$. We prefer the first option as it treats $\kappa$ and $\lambda$ on the same footing, but we stress that both points of view are equally valid.  
 
\subsection{Two and four-point functions}

We denote $\Gamma^{(2)}_{\mba\mbb}$ and $\Gamma^{(4)}_{\mba \mbb \mbc \mbd}$ the one-particle irreducible two and four-point functions at zero external momentum.\footnote{In the counterterm picture $\Gamma^{(2)}_{\mba\mbb}$ is the two-point function with the local power divergence subtracted.} We compute them up to three loops using the bare expansion in terms of connected amputated one-particle irreducible Feynman diagrams $\mathcal{G}$, whose amplitude in Schwinger parametrization reads: 
\be\label{eq:amp_final}
\mathcal{A}(\mathcal{G} ) =  \mu^{ (d-4\zeta)(V-1)} \; \mathcal{\hat{A}}(\mathcal{G}) \,, \quad
\mathcal{\hat{A}}(\mathcal{G}) =
 \frac{1} { 
  \big[ (4\pi)^{d/2} \Gamma(\zeta)^2 \big]^{V-1} }
\int_0^{\infty}
\prod_{e \in \mathcal{G}} d a_e
\;\;
\frac{\prod_{e \in \mathcal{G}} a_e^{\zeta-1} \; e^{-\sum_{e \in \mathcal{G}} a_e}}
{\big(\sum_{\cT \in \mathcal{G}  } \prod_{e \notin \cT } a_e\big)^{d/2} } \,,
\ee
where $V$ denote the numbers of vertices of $\mathcal{G}$, $e \in \mathcal{G}$ runs over the edges of $\mathcal{G}$, and $\cT$ denotes the spanning trees in $\mathcal{G}$ (e.g.\ \cite{Rivasseau:1991ub}). Note that we used the fact that we only deal with four-point graphs with quartic vertices, as these are sufficient to describe the divergent graphs described above.

\begin{figure}[htbp]
\centering
\captionsetup[subfigure]{labelformat=empty}
\subfloat[$D$]{\includegraphics[scale=1]{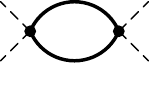}} \hspace{20pt} 
\subfloat[$D^2$]{\includegraphics[scale=1]{D2.pdf}} \hspace{20pt} 
\subfloat[$S$]{\includegraphics[scale=1]{S.pdf}} 
\caption{One- and two-loop contributions to the bare expansion.}
\label{fig:1_2_loops}
\end{figure}

\begin{figure}[htbp]
\centering
\captionsetup[subfigure]{labelformat=empty}
\begin{tabular}{cccc}
\subfloat[$D^3$]{\includegraphics[scale=1]{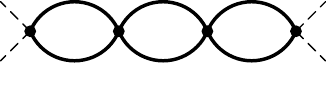}} &
\subfloat[$DS$]{\includegraphics[scale=1]{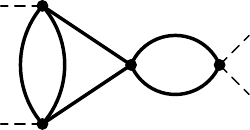}} &
\subfloat[$U$]{\includegraphics[scale=1]{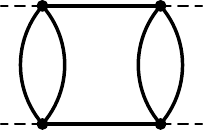}} &
\subfloat[$T$]{\includegraphics[scale=1]{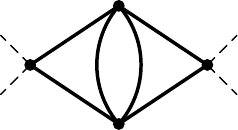}}  \\
\subfloat[$I_1$]{\includegraphics[scale=1]{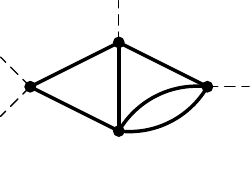}} & 
\subfloat[$I_2$]{\includegraphics[scale=1]{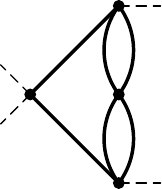}} &
\subfloat[$I_3$]{\includegraphics[scale=1]{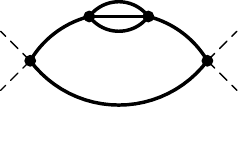}} &
\subfloat[$I_4$]{\includegraphics[scale=1]{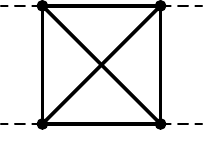}} 
\end{tabular}
\caption{Three-loop contributions.}
\label{fig:3_loops}
\end{figure}

\paragraph{The four-point function.}
There is only one diagram contributing at one loop, two diagrams at two loops (figure~\ref{fig:1_2_loops}), and eight at three loops (figure~\ref{fig:3_loops}). We call $D,S,T,U,I_1,I_2,I_3,I_4$ the amplitudes 
$\mathcal{\hat{A}}(G)$ of these diagrams 
(see figure~\ref{fig:1_2_loops} and \ref{fig:3_loops} for the detailed notation),\footnote{The choice of letters has no particular meaning.} and we use the fact that the amplitude of a one-vertex reducible diagram (that is, a diagram that disconnects by deleting a vertex) factors into the product of amplitudes of its one-vertex irreducible parts.

One has to be careful to conserve the permutation symmetry of the four-point function in its indices. To this end one should completely symmetrize over the external indices, but due to specific invariances of the diagrams under relabelling, some of the symmetrized terms are trivially equal. Grouping together the terms in classes of explicitly equal terms we get:
\begin{align} \label{eq:bare_4pt}
& \Gamma^{(4)}_{\mba \mbb \mbc \mbd} \, = \, 
\lambda_{\mba \mbb \mbc \mbd}-\frac{1}{2} \big(\lambda_{\mba \mbb \mbe \mbf}\lambda_{\mbe \mbf \mbc \mbd} + 2 \textrm{ terms} \big) \, \mu^{-\epsilon} D \crcr
& 
+\frac{1}{4} \big( \lambda_{\mba \mbb \mbe \mbf}\lambda_{\mbe \mbf \mbg \mbh}\lambda_{\mbg \mbh \mbc \mbd}+ 2 \textrm{ terms} \big) \, \mu^{-2\epsilon} D^2  +\frac{1}{2} (\lambda_{\mba \mbb \mbe \mbf}\lambda_{\mbe \mbg \mbh \mbc}\lambda_{\mbf \mbg \mbh \mbd}+ 5 \textrm{ terms}) \, \mu^{-2\epsilon} S \crcr
& 
- \, \frac{1}{8}  \big(\lambda_{\mba \mbb \mbe \mbf} \lambda_{\mbe \mbf \mbg \mbh} \lambda_{\mbg \mbh \mbm \mbn} \lambda_{\mbm \mbn \mbc \mbd} + 2 \textrm{ terms} \big) \,\mu^{-3\epsilon}  D^3  
- \, \frac{1}{4} \big(\lambda_{\mba \mbb \mbe \mbf} \lambda_{\mbe \mbf \mbg \mbh} \lambda_{\mbg \mbm \mbn \mbc} \lambda_{\mbh \mbm \mbn \mbd} + 5 \textrm{ terms} \big) \, \mu^{-3\epsilon}   SD \crcr
& 
- \, \frac{1}{4} \big(\lambda_{\mba \mbe \mbf \mbg} \lambda_{\mbb \mbe \mbf \mbh} \lambda_{\mbg \mbm \mbn \mbc} \lambda_{\mbh \mbm \mbn \mbd}  + 5 \textrm{ terms} \big) \, \mu^{-3\epsilon} U - \, \frac{1}{4} \big(\lambda_{\mba \mbb \mbe \mbf} \lambda_{\mbe \mbg \mbh \mbm} \lambda_{\mbf \mbg \mbh \mbn} \lambda_{\mbm \mbn \mbc \mbd} + 2 \textrm{ terms} \big) \, \mu^{-3\epsilon} T \crcr
& - \, \frac{1}{2} \big(\lambda_{\mba \mbb \mbe \mbf} \lambda_{\mbe \mbg \mbh \mbm} \lambda_{\mbf \mbg \mbn \mbc} \lambda_{\mbh \mbm \mbn \mbd} + 11 \textrm{ terms} \big) \, \mu^{-3\epsilon} I_1 - 
\,  \frac{1}{4} \big(\lambda_{\mba \mbb \mbe \mbf} \lambda_{\mbe \mbg \mbh \mbc} \lambda_{\mbf \mbm \mbn \mbd} \lambda_{\mbg \mbh \mbm \mbn} + 5 \textrm{ terms} \big) \, \mu^{-3\epsilon}  I_2  \crcr  
&  
- \, \frac{1}{6} \big(\lambda_{\mba \mbb \mbe \mbf}\lambda_{\mbh \mbm \mbn \mbf}\lambda_{\mbh \mbm \mbn \mbg}\lambda_{\mbg \mbe \mbc \mbd} +2 \textrm{ terms} \big)\, \mu^{-3\epsilon} I_3  
-  \,\big(\lambda_{\mba \mbe \mbm \mbh}\lambda_{\mbb \mbe \mbf \mbn}\lambda_{\mbc \mbf \mbm \mbg}\lambda_{\mbd \mbg \mbn \mbh}  \big) \, \mu^{-3\epsilon} I_4  \, .
\end{align}
where the ``$+  \dots \,\rm{terms}$'' notation designates a sum over terms obtained by permuting the external indices 
in non-equivalent ways. For instance, the 3 terms in the first line can be seen as the choice of the index $\mbb$, $\mbc$, or $\mbd$, paired with $\mba$ on the same coupling constant. The $I_1$ diagram is the least symmetric in our lot, only being invariant under exchange of the two external indices attached to the same vertex, thus giving 12 inequivalent terms out of the 24 permutations.

The integrals $D,S,T,U,I_1,I_2,I_3,I_4$ are computed in appendix \ref{app:integrals}.

\paragraph{The two-point function.} The UV-divergent diagrams contributing to
$\Gamma^{(2)}_{\mba\mbb}$ are a subset of those contributing to the four-point function. More specifically, they are the diagrams  in  figure~\ref{fig:1_2_loops} and \ref{fig:3_loops} with at least one vertex having two external half-edges, hence $U$ and $I_4$ do not contribute. We then substitute one of the vertices  having two external half-edges with a $\kappa$ vertex, and we get:
\begin{align} \label{eq:bare_2pt}
 \Gamma^{(2)}_{\mbc \mbd} \, = & \, 
\kappa_{ \mbc \mbd}-\frac{1}{2} \big( \kappa_{\mbe \mbf}\lambda_{\mbe \mbf \mbc \mbd} \big) \, \mu^{-\epsilon} D 
+\frac{1}{4} \big( \kappa_{\mbe \mbf} \lambda_{\mbe \mbf \mbg \mbh}\lambda_{\mbg \mbh \mbc \mbd} \big) \, \mu^{-2\epsilon} D^2 
+\frac{1}{2} \big( \kappa_{\mbe \mbf}\lambda_{\mbe \mbg \mbh \mbc}\lambda_{\mbf \mbg \mbh \mbd} \big) \, \mu^{-2\epsilon} S \crcr 
&
- \, \frac{1}{8}  \big( \kappa_{ \mbe \mbf} \lambda_{\mbe \mbf \mbg \mbh} \lambda_{\mbg \mbh \mbm \mbn} \lambda_{\mbm \mbn \mbc \mbd} \big) \,\mu^{-3\epsilon}  D^3  
- \, \frac{1}{4} \big( \kappa_{ \mbe \mbf} \lambda_{\mbe \mbf \mbg \mbh} \lambda_{\mbg \mbm \mbn \mbc} \lambda_{\mbh \mbm \mbn \mbd} \big) \, \mu^{-3\epsilon}   SD \crcr 
& 
- \, \frac{1}{4} \big( \kappa_{ \mbe \mbf} \lambda_{\mbe \mbg \mbh \mbm} \lambda_{\mbf \mbg \mbh \mbn} \lambda_{\mbm \mbn \mbc \mbd} \big) \, \mu^{-3\epsilon} T - \, \frac{1}{2} \big( \kappa_{ \mbe \mbf} \lambda_{\mbe \mbg \mbh \mbm} \lambda_{\mbf \mbg \mbn \mbc} \lambda_{\mbh \mbm \mbn \mbd}  + 1 \textrm{ term} \big) \, \mu^{-3\epsilon} I_1 \crcr 
&
- \,  \frac{1}{4} \big( \kappa_{ \mbe \mbf} \lambda_{\mbe \mbg \mbh \mbc} \lambda_{\mbf \mbm \mbn \mbd} \lambda_{\mbg \mbh \mbm \mbn} \big) \, \mu^{-3\epsilon}  I_2  
- \, \frac{1}{6} \big(\kappa_{ \mbe \mbf}\lambda_{\mbh \mbm \mbn \mbf}\lambda_{\mbh \mbm \mbn \mbg}\lambda_{\mbg \mbe \mbc \mbd}   \big)\, \mu^{-3\epsilon} I_3  \, .
\end{align}
In general $\Gamma^{(2)}_{\mba\mbb}$ receives contribution also from diagrams with more insertions of $\kappa$ vertices, which are UV-convergent (but IR-divergent); however, we are interested in perturbative fixed-point theories with $\kappa=0$, hence we will ignore such contributions. For the same reason we have not included any $\kappa$ contribution in $\Gamma^{(4)}_{\mba \mbb \mbc \mbd}$. We are only introducing the quadratic operator perturbations as a means to obtain their scaling dimension at the fixed point, for which the linear terms in $\kappa$ suffice.
An equivalent way to rephrase this is to only consider the theory with $\kappa=0$, and in addition study the renormalization of the composite mass quadratic operator $[\phi^2]$ as in \cite{Pelissetto:2000ek,Brezin:1974eb,Brezin:1974-add}.

Observe that the overall combinatorial coefficients in \eqref{eq:bare_2pt} are the same ones as in \eqref{eq:bare_4pt}, but all the symmetries are broken, that is, for each term only one class is selected among the distinct classes involved in the four-point function ($I_1$ is special as its contribution to $\Gamma^{(2)}_{\mbc\mbd}$ is not a priori symmetric in the indices $\mbc\mbd$, hence one still gets a sum over the two terms). 


\subsection{The beta functions}
\label{sec:beta_func}

In the BPHZ subtraction with IR regulator 
\cite{Rivasseau:1991ub} (which is equivalent to the Wilsonian picture) we identify the dimensionless four-point function at zero external momenta with the running coupling:
\be \label{eq:bphz}
g_{\mba \mbb \mbc \mbd} = \, \mu^{- \epsilon} \, \Gamma^{(4)}_{\mba \mbb \mbc \mbd} \,, \qquad
 r_{\mbc \mbd} = \, \mu^{-(d-2\Delta_{\phi} )} \, \Gamma^{(2)}_{\mbc \mbd} \,.
\ee
The beta functions are the scale derivatives of the running coupling at fixed bare couplings:
\begin{equation}\label{eq:bseries}
\beta^{(4)}_{\mba \mbb \mbc \mbd}=\mu \partial_{\mu} g_{\mba \mbb \mbc \mbd}\, ; \qquad 
 \beta^{(2)}_{\mbc \mbd} = \mu \partial_{\mu} r_{\mbc\mbd} \, .
\end{equation}
We rescale the couplings as\footnote{Notice that $\left(4\pi\right)^{d/2}\Gamma(d/2) =2 (2\pi)^d/{\rm Vol}(S^{d-1})$, where ${\rm Vol}(S^{d-1})= 2\pi^{d/2}/\Gamma(\tfrac{d}{2})$ is the volume of the $(d-1)$-dimensional unit sphere. Our rescaling thus differs by a factor two from the one used by some other authors.}
$g_{\mba \mbb \mbc \mbd}= \left(4\pi\right)^{d/2}\Gamma(d/2) \, \tilde{g}_{\mba \mbb \mbc \mbd} $ and 
$r_{\mba \mbb }= \left(4\pi\right)^{d/2}\Gamma(d/2) \, 
\rt _{\mba \mbb} $, and we denote:  
\begin{align}
\alpha_{D} \, & =  \, \epsilon (4\pi)^{d/2} \,
\Gamma( \tfrac{d}{2}) \frac{D}{2} \, , &
\alpha_{S} \, &= \, \epsilon (4\pi)^{d}\Gamma( \tfrac{d}{2})^2 \, \frac{(D^2 - 2 S)}{2} \, , \crcr
\alpha_{U} \, & =\, \epsilon (4\pi)^{3d/2}\Gamma(\tfrac{d}{2})^3 \, \frac{(D^3-4DS+3U)}{4} \, , &
\alpha_{T} \,  &=\, \epsilon (4\pi)^{3d/2}\Gamma(\tfrac{d}{2})^3 \, \frac{(3T-2DS)}{4} \, , \;\; \crcr
\alpha_{I_1} \, &   =\, \epsilon (4\pi)^{3d/2}\Gamma(\tfrac{d}{2})^3 \, \frac{(D^3-3DS+3I_1)}{2} \, , &
\alpha_{I_2} \,  & = \, \epsilon (4\pi)^{3d/2}\Gamma(\tfrac{d}{2})^3\frac{(D^3-4DS+3I_2)}{4} \, , \crcr
\alpha_{I_3} \, &  =\, \epsilon (4\pi)^{3d/2}\Gamma(\tfrac{d}{2})^3 \, \frac{I_3}{2} \, , &
\alpha_{I_4} \, & =\, \epsilon  (4\pi)^{3d/2}\Gamma(\tfrac{d}{2})^3 \,  3I_4 \, \,.
\end{align}
Using appendix \ref{app:renseries}, we get:
\begin{align}
\beta^{(4)}_{\mba \mbb \mbc \mbd} &= -\epsilon \gt_{\mba \mbb \mbc \mbd} + \alpha_{D}\left(\gt_{\mba \mbb \mbe \mbf}\gt_{\mbe \mbf \mbc \mbd} + 2 \textrm{ terms} \right)  + \alpha_{S}\left(\gt_{\mba \mbb \mbe \mbf}\gt_{\mbe \mbg \mbh \mbc}\gt_{\mbf \mbg \mbh \mbd}+ 5 \textrm{ terms}\right) \crcr 
& \quad  \, + \, \alpha_{U} (\gt_{\mba \mbe \mbf \mbg} \gt_{\mbb \mbe \mbf \mbh} \gt_{\mbg \mbm \mbn \mbc} \gt_{\mbh \mbm \mbn \mbd} + 5 \textrm{ terms} )  + \, \alpha_{T} (\gt_{\mba \mbb \mbe \mbf} \gt_{\mbe \mbg \mbh \mbm} \gt_{\mbf \mbg \mbh \mbn} \gt_{\mbm \mbn \mbc \mbd} + 2 \textrm{ terms} )   \crcr
& \quad  \,  + \, \alpha_{I_1} (\gt_{\mba \mbb \mbe \mbf} \gt_{\mbe \mbg \mbh \mbm} \gt_{\mbf \mbg \mbn \mbc} \gt_{\mbh \mbm \mbn \mbd} + 11 \textrm{ terms} ) + \,\alpha_{I_2}(\gt_{\mba \mbb \mbe \mbf} \gt_{\mbe \mbg \mbh \mbc} \gt_{\mbf \mbm \mbn \mbd} \gt_{\mbg \mbh \mbm \mbn} + 5 \textrm{ terms} )  \crcr
& \quad  + \, \alpha_{I_3} (\gt_{\mba \mbb \mbe \mbf}\gt_{\mbh \mbm \mbn \mbf}\gt_{\mbh \mbm \mbn \mbg}\gt_{\mbg \mbe \mbc \mbd} +2 \textrm{ terms} ) + \, \alpha_{I_4} ( \gt_{\mba \mbe \mbm \mbh}\gt_{\mbb \mbe \mbf \mbn}\gt_{\mbc \mbf \mbm \mbg}\gt_{\mbd \mbg \mbn \mbh} ) 
\label{eq:beta_abcd_alpha3}
\end{align}
\begin{align}
\beta^{(2)}_{\mbc\mbd}
&= - (d-2\Delta_{\phi} ) \rt_{\mbc \mbd} +\alpha_{D}  \big( \rt_{ \mbe \mbf}\gt_{\mbe \mbf \mbc \mbd} \big) + \alpha_{S} \big(\rt_{\mbe \mbf} \gt_{\mbe \mbg \mbh \mbc}\gt_{\mbf \mbg \mbh \mbd} \big) + \, \alpha_{T}  
\big(\rt_{\mbe \mbf} \gt_{\mbe \mbg \mbh \mbm} \gt_{\mbf \mbg \mbh \mbn} \gt_{\mbm \mbn \mbc \mbd} \big)  \crcr
& \quad 
 + \, \alpha_{I_1}
(\rt_{\mbe \mbf} \gt_{\mbe \mbg \mbh \mbm} \gt_{\mbf \mbg \mbn \mbc} \gt_{\mbh \mbm \mbn \mbd} + 1 \textrm{ term} )  + \, \alpha_{I_2}
\big(\rt_{ \mbe \mbf} \gt_{\mbe \mbg \mbh \mbc} \gt_{\mbf \mbm \mbn \mbd} \gt_{\mbg \mbh \mbm \mbn} \big)  \crcr 
& \quad + \, \alpha_{I_3}
\big( \rt_{\mbe \mbf} \gt_{\mbh \mbm \mbn \mbf} \gt_{\mbh \mbm \mbn \mbg} \gt_{\mbg \mbe \mbc \mbd} \big) \,. 
\label{eq:beta2_abcd_alpha3}
\end{align}
The main result of this chapter is the determination of the constants $\alpha$ in appendix \ref{app:integrals}:
\begin{align}
& \alpha_{D} \, = \, 1 +\frac{\epsilon}{2}\big[\psi(1)-\psi(\tfrac{d}{2}) \big]+\frac{\epsilon^2}{8}\left[\left(\psi(1)-\psi(\tfrac{d}{2})\right)^2+ \psi_1(1)-\psi_1(\tfrac{d}{2})\right] \,, \crcr
&\alpha_{S} \, = \,  2\psi( \tfrac{d}{4} ) - \psi( \tfrac{d}{2})-\psi(1)   +\frac{\epsilon}{4}\Big[\left[2\psi(\tfrac{d}{4})-\psi(\tfrac{d}{2})-\psi(1)\right]
\left[3\psi(1)-5\psi(\tfrac{d}{2})+2\psi(\tfrac{d}{4})\right]   \crcr
& \qquad \; + 3\psi_1(1) + 4\psi_1(\tfrac{d}{4})-7\psi_1(\tfrac{d}{2})  -4 J_0(\tfrac{d}{4}) \Big] \, , \crcr
& \alpha_{U} = \alpha_{I_2} =\,  - \psi_1(1)-\psi_1(\tfrac{d}{4})+2\psi_1( \tfrac{d}{2})
+ J_0(\tfrac{d}{4})\, , \crcr
& \alpha_{T} \, = \, \frac{1}{2}\Big[2\psi(\tfrac{d}{4}) - \psi(\tfrac{d}{2})-\psi(1) \Big]^2 + \frac{1}{2}  \psi_1(1)+ \psi_1(\tfrac{d}{4}) - \frac{3}{2} \psi_1(\tfrac{d}{2}) - \, J_0(\tfrac{d}{4}) \, , \crcr
& \alpha_{I_1} \, = \, \frac{3}{2}\left[2\psi( \tfrac{d}{4} ) - \psi( \tfrac{d}{2})-\psi(1)\right]^2
 + \frac{1}{2} \psi_1(1) 
-\frac{1}{2}\psi_1(\tfrac{d}{2})
\crcr
& \alpha_{I_3} \, = \,\frac{\Gamma(-\tfrac{d}{4})\Gamma(\tfrac{d}{2})^2}{3 \, \Gamma( \tfrac{3d}{4})}\, , 
\crcr
& \alpha_{I_4} \, = \,\frac{\, \Gamma(1 + \tfrac{d}{4})^3\Gamma(- \tfrac{d}{4})}{
 \, \Gamma(\tfrac{d}{2} )} \; 6 \, \Big[  \psi_1(1) -  \psi_1(\tfrac{d}{4})  \Big]  \,,
 \label{eq:alphas3}
\end{align}
with $\psi_i$ the polygamma functions of order $i$ and $J_0$ the sum:
\begin{equation}
J_0(\tfrac{d}{4})=\frac{1}{\Gamma(\tfrac{d}{4})^2}\sum_{n \geq 1}\frac{\Gamma(n+\tfrac{d}{2})\Gamma(n+ \tfrac{d}{4})^2}{n(n!)\Gamma(\tfrac{d}{2}+2n)}\Big[2\psi(n+1)-\psi(n)-2\psi(n+\tfrac{d}{4})-\psi(n+\tfrac{d}{2})+2\psi(\tfrac{d}{2}+2n)\Big] \,.
\end{equation}
As we are interested in Wilson-Fisher-like fixed points, with $\gt$ of order $\epsilon$, we have expanded the constants $\alpha$ in $\epsilon$ up to a consistent order (such that the beta functions are written up to order $\epsilon^p \gt^q$ with $p+q= 4$).

Notice that $\alpha_{I_3} $ is the only constant blowing up for $d\to 4$. This is because it contains a (melonic/sunset) two-point subgraph that is finite for $d<4$, and thus not renormalized by a counterterm; however, in $d=4$ such subgraph is divergent, hence the singularity in $\alpha_{I_3} $.

\section{Applications}
\label{sec:appl}
The most general model \eqref{eq:actionms} has only $\mathbb{Z}_2$ symmetry, that is, invariance under simultaneous sign flip of all the fields.
However, we do not know how to solve the fixed point equations in full generality, and all the known interacting fixed points with $\cN\geq 2$ have a symmetry group strictly larger than $\mathbb{Z}_2$ \cite{Rychkov:2018vya}.
In this section, using the results obtained in the previous section, we will study the fixed-points of some specific models, characterized by their invariance under different symmetry groups. In particular we will only consider some of the models  which have been studied the most, at least in their short-range version, because of their physical interest.

\subsection{The long-range Ising model}
\label{sec:ising model}
The Ising model is the special case $N=1$ of the $O(N)$ vector model which we will discuss in the next subsection. We prefer to  discuss it separately
 in this subsection because of its physics and historical importance, and because it is the long-range model for which more results are available in the literature.

Setting $\cN=1$, $\gt_{\mba \mbb \mbc \mbd} =  \gt $ and $\rt_{\mba \mbb}=\rt$ in  \eqref{eq:beta_abcd_alpha3} and \eqref{eq:beta2_abcd_alpha3},  we find the beta functions:
	\begin{align}
		\beta^{(4)} \, &= \, - \epsilon \, \gt \, + \, 3 \, \alpha_D \, \gt^2 \, + \, 6 \, \alpha_S \, \gt^3
		\, + \, (3\alpha_T + 6 \alpha_U + 12\alpha_{I_1} + 6 \alpha_{I_2} + 3\alpha_{I_3} + \alpha_{I_4}) \gt^4 \, , \\
		\beta^{(2)} \, &= \, - (d-2\Delta_{\phi} ) \, \rt \, + \, \alpha_{D} \, \rt \, \gt \, + \, \alpha_{S} \, \rt \, \gt^2 \, + \, (\alpha_{T} + 2\alpha_{I_1} + \alpha_{I_2} + \alpha_{I_3}) \, \rt \, \gt^3  \, .
	\end{align}
Parametrizing the coefficients of the $\epsilon$ expansion  of the one- and two-loop constants $\alpha$ as: 
	\begin{align}
		\alpha_D \, &= \, 1\, + \, \alpha_{D,1} \, \epsilon \, + \, \alpha_{D,2} \, \epsilon^2 \, + \, \mathcal{O}(\epsilon^3) \, ,\crcr
		\alpha_S \, &= \, \alpha_{S,0} \, + \, \alpha_{S,1} \, \epsilon \,  + \, \mathcal{O}(\epsilon^2) \, ,
	\label{eq:alpha_param}
	\end{align}
we can solve for the fixed point coupling perturbatively in $\epsilon$, obtaining $\rt_\star=0$ and:
	\begin{align}
		\gt_{\star} \, &= \, \frac{\epsilon}{3} \, - \, \left( \frac{3\alpha_{D,1} + 2\alpha_{S,0}}{9} \right) \, \epsilon^2
		\, - \, \frac{\epsilon^3}{81} \bigg[ 27 \alpha_{D,2} -24 \alpha_{S,0}^2 + 9 (2\alpha_{S,1} - 3\alpha_{D,1}^2) \crcr
		&\hspace{80pt} - 54\alpha_{D,1} \alpha_{S,0} + 3\alpha_T + 6 \alpha_U + 12\alpha_{I_1} + 6 \alpha_{I_2} + 3\alpha_{I_3} + \alpha_{I_4} \bigg]
		\, + \, \mathcal{O}(\epsilon^4) \, .
	\end{align}
As the stability matrix is triangular, the stability exponents are simply given by:
\begin{align}
		\partial_{\gt}\beta^{(4)}(\gt_{\star}) \, &= \, \epsilon \, + \, \frac{2\alpha_{S,0}}{3} \, \epsilon^2
		\, + \, \frac{2\epsilon^3}{27} \bigg[ -12 \alpha_{S,0}^2 + 9 \alpha_{S,1} - 18\alpha_{D,1} \alpha_{S,0}\crcr
		&\hspace{120pt} + 3\alpha_T + 6 \alpha_U + 12\alpha_{I_1} + 6 \alpha_{I_2} + 3\alpha_{I_3} + \alpha_{I_4} \bigg]
		\, + \, \mathcal{O}(\epsilon^4) \, , \\
		\partial_{\rt}\beta^{(2)}(\gt_{\star}) \, &= \, - (d-2\Delta_{\phi} ) \, + \, \frac{\epsilon}{3} \, - \, \frac{\alpha_{S,0}}{9} \, \epsilon^2 \crcr
		&\quad \, - \, \frac{\epsilon^3}{81} \bigg[ -12 \alpha_{S,0}^2 + 9 \alpha_{S,1} 
		- 18 \alpha_{D,1} \alpha_{S,0} + 6 \alpha_U + 6\alpha_{I_1} + 3 \alpha_{I_2} + \alpha_{I_4} \bigg] \, + \, \mathcal{O}(\epsilon^4) \, .
\end{align}

As usual, the stability exponents are related to the critical exponents, which describe universal properties of  critical phenomena. The anomalous dimension $\eta$, the susceptibility exponent $\gamma$, and the correlation length exponent $\nu$ satisfy the scaling relation $\gamma= (2-\eta) \nu$, and  for the long-range models we have also $2-\eta = 2\zeta$.
Therefore, it suffices to consider $\nu$, which is given by:
	\begin{equation}
		\nu^{-1} \, = \, - \, \partial_{\rt}\beta^{(2)}(\gt_{\star}) \, ,
	\label{eq:nu}
	\end{equation}
and the correction-to-scaling exponent $\omega$, given by:
\be
\omega = \partial_{\gt}\beta^{(4)}(\gt_{\star}) \,.
\ee
Using \eqref{eq:alphas3} and $d-2\Delta_{\phi}=(d+\epsilon)/2=2\zeta$, the exponents are obtained as:
\begin{align}
	\omega\, &= \, \epsilon \,+ \, \frac{2}{3} \epsilon ^2   \Big[ 2\psi(\tfrac{d}{4}) - \psi(\tfrac{d}{2}) -\psi(1) \Big] \crcr
	& \quad +\, \frac{1}{18} \epsilon ^3 \bigg[ 13 \left(
   2\psi(\tfrac{d}{4})-\psi(1)  -\psi(\tfrac{d}{2})\right)^2 +3\left(\psi_1(1)- \psi_1(\tfrac{d}{2})\right) \crcr
   & \qquad  + \frac{8 \Gamma \left(-\frac{d}{4}\right) \left(\Gamma \left(\frac{d}{2}\right)^3+\Gamma \left(\frac{3
   d}{4}\right) \Gamma \left(\frac{d}{4}+1\right)^3 \left(\psi_1(1)- \psi_1(\tfrac{d}{4})\right)\right)}{\Gamma
   \left(\frac{d}{2}\right) \Gamma \left(\frac{3 d}{4}\right)} \bigg] \, + \, \mathcal{O}(\epsilon^4) \,, 
    \label{eq:omega-Ising}\\
		\nu^{-1} \, &= \, \frac{d}{2} \, + \, \frac{\epsilon}{6} \, + \, \frac{\epsilon^2}{9} \Big[ 2\psi(\tfrac{d}{4}) - \psi(\tfrac{d}{2}) -\psi(1) \Big] \,  \crcr
		&\quad + \, \frac{\epsilon^3}{108} \bigg[\psi_1(1)-\left(
   2\psi(\tfrac{d}{4})-\psi(1)  -\psi(\tfrac{d}{2})\right)^2-\psi_1(\tfrac{d}{2})\crcr 
		& \qquad +8\frac{\Gamma(1+\tfrac{d}{4})^3\Gamma(-\tfrac{d}{4})\left(\psi_1(1)-\psi_1(\tfrac{d}{4})\right)}{\Gamma(d/2)}\bigg] \, + \, \mathcal{O}(\epsilon^4) \, .
  \label{eq:nu-Ising}
\end{align}

Notice that $\alpha_{I_3}$ does not contribute to $\nu$, but it appears in $\omega$.
As mentioned earlier, $\alpha_{I_3}\to-\infty$ as we approach $d=4$, and therefore, while the coefficients of the $\epsilon$ expansion of $\nu$ remain small at this order, the three-loop coefficient for the dimension of the quartic operator becomes large in this limit. It has been argued in \cite{Honkonen:1990} that such growing coefficients provide an explanation of the transition from long-range to short-range behavior, which is supposed to happen when $2\zeta = 2-\eta_{\rm SR}$, where $\eta_{\rm SR}$ stands for the anomalous dimension of the short-range Ising model \cite{Sak:1973}.

In integer dimensions, $\nu^{-1}$ evaluates to:\footnote{We stress again that while short-range models cannot undergo a phase transition at $d=1$, the existence of a phase transition in the one-dimensional long-range Ising model with $0< 2\z < 1$ has been proved rigorously in \cite{Dyson:1968up}.}
\begin{align}
\nu^{-1} \, & = \, 1.5 + 0.1667\, \epsilon - 0.1812\, \epsilon^2 +  0.2633\, \epsilon^3 \,, & \text{at } \; d=3\,, \crcr
\nu^{-1} \, & = \,  1 + 0.1667\, \epsilon - 0.3081\, \epsilon^2 +  0.5301\, \epsilon^3 \,, & \text{at } \; d=2\,, \crcr
\nu^{-1} \, & = \,  0.5 + 0.1667\, \epsilon- 0.6571\, \epsilon^2 + 2.018\, \epsilon^3 \,, & \text{at } \; d=1\,.
\end{align}

In Table~\ref{tab:critexp} we compare the numerical values of $\nu$ in one dimension at different values of $\epsilon$ obtained from our loop expansion with those from numerical simulations \cite{Glumac:1989,Luijten:1997-thesis,uzelac2001critical,tomita2009monte}. Other numerical results can also be found in \cite{Luijten:1997} for the mean-field regime, and in \cite{Loscar:2018,rodriguez2009study,Xu:1993} for other values of $\z$ in one and two dimensions.
As the perturbative series is only asymptotic, a summation method must be employed for its quantitative use (as for example in \cite{Guida:1998bx}). We have used the most basic method, the Pad\'e-Borel summation (e.g.\ \cite{Kleinert:2001ax} and references therein), with which we have found an improvement with respect to the naive series (which we report for comparison), especially at larger values of $\epsilon$.
In Table~\ref{tab:critexp-omega} we provide a similar comparison for $\omega$ in one dimension with available numerical simulations \cite{Luijten:1997-thesis}.
\begin{table}[htb]
\begin{center}
\begin{tabular}{|c|c||c|c|c|c|c|c|c|c|}\hline
$\epsilon$ &  $2\zeta$  &  mean-field      &     three-loop   &    PB $[2/1]$ &   Ref.~\cite{Glumac:1989}        &     Ref.~\cite{Luijten:1997-thesis}  & Ref.~\cite{uzelac2001critical} & Ref.~\cite{tomita2009monte}                  \\ \hline
0.2 & 0.6 &1.6667    &     1.9015    &  1.937(16)   &     2.16        &     1.98      & 2.00(16) & 1.80(21)    \\
0.4 & 0.7 & 1.4286   &      1.5559        &  1.94(6)  &    2.123       &      2.01 & 1.96(15) & 1.83(17)               \\
0.6 & 0.8 & 1.25   &    0.4881  &  1.98(12)  &     2.208        &       2.17 & 2.13(18) & 1.89(5)              \\
\hline
\end{tabular}\end{center}
\caption{The critical exponent $\nu$ for the long-range Ising model at $d=1$, as computed in mean-field ($\nu=1/(2\z)$), by the naive three-loop series, and by a Pad\'e-Borel (PB) summation with $[2/1]$ approximant. The error in the latter is estimated by the difference with the PB summation of the two-loop series with $[1/1]$ approximant. The last four columns report numerical results from the literature for comparison.
}
\label{tab:critexp}
\end{table}
\begin{table}[htb]
\begin{center}
\begin{tabular}{|c|c||c|c|c|c|}\hline
$\epsilon$ &  $2\zeta$ &   one-loop         &     three-loop    &    PB $[2/1]$       &     Ref.~\cite{Luijten:1997-thesis}                    \\ \hline
0.2 & 0.6 & 0.2          &     0.35    &  0.139(17)          &     0.15         \\
0.4 & 0.7 & 0.4      &     2.20      &  0.24(6)         &      0.23                 \\
0.6 & 0.8 & 0.6       &    7.39   &  0.33(11)       &       0.25              \\
\hline
\end{tabular}\end{center}
\caption{The critical exponent $\omega$ for the long-range Ising model at $d=1$, as computed by one- and three-loop truncations and by a Pad\'e-Borel summation of the three-loop series with $[2/1]$ approximant, with errors estimated as before. The last column reports numerical results from \cite{Luijten:1997-thesis}  for comparison.
}
\label{tab:critexp-omega}
\end{table}

In Table \ref{tab:critexp2} we compare numerical values of $\nu$ in two dimensions at different values of $\epsilon$ obtained with our loop expansion and summation method with numerical simulations \cite{Angelini:2014}. The value $\epsilon=1.5$ is the maximum value of $\epsilon$ we consider because it corresponds to the transition between long-range and short-range behavior happening at $2\zeta=2-\eta_{SR}$ (and we indeed find a value of $\n$ consistent with the exact result in two dimensions, $\n_{SR}=1$). 

\begin{table}[htb]
\begin{center}
\begin{tabular}{|c|c||c|c|c|c|}\hline
$\epsilon$ &  $2\zeta$  & mean-field &     three-loop    &   PB $[2/1]$ &   Ref.~\cite{Angelini:2014}                         \\ \hline
0.4 & 1.2 &0.8333 &0.9463  & 0.966(6) &   0.977(34)     \\
0.61812 & 1.30906 & 0.7640& 0.8748 & 0.962(14) & 0.986(33)     \\
1.2 & 1.6 & 0.625& 0.1821 & 0.98(5) &  1.004(34)   \\
1.5 & 1.75 &0.5714 & -0.6457  & 0.99(7) &  1.02 (12) \\
\hline
\end{tabular}\end{center}
\caption{The critical exponent $\nu$ for the long-range Ising model at $d=2$, as computed in mean-field ($\nu=1/(2\z)$), by the naive three-loop series, and by a Pad\'e-Borel (PB) summation with $[2/1]$ approximant (with error estimated by the difference with the PB summation of the two-loop series with $[1/1]$ approximant). The last column reports numerical results from the literature for comparison.}
\label{tab:critexp2}
\end{table}

\paragraph{On the conjectured relation between short-range and long-range Ising models.}
The peculiar value $\epsilon=0.61812$ was considered in \cite{Angelini:2014} (and hence in our  Table \ref{tab:critexp2}) in order to test a conjecture relating the long-range Ising model at given dimension $d$, to the short-range Ising model at a different dimension $d_{SR}$ \cite{Banos:2012,Angelini:2014,Defenu:2014}. According to the conjecture, the effective dimension or effective $\z$ should be found from the relation:
 \be \label{eq:LR-SR-1}
 \f{2\zeta}{d}= \f{2-\eta_{SR}(d_{SR})}{d_{SR}}\,,
 \ee
 and one should then also see other relations among the critical exponents, such as: 
 \be \label{eq:LR-SR-2}
 d\,\nu(\z,d)=d_{SR}\, \nu_{SR}(d_{SR})\,, \quad \omega(\z,d)/d=\omega_{SR}(d_{SR})/d_{SR}\,.
 \ee
Therefore, the short-range model at $d_{SR}=3$ should be related to the long-range model at $d=2$ if one takes  $2\zeta=1.30906$, or equivalently $\epsilon=0.61812$.
It should be said that there is no compelling evidence in favor of this conjecture and actually we will now show that it is only an approximate relation valid close to the upper critical dimensions ($d_{SR}\simeq 4$ and $d\simeq 4\z$), as also observed numerically in \cite{Angelini:2014} and remarked in \cite{Paulos:2015jfa,Behan:2017emf}.
Close to the upper critical dimensions, equation \eqref{eq:LR-SR-1} can be solved perturbatively.
Consider indeed the long-range Ising model at arbitrary $d<4$ and with long-range exponent \eqref{eq:zeta-eps}, for small $\epsilon$; next, consider the short-range Ising model in $d_{SR}=4-\epsilon_{SR}$, for small $\epsilon_{SR}$. Using the known expansion of $\eta_{SR}(4-\epsilon_{SR})$ (e.g.\ \cite{Kleinert:2001ax}), from \eqref{eq:LR-SR-1} we obtain:
\be
\f{\epsilon}{d} = \f{\epsilon_{SR}}{4} + \f{23 \epsilon_{SR}^2}{432} + \f{ 185 \epsilon_{SR}^3}{46656} +\cO(\epsilon_{SR}^4) \,.
\ee
Plugging this expression into \eqref{eq:omega-Ising} and \eqref{eq:nu-Ising}, and using  \eqref{eq:LR-SR-2} in order to determine $\nu_{SR}(4-\epsilon_{SR})$ and $\omega_{SR}(4-\epsilon_{SR})$, we find that they agree with the known short-range exponents up to order $\epsilon_{SR}$, but they disagree already at order $\epsilon_{SR}^2$, with in particular some surviving polygamma functions that should not appear in the short-range case.
We conclude that the relations \eqref{eq:LR-SR-1} and \eqref{eq:LR-SR-2} can at best be approximate and qualitative, in particular they can be trusted as long as one can trust the one-loop approximation, for which they are exact.
In fact, from a numerical point of view, the results of the $d=2$ long-range Ising model with $\epsilon=0.61812$ and the $d_{SR}=3$ short-range one are quite close but not in perfect agreement.
Comparing the short-range results from the literature \cite{El-Showk:2014dwa}, $3\nu_{SR}(3)\simeq 1.88997(15)$, with our resummed result $2\nu_{LR}\simeq 1.924(28)$, we find that the values are approximately compatible with equation \eqref{eq:LR-SR-2}, within the given errors. 
On the other hand, from the Pad\'e-Borel summation of our three-loop result  at $d=2$ we find that $\omega/2\simeq 0.20(4)$, which is slightly off from  $\omega_{SR}(3)/3 \simeq 0.2767(6)$ obtained again from \cite{El-Showk:2014dwa}.

\subsection{The long-range $O(N)$ vector model}
\label{sec:vector model}

For general $\cN=N$, the largest possible symmetry group is $O(N)$, which reduces the number of couplings 
to just one, corresponding to the unique quartic invariant $(\phi_{\mba}\phi_{\mba})^2$.
The long-range quartic $O(N)$ model is defined by the choice of couplings:
	\begin{equation} \label{eq:g-vector}
		\gt_{\mba \mbb \mbc \mbd} \, = \, \frac{\gt}{3} \, \big( \delta_{ab} \delta_{cd} + \delta_{ac} \delta_{bd} + \delta_{ad} \delta_{bc} \big) \, , \qquad
		\rt_{\mba \mbb} \, = \, \rt \, \delta_{ab}  \, ,
	\end{equation}
where we have identified $\mba=a=1,\ldots ,N$, and so on.
By substituting \eqref{eq:g-vector} in \eqref{eq:beta_abcd_alpha3} and \eqref{eq:beta2_abcd_alpha3}, we find the beta functions:
	\begin{align}
		\beta^{(4)} \, &= \, - \epsilon \, \gt \, + \, \frac{\alpha_{D}}{3} (N+8) \gt^2 \, + \, \frac{2\alpha_{S}}{9} (5N + 22) \gt^3
		\,  \crcr
		&\quad+ \, \Big[ (3N^2 +22N + 56)(2\alpha_{I_2}+ \alpha_{T}) + 2(N^2+20N+60)(2\alpha_{I_1}+ \alpha_{U})\crcr
		&\qquad  + 3( N+8)(N+2) \alpha_{I_3} + (5N +22) \alpha_{I_4} \Big] \frac{\gt^4}{27} \, , \\
		\beta^{(2)} \, &= \, - (d-2\Delta_{\phi} ) \, \rt \, + \, \frac{\alpha_{D}}{3} \, (N+2) \, \rt \, \gt \, + \, \frac{\alpha_{S}}{3} \, (N+2) \, \rt \, \gt^2\crcr
		&\quad + \, \frac{(N+2)}{27} \, \Big( 3(N+2) (\alpha_{T} + \alpha_{I_3}) + (N+8) (2\alpha_{I_1} + \alpha_{I_2} ) \Big) \, \rt \, \gt^3  \, .
	\end{align}
Besides the trivial fixed point $\gt_{\star}=\rt_\star=0$, we also find a (perturbative in $\epsilon$) non-trivial fixed point with $\rt_\star=0$ and:
	\begin{align}
		\gt_{\star} \, &= \, \frac{3\epsilon}{N+8}
		\, - \, \frac{3\epsilon^2}{(N+8)^3} \, \Big[ (N+8)^2\alpha_{D,1} + 2(5N+22)\alpha_{S,0} \Big] \crcr
		&\quad + \, \frac{3\epsilon^3}{(N+8)^5} \bigg[ (N+8)^2 \Big( (N+8)^2\alpha_{D,1}^2 - 2(5N +22)( \alpha_{S,1} -3 \alpha_{S,0} \alpha_{D,1})-3(N+2)\alpha_{I_3} \Big)\crcr
		&\hspace{60pt} - (N+8) \Big( (3N^2 +22N + 56) (2\alpha_{I_2}+\alpha_{T}) + 2(N^2+20N+60)(2\alpha_{I_1}+ \alpha_{U}) \crcr
		&\hspace{60pt}   + (5N +22) \alpha_{I_4} \Big)  -(N+8)^4 \alpha_{D,2} + 8(5N +22)^2 \alpha_{S,0}^2  \bigg]
		\, + \, \mathcal{O}(\epsilon^4) \, ,
	\end{align}
where we used the parametrization \eqref{eq:alpha_param}.

This is the long-range version of the Wilson-Fisher fixed point, describing several important universality classes: self-avoiding walks ($N=0$), Ising  ($N=1$), XY  ($N=2$), Heisenberg  ($N=3$), and spherical  ($N=\infty$); see for example  \cite{Pelissetto:2000ek} for a review of their short-range version.

The long-range critical exponents are given by:
	\begin{align} \label{eq:omega-ON}
		\omega \equiv \partial_{\gt}\beta^{(4)}(\gt_{\star}) \, &= \, \epsilon \, + \, \frac{2(5N +22)\alpha_{S,0}}{(N+8)^2} \, \epsilon^2
		\, + \, \frac{2\epsilon^3}{(N+8)^4} \bigg[ -4 (5N +22)^2\alpha_{S,0}^2 \crcr
	         & \qquad+ (N+8)^2 (5N +22) (\alpha_{S,1} -2 \alpha_{D,1}\alpha_{S,0}) \crcr
		&\qquad + (N+8) \Big( (3N^2 +22N + 56)(2\alpha_{I_2}+ \alpha_{T}) \crcr
		& \hspace{70pt} + 2(N^2+20N+60)(2\alpha_{I_1}+ \alpha_{U} )\crcr
		&\hspace{70pt} + 3(N+8)(N+2) \alpha_{I_3} + (5N +22) \alpha_{I_4} \Big) \bigg]
		\, + \, \mathcal{O}(\epsilon^4)\, , \\
		 \label{eq:nu-ON}
		\nu^{-1}\equiv - \partial_{\rt}\beta^{(2)}(\gt_{\star}) \, &= \, 
		 2\zeta \, - \, \frac{(N+2)}{N+8} \, \epsilon \, + \, \frac{(N+2)(7N+20)\alpha_{S,0}}{(N+8)^3} \, \epsilon^2 \crcr
		&\quad + \, \frac{(N+2)\epsilon^3}{(N+8)^5} \bigg[ -4(5N+22)(7N+20) \alpha_{S,0}^2 \crcr
		& \quad \qquad + (N+8)^2(7N+20) (\alpha_{S,1} -2 \alpha_{D,1} \alpha_{S,0}) \crcr
		& \quad \qquad+ (N+8) \Big(-8(N-1) \alpha_{T} + 2(N^2+20N+60) \alpha_U \crcr
		& \hspace{70pt}+ 2(N^2+24N+56) \alpha_{I_1} \crcr
		&\hspace{70pt}+ (5N^2+28N+48) \alpha_{I_2} + (5N +22) \alpha_{I_4} \Big) \bigg] \, + \, \mathcal{O}(\epsilon^4)
		 \, . 
	\end{align}

We can compute the critical value of $N$, $N_c$, at which $\omega$ vanishes and the Wilson-Fisher fixed point becomes marginal. At first order in $\epsilon$, we find:
\begin{equation}
N_c=-8 \pm 6\sqrt{2|\alpha_{S,0}|}\epsilon^{1/2}+\mathcal{O}(\epsilon) \; .
\end{equation}
At small $\epsilon$ this corresponds to a negative value of $N$, hence for the bosonic model the quartic operator never becomes relevant. However,  it can cross marginality for symplectic fermions, whose perturbative series is related to the bosonic one by changing sign of $N$ (see for example \cite{Giuliani:2020aot} where the renormalization group of a long-range model of symplectic fermions was studied).

The two-loop results for $\nu$ and $\omega$ agree with those first reported in \cite{Fisher:1972zz} and \cite{Yamazaki:1977pt}, respectively, while the three-loop contributions are new. At low $N$ and integer $d$, $\nu^{-1}$ is:
\begin{align}
\nu^{-1}&=0.5+0.1\epsilon-0.8043\epsilon^2+2.002\epsilon^3 \,, \quad (d=1, N=2 ) \,, \crcr
\nu^{-1}&=1+0.1\epsilon-0.3771\epsilon^2+0.5306\epsilon^3 \,, \quad ( d=2, N=2 ) \,, \crcr
\nu^{-1}&=1.5+0.1\epsilon-0.2218\epsilon^2+0.2672\epsilon^3 \,, \quad ( d=3 , N=2 ) \,, \crcr
\nu^{-1}&=0.5+0.04545\epsilon-0.9109\epsilon^2+1.963\epsilon^3 \,, \quad ( d=1, N=3 ) \,, \crcr
\nu^{-1}&=1+0.04545\epsilon-0.4270\epsilon^2+0.5212\epsilon^3 \,, \quad ( d=2, N=3 ) \,, \crcr
\nu^{-1}&=1.5+0.04545\epsilon-0.2512\epsilon^2+0.2632\epsilon^3 \,, \quad ( d=3, N=3 ) \,.
\end{align}

In Table \ref{tab:critvect22} we report the numerical values of $\nu$ and $\omega$ in two dimensions at $N=2$ and in three dimensions at $N=2,3$ for different values of $\epsilon$. 

\begin{table}[htb]
\begin{center}
\begin{tabular}{|c|c|c||c|c|c| |c|c|c| }\hline
\multirow{2}{*}{$(d,N)$ } & \multirow{2}{*} {$\epsilon$} &  \multirow{2}{*}{$2\zeta$} &  \multicolumn{3}{c||}{$\nu$}  
& 
 \multicolumn{2}{c|}{ $\omega$ }\\
& & & mean-field  &     three-loop    &    PB $[2/1]$  & three-loop &  PB $[2/1]$\\ \hline
\multirow{3}{*}{(2,2)}
& 0.2 & 1.1 &0.9091     &     0.9906   &  0.992(4)  & 0.1945 & 0.160(8)\\
& 0.4 & 1.2 & 0.8333    &   0.9831      &  1.000(18) & 0.6399 & 0.287(35)\\
& 0.6 & 1.3 & 0.7692  &    0.9482   &  1.02(4)  & 1.729 & 0.40(7)\\
\hline
\multirow{3}{*}{(3,2)} 
& 0.2 & 1.6 &0.625  &     0.6608   &  0.6610(7)  & 0.1835 & 0.173(5) \\
& 0.4 & 1.7 & 0.5882 &      0.6567      &  0.6600(35)  & 0.4350 & 0.317(25)\\
& 0.6 & 1.8 & 0.5556  &    0.6480   &  0.662(8)   &  0.9061 & 0.45(6)\\
\hline
\multirow{3}{*}{(3,3)} 
& 0.2 & 1.6 &0.625    &     0.6661   &  0.6663(18)   & 0.1832 & 0.174(5)\\
& 0.4 & 1.7 & 0.5882 &      0.6686      &  0.671(8)  & 0.4251 & 0.319(24)\\
& 0.6 & 1.8 & 0.5556 &    0.6682   &  0.680(17)   & 0.8642  &  0.45(5) \\
\hline
\end{tabular}\end{center}
\caption{The critical exponents $\nu$ and $\omega$ for the long-range vector model at various $d$ and $N$ computed in mean-field ($\nu=1/(2\z)$), by the naive three-loop series, and by a Pad\'e-Borel (PB) summation with $[2/1]$ approximant (with error estimated by the difference with the PB summation of the two-loop series with $[1/1]$ approximant).}
\label{tab:critvect22}
\end{table}

\paragraph{Large-$N$.}
Next, we consider the $1/N$ expansion, which can be obtained either by rescaling the coupling by $\gt \to \bar{g}/N$ and following the usual large-$N$ analysis from the beginning, or by expanding the finite-$N$ result \eqref{eq:omega-ON} and \eqref{eq:nu-ON}.
Up to order $\mathcal{O}(N^{-2})$, the critical exponents are given by:
	\begin{align}
		\omega \, &= \, \epsilon   \, + \, \frac{2\epsilon^2}{ N} \Big[ 5 \alpha_{S,0}
		+ (3\alpha_{T} + 2\alpha_{U} + 4\alpha_{I_1} + 6\alpha_{I_2} + 3 \alpha_{I_3}+5(\alpha_{S,1}-2\alpha_{D,1}\alpha_{S,0}) ) \epsilon \Big] \crcr
		&\quad - \frac{2\epsilon^2}{N^2}\Big[ 58\alpha_{S,0}+\big(16\alpha_{I_1}+100\alpha_{I_2}+42\alpha_{I_3}-5\alpha_{I_4}+58(\alpha_{S,1}-2\alpha_{D,1}\alpha_{S,0})\crcr
& \qquad \quad +100\alpha_{S,0}^2+50\alpha_T+8\alpha_U\big)\epsilon\Big] \, + \, \mathcal{O}(N^{-3},\epsilon^4)  \, , \\
		\nu^{-1}\, & = \, 2\Delta_{\phi}
		\, + \, \frac{\epsilon}{ N} \, \Big[ 6 + 7 \alpha_{S,0} \epsilon
		+ \big(2\alpha_{U} + 2\alpha_{I_1} + 5\alpha_{I_2}+7(\alpha_{S,1}-2\alpha_{D,1}\alpha_{S,0})\big) \epsilon^2\Big]\crcr
		& \quad - \frac{\epsilon}{N^2}\,\Big[48+134\alpha_{S,0}\epsilon +\big(12\alpha_{I_1}+122\alpha_{I_2}-5\alpha_{I_4}+134(\alpha_{S,1}-2\alpha_{D,1}\alpha_{S,0})\crcr
&\qquad \quad +140\alpha_{S,0}^2+8\alpha_{T}+20\alpha_U\big)\epsilon^2 \Big] \, + \, \mathcal{O}(N^{-3},\epsilon^4) \, .
	\end{align}
The leading-order term of $\nu^{-1}$ (where we used $2\z-\epsilon =d-2\Delta_{\phi} -\epsilon =  2\Delta_{\phi}$) gives the spherical model result $\g= 2\z \nu = 2\zeta/(d-2\z)$, obtained first in \cite{Joyce:1966}.
The order $1/N$ contribution to $\nu$ was computed to all orders in $\epsilon$ in  \cite{Fisher:1972zz} and we reproduce it up to three loops.
The order $1/N^2$ contribution is new.

\subsection{The long-range cubic model}
\label{sec:cubic}

The next model we consider is obtained by breaking explicitly the $O(N)$ symmetry with an interaction of the form $\sum_{\mba} \phi_{\mba}^4$.
As the four fields share the same index, the continuous symmetry is completely broken; however, we can still independently flip the sign of any component, or permute them. In other words, we are left with the (hyper-)cubic symmetry group $(\mathbb{Z}_2)^N \rtimes S_N$.
For reviews of the short-range model with the same symmetry group, see \cite{Aharony:1976br,Pelissetto:2000ek,Kleinert:2001ax}; for some of the most recent results, using renormalization or bootstrap methods, see \cite{Stergiou:2018gjj,Kousvos:2018rhl,Kousvos:2019hgc,Adzhemyan:2019gvv,Antipin:2019vdg} and references therein. 
The long-range version of the model has been mostly unexplored, and we are only aware of the two-loop calculations of \cite{Yamazaki:1978-cubic,Yamazaki:1981-cubic,Chen:2001}.

The model is defined by the following choice of coupling $g_{\mba\mbb\mbc\mbd}$:
\begin{equation}
g_{\mba\mbb\mbc\mbd}=\frac{g_d}{3}\left(\delta_{ab}\delta_{cd}+\delta_{ac}\delta_{bd}+\delta_{ad}\delta_{bc}\right)
+g_c \delta_{ab}\delta_{ac}\delta_{ad} \,.
\label{eq:cubic_coupling}
\end{equation}
Substituting \eqref{eq:cubic_coupling} into \eqref{eq:beta_abcd_alpha3} and \eqref{eq:beta2_abcd_alpha3}, we find the beta functions up to three loops:
\begin{align}
\beta^{(4)}_d= & -\epsilon \tilde{g}_d+\frac{\alpha_{D}}{3}\left[6\tilde{g}_c+(N+8)\tilde{g}_d\right]\tilde{g}_d+\frac{2\alpha_{S}}{9}\left[9\tilde{g}_c^2+36\tilde{g}_c\tilde{g}_d+(5N+22)\tilde{g}_d^2\right]\tilde{g}_d\crcr
&  +2\Big[2\alpha_{I_1}+\alpha_{I_2}+\alpha_{I_3}+\alpha_{T}\Big]\tilde{g}_d\tilde{g}_c^3 +\frac{1}{3}\Big[52\alpha_{I_1}+30\alpha_{I_2}+(N+20)\alpha_{I_3}\crcr
&  +(N+18)\alpha_{T}+2\alpha_{I_4}+16\alpha_{U}\Big]\tilde{g}_d^2\tilde{g_c}^2  + \frac{2}{9}\Big[4(2N+27)\alpha_{I_1}+(9N+52)\alpha_{I_2}\crcr
&  +6(N+5)\alpha_{I_3}+8\alpha_{I_4}+2(3N+13)\alpha_{T}+4(N+12)\alpha_{U}\Big]\tilde{g}_d^3\tilde{g}_c \crcr
&+\frac{1}{27}\Big[2(N^2+20N+60)(2\alpha_{I_1}+\alpha_{U})+(3N^2+22N+56)(2\alpha_{I_2}+\alpha_{T})\crcr
& \qquad +3(N+2)(N+8)\alpha_{I_3}+(5N+22)\alpha_{I_4}  \Big]\tilde{g}_d^4  \, ,\\
\beta^{(4)}_c=& -\epsilon \tilde{g}_c+\alpha_{D}\left[3\tilde{g}_c+4\tilde{g}_d\right]\tilde{g}_c+\frac{2\alpha_{S}}{3}\left[9\tilde{g}_c^2+24\tilde{g}_c\tilde{g}_d+(N+14)\tilde{g}_d^2\right]\tilde{g}_c \crcr
& +\Big[12\alpha_{I_1}+6\alpha_{I_2}+3\alpha_{I_3}+\alpha_{I_4}+3\alpha_{T}+6\alpha_{U}\Big]\tilde{g}_c^4\crcr
& +2\Big[22\alpha_{I_1}+11\alpha_{I_2}+5\alpha_{I_3}+2\alpha_{I_4}+5\alpha_{T}+12\alpha_{U}\Big]\tilde{g}_c^3\tilde{g}_d \crcr
& +\frac{1}{3}\Big[4(N+40)\alpha_{I_1}+6(N+12)\alpha_{I_2}+3(N+10)\alpha_{I_3}\crcr
& \qquad +16\alpha_{I_4}+(N+34)\alpha_{T}+4(N+22)\alpha_{U}\Big]\tilde{g}_c^2\tilde{g}_d^2 \crcr
&+\frac{2}{27}\Big[12(3N+22)\alpha_{I_1}+3(N^2+6N+40)\alpha_{I_2}+18(N+2)\alpha_{I_3}\crcr
& \qquad +2(N+14)\alpha_{I_4}+6(N+10)\alpha_{T}+24(N+6)\alpha_{U}\Big]\tilde{g}_c\tilde{g}_d^3 \, , \\
\beta^{(2)}=& -(d-2\Delta_{\phi})\tilde{r}+\alpha_{D}\Big[\tilde{g}_c+\frac{\tilde{g}_d}{3}(N+2)\Big]\tilde{r}+\frac{\alpha_{S}}{3}\Big[3\tilde{g}_c^2+6\tilde{g}_c\tilde{g}_d+(N+2)\tilde{g}_d^2\Big]\tilde{r}\crcr
&+\Big[2\alpha_{I_1}+\alpha_{I_2}+\alpha_{I_3}+\alpha_{T}\Big]\tilde{g}_c^3\tilde{r}+\frac{1}{3}\Big[18\alpha_{I_1}+9\alpha_{I_2}+(N+8)(\alpha_{I_3}+\alpha_{T})\Big]\tilde{g_c^2}\tilde{g}_d\tilde{r} \crcr
& +\Big[(N+8)(2\alpha_{I_1}+\alpha_{I_2})+3(N+2)(\alpha_{I_3}+\alpha_{T})\Big]\Big[\frac{N+2}{27}\tilde{g}_d+\frac{1}{3}\tilde{g}_c\Big]\tilde{g}_d^2\tilde{r} \, .
\end{align}
Using \eqref{eq:alpha_param} and expanding in $\epsilon$, we find three non-trivial fixed points.

 \paragraph{Heisenberg fixed point.} The first fixed point, with $g_c=0$, is the $O(N)$-symmetric Heisenberg one:
\begin{align}
\tilde{g}_d^{\star,H}&=\frac{3\epsilon}{N+8}-\frac{3\epsilon^2}{(N+8)^3}\left[(N+8)^2\alpha_{D,1}+2(5N+22)\alpha_{S,0}\right] \crcr
&+\frac{3\epsilon^3}{(N+8)^4}\Big[-2(N^2+20N+60)(2\alpha_{I_1}+\alpha_{U})-(3N^2+22N+56)(2\alpha_{I_2}+\alpha_{T}) \crcr
&-3(N+8)(N+2)\alpha_{I_3}-(5N+22)\alpha_{I_4} +(N+8)^3(\alpha_{D,1}^2-\alpha_{D,2})\crcr
&-2(N+8)(5N+22)(\alpha_{S,1}-3\alpha_{D,1}\alpha_{S,0})+8\frac{(5N+22)^2}{N+8}\alpha_{S,0}^2\Big]+\mathcal{O}(\epsilon^4)\,,\crcr
\tilde{g}_c^{\star,H}&=0 \,. 
\end{align}
\paragraph{Ising fixed point.} The second one, with $g_d=0$, is the Ising fixed point:
\begin{align}
\tilde{g}_d^{\star,I}&=0 \,, \crcr
 \tilde{g}_c^{\star,I}&=\frac{\epsilon}{3}-\frac{\epsilon^2}{9}\Big[3\alpha_{D,1}+2\alpha_{S,0}\Big] +\frac{\epsilon^3}{81}\Big[-3(4\alpha_{I_1}+2\alpha_{I_2}+\alpha_{I_3})+27(\alpha_{D,1}^2-\alpha_{D,2}+2\alpha_{D,1}\alpha_{S,0})\crcr
 & \qquad \quad-\alpha_{I_4}+6(4\alpha_{S,0}^2-3\alpha_{S,1})-3(\alpha_{T}+2\alpha_{U})\Big]+ \mathcal{O}(\epsilon^4)  \,.
\end{align}
\paragraph{Cubic fixed point.} The last fixed point, with both couplings non-zero, is the cubic fixed point:
\begin{align}
\tilde{g}_d^{\star,C}&=\frac{\epsilon}{N}+\frac{\epsilon^2}{3N^3}\Big[2(N-1)(N-6)\alpha_{S,0}-3N^2\alpha_{D,1}\Big]  \crcr
& -\frac{\epsilon^3}{27N^5}\Big[-12N(N^3-6N^2+2N+4)\alpha_{I_1}-6N(N^3-9N^2+14N-8)\alpha_{I_2}\crcr
& +3N^2(N-1)(N+2)\alpha_{I_3}-N(2N^3-6N^2-7N+14)\alpha_{I_4}-27N^4(\alpha_{D,1}^2-\alpha_{D,2})\crcr
&+18N^2(N-1)(N-6)(3\alpha_{D,1}\alpha_{S,0}-\alpha_{S,1})+24(N-1)(N^3+5N^2-40N+36)\alpha_{S,0}^2\crcr
& +3N(N^3+3N^2-10N+8)\alpha_{T}-6N(2N^3-9N^2+4N+4)\alpha_{U}\Big] + \mathcal{O}(\epsilon^4) \,, \crcr
 \tilde{g}_c^{\star,C}&=\frac{\epsilon(N-4)}{3N}-\frac{\epsilon^2}{9N^3}\left(3N^2(N-4)\alpha_{D,1}+2(N-1)(N^2+6N-24)\alpha_{S,0}\right) \crcr
 &-\frac{\epsilon^3}{81N^5}\Big[12N(N^2-2N-2)(N^2+6N-8)\alpha_{I_1}+6N(N^4+7N^3-46N^2+64N-32)\alpha_{I_2}\crcr
 & +3N^2(N-1)(N+2)(N-4)\alpha_{I_3}+N(N+2)(N^3+6N^2-32N+28)\alpha_{I_4}\crcr
 & -27N^4(N-4)(\alpha_{D,1}^2-\alpha_{D,2})-18N^2(N-1)(N^2+6N-24)(3\alpha_{D,1}\alpha_{S,0}-\alpha_{S,1})\crcr
 & -24(N-1)(N^4+5N^3+28N^2-172N+144)\alpha_{S,0}^2\crcr
 & +3N(N^4-5N^3-10N^2+40N-32)\alpha_{T}\crcr
 & +6N(N^4+10N^3-40N^2+16N+16)\alpha_{U}\Big] + \mathcal{O}(\epsilon^4) \,. 
\end{align}

In the limit $N \rightarrow \infty$, the cubic and Ising fixed points are equal. For $N=2$, the cubic and Ising fixed points verify:
\begin{equation}
\tilde{g}_d^{\star,C}=\tilde{g}_d^{\star,I}+\frac{3}{2}\tilde{g}_c^{\star,I} \,, \; ~~ \tilde{g}_c^{\star,C}=-\tilde{g}_c^{\star,I} \, ,
\end{equation}
which was first noticed  in the short-range case in \cite{Kleinert:2001ax}.

The critical exponents are the eigenvalues of the stability matrix. As the latter has a block triangular structure, the correction-to-scaling exponents $\omega_d$ and $\omega_c$ are the eigenvalues of the reduced stability matrix $\partial (\beta_s, \beta_d)/\partial (\gt_s, \gt_d)|_{\gt=\gt_\star}$, while the correlation length exponent is simply obtained from $\partial_{\rt} \beta^{(2)}|_{\gt=\gt_\star,\rt=0}$. 
They are given by the following expressions for the three non-trivial fixed points.

 \paragraph{Heisenberg fixed point.} For the Heisenberg fixed point we have:
\begin{align}
\omega_d^H=\, &\epsilon+\frac{2\epsilon^2(5N+22)\alpha_{S,0}}{(N+8)^2}\crcr
&+\frac{2\epsilon^3}{(N+8)^3}\Big[2(N^2+20N+60)(2\alpha_{I_1}+\alpha_{U})+(3N^2+22N+56)(2\alpha_{I_2}+\alpha_{T})\crcr
&\quad+3(N+8)(N+2)\alpha_{I_3}+(5N+22)\alpha_{I_4}-(N+8)(5N+22)(2\alpha_{D,1}\alpha_{S,0}-\alpha_{S,1})\crcr
&\quad-4\frac{(5N+22)^2}{N+8}\alpha_{S,0}^2\Big]+\mathcal{O}(\epsilon^4)\,, \crcr
\omega_c^H=\, &-\epsilon\frac{N-4}{N+8}+\frac{6\epsilon^2(N^2+2N+24)\alpha_{S,0}}{(N+8)^3} \crcr
& +\frac{2\epsilon^3}{(N+8)^4}\Big[12(N^2+6N+56)\alpha_{I_1}+3(N^3+2N^2+96)\alpha_{I_2}+2(N^2+7N+46)\alpha_{I_4}\crcr 
&\quad-12\frac{(5N+22)(N^2+2N+24)}{N+8}\alpha_{S,0}^2+3(N+8)(N^2+2N+24)(\alpha_{S,1}-2\alpha_{D,1}\alpha_{S,0})\crcr
&\quad-12(N^2+2N-12)\alpha_{T}+12(N^2+8N+36)\alpha_{U}\Big] +\mathcal{O}(\epsilon^4) \,, \crcr
\nu_H^{-1}=\, & 2\zeta 
-\epsilon\frac{N+2}{N+8}+\epsilon^2\frac{(N+2)(7N+20)\alpha_{S,0}}{(N+8)^3}\crcr
&-\frac{\epsilon^3(N+2)}{(N+8)^4}\Big[2(N^2+24N+56)\alpha_{I_1}+(5N^2+28N+48)\alpha_{I_2}+(5N+22)\alpha_{I_4}\crcr
&\quad -\frac{4(5N+22)(7N+20)}{N+8}\alpha_{S,0}^2+(7N+20)(\alpha_{S,1}-2\alpha_{D,1}\alpha_{S,0})-8(N-1)\alpha_{T}\crcr
&\quad +2(N^2+20N+60)\alpha_{U} \Big] +\mathcal{O}(\epsilon^4)\,.  
\end{align}
\paragraph{Ising fixed point.} For the Ising fixed point we have:
\begin{align}
\omega_d^I=\, &-\frac{\epsilon}{3}-\frac{2\epsilon^2\alpha_{S,0}}{9}-\frac{2\epsilon^3}{81}\Big[6\alpha_{I_1}+3\alpha_{I_2}+\alpha_{I_4}-6\alpha_{S,0}(3\alpha_{D,1}+2\alpha_{S,0})\crcr
& \qquad +9\alpha_{S,1}+6\alpha_{U}\Big]+\mathcal{O}(\epsilon^4) \,, \crcr 
\omega_c^I=\, &\epsilon+\frac{2\epsilon^2\alpha_{S,0}}{3} +\frac{2\epsilon^3}{27}\Big[12\alpha_{I_1}+6\alpha_{I_2}+3\alpha_{I_3}+\alpha_{I_4}-6\alpha_{S,0}(3\alpha_{D,1}+2\alpha_{S,0})\crcr
& \qquad +9\alpha_{S,1}+3\alpha_{T}+6\alpha_{U}\Big] +\mathcal{O}(\epsilon^4) \,,\crcr 
\nu_I^{-1}=\, & 2\zeta 
-\frac{\epsilon}{3}+\epsilon^2\frac{\alpha_{S,0}}{9}-\frac{\epsilon^3}{81}\Big[-6\alpha_{I_1}-3\alpha_{I_2}-\alpha_{I_4} +6\alpha_{S,0}(3\alpha_{D,1}+2\alpha_{S,0})\crcr
& \qquad -9\alpha_{S,1}-6\alpha_{U}\Big] +\mathcal{O}(\epsilon^4) \,. 
\end{align}
\paragraph{Cubic fixed point.} For the cubic fixed point we have:
\begin{align}
\omega_d^C=\, &\epsilon+\frac{2\epsilon^2(N-1)(N^2+12)\alpha_{S,0}}{3N^2(N+2)}\crcr
& +\frac{2\epsilon^3}{27N^2(N+2)}\Big[12(N^4-2N^3+14N^2-8N-8)\alpha_{I_1}\crcr
& +6(N^4+N^3+8N^2-20N+16)\alpha_{I_2}\crcr
&\quad +3N(N+2)^2(N-1)\alpha_{I_3}+(N^4-4N^3+16N^2+6N-28)\alpha_{I_4}\crcr
&\quad+9N(N-1)(N^2+12)(\alpha_{S,1}-2\alpha_{D,1}\alpha_{S,0})\crcr
&\quad -12\frac{(N-1)(N^6+8N^5-14N^4+4N^3+384N^2-176N-288)}{N(N+2)^2}\alpha_{S,0}^2\crcr
&\quad +3(N^4+N^3+8N^2-20N+16)\alpha_{T} +6(N^4-2N^3+14N^2-8N-8)\alpha_{U}\Big]+\mathcal{O}(\epsilon^4) \,,\crcr
\omega_c^C=\, &\epsilon\frac{N-4}{3N}+\frac{2\epsilon^2(N-1)(N^3-4N^2-36N+48)\alpha_{S,0}}{9N^3(N+2)} \crcr
& +\frac{2(N-1)\epsilon^3}{81N^4(N+2)}\Big[6(N^4+2N^3-68N^2+72N+32)\alpha_{I_1}\crcr
&\quad +3(N^4-4N^3-44N^2+96N-64)\alpha_{I_2} +(N^4-N^3-44N^2+24N+56)\alpha_{I_4}\crcr
&\quad +9N(N^3-4N^2-36N+48)(\alpha_{S,1}-2\alpha_{D,1}\alpha_{S,0})\crcr
&\quad -12\frac{N^7+22N^5-180N^4-864N^3+1616N^2+800N-1152}{N(N+2)^2}\alpha_{S,0}^2\crcr
&\quad +12(N-2)(N^2-2N+4)\alpha_{T} +6(N^4-2N^3-40N^2+40N+16)\alpha_{U}\Big]+\mathcal{O}(\epsilon^4) \,,\crcr
\nu_C^{-1}=\, & 2\zeta 
-2\epsilon\frac{N-1}{3N}-\epsilon^2\frac{(N-1)(N^2-18N+24)\alpha_{S,0}}{9N^3}\crcr
& -\frac{\epsilon^3(N-1)}{81N^5}\Big[6N(N^3-14N^2+8N+16)\alpha_{I_1}+3N(N^3-26N^2+56N-32)\alpha_{I_2}\crcr
&\quad +N(N+2)(N^2-11N+14)\alpha_{I_4}+9N^2(N^2-18N+24)(\alpha_{S,1}-2\alpha_{S,0}\alpha_{D,1})\crcr
&\quad -12(N-1)(N^3+4N^2-100N+144)\alpha_{S,0}^2+24N(N-2)\alpha_{T}\crcr
&\quad +6N(N^3-14N^2+12N+8)\alpha_{U}\Big]+\mathcal{O}(\epsilon^4) \,,
\end{align}
with which we recover at two-loop the results of \cite{Yamazaki:1978-cubic,Yamazaki:1981-cubic,Chen:2001}.

The Gaussian fixed point is doubly unstable, while at the Ising fixed-point one eigenvalue is negative ($\omega_d^I$) and the other is positive. 
The stability of the Heisenberg and cubic fixed points are related. Indeed, there exists a critical value $N=N_c$ for which the cubic and Heisenberg fixed points collapse and exchange stability, as in the short-range model \cite{Kleinert:2001ax}. The Heisenberg fixed point is stable for $N<N_c$ and the cubic fixed point is stable for  $N>N_c$. Using the condition $\tilde{g}_c^{\star,C}=0$ (or equivalently $\omega_c^C=0$), we find the following $\epsilon$ expansion for $N_c$:
\begin{equation}
N_c= 4+2\epsilon\alpha_{S,0}+\frac{\epsilon^2}{6}\left(8\alpha_{I_1}+4\alpha_{I_2}+\frac{5}{4}\alpha_{I_4}+12(\alpha_{S,1}-2\alpha_{D,1}\alpha_{S,0})-13\alpha_{S,0}^2-\alpha_{T}+7\alpha_{U}\right) \,. 
\end{equation}
We notice that the two-loop result (order $\epsilon$) coincides with the similar short-range result \cite{Kleinert:2001ax} upon taking $\alpha_{S,0}\to -1$, which corresponds to the value at $d=4$. The three-loop results are instead not related in a similar way.

Physically, the value $N=3$ is very interesting because the $O(3)$-symmetric fixed point characterizes the critical behavior of the Heisenberg model of magnetism. It is thus important to know whether $N_c$ is below or above $3$. Indeed, if $N_c$ is greater than $3$ all magnetic systems with cubic symmetry will have a $O(3)$-symmetric critical behavior as the Heisenberg fixed point will be the relevant one. However, if $N_c$ is smaller than $3$ the cubic fixed point will be the relevant one and will govern the critical behavior of magnetic systems with cubic symmetry. Table \ref{tab:nc} gives values of $N_c$ at $d=3$ for different values of $\epsilon$ obtained at different loop-orders and with the Pad\'e-Borel summation method. 
For $\epsilon=0.2,0.4$ we find $N_c$ greater than $3$ whereas it is smaller than $3$ for $\epsilon=0.6$. However, for short-range cubic models, results at three loops (in $d=4-\epsilon_{SR}$, extrapolated to $\epsilon_{SR}=1$) gave $N_c$ above $3$ while higher-loop computations \cite{Adzhemyan:2019gvv,Kleinert:2001ax} gave values of $N_c$ below three. In order to accurately conclude on the value of $N_c$ in the long-range model for $\epsilon$ close to $1$ we need higher-loop results.

\begin{table}[htb]
\begin{center}
\begin{tabular}{|c||c|c|c|}\hline
$\epsilon$ &   one-loop          &     three-loop    &    PB $[1/1]$  \\ \hline
0.2 & 4 &        3.5712   &  3.500(5)  \\
 0.4 & 4 &      3.5897      &  3.171(13)                 \\
 0.6 & 4 &      4.0553  &  2.926(21)   \\
\hline
\end{tabular}\end{center}
\caption{The critical value $N_c$ for the long-range cubic model at $d=3$, as computed by a one-loop and three-loop truncation and by a Pad\'e-Borel summation of the three-loop series with $[1/1]$ approximant (the error is estimated by the difference with the PB summation of the two-loop series with $[0/1]$ approximant).}
\label{tab:nc}
\end{table}

We can nonetheless study the numerics of the critical exponents $\nu_C$ and  $\nu_H$ of the cubic and Heisenberg fixed points at $N=3$ for different values of $\epsilon$. The exponent $\nu_H$ at $d=3,N=3$ is identical with the exponent $\nu$ reported in table \ref{tab:critvect22}. For comparison, the corresponding $\nu_C$ is displayed in table \ref{tab:nucubic}. 

\begin{table}[h]
\begin{center}
\begin{tabular}{|c|c||c|c|c|}\hline
$\epsilon$& $2\zeta$ &  mean-field &   three-loop     &   PB $[2/1]$      \\ \hline
0.2 & 1.6 &   0.625 &     0.6657   &  0.6659(20)   \\
0.4 & 1.7 & 0.5882  &      0.6681      &  0.671(8)                 \\
0.6 & 1.8 &  0.5556 &    0.6674   &  0.681(19)   \\
\hline
\end{tabular}\end{center}
\caption{The critical exponents $\nu_C$  for the long-range cubic model at $d=3$ and $N=3$, as computed by a one-loop and three-loop truncation and by a Pad\'e-Borel summation of the three-loop series with $[2/1]$ approximant (with error estimated by the difference with the PB summation of the two-loop series with $[1/1]$ approximant).}
\label{tab:nucubic}
\end{table}
The results of the Pad\'e-Borel summation show that $\nu_H$ and $\nu_C$ lie very close to each other. This is due to the fact that $N_c\simeq 3$, so that for $N=3$ the two fixed points are very close to each other. As a consequence, for the physically interesting case $N=3$, the Heisenberg and the cubic critical behavior are practically indistinguishable, as noticed in \cite{Kleinert:1996hy}.

\subsection{The long-range $O(M) \times O(N)$ bifundamental model}
\label{sec:bifundamental}

The last special case of the general multi-scalar model we discuss is the $O(M)\times O(N)$ model. We consider two integers $M$ and $N$, such that $\cN=MN$, and we impose $O(M)\times O(N)$ symmetry.\footnote{In order to have a faithful action of the symmetry group, we should take the latter to be $O(M)\times O(N)/\mathbb{Z}_2$. In the rest of the chapter, for conciseness we will simply refer to $O(M)\times O(N)$ symmetry, as very common in the literature.}
The resulting model, called \emph{bifundamental model} in \cite{Rychkov:2018vya}, has two quartic couplings, and it has been extensively studied in its short-range version (e.g.\ \cite{Kawamura:1988,Kawamura:1990,Pelissetto:2001fi,Gracey:2002pm,Delamotte:2003dw,Kompaniets:2020,Henriksson:2020fqi}).
However, we are not aware of any work on its long-range version.
The model is obtained by the substitution:
	\begin{align} \label{eq:g-bifund}
		g_{\mba \mbb \mbc \mbd} \, &= \, 
		\frac{g_s}{6} \big( \delta_{a_1 b_1} \delta_{c_1 d_1} ( \delta_{a_2 c_2}\delta_{b_2 d_2}+ \delta_{a_2 d_2}\delta_{b_2 c_2})\, + \, 
		 \delta_{a_1 c_1} \delta_{b_1 d_1} (\delta_{a_2 b_2} \delta_{c_2 d_2} +\delta_{a_2 d_2} \delta_{c_2 b_2})\crcr
		&\qquad\quad + \,
		 \delta_{a_1 d_1} \delta_{c_1 b_1} ( \delta_{a_2 c_2}\delta_{b_2 d_2}+ \delta_{a_2 b_2}\delta_{d_2 c_2}) \big) \crcr
		&\quad \, + \, \frac{g_d}{3} \big( \delta_{a_1b_1} \delta_{a_2b_2} \delta_{c_1d_1} \delta_{c_2d_2} \, + \, \delta_{a_1c_1} \delta_{a_2 c_2} \delta_{b_1d_1} \delta_{b_2d_2} \,+\, \delta_{a_1d_1} \delta_{a_2 d_2} \delta_{c_1b_1} \delta_{c_2b_2}\big) \, , \crcr
		r_{\mba \mbb} \, &= \, r \, \delta_{a_1b_1} \delta_{a_2b_2} \, ,
	\end{align}
where each boldface index is split in a pair of indices, $\mba=(a_1, a_2)$ and so on,
where the first index corresponds to the $O(M)$ group ($a_1 = 1, \ldots, M$) while the second index corresponds to the $O(N)$ group ($a_2 = 1, \ldots, N$). This model can be viewed as a modification of the $O(\cN)$ model by the $g_s$ term, which  explicitly breaks the $O(MN)$ symmetry down to $O(M)\times O(N)$.
It can of course also be viewed as a rectangular matrix field theory, with $g_s$ being associated to the single-trace interaction $\Tr[\phi \phi^t \phi \phi^t]$ and $g_d$ to the double-trace interaction $\Tr[\phi \phi^t]^2$, thus explaining our choice of subscripts.

Plugging \eqref{eq:g-bifund} into \eqref{eq:beta_abcd_alpha3} and \eqref{eq:beta2_abcd_alpha3}, we obtain the beta functions:\footnote{Since the three-loop contributions to the beta functions are too lengthy, we do not display them here. They will however be taken into account for the analysis of the fixed points.}
	\begin{align}
		\beta^{(4)}_s \, &= \, - \epsilon \, \gt_s \, + \, \frac{\alpha_{D}}{3} \Big[ (M+N+4)\gt_s + 12\gt_d \Big] \gt_s \crcr
		&\quad + \, \frac{\alpha_{S}}{9} \Big[ (2MN+5M+5N+27) \gt_s^2 + 6(MN+ 14) \gt_d^2 + 12 (2M+2N+5) \gt_d\gt_s \Big] \gt_s \, , \crcr
		\beta^{(4)}_d \, &= \, - \epsilon \, \gt_d \, + \, \frac{\alpha_{D}}{3} \Big[ 3\gt_s^2 +2 (M+N+1) \gt_s \gt_d + (MN + 8) \gt_d^2 \Big] \crcr
		&\quad + \, \frac{\alpha_{S}}{9} \Big[ 3(M+N+3) \gt_s^3 + 3 (MN + M + N + 15) \gt_s^2 \gt_d \crcr
		&\qquad\quad + 24(M + N + 1) \gt_s \gt_d^2 + 2 (5MN+22) \gt_d^3 \Big] \, , \crcr
		\beta^{(2)} \, &= \, - (d-2\Delta_{\phi} ) \, \rt \, + \, \frac{\alpha_{D}}{3} \, \Big[ (M+N+1) \gt_s + (MN+2) \gt_d \Big] \, \rt \crcr
		&\quad + \, \frac{\alpha_S}{6} \, \Big[ (MN+M+N+3) \gt_s^2 + 4 (M+N+1) \gt_s \gt_d + 2(MN+2) \gt_d^2 \Big] \, \rt \, . 
	\end{align}
At order $\epsilon$, the critical couplings are given by
	\begin{align}
		(\tilde{g}_s^{\star}, \tilde{g}_d^{\star})& \, = \,  (0,0) \, , \, \left(0, \, \frac{3\epsilon}{MN+8} \right) \, , \crcr
		 & \bigg( \frac{(12-3MN) \epsilon}{4+10(M+N)-MN(M+N+4) \pm 6 \sqrt{Q} } \, , \,  \crcr
		&   \, \frac{-3(-80+2M+2N+M^2+N^2+2MN\mp 4(M+N+4)\sqrt{Q}) \epsilon}{2\left(464-56(M+N)-16(M^2+N^2+MN)+8MN(M+N) +MN(M+N)^2\right)} \bigg)  ,		
	\end{align}
where
	\begin{equation}
		Q \, = \, 52 - 4 (M+N) + (M^2 - 10MN +N^2) \, .
	\end{equation} 
The first solution is the trivial one, and the second solution is the Heisenberg  fixed-point with  $O(MN)$ symmetry.
The third and fourth solutions are the chiral and anti-chiral fixed-points \cite{Kawamura:1988}. When $\tilde{g}_s^{\star}<0$, the latter are also called sinusoidal and anti-sinusoidal fixed points.

There are four regimes of criticality at fixed $M$  depending on the stability of the Heisenberg and chiral fixed points:
\begin{itemize}
\item If $N> N_{c+}$ there are four real fixed points, and the chiral one is stable.
\item If $N_{c-}<N< N_{c+}$, only the Gaussian and the Heisenberg fixed points are real, and they are both unstable.
\item If $N_H <N < N_{c-}$, there are again four real fixed points, and the chiral (or sinusoidal) one is stable.
\item If $N<N_H$, there are still four real fixed points, but the Heisenberg one is stable. 
\end{itemize}

To compute $N_{c\pm}$ and $N_H$, we use the following ansatz:
	\begin{equation}
		N \, = \, N_0 \, + \, N_1\, \epsilon \,+ \, N_2 \, \epsilon^2 \, + \, \mathcal{O}(\epsilon^3) \, ,
	\end{equation} 
and we solve:
\begin{equation}
		{\rm det} \left| \frac{\partial (\beta_s, \beta_d)}{\partial (\gt_s, \gt_d)} \right|_{\gt=\gt^\star} \, = \, 0 \, .
	\end{equation} 
This leads to $N_{c\pm}=N_{c\pm,0}+N_{c\pm,1}\epsilon+N_{c\pm,2}\epsilon^2+\mathcal{O}(\epsilon^3)$ with:
	\begin{equation}
		N_{c\pm,0} \, = \, 2 + 5M \pm 2 \sqrt{6(M+2)(M-1)} \, ,
	\end{equation} 
	\begin{equation}
		N_{c\pm,1}\, = \, (5M+2)\alpha_{S,0}\pm \frac{\alpha_{S,0}}{2s}\Big[25M^2+22M-32\Big] \, ,
	\end{equation} 
\begin{align}
N_{c\pm,2}\,&= \, \frac{\alpha_{I_1}}{Q_2}\Big[\frac{5M^5+14M^4-277M^3-530M^2+496M+400}{3}\crcr
& \qquad \quad \pm\frac{s(25M^4+59M^3-1434M^2-1900M+2944)}{36}\Big] \crcr
& +\frac{\alpha_{I_2}}{Q_2}\Big[\frac{2(5M^5+11M^4-283M^3-341M^2+418M+136)}{3}\crcr
& \qquad \quad\pm\frac{s(25M^4+41M^2-1398M^2-946M+1216)}{18}\Big]\crcr
&+\frac{\alpha_{I_4}}{Q_2}\Big[\frac{8M^5+31M^4-426M^3-1376M^2+1184M+1632}{96}\crcr
& \qquad \quad\pm\frac{s(20M^4+73M^3-1230M^2-2960M+6176)}{576}\Big] \crcr
& +\frac{\alpha_{T}}{Q_2}\Big[\frac{5M^5+8M^4-289M^3-152M^2+340M-128}{12}\crcr
& \qquad \quad\pm\frac{s(25M^4+23M^3-1362M^2+8M-512)}{144}\Big] \crcr
& +\frac{\alpha_{U}}{Q_2}\Big[\frac{25M^5+64M^4-1397M^3-2272M^2+2324M+1472}{24}\crcr
& \qquad \quad\pm\frac{s(125M^4+259M^3-7098M^2-7592M+11264)}{288}\Big] \crcr
& +\left(\alpha_{S,1}-2\alpha_{S,0}\alpha_{D,1}\right)\Big[5M+2\mp\frac{3(M-6)}{2s}\pm\frac{25s}{12}\Big] \crcr
& -\frac{\alpha_{S,0}^2}{Q_1}\Big[\frac{7M^7+32M^6-744M^5-2882M^4+21608M^3+62520M^2-61952M-62464)}{8} \crcr
& \qquad \quad \pm\frac{1}{96s}\Big(235M^8+1180M^7-26243M^6-108344M^5+791476M^4\crcr
& \qquad \qquad \qquad  +2530384M^3-3402944M^2-6391808M+6897664\Big)\Big] \,,
\end{align}
with
	\begin{align}
s \, &= \, \sqrt{6(M+2)(M-1)} \crcr
Q_1 \, &=\, (M+8)^2(M-7)^2(M+2)(M-1) \crcr
Q_2\, &=\, (M+8)(M-7)(M+2)(M-1) \, .
	\end{align} 
	
For $N_H$, we find:
\begin{align}
N_H=&\frac{4}{M}+\frac{2\alpha_{S,0}}{M}\epsilon \crcr
& +\frac{4(8\alpha_{I_1}+4\alpha_{I_2}-\alpha_T)+5\alpha_{I_4}+48(\alpha_{S,1}-2\alpha_{D,1}\alpha_{S,0})-52\alpha_{S,0}^2+14\alpha_{U}}{24M}\epsilon^2+\mathcal{O}(\epsilon^3)\,.
\end{align}

We notice again that the two-loop results for the critical values of $N$ (order $\epsilon$) coincide with the corresponding short-range results \cite{Pelissetto:2001fi} upon taking $\alpha_{S,0}\to -1$.  The three-loop results are instead not related in a similar way.

We can look at the numerical values of $N_H$ and $N_{c\pm}$ at $d=3$ and $M=2$. Table \ref{tab:ncp} 
gives values of $N_{c\pm}$ and $N_H$ for different values of $\epsilon$ with either a three-loop truncation or a Pad\'e-Borel summation method. 
The table indicates that for $d=3$ and $M=N=2$ the chiral (or sinusoidal) fixed point might exist and be stable for sufficiently small $\epsilon$. However, at $N=3$ the chiral fixed point is not present, and the Heisenberg one is not stable.  
\begin{table}[h!]
\begin{center}
\begin{tabular}{|c|c||c|c|c|}\hline
$\epsilon$ &  & one-loop &        three-loop   &    PB $[1/1]$ \\ \hline
\multirow{3}{*}{0.2} & $N_{c+}$ & 21.8 &   16.36  &  15.7(12) \\
                     & $N_{c-}$ & 2.202 &  2.120  &  2.076(35) \\
                     & $N_{H}$ &   2  &  1.806  &  1.759(12)  \\
\hline
\multirow{3}{*}{0.4} & $N_{c+}$ & 21.8 &   15.35  &  11.6(29) \\
                     & $N_{c-}$ & 2.202 &  2.245  &   2.01(9) \\
                     & $N_{H}$ &   2  &   1.875      &  1.608(29)  \\
\hline
\multirow{3}{*}{0.6} & $N_{c+}$ & 21.8 &  18.76   &  8(4)     \\
                     & $N_{c-}$ & 2.202 & 2.578 &  1.96(14) \\
                     & $N_{H}$ &   2  &  2.207  &  1.50(5) \\
\hline
\end{tabular}\end{center}
\caption{The critical values $N_{c\pm}$ and  $N_{H}$ for the long-range bifundamental model at $d=3$ and $M=2$, as computed by a one-loop and three-loop truncation and by a Pad\'e-Borel summation of the three-loop series with $[1/1]$ approximant (with error estimated by the difference with the PB summation of the two-loop series with $[0/1]$ approximant).}
\label{tab:ncp}
\end{table}


\section{Conclusions}
\label{sec:concl3}

Long-range models have several intriguing properties and they provide an interesting playground for statistical physics methods.
Nevertheless, they have been much less explored than their short-range counterparts. In particular, fewer models have been considered, and all perturbative results to date had been limited to two loops.
In this chapter we contributed in two ways to improving the situation in the case of long-range multi-scalar models: first, we computed the renormalization group beta functions for general quartic interaction up to three loops; second, we used them to provide higher-order results in the long-range Ising, $O(N)$, cubic, as well as $O(M)\times O(N)$ models.
Along the way, we showed that the hypothetical relations between the long-range Ising model at given dimension $d$ and the short-range Ising model at a different dimension $d_{SR}$ \cite{Banos:2012,Angelini:2014,Defenu:2014} only hold up to first order in $\epsilon_{SR}=4-d_{SR}$, failing at second order.

It is instructive to compare our computations to the analogue three-loop computations for short-range multi-scalar models by Brezin et al.\ \cite{Brezin:1974eb,Brezin:1974-add,Brezin:1973jt}. The setting is very similar to ours, as we do not use the minimal subtraction scheme, relying instead on renormalization conditions at a subtraction point. On the latter we differ, as for the four-point function they adopted a non-zero symmetric subtraction point in momentum space, preserving the massless propagator while avoiding IR divergences in the renormalization condition. Unfortunately, for our integrals this option turned out to be unfeasible: we have of course the same topologies of Feynman diagrams as they do, but with propagators with an essentially arbitrary power of $1/p^2$. While in the short-range case the three-loop integrals can be performed for example in Feynman parametrization, the same representation includes in our case extra factors $u_i^{\z-1}$ 
subject to the constraint $\sum_i u_i=1$: the constraint makes it difficult to perform the explicit integration over the parameters $u_i$. Similarly, the Schwinger parametrization with symmetric subtraction point leads to inconvenient exponents of rational functions.
Therefore, we opted for a subtraction point at zero external momenta, and were forced to introduce an IR regulator in the propagator, \eqref{eq:paramcov}. We then used a Mellin-Barnes representation for the Schwinger parametrization of the amplitudes, with the sole exception of the $I_4$ integral of appendix \ref{app:I_4}, for which we used the Gegenbauer polynomial technique \cite{Chetyrkin:1980pr} applied directly to the integral in  momentum space. All these details are presented in the appendices \ref{sec:MB} and \ref{app:integrals}. 
We want to emphasize that despite the vast literature on Feynman integral calculus (e.g.\ \cite{Smirnov:2006ry,Weinzierl:2006qs,Panzer:2014kia,Kotikov:2018wxe} and relative encyclopedic work \cite{Bogner:2017xhp}) we were unable to find directly applicable methods besides the ones presented here. Of course this could be due to our limitations, and it would be interesting to further explore the evaluation of long-range Feynman integrals with other methods, possibly allowing a massless renormalization scheme or higher-loop computations.

We conclude with one comment about the crossover from long-range to short-range critical behavior for more general models than Ising. For the latter, such crossover has attracted quite some interest \cite{Sak:1973,Blanchard:2012xv,Angelini:2014,Brezin:2014,Defenu:2014,Behan:2017dwr,Behan:2017emf}, with the emerging picture being that as one varies $\z$ in the $(0,1)$ interval, at fixed $d$, three regimes are met: the long-range mean-field regime for $0<\z<d/4$; the long-range non-trivial critical regime for $d/4<\z<\z^\star$; and the short-range critical regime for $\z>\z^\star$. The value $\z^\star$ is such that $2\z^\star = 2 -\eta_{\rm SR}$, where $\eta_{\rm SR}$ stands for the anomalous dimension of the short-range Ising model in $d$ dimensions. The picture found in \cite{Behan:2017dwr,Behan:2017emf} actually suggests that the crossover at $\z=\z^\star$ happens not just to the short-range fixed point, but rather to the short-range fixed point plus a decoupled Gaussian field. It would be interesting to extend such picture to the general multi-scalar models.


\begin{subappendices}
\section{The renormalized series} 
\label{app:renseries}
The inversion of the bare series is immediate using the Bogoliubov-Parasuk recursion given in \eqref{eq:BPrecursion}. We have to adapt it to take into account the fact that the coupling has now four indices:
\be
\begin{split}
 g_{\mba \mbb\mbc\mbd} 
 & = \mu^{-\epsilon} \lambda_{\mba \mbb\mbc\mbd} - \sum_{\mathcal{G}}
 s(\mathcal{G})( - 1)^V \mathbf{\mathcal{G}}( \mu^{-\epsilon}\lambda)_{\mba\mbb\mbc\mbd} \; {\cal \hat A}(\mathcal{G}) \,, \crcr
  \mu^{-\epsilon} \lambda_{\mba \mbb\mbc\mbd} 
 & = g_{\mba \mbb\mbc\mbd} - \sum_{\mathcal{G}}
 s(\mathcal{G})(-1)^V \mathbf{\mathcal{G}}( g )_{\mba\mbb\mbc\mbd} \; K_G \,,
\end{split}
\ee
where the sums run over one particle irreducible four-point graphs $\mathcal{G}$ and $ \mathbf{\mathcal{G}}( \cdot )_{\mba\mbb\mbc\mbd} $ is the contraction of coupling constants associated to $\mathcal{G}$, which in our case has four external points with four associated external indices $\mba\mbb\mbc\mbd$. 
The counterterms $K_G$ are defined recursively as in \eqref{eq:rec_counter}.

At one loop $D$ has no four-point subgraphs, hence 
$ K_D = -D$. At two loops, $D^2$ has two subgraphs $D$ (sharing a vertex), while $S$ has one subgraph $D$, hence:
\be
K_{D^2} = - D^2  - 2 D K_D = D^2 
 \,,\qquad 
 K_S = - S - D  K_D = D^2 - S
 \,,
\ee
and a short computation yields at three loops:
\be
\begin{split}
& K_{D^3} = -D^3 \,, \quad
K_{DS} = -D (D^2-S) \,, \quad
K_{U} = -U + 2SD - D^3 \,,\quad
 K_{T} = -T + 2DS - D^3\,, \crcr
& K_{I_1}  = -I_1 + 2SD -D^3 \,, \quad 
K_{I_2} = -I_2 + 2DS -D^3 \,, 
\quad K_{I_3} = -I_3 \,,\quad K_{I_4} = -I_4 \,.
 \end{split}
\ee
Observe that $I_3$ has four-point subgraphs, but upon contracting we obtain a graph with a tadpole hence with zero bare amplitude. Therefore, the inverse of \eqref{eq:bphz}, with right-hand sides given in \eqref{eq:bare_4pt} and \eqref{eq:bare_2pt}, are:
\begin{align}
& \mu^{-\epsilon}\lambda_{\mba \mbb \mbc \mbd}
\, = \, g_{\mba \mbb \mbc \mbd}+\frac{1}{2} \big(g_{\mba \mbb \mbe \mbf}g_{\mbe \mbf \mbc \mbd} + 2 \textrm{ terms} \big) \, D \crcr
& \quad
+ \, \frac{1}{4} \big(g_{\mba \mbb \mbe \mbf} g_{\mbe \mbf \mbg \mbh} g_{\mbg \mbh \mbc \mbd} + 2 \textrm{ terms} \big) \,D^2 \, 
+ \,\frac{1}{2} \big(g_{\mba \mbb \mbe \mbf} g_{\mbe \mbg \mbh \mbc} g_{\mbf \mbg \mbh \mbd} + 5 \textrm{ terms} \big)  \, (D^2 - S)   \crcr
&\quad
+ \, \frac{1}{8} \big(g_{\mba \mbb \mbe \mbf} g_{\mbe \mbf \mbg \mbh} g_{\mbg \mbh \mbm \mbn} g_{\mbm \mbn \mbc \mbd} 
+ 2 \textrm{ terms} \big) \, D^3  \, 
+ \, \frac{1}{4} \big(g_{\mba \mbb \mbe \mbf} g_{\mbe \mbf \mbg \mbh} g_{\mbg \mbm \mbn \mbc} g_{\mbh \mbm \mbn \mbd} 
+ 5 \textrm{ terms} \big) \, D(D^2-S)   \crcr
& \quad 
+ \, \frac{1}{4} \big(g_{\mba \mbe \mbf \mbg} g_{\mbb \mbe \mbf \mbh} g_{\mbg \mbm \mbn \mbc} g_{\mbh \mbm \mbn \mbd} + 5 \textrm{ terms} \big) \; (D^3-2DS+ U)   \crcr
& \quad   + \, \frac{1}{4} \big(g_{\mba \mbb \mbe \mbf} g_{\mbe \mbg \mbh \mbm} g_{\mbf \mbg \mbh \mbn} g_{\mbm \mbn \mbc \mbd} + 2 \textrm{ terms} \big) (D^3-2DS+ T)  \crcr
& \quad   + \,\frac{1}{2} \big(g_{\mba \mbb \mbe \mbf} g_{\mbe \mbg \mbh \mbm} g_{\mbf \mbg \mbn \mbc} g_{\mbh \mbm \mbn \mbd}  + 11 \textrm{ terms} \big) \, (D^3-2DS+I_1)   \crcr
& \quad
+ \,  \frac{1}{4} \big(g_{\mba \mbb \mbe \mbf} g_{\mbe \mbg \mbh \mbc} g_{\mbf \mbm \mbn \mbd} g_{\mbg \mbh \mbm \mbn} 
+ 5 \textrm{ terms} \big) (D^3-2DS+ I_2)    \crcr
& \quad  + \, \frac{1}{6} \big(g_{\mba \mbb \mbe \mbf}g_{\mbh \mbm \mbn \mbf}g_{\mbh \mbm \mbn \mbg}g_{\mbg \mbe \mbc \mbd} +2 \textrm{ terms} \big) \, I_3  + \,  \big( g_{\mba \mbe \mbm \mbh}g_{\mbb \mbe \mbf \mbn}g_{\mbc \mbf \mbm \mbg}g_{\mbd \mbg \mbn \mbh}  \big) \, I_4\, ,
\label{eq:inverse}
\end{align}
while for the two-point coupling we get:
\begin{align}
& \mu^{-(d-2\Delta_{\phi})}\kappa_{\mbc \mbd} \, =  \, 
r_{ \mbc \mbd} + \frac{1}{2} \big( r_{\mbe \mbf} g_{\mbe \mbf \mbc \mbd} \big) \, D 
+\frac{1}{4} \big( r_{\mbe \mbf} g_{\mbe \mbf \mbg \mbh}g_{\mbg \mbh \mbc \mbd} \big) \, D^2 
+\frac{1}{2} \big( r_{\mbe \mbf} g_{\mbe \mbg \mbh \mbc}g_{\mbf \mbg \mbh \mbd} \big) \, (D^2-S) \crcr 
&
+ \, \frac{1}{8}  \big( r_{ \mbe \mbf} g_{\mbe \mbf \mbg \mbh} g_{\mbg \mbh \mbm \mbn} g_{\mbm \mbn \mbc \mbd} \big) \, D^3  
+ \, \frac{1}{4} \big( r_{ \mbe \mbf} g_{\mbe \mbf \mbg \mbh} g_{\mbg \mbm \mbn \mbc} g_{\mbh \mbm \mbn \mbd} \big) \,  
D(D^2-S) \crcr 
& + \, \frac{1}{4} \big( r_{ \mbe \mbf} g_{\mbe \mbg \mbh \mbm} g_{\mbf \mbg \mbh \mbn} g_{\mbm \mbn \mbc \mbd} \big) \, 
( D^3 -2DS  + T ) + \, \frac{1}{2} \big( r_{ \mbe \mbf} g_{\mbe \mbg \mbh \mbm} g_{\mbf \mbg \mbn \mbc} g_{\mbh \mbm \mbn \mbd}  + 1 \textrm{ term} \big) \,(D^3 - 2DS + I_1 )\crcr 
&
+ \,  \frac{1}{4} \big( r_{ \mbe \mbf} g_{\mbe \mbg \mbh \mbc} g_{\mbf \mbm \mbn \mbd} g_{\mbg \mbh \mbm \mbn} \big) \, (D^3 - 2DS +  I_2 ) 
+ \, \frac{1}{6} \big(r_{ \mbe \mbf}\lambda_{\mbh \mbm \mbn \mbf}\lambda_{\mbh \mbm \mbn \mbg}\lambda_{\mbg \mbe \mbc \mbd}   \big)\,   I_3  \, .
\end{align}

In order to compute the $\beta$ functions in practice,
we can for instance derive the bare series in \eqref{eq:bare_4pt} and \eqref{eq:bare_2pt} with respect to $\mu$ and then substitute the bare constants in terms of the renormalized ones using the renormalized series. We get:
\begin{align}
\beta^{(4)}_{\mba \mbb \mbc \mbd} 
&= -\epsilon g_{\mba \mbb \mbc \mbd} +\frac{\epsilon}{2}  D\, \big(g_{\mba \mbb \mbe \mbf}g_{\mbe \mbf \mbc \mbd} + 2 \textrm{ terms} \big)  + 
\frac{\epsilon}{2} \left(D^2-2S\right) \,
\big(g_{\mba \mbb \mbe \mbf}g_{\mbe \mbg \mbh \mbc}g_{\mbf \mbg \mbh \mbd}+ 5 \textrm{ terms}\big) \crcr 
& \quad  \, + \, \frac{\epsilon}{4} (D^3-4DS+3U) \,
\big( g_{\mba \mbe \mbf \mbg} g_{\mbb \mbe \mbf \mbh} g_{\mbg \mbm \mbn \mbc} g_{\mbh \mbm \mbn \mbd} + 5 \textrm{ terms} \big)  \crcr
& \quad  \, + \, \frac{\epsilon}{4}(3T-2DS) \, \big(g_{\mba \mbb \mbe \mbf} g_{\mbe \mbg \mbh \mbm} g_{\mbf \mbg \mbh \mbn} g_{\mbm \mbn \mbc \mbd} + 2 \textrm{ terms} \big)  \crcr
& \quad  \, + \, \frac{\epsilon}{2}(D^3-3DS+3I_1) \, \big(g_{\mba \mbb \mbe \mbf} g_{\mbe \mbg \mbh \mbm} g_{\mbf \mbg \mbn \mbc} g_{\mbh \mbm \mbn \mbd} + 11 \textrm{ terms} \big)  \crcr
& \quad  \, + \, \frac{\epsilon}{4}(D^3-4DS+3I_2) \,
\big(g_{\mba \mbb \mbe \mbf} g_{\mbe \mbg \mbh \mbc} g_{\mbf \mbm \mbn \mbd} g_{\mbg \mbh \mbm \mbn} + 5 \textrm{ terms} \big)  \crcr
& \quad  + \, \frac{\epsilon}{2} I_3 \, \big(g_{\mba \mbb \mbe \mbf}g_{\mbh \mbm \mbn \mbf}g_{\mbh \mbm \mbn \mbg}g_{\mbg \mbe \mbc \mbd} +2 \textrm{ terms} \big) \,  + \, 3\epsilon I_4 \,\big( g_{\mba \mbe \mbm \mbh}g_{\mbb \mbe \mbf \mbn}g_{\mbc \mbf \mbm \mbg}g_{\mbd \mbg \mbn \mbh}  \big) \,,
\end{align}
while for the quadratic coupling we get
\begin{align}
\beta^{(2)}_{\mbc \mbd} 
&= - (d-2\Delta_{\phi} ) r_{\mbc \mbd} +\frac{\epsilon}{2}  D\big( r_{ \mbe \mbf}g_{\mbe \mbf \mbc \mbd} \big) + \frac{\epsilon}{2} \left(D^2-2S\right)
 \big(r_{\mbe \mbf}g_{\mbe \mbg \mbh \mbc}g_{\mbf \mbg \mbh \mbd} \big) \crcr 
& \quad  \,  + \, \frac{\epsilon}{4}(3T-2DS) 
\big(r_{\mbe \mbf} g_{\mbe \mbg \mbh \mbm} g_{\mbf \mbg \mbh \mbn} g_{\mbm \mbn \mbc \mbd} \big) + \, \frac{\epsilon}{2}(D^3-3DS+3I_1) 
(r_{\mbe \mbf} g_{\mbe \mbg \mbh \mbm} g_{\mbf \mbg \mbn \mbc} g_{\mbh \mbm \mbn \mbd} + 1 \textrm{ term} )  \crcr
& \quad  \,  + \, \frac{\epsilon}{4}(D^3-4DS+3I_2) 
\big(r_{ \mbe \mbf} g_{\mbe \mbg \mbh \mbc} g_{\mbf \mbm \mbn \mbd} g_{\mbg \mbh \mbm \mbn} \big) 
 + \, \frac{\epsilon}{2} I_3 
\big( r_{\mbe \mbf} g_{\mbh \mbm \mbn \mbf}g_{\mbh \mbm \mbn \mbg}g_{\mbg \mbe \mbc \mbd} \big) \,.
\label{eq:gamma_ab}
\end{align}

\section{Mellin-Barnes representation}
\label{sec:MB}

We briefly review the Mellin-Barnes representation used repeatedly in appendix \ref{app:integrals}. The Mellin transform of a function $f$ and the inverse Mellin transform are:
 \be
  \phi(s) = \int_0^{\infty} dx  \; x^{s-1} f(x)  \,, \qquad
  f(x) = \int_{c-\imath \infty}^{c+\imath \infty} \frac{ds}{2\pi \imath}\; x^{-s}\phi(s) \,,
 \ee
with $c$ such that $\phi(s)$ is analytic and decreases at infinity in a strip around $c$. In particular, changing variables in the inverse Mellin transform we have:
\be
 \Gamma(s) = \int_0^{\infty} dx\; x^{s-1} \, e^{-x} \,, \qquad
  e^{-x} = \int_{0^--\imath \infty}^{0^-+\imath \infty}
    \frac{ds}{2\pi \imath} \; x^{s} \, \Gamma(-s) \,,
\ee
that is, the $\Gamma$ function is the Mellin transform of the exponential.\footnote{The first expression is the definition of the $\Gamma$ function while the second one is obtained by going around the poles of $\Gamma(-s)$ located at $s=n$, taking into account that the contours are negatively oriented and that 
$
 {\rm Res} \big( \Gamma(-s) ,n \big) = \lim_{s\to  n} (s - n) \Gamma(- s ) = - \lim_{z\to -n} (z + n) \Gamma(z) = - (-1)^n / n!$.} 
 For ${\rm Re}(u) > 0 $ we have:
\begin{align}
   \frac{ \Gamma(u) }{(A+B)^u} & = \int_{0}^{\infty} dx \int_{0^- -\imath \infty}^{0^- +\imath \infty}
    \frac{dz }{2\pi \imath}
    \;x^{u-1} (x A)^{z} \Gamma(-z ) e^{-xB } \crcr
   &  =\int_{0^--\imath \infty}^{0^-+\imath \infty} \; 
    \frac{dz }{2\pi \imath} \; \Gamma(-z)  \Gamma(u+z) A^z  B^{-u-z} \,.
\end{align}
Denoting $[dz] = \frac{dz}{2\pi \im}$ we get:
 \be
 \begin{split}
 &  \frac{1}{(A_1 + \dots + A_{q+1})^{u}}  = \crcr 
 & \qquad \int_{0^- -\imath \infty}^{0^- +\imath \infty} [dz] \;
   \frac{\Gamma(-z_1) \dots \Gamma(-z_q)  \Gamma(z_1 + \dots +z_q +u)  }{\Gamma(u)}  \;  A_1^{z_1} \dots A_q^{z_q} 
   A_{q+1}^{-z_1 - \dots -z_q -u} \,,
 \end{split}
 \label{eq:MB_rep}
 \ee
which is the main formula we use in appendix \ref{app:integrals}. The only a priori restriction we have on the Mellin contour is that ${\rm Re} (z_i) <0$.

\paragraph{A general remark.} 
Let us assume we are interested in the $\epsilon \to 0$ limit of an integral of the type:
\[
 \int_{C}[dz] \; \frac{H_{\epsilon}(z)}{(z-z_0)(z-z_0-\epsilon)} \,,
\]
where $H_{\epsilon}(z)$ depends parametrically on $\epsilon$ and the two poles (of order one) at $z_0$ and $z_0+\epsilon$ are the only poles inside the contour $C$. In the $\epsilon \to 0$ limit the two poles collapse. The integral has a well-defined $\epsilon \to 0$ limit and can be computed by the residue at the double pole if the function is analytic around $z_0$ uniformly in $\epsilon$ (that is the Taylor series of the function around $z_0$ has a radius of convergence which does not depend on $\epsilon$). To see this, observe that:
\begin{equation*}
\begin{split}
 \int_{C}[dz] \; \frac{H_{\epsilon}(z)}{(z-z_0)(z-z_0-\epsilon)}
& = H_{\epsilon}(z_0) \frac{1}{-\epsilon} + H_{\epsilon}(z_0+\epsilon) \frac{1}{\epsilon} = H_{\epsilon}'(z_0) + O(\epsilon)\\
& = \lim_{z\to z_0} \frac{d}{dz}\left[ (z-z_0)^2 H_0(z)\right] + O(\epsilon)\,,
\end{split}
\end{equation*}
where we recall that the residue at a pole of order $n$ is ${\rm Res}(f,c) = 
\frac{1}{(n-1)!} \lim_{z\to c} [(z-c)^n f]^{(n-1)} $. 
The $\epsilon \to 0$ limit does not exist if 
$H_\epsilon(z_0)$ or its derivative diverges in the $\epsilon\to 0$ limit.

\section{The integrals}
\label{app:integrals}

In this appendix we compute the integrals appearing in the three loops beta function.

\subsection{One-loop integral $D$}

First, let us compute the amplitude $D$ of the one-loop graph. From \eqref{eq:amp_final} this is:
\begin{equation}
D=\frac{1}{(4\pi)^{d/2}\Gamma(\zeta)^2}\int_{0}^{\infty} 
da_1da_2 \; \frac{(a_1 a_2)^{\zeta-1} e^{-\sum a}}{ (a_1 + a_2)^{d/2}}  \,.
\end{equation}
We will repeatedly use below the integral:
\be\label{eq:I} 
  \int_{0}^{\infty}[da] \;  
 \frac{ (a_1 a_2)^{ u -1}}{(a_1+a_2)^{ \gamma }} \;
 e^{- (a_1+a_2) } =  \frac{\Gamma(u)^2 \Gamma(2u-\gamma)}{\Gamma(2u)} \,,
\ee
which is convergent for $2{\rm Re}( u ) > {\rm Re}(\gamma)$ and ${\rm Re}(u)>0$. $D$ is the particular case $u = \zeta, \gamma = d/2$:
\begin{equation}
\label{eq:D_tot}
\boxed{
  D  =  \frac{ 1 }{(4\pi)^{d/2}\Gamma(2\zeta)} \Gamma(\tfrac{\epsilon}{2} )  \,. }
\end{equation}
At the relevant orders in $\epsilon$ this is: 
\begin{align}
D&=\frac{1}{\left(4\pi\right)^{d/2}\Gamma(\tfrac{d}{2})}\left(\frac{2}{\epsilon}-\psi(\tfrac{d}{2})+\psi(1) +\frac{\epsilon}{24}(6(\psi(1)-\psi(\tfrac{d}{2}))^2+\pi^2-6\psi_1(\tfrac{d}{2}))\right) +\mathcal{O}(\epsilon^2) \,,\crcr
 \alpha_{D}& = \epsilon(4\pi)^{d/2}\Gamma(\tfrac{d}{2}) \; \frac{D}{2}\crcr
  &= 1 +\frac{\epsilon}{2}\left(\psi(1)-\psi(\tfrac{d}{2})\right)+\frac{\epsilon^2}{8}\left(\left(\psi(1)-\psi(\tfrac{d}{2})\right)^2+\psi_1(1)-\psi_1(\tfrac{d}{2})\right)
\,,
\label{eq:D_exp}
\end{align}
where we used $\psi_1(1)=\pi^2/6$.

\subsection{Two-loop integral $S$}

For the $S$ graph in figure~\ref{fig:1_2_loops}, \eqref{eq:amp_final} yields:
\begin{equation}
S=\frac{1}{(4\pi)^d\Gamma(\zeta)^4}\int_{0}^{\infty} [dadb] \; \frac{(a_1 a_2 b_1 b_2)^{\zeta-1}}{ \big[(a_1 + a_2)(b_1 + b_2 ) + b_1 b_2 \big] ^{d/2 }} \;e^{-(a_1+a_2+b_1+b_2)} \,.
\end{equation}
Using Mellin parameters (see appendix \ref{sec:MB}) we  write:
\be
 \frac{1}{\big[(a_1 + a_2)(b_1 + b_2 ) + b_1 b_2 \big]^{d/2}}
  =  \frac{1}{\Gamma( \tfrac{d}{2}) } \int_{0^-}[dz]  \Gamma(-z)
  \Gamma(\tfrac{d}{2}+z)  
   \;  \frac{ (b_1b_2)^z }{(b_1+b_2)^{\tfrac{d}{2} + z} } \;\frac{1}{(a_1+a_2)^{ \tfrac{d}{2} + z }} \,,
\ee
and we integrate $a$ and $b$ using \eqref{eq:I} to obtain:
\be
 S = \frac{1}{(4\pi)^d\Gamma(2\zeta)
 \Gamma(\tfrac{d}{2})\Gamma(\zeta)^2 } 
\int_{0^-}[dz] \;  \Gamma(-z) \Gamma( \tfrac{d}{2}+z)  
  \frac{\Gamma(z+\zeta)^2}{ \Gamma(2z+2\zeta ) }
\Gamma( \tfrac{\epsilon}{2} + z)  \; 
   \Gamma(\tfrac{\epsilon}{2} -z) \,.
\ee

In the right half complex plane the integrand has poles at $z = n$ and $ n +\epsilon / 2$, for $n\in \mathbb{N}_0$. Passing to the right of the first two poles we get:
\be\label{eq:S1full}
\boxed{
\begin{split}
  S = & \, D^2 + 
  \; \frac{ 1 }{(4\pi)^d \Gamma(2\zeta) \Gamma(\zeta)^2\Gamma( \tfrac{d}{2})  }  \Gamma(\tfrac{d}{2} + \tfrac{ \epsilon}{2})\frac{\Gamma( \tfrac{\epsilon}{2}+\zeta)^2}{ \Gamma(\epsilon+2\zeta ) } \Gamma(-\tfrac{ \epsilon}{2} ) \Gamma(\epsilon) + \frac{J_{\epsilon}(\zeta)}{(4\pi)^d \Gamma(\frac{d}{2})^2 } \,, \crcr
  J_{\epsilon}(\zeta)  = & \, \frac{\Gamma(\tfrac{d}{2})}{\Gamma(\zeta)^2\Gamma(2\zeta) } 
\int_{1^-}[dz] \;  \Gamma(-z) \Gamma(\tfrac{d}{2}+z)  
  \frac{\Gamma(z+\zeta)^2}{ \Gamma(2z+2\zeta ) }
\Gamma( \tfrac{\epsilon}{2} + z)  \; 
   \Gamma(\tfrac{\epsilon}{2} -z) \,. 
\end{split} }
\ee
The crucial point is that $J_{\epsilon}(\zeta)$ has a finite limit for $\epsilon \to 0$:
\be
 J_0( \tfrac{d}{4}) = \frac{1}{\Gamma( \tfrac{d}{4})^2} 
\int_{1^-}[dz] \;  \Gamma( \tfrac{d}{2}+z)  
  \frac{\Gamma(z+ \tfrac{d}{4})^2}{ \Gamma(2z+ \tfrac{d}{2} ) }
\Gamma(  z)  \; 
   \Gamma( -z)^2 \,,
\ee
the integrand having poles of order $2$ with finite residues at the positive integers. $J_0(\tfrac{d}{4})$ can also be expressed as an infinite sum, which we will use for numerical estimates:
\begin{equation}
J_0(\tfrac{d}{4})=\frac{1}{\Gamma(\tfrac{d}{4})^2 }\sum_{n \geq 1}\frac{\Gamma(n+ \tfrac{d}{2})\Gamma(n+ \tfrac{d}{4})^2}{n(n!)\Gamma( \tfrac{d}{2}+2n)} \Big(2\psi(n+1)-\psi(n)-2\psi(n+\tfrac{d}{4})-\psi(n+\tfrac{d}{2})+2\psi(\tfrac{d}{2}+2n)\Big) \,.
\end{equation}
At the relevant order, we obtain for $S$:
\begin{align}
S=& \frac{1}{\left(4\pi\right)^{d}\Gamma(\tfrac{d}{2})^2}\left(\frac{2}{\epsilon^2}+\frac{1}{\epsilon}\Big[3\psi(1)-\psi( \tfrac{d}{2})-2\psi(\tfrac{d}{4})\Big] +\frac{7}{4}\psi(1)^2-\frac{\pi^2}{24}-\psi(1)\psi(\tfrac{d}{4})-\psi(\tfrac{d}{4})^2 \right. \crcr
& \; \left. -\frac{5}{2}\psi(1)\psi(\tfrac{d}{2})+3\psi(\tfrac{d}{4})\psi(\tfrac{d}{2})-\frac{\psi(\tfrac{d}{2})^2}{4}-\psi_1(\tfrac{d}{4})+\frac{5}{4}\psi_1(\tfrac{d}{2})\right)+ \frac{J_0( \tfrac{d}{4})}{(4\pi)^d \Gamma(\tfrac{d}{2})^2} +\mathcal{O}(\epsilon)
\label{eq:S_exp} \,.
\end{align}
We are interested in $\alpha_{S}  = \epsilon (4\pi)^d \Gamma(\tfrac{d}{2})^2\frac{(D^2-2S)}{2}$ for which we get:
\begin{align}
\alpha_{S}  = & \,  2\psi( \tfrac{d}{4} ) - \psi( \tfrac{d}{2})-\psi(1)\crcr
& \, +\frac{\epsilon}{4}\Big[\left(2\psi(\tfrac{d}{4})-\psi(\tfrac{d}{2})-\psi(1)\right)\left(3\psi(1)-5\psi(\tfrac{d}{2})+2\psi(\tfrac{d}{4})\right)  \crcr
& \qquad  +3\psi_1(1)+4\psi_1(\tfrac{d}{4})-7\psi_1(\tfrac{d}{2})-4J_0(\tfrac{d}{4})\Big]   
\,.
\label{eq:alpha_2}
\end{align}

\subsection{Three-loop integrals}

We will treat $I_4$ separately at the end. Using \eqref{eq:amp_final}, the integrals for the other three loops graphs of 
figure~\ref{fig:3_loops} are:
\begin{align}\label{eq:SDTU3}
 T = & \frac{1}{(4\pi)^{3d/2}\Gamma(\zeta)^6}\int_{0}^{ \infty }  \frac{  (a_1 a_2 b_1 b_2 c_1  c_2)^{\zeta-1}  e^{-\sum a}  }
  { \big[ (a_1 + a_2)(b_1 + b_2) (c_1 + c_2) +  b_1b_2(a_1+a_2+c_1+c_2) \big]^{d/2}  }\,, \crcr
 U = & \frac{1}{(4\pi)^{3d/2}\Gamma(\zeta)^6} \int_{0}^{\infty}   \; \frac{  (a_1 a_2 b_1 b_2 c_1  c_2)^{\zeta-1}  e^{-\sum a}  }
  { \big[ (a_1 + a_2)(b_1 + b_2) (c_1 + c_2) + (a_1 + a_2) c_1c_2 + a_1a_2(c_1+c_2)\big]^{d/2}  } \,, \crcr
I_1 = & \frac{1}{(4\pi)^{3d/2}\Gamma(\zeta)^6}\int_0^{\infty} 
 \frac{ (a_1a_2 b c_1 c_{2'}c_{2''})^{\zeta-1} e^{-\sum a}}
 { \big[ (a_1+a_2)[ c_1( c_{2'} +c_{2''} ) +c_{2'}c_{2''} ] + b [ c_{2'}c_{2''} + ( c_{2'} + c_{2''}) ( a_1 + a_2 + c_1 ) ] \big]^{d/2}} \,,
\crcr
I_2  = & \frac{1}{(4\pi)^{3d/2}\Gamma(\zeta)^6}\int_0^{\infty} 
 \frac{  (a_1a_2b_{1'} b_{1''} b_{2'} b_{2''})^{\zeta-1} e^{-\sum a} }{\big[ (a_1+a_2) ( b_{1'} + b_{1''} )(b_{2'} + b_{2''}  )  + b_{1'} b_{1''} (b_{2'} + b_{2''}) + b_{2'} b_{2''} (b_{1'} + b_{1''}) \big]^{d/2} } \, , \crcr
I_3 = & \frac{1}{(4\pi)^{3d/2}\Gamma(\zeta)^6}\int_0^{\infty} 
 \frac{  ( a_1a_2a_3 b_1 b_2 b_3 )^{\zeta-1} e^{-\sum a} }{\big[ (a_1+a_2+a_3) ( b_1 b_2 + b_1b_3 + b_2b_3) + b_1 b_2 b_3  \big]^{d/2} } \, , 
 \end{align}
where $\sum a$ is an abusive notation which signifies the sum over all the Schwinger parameters.

Notice that $U=I_2$. We thus have only five more integrals to compute. To simplify the notation below, we will call $\tilde{D}=(4\pi)^{d/2}D$, $\tilde{S}=(4\pi)^d S$, $\tilde{T} =(4\pi)^{3d/2} T$ and so on.


\subsubsection{The $T$ integral}
We start with the $T$ integral. We split the denominator using Mellin parameters:
\begin{align}
&\frac{1}
 { [(a_1+a_2)(b_1+b_2)(c_1 + c_2)
+ b_1b_2(a_1+a_2+c_1+c_2) ]^{d/2} }  \crcr
& \quad = \int_{0^-}[dz_1 ]  \int_{0^-}[ dz_2]
 \frac{\Gamma(-z_1) \Gamma(-z_2) \Gamma( \tfrac{d}{2} +z_1+z_2)}{\Gamma(\tfrac{d}{2})} 
 \frac{ [b_1b_2(a_1+a_2)]^{z_1} [b_1b_2(c_1+c_2)]^{z_2} }
{ [(a_1+a_2)(b_1+b_2)(c_1 + c_2)]^{\tfrac{d}{2} + z_1+z_2} } \,,
\end{align}
and integrating out $a,b,c$ we get:
\begin{align}
 \tilde{T}= \frac{1}{\Gamma(\zeta)^2\Gamma(2\zeta)^2 \Gamma(\tfrac{d}{2})} & \int_{0^-}[dz_1 ]  \int_{0^-}[ dz_2]  \;
  \Gamma(\tfrac{d}{2} +z_1+z_2) 
\;\; \frac{ \Gamma(\zeta + z_1+z_2)^2 } 
{\Gamma(2\zeta + 2z_1+2z_2)}  \Gamma( \tfrac{\epsilon}{2} + z_1 +z_2) \crcr
 & \qquad \qquad \times \Gamma(-z_1) \Gamma(-z_2)   \Gamma( \tfrac{\epsilon}{2}-z_2)    \Gamma( \tfrac{\epsilon}{2}-z_1)
 \,.
\end{align} 
We deform both contours to the right. The poles in $z_1$ and $z_2$ are completely independent, that is for any $z_1$ to the right of $0^-$, the poles in $z_2$ are always located at the values  $n_2$, $n_2+\epsilon/2$. 
We can then push first the contour of $z_2$, pick up the poles in $z_2$ at $z_1$ fixed and then push the contour of $z_1$. Only the poles at $(0,0),(0,\epsilon/2), (\epsilon/2,0)$ and $(\epsilon/2,\epsilon/2)$ give singular contributions. We then obtain: 
\begin{align}
  \tilde{T}  = &
 \frac{1 }{\Gamma(2\zeta)^3 } \Gamma( \tfrac{\epsilon}{2})^3 
 + 2\frac{ \Gamma( \tfrac{d}{2} + \tfrac{\epsilon}{2})\Gamma(\zeta + \tfrac{\epsilon}{2})^2 }{ \Gamma(\zeta)^2\Gamma(2\zeta)^2 \Gamma( \tfrac{d}{2} ) 
 \Gamma(2\zeta +\epsilon) }  \Gamma(\epsilon) \Gamma(-\tfrac{\epsilon}{2}) \Gamma(\tfrac{\epsilon}{2}) \crcr
 &  + \frac{\Gamma( \tfrac{d}{2} +\epsilon)}{ \Gamma(\zeta)^2\Gamma(2\zeta)^2 \Gamma( \tfrac{d}{2} )}  
 \,  \frac{\Gamma(\zeta +\epsilon)^2}{\Gamma(2\zeta +2\epsilon)} \Gamma( \tfrac{3\epsilon}{2})\Gamma(-\tfrac{\epsilon}{2})^2 
 \crcr
 &+  \frac{2}{\Gamma(\zeta)^2\Gamma(2\zeta)^2 \Gamma(\tfrac{d}{2})}  \int_{1^-}[dz_1 ] 
 \;  \Gamma(-z_1)  \Gamma(\tfrac{\epsilon}{2}-z_1)
  \bigg[
  \Gamma( \tfrac{d}{2}+ z_1) 
\;\; \frac{ \Gamma(\zeta + z_1)^2 } 
{\Gamma(2\zeta + 2z_1)}   \Gamma( \tfrac{\epsilon}{2}+z_1)  \Gamma( \tfrac{\epsilon}{2} )\crcr
 & \qquad \qquad \qquad \qquad \qquad  +  
 \Gamma( \tfrac{d}{2}+ \tfrac{\epsilon}{2} +z_1) 
\;\; \frac{ \Gamma(\zeta + \tfrac{\epsilon}{2} + z_1 )^2 } 
{\Gamma(2\zeta +\epsilon + 2z_1)} 
\Gamma(\epsilon+z_1) \Gamma( - \tfrac{ \epsilon}{2} ) \bigg] 
 \crcr
 &+ \frac{1}{\Gamma(\zeta)^2\Gamma(2\zeta)^2 \Gamma(\tfrac{d}{2})}   \int_{1^-}[dz_1 ]  \int_{1^-}[ dz_2]  \; \Gamma( \tfrac{d}{2} +z_1+z_2) 
\;\; \frac{ \Gamma(\zeta + z_1+z_2)^2 } 
{\Gamma(2\zeta + 2z_1+2z_2)}  \Gamma( \tfrac{\epsilon}{2} + z_1 +z_2) \crcr
 & \qquad \qquad  \qquad \qquad \qquad \times \Gamma(-z_1) \Gamma(-z_2)   \Gamma( \tfrac{\epsilon}{2}-z_2)    \Gamma( \tfrac{\epsilon}{2}-z_1) \,.
\end{align}
The first two lines are explicit and the single and double integrals are of order $O(\epsilon^0)$.
Overall we get: 
\be
\begin{split}
 T\, &= \, D^3 
 +\frac{\Gamma(-\tfrac{\epsilon}{2})}{(4\pi)^{3d/2}\Gamma(\zeta)^2\Gamma(2\zeta)^2\Gamma( \tfrac{d}{2})} \Bigg[ 
 \frac{2\Gamma( \tfrac{d}{2} +  \tfrac{\epsilon}{2})\Gamma(\zeta + \tfrac{\epsilon}{2} )^2 \Gamma(\epsilon)  \Gamma( \tfrac{\epsilon}{2} )}{\Gamma(2\zeta +\epsilon) } \crcr
&\hspace{120pt} \, + \, \frac{\Gamma( \tfrac{d}{2} + \epsilon)\Gamma(\zeta +\epsilon)^2\Gamma(\tfrac{3\epsilon}{2} )\Gamma(- \tfrac{\epsilon}{2} )}{\Gamma(2\zeta +2\epsilon)} \Bigg] +O(\epsilon^0) \,.
\end{split}
\label{eq:T1_exp}
\ee

At the relevant order in $\epsilon$, this is:

\begin{equation}
\boxed{
\begin{split}
T=\frac{1}{3(4\pi)^{3d/2}\Gamma(d/2)^3}&\left[\frac{8}{\epsilon^3}+\frac{8}{\epsilon^2}\left(2\psi(1)-\psi(\tfrac{d}{4})-\psi(\tfrac{d}{2})\right)\right. \\
& \left. +\frac{1}{3\epsilon}\left(\pi^2+12\left(2\psi(1)-\psi(\tfrac{d}{4})-\psi(\tfrac{d}{2})\right)^2-6\psi_1(\tfrac{d}{2})\right)\right]+\mathcal{O}(\epsilon^0) \,.
\end{split}
}
\end{equation}

The beta function coefficient is
$ \alpha_{T} =\epsilon (4\pi)^{3d/2}\Gamma(\tfrac{d}{2})^3 \frac{(3T-2DS)}{4} $, for which we obtain:
\begin{equation}
\alpha_{T}    = \,  \frac{1}{2}\Big[2\psi(\tfrac{d}{4}) - \psi(\tfrac{d}{2})-\psi(1) \Big]^2 
+\frac{1}{2}\psi_1(1)+  \psi_1(\tfrac{d}{4}) - \frac{3}{2} \psi_1(\tfrac{d}{2}) - \, J_0(\tfrac{d}{4}) \, . 
\label{eq:alpha_5}
\end{equation}

\subsubsection{The $U$ integral}

The $U$ integral is more complicated. We use:
\begin{align}
&\frac{1}
 { [(a_1+a_2)(b_1+b_2)(c_1 + c_2)
+ c_1c_2(a_1+a_2) + a_1a_2 (c_1+c_2) ]^{d/2} }  \crcr
& \quad = \int_{0^-}[dz_1] \int_{0^-}[dz_2] 
 \frac{\Gamma(-z_1) \Gamma(-z_2) \Gamma(\tfrac{d}{2}+z_1+z_2)}{\Gamma( \tfrac{d}{2})} 
 \frac{ [c_1c_2(a_1+a_2)]^{z_1} [a_1a_2(c_1+c_2)]^{z_2} }
{ [(a_1+a_2)(b_1+b_2)(c_1 + c_2)]^{\tfrac{d}{2} + z_1+z_2} } \,.
\end{align}
Integrating out $a,b,c$ we end up with:
\begin{align}
 \tilde{U}= 
 \frac{1}{\Gamma(\zeta)^4\Gamma(2\zeta)\Gamma(\tfrac{d}{2})} &
 \int_{0^-}[dz_1] \int_{0^-}[dz_2]  \; 
 \Gamma(\tfrac{d}{2}+z_1+z_2) 
 \frac{\Gamma(\zeta + z_1)^2\Gamma(\zeta + z_2)^2}{\Gamma(2\zeta + 2z_1)\Gamma(2\zeta + 2z_2) } \crcr
 & \times \Gamma(-z_1) \Gamma(-z_2) 
 \Gamma(\tfrac{\epsilon}{2} -z_1-z_2) \Gamma(\tfrac{\epsilon}{2} +z_1 )
 \Gamma( \tfrac{\epsilon}{2}+z_2) \,. 
\end{align}
The problem now is that, due to the $\Gamma(\epsilon/2 - z_1-z_2)$ factor, the poles in $z_2$ in the right half complex plane depend on $z_1$. This makes the integral quite tricky. The first poles in $z_2$ are located at 
$0$ and $\epsilon/2-z_1$, hence:
\begin{align}
\tilde{U}= &  \frac{1}{\Gamma(\zeta)^4\Gamma(2\zeta)\Gamma( \tfrac{d}{2})} 
 \int_{0^-}[dz_1] \bigg(
 \Gamma( \tfrac{d}{2}+z_1) 
 \frac{\Gamma(\zeta + z_1)^2\Gamma(\zeta )^2}{\Gamma(2\zeta + 2z_1)\Gamma(2\zeta ) } 
 \Gamma(-z_1) \Gamma( \tfrac{\epsilon}{2} -z_1) 
 \Gamma( \tfrac{\epsilon}{2} +z_1 ) \Gamma(\tfrac{\epsilon}{2}) \crcr 
& \qquad    + 
    \Gamma( \tfrac{d}{2}+ \tfrac{\epsilon}{2}) 
 \frac{\Gamma(\zeta + z_1)^2\Gamma(\zeta + \tfrac{\epsilon}{2}-z_1)^2}{\Gamma(2\zeta + 2z_1)\Gamma(2\zeta + \epsilon-2z_1) } \Gamma(-z_1) 
 \Gamma( - \tfrac{\epsilon}{2} +z_1) 
 \Gamma( \tfrac{\epsilon}{2} +z_1 )\Gamma(\epsilon-z_1)  \bigg)
  \crcr
& +  \frac{1}{\Gamma(\zeta)^4\Gamma(2\zeta)\Gamma(\tfrac{d}{2})} 
 \int_{0^-}[dz_1]\int_{1^-}[dz_2]  \; 
 \Gamma( \tfrac{d}{2}+z_1+z_2) 
 \frac{\Gamma(\zeta + z_1)^2\Gamma(\zeta + z_2)^2}{\Gamma(2\zeta + 2z_1)\Gamma(2\zeta + 2z_2) }\crcr
 &  \qquad \qquad \times \Gamma(-z_1) \Gamma(-z_2) 
 \Gamma(\tfrac{\epsilon}{2} -z_1-z_2) \Gamma(\tfrac{\epsilon}{2} +z_1 )
 \Gamma(\tfrac{\epsilon}{2}+z_2) \,. 
 \label{eq:tildeU1}
\end{align}
We aim to compute this up to order $1/\epsilon$. The first term is $\tilde D \tilde S$. The second term  has poles at $z_1=0,z_1=\epsilon$ and $z_1=\zeta+\epsilon/2$. The residue at the last pole gives a convergent contribution, thus the divergent part is at most:
\begin{align}
&\frac{\Gamma(\tfrac{d}{2}+ \tfrac{\epsilon}{2})}{\Gamma(\zeta)^4\Gamma(2\zeta)\Gamma( \tfrac{d}{2})} \left [ \frac{\Gamma(\zeta+ \tfrac{\epsilon}{2})^2\Gamma(\zeta)^2\Gamma( \tfrac{\epsilon}{2})\Gamma(\epsilon)\Gamma(-\tfrac{\epsilon}{2})}{\Gamma(2\zeta+\epsilon)\Gamma(2\zeta)} +\frac{\Gamma(\zeta- \tfrac{\epsilon}{2})^2\Gamma(\zeta+\epsilon)^2\Gamma( \tfrac{3\epsilon}{2})\Gamma(-\epsilon)\Gamma(\tfrac{\epsilon}{2})}{\Gamma(2\zeta-\epsilon)\Gamma(2\zeta+2\epsilon)} \right. \crcr
& \qquad \qquad \qquad \qquad \quad\left. - \frac{\Gamma(\zeta)^2\Gamma(\zeta+ \tfrac{\epsilon}{2})^2\Gamma(\tfrac{d}{2}+\tfrac{\epsilon}{2})\Gamma(\epsilon)\Gamma( \tfrac{\epsilon}{2})\Gamma(-\tfrac{\epsilon}{2})}{\Gamma(2\zeta)\Gamma(2\zeta+\epsilon)}\right. \crcr
& \qquad \quad + \int_{1^-}[dz_1] \; 
 \frac{\Gamma(\zeta + z_1)^2\Gamma(\zeta + \tfrac{\epsilon}{2}-z_1)^2}{\Gamma(2\zeta + 2z_1)\Gamma(2\zeta + \epsilon-2z_1) } \Gamma(-z_1) \Gamma( - \tfrac{\epsilon}{2} +z_1) 
 \Gamma( \tfrac{\epsilon}{2} +z_1 )\Gamma(\epsilon-z_1) \bigg] \,, 
\end{align}
but, using the remark in appendix~\ref{sec:MB} the last integral has a finite limit for $\epsilon \to 0$.

The double integral in \eqref{eq:tildeU1} is also tractable. First, we reduce it to:
\begin{align}
&  \tilde{D} \frac{ J_{\epsilon}(\zeta) }{\Gamma(\tfrac{d}{2})^2} -   \frac{1}{\Gamma(\zeta)^4\Gamma(2\zeta)\Gamma(\tfrac{d}{2})} 
 \int_{1^-}[dz_2]  \; 
 \Gamma( \tfrac{d}{2} +1+ \tfrac{\epsilon}{2} ) 
 \frac{\Gamma(\zeta + 1+ \tfrac{\epsilon}{2} -z_2 )^2\Gamma(\zeta + z_2)^2}{\Gamma(2\zeta + 2 +\epsilon -2z_2)\Gamma(2\zeta + 2z_2) }\crcr
 &\qquad  \qquad \times \Gamma(-1 - \tfrac{\epsilon}{2}+z_2) \Gamma(-z_2) 
  \Gamma(\epsilon +1-z_2 ) \Gamma(\tfrac{\epsilon}{2}+z_2)  \crcr
  & +  \frac{1}{\Gamma(\zeta)^4\Gamma(2\zeta)\Gamma(\tfrac{d}{2} ) } 
 \int_{1^-}[dz_1]\int_{1^-}[dz_2]  \; 
 \Gamma( \tfrac{d}{2}+z_1+z_2) 
 \frac{\Gamma(\zeta + z_1)^2\Gamma(\zeta + z_2)^2}{\Gamma(2\zeta + 2z_1)\Gamma(2\zeta + 2z_2) }\crcr
 &  \qquad \qquad \times \Gamma(-z_1) \Gamma(-z_2) 
 \Gamma( \tfrac{\epsilon}{2} -z_1-z_2) \Gamma(\tfrac{\epsilon}{2} +z_1 )
 \Gamma( \tfrac{\epsilon}{2}+z_2) \,. 
\end{align}
Again the single integral is $O(\epsilon^0)$.
A detailed study of the remaining double integral shows that only the poles at $z_2=\epsilon/2-z_1+n$ with $n\geq 1$ and $z_1=m_1$ and $z_1=\epsilon+m_1$ with 
$1\le m_1\le n$ respectively $z_1=n+\epsilon/2-m_2 $ and $0\leq m_2 \leq n-1$ can contribute to the singular part but their summed contribution is in fact $O(\epsilon^0)$.

We finally have:
\be
\begin{split}
U \, = \, DS + \frac{\Gamma(\tfrac{d}{2}+\tfrac{\epsilon}{2})\Gamma(-\epsilon)\Gamma(\tfrac{\epsilon}{2})\Gamma(\tfrac{3\epsilon}{2})\Gamma(\zeta+\epsilon)^2\Gamma(\zeta-\tfrac{\epsilon}{2})^2}{(4\pi)^{3d/2}\Gamma(\zeta)^4\Gamma(2\zeta)\Gamma(\tfrac{d}{2})\Gamma(2\zeta+2\epsilon)\Gamma(2\zeta-\epsilon)} 
 \, + \, D \frac{J_{\epsilon}(\zeta)}{(4\pi)^d
 \Gamma(\tfrac{d}{2})^2}+\mathcal{O}(\epsilon^0)
\end{split} \,, 
\label{eq:U2_exp}
\ee

which is at the relevant order:
\begin{equation}
\boxed{
\begin{split}
U=\frac{1}{3(4\pi)^{3d/2}\Gamma(\tfrac{d}{2})^3}&\Bigg[\frac{8}{\epsilon^3}+\frac{4}{\epsilon^2}\left(5\psi(1)-4\psi(\tfrac{d}{4})-\psi(\tfrac{d}{2})\right)\crcr
& + \frac{1}{\epsilon}\Bigg(-\frac{7\pi^2}{6}-12\psi_1(\tfrac{d}{4})+19\psi_1(\tfrac{d}{2})+12J_0(\tfrac{d}{4}) +19\psi(1)^2-5\psi(\tfrac{d}{2})^2 \crcr
&  \qquad +32\psi(\tfrac{d}{2})\psi(\tfrac{d}{4})-8\psi(\tfrac{d}{4})^2-2\psi(1)(11\psi(\tfrac{d}{2})+8\psi(\tfrac{d}{4})) \Bigg) \Bigg] +\mathcal{O}(\epsilon^0) \,.
\end{split}
}
\end{equation}

In the beta function we are interested in the combination $\alpha_{U}=\epsilon(4\pi)^{3d/2}\Gamma(d/2)^3\frac{(D^3-4DS+3U)}{4}$:
\begin{equation}
\alpha_{U}=-\psi_1(1)-\psi_1(\tfrac{d}{4})+2\psi_1(\tfrac{d}{2})+J_0(\tfrac{d}{4}) \, .
\label{eq:alpha_3}
\end{equation}

\subsubsection{The $I_1$ integral}

Let us now compute $I_1$. First we can simplify it by introducing $a_1=a\beta, a_2=a(1-\beta)$ and integrating out $\beta$. We obtain:
\begin{equation}
\tilde{I}_1 = \frac{1}{\Gamma(\zeta)^4\Gamma(2\zeta)} \int_0^{\infty} 
 \frac{ a^{2\zeta-1}( b c_1 c_{2'}c_{2''})^{\zeta-1} e^{-(a+b+c_1+c_{2'}+c_{2''})}}
 { \big[ a(b+ c_1)( c_{2'} +c_{2''} ) + bc_1( c_{2'} + c_{2''}) +ac_{2'}c_{2''}  + b c_{2'}c_{2''}   \big]^{d/2}} \,.
\end{equation}
We split the denominator using three Mellin parameters:
\begin{align}
&\frac{1}{\big[ a(b+ c_1)( c_{2'} +c_{2''} ) + bc_1( c_{2'} + c_{2''}) +ac_{2'}c_{2''}  + b c_{2'}c_{2''}   \big]^{d/2}}
  = \int_{0^-}[dz_1] \int_{0^-}[dz_2] \int_{0^-}[dz_3] 
\crcr
& \qquad
 \frac{\Gamma(-z_1) \Gamma(-z_2)\Gamma(-z_3) 
 \Gamma(\tfrac{ d}{2} +z_1+z_2+z_3)}{\Gamma( \tfrac{d}{2})} \; 
 \frac{ [ac_{2'}c_{2''}]^{z_1} [bc_{2'}c_{2''})]^{z_2}[bc_1(c_{2'}+c_{2''})]^{z_3} }
{ [a(b+c_1)(c_{2'} + c_{2''})]^{d/2 + z_1+z_2+z_3} } \; 
 \,,
\end{align}
and integrating out the Schwinger parameters we find:
\begin{align}
& \tilde{I}_1=\frac{1}{\Gamma(\zeta)^4\Gamma(2\zeta)\Gamma(\tfrac{d}{2})}\int_{0^-}[dz_1] \int_{0^-}[dz_2] \int_{0^-}[dz_3]  \Gamma(-z_1)\Gamma(-z_2)\Gamma(-z_3)\Gamma(\tfrac{d}{2}+z_1+z_2+z_3) \crcr
& \qquad \qquad \times
\frac{\Gamma(\zeta+z_1+z_2)^2\Gamma(\tfrac{\epsilon}{2}+z_1+z_2)}{\Gamma(2\zeta+2z_1+2z_2)}  \frac{\Gamma(\zeta+z_2+z_3)\Gamma(\zeta+z_3)}{\Gamma(2\zeta+2z_3+z_2)} \crcr
&\qquad \qquad \qquad \times\Gamma(\tfrac{\epsilon}{2}-z_2-z_3)\Gamma(\tfrac{\epsilon}{2}+z_3-z_1) \,.
\end{align}
Again, we have a mixing between the poles of the $z_i$. The first poles of $z_3$ are located at $0$ and $\epsilon/2-z_2$ and we have:
\begin{align}
 \tilde{I}_1&= \frac{1}{\Gamma(\zeta)^4\Gamma(2\zeta)\Gamma( \tfrac{ d}{2})}\int_{0^-}[dz_1] \int_{0^-}[dz_2]
\Gamma(-z_1)\Gamma(-z_2) \crcr
& \times \bigg[  \Gamma(\tfrac{ d}{2} +z_1+z_2) \frac{\Gamma(\zeta+z_1+z_2)^2\Gamma(\tfrac{\epsilon}{2}+z_1+z_2)}{\Gamma(2\zeta+2z_1+2z_2)} \; \frac{\Gamma(\zeta+z_2)\Gamma(\zeta)}{\Gamma(2\zeta+z_2)}\Gamma(\tfrac{\epsilon}{2}-z_2)\Gamma(\tfrac{\epsilon}{2}-z_1) \crcr
& \;\;+  \Gamma(z_2-\tfrac{\epsilon}{2})\Gamma(\tfrac{ d}{2} +z_1+\tfrac{\epsilon}{2})  \; \frac{\Gamma(\zeta+z_1+z_2)^2\Gamma(\tfrac{\epsilon}{2}+z_1+z_2)}{\Gamma(2\zeta+2z_1+2z_2)}  \;  \frac{\Gamma(\zeta+\tfrac{\epsilon}{2})\Gamma(\zeta+\tfrac{\epsilon}{2}-z_2)}{\Gamma(2\zeta+\epsilon-z_2)}\Gamma(\epsilon-z_1-z_2) \bigg] \crcr
&+ \frac{1}{\Gamma(\zeta)^4\Gamma(2\zeta)\Gamma(\tfrac{d}{2} )}\int_{0^-}[dz_1] \int_{0^-}[dz_2] \int_{1^-}[dz_3] \Gamma(-z_1)\Gamma(-z_2)\Gamma(-z_3)\Gamma(\tfrac{d}{2}+z_1+z_2+z_3) \crcr
& \qquad \times \frac{\Gamma(\zeta+z_1+z_2)^2\Gamma(\tfrac{\epsilon}{2}+z_1+z_2)}{\Gamma(2\zeta+2z_1+2z_2)}  \; \frac{\Gamma(\zeta+z_2+z_3)\Gamma(\zeta+z_3)}{\Gamma(2\zeta+2z_3+z_2)}\crcr
&\qquad \qquad  \times\Gamma(\tfrac{\epsilon}{2}-z_2-z_3)\Gamma(\tfrac{\epsilon}{2}+z_3-z_1) \,.
\label{eq:I1_double_int}
\end{align}

Let us call $\tilde{I}_{1,1}$ the first double integral. 
The computation of this integral is very similar to the one of $T$. The poles of $z_1$ and $z_2$ are completely independent. Only the poles at $(0,0),(0,\epsilon/2);(\epsilon/2,0),(\epsilon/2,\epsilon/2)$ give singular contributions and we get:
\begin{align}
\tilde{I}_{1,1}=&\frac{\Gamma(\tfrac{\epsilon}{2})^3}{\Gamma(2\zeta)^3}+\frac{\Gamma(\zeta+\tfrac{\epsilon}{2})^3\Gamma(\tfrac{d}{2}+\tfrac{\epsilon}{2})}{\Gamma(\zeta)^3\Gamma(2\zeta)\Gamma(\tfrac{d}{2})\Gamma(2\zeta+\epsilon)\Gamma(2\zeta+\tfrac{\epsilon}{2})}\Gamma(-\tfrac{\epsilon}{2})\Gamma(\epsilon)\Gamma(\tfrac{\epsilon}{2}) \crcr
&+ \frac{\Gamma(\zeta+\tfrac{\epsilon}{2})^2\Gamma(\tfrac{d}{2}+\tfrac{\epsilon}{2})}{\Gamma(\zeta)^2\Gamma( \tfrac{d}{2})\Gamma(2\zeta)^2\Gamma(2\zeta+\epsilon)}\Gamma(-\tfrac{\epsilon}{2})\Gamma(\epsilon)\Gamma(\tfrac{\epsilon}{2}) \crcr
& + \frac{\Gamma(\zeta+\epsilon)^2\Gamma(\zeta+\tfrac{\epsilon}{2})\Gamma(\tfrac{d}{2}+\epsilon)}{\Gamma(\zeta)^3\Gamma(d/2)\Gamma(2\zeta)\Gamma(2\zeta+2\epsilon)\Gamma(2\zeta+\tfrac{\epsilon}{2})}\Gamma(-\tfrac{\epsilon}{2})^2\Gamma(\tfrac{3\epsilon}{2}) +\mathcal{O}(\epsilon^0) \,.
\end{align}
Let us now call $\tilde{I}_{1,2}$ the second term in \eqref{eq:I1_double_int}.
The poles in $z_1$ and $z_2$ mix and the computation is similar to the one of $U$.  The first poles in $z_2$ are located at $z_2=0$, $z_2=\epsilon/2$ and $z_2=\epsilon-z_1$. Pushing the integral over $z_2$ past these poles, we obtain singular contributions only from the single integrals (and in particular from the poles in $z_1$ closest to zero). In the end we obtain:
\begin{align}
 \tilde{I}_{1,2}=\frac{1}{\Gamma(\zeta)^4\Gamma(2\zeta)\Gamma(\tfrac{d}{2})}&\bigg[  \frac{\Gamma(\zeta)^2\Gamma(\zeta+\tfrac{\epsilon}{2})^2\Gamma(\tfrac{d}{2}+\tfrac{\epsilon}{2})}{\Gamma(2\zeta)\Gamma(2\zeta+\epsilon)}\Gamma(-\tfrac{\epsilon}{2})\Gamma(\tfrac{\epsilon}{2})\Gamma(\epsilon) \crcr
& - \frac{\Gamma(\zeta)\Gamma(\zeta+\tfrac{\epsilon}{2})^3\Gamma(\tfrac{d}{2}+\tfrac{\epsilon}{2})}{\Gamma(2\zeta+\tfrac{\epsilon}{2})\Gamma(2\zeta+\epsilon)}\Gamma(-\tfrac{\epsilon}{2})\Gamma(\tfrac{\epsilon}{2})\Gamma(\epsilon) \crcr
&+ \frac{\Gamma(\zeta+\epsilon)^2\Gamma(\zeta+\tfrac{\epsilon}{2})\Gamma(\zeta-\tfrac{\epsilon}{2})\Gamma(\tfrac{d}{2}+\tfrac{\epsilon}{2})}{\Gamma(2\zeta)\Gamma(2\zeta+2\epsilon)}\Gamma(\tfrac{\epsilon}{2})\Gamma(\tfrac{3\epsilon}{2})\Gamma(-\epsilon)   \bigg] \crcr
& + \frac{\Gamma(\tfrac{3\epsilon}{2})}{\Gamma(\tfrac{d}{4})\Gamma(\tfrac{d}{2})^3}\int_{1^-} [dz_1] \Gamma(-z_1)^2\Gamma(z_1)\Gamma(\tfrac{d}{4}+z_1)+\mathcal{O}(\epsilon^0)\, .
\end{align}
The triple integral in \eqref{eq:I1_double_int}, which we call $\tilde{I}_{1,3}$ has poles in $z_2$ between $0^-$ and $1^-$ located at $0$ and $1+\epsilon/2-z_3$. Pushing the contour of  $z_2$ past these poles, a very lengthy but straightforward computation gives:
\begin{equation}
\tilde{I}_{1,3}= \tilde{D} \;\frac{ J_{\epsilon}(\zeta) }{\Gamma(\tfrac{d}{2} )^2}+ \mathcal{O}(\epsilon^0)\, .
\end{equation}
Gathering all the terms from $\tilde{I}_{1,1}$, $\tilde{I}_{1,2}$ and $\tilde{I}_{1,3}$, we have:
\begin{align}
I_1=&\frac{\Gamma(\tfrac{\epsilon}{2})^3 }{(4\pi)^{\tfrac{3d}{2}}\Gamma(2\zeta)^3} +\frac{\Gamma(\zeta+\epsilon)^2\Gamma(\zeta+\tfrac{\epsilon}{2})\Gamma(\zeta-\tfrac{\epsilon}{2})\Gamma(\tfrac{d}{2}+\tfrac{\epsilon}{2})}{(4\pi)^{\tfrac{3d}{2}}\Gamma(\zeta)^4\Gamma(2\zeta)^2\Gamma(\tfrac{d}{2})\Gamma(2\zeta+2\epsilon)}\Gamma(\tfrac{\epsilon}{2})\Gamma(\tfrac{3\epsilon}{2})\Gamma(-\epsilon) \crcr
&+ \frac{\Gamma(\zeta+\epsilon)^2\Gamma(\zeta+\tfrac{\epsilon}{2})\Gamma(\tfrac{d}{2}+\epsilon)}{(4\pi)^{\tfrac{3d}{2}}\Gamma(\zeta)^3\Gamma(\tfrac{d}{2})\Gamma(2\zeta)\Gamma(2\zeta+2\epsilon)\Gamma(2\zeta+\tfrac{\epsilon}{2})}\Gamma(-\tfrac{\epsilon}{2})^2\Gamma(\tfrac{3\epsilon}{2}) \crcr
& +\frac{2\Gamma(\zeta+\tfrac{\epsilon}{2})^2\Gamma(\tfrac{d}{2}+\tfrac{\epsilon}{2})}{(4\pi)^{\tfrac{3d}{2}}\Gamma(\zeta)^2\Gamma(\tfrac{d}{2})\Gamma(2\zeta)^2\Gamma(2\zeta+\epsilon)}\Gamma(-\tfrac{\epsilon}{2})\Gamma(\epsilon)\Gamma(\tfrac{\epsilon}{2})    \crcr
& + \Gamma(\tfrac{3\epsilon}{2})\frac{1}{(4\pi)^{\tfrac{3d}{2}}\Gamma(\tfrac{d}{4})\Gamma(\tfrac{d}{2})^3}\int_{1^-} [dz_1] \Gamma(-z_1)^2\Gamma(z_1)\Gamma(\tfrac{d}{4}+z_1) + D \; \frac{ J_{\epsilon}(\zeta) }{\Gamma(\tfrac{d}{2})^2}+\mathcal{O}(\epsilon^0) \,.
\end{align}
The last step is to compute the finite integral in this equation. We close the contour to the left and pick up a pole of order $3$ at $z=0$ and poles of order one at $z=-n_1$ and $z=-d/4-n_2$ with $n_1\geq 1$ and $n_2\geq 0$. The sums over $n_1$ and $n_2$ can be computed in terms of gamma and polygamma functions and we find:
\begin{align}
& \frac{1}{\Gamma( \tfrac{d}{4} )\Gamma(\tfrac{d}{2})^3}\int_{1^-} [dz] \Gamma(-z)^2\Gamma(z)\Gamma(\tfrac{d}{4}+z) \crcr
& = \frac{1}{\Gamma(\tfrac{d}{2})^3}\left(-\pi^2\csc(\tfrac{d\pi}{4})^2+\psi_1(1-\tfrac{d}{4})+\frac{1}{2}\psi_1(\tfrac{d}{4})+\frac{1}{2}\psi(1)^2+\frac{\pi^2}{4}-\psi(1)\psi(\tfrac{d}{4})+\frac{1}{2}\psi(\tfrac{d}{4})^2\right).
\end{align}

We finally have for $I_1$:

\begin{equation}
\boxed{
\begin{split}
I_1=\frac{1}{(4\pi)^{3d/2}\Gamma(d/2)^3}\Bigg[&\frac{4}{3\epsilon^3}-\frac{4}{\epsilon^2}\left(\psi(\tfrac{d}{4})-\psi(1)\right) \crcr
& +\frac{1}{\epsilon}\Bigg(2J_0(\tfrac{d}{4})-\frac{\pi^2}{9}+5\psi(1)^2-\psi(\tfrac{d}{2})^2+4\psi(\tfrac{d}{4})\psi(\tfrac{d}{2})+2\psi(\tfrac{d}{4})^2\crcr
&  -2\psi(1)(\psi(\tfrac{d}{2})+4\psi(\tfrac{d}{4}))-2\psi_1(\tfrac{d}{4})+\frac{8}{3}\psi_1(\tfrac{d}{2}) \Bigg) \Bigg] +\mathcal{O}(\epsilon^0) \,,
\end{split}
}
\end{equation}

and finally we get for $\alpha_{I_1}$:
\begin{equation}
\alpha_{I_1}=\frac{3}{2}\left[2\psi(\tfrac{d}{4})-\psi(\tfrac{d}{2})-\psi(1)\right]^2+\frac{1}{2}\psi_1(1)-\frac{1}{2}\psi_1(\tfrac{d}{2}) \, .
\end{equation}

\subsubsection{The $I_3$ integral}

For the $I_3$ diagram, one needs to take into account the subtraction of the local part of the two-point insertion. After some trivial integrals, the subtracted $I_3$ writes 
using a Taylor expansion with integral rest:
\begin{equation}
I_3=\frac{ \tfrac{-2}{d} }{(4\pi)^{3d/2}\Gamma(3\zeta)\Gamma(\zeta)^3}\int_0^1 dt\int [dadb] \frac{a^{3\zeta-1}(b_1b_2b_3)^{\zeta}e^{-a-b_1-b_2-b_3}}{\left[a(b_1b_2+b_1b_3+b_2b_3)+tb_1b_2b_3\right]^{d/2+1}} \,.
\end{equation}
We then introduce Mellin parameters via:
\begin{align}
&\frac{1}{\left[a(b_2b_3+b_1b_2+b_1b_3)+tb_1b_2b_3\right]^{1+d/2}}\crcr
& =\int_{0^-}[dz_1] \int_{0^-} [dz_2]\frac{\Gamma(-z_1)\Gamma(-z_2)\Gamma(z_1+z_2+d/2+1)}{\Gamma(1+d/2)}\frac{(tb_1b_2b_3)^{z_1}(a(b_2b_3))^{z_2}}{\left[ab_1(b_2+b_3)\right]^{z_1+z_2+d/2+1}} \,,
\end{align}
and integrate out the Schwinger parameters and $t$  to obtain:
\begin{align}
I_3=&\frac{ \tfrac{-2}{d} }{(4\pi)^{3d/2}\Gamma(3\zeta)\Gamma(\zeta)^3\Gamma(1+\tfrac{d}{2})}\int_{(\tfrac{d+3\epsilon}{4}-1)^-} [dz_1] \int_{(\tfrac{-d+\epsilon}{4})^-} [dz_2]  \Gamma(-z_1)\Gamma(-z_2)\crcr 
&\qquad \quad \times \Gamma(3\zeta-\tfrac{d}{2}-1-z_1)\Gamma(\zeta-\tfrac{d}{2}-z_2) \Gamma(1+\tfrac{d}{2}+z_1+z_2)\crcr
& \qquad \quad \times \frac{\Gamma(\zeta+1+z_1+z_2)^2}{\Gamma(2\zeta+2+2z_1+2z_2)(z_1+1)}\Gamma(\tfrac{\epsilon}{2}+1+z_1+z_2) ,
\end{align}
where we moved the contours such that all gamma functions have positive arguments (i.e. the integrals over the Schwinger parameters are convergent) and the poles of the first four gamma functions are separated from the ones of the other gamma functions by the contours. The pole in $z_1$ and $z_2$ are independent, and the only one contributing at order ${\cal O}(\epsilon^{-1})$ is located at $z_1=3\zeta-d/2-1$ and $z_2=\zeta-d/2$.  We finally obtain:
\begin{equation}
\boxed{ I_3=\frac{2}{3\epsilon}\frac{\Gamma(- \tfrac{d}{4})}{(4\pi)^{3d/2}\Gamma( \tfrac{3d}{4} )\Gamma( \tfrac{d}{2})}+\mathcal{O}(\epsilon^0)\,, }
\end{equation}
and for $\alpha_{I_3}= \epsilon (4\pi)^{3d/2}\Gamma(\tfrac{d}{2})^3 \frac{I_3}{2}$ we get:
\begin{equation}
\alpha_{I_3}=\frac{\Gamma(- \tfrac{d}{4})
\Gamma( \tfrac{d}{2} )^2}{3 \, \Gamma( \tfrac{3d}{4} )} \,. 
\end{equation}

\subsubsection{The $I_4$ integral}
\label{app:I_4}

We will compute the $I_4$ integral differently.
The Feynman integral corresponding to the tetrahedron diagram at zero external momenta is:
\begin{align}
& \mu^{-3\epsilon}I_4= \int \, \frac{d^d q_1}{(2\pi)^d} \frac{d^d q_2}{(2\pi)^d}  \frac{d^d q_3}{(2\pi)^d}  \crcr 
& \frac{1}{(q_1^2+\mu^2)^{\zeta}(q_2^2+\mu^2)^{\zeta}(q_3+\mu^2)^{\zeta}((q_1-q_2)^2+\mu^2)^{\zeta}((q_3-q_1)^2+\mu^2)^{\zeta}((q_3-q_2)^2+\mu^2)^{\zeta}} \,.
\end{align}
This is a much harder diagram to compute with the Schwinger parametrization and Mellin-Barnes representation, so we will adopt a different approach, following the method discussed in \cite{Kreimer:1996js} for the case $\zeta=1$, and based on the Gegenbauer polynomial technique \cite{Chetyrkin:1980pr}.
As this diagram does not contain divergent subgraphs, it diverges as a simple pole in $\epsilon$ and we are only interested in determining the residue at the pole. The ultraviolet divergence arises when all three-loop momenta are large, and its coefficient is independent of the chosen IR regularization: we can set $\mu=0$ and deal with the infrared divergence in a simpler fashion.
The main trick is to use the following expansion, valid  for $p>q$:
\be
\frac{1}{(p-q)^{2\zeta}} = \frac{1}{p^{2\zeta}} \sum_{n= 0}^{+\infty} C_n^{\zeta}(\hat{p}\cdot\hat{q}) \left(\frac{q}{p}\right)^n \,,
\ee
where $p=\sqrt{p^2}$ and $\hat{p}_\mu = p_\mu/p$. The expansion coefficients $C_n^\zeta(x)$ are the Gegenbauer polynomials, satisfying
\be
C_n^{\zeta}(1)=  \frac{\Gamma(n+2\zeta)}{\Gamma(2\zeta) n!} \,, \qquad
\int d\hat{q}\; C_n^{\zeta}(\hat{p}\cdot\hat{q}) C_{n'}^{\zeta}(\hat{p}'\cdot\hat{q}) = \frac{\zeta}{n+\zeta} \delta_{nn'} C_n^{\zeta}(\hat{p}\cdot\hat{p}') \,,
\ee
where the angular integral is normalized such that $\int d\hat{q} =1$.

Noticing that $I_4$ is totally symmetric in the three-loop momenta, we can choose $q_1<q_2<q_3$ and write:
\begin{align}
I_4 &=6\mu^{3\epsilon} \int_{q_1<q_2<q_3} \, \frac{d^d q_1}{(2\pi)^d} \frac{d^d q_2}{(2\pi)^d}  \frac{d^d q_3}{(2\pi)^d}  \frac{1}{q_1^{2\zeta}q_2^{2\zeta}q_3^{2\zeta}(q_1-q_2)^{2\zeta}(q_3-q_1)^{2\zeta}(q_3-q_2)^{2\zeta}} \crcr
&= 6 \mu^{3\epsilon} \int_{q_1<q_2<q_3} \, \frac{d^d q_1}{(2\pi)^d} \frac{d^d q_2}{(2\pi)^d}   \frac{d^d q_3}{(2\pi)^d}   \frac{1}{q_1^{2\zeta}q_2^{4\zeta}q_3^{6\zeta}}\crcr
&\quad\qquad \times \sum_{n_1,n_2,n_3} C_{n_1}^{\zeta}(\hat{q_1}\cdot\hat{q_2}) \left(\frac{q_1}{q_2}\right)^{n_1} C_{n_2}^{\zeta}(\hat{q_1}\cdot\hat{q_3}) \left(\frac{q_1}{q_3}\right)^{n_2} C_{n_3}^{\zeta}(\hat{q_3}\cdot\hat{q_2}) \left(\frac{q_2}{q_3}\right)^{n_3} \,.
\end{align}
Separating into radial and angular integrals, this becomes:
\begin{align}
I_4&= \frac{6\mu^{3\epsilon}  {\rm Vol}(S^{d-1})^3}{(2\pi)^{3d}} \int_{q_1<q_2<q_3} \, d q_1d q_2 d q_3  \, q_1^{d-1-2\zeta}q_2^{d-1-4\zeta}q_3^{d-1-6\zeta}\crcr
& \times \sum_{n_1,n_2,n_3} \int d \hat{q}_1d \hat{q}_2 d \hat{q}_3 \, C_{n_1}^{\zeta}(\hat{q_1}\cdot\hat{q_2}) \left(\frac{q_1}{q_2}\right)^{n_1} C_{n_2}^{\zeta}(\hat{q_1}\cdot\hat{q_3}) \left(\frac{q_1}{q_3}\right)^{n_2} C_{n_3}^{\zeta}(\hat{q_3}\cdot\hat{q_2}) \left(\frac{q_2}{q_3}\right)^{n_3} ,
\end{align}
and using the orthogonality relation of the Gegenbauer polynomials we get:
\begin{align}
I_4 
&= \frac{6 \mu^{3\epsilon} {\rm Vol}(S^{d-1})^3}{(2\pi)^{3d}} \int_{q_1<q_2<q_3} \, d q_1d q_2 d q_3  \, q_1^{d-1-2\zeta}q_2^{d-1-4\zeta}q_3^{d-1-6\zeta}\crcr
&\quad\qquad \times \sum_{n_1,n_2}  \frac{\zeta}{n_1+\zeta} \int d \hat{q}_1  d \hat{q}_3 \, C_{n_1}^{\zeta}(\hat{q_1}\cdot\hat{q_3}) C_{n_2}^{\zeta}(\hat{q_1}\cdot\hat{q_3}) \left(\frac{q_1}{q_3}\right)^{n_1+n_2} \crcr
&= \frac{6 \mu^{3\epsilon} {\rm Vol}(S^{d-1})^3}{(2\pi)^{3d}} \int_{\mu}^{+\infty} d q_3 \int_{\mu}^{q_3} d q_2  \int_{\mu}^{q_2}  d q_1  \, q_1^{d-1-2\zeta}q_2^{d-1-4\zeta}q_3^{d-1-6\zeta} \crcr 
& \qquad \qquad \times \sum_{n\geq 0}  \left(\frac{\zeta}{n+\zeta}\right)^2 \frac{\Gamma(n+2\zeta)}{\Gamma(2\zeta) n!}  \left(\frac{q_1}{q_3}\right)^{2 n} \crcr
&= \frac{6  {\rm Vol}(S^{d-1})^3}{(2\pi)^{3d}} \frac{1}{\epsilon} \frac{ (\tfrac{d}{4})^2}{12 
\Gamma(\tfrac{d}{2})} \sum_{n\geq 0}  \frac{1}{(n+
\tfrac{d}{4} )^4} \frac{\Gamma(n+ \tfrac{d}{2})}{n!}  + O(\epsilon^0) \,.
\end{align}
We finally obtain $I_4$ remembering that ${\rm Vol}(S^{d-1})= 2\pi^{d/2}/\Gamma(\tfrac{d}{2})$,  and noting that 
the last sum can be written in terms of polygamma functions:
\be
\boxed{ I_4 = \frac{1}{\epsilon} \frac{d^2 }{48 (4\pi)^{3d/2}}   
\frac{\Gamma(\tfrac{d}{4})^3\Gamma(1-\tfrac{d}{4})}{\Gamma(\tfrac{d}{2})^4}(6\psi_1(\tfrac{d}{4})-\pi^2) + O(\epsilon^0) \,. }
\ee
Notice that for $d=4$ this reduces to $I_4 = \frac{{\rm Vol}(S^{d-1})^3}{2 (2\pi)^{3d} \epsilon} \zeta(3)$, in agreement with \cite{Brezin:1974-add}.

In the beta functions we are interested in $\alpha_{I_4}=\epsilon(4\pi)^{3d/2}\Gamma(\tfrac{d}{2})^3 3I_4$. We obtain:

%

\begin{equation}
\alpha_{I_4} \, = \,\frac{\, \Gamma(1 + \tfrac{d}{4})^3\Gamma(- \tfrac{d}{4})}{
 \, \Gamma(\tfrac{d}{2} )} \; 6 \, \Big[  \psi_1(1) -  \psi_1(\tfrac{d}{4})  \Big] \, .
\end{equation}

\end{subappendices}

\chapter{The large-$N$ $O(N)^3$ model}
\label{chap:CTKT}

\section{Line of fixed points in long range}
\label{sec:line}
In this chapter, we study rigorously the $O(N)^3$ bosonic long-range model which is a special case of the multi-scalar long-range model we considered in the previous chapter. In order to tame rigorously the divergences appearing in the two- and four-point functions we use here a different scheme than in the previous chapter, namely Wilsonian renormalization with IR and UV cutoffs. 
In the first part of the chapter, we study the renormalization group and fixed point of the $O(N)^3$ bosonic long-range model. 
In sections \ref{sec:modelCTKT} and \ref{sec:model-graphs} we introduce in detail the model, its expansion in Feynman graphs, and the 2PI formalism, which neatly captures the resummed $n$-point functions at  large $N$. In section \ref{sec:RG} we review the Wilsonian renormalization group formalism that is the backbone of our construction. In sections \ref{sec:2point} and \ref{sec:4point} we construct and renormalize the two- and four-point functions, thus obtaining the beta functions. In section \ref{sec:divergences} we discuss in detail the coefficients of the beta functions to all orders in $g$.  
In the second part of the chapter, we study the property of the CFT at the IR fixed point.
In section \ref{sec:OPE}, we compute the dimensions of the bilinear primary operators and the corresponding OPE coefficients for $d\ne 1, 2$.
In section \ref{sec:d=3}, we detail the case $d=3$.
As the cases $d=2$ and $d=1$ are special, we study them in section \ref{sec:d=2}.
In section \ref{sec:character}, we use representation theory to derive the spectrum of bilinear primary operators in the free theory for integer dimension $d=3,2,1$. We summarize our results and conclude in section \ref{sec:conclusions}.
In appendix \ref{app:melon}, we detail the computation of the melon integral. In appendices \ref{ap:eigenvalues} and  \ref{app:measure}, we give some technical details on CFT data. In appendix \ref{app:free}, we further comment on the free theory.
Lastly, in appendix \ref{app:original syk}, we review the OPE coefficients of the original SYK model and those of the conformal SYK model of Gross and Rosenhaus \cite{Gross:2017vhb}.

\subsection{The bosonic CTKT model}
\label{sec:modelCTKT}

We will deal in this chapter with a modified version of the $O(N)^3$ model of Giombi, Klebanov and Tarnopolsky \cite{Giombi:2017dtl} which itself is a bosonic version in $d$ dimension of the CTKT model presented in section \ref{sec:other_melonic}.
We consider an uncolored model with real tensor field of rank $3$, $\phi_{a_1a_2 a_3}(x)$ with $O(N)^3$ symmetry. We denote $\mba = (a_1,a_2,a_3)$.
The action of the model is:
\be\label{eq:actionCTKT} 
\begin{split}
    S[\phi]  & =   \frac{1}{2} \int d^dx \;   \phi_{\mba}(x) (   - \Delta)^{\zeta}\phi_{\mba}(x) + S^{\rm int}[\phi]\;,\crcr
   S^{\rm int}[\phi]  & = \frac{ m^{2\zeta}}{2} \int d^dx \;   \phi_{\mba}(x) \delta_{\mba \mbb} \phi_{\mbb}(x)  + 
   \frac{ \lambda }{4 N^{3/2}} \int d^d x \;   \delta^t_{\mba \mbb\mbc\mbd} \; \phi_{\mba}(x) \phi_{\mbb}(x)  \phi_{\mbc}(x) \phi_{\mbd }(x)\crcr
       & \qquad +   \int d^d x \;  \left( \frac{ \lambda_p }{4 N^{2}} \; \delta^p_{\mba\mbb; \mbc\mbd} +  \frac{ \lambda_d }{4 N^{3}}  \; \delta^d_{\mba\mbb; \mbc\mbd } \right) \; \phi_{\mba}(x) \phi_{\mbb}(x)  \phi_{\mbc}(x) \phi_{\mbd }(x) 
     \; ,     
\end{split}
\ee 
where $\Delta=  \partial_{\mu}\partial^{\mu}$,  $\delta_{\mba \mbb}  = \prod_{i=1}^3 \delta_{a_i b_i} $ and:
\begin{equation}
\begin{split}
    \delta^t_{\mba \mbb\mbc\mbd} & = \delta_{a_1 b_1}  \delta_{c_1 d_1} \delta_{a_2 c_2}  \delta_{b_2 d_2 } \delta_{a_3 d_3}   \delta_{b_3 c_3} \;  , \quad
  \delta^p_{\mba\mbb; \mbc\mbd }= \frac{1}{3} \sum_{i=1}^3  \delta_{a_ic_i} \delta_{b_id_i} \prod_{j\neq i}  \delta_{a_jb_j}  \delta_{c_jd_j} \;,
 \crcr
  \delta^d_{\mba\mbb; \mbc\mbd } & = \delta_{\mba \mbb}  \delta_{\mbc \mbd} \;, 
  \end{split}
  \label{eq:delta_invariants}
\end{equation}
where $t$ stands for \emph{tetrahedron}, $d$ for \emph{double-trace} and $p$ for \emph{pillow} pattern of contraction. The interaction invariants are graphically represented in figure~\ref{fig:trace_invariants}. Because it plays a special role below, we have distinguished the coupling $\lambda$ of the tetrahedral invariant and did not assign any subscript to it.

The CTKT model is obtained for $\zeta=1$, but we will allow here a non-trivial power of the Laplacian $\frac{d}{4} \leq \zeta \leq 1$, making the model long-range. Unlike the fermionic CTKT model in one dimension (see \eqref{eq:fermionCTKT}), where one keeps only the tetrahedral interaction, in higher dimensions we have to include all the terms demanded by
perturbative renormalizability, hence the mass, pillow, and double-trace terms in \eqref{eq:actionCTKT}.

To simplify the notation, we sometimes denote $A=(\mba,x)$, $\delta_{AB} = \delta_{(\mba,x) (\mbb,y)} = \delta_{\mba \mbb} \delta(x-y)$ and $\delta(x-y) = \delta_{xy}$.
We denote bilocal operators by bold face. For instance the covariance of the theory $ \mbC$ is $\mbC_{AB} = \mbC_{\mba \mbb}(x,y)  \equiv \delta_{\mba \mbb} \;  C(x,y)$ with $C(x,y)$ given in \eqref{eq:cov3} as in the previous chapter. 

At large $N$ the theory simplifies significantly: the partition function and correlations admit a $1/N$ expansion, as we will now recall.

\subsection{Feynman graphs and large-$N$ expansion}
\label{sec:model-graphs}

\paragraph{Feynman graphs}

The free energy (and the connected $n$-point functions) of the theory can be expanded in connected Feynman graphs $\cG$. 
We will actually use two types of graphs: 4-colored graphs and ordinary Feynman graphs.

\begin{figure}[ht]
\begin{center}
\includegraphics[width=0.3\textwidth]{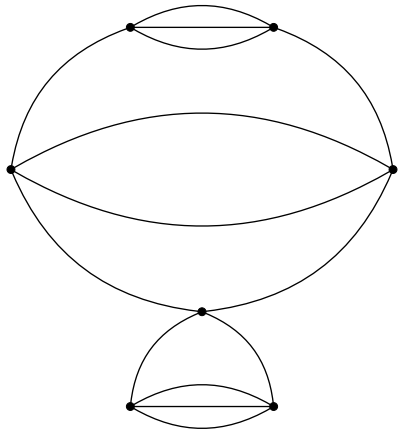}
 \caption{A melon tadpole graph, where all the invariants have been shrunk to point-like vertices.} \label{fig:melontadpoles}
 \end{center}
\end{figure}

The ordinary Feynman graphs are obtained by shrinking each interaction bubble to a point (appropriately colored in order to still distinguish the different interaction bubbles, if necessary). An example is given in figure~\ref{fig:melontadpoles}, where however we omit the colors of the vertices.

While the ordinary Feynman graphs are simpler, and they are sufficient for representing Feynman integrals (which we will do later), the 4-colored graphs are useful for identifying the correct powers of $N$.
In fact, in a 4-colored graph, as explained in section \ref{sec:uncolored}, there is a free sum, that is a factor $N$, per face. 
We denote $n_t(\cG)$, $n_p(\cG)$ and $n_d(\cG)$ the numbers of tetrahedral, pillow, and double-trace bubbles, and $F(\cG)$ the number of faces of $\cG$. We associate a variable $x_v$ to each bubble in $\cG$. The free energy of the model is:
\begin{align}
 {\mathcal F} & =-  \ln \bigg\{  \int [d\phi]\; e^{-S[\phi]}\bigg\} \crcr
 & = \sum_{\cG} N^{F -\frac{3}{2}n_t - 2 n_p - 3n_d}  \frac{\lambda^{n_t}}{ n_t! 4^{n_t}} \frac{\lambda_p^{n_p}}{ n_p! 12^{n_p}}   \frac{\lambda_d^{n_d}}{ n_d! 4^{n_d}} (-1)^{n_t+n_p+n_d+1} A(\cG) \int_x 1  \;, \crcr
 A(\cG)& =  
 \int \prod_{ v\neq v_0 } dx_{v} \prod_{e\in \cG } C(x_e,y_e) \;,
\end{align}
where $\cG$ runs over connected vacuum 4-colored graphs with labeled tensor vertices, $v_0$ is an arbitrary root vertex, and $x_e$ and $y_e$ denote the positions of the end vertices of the edge $e$. 

\begin{figure}[htbp]
\centering
\captionsetup[subfigure]{labelformat=empty}
\subfloat[]{\includegraphics[scale=0.75]{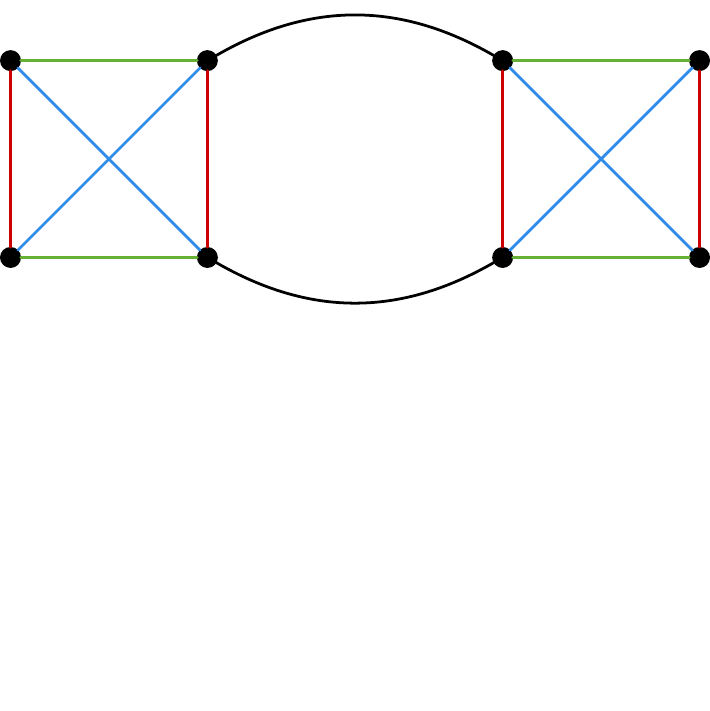}}
\hspace{1cm}
\subfloat[]{\includegraphics[scale=0.75]{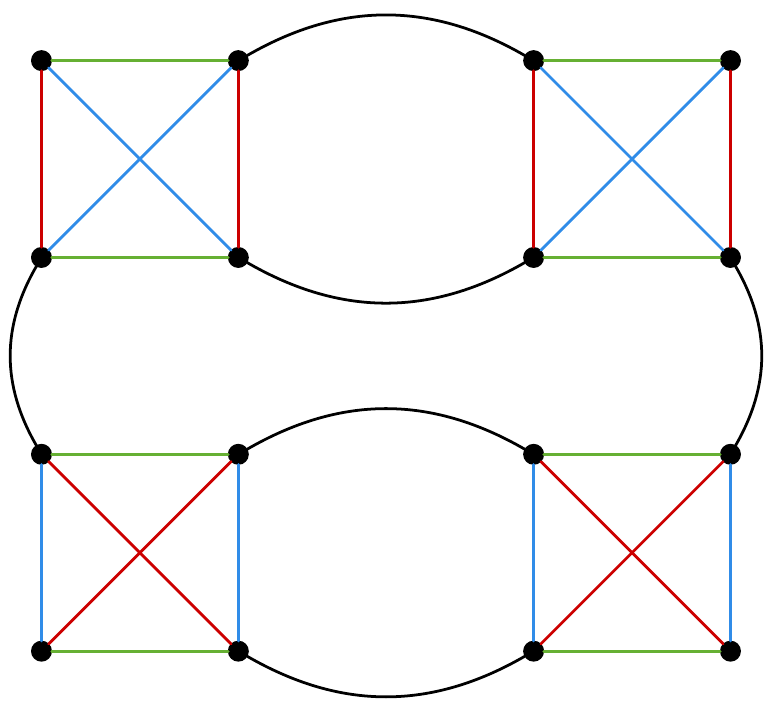}}
\caption{Two Feynman graphs, with external tensor contractions equivalent to the pillow (left)
and double-trace (right) invariants.}
\label{fig:radiative_corr}
\end{figure}

\paragraph{The $1/N$ expansion.}
The model has a $1/N$ expansion\cite{Carrozza:2015adg,Klebanov:2016xxf}. The simplest way to see this is to observe that pillow and double-trace vertices can be obtained as radiative corrections from the tetrahedral vertex: the pillow is a rung (figure \ref{fig:radiative_corr}, left), and the double-trace 
is a ladder made out of two rungs with different color inside their loop (figure \ref{fig:radiative_corr}, right). Replacing the pillow and double-trace vertices in a graph by their minimal resolution in terms of tetrahedral vertices one associates to any graph $\cG$ 
a graph $\hat \cG$ having \emph{only} tetrahedral vertices but the same scaling in $N$:
\[
 F(\cG) -\frac{3}{2}n_t(\cG) - 2 n_{p}(\cG) - 3n_{d}(\cG) =  F(\hat \cG) -\frac{3}{2}n_t( \hat \cG) \;.
\]

We then showed in section \ref{sec:uncolored}, that $\hat \cG$ has degree zero if and only if it is melonic. The original graph $\cG$ has then also degree zero if and only if $\hat \cG$ is melonic.
That is the leading order graphs are melonic \emph{after} substituting all the pillows and double-trace vertices by their minimal realizations in terms of 
the tetrahedral vertex. In terms of the original interactions in $\cG$, one gets \emph{melon tadpole} \cite{Benedetti:2017qxl} graphs, that is 
graphs obtained by iterated insertions of melons or tadpoles into melons or tadpoles,
see figure~\ref{fig:melontadpoles}. Observe that all the tadpoles are based on
either pillow or double-trace vertices, while the end vertices of the melons are tetrahedral.

\subsubsection{The 2PI effective action}
\label{sec:model-2PI}

The two-particle irreducible (2PI) effective action formalism is particularly well adapted to the tensor $1/N$ expansion \cite{Benedetti:2018goh}.\footnote{In fact, the $1/N$ expansion offers a controlled way of implementing as a proper expansion scheme the so-called $\Phi$-derivable truncations studied for example in \cite{Blaizot:2003br,Berges:2005hc,Blaizot:2010zx}.}
First of all, observe that $S[- \phi] = S[\phi]$, hence the odd-point functions are zero in the absence of spontaneous symmetry breaking, which we will assume in the following. 
We define the generating function with bilocal source $\mbK_{AB}= \mbK_{\mba \mbb}(x,y)$:\footnote{We omit the source term linear in the fields in order to keep the presentation concise, as the bilinear term is enough for our purposes. For the general construction see \cite{Benedetti:2018goh}.}
\be \label{eq:W-2PI}
 e^{ W [\mbK]} = \int d\mu_{ \mbC  }( \phi )\; e^{-S^{\rm int}[\phi] + \frac{1}{2} \phi_A \mbK_{AB} \phi_B}   \;, 
\ee
where $d\mu_{ \mbC  }$ denotes the normalized Gaussian measure with covariance $\mbC$.
The source effectively shifts the inverse covariance: $\mbC^{-1}\to\mbC^{-1}-\mbK$.
Taking into account that the odd-point functions are still zero in the presence of the source, 
the derivatives of $W$ write in terms of the connected two and four-point functions with source $\mbK$:\footnote{Note that the derivative of a symmetric function $\mbK_{AB} = \mbK_{BA}$ with respect to itself is the projector on symmetric functions
\[ \frac{\delta \mbK_{AB}}{\delta \mbK_{EF}} = {\cal S}_{AB;EF} = \frac{1}{2} (\delta_{AE} \delta_{BF} + \delta_{AF} \delta_{BE}) \;.\]}

\begin{equation}
\begin{split}
2 \frac{\delta  W}{ \delta \mbK_{AB}}  = & \braket{\phi_{A} \phi_{B} }^c_{\mbK} \;, \crcr
4 \frac{\delta^2  W}{ \delta \mbK_{AB} \delta \mbK_{EF} }  = & \braket{\phi_{A}  \phi_{B} \phi_{E} \phi_{F} }^c_{\mbK}  + 
  \braket{\phi_{A}  \phi_{E} }^c_{\mbK} \braket{  \phi_{B} \phi_{F} }^c_{\mbK} +
  \braket{\phi_{A }   \phi_{F} }^c_{\mbK} \braket{  \phi_{B } \phi_{E} }^c_{\mbK} \;.
  \end{split}
\label{eq:Wderiv}
\end{equation}

Setting $\mbK=0$, one recovers the connected two- and four-point functions of the original theory.
Inverting $ 2 \frac{\delta  W}{ \delta \mbK_{AB}}  = \mbG_{AB} $ yields the source $\mbK[\mbG] $ which ensures that the connected two-point function is exactly $\mbG_{AB}$. 
The Legendre transform of $W$ is:
\be \label{eq:GammaG}
  \Gamma [ \mbG]   =    \bigg\{ -   W[\mbK] + \frac{1}{2} \Tr[\mbG \mbK] \bigg\}_{ \mbK = \mbK[\mbG]}  \;,
\ee
where $\Tr$ denotes a trace over both the indices and the positions. The derivatives of $\Gamma$ are:
\begin{align*}
 &  \frac{\delta  \Gamma}{\delta \mbG_{AB}} = \frac{1}{2} \mbK_{AB} 
 \;, \qquad \frac{\delta^2   \Gamma}{\delta \mbG_{AB} \delta \mbG_{EF}}   =   \frac{1}{2} \frac{\delta \mbK_{AB}}{\delta \mbG_{EF}} 
 = \frac{1}{2} \left(  \frac{\delta \mbG  }{\delta K } \right)^{-1}  = M^{-1} \;,
  \crcr
 & M_{ (AB) ; (EF)}   = 4 \frac{\delta^2   W}{ \delta \mbK_{AB}\delta \mbK_{EF}}  
 =  \braket{\phi_{A}  \phi_{B} \phi_{E} \phi_{F} }^c_{\mbK[\mbG]}  + 
  \mbG_{AE} \mbG_{BF}  + \mbG_{AF} \mbG_{BE} \;. 
\end{align*} 
The field equations $ \frac{\delta \Gamma}{\delta \mbG} =  0$ are equivalent to $\mbK =0$,   and we denote their solution $\bar \mbG$. 
The on-shell two-point function is diagonal in the tensor indices 
$ \bar \mbG_{AB} = \delta_{\mba \mbb} \bar G(x,y)$. 

Let us denote $ - \Gamma^{2PI}[\mbG]$ the sum of non-trivial vacuum 2PI graphs (i.e. which do not disconnect by cutting two edges) with vertices defined by $ S[\phi]$ and with propagators given by $\mbG$. 
The self-energy $\mbS$ (the sum of non-trivial one-particle irreducible two-point graphs with propagator $\mbC$) can be obtained as:
\be
\mbS_{AB}[\mbG] =  - 2 \frac{\delta \Gamma^{2PI}[\mbG]}{\delta \mbG_{AB}}  \;, 
\ee
where the derivative selects and cuts an edge and the factor $2$ counts the ways to attach it to the external points.
The derivative of the self-energy with respect to the two-point function yields the amputated 2PI four-point kernel \cite{Berges:2004yj}. The 2PI irreducible kernel amputated to the right only is:
\be
 {\cal K}_{A'B' ; EF} =  \mbG_{A' A} \mbG_{ B' B} \frac{\delta \mbS_{AB}}{ \delta \mbG_{EF} } \;.
\ee

The full two-point function obeys the Schwinger-Dyson equation $\mbG^{-1} = \mbC^{-1} - \mbK[\mbG] - \mbS[\mbG]$.
Solving for $\mbK$, we get $
 \frac{\delta  \Gamma}{\delta \mbG} = \frac{1}{2} \mbK  = \frac{1}{2} \mbC^{-1} -  \frac{1}{2}  \mbG^{-1} + \frac{\delta \Gamma^{2PI}}{\delta \mbG} \;,
$
and:
\begin{align}
  \Gamma[\mbG] & = \frac{1}{2} \Tr[ \mbC^{-1} \mbG ] -\frac{1}{2} \Tr\ln(\mbG) +  \Gamma^{2PI}[ \mbG ] \;, \\
 \frac{\delta^2   \Gamma}{\delta \mbG_{AB} \delta \mbG_{EF}}  & =   \frac{1}{2} \frac{\delta \mbK_{AB}}{\delta \mbG_{EF}}  = \frac{1}{2} \mbG^{-1}_{AA'} \mbG^{-1}_{BB'} \bigg( {\cal S} - {\cal K} \bigg)_{A'B'; EF} \;,
\end{align}
with ${\cal S}$ the projector on symmetric functions. Now, as the kernel ${\cal K}_{A'B';EF}$ is symmetric in $A'B'$ (and in $EF$), we have ${\cal K} = {\cal S} {\cal K}$,
and using \eqref{eq:Wderiv} we get:
\begin{equation} \label{eq:4point}
 \braket{\phi_{A}  \phi_{B} \phi_{E} \phi_{F} }^c_{\mbK[\mbG]}  =2 \left( \frac{{\cal K}}{1 - {\cal K}} {\cal S} \right)_{ AB; E'F' }  G_{E'E}  G_{F'F} \;.
\end{equation}

The terms of the 2PI action can be organized in powers of $1/N$. The scaling in $N$ of a term is obtained by substituting for the two-point function its on-shell value $\delta_{\mba \mbb} \bar G(x,y)$.
At leading and next-to-leading order in $N$, the combination of the $1/N$ expansion and the 2PI condition leads to a finite number of graphs:
\begin{itemize}
 \item leading order ($N^3$): a graph with a mass two-valent vertex and one edge, a melon with two tetrahedral vertices, a double-tadpole with the pillow vertex and one with the double-trace vertex,
\item  next-to-leading order ($N^{5/2}$): three double-tadpoles with the tetrahedral vertex (the three possible choices for closing a tadpole are distinguished by the coloring of the tetrahedron).
\end{itemize}
Thus at leading and  next-to-leading order we get \cite{Benedetti:2018goh}:
\begin{equation}
\begin{split}
 - \Gamma^{2PI}[\mbG] = & - \frac{m^{2\zeta}}{2} \Tr[\mbG]  - \frac{\lambda_p}{4N^2} \int_x \mbG_{( \mba,x)(\mbb,x)} \delta^p_{\mba\mbb ; \mbc\mbd} \mbG_{(\mbc,x)(\mbd,x)} - 
  \frac{\lambda_d}{4N^3} \int_x \mbG_{( \mba,x)(\mbb,x)} \delta^d_{\mba\mbb ; \mbc\mbd} \mbG_{(\mbc,x)(\mbd,x)} \\
  & + \frac{1}{2} \left( \frac{\lambda}{4 N^{3/2}}\right)^2 4 \int_{x,y} \delta^t_{\mba\mbb \mbc\mbd} \delta^t_{\mba'\mbb' \mbc'\mbd'}
  \mbG_{( \mba, x)(\mba',y)} \mbG_{( \mbb, x)(\mbb',y)}  \mbG_{( \mbc, x)(\mbc',y)} \mbG_{( \mbd, x)(\mbd',y)}  \\
  & -  \frac{\lambda}{4N^{3/2}}   \int_x \mbG_{( \mba,x)(\mbb,x)} \mbG_{(\mbc,x)(\mbd,x)} \bigg(  \delta^t_{\mba\mbb   \mbc\mbd} +  \delta^t_{\mba  \mbc \mbb  \mbd}  +  \delta^t_{\mba \mbc \mbd   \mbb}  \bigg)\;,
  \end{split}
\end{equation}
where the first two lines are leading order and the last one is next-to-leading order. The self-energy is:
\begin{equation}
\begin{split}
  \mbS_{(\mba,x)(\mbb,y)}  = & -m^{2\zeta} \delta_{\mba \mbb}\delta_{xy} - \frac{\lambda_p}{N^2}  \delta_{xy} \delta^p_{\mba\mbb ; \mbc\mbd} \mbG_{(\mbc,x)(\mbd,x)} 
 - \frac{\lambda_d}{N^3}  \delta_{xy} \delta^d_{\mba\mbb ; \mbc\mbd} \mbG_{(\mbc,x)(\mbd,x)} \crcr
    & + \frac{\lambda^2}{N^3} \delta^t_{\mba\mbc_1 \mbc_2\mbc_3} \delta^t_{\mbb \mbd_1 \mbd_2\mbd_3} \mbG_{( \mbc_1, x)(\mbd_1,y)}  \mbG_{( \mbc_2, x)(\mbd_2,y)} \mbG_{( \mbc_3, x)(\mbd_3,y)} \crcr
    & -  \frac{\lambda}{N^{3/2}} \delta_{xy}\bigg(  \delta^t_{\mba\mbb   \mbc\mbd} +  \delta^t_{\mba  \mbc \mbb  \mbd}  +  \delta^t_{\mba \mbc \mbd   \mbb}  \bigg)  \mbG_{(\mbc,x)(\mbd,x)} \;.
\end{split} 
\end{equation}

Finally, the four-point kernel at leading and next-to-leading order is:
\begin{equation} \label{eq:kernel}
\begin{split}
& {\cal K}_{ (\mba',x')(\mbb',y') ; (\mbc,z)(\mbd,t) }  =  \mbG_{(\mba',x') (\mba,x)  }\mbG_{(\mbb',y') (\mbb,y)  } \bigg[
    - \frac{\lambda_p}{N^2}  \delta_{xy} \delta_{xz} \delta_{xt} \delta^p_{\mba\mbb ; \mbc\mbd} - \frac{\lambda_d}{N^3}  \delta_{xy} \delta_{xz} \delta_{xt} \delta^d_{\mba\mbb ; \mbc\mbd}  \crcr
 & \qquad + \frac{\lambda^2}{N^3} \delta^t_{\mba\mbc_1 \mbc_2\mbc_3} \delta^t_{\mbb \mbd_1 \mbd_2\mbd_3}    \sum_{i=1}^3
     \left( \frac{1}{2}\delta_{xz}\delta_{yt} \delta_{\mbc_i \mbc} \delta_{\mbd_i \mbd} + \frac{1}{2}\delta_{xt}\delta_{yz}  \delta_{\mbc_i \mbd} \delta_{\mbd_i \mbc}  \right) 
    \prod_{j\neq i} 
    \mbG_{( \mbc_j, x)(\mbd_j,y)} \crcr
&  \qquad -  \frac{\lambda}{N^{3/2}} \delta_{xy}  \delta_{xz} \delta_{xt} 
    \frac{ \bigg(  \delta^t_{\mba\mbb   \mbc\mbd} +\delta^t_{\mba\mbb   \mbd\mbc} +  \delta^t_{\mba  \mbc \mbb  \mbd} +  \delta^t_{\mba  \mbd \mbb  \mbc} 
    +  \delta^t_{\mba \mbc \mbd   \mbb} + \delta^t_{\mba \mbd \mbc   \mbb}  \bigg) }{2}
 \bigg] \;.
\end{split} 
\end{equation}

Below we will be interested in evaluating the two- and four-point functions on-shell where $ \bar \mbG_{AB} = \delta_{\mba \mbb} \bar G(x,y)$.
We will drop the bar on $G(x,y)$ in order to simplify the notation.

\subsection{Renormalization}
\label{sec:RG}
 
\paragraph{Motivation.} We consider $d < 4$. According to \cite{Klebanov:2016xxf,Giombi:2017dtl}, the model in \eqref{eq:actionCTKT} with $\zeta =1$ should have a non-trivial conformal infrared limit.
We aim to study rigorously this putative conformal infrared limit. In the IR, the full two-point function is expected to acquire a non-trivial scaling behavior $G(p) \sim p^{- d/2 }$. 
Using the full two-point function as propagator, the theory with interaction $S^{\rm int}$ in \eqref{eq:actionCTKT} exhibits \emph{ultraviolet} divergences 
in any $d$: the two-point\footnote{As we use the resummed two-point function, the graphs of the effective theory do not have any two-point subgraphs. However, the full two-point function must satisfy the Schwinger-Dyson equation which does exhibit ultraviolet divergences.} and four-point graphs are ultraviolet divergent.
In order to make sense of the infrared theory one has two options:
\begin{itemize}
 \item set from the beginning  $ \zeta = d / 4 $ that is start from a bare covariance
 that reproduces the infrared scaling of the two-point function. In the SYK model in one dimension this has been studied by
 Gross and Rosenhaus \cite{Gross:2017vhb}. In $d=3$ (with no tensor indices) the choice $\zeta = 3/4 + \epsilon$ yields the Brydges-Mitter-Scoppola model 
 \cite{Brydges:2002wq,Abdesselam:2006qg}.
 \item argue that the ultraviolet divergences are just an artifact of using the infrared ansatz for the two-point function: for $d < 4$ the free covariance dominates in the ultraviolet, hence the effective two-point function will behave at large momentum as $p^{-2}$. 
\end{itemize}

The second option is very difficult to implement (even non-rigorously) in practice. One would have to consider that the infrared scaling $G(p) \sim p^{-d/2}$ is a good approximation up to some momentum scale $\Lambda$. 
Neglecting the higher momenta makes all the correlation functions depend on (and in fact diverge with) the non-physical dimensionful parameter $\Lambda$. In order to eliminate this dependence\footnote{The same situation arises in quantum electrodynamics. Although we all agree that QED is not a UV complete theory and in the UV one needs to take into account the rest of the standard model, it still makes sense in the infrared to study QED with a cutoff $\Lambda$ and renormalize it. This leads at low energy to some reasonably accurate predictions, like the anomalous magnetic moment of the electron.} one still needs to subtract these divergences using a bare theory at scale $\Lambda$ with bare covariance $C(p) \sim p^{ - d/2}$.

\bigskip

We choose the first option, and from now on we assume $\zeta = d/4$, which corresponds to the marginal scaling of the long-range propagator, although we will keep $\zeta$ arbitrary in most formulas, for convenience and generality. The power counting is then the same as in the previous chapter (see \eqref{eq:power_counting}). The two-point graphs are power divergent  ($\text{deg}(\mathcal{G})=\frac{d}{2}$) in the UV and the four-point graphs are logarithmically divergent in the UV. Graphs with more than six external points are naively UV convergent.

 
\subsubsection{Wilsonian renormalization group} 

In order to access the infrared limit one needs to study the renormalization of the theory. We want to treat rigorously power divergences appearing in the two-point functions, therefore the dimensional renormalization scheme at $\epsilon >0$ of chapter \ref{chap:3loops} is not adapted here. We will instead rely on the Wilsonian renormalization group framework presented in section \ref{subsec:WR}. 

However, as we work here with the 2PI effective action with a non-trivial power of the Laplacian $\zeta$, we will review some essentials in order to clarify our logic and to highlight some subtleties that arise in this case.\footnote{Our presentation essentially follows the formulation of \cite{Morris:1993qb}; see also \cite{Blaizot:2010zx,Carrington:2012ea,Carrington:2014lba} for the functional RG of the 2PI effective action.} We start from \eqref{eq:W-2PI} with an explicit UV cutoff $\Lambda$:
\be 
\begin{split}
 e^{ W [\mbK]} & = \int d\mu_{ C^{\Lambda}}  (\phi) \; e^{-S_{\Lambda}[\phi]+ \frac{1}{2} \phi_A \mbK_{AB} \phi_B} \;,\\
C^{\Lambda}(p) & = \frac{1}{p^{2\zeta}} \, \Theta\left(\frac{p^2}{\Lambda^2}\right)  \; , \qquad C^{\Lambda}(x) =  \int_p \; \frac{e^{-\imath p x}}{ p^{2\zeta} } \,\Theta\left(\frac{p^2}{\Lambda^2}\right)  \; ,
\end{split}
\ee
where $d\mu_{ C^{\Lambda}  }$ denotes the normalized Gaussian measure with covariance $C^{\Lambda}$. We denote by convention $S_{\Lambda}[\phi] \equiv S^{\rm int}[\phi]$ the bare potential of our model \eqref{eq:actionCTKT}.
The ultraviolet divergences are regularized by the multiplicative 
cutoff function $\Theta(p^2/\Lambda^2)$.
While the specific choice of the cutoff function should not affect the main results, we will choose once and for all to use
a normalized upper incomplete Euler gamma function:
\be
  \Theta\left(\frac{p^2}{\Lambda^2}\right) = \frac{\Gamma \left(\zeta; \frac{p^2}{\Lambda^2} \right)}{\Gamma(\zeta)} = \frac{1}{\Gamma(\zeta)}\int_{\frac{p^2}{\Lambda^2}}^{\infty} d\alpha \; \alpha^{\zeta-1} e^{-\alpha} \;,
\ee
which implements a parametric cutoff for the Schwinger parameter $\alpha$, and which, for $\zeta=1$, reduces to the standard exponential cutoff.

Let $k\le \Lambda$ be an infrared scale.
The Wilsonian RG transformation consists in integrating out the modes with momenta between $\Lambda$ and $k$, and then rescaling the momenta by $\Lambda/k$ and the fields by their wave function renormalization in order to re-establish the original free covariance of the leftover modes. 

To be more precise, we introduce the slice cutoff function $\chi^{\Lambda}_k(p) = \Theta\left( p^2/\Lambda^2 \right)- \Theta\left( p^2/k^2 \right) $
and we split the covariance as $C^{\Lambda} = C^{k} + C^{\Lambda}_k$, where $C^{k} $ is the covariance with UV cutoff $k $ and $C^{\Lambda}_k$ is the covariance of the fluctuations (the modes with momenta between $\Lambda$ and $k$):
\be
 C^{\Lambda}_k (p) = \frac{1}{p^{2\zeta}} \,\chi^{\Lambda}_k(p) =  \frac{1}{\Gamma(\zeta)}\int_{\Lambda^{-2}}^{k^{-2}}  d\alpha \; \alpha^{\zeta-1} e^{-\alpha p^2 } \; .
\ee

Associated to the split of the covariance, the Gaussian integral also splits as:
\be \label{eq:1stRGstep}
\begin{split}
 e^{ W [\mbK]}   & = \int d\mu_{C^{\Lambda}}(\phi )  \;e^{ - S_{\Lambda}[\phi] + \frac{1}{2} \phi_A \mbK_{AB} \phi_B } \\
 & = 
\int d\mu_{C^{ k } }(\phi)  \int d\mu_{C^{\Lambda}_k}(\chi) \; 
e^{ - S_{\Lambda} [\phi + \chi ] + \frac{1}{2} (\phi_A+\chi_A) \mbK_{AB} (\phi_B+\chi_B)} \\
&= \int d\mu_{C^{ k }} (\phi ) \;  e^{   W_k [\mbK;\phi ]}  \;,
\end{split}
\ee
where by a change of variables we can write
\be \label{eq:Wk}
e^{   W_k [\mbK;\phi ]} = e^{-\frac{1}{2}\phi_A (\mbC^{\Lambda}_k)^{-1}_{AB}\phi_B } \int d\mu_{C^{\Lambda}_k}(\chi) \; 
e^{ - S_{\Lambda} [ \chi ] + \frac{1}{2} \chi_A \mbK_{AB} \chi_B + \phi_A (\mbC^{\Lambda}_k)^{-1}_{AB} \chi_B } \;.
\ee
Defining a new field $J_A =(C^{\Lambda}_k)^{-1}_{AB} \phi_B$, we recognize 
\be
\hat{W}_k [\mbK,J ] =W_k [\mbK;\mbC^{\Lambda}_k J ] + \frac{1}{2} J_A (\mbC^{\Lambda}_k)_{AB} J_B \;,
\ee
to be the connected generating functional with local and bilocal sources and covariance $C^{\Lambda}_k$.

Performing a double Legendre transform we define the full 2PI effective action (equation \eqref{eq:GammaG} being the special case $J=\phi=0$):\footnote{We distinguish the various effective actions by their arguments: $\Gamma[\phi,\mbG]$ is the full 2PI effective action, while $\Gamma[\mbG]=\Gamma[\phi=0,\mbG]$ is the reduced action presented in section \ref{sec:model-2PI}; and $\Gamma[\phi]=\Gamma[\phi,\bar\mbG]$ is the 1PI effective action. Lastly, we use the subscript $k$ to denote the presence of the Wilsonian cutoff.}
\be \label{eq:GammaGphi}
\Gamma_{k}[\phi,\mbG]  =  - \hat{W}_{k}[\mbK, J ] +   J_A  \phi_A + \frac{1}{2}  \phi_A \mbK_{AB}  \phi_B + \frac{1}{2} \Tr[G \mbK ] 
\; ,
\ee
where on the right-hand side $J$ and $\mbK$ satisfy $\delta \hat{W}/\delta J_A =\phi_A$ and $\delta \hat{W}/\delta\mbK_{AB} = \frac{1}{2} (\phi_A \phi_B + \mbG_{AB})$.

Acting on $\hat{W}_{k}[\mbK, J ]$ with a $k$-derivative, and using \eqref{eq:Wk}, we obtain:
\be
k \partial_k  \hat{W}_{k}[\mbK, J ] = - \Tr\left[ k \partial_k (\mbC_{k}^{\Lambda})^{-1}  \frac{\delta \hat{W}_{k}}{\delta \mbK}\right] \;.
\ee
Next, acting with a $k$-derivative on \eqref{eq:GammaGphi}, we find that the 2PI effective action satisfies the flow equation:
\be  \label{eq:WE} 
 k \partial_k \Gamma_{k}[\phi,G] =  \frac{1}{2}  \Tr\left[ k \partial_k (C_{k}^{\Lambda})^{-1}  G\right]+ \frac{1}{2} \phi_A k \partial_k (C_{k}^{\Lambda})^{-1}_{AB}  \phi_B \; .
\ee
Going on-shell for the two-point function, i.e.\ setting $G=G_k^{\Lambda}$ directly in the flow equation (which is a valid operation since by definition $\delta\Gamma_k/\delta G|_{G=G_k^{\Lambda}} =0$), one recovers the flow equation for the 1PI effective action $\Gamma_{k}[\phi]$ of \cite{Wetterich:1992yh} (see also \cite{Ellwanger:1993mw,Morris:1993qb}).

Expanding the 1PI effective action in powers of the field, and using translation invariance, we write:
\be
\begin{split}
\Gamma_{k}[\phi] & = \sum_{n\geq 0} \frac{1}{n!} \int_{x_1,\ldots, x_n} \Gamma_{k}^{(n)}(x_1,\ldots,x_n) \phi(x_1) \cdots \phi(x_n)\\
&= \sum_{n\geq 0} \frac{(2\pi)^d}{n!} \int_{p_1,\ldots, p_n} \Gamma_{k}^{(n)}(p_1,\ldots,p_n) \phi(p_1) \cdots \phi(p_n) \delta(p_1+\ldots+p_n) \;.\end{split}
\ee

A hierarchy of equations for the $n$-point functions $\Gamma_{k}^{(n)}(x_1,\ldots,x_n)$ is obtained by acting with derivatives on \eqref{eq:WE}, and using the fact that
\be
G_{k}^{\Lambda} = \left(\frac{ \delta^2 \Gamma_{k}[\phi] }{ \delta \phi \delta \phi } \right)^{-1} \;,
\ee
and hence
\be
\frac{ \delta G_{k}^{\Lambda}(x,y)}{ \delta \phi(z) } = - \int_{u,v} G_{k}^{\Lambda}(x,u) \frac{\delta^3 \Gamma_{k}[\phi] }{ \delta \phi(u) \delta \phi(v) \delta \phi(z) } G_{k}^{\Lambda}(v,y) \;.
\ee

Assuming as above that the theory has $\mathbb{Z}_2$ invariance, and thus that the $n$-point functions vanish for odd $n$, we obtain
\be \label{eq:Gamma2-FRG}
k\partial_k \Gamma_{k}^{(2)} = - \frac12  \Tr\left[k\partial_k (C_{k}^{\Lambda})^{-1}  G_{k}^{\Lambda} \Gamma_{k}^{(4)}  G_{k}^{\Lambda}\right]+ k\partial_k (C_{k}^{\Lambda})^{-1} \; ,
\ee
\be \label{eq:Gamma4-FRG}
k\partial_k \Gamma_{k}^{(4)} = - \frac12  \Tr\left[k\partial_k (C_{k}^{\Lambda})^{-1}  G_{k}^{\Lambda} \Gamma_{k}^{(6)}  G_{k}^{\Lambda}\right]+ \sum_{``s,t,u\; {\rm channels}''}   \Tr\left[k\partial_k (C_{k}^{\Lambda})^{-1}  G_{k}^{\Lambda} \Gamma_{k}^{(4)}  G_{k}^{\Lambda} \Gamma_{k}^{(4)}  G_{k}^{\Lambda}\right]\; ,
\ee
and so on.

The second term in \eqref{eq:Gamma2-FRG} is often eliminated by redefining $\Gamma_{k}[\phi] =  \frac{1}{2} \phi_A  (C_{k}^{\Lambda})^{-1}_{AB}  \phi_B + \hat\Gamma_{k}[\phi]$. 
The new two-point vertex can be identified with the self-energy,  $\hat\Gamma_{k}^{(2)}(p) = -\Sigma_k(p)$, and we expect for $p\ll k$ the self-energy to contain a term proportional to the kinetic term of the model: $ \Sigma_k(p) = - (Z_k^{-1} - 1) p^{2\zeta} + \dots$. 
Therefore, rescaling by $Z_k$ (together with a rescaling of momenta) is necessary in order to restore the covariance $C^{\Lambda}$ in the last line of \eqref{eq:1stRGstep}, as demanded by the second step of the Wilsonian RG.
The rescaling will not play an important role for us, as our couplings are dimensionless, but we should remember to multiply $\Gamma_{k}^{(n)}$ by $Z_k^{n/2}$.

For tensor models at large $N$, the hierarchy of equations can be closed,\footnote{A closure of this type was considered in \cite{Blaizot:2010zx} but lacking a melonic $1/N$ expansion, that was the result of an arbitrary truncation rather than a controlled expansion.} as all the $n$-point functions with $n>4$ can be expressed in terms of $ \Gamma^{(2)}_k $ and $ \Gamma^{(4)}_k $, and one could attempt to solve the resulting system of (two) equations. 
In particular, the 1PI four-point vertex in the absence of odd vertices is given by (minus) the amputated connected four-point function  \eqref{eq:4point}, i.e.:
\be \label{eq:4point1PI}
\Gamma^{(4)}{}_{ABEF} = - 2 \left( \frac{{\cal K}}{1 - {\cal K}} {\cal S} \right)_{ AB; E'F' }  G^{-1}_{E'E}  G^{-1}_{F'F} \;.
\ee
which, combined with \eqref{eq:Gamma2-FRG}, gives
\be \label{eq:Gamma2-FRG-redux}
k\partial_k \Gamma_{k}^{(2)} =   \Tr\left[k\partial_k (C_{k}^{\Lambda})^{-1} \left( \frac{{\cal S}}{1 - {\cal K}}  \right)\right] \;.
\ee
Such expression is actually generic, but in the melonic large-$N$ limit the kernel $\cal K$ has the closed expression \eqref{eq:kernel} in terms of the full propagator, rather than being an infinite sum over all kernel diagrams. More importantly $ \Gamma^{(6)}_k $ can be expressed in a closed form in terms of the full propagator and $ \Gamma^{(4)}_k $ (by the type of contact and planar diagrams encountered in \cite{Gross:2017aos}), thus closing equation \eqref{eq:Gamma4-FRG}.

However, such expressions for $ \Gamma^{(4)}_k $ and $ \Gamma^{(6)}_k $ are of limited use here: first, they require renormalization; second, they involve a summation over an infinite series of diagrams.
For these reasons, we will not use explicitly such equations, although it is useful to keep in mind that our construction is implicitly related to them.

First, we will deal with the Schwinger-Dyson equation $\Gamma_{k}^{(2)} = (C_{k}^{\Lambda})^{-1}-\Sigma[G_k^{\Lambda}]$. Once we have obtained the full renormalized two-point function, we will construct and renormalize the four-point vertex \eqref{eq:4point1PI}, from which we will define the effective couplings, and lastly their beta functions.

Before embarking into that, we should further comment on two non-trivial issues raised by the RG framework.

\paragraph{Wave function renormalization.}

We explained in chapter \ref{chap:3loops}, that for long-range models there is no anomalous field dimension. Nevertheless, the RG flow generates a finite wave function renormalization $Z_k$.
For a theory with $\zeta=1$, the wave function is obtained by a Taylor expansion in $p$ of the two-point function, but this can not work in our case, $\zeta=d/4<1$, because Taylor expansions generate only integer powers of momenta. We will see in section~\ref{sec:2point} how the wave function comes about for non-integer powers of momenta.

\paragraph{Subtraction at zero momentum.} Although our theory is massless, we perform a subtraction at zero momentum: the effective dimensionless four-point coupling is 
$ Z_k^{2}  \Gamma^{(4)}_k \left( 0 ,0 ,0,0 \right)  $. In the Wilsonian picture described above, this is built in, as explained in section \ref{subsec:WR}.
 
%

However, this comes at a price: some facts one usually takes for granted when it comes to renormalization need to be revisited.  In particular, one expects that the coefficients of the beta function have finite limits when the UV cutoff is lifted to infinity. These coefficients turn out to be sums over amplitudes of graphs renormalized by the BPHZ subtraction \cite{bogoliubow1957multiplikation,Hepp:1966eg,Zimmermann:1969jj} operator and  the fact that they are finite is simply the statement of the 
BPHZ theorem. However, the  BPHZ theorem does not apply as it only works for massive theories, and Lowenstin's extension \cite{Lowenstein:1974qt,Lowenstein:1975rg} can't be used either as it does not subtract at zero momentum. The zero momentum subtraction in massless theories is much more involved: one needs to use multiscale analysis \cite{Rivasseau:1991ub} and the classification of inclusion forests in order to show that the subtracted amplitudes are indeed finite.

\subsection{The two-point function}
\label{sec:2point}

We first consider (formally) the theory without cutoffs. The covariance,
on-shell two-point function and on-shell self-energy are diagonal in the tensor indices $\mbC_{AB} = \delta_{\mba \mbb} C(x,y)$,
$ \bar \mbG_{AB} = \delta_{\mba \mbb} G(x,y)$ and $\mbS_{AB} = \delta_{\mba \mbb} \Sigma(x,y)$. 
The on-shell Schwinger-Dyson equation becomes at leading and next-to-leading order in $1/N$: 
\be
 \Sigma(x,y) = -m^{2\zeta} \delta_{xy}  - (\lambda_p +\lambda_d) \delta_{xy} G(x,x) + \lambda^2 G(x,y)^3 - 3 \frac{\lambda}{N^{1/2}} \delta_{xy} G(x,x) \;, \quad G^{-1} = C^{-1} - \Sigma \;.
\ee
Taking $N\to \infty$, this becomes in momentum space:
\be\label{eq:SDECTKT}
\begin{split}
  \Sigma(p) & = - m^{2\zeta}    +     \lambda^2    \int_{q_1,q_2}    G(q_1)  G(q_2)  G( p +q_1 + q_2  )    - ( \lambda_p + \lambda_d ) \int_{q}    G(q) \;,
\crcr
   G(p)^{-1} &  =   C(p)^{-1} -  \Sigma(p) \; .
\end{split}
\ee 
For $C(p)^{-1}=p^2$ (i.e.\ for $\zeta=1$), a simple power counting argument indicates that the solution admits two regimes \cite{Klebanov:2016xxf}: a free scaling regime in the ultraviolet $G(p)^{-1}\sim p^2$ (with $C(p)^{-1}$ dominating over $\Sigma(p)$), and an anomalous scaling regime in the infrared $G(p)^{-1}\sim p^{d/2}$ (with $\Sigma(p)$ dominating over $C(p)^{-1}$).
For the reasons discussed above, we choose $\zeta\neq 1$ to match the infrared conformal behavior.
In fact, with $C(p)^{-1}=p^{2\zeta}$, the Schwinger-Dyson equation is formally solved by $ G(p)^{-1} =  p^{2\zeta}/\mathcal{Z}$\footnote{Throughout this thesis we will denote $\mathcal{Z}$ the coefficient of the full two-point function and $Z$ the usual wave function renormalization.} with $\zeta = d/4$:
\be\label{eq:formal}
  \mathcal{Z}^{-1} p^{2\zeta} = p^{2\zeta} +  m^{2\zeta} - \mathcal{Z}^3\lambda^2   \int_{q_1,q_2}   \;\frac{1}{q_1^{2\zeta}} \; \frac{1}{q_2^{2\zeta}} \; \frac{1}{ ( p +q_1 + q_2  )^{2\zeta} } +  \mathcal{Z}\left(  \lambda_p + \lambda_d \right)   \int_{q} \;\frac{1}{ q^{2\zeta} }  \; ,
\ee 
as the double integral (which we call the  melon integral) gives, after a rescaling of $q_1$ and $q_2$ by $|p|$, a global $|p|^{2d - 6\zeta} = |p|^{2\zeta}$. Differently from \cite{Klebanov:2016xxf}, there is only one regime: $\Sigma(p)$ and $C(p)^{-1}$ are of the same order in $p$.
The problem is that both integrals in \eqref{eq:formal} are divergent, thus we need regularization and renormalization.

Using the slice propagator\footnote{Thus $\lambda, \lambda_{p},\lambda_{d}$ and $m^{2\zeta}$ become the bare couplings and mass at scale $\Lambda$.} $C^{\Lambda}_k(p) = C(p)\chi^{\Lambda}_k(p)$ and denoting the self-energy and the two-point  function with cutoffs $\Sigma^{\Lambda}_k(p)$ and $G^{\Lambda}_k(p)$, the Schwinger-Dyson equation with cutoffs becomes:
\begin{equation}
\begin{split}\label{eq:SDEcutoff}
 G^{\Lambda}_k(p) & = \frac{1}{ C(p)^{-1} - \chi^{\Lambda}_k(p) \Sigma^{\Lambda}_k(p) } \, \chi^{\Lambda}_k(p)  \equiv
 G\left( p ; \Lambda , k \right) \chi^{\Lambda}_k(p) \;, \\
  G\left( p ; \Lambda , k \right) ^{-1} & =  C(p)^{-1} - \chi^{\Lambda}_k(p) \bigg[ - m^{2\zeta}  -  \left( \lambda_{p} + \lambda_{d}  \right)  \int_{q}  G^{\Lambda}_k(q_1)   \\
  & \qquad \quad   +  \lambda^2  \int_{q_1,q_2} G^{\Lambda}_k(q_1) G^{\Lambda}_k(q_2)   G^{\Lambda}_k( p +q_1 + q_2  )   \bigg] \; . 
\end{split} 
\end{equation}
The first equation shows that the two-point function is proportional to the cutoff.

Let us step back once more for a moment and consider again the case $C(p)^{-1}=p^2$: the textbook observation is that at fixed 
$\Lambda$ and $k$, $ G\left( p ;\Lambda,k \right)^{-1}$  is analytic around $p=0$, hence:
\[
 G\left( p ;\Lambda,k \right)^{-1} = G\left( 0 ;\Lambda,k \right)^{-1} +  Z_k p^2 + O(p^4) \;,
\]
and one can extract the wave function renormalization $Z_k$.
Such Taylor expansion (known as the derivative expansion)
has a finite radius of convergence in $p/k$, hence it fails in the $k\to 0$ limit,
but on general grounds (see for example \cite{Delamotte:2007pf}) we expect that:
\begin{itemize}
 \item for $ k \ll p$  the inverse two-point function behaves like $G^{\Lambda}_k(p)^{-1} \sim p^{2-\eta}$,
 \item for $ p \ll k$  the inverse two-point function behaves like $G^{\Lambda}_k(p)^{-1} \sim k^{-\eta} p^2 $,
\end{itemize}
where $\eta$ is the anomalous field dimension. Therefore, in order to extract $\eta$ it is typically enough to obtain the scaling behavior of $Z_k$ with $k$.
However, this is \emph{not} how we are going to treat the two-point function, for the following two reasons. First, as explained before  we are interested in the anomalous scaling of the propagator, to be used in the SDE (we want to do more than the usual RG analysis, we want to show the appearance of the anomalous scaling in the SDE), in the four-point function, and so on. Second, with $C(p)^{-1}=p^{2\zeta}$, $\zeta\neq 1$, we have a non-analytic behavior from the start, and we cannot obtain the wave function renormalization from a Taylor expansion. 

It is unfortunately too difficult to solve the SDE with cutoffs analytically,
therefore we aim to have an ansatz for the two-point function with cutoffs $ G^{\Lambda}_k(p)$ which explicitly exhibits a conformal behavior in the $\Lambda\to \infty, k\to 0$ limit.
We take the ansatz:
\be \label{eq:Gansatz}
 G^{\Lambda}_k(p)=  \frac{\mathcal{Z}}{ p^{2\zeta}} \; \chi^{\Lambda}_k(p) \;,
\ee
which reproduces the expected infrared scaling for $k \ll p$ 
with\footnote{The scaling for $p\ll k$ is recovered by observing 
that $\chi^{\infty}_k(p)$ has a series expansion:
\[
\chi^{\infty}_k(p) = \left(\frac{p^2}{k^2}\right)^{\zeta} \frac{1}{\Gamma(\zeta)} \left(\frac{1}{\zeta}  -\frac{1}{1+\zeta} \frac{p^2}{k^2}  +O\left(\left(\frac{p^2}{k^2}\right)^{2}\right) \right) \;,
\]
hence at small $p$ we get:
\[
\hat G^{\infty}_k(p)^{-1} = \frac{1}{p^{2\zeta}} \; \chi^{\infty}_k(p) \simeq
k^{2\zeta} \Gamma(\zeta) \zeta \left(1 +\frac{\zeta}{1+\zeta} \frac{p^2}{k^2}   \right) \; ,
\]
consistent with an anomalous field dimension $\eta = 2 -2\zeta$.} $\eta = 2-2\zeta$.

This ansatz is a solution of the SDE with cutoffs up to terms that are suppressed in the limit $k\to 0$ provided that the mass is tuned to criticality.
To see this, let us denote the cutoffed tadpole and  melon integrals $T^{\Lambda}_k$ and $M^{\Lambda}_k(p)$:
\begin{equation}
T^{\Lambda}_k = \int_q G^{\Lambda}_k(q)\; , \qquad 
M^{\Lambda}_k(p) =   \int_{q_1,q_2}  G^{\Lambda}_k(q_1)   G^{\Lambda}_k(q_2)     G^{\Lambda}_k( p +q_1 + q_2  ) \; ,
\end{equation}
where, using Schwinger parameters, the melon integral writes:
\begin{equation}
M^{\Lambda}_k(p) = \frac{\mathcal{Z}^3}{(4\pi)^d \Gamma(\zeta)^3 }    \int_{\Lambda^{-2}}^{k^{-2}}  d\alpha_1 d\alpha_2 d\alpha_3  \; 
   \frac{ (\alpha_1\alpha_2 \alpha_3)^{\zeta-1}   }{ ( \alpha_1 \alpha_2 + \alpha_1 \alpha_3 + \alpha_2 \alpha_3)^{d/2} }
   e^{ - p^2 \frac{ \alpha_1 \alpha_2 \alpha_3}{ \alpha_1 \alpha_2 + \alpha_1 \alpha_3 + \alpha_2 \alpha_3 } } \; .
\end{equation}
The Schwinger-Dyson equation with cutoffs then becomes:
\begin{equation}
 \label{eq:substSDE}
 \frac{p^{2\zeta}}{\mathcal{Z}} =  p^{2\zeta} + \bigg[ m_k^{2\zeta}    - \lambda^2    \big( M^{\Lambda}_k(p) - M^{\Lambda}_k(0) \big) \bigg] \chi^{\Lambda}_k(p) \;, 
\end{equation}
where all the $p$-independent contributions in the square bracket have been absorbed in the renormalized mass
\be
m_k^{2\zeta} =  m^{2\zeta}  +     \left( \lambda_{p} + \lambda_{d}  \right) T^{\Lambda}_k  - 
  \lambda^2  M^{\Lambda}_k(0)\;.
\ee 
We can tune the UV mass so as to both cancel the ultraviolet mass divergences in the SDE for the two-point function
 and ensure that the renormalized mass goes to zero in the $k=0$ limit. 
 
 \begin{proposition}
There exists $m^{2\zeta}$ depending only on $\Lambda$ and the bare coupling constants $ \lambda_{p} , \lambda_{d} , \lambda$ such that 
\[
 \lim_{k\to 0}m_k^{2\zeta} =0 \;.
\]
\end{proposition}

\begin{proof}
 
 The tadpole integral is:
 \[
 \begin{split}
  T^{\Lambda}_k  = \int_{q}   G^{\Lambda}_k(q) & =  \mathcal{Z} \int_{\Lambda^{-2}}^{k^{-2}}  d\alpha  \int_q  \; \frac{ \alpha^{\zeta-1}}{\Gamma(\zeta)} e^{ -\alpha q^2} = \frac{\mathcal{Z}}{   (4\pi)^{d/2} }\; \frac{-4}{\Gamma(d/4)d} \left( (k^2)^{\frac{d}{4}} - (\Lambda^2)^{\frac{d}{4}} \right) \;,
 \end{split}
 \]
 where we used $\zeta = \frac{d}{4}$. Combining this with the $M^{\Lambda}_k(0)$ integral computed in Appendix \ref{app:melon} we get
 $m_k^{2\zeta} = m^{2\zeta} + A(\lambda_{p},\lambda_{d},\lambda) ( \Lambda^{d/2} 
 - k^{d/2} ) $ with:
 \[
A(\lambda_{p},\lambda_{d},\lambda)=\frac{4\mathcal{Z}\left( \lambda_{p } + \lambda_{d}  \right)}{(4\pi)^{d/2}\Gamma(d/4)d}-\mathcal{Z}^3\lambda^2\frac{24}{d(4\pi)^d\Gamma(d/4)^3}\int_1^{\infty}dx \int_1^{\infty}dy \frac{x^{-1}y^{d/4-1}}{\left(1+y+xy\right)^{d/2}} \;,
 \]
Choosing 
\be \label{eq:baremass}
m^{2\zeta} = - A(\lambda_{p},\lambda_{d},\lambda) \Lambda^{d/2} \;,
\ee
we obtain:
\[
 m^{2\zeta}_k =- k^{d/2} A\left( \lambda_p, \lambda_d, \lambda \right) \;,
\]
which goes to 0 when sending $k\to 0$. 
\end{proof}

 \begin{proposition}
Choosing $m^{2\zeta}$ as in \eqref{eq:baremass}, the Schwinger-Dyson equation \eqref{eq:SDEcutoff} is solved by the ansatz \eqref{eq:Gansatz}, with $\zeta= \frac{d}{4}$, and with $\mathcal{Z}$ satisfying
\be\label{eq:wave}
1 -\mathcal{Z}  =   \mathcal{Z}^4\lambda^2 \frac{1}{(4\pi)^d } \;\frac{\Gamma \left( 1 -\frac{d}{4} \right) }{ \frac{d}{4}\Gamma\left( 3 \frac{d}{4}\right)} \;.
\ee
\end{proposition}

\begin{proof}

We show in Appendix \ref{app:melon} that:
\[
M^{\Lambda}_k(p)
   =M^{\Lambda}_k(0)- \frac{\mathcal{Z}^3p^{2d-6\zeta} f\left( \frac{ k^2 }{p^2}, \frac{p^2}{\Lambda^2} \right)}{(4\pi)^d \Gamma(\zeta)^3} \;, \qquad
 f(0,0)  =\frac{\Gamma(1-d+3\zeta) \Gamma(d/2-\zeta)^3 }{( d-3\zeta ) \Gamma(3d/2-3\zeta)}\;.
\]
Choosing $m^{2\zeta}$ as in  \eqref{eq:baremass} to exactly cancel the UV divergent pieces arising from $T^{\Lambda}_k $ and $M^{\Lambda}_k (0)$,
we obtain a renormalized mass $m^{2\zeta}_k $ which is tuned to criticality
$\lim_{k\to 0} m_k^{2\zeta} = 0 $. With this choice, 
we can take $\Lambda \to \infty, k\to 0$ in \eqref{eq:substSDE} and obtain:
\[
\begin{split}
 (1-\mathcal{Z}) p^{2\zeta} =   \mathcal{Z}^4 \lambda^2p^{2d-6\zeta} \frac{ f(0,0) }{(4\pi)^{d} \Gamma(\zeta)^{3} }  \; ,
\end{split}
\]
which is solved by $\zeta= \frac{d}{4}$ and
\[
1 - \mathcal{Z}  =   \mathcal{Z}^4\lambda^2 \frac{1}{(4\pi)^d } \;\frac{\Gamma \left( 1 -\frac{d}{4} \right) }{ \frac{d}{4}\Gamma\left( 3 \frac{d}{4}\right)} \;.
\]
\end{proof}

Notice that dropping the $-\mathcal{Z}$ term in \eqref{eq:wave} (which comes from the inverse free covariance) we recover the result of \cite{Klebanov:2016xxf}.

\subsection{The four-point couplings}
\label{sec:4point}
 
We denote $\hat \delta^{t}_{\mba \mbb  \mbc \mbd} = \frac{1}{N^{3/2}} \delta^{t}_{\mba \mbb  \mbc \mbd}$,  $\hat \delta^{p}_{\mba \mbb ; \mbc \mbd} = \frac{1}{N^2} \delta^{p}_{\mba \mbb ; \mbc \mbd}$ and $\hat \delta^d_{\mba \mbb ; \mbc \mbd} = \frac{1}{N^3} \delta^{d}_{\mba \mbb ; \mbc \mbd}$
the rescaled tetrahedron, pillow and double-trace contraction operators. The four-point function:
\begin{equation}\label{eq:2.0}
 \braket{\phi_{A}  \phi_{B} \phi_{C} \phi_{D} }^c  =2 \left( \frac{{\cal K}}{1 - {\cal K}} {\cal S} \right)_{ AB; C'D' }  G_{C'C}  G_{D'D}  \; ,
\end{equation}
is computed in terms of the four-point kernel ${\cal K}$ which at leading and next-to-leading order in $1/N$ is, using the shorthand notation $G_{xy} = G(x,y)$:
\begin{align}
  {\cal K}_{ (\mba,x')(\mbb,y') ; (\mbc,z)(\mbd,w) } &=  G_{x'x} G_{y'y}  \bigg[  
    -  \lambda_p   \delta_{xy} \delta_{xz} \delta_{xw} \hat \delta^p_{\mba\mbb ; \mbc\mbd} -  \lambda_d  \delta_{xy} \delta_{xz} \delta_{xw} \hat \delta^d_{\mba\mbb ; \mbc\mbd}   
    \crcr
    & + 3 \lambda^2 G_{xy}^2 \delta_{xz}\delta_{yw} \hat \delta^p_{\mba\mbb ; \mbc\mbd} \crcr
    & -   \frac{\lambda}{N^{3/2}} \delta_{xy}  \delta_{xz} \delta_{xw}   \left( \frac{  \delta^t_{\mba\mbb   \mbc\mbd} +\delta^t_{\mba\mbb   \mbd\mbc} +  \delta^t_{\mba  \mbc \mbb  \mbd} +  \delta^t_{\mba  \mbd \mbb  \mbc} 
    +  \delta^t_{\mba \mbc \mbd   \mbb} + \delta^t_{\mba \mbd \mbc   \mbb} }{2} \right)
 \bigg] \, .
\end{align}

Due to the $O(N)^3$ symmetry as well as the color permutation symmetry, minus the amputated four-point function
$\Gamma^{(4)}_{ (\mba,x) (\mbb,y)  (\mbc,z) (\mbd,w) } $ is a sum of three classes of terms.

\paragraph{Tetrahedral terms.} We have six tetrahedral terms:
    \[
   \sum_{ (\mbb,y) (\mbc,z) (\mbd,w) }
      \delta^{t}_{\mba \mbb\mbc\mbd} \Gamma^{(4,t)}_{xyzw}  \;, 
     \qquad \Gamma^{(4,t)}_{xyzw}  =  \frac{  \lambda }{4 N^{3/2}} \delta_{xy} \delta_{xz} \delta_{xw} \;,
    \]
where the sum runs over the six permutations of the couples $(\mbb,y) (\mbc,z) (\mbd,w) $, coming from the next-to-leading order contribution to the kernel.
The effective tetrahedral coupling is then exactly 
\be\label{eq:tetra}
  g =k^{d-4\zeta} \mathcal{Z}^2\lambda \;.
\ee

The tetrahedral coupling has a finite flow: in the $\Lambda\to \infty, \,k\to 0$ limit the effective coupling is just a rescaling of the bare one by the wave function constant.
  In particular, denoting 
  \be \label{eq:gc}
  g_c^{-2} = \Gamma \left( 1 -\frac{d}{4} \right) \left[ (4\pi)^d  \, \frac{d}{4}\Gamma\left( 3 \frac{d}{4}\right) \right]^{-1} \;,
  \ee
 the coefficient $\mathcal{Z}$ and the bare tetrahedral coupling write in terms of the renormalized one as:
  \[
   \mathcal{Z}= 1 - \frac{g^2}{g_c^2} \; \qquad \lambda = \frac{g}{\mathcal{Z}^2} \;.
  \]
\begin{figure}[ht]
\begin{center}
\includegraphics[height=3cm]{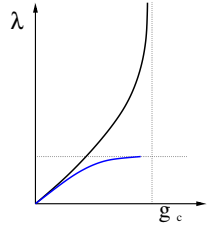} 
 \caption{The bare coupling as a function of the renormalized one. We represented in black the case $\lambda$ real, and
 in blue the absolute value in the case $\lambda$ purely imaginary.} \label{fig:tetra}
 \end{center}
\end{figure}
We distinguish two cases (see figure~\ref{fig:tetra}): $\lambda$ real and $\lambda$ purely imaginary:
   \begin{itemize}
    \item \emph{$\lambda$ (and $g$) real.} In this case $\lambda(g)$ is invertible to $g(\lambda)$ for any $\lambda $, $g(\lambda) <g_c$, $g$ asymptotes to $g_c$ and $\mathcal{Z}$ goes to zero when $\lambda\to \infty$ ($g\to g_c$) . 
    \item \emph{$\lambda$ (and $g$) imaginary.} In this case $\lambda(g)$
    is invertible to $g(\lambda)$ for $|\lambda |< 3^{3/2}2^{-4}g_c$, corresponding to  $| g | < 3^{-1/2}g_c$, and $\mathcal{Z}$ is bounded.
   \end{itemize}
 
\paragraph{Pillow and double-trace terms.} We have three pillow and three double-trace terms, corresponding to the three channels $(\mba,x) (\mbb,y)  \to (\mbc,z) (\mbd,w) $,
$(\mba,x) (\mbc,z)   \to (\mbb,y) (\mbd,w) $ and $(\mba,x) (\mbd,w)  \to (\mbb,y)   (\mbc,z)  $. We write minus the pillow 1PI four-point function as:
    \[
       2  \left( \hat \delta^{p}_{\mba\mbb; \mbc\mbd} \Gamma^{(4,p)}_{xy; zw} +   \hat \delta^{p}_{\mba\mbc; \mbb\mbd} \Gamma^{(4,p)}_{xz; yw}  +   \hat \delta^{p}_{\mba\mbd; \mbb\mbc} \Gamma^{(4,p)}_{xw; yz} \right) \;,
    \]
where the factor 2 is conventional. The double-trace contribution is obtained by changing the superscript $p$ to $d$. At leading order in $1/N$ the sum of the pillow and double-trace contributions in one channel is:
\be
\begin{split}
   -  \hat \delta^{p}_{\mba\mbb; \mbc\mbd} \Gamma^{(4,p)}_{xy; zw}  -  \hat \delta^{d}_{\mba\mbb; \mbc\mbd} \Gamma^{(4,d)}_{xy; zw}    =
      G^{-1}_{xx'} G^{-1}_{yy'}  \left( \frac{K}{1-K} \right)_{x'y';zw} \;,
\end{split}
\ee
where $K$ is the on-shell leading order four-point kernel:
\[
 K_{ (\mba,x')(\mbb,y') ; (\mbc,z)(\mbd,w) } =   G_{x'x} G_{y'y}  \bigg[  
    -  \lambda_p   \delta_{xy} \delta_{xz} \delta_{xw} \hat \delta^p_{\mba\mbb ; \mbc\mbd} -  \lambda_d  \delta_{xy} \delta_{xz} \delta_{xw} \hat \delta^d_{\mba\mbb ; \mbc\mbd}   
    + 3 \lambda^2 G_{xy}^2 \delta_{xz}\delta_{yw} \hat \delta^p_{\mba\mbb ; \mbc\mbd} 
 \bigg] \; .
\]
In momentum space, we have $\Gamma^{(4,d) }_{p_1p_2p_3p_4} = (2\pi)^d \delta(p_1+p_2+p_3+p_4 )\Gamma^{(4,d)}(p_1,p_2,p_3,-p_1-p_2-p_3)$,
and the four-point function in the channel $\mba \mbb  \to \mbc \mbd $ is:
\be\label{eq:fourpointdef}
  - \hat \delta^{p}_{\mba\mbb; \mbc\mbd} \Gamma^{(4,p)}_{p_1p_2;  r_1   r_2  } - \hat \delta^{d}_{\mba\mbb; \mbc\mbd } \Gamma^{(4,d)}_{p_1p_2;  r_1  r_2  } 
 =   \frac{1}{  G(p_1)  G(p_2)   } \left(  \frac{K}{1-K} \right)_{\mba\mbb; \mbc\mbd}( p_1, p_2 ; r_1 , r_2 ) \; ,
\ee
with $1$ the identity operator on bilocal functions $1 = (2\pi)^{2d} \delta(p_1-q_1) \delta(p_2-q_2)$ and the four-point kernel $K$ in momentum space:
\be
\begin{split}\label{eq:fourpointkernel}
  K_{ p_1,p_2; q_1,q_2 } = & (2\pi)^d \delta(p_1+p_2 - q_1 -q_2)   G(p_1) G(p_2)  \\
 & \qquad  \times \bigg[ \hat \delta^{p }  \; 3  \lambda^2 \int_q  G(q) G(q + p_1 -q_1)     -   \hat \delta^{p }  \lambda_{p}   - \hat \delta^{d }  \lambda_{d}  \bigg] \; .
\end{split}
\ee

When expanding the geometric series in $K$ we need to deal with powers of $\hat \delta^p$ and $\hat \delta^d$. This is slightly unpleasant as $\hat \delta^p \hat \delta^p =  \frac{1}{3} \hat \delta^p + \frac{2}{3} \hat \delta^d$, 
$\hat \delta^p \hat \delta^d =  \hat \delta^d$ and $\hat \delta^d \hat \delta^d = \hat \delta^d$, that is $\hat \delta^p$ and 
$\hat \delta^d$ are not mutually orthogonal. In particular, the pillow and double-trace couplings mix. It is convenient to parameterize the interaction in terms of two independent couplings which do not mix. The operators:
\be
 \hat{P}^{(1)}_{\mba\mbb; \mbc\mbd} = 3( \hat \delta^p_{\mba\mbb; \mbc\mbd} - \hat \delta^d_{\mba\mbb; \mbc\mbd} ) \; ,\qquad \hat{P}^{(2)}_{\mba\mbb; \mbc\mbd} = \hat \delta^d_{\mba\mbb; \mbc\mbd} \;,
 \label{eq:proj_ortho}
\ee
are two mutually orthogonal projectors which span the interaction space.\footnote{This corresponds to the traceless-trace decomposition in the intermediate field representation of the pillow and double-trace interactions \cite{Benedetti:2018ghn}.} We parametrize the interaction in terms of  
$\lambda_1 = \frac{\lambda_p}{3}$ and $\lambda_2 = \lambda_d + \lambda_p$.
Thus, the four-point kernel in momentum space becomes:
\begin{equation}
\begin{split}\label{eq:newker}
  & K_{ p_1,p_2; q_1,q_2 } =  (2\pi)^d \delta(p_1+p_2 - q_1 -q_2)   G(p_1) G(p_2)  \\
&\qquad \times \bigg[ \bigg( \lambda^2 \int_q G(q)G(q+p_1-q_1)       -\lambda_1 \bigg) \hat{P}^{(1)} + \bigg( 3\lambda^2 \int_q G(q)G(q+p_1-q_1) -  \lambda_2 \bigg) \hat{P}^{(2)} 
\bigg]\; ,
\end{split}
\end{equation}
and the effective four-point function (where 
$\Gamma^{(4;1)}  = \frac{1}{3}\Gamma^{(4,p)} $ and $\Gamma^{(4;2)}  = 
\Gamma^{(4,d)} +  \Gamma^{(4,p)} $):
\be\label{eq:reparam}
- \Gamma^{(4;1)} \hat{P}^{(1)} - \Gamma^{(4;2)} \hat{P}^{(2)} = G^{-1}G^{-1} \frac{K}{1-K} \; .
\ee

\subsubsection{The bare expansion}

\eqref{eq:reparam} and \eqref{eq:newker} are used to obtain the bare expansion of the running couplings:
\be
g_{1} = k^{d-4\zeta} \mathcal{Z}^2\Gamma_k^{(4;1)}(0,0,0,0) \;,\qquad
g_{2} = k^{d-4\zeta} \mathcal{Z}^2\Gamma_k^{(4;2)}(0,0,0,0) \; ,
\ee 
as decoupled series in the bare couplings. The two cases are identical, up to replacing $\lambda^2$ by $3\lambda^2$, hence we discuss below only $g_1$. 
\begin{figure}[ht]
\begin{center}
\includegraphics[scale=1.3]{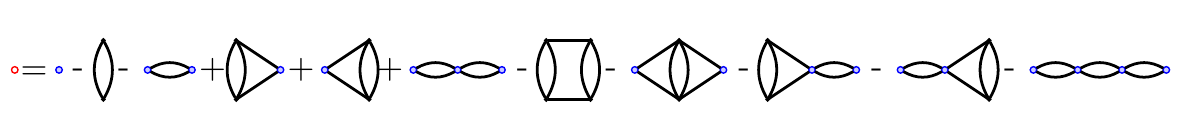} 
 \caption{The bare series up to quartic order. For $g_1$ the blue vertices represent $\lambda_1$ and the rungs contribute $\lambda^2$, while for $g_2$
 the blue vertices represent $\lambda_2$ and the rungs 
 contribute $3\lambda^2$.} \label{fig:bare1}
 \end{center}
\end{figure}

The bare series is a sum over connected amputated chain graphs $\cG$ depicted in figure~\ref{fig:bare1}.  A connected amputated chain graph $\cG$ is a sequence of irreducible pieces connected one to another by pairs of parallel horizontal edges. The irreducible pieces are either vertical ladder rungs with two tetrahedral couplings or bare vertices $\lambda_1$. There are $2^{n}$ chain graphs with $n$ irreducible parts (that is vertical rungs or bare vertices). To each graph we associate its amplitude in Schwinger parametrization. The formula is similar to \eqref{eq:amp_final} but with cutoffs:
\be\label{eq:ampli1}
 A(\cG) = \frac{1}{\Gamma(\zeta)^{2V(\mathcal{G})}(4\pi)^{d(V-1)/2}}\bar{A}(\cG) \; , \quad \bar{A}(\cG)= \int_{\Lambda^{-2}}^{k^{-2}} 
 \left( \prod_{e\in \cG} d\alpha_e \;  \alpha_e^{\zeta-1}\right) \; 
 \frac{ 1 }{ \big[ \sum_{ {\cal T} \subset \cG} \prod_{e\notin {\cal T} } \alpha_e \big]^{d/2} } \;,
\ee
where $e\in \cG$ denotes the edges of $\cG$, $V(\mathcal{G})$ its vertices and ${\cal T}$ runs over the spanning trees in $\cG$.
For instance, the amplitudes $\bar{A}(\cG)$ of the graphs depicted on the right hand side in figure~\ref{fig:bare1} are, from left to right:
\[
1 \, , \;\;  U_1 \, , \;\;  T_0 \, , \;\;  S_1 \, , \;\; S_1 \, , \;\; T_0^2 \, , \;\; 
U_2 \, , \;\; T_1 \, , \;\; S_1T_0\, , \;\;  T_0S_1 \, , \;\; T_0^3 \;,
\]
where $S_1$, $T_0,T_1$ and $U_1,U_2$ denote the integrals (to simplify the notation we suppress the measure):
\begin{equation}\label{eq:SDTU}
\begin{split}
 U_1 = & T_0 =   \int_{\Lambda^{-2}}^{k^{-2}} \frac{(a_1 a_2)^{\zeta-1}}{ (a_1 + a_2)^{d/2}} \equiv D  \; , \qquad
 S_1 =  \int_{\Lambda^{-2}}^{k^{-2}} \frac{(a_1 a_2 b_1 b_2)^{\zeta-1}}{ \big[(a_1 + a_2)(b_1 + b_2 ) + a_1 a_2 \big] ^{d/2}}  \;,  \crcr
 U_2 = & \int_{\Lambda^{-2}}^{k^{-2}} \frac{  (a_1 a_2 b_1 b_2 c_1  c_2)^{\zeta-1}   }
  { \big[ (a_1 + a_2)(b_1 + b_2) (c_1 + c_2) + (a_1 + a_2) c_1c_2 + a_1a_2(c_1+c_2)\big]^{d/2}  }  \;, \crcr
 T_1 = & \int_{\Lambda^{-2}}^{k^{-2}} \frac{  (a_1 a_2 b_1 b_2 c_1  c_2)^{\zeta-1}   }
  { \big[ (a_1 + a_2)(b_1 + b_2) (c_1 + c_2) +  b_1b_2(a_1+a_2+c_1+c_2) \big]^{d/2}  }  \;.
\end{split}
\end{equation}
 Observe that (setting again $\zeta=d/4$) the amplitude of a graph diverges like some power of 
$ \ln (\Lambda^2/ k^2)$.\footnote{Those integrals were computed in dimensional regularization in chapter \ref{chap:3loops}. Divergences in $ \ln (\Lambda^2/ k^2) $ then correspond to poles in $1/\epsilon$.} 
 
The chain graph consisting in a bare vertex has amplitude $1$. We denote 
$\mathfrak{G}$ the set of connected chain graphs with at least two internal vertices. The number of tetrahedral vertices of $\cG\in\mathfrak{G}$, $n_t(\cG)$, is always even. We denote $n_1(\cG)$ the numbers of blue vertices of $\cG$. The graphs $\cG\in\mathfrak{G}$ are such that $n_t(\cG)+n_1(\cG) \ge 2$.
We rescale the bare and effective coupling as 
$\bar{g}_1 = (4\pi)^{ - d/2} \Gamma(\zeta)^{-2} g_1$ and so  on. Recalling that $g = \lambda \mathcal{Z}^2$ the bare expansion writes: 
\begin{align*}
  \bar{g}_1(\bar{\lambda}_1, \bar{g})  =   \mathcal{Z}^2\bar{\lambda}_1  
  +  \sum_{\cG\in \mathfrak{G}} (-1)^{1 + n_1(\cG) } 
      \bar{g}^{n_{t}(\cG)} 
 \left( \mathcal{Z}^2 \bar{\lambda}_1  \right)^{n_1(\cG)} 
 \bar{A}(\cG)\; .
\end{align*}
The same formula holds for $\bar{g}_2$ by replacing $\bar{g}^2$ by $3\bar{g}^2$.
Up to total degree $4$ in the coupling constants the bare expansion is:
\be\label{eq:Bare}
\begin{split}
  \bar{g}_1(\bar{\lambda}_1,\bar{g})  = &  \left(\mathcal{Z}^2 \bar{\lambda}_1\right)- \bar{g}^2 U_1  - 
  \left(\mathcal{Z}^2\bar{\lambda}_1\right)^2T_0 + 2\bar{g}^2 \left(\mathcal{Z} \bar{\lambda}_1\right) S_1 + \left(\mathcal{Z}^2 \bar{\lambda}_1\right)^3  T_0^2 \crcr
  &\qquad  
   -\bar{g}^4 U_2 -\bar{g}^2 \left(\mathcal{Z}^2 \bar{\lambda}_1\right)^2 T_1 - 2\bar{g}^2 \left(\mathcal{Z}^2 \bar{\lambda}_1\right)^2 S_1 T_0 - \left(\mathcal{Z}^2 \bar{\lambda}_1\right)^4 T_0^3 \;.
\end{split}
\ee 

 \paragraph{One vertex reducible graphs.}
\begin{figure}[ht]
\begin{center}
 \includegraphics[scale=1]{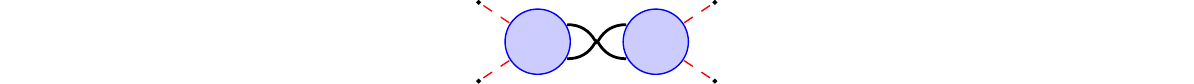} 
 \caption{One vertex reducible graph. We represented in dashed red the amputated external edges.} \label{fig:graphs3}
 \end{center}
\end{figure}

A graph is called one vertex reducible (1VR) if it disconnects into two non-trivial graphs (i.e graphs having internal vertices) by cutting a vertex (see figure~\ref{fig:graphs3}).
As there are no two-point subgraphs (we are using the full propagator), any 1VR four-point graph disconnects into two four-point graphs by cutting the vertex ``vertically'' and adding a pair of external edges on each resulting ``half-vertex''. We write $\cG = \cG_1 \cG_2$. 
By this procedure, any four-point graph can be decomposed as a chain $\cG = \cG_1\dots \cG_q$ where $\cG_i$ are one vertex irreducible (1VI). The amplitude factors on the 1VI components: $\bar{A}(\cG) = \bar{A}(\cG_1)\dots \bar{A}(\cG_q)$. We classify the 1VI graphs into three families:
\begin{itemize}
 \item The pure \emph{ladders}, depicted in figure~\ref{fig:pureladders}, consisting in a non-empty sequence of vertical ladders with tetrahedral vertices. We denote $U_r$ the amplitude of the ladder graph with $r$ rungs (by some abuse of notation we will denote the graph itself also $U_r$ when no confusion can arise). The sum over the ladders is:
 \be
    U(\bar{g}) =  \sum_{r\ge 1} \bar{g}^{2r} U_r  = \bar{g}^2 D + 
  \bar{g}^4 U_2 + \dots \;.
 \ee
\begin{figure}[H]
\begin{center}
\includegraphics[scale=1]{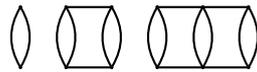} 
 \caption{The pure ladders $U_1,U_2$ and $U_3$.} \label{fig:pureladders}
 \end{center}
\end{figure}
 
 \item The ``v-ladders'' or \emph{caps}, that is ladders having a blue bare vertex at one end, depicted in figure~\ref{fig:pladders}. They consist in a blue bare vertex followed by a non-empty sequence of vertical ladder rungs with tetrahedral vertices. 
 We denote $S_r$ the amplitude of the cap with $r$ rungs (and the graph itself also $S_r$). The sum over the caps is:
 \be 
 S(\bar{g}) =  \sum_{r \ge 1 }\bar{g}^{2r} S_r 
        = \bar{g}^2 S_1 + \dots \;.
 \ee
\begin{figure}[H]
\begin{center}
\includegraphics[scale=1]{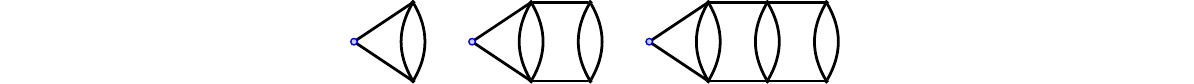} 
 \caption{The caps $S_1,S_2$ and $S_3$.} \label{fig:pladders}
 \end{center}
\end{figure}
 \item The ``vv-ladders'' or \emph{double caps} having a blue bare vertex at each end, depicted in figure~\ref{fig:pladdersp}. They consist in a blue bare vertex followed by a \emph{possibly empty} sequence of vertical ladder rungs with tetrahedral vertices followed by a blue bare vertex. 
   We denote $T_r$ the amplitude of the double cap graph with $r$ rungs (and the graph itself also $T_r$). The sum over the double caps is:
   \be
  T (\bar{g}) =  \sum_{r\ge 0} \bar{g}^{2r } T_r
      = D + \bar{g}^2 T_1 + \dots \;,
  \ee
\begin{figure}[H]
\begin{center}
\includegraphics[scale=1]{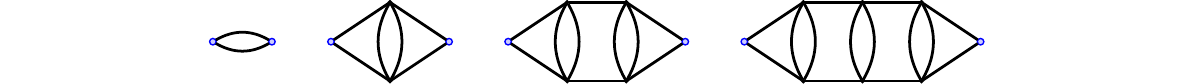} 
 \caption{Some double caps.} \label{fig:pladdersp}
 \end{center}
\end{figure}
\end{itemize}
Observe that in the generating functions $S(\bar{g})$ and $T(\bar{g})$ we have \emph{not} included any coupling constants for the blue vertices.
By counting the number of reducibility vertices in a graph the bare series is simply:
\begin{equation}
\label{eq:bare1VR}
\begin{split}
 \bar{g}_1 =&  -U(\bar{g}) + \mathcal{Z}^2\bar{\lambda}_1 \big[ 1 + S(\bar{g}) \big] \bigg[ \sum_{q\ge 0}
 \left(- \mathcal{Z}^2 \bar{\lambda}_1 \right)^q \sum_{r_1,\dots r_{q} \ge 0 }  \prod_{i=1}^q \bar{\lambda}^{2r_i} T_{r_i} \bigg]
 \big[ 1 + S(\bar{g})\big] \crcr 
= &- U(\bar{g}) + \left(- \mathcal{Z}^2 \bar{\lambda}_1 \right) 
\frac{\big[ 1+S(\bar{g}) \big]^2}
 { 1 + \mathcal{Z}^2 \bar{\lambda}_1  T( \bar{g} ) } \;. 
\end{split}
\end{equation}

A similar expression holds for the usual $\phi^4$ model (with of course other types of graphs contributing too): $U,S,T$ can be computed directly starting from the four-point kernel, by separating the contribution of the bare vertex from the rest. The important difference is that in general $U,S$ and $T$ depend on $\bar{\lambda}_1$, whereas in our case they depend only on the parameter $\bar{g}$. This makes the $\beta$ function in our case particularly simple: as we will see below, the all orders $\beta$ function is only quadratic in the running coupling.

\begin{figure}[ht]
\begin{center}
\includegraphics[scale=1.3]{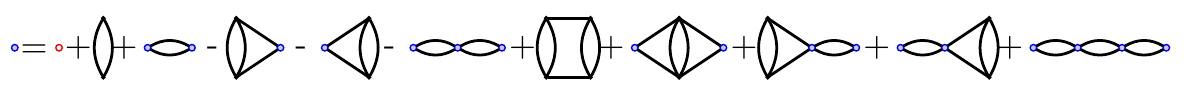} 
 \caption{The renormalized series by iterated substitutions. On the right hand side the 
 vertices $\bar{\lambda}_1$ (in blue) should be iteratively substituted with the right hand side itself.} \label{fig:graphs2}
 \end{center}
\end{figure}

\paragraph{The renormalized expansion.}
The bare series can be inverted to yield the renormalized series. 
This is usually done by iterative substitutions:
\be
\mathcal{Z}^2 \bar{\lambda}_1  = \bar{g}_1 + \bigg[ \sum_{\cG\in \mathfrak{G}} (-1)^{n_1(\cG) } \bar{g}^{n_t(\cG)} \left(\mathcal{Z}^2\bar{\lambda}_1 \right)^{n_1(\cG)}\bar{A}(\cG)\bigg]_{ \mathcal{Z}^2\bar{\lambda}_1 = 
\bar{g}_1 + \sum_{\cG\in \mathfrak{G}} (-1)^{n_1(\cG) } \bar{g}^{n_t(\cG)}
\left(\mathcal{Z}^2\bar{\lambda}_1 \right)^{n_1(\cG)}\bar{A}(\cG)  } \; ,
\ee
as represented in figure \ref{fig:graphs2}. It can also be done by using the Bogoliubov-Parasuk recursion as in \eqref{eq:BPrecursion}.

In the most general case the result of iterated insertions can be written compactly using Zimmermann forests \cite{Zimmermann:1969jj}.
However, in our case we can invert the bare series directly:
 \be\label{eq:renodirect}
  \mathcal{Z}^2\bar{\lambda}_1 = 
  \frac{\bar{g}_1+U(\bar{g})}{ \big[1+S(\bar{g}) \big]^2 - U(\bar{g})T(\bar{g}) - \bar{g}_1 T(\bar{g})}  \;.
\ee
For the usual $\phi^4$, the 1VI expansion can not be inverted so simply:
while in our case \eqref{eq:renodirect} is a series in $\bar{g}_1$ whose coefficients are series in the parameter $\bar{g}$, in the usual $\phi^4$ the coefficients themselves depend on $\bar{g}_1$.
The renormalized expansion in our case up to quartic order is:
\begin{align*}
 \mathcal{Z}^2 \bar{\lambda}_1 = & \bar{g}_1 +  \bar{g}^2 U_1 + \bar{g}_1^2 T_0  -
 2 \bar{g}^2 \bar{g}_1 (S_1-U_1T_0)  
              +  \bar{g}_1^3  T_0^2 +    \bar{g}_1^4  T_0^3 \crcr
          &    + \bar{g}^4 \big[ U_2 - 2 S_1 U_1  + U_1^2T_0 \big] 
              + \bar{g}_1^2 \bar{g}^2 \big[ T_1 - 4S_1T_0 + 3U_1T_0^2  \big] \; .
\end{align*}

\subsubsection{Beta functions} 
\label{sec:4point-beta}

Let us denote $ \pmb{\partial} = k\partial_k $.
The $\beta$ function can be computed in two ways. Usually one starts from the renormalized series $\bar{\lambda}_1(k,\bar{g}_1)$, where $k$ denotes the explicit dependence of the coefficients on the IR cutoff and  equates to zero the scale derivative of the bare coupling. This has the advantage that the $\beta$ function is directly written in terms of the running
coupling. In our case however, it is more convenient to derive directly the bare expansion in \eqref{eq:bare1VR} and rewrite the derivative as:
\be\label{eq:beta}
 \beta_{\bar{g}_1} = \pmb{\partial} \bar{g}_1
 = \beta_0^{\bar{g}}
  - 2  \beta_1^{\bar{g}} \bar{g}_1  + \beta_2^{\bar{g}} \bar{g}_1^2\;,
\ee
with the coefficients of the $\beta$ function given by:
\begin{equation}\label{eq:coef}
\begin{split}
 \beta_0^{\bar{g}}  &= - \pmb{\partial} U + 2 \frac{U}{1+S} \pmb{\partial} S - \frac{U^2}{(1+S)^2} \pmb{\partial} T  
 = - \bar{g}^2 \pmb{\partial} U_1 -
        \bar{g}^4 \big[ \pmb{\partial}  U_2 - 2 U_1 \pmb{\partial} S_1 + U_1^2 \pmb{\partial} T_0 \big]  + O(\bar{g}^6)
 \; ,\\
 \beta_1^{\bar{g}} &= - \frac{1}{1+S} \pmb{\partial}S + \frac{U}{(1+S)^2} \pmb{\partial} T = -
 \bar{g}^2 ( \pmb{\partial} S_1 - U_1 \pmb{\partial} T_0 ) 
     + O(\bar{g}^4)
 \;, \\ 
 \beta_2^{g} &= - \frac{1}{(1+S)^2} \pmb{\partial} T
 = - \pmb{\partial} T_0 - \bar{g}^2 ( \pmb{\partial} T_1 -2
          S_1 \pmb{\partial} T_0 ) + O(\bar{g}^4)  \; .
\end{split}
\end{equation}
We will discuss these coefficients further in section \ref{sec:divergences}.
At three loops the $\beta$ function is:
\be\label{eq:beta3loops}
\begin{split}
 \beta_{\bar{g}_1} =& - \bar{g}^2 \pmb{\partial} U_1 -  \bar{g}^4 
 \big[ \pmb{\partial}  U_2 - 2  U_1 \pmb{\partial} S_1 + U_1^2 \pmb{\partial} T_0 \big]\crcr
 &  + 2\bar{g}_1 \bigg[ \bar{g}^2 ( \pmb{\partial} S_1 - U_1 \pmb{\partial} T_0 )\bigg]- \bar{g}_1^2 \bigg[\pmb{\partial} T_0 + \bar{g}^2 
 ( \pmb{\partial} T_1 -2 S_1 \pmb{\partial} T_0 ) \bigg] \;,
 \end{split}
\ee
The $\beta$ functions of the original pillow and double-trace couplings $\bar{\lambda}_p = 3\bar{\lambda}_1,\bar{\lambda}_d = \bar{\lambda}_2 - 3\bar{\lambda}_1$ can be reconstructed as: 
\begin{align*}
  \beta_{p} =&  3 \beta_0^{\bar{g}} + 2\beta_1^{\bar{g}} \, \bar{g}_p + \frac{1}{3} 
  \beta_2^{\bar{g}}\, \bar{g}_p^2  \;, \crcr
\beta_{d} =&   \big[ \beta_0^{\sqrt{3}\bar{g}}  - 3 \beta_0^{\bar{g}} \big]
 + 2 \beta_1^{\sqrt{3}\bar{g}} \; \bar{g}_d + 2 
 \big[\beta_1^{\sqrt{3}\bar{g}} - \beta_1^{\bar{g}} \big] \bar{g}_p  + 
 \beta_2^{ \sqrt{3}\bar{g} }  \, \bar{g}_d ^2 
+ 2 \beta_2^{ \sqrt{3}\bar{g} } \bar{g}_d  \bar{g}_p +
 \big[ \beta_2^{\sqrt{3}\bar{g}} - \frac{1}{3}  \beta_2^{ \bar{g}} \big] \bar{g}_p^2  
 \; ,
\end{align*}
which is at two loops, using $U_1 = T_0 = D$:
\begin{align*}
 \beta_{p} = & - 3\bar{g}^2 \pmb{\partial} D + 2 \bar{g}^2 ( \pmb{\partial} S_1 - D \pmb{\partial} D ) \bar{g}_p - \frac{1}{3}\pmb{\partial} D \bar{g}_p^2  \; , \crcr
 \beta_d = & 6 \bar{g}^2 ( \pmb{\partial} S_1 - D \pmb{\partial} D ) \bar{g}_d
  + 4 \bar{g}^2 ( \pmb{\partial} S_1 - D \pmb{\partial} D ) \bar{g}_p
   - \bigg(  \bar{g}_p^2 + 2\bar{g}_p\bar{g}_d + \frac{2}{3} \bar{g}_p^2\bigg)\pmb{\partial} D \;.
\end{align*} 
Notice that the $\bar{g}_p$- and $\bar{g}_d$-independent part of $ \beta_d$ starts at order $\bar{g}^4$, as expected from the minimal resolution of the double-trace bubble in terms of tetrahedra (see figure~\ref{fig:graph}).

Lastly we can explicitly check that the lowest order coefficients of $\beta_{0,1,2}^{\bar{g}}$ are convergent and given by:\footnote{Contrary to the results of chapter \ref{chap:3loops}, using cutoffs we obtain analytic results only for the one loop integrals.}
\be
 \pmb{\partial} D = -4 \int_{k^2\Lambda^{-2}}^{1}
 d\alpha \;\frac{\alpha^{\zeta-1}}{(1+\alpha)^{d/2}} \to_{\Lambda\to\infty}
    - 2 \frac{\Gamma\left( \frac{d}{4}\right)^2}{ \Gamma \left( \frac{d}{2}\right)}  \;,
\ee
\be
 -  \frac{1}{4} (\pmb{\partial S} - D \pmb{\partial}D ) 
  = I_1 +I_2 \;,
\ee
with
\begin{align*}
 I_1 & = \int_0^1 \frac{ (a_1a_2 b_2)^{\zeta-1}}{[ (a_1+a_2)(1+b_2) + b_2]^{d/2}}  \; ,\crcr
 I_2 & = -\frac{d}{2} \int_0^1 du \int_0^1 
 \frac{ (a_2 b_1b_2)^{\zeta-1} b_1 b_2}{[ (1+a_2)(b_1+b_2) + u b_1b_2]^{d/2+1}} \; .
\end{align*}

\subsubsection{Flow and fixed points} 
\label{sec:4point-FP}

Being quadratic in $\bar{g}_1$, the beta function \eqref{eq:beta} admits  two fixed points:
 \begin{equation}
 \bar{g}_{1\pm} =  \frac{
    \beta_1^{\bar{g}} 
    \pm \sqrt{ (\beta_1^{\bar{g}})^2 -\beta_0^{\bar{g}}\beta_2^{\bar{g}} }
    }{\beta_2^{\bar{g}}}  = \pm\sqrt{-\bar{g}^2} + O(\bar{g}^2) \; .
    \label{eq:FP-ON3}
   \end{equation}  
In fact, we can even solve the full flow, in terms of the beta function coefficients \eqref{eq:coef}.
Taking $\bar{g}$ to be purely imaginary and small, so that $(\beta_1^{\bar{g}})^2 -\beta_0^{\bar{g}}\beta_2^{\bar{g}} >0$, we find
\be \label{eq:flow}
\bar{g}_1(k) = \frac{
    \beta_1^{\bar{g}} 
    - \sqrt{ (\beta_1^{\bar{g}})^2 -\beta_0^{\bar{g}}\beta_2^{\bar{g}} } \,\tanh\left( \sqrt{ (\beta_1^{\bar{g}})^2 -\beta_0^{\bar{g}}\beta_2^{\bar{g}} } \ln(k/k_0) + c  \right)
    }{\beta_2^{\bar{g}}} \;,
\ee
where $c$ is an integration constant to be fixed by the initial condition $\bar{g}_1(k_0)=\bar{g}_1^{(0)}$:
\be
c = {\rm arctanh}\left( \frac{\beta_1^{\bar{g}} - \beta_2^{\bar{g}} \bar{g}_1^{(0)} }{ \sqrt{ (\beta_1^{\bar{g}})^2 -\beta_0^{\bar{g}}\beta_2^{\bar{g}} } }  \right) \;.
\ee
We then see that $\bar{g}_{1+}$ is an IR fixed point (reached for $k\to 0$) and $\bar{g}_{1-}$ is a UV fixed point (reached for $k\to \infty$).

\begin{figure}[htbp]
\centering
\includegraphics[scale=1]{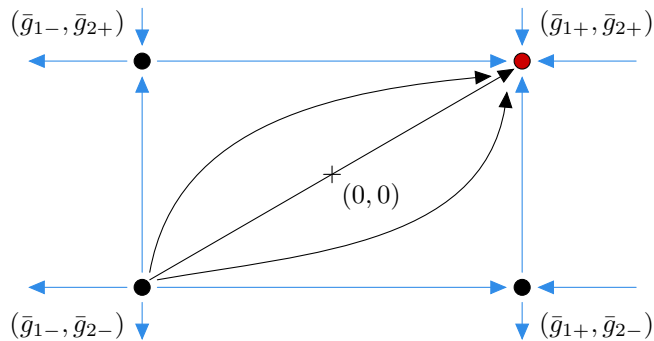}
\caption{Trajectory between the fixed points in the space $(\bar{g}_1,\bar{g}_2)$. The red dot is the IR stable fixed point.}
\label{fig:trajectory}
\end{figure}

The corresponding critical exponents are:
   \be
  \beta'_{\bar{g}_1}( \bar{g}_{1\pm}) = \pm 2\sqrt{ (\beta_1^{\bar{g}})^2 -\beta_0^{\bar{g}}\beta_2^{\bar{g}} }
     = \pm \sqrt{-\bar{g}^2} \left( 4\frac{\Gamma(\frac{d}{4})^2}{ \Gamma(\frac{d}{2})}\right) + O(\bar{g}^3)\;.
 \ee
The beta function $\beta_{\bar{g}_2}$ admits two fixed points and critical exponents of the same form, with $\bar{g}\to\sqrt{3}\bar{g}$.
We illustrate the trajectories between the fixed points in figure \ref{fig:trajectory}.

\subsection{The Beta function coefficients}
\label{sec:divergences}
  
In this section we study the coefficients $\beta_0^{\bar{g}}, \beta_1^{\bar{g}}$ and $\beta_2^{\bar{g}}$. 
As presented in \eqref{eq:coef} they are ratios of sums of amplitudes of graphs which can be arbitrarily UV divergent. We will now show that:
\[
 - \beta_0^{\bar{g}} = R \{ \pmb{\partial}U \} \;, \qquad 
 - \beta_1^{\bar{g}} = R \{ \pmb{\partial}S \}  \;, \qquad
 - \beta_2^{\bar{g}} = R \{ \pmb{\partial}T \} \;,
\]
with $R$ the BPHZ \cite{bogoliubow1957multiplikation,Hepp:1966eg,Zimmermann:1969jj} subtraction operator. 

\paragraph{The scale derivative.}
 
Let us consider an amputated graph $\cG$, and let $E(\cG)$, $V(\cG)$, $n(\cG)$ be the numbers of edges, vertices, and external half-edges of $\cG$, respectively. Below, we are interested in 
chain graphs with $n(\cG)=4$. We have $ 2 E(\cG) =4 V(\cG) -n(\cG)$, and the amplitude of $\cG$ from \eqref{eq:ampli1} is:
\be
 \bar{A}(\cG) = \int_{\Lambda^{-2}}^{k^{-2}} 
  \left( \prod_{e\in \cG} d\alpha_e \, \alpha_e^{\zeta-1}\right) \; 
 \frac{ 1 }{ \big[ U_{\cG}(\alpha)\big]^{d/2} }  \;, \qquad U_{\cG}(\alpha) = \sum_{ {\cal T} \subset \cG} \prod_{e\notin {\cal T} } \alpha_e \;,
\ee
with $U_{\cG}(\alpha)$ the first Symanzik polynomial (see for instance \cite{Rivasseau:1991ub,Krajewski:2008fa}) of $\cG$. $U_{\cG}(\alpha)$ is a sum of positive terms and each monomial has the global degree $E(\cG)- V(\cG) +1$ in the variables $\alpha$.
We are interested in the scale derivative of the amplitude. After uniformly rescaling all the parameters by $k^{-2}$ and using $n(\cG)=4$, we have:
\be\label{eq:amplirescale}
  \pmb{\partial} \bar{A}(\cG) =   -2 \sum_{e_0 \in \cG} 
  \int_{k^2\Lambda^{-2}}^{1} 
  \left(  \prod_{e\in \cG}^{e\neq e_0} d\alpha_e \, \alpha_e^{\zeta-1}\right)\; 
  \frac{ 1 }{ \big[ U_{\cG}(\alpha)
  \big]_{\alpha_{e_0}=1}^{d/2} } \; ,  
\ee
that is the scale derivative is a sum over graphs where one of the (rescaled) parameters has been set to its maximal value 1. 
We call this edge \emph{the marked edge} and we represent it as dashed.
Below we will need to consider the cap with zero rungs $\bullet$ which consists in a vertex (a blue vertex associated to $\bar{\lambda}_1$) joined by two edges to the rest of the graph.

\begin{figure}[ht]
\begin{center}
\includegraphics[scale=1]{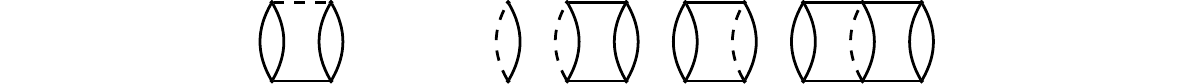} 
 \caption{From left to right, the graphs $ [ U_1 H U_1]$ and 
 $ V , [V U_1] ,  [U_1V ]$ and $[U_1VU_1]$.} \label{fig:Uh}
 \end{center}
\end{figure}

The derivatives can act on the horizontal or the vertical edges in a graph.
We call $H$ (for horizontal) the subgraph with two parallel horizontal edges, one of which is marked and $V$ (for vertical) the vertical rung with one marked edge. 
Correspondingly, we denote
$[ U_p H U_q ]$  the amplitude of a graph consisting in a ladder with $p\ge 1$ rungs followed by two horizontal edges, one of which is marked, followed by a ladder with $q\ge 1$ rungs
 and $[U_p V U_q]$ the amplitude of a graph consisting in a ladder with $p\ge 1$ rungs followed by a rung with a marked edge, followed by a ladder with $q\ge 1$ rungs. We denote their generating functions:
 \[ UHU = \sum_{p,q\ge 1}  \bar{g}^{2p+2q} [ U_pHU_q ] \;, \qquad g^2 U VU  = \sum_{p,q\ge 1}  \bar{g}^{2p+2q+2} [ U_p V U_q] \;. \] 
Several examples are depicted in figure~\ref{fig:Uh}. 
 
\begin{figure}[ht]
\begin{center}
\includegraphics[scale=1]{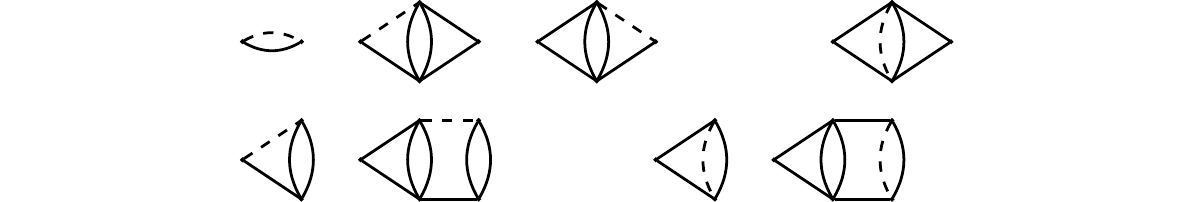} 
 \caption{Top row, from left to right the graphs $[\bullet H\bullet ] , 
 [\bullet H S_1], [S_1 H \bullet]  $ and $[\bullet  V \bullet]$.
 Bottom row, from left to right $[\bullet HU_1],[S_1HU_1]$ and
 $[\bullet V ],[S_1V ]$.
 } \label{fig:Th}
 \end{center}
\end{figure}

The scale derivative of $U_r$ is:
\begin{align*}
&\pmb{\partial} U_1 = (-4) V \;,\\ 
&\pmb{\partial} U_r = 
  (- 4) \bigg[  \sum_{p,q\ge 1}^{p+q = r}  [ U_p H U_q ]  + \sum_{p,q\ge 1}^{p+q = r-1}   [ U_p V U_q ] + [ V U_{r-1} ] + [ U_{r-1} V ] \bigg] \;,
   \;\;\;\; \forall r\ge 1\;,
\end{align*}
which for generating functions becomes:
\be\label{eq:derivU}
 \pmb{\partial} U = ( -4 ) \bigg\{ \bar{g}^2  V   + 2 \bar{g}^2 VU  + U ( H + \bar{g}^2 V ) U  \bigg\}\;.
\ee
Similarly, we get, using obvious notation, (see figure~\ref{fig:Th}):
 \begin{align}\label{eq:derivT}
 & \pmb{\partial} T  = (-4 ) \bigg\{
  ( \bullet + S) (H +\bar{g}^2 V ) ( \bullet + S ) \bigg\} \;,  \\ \label{eq:derivS}
& \pmb{\partial } S   =  (-4) \bigg\{
 (\bullet +S) \big[ \bar{g}^2  V + ( H + \bar{g}^2 V) U \big] \bigg\} \;.
 \end{align}

\paragraph{Taylor operators.}
A connected amputated subgraph $\gamma \subset \cG$ is a subset of the edges 
of $\mathcal{G}$. Like $\cG$, $\gamma$ contains all the end vertices of its edges. The external half-edges (or legs) of $\gamma$ are either half-edges of $\cG$ or come from the edges of $\cG$ which do not belong to $\gamma$ but are incident to vertices in $\gamma$. 

For any subgraph $\gamma$ of $\cG$, we denote $\tau_{\gamma}$ the ``localization'' operator acting on $\bar{A}(\cG)$ which on the one hand separates the subgraph $\gamma$ and on the other, in $\cG$, it shrinks $\gamma$ to a vertex to give the graph $\cG/\gamma$:
\[
 \tau_{\gamma} \bar{A}(\cG) = \bar{A}(\cG/ \gamma) \bar{A}(\gamma) \;. 
\]
The edges of $\cG$ are partitioned between the edges of $\gamma$ and the ones of $\cG/\gamma$. We denote $ \alpha_{\cG/\gamma} = \{ \alpha_e| e\in \cG/\gamma \} $ and $\alpha_{\gamma} = \{ \alpha_e| e\in \gamma \}$. Observe that 
$E(\cG) = E(\gamma) + E(\cG /\gamma)$ and $V(\cG) = V(\gamma) + V(\cG /\gamma) - 1$. Any spanning tree in $\cG$ is:
\begin{itemize}
 \item either the union of a spanning tree in $\gamma$ with a spanning tree in $\cG/\gamma$, in which case the global degree in  $\alpha_{\gamma}$ of the corresponding term in $U_{\cG}(\alpha)$ is exactly $ E(\gamma) - V(\gamma) +1$. Any tree in $\gamma$ and any tree in $\cG/\gamma$ lead to exactly one tree in $\cG$.
 \item or not, in which case the global degree in $\alpha_{\gamma}$ of the corresponding term in $U_{\cG}(\alpha)$ is at least $ E(\gamma) - V(\gamma) +2$.
\end{itemize}
It follows that under a uniform rescaling of $\alpha_{\gamma}$ by $u$ we have:
\[
 U_{\cG}(\alpha)\big{|}_{\alpha_{\gamma} = u\alpha_{\gamma}}  = 
 u^{   E(\gamma) - V(\gamma) + 1  }  
 \bigg{[} U_{\cG/\gamma}(\alpha_{\cG/\gamma} )U_{\gamma}(\alpha_{\gamma}) 
    + \sum_{q\ge 1} u^q 
    [ \alpha_{\gamma} ]^{E(\gamma) - V(\gamma) + 1 + q}  
    [\alpha_{\cG/\gamma}]^{ E(\cG/\gamma) - V(\cG/\gamma) +1 -q }
 \bigg{]} \;,
\]
where we indicated the global scaling with $\alpha_{\gamma}$
and $\alpha_{\cG/\gamma} $ of the corrections.
The localization operator $\tau_{\gamma}$ can be implemented as a Taylor operator
\cite{Bergere:1974zh,Bergere:1977ft,Bergere:1980sm}  acting on the integrand:
\begin{equation}\label{eq:taylor}
\begin{split}
 \tau_{\gamma}\frac{ 1 }{ [U_{\cG}(\alpha)]^{d/2}}  & = 
                            \frac{ u^{\frac{d}{2} \left[ E(\gamma) - V(\gamma) + 1\right] }  }{ \big[U_{\cG}(\alpha) \big{|}_{\alpha_{\gamma} = u\alpha_{\gamma}} \big]^{d/2}} \bigg|_{u\to 0} 
       = \frac{1}{[U_{\cG/\gamma}(\alpha_{\cG/\gamma})]^{d/2}} \;
       \frac{1}{[U_{\gamma}(\alpha_{\gamma})]^{d/2}}    \;, \crcr
(1 - \tau_{\gamma} )\frac{ 1 }{ [U_{\cG}(\alpha)]^{d/2}} & = 
 \int_0^1 du\; \frac{d}{du} \bigg\{     u^{\frac{d}{2} \left[ E(\gamma) - V(\gamma) + 1\right] }   \frac{ 1 }{ \big[U_{\cG}(\alpha) \big{|}_{\alpha_{\gamma} = u\alpha_{\gamma}} \big]^{d/2}}      \bigg\} \;.
\end{split}
\end{equation}

\paragraph{Subtraction operator.}
We call the one particle irreducible four-point subgraphs\footnote{As we deal with graphs with no two-point subgraphs, all the connected four-point subgraphs are automatically one particle irreducible.} of 
$\cG$  \emph{dangerous}. 
An inclusion forest of dangerous subgraphs is a set of dangerous subgraphs which are either \emph{nested} or \emph{totally disjoint} (that is they do not have any vertex in common):
\[
 \bigg\{ \gamma \subset \cG \;, n(\gamma) =4 \; \bigg{|} \;\; 
  \forall \gamma_1,\gamma_2 \; \text{ either } \gamma_1 \subset \gamma_2\, \text{ or } \gamma_2 \subset \gamma_1 \, \text{ or }
   \gamma_1\cap \gamma_2 =\emptyset \bigg\} \;.
\]
We denote $\pmb{F}(\cG)$ the set of all the inclusion forests of dangerous subgraphs of $\cG$, including the empty forest. The BPHZ subtraction operator \cite{bogoliubow1957multiplikation,Hepp:1966eg,Zimmermann:1969jj}
is:
\[
 R  =\sum_{F \in \pmb{F}(\cG)} \prod_{\gamma \in F} (-\tau_{\gamma}) \;.
\]
The operator is well defined because the localization operators of graphs in a forest commute. 

We are concerned here with ladders, caps and double caps $U,S$ and $T$. In all these cases the dangerous proper subgraphs (i.e. different from the graph itself) have a  particularly simple structure.

\begin{lemma}
Any proper four-point subgraph $\gamma \subset \cG$ consists in a sequence of vertical rungs connected by horizontal edges. $\gamma$ can reach one end of the graph or not (the end is either a vertex with two external points for $T$ and for one end of $S$, or a rung with two external vertices for $U$ and for the other end of $S$).
\end{lemma}

\begin{proof}
 Consider $\cG$ a ladder, cap or double cap and a four-point proper subgraph $\gamma\subset \cG$ (that is $\gamma$ is not $\cG$ itself):
\begin{itemize}
\item assume $\gamma$ does not contain any internal vertex of $\cG$ (i.e. a vertex which is not incident to external half-edges of $\cG$). This is impossible for $S$. For $U$, $\gamma$ must consist in 
a vertical rung connecting two of the boundary vertices. For $T$, 
only $T_0$ has such a subgraph, $\gamma = T_0$, which is not proper.

\item assume $\gamma$ contains an internal vertex $v$. Then this vertex is part of a rung with edges $e_1,e_2$ connecting $v$ with $v'$.
    \begin{itemize}
     \item assume that neither $e_1$ nor $e_2$ belong to $\gamma$. Then both the horizontal edges incident at $v$ must belong to $\gamma$ (otherwise $\gamma$ is 1PR), and $v'$ must also belong to $\gamma$ (otherwise again $\gamma$ is 1PR). But $v$ and $v'$ already support four external points for $\gamma$, hence $\gamma$ has no additional external points, which is impossible.
     \item assume only $e_1$ belongs to $\gamma$ but $e_2$ does not.
     Consider the four horizontal edges incident to $v$ and $v'$
        \begin{itemize}
         \item  if neither one of them or only one of them belongs to $\gamma$ then $\gamma$ has more than four external points.
         \item  if exactly two of them belong to $\gamma$ then 
         there are already four external points supported by $v$ and $v'$, hence $\gamma$ cannot have any additional external points which is impossible. 
         \item if only three of them belong to $\gamma$ then $\gamma$ is 1PR. 
         \item if all the four belong to $\gamma$, then $\gamma$ splits into a left part (to the left of the rung $e_1,e_2$) and a  right part (to the right of $e_1,e_2$). Each part brings at least two additional external points, which makes $\gamma$ at least a six-point graph.
        \end{itemize}
    \end{itemize}
\end{itemize}

It follows that $\gamma$ must contain a rung with edges $e_1e_2$ connecting two internal vertices $v$ and $v'$. The iteration is now simple: 
\begin{itemize}
 \item either $\gamma$ consists in only this rung,
 \item or  (as $\gamma$ is 1PI) it contains the pair of horizontal edges incident to $v$ and $v'$ pointing to the right (or the pair pointing to the left or both). These edges either:
     \begin{itemize}
      \item reach the end of the graph which is either an external vertex with two external points ($T$ or one end of $S$) or a rung with two external points ($U$ or the other end of $S$). In the second case the rung must belong to $\gamma$, as two external points must come from the left of $\gamma$.
      \item reach a pair of internal vertices $w$ and $w'$ connected by a rung $e_1',e_2'$. But then the entire rung $e_1',e_2'$ belongs to $\gamma$ and we iterate.
     \end{itemize}
\end{itemize}
\end{proof}

Now, the graphs contributing to $\pmb{\partial}T,~\pmb{\partial}S$ and 
$\pmb{\partial}U$ have the additional marked edge whose parameter is set to $1$. In this case, we restrict to dangerous subgraphs which do not contain the marked edge. Then the dangerous subgraphs are 
confined to live either to the left or to the right of the marked edge,
and the subtraction operator factors into the sum over the forests of left subgraphs and the sum over forests of right subgraphs:
\[
  R = R^{\rm left} R^{\rm right } \;,\qquad 
 R^{\rm left}  =\sum_{F^{\rm l} \in \pmb{F}^{\rm l}(\cG)} \prod_{\gamma^{\rm l} \in F^{\rm l}} (-\tau_{\gamma^{\rm l}}) \; ,\;\;
 R^{\rm right}=    \sum_{F^{\rm r} \in \pmb{F}^{\rm r}(\cG)} 
   \prod_{\gamma^{\rm r} \in F^{\rm r}} (-\tau_{\gamma^{\rm r}}) \;,
\]
where the forests in $\pmb{F}^{\rm l}(\cG)$ (respectively 
$\pmb{F}^{\rm r}(\cG)$) contain only graphs at the left (resp. right) 
of the marked edge. Let us discuss the forest of left subgraphs. We denote the rungs to the left of the marked edge by $1,2\dots, p$ ($p$ being the closest to the marked edge). We treat the case in which the left end of the graph is a cap (a similar reasoning works for the case of an end rung). We denote $S_r$ the subgraph starting at the left end of $\cG$ and having $r$ rungs. Any left forest can be obtained from a forest which does not contain $S_p$ by adding or not $S_p$. Thus:
\[
 R^{\rm left} = (1-\tau_{S_p} )
 \sum_{F^{\rm l} \in \pmb{F}^{\rm l}(\cG)}^{S_p \notin F^{\rm l}} \prod_{\gamma^{\rm l} \in F^{\rm l}} (-\tau_{\gamma^{\rm l}}) \; .
\]
Now, among the graphs $\gamma^{\rm l}$ in the forest $F^{\rm l}$ some, denoted
$\gamma^{\rm l}_{\supset p}$, contain the rung $p$, and the rest do not. We have:
\[
 (1-\tau_{S_p} ) \tau_{ \gamma^{\rm l}_{\supset p} } =0 \;,
\]
therefore the sum truncates to the forests such that none of the graphs in the forest contains the last rung. Iterating we find:
\[
 R^{\rm left} = \prod_{i=1}^p (1-\tau_{S_i} ) \;.
\]
Let us define
\[
 R^S = \prod_{i\ge 1} (1-\tau_{S_i} ) \;, \qquad
  R^U = \prod_{j\ge 1} (1-\tau_{U_j} ) \;, 
\]
where $\tau_{S_p}\bar{A}(\cG)$ (resp. $\tau_{U_p}\bar{A}(\cG)$) is  zero if $S_p$ (resp.  $U_p$) is not a subgraph of $\cG$.
We can now state the main result of this section.
\begin{theorem}\label{thm:renorm}
 The coefficients $\beta^{\bar{g}}_0$, $\beta^{\bar{g}}_1 $ and $\beta^{\bar{g}}_2$ are minus the renormalized scale derivatives of the ladders, caps and double caps generating functions:
 \begin{align*}
-  \beta_0^{\bar{g}} &=  \pmb{\partial} U - 2 \frac{U}{1+S} \pmb{\partial} S + \frac{U^2}{(1+S)^2} \pmb{\partial} T   =R \big\{ \pmb{\partial} U  \big\} =
   \big( R^U\big)^{\rm left} \big(R^U\big)^{\rm right} \big\{ \pmb{\partial} U  \big\}
 \; , \crcr
-  \beta_1^{\bar{g}} &= \frac{1}{1+S} \pmb{\partial}S - \frac{U}{(1+S)^2} \pmb{\partial} T    = R \big\{ \pmb{\partial} S  \big\}  =    \big( R^S\big)^{\rm left} \big(R^U\big)^{\rm right}  \big\{\pmb{\partial} S\big\}
 \;,  
 \crcr 
 -  \beta_2^{\bar{g}} &=  \frac{1}{(1+S)^2} \pmb{\partial} T =R \big\{ \pmb{\partial} T \big\}
 =   \big( R^S\big)^{\rm left} \big(R^S\big)^{\rm right}  \big\{ \pmb{\partial} T \big\}
 \; .
 \end{align*}
\end{theorem}

\begin{proof} Recalling the scale derivatives from 
equations \eqref{eq:derivU}, \eqref{eq:derivT}, and \eqref{eq:derivS}, 
the theorem follows provided that, for $\cG = H,V$, we have:
\[
 R^S \big\{ (  \bullet +S ) \cG  \big\}  = 
 \left( \frac{1}{1+S}  \right)
    \big[ ( \bullet +S ) \cG   \big]
 \;,\qquad  R^U \big\{  \cG  U  \big\} = 
    \cG U  -  
    \left( \frac{U}{1+S} \right) \big[ \cG ( \bullet +S )  \big] \;.
\]
Let us first check $R^S$. We have:
\be
\begin{split}
  R^S \big\{ ( \bullet +S ) \cG  \big\}  & =  \bullet \cG  +  \sum_{r\ge 1} \bar{g}^{2r} \left[ \prod_{i=1}^r
  (1-\tau_{S_i}) \right] [ S_r\cG ] \crcr
 & =  \bullet \cG   +  S  \cG  + 
  \sum_{r\ge 1} \bar{g}^{2r}  \sum_{q=1}^r 
   \sum_{1\le i_1<\dots <i_q \le r}
 \left[ \prod_{s=1}^q   (-\tau_{S_{i_s}} )  \right] [S_r\cG] \;.
 \end{split}
\ee
Taking into account the action of the localization operators on the cap, the sum over $r$ becomes:
\begin{align*}
&  \sum_{r\ge 1} \bar{g}^{2r}  \sum_{q=1}^r (-1)^q
   \sum_{1\le i_1<\dots <i_q \le r}
  S_{i_1} S_{i_2 - i_1} \dots S_{i_{q}-i_{q-1}}
 [ S_{r-i_q }\cG] \crcr
 & = \sum_{r\ge 1} \bar{g}^{2r}  \sum_{q=1}^r (-1)^q
   \sum_{d_1, \dots d_{q} \ge 1}^{d_1 + \dots + d_{q} \le r}
  S_{d_1}  \dots S_{d_{q}} [ S_{r- d_1 \dots - d_q  } \cG ] \crcr
  & =  
    \sum_{q\ge 1} (-1)^q
  \left( \sum_{d\ge 1} \bar{g}^{2d} S_{d} \right)^q  
  \left(   \bullet \cG   +\sum_{p\ge 1} \bar{g}^{2p} [S_p \cG]  \right)
   = \left( \frac{1}{1+S} -1 \right)
   \big[  ( \bullet + S) \cG \big] \;.
\end{align*}
Concerning $R^U$, we have:
 \begin{align*}
  R^U \big\{  \cG U   \big\} &=     \cG U  
  + \sum_{r\ge 1} \bar{g}^{2r} \sum_{q=1}^r (-1)^q \sum_{1\le i_1<\dots <i_q \le r}
 \left[ \prod_{s=1}^q   ( \tau_{U_{i_s}} )  \right] [ \cG U_r]
 \crcr
 & = \cG U 
  + \sum_{r\ge 1} \bar{g}^{2r} \sum_{q=1}^r (-1)^q \sum_{1\le i_1<\dots <i_q \le r}
  U_{i_1} S_{i_2-i_1} \dots S_{i_q - i_{q-1}} [ \cG S_{r-i_q} ] \crcr
  & = \cG U + \sum_{q\ge 1} (-1)^q 
  \left(  \sum_{d\ge 1} \bar{g}^{2d} U_d \right)
   \left( \sum_{d\ge 1} \bar{g}^{2d}S_d \right)^{q-1} \left( 
    \cG\bullet  + \sum_{p\ge 1} \bar{g}^{2p} [ \cG S_p ]  \right) \crcr
   & = \cG U - \frac{U}{1+S} \big[  \cG ( \bullet +S) \big] \;.
 \end{align*}

\end{proof}

\begin{figure}[ht]
\begin{center}
\includegraphics[scale=1]{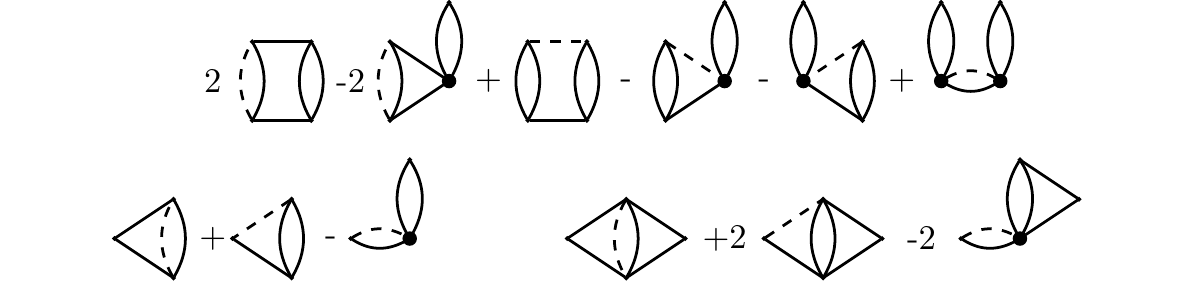} 
 \caption{First non-trivial subtractions. We represented the shrunk vertex as thickened.} \label{fig:betaeq}
 \end{center}
\end{figure}

At first orders the bare amplitudes are: 
\begin{align*}
\left( - \frac{1}{4} \right)  \pmb{\partial} U =&  
 \bar{g}^2 V  + 2\bar{g}^4 VU_1 + \bar{g}^4 U_1 H U_1 + O(\bar{g}^6) \;,\crcr
 \left( - \frac{1}{4} \right)  \pmb{\partial} S = & 
 \bar{g}^2 \bullet V + \bar{g}^2 \bullet H U_1 \;, 
 \qquad \left( - \frac{1}{4} \right)  \pmb{\partial} T = \bullet H \bullet 
 + \bar{g}^2 \bullet V \bullet +2 \bar{g}^2  \bullet H S_1 \; ,
\end{align*}
and the first non-trivial renormalized amplitudes are (see figure~\ref{fig:betaeq}):
\begin{align*}
    \frac{ (- 1) }{4}  R[\pmb{\partial}U] \bigg{|}_{\bar{g}^4} & = 
    2(1-\tau_{U_1}) [VU_1] + (1-\tau_{U_1}) (1-\tau_{U_1})  [ U_1 H U_1 ] \;,
 \crcr
  \frac{ (- 1) }{4}  R[\pmb{\partial}S] \bigg{|}_{\bar{g}^2}& =
  \bullet V + (1-\tau_{U_1} ) \bullet [HU_1] \;, \qquad
  \frac{ (- 1) }{4}  R[\pmb{\partial}T] \bigg{|}_{\bar{g}^2}  = 
  \bullet V \bullet +2  (1 - \tau_{S_1}) \bullet H S_1 \;.
\end{align*}
This should be compared to the coefficients in \eqref{eq:beta3loops}.

\paragraph{Convergence.} It remains to prove that the subtracted amplitudes are convergent. This might seem obvious at first sight: this is the whole 
``raison d'\^etre'' of the subtraction operator. However, it turns out that for massless theories subtracted a zero momentum the usual proofs do not work. What works is the
proof in \cite{Rivasseau:1991ub} chapter II.3 relying on multiscale analysis and the classification of forests.
The analysis is however quite involved and we will not reproduce it here.

\section{Unitarity}
\label{sec:unitarity}
In this section we are now interested in studying the CFT at the large-$N$ infrared fixed point.

\subsection{Primary operators and OPE coefficients}
\label{sec:OPE}
We start by computing the dimensions of the bilinear primaries of arbitrary spin and their OPE coefficients as defined in section \ref{sec:CFT}.

\paragraph{Dimensions of primaries.}
Our model has $\Delta_{\phi}=d/4$ and the eigenvalues of the two-particle irreducible four-point kernel are:
\begin{equation}
k(h,J)=3\bar{g}^2\Gamma(d/4)^4
 \frac{\Gamma(-\frac{d}{4}+\frac{h+J}{2})\Gamma(\frac{d}{4}-\frac{h-J}{2})}{\Gamma(\frac{3d}{4}-\frac{h-J}{2})\Gamma(\frac{d}{4}+\frac{h+J}{2})} \;.
\label{eq:k(h,J)}
\end{equation}
We give a detail computation of the eigenvalues in appendix \ref{ap:eigenvalues}.

We are interested in the solutions $h_{m,J}$ of the equation $k(h,J) = 1 $ with ${\rm Re}(h_{m,J}) \ge d/2$ for small $\bar{g}$. Such solutions correspond to values of $h$ for which the ratio of gamma functions diverges. As the gamma function does not have any zeros in the complex plane ($1/\Gamma(z)$ is an entire function) divergences only arise near the poles of 
the numerator $ \Gamma(-\frac{d}{4}+\frac{h+J}{2})\Gamma(\frac{d}{4}-\frac{h-J}{2}) $. We are only interested in the poles in the region ${\rm Re}(h) \ge d/2$, therefore the relevant poles of the numerator are located at the classical dimensions of the bilinear spin $J$ operators:
\begin{equation}
h^{\rm classical} = d/2 + J + 2m  \; , \qquad m\ge 0 \;.
\end{equation}
The poles of $ \Gamma(-\frac{d}{4}+\frac{h+J}{2})$ are located at $h = d/2-J-2m$ which does not obey ${\rm Re}(h) \ge d/2$ for $(m,J)\neq (0,0)$. In the neighborhood  of each pole we parametrize $h = d/2 + J + 2m + 2z$ with $z\sim o(\bar{g})$. Then $k(h,J)$ becomes:  
\begin{equation}
k(d/2 + J + 2m + 2z  ,J) = 3\bar{g}^2 \Gamma(d/4)^4 
    \frac{ \Gamma(J+m+z) \Gamma(-m-z)  }
    { \Gamma( \frac{d}{2} -m-z ) \Gamma( \frac{d}{2} + J +m + z) } \; .
\end{equation}

For any $d$, the pole $(m,J) = (0,0) $ is special because both the $\Gamma$ functions in the numerator are singular, while for the poles $(m,J)\neq(0,0)$ only one of  them is. Moreover the case $d=2$ is special as $\Gamma( \frac{d}{2} -m-z ) $ diverges for $m\ge 1$ at small $z$. We take $d\neq 2$ and we will deal with $d=2$ in section \ref{sec:d=2}.
The function $k(h,J)$ close to the pole $(m,J)$ takes the form:
\begin{equation}
\label{eq:2ks}
\begin{split}
  k_{(0,0)}(z) \, &= \, 3\bar{g}^2 \Gamma(d/4)^4  \; 
   \frac{ \Gamma(1+z) \Gamma(1-z)  }{  (- z^2) \Gamma(\frac{d}{2}-z) \Gamma(\frac{d}{2}+z) }  \;, \crcr
k_{(m,J)} (z) \, & \xlongequal{(m,J) \neq (0,0)} \,
  3\bar{g}^2 \Gamma(d/4)^4
    \frac{\Gamma(J+m+z)}{ \Gamma(\frac{d}{2}+J+m+z) }
     \; \frac{ \Gamma(1+z) \Gamma(1-z) }{ z \Gamma(m+1 + z) }
     \;\frac{ \Gamma( m+1 - \frac{d}{2} +z )}{  
       \Gamma(z-\frac{d}{2}) \Gamma(\frac{d}{2} +1 -z)}  \;,  
\end{split}
\end{equation}
where in the second line we used $\Gamma(-m + a) = (-1)^{m+1} \Gamma(-a) \Gamma(1+a) /\Gamma(m+1-a)$. The dimensions of the physical operators in the interacting theory are the solutions of the equation $k(h_{m,J},J)=1$, that is:
\begin{equation}
   h_{m,J} = \frac{d}{2} + J +2 m + 2z_{m,J}\;, \qquad
     k_{(m,J)} ( z_{m,J} ) = 1 \;.
\end{equation}

The solutions $z_{m,J}$ (which are the anomalous scalings of the bilinear primaries at the fixed point) are obtained as follows.

\paragraph{\it The case $(m,J) = (0,0)$} The anomalous dimension $z_{0,0}$ is the solution of:
  \begin{equation}\label{eq:z00}
     (- z^2) \frac{ \Gamma(\frac{d}{2}-z) \Gamma(\frac{d}{2}+z) }  
   { \Gamma(1+z) \Gamma(1-z)  }
    =    3\bar{g}^2 \Gamma(d/4)^4  \; ,
  \end{equation}
  where only the solutions with ${\rm Re}(z)\ge 0$ are picked up.
  Observe that the left-hand side of \eqref{eq:z00} is a series 
  in $z^2$ with real coefficients, therefore $z_{0,0}^2$ is a series in $\bar{g}^2$ with real coefficients starting at first order which implies:
  \begin{equation}
    z_{0,0} =  \pm \frac{  \Gamma(d/4)^2  }{ \Gamma(d/2) } \sqrt{   - 3\bar{g}^2  } 
       \left( 1 + \sum_{q\ge 0} C_q \bar{g}^{2q}\right) \;, \qquad 
         C_q \in \mathbb{R} \;.
  \end{equation}
  At first order in $\bar{g}$ we have:
  \begin{equation}
     z_{0,0} = \pm \frac{  \Gamma(d/4)^2  }{ \Gamma(d/2) } \sqrt{   - 3\bar{g}^2  } + {\cal O}(\bar{g}^3) \;, \qquad       h_{\pm} = \frac{d}{2} \pm 2 \frac{  \Gamma(d/4)^2  }{ \Gamma(d/2) } \sqrt{   - 3\bar{g}^2  } + {\cal O}(\bar{g}^3) \;.
  \end{equation}  
  
\paragraph{\it The case $(m,J)\neq (0,0)$} The remaining anomalous dimensions are the solutions of:
   \begin{equation}
    z \;  \frac{ \Gamma(\frac{d}{2}+J+m+z)  \Gamma(m+1 + z) \Gamma(z-\frac{d}{2}) \Gamma(\frac{d}{2} +1 -z)} 
      {\Gamma(J+m+z) \Gamma(1+z) \Gamma(1-z)
      \Gamma( m+1 - \frac{d}{2} +z )
      }
      =   3\bar{g}^2 \Gamma(d/4)^4 \;.
   \end{equation}
   Clearly $z_{m,J}$ are series in $\bar{g}^2$ with real coefficients and at first order in $\bar{g}$ we have:
    \begin{equation}
    \begin{split}
      z_{m,J} & =  3\bar{g}^2 \Gamma(d/4)^4
        \frac{\Gamma(J+m) \Gamma( m+1 - \frac{d}{2} ) }
        { \Gamma(\frac{d}{2}+J+m)  \Gamma(m+1 ) \Gamma(-\frac{d}{2}) \Gamma(\frac{d}{2} +1 )} + 
     {\cal O}(\bar{g}^4) \crcr
         h_{m,J}& = \frac{d}{2} + J + 2m + 2 \frac{ 3\bar{g}^2 \Gamma(d/4)^4 
     \Gamma(m + J)  \Gamma(m+1 - \frac{d}{2}) \sin\left( -\frac{\pi d}{2} \right)  }{ \Gamma(\frac{d}{2} + J + m ) \Gamma(m+1) \; \pi } + 
     {\cal O}(\bar{g}^4) \;.
     \end{split}
    \label{eq:hmJSol}
    \end{equation}

 \paragraph{The OPE coefficients.} 
In appendix \ref{app:measure}, we give a detailed computation of the measure and residue factors which are needed for the OPE coefficients \eqref{eq:OPE coef}.
Here we simply present the final result. Putting all factors together the OPE coefficients are:
\begin{equation}\begin{split}
 c_{0,0}^2 & = 
   2 
-   \, 4\,  z_{0,0}\bigg[   \psi(d/2) +  \psi(1) - 2\psi(d/4)  \bigg]  
+ O(z_{0,0}^2) \crcr
& =   2 \pm  \sqrt{ -3\bar{g}^2 }  1 \frac{4\, \Gamma(d/4)^2   }{\Gamma(\frac{d}{2})} \bigg[ 2\psi(d/4)  -  \psi(d/2) -  \psi(1)  \bigg]  
+ {\cal O}(\bar{g}^3) \;,
\end{split}
\end{equation}
and for $(m,J) \neq (0,0)$:
\begin{equation}
\begin{split}
 c_{m,J}^2 & = \frac{2\Gamma(J+\tfrac{d}{2})   \Gamma(J+m )
    \Gamma(\frac{d}{2} + J + 2m  -1)}{\Gamma(J+1) \Gamma(m+1 )
  \Gamma( 1 + 2m - \frac{d}{2}  )\Gamma( \frac{d}{2} + J + m)}  \crcr
 &  \quad \times \frac{\Gamma( 1 + m - \frac{d}{2}  )\Gamma(\frac{d}{4} + J + m  )^2}{\Gamma(J + 2m  )\Gamma(\frac{d}{2} + 2J + 2m   -1)
  \Gamma( \frac{d}{4} -m  )^2} + {\cal O}(\bar{g}^2) \;.
 \end{split}
 \label{eq:OPE-coef-generalmJ}
\end{equation}

\paragraph{Summary of results.} The conclusions of the computation of the OPE coefficients at small $\bar{g}$ are:
\begin{itemize}
 \item[-] at $\bar{g}=0$ we recover the classical dimensions $h^{\rm classical}=d/2 +J+2m$.
 \item[-] at $\bar{g}\neq 0$ we get the dimensions $h_{m,J}= d/2 + J +2m +2z_{m,J}$ with $z_{0,0}\sim \sqrt{-\bar{g}^2}$ and $z_{m,J} \sim \bar{g}^2$ for $(m,J)\neq (0,0)$. For $(m,J)\neq (0,0)$, $z_{m,J}$ is always real. $z_{0,0}$
 is real for purely imaginary coupling and purely imaginary for real coupling. This is true at all orders in $\bar{g}$.
 \item[-] at order $\bar{g}^0$ all the OPE coefficients $c_{m,J}$ are real,  $c_{m,J}^2>0$. This is reassuring as it means that  the free theory is unitary, which it is (from OS positivity).
 \item[-] the OPE coefficients $c_{m,J}$ with $(m,J)\neq (0,0)$ are always real, $c_{m,J}^2>0$, at all orders in $\bar{g}$  because they are series with real coefficients in $z_{m,J}$ which in turn is a series with real coefficients in $\bar{g}^2$.
 \item[-] the OPE coefficient $c_{0,0}$ is: 
    \begin{itemize}
     \item complex ($c_{0,0}^2$ has a non zero imaginary part) at all orders in $\bar{g}$ for $\bar{g}$ real,
     \item real ($c_{0,0}^2>0$) at all orders in $\bar{g}$ for $\bar{g}$ purely imaginary,
    \end{itemize}
   this is because $c_{0,0}$ is a series with real coefficients in $z_{0,0}$.
\end{itemize}

\subsubsection{The $d=3$ case}
\label{sec:d=3}
In this subsection, we focus on the $d=3$ case.
Setting $\Delta_{\phi}=d/4$ with $d=3$, we obtain the eigenvalues:
	\begin{equation}
		k(h,J) \, =  \, 3\bar{g}^2 \, \Gamma(3/4)^4 \, \frac{\Gamma(\frac{3}{4}+\frac{J}{2}-\frac{h}{2})\Gamma(\frac{h}{2}+\frac{J}{2}-\frac{3}{4})}
		{\Gamma(\frac{9}{4}+\frac{J}{2}-\frac{h}{2})\Gamma(\frac{h}{2}+\frac{J}{2}+\frac{3}{4})} \, ,
	\end{equation}
and the measure:
	\begin{equation}
	\begin{split}
		\mu_{3/4}^3(h,J) \, 
		&= \, \left( \frac{1+(-1)^J}{2} \right) \frac{ \Gamma(h-1)\Gamma(J+\frac{3}{2})\Gamma(3-h+J)\Gamma(\frac{h+J}{2})^2}
		{\Gamma(h-\frac{3}{2})\Gamma(J+1)\Gamma(\frac{3-h+J}{2})^2\Gamma(h+J-1)} \\
		&\hspace{80pt} \times \frac{\Gamma(\frac{h}{2}+\frac{J}{2}-\frac{3}{4})\Gamma(\frac{3}{4}-\frac{h}{2}+\frac{J}{2})}
		{\Gamma(\frac{3}{4}+\frac{h}{2}+\frac{J}{2})\Gamma(\frac{9}{4}-\frac{h}{2}+\frac{J}{2})} \, .
	\end{split}
	\end{equation}

\begin{figure}[htb]
	\begin{center}
		\scalebox{0.55}{\includegraphics{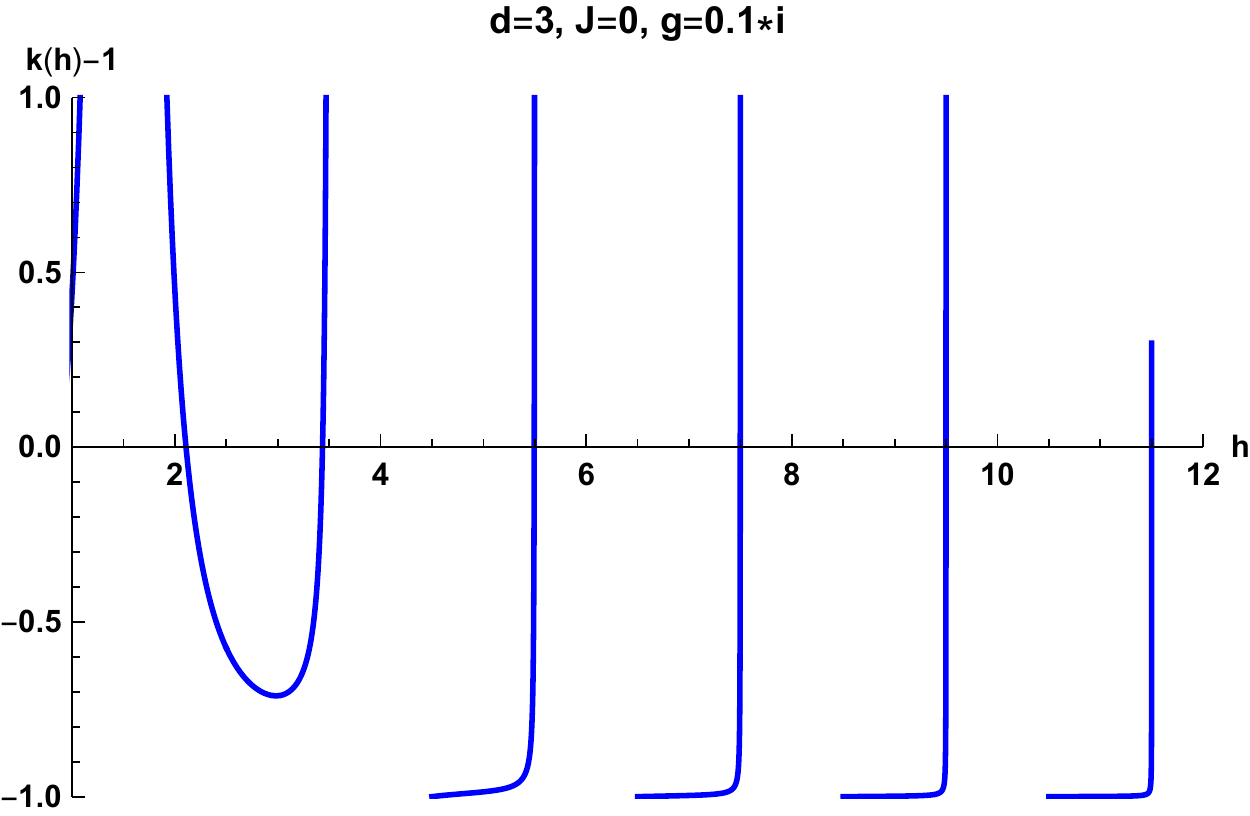}} \qquad 
		\scalebox{0.55}{\includegraphics{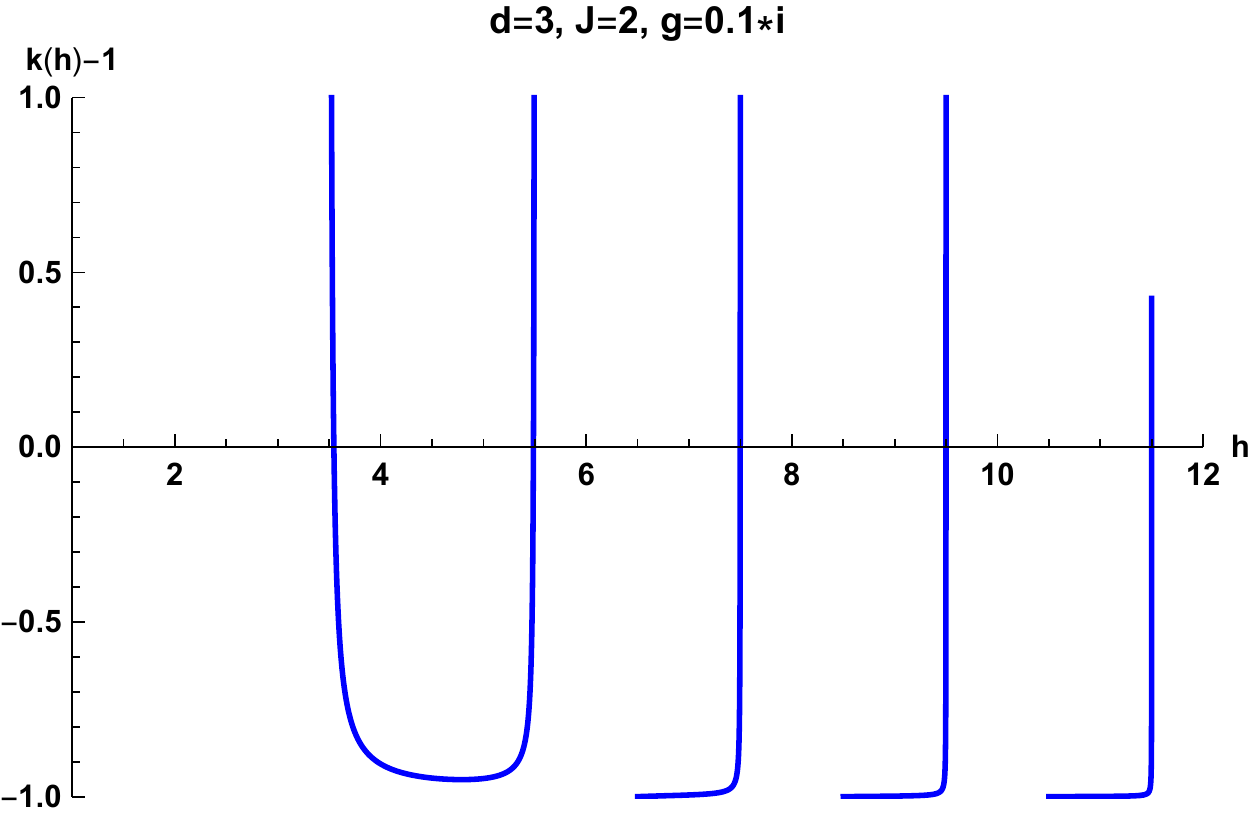}} \\ \, \vspace{30pt}
		\scalebox{0.55}{\includegraphics{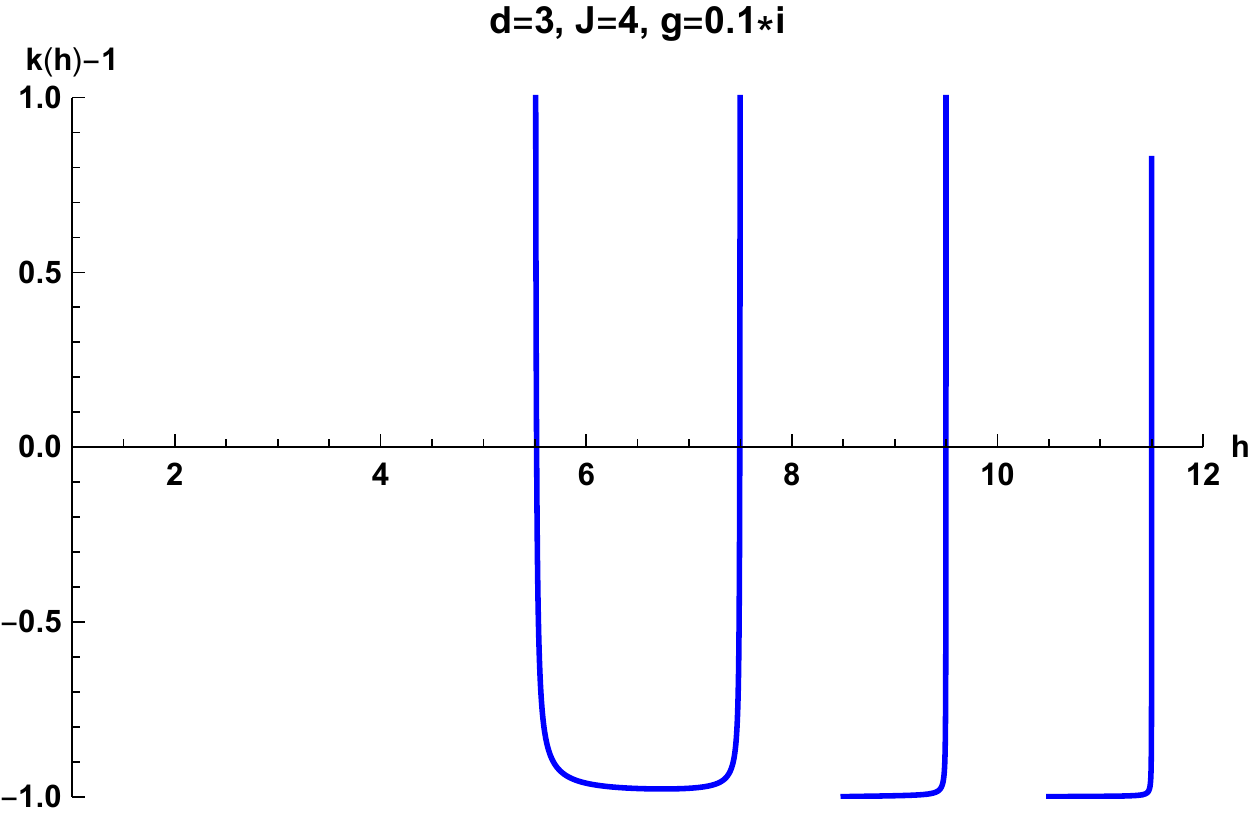}}
	\end{center}
	\caption{Plots of $k(h)-1$ for $d=3$, $\bar{g}=0.1i$, and $J=0,2,4$.}
	\label{fig:d=3}
\end{figure}

The plots in Figure \ref{fig:d=3} graphically show that we can find one solution for the conformal dimension close to $h_{m,J}=3/2+J+2m$
for each non-negative integers $m$ and $J$.
For the $(m,J)=(0,0)$ case, there is a rather bigger deviation from $h_{0,0}=3/2$.

\paragraph{\it The case $(m,J)=(0,0)$.}
Expanding for small coupling constant $g$, we find the physical conformal dimensions:
	\begin{equation}
		h_{\pm} \, = \, \frac{3}{2} \, \pm \, 4 \sqrt{- \frac{3\bar{g}^2}{\pi}} \, \Gamma(3/4)^2 \, + \, \mathcal{O}(\bar{g}^3) \, ,
	\end{equation}
and associated OPE coefficients:
	\begin{equation}
		c_{\pm}^2 \, = \, 2 \, \pm \,   \frac{8}{\pi}(\pi - 2 - 4 \log 2)\Gamma(3/4)^2 \sqrt{-3\pi \bar{g}^2} \, + \, \mathcal{O}(\bar{g}^3) \, .
	\end{equation}
For a real value of the coupling constant $\bar{g}$, both solutions $h_{\pm}$ are on the contour integral ${\rm Re}(h) = 3/2$  and  extra care is needed. In this case both $c_{\pm}^2$ are not real.

For a purely imaginary value of the coupling constant, $h_+$ is at the right of the contour while $h_-$ is at the left.
Therefore in this case only $h_+$ is in the spectrum of the model.
The associated OPE coefficient is $c_{+}^2=2 + \cdots>0$ for small coupling $\bar{g}$.

\paragraph{\it The case $(m,J)\ne(0,0)$.}
The other solutions are also obtained by small coupling expansion as: 
	\begin{equation}
		h_{m,J} \, = \, \frac{3}{2} + J + 2m \, + \, \frac{6 \Gamma(\tfrac{3}{4})^4 \Gamma(m-\frac{1}{2}) \Gamma(m+J) }{\pi \, \Gamma(m+1)\Gamma(m+J+\frac{3}{2})}\bar{g}^2
		\, + \, \mathcal{O}(\bar{g}^4) \, , 
	\end{equation}
for non-negative integers $m$ and $J$ excluding the $(m,J)=(0,0)$ case.
The associated OPE coefficients are given by:
	\begin{equation}
		c_{m,J}^2 \, = \frac{ \Gamma(m-\frac{1}{2})\Gamma(m+\frac{1}{4})\Gamma(J+\frac{3}{2})\Gamma(m+J)\Gamma(m+J+\frac{3}{4})\Gamma(2m+J+\frac{1}{2})}
		{4^{2m+J-1}\, \pi\,\Gamma(m+1)\Gamma(J+1)\Gamma(m-\frac{1}{4})\Gamma(2m+J)\Gamma(m+J+\frac{1}{4})\Gamma(m+J+\frac{3}{2})}
		\, + \, \mathcal{O}(\bar{g}^2) \, .
	\end{equation}
The zeroth order (i.e. $\bar{g}^0$) contributions are real and positive for any $m$ and $J$.
Hence, the OPE coefficients are real for all $m$ and $J$ in the small coupling regime.
This is a strong indication of unitarity of the model for $d=3$.

\subsubsection{The $d=2$ case}
\label{sec:d=2}
We  now focus on the $d=2$ case. We first set $d=2$ directly in the above equations. With $\Delta_{\phi}=d/4$, we have for the eigenvalues of the four-point kernel: 
	\begin{equation}
		k(h,J) \, = \, - \, \frac{12 \pi^2 \bar{g}^2}{(h+J-1)(h-J-1)} \, ,
	\end{equation}
therefore the solutions of $k(h,J)=1$ are given by:
	\begin{equation}
		h_{\pm} \, = \, 1 \, \pm \, \sqrt{J^2-12 \pi^2 \bar{g}^2} \, .
	\end{equation}
A physical dimension needs to have real part at least $1$, therefore these solutions exist in the spectrum only in the following range of the coupling constant:
	\begin{align}
		&{\rm For\ \, } h_+: \quad \bar{g}^2 \, \le \, \frac{J^2}{12 \pi^2} \, , \\
		&{\rm For\ \, } h_-: \quad \frac{J^2}{12 \pi^2} \, \le \, \bar{g}^2 \, .
	\end{align}
From this we can see that $J=0$ is a special case where the $\bar{g}^2 \to 0$ limit is well-defined for both $h_{\pm}$.
On the other hand, for $J>0$, the weak coupling limit $\bar{g}^2 \to 0$ is only well-defined for $h_+$ and in the weak coupling limit $h_-$ does not exist in the spectrum.
Therefore, the $h_+$ solution in $J>0$ corresponds to the $h_{m,J}$ solution in \eqref{eq:hmJSol} with $m=0$.
The measure in $d=2$ is given by:
	\begin{equation}
		\mu_{1/2}^2(h,J) \, 
		= \, \left( \frac{1+(-1)^J}{2^{2h-2}} \right) \frac{\Gamma(\frac{h+J}{2})\Gamma(\frac{1-h+J}{2})}{\Gamma(\frac{h+J+1}{2})\Gamma(\frac{2-h+J}{2})} \, .
	\end{equation}

\paragraph{\it Spin $J=0$.}
The OPE coefficients for the $(m,J)=(0,0)$ case are:
	\begin{equation}
		c_{\pm}^2 \, = \,  2 \, \mp \sqrt{-3\bar{g}^2}  \, 16 \pi \log2 \, + \, \mathcal{O}(\bar{g}^2) \, ,
	\end{equation}
that is these OPE coefficients are real for small pure imaginary coupling $\bar{g}$.

\paragraph{\it Spin $J>0$.}
The OPE coefficients for the $J>0$ case are: 
	\begin{equation}
		c_{0,J}^2 \, = \, \frac{2^{ 2-2J} \, \Gamma(J+\frac{1}{2})}{\sqrt{\pi}\, \Gamma(J+1)} \, + \, \mathcal{O}(\bar{g}^2) \, .
	\end{equation}
The zeroth order $\mathcal{O}(\bar{g}^0)$ contributions are real and positive for any $J$.
Hence, the OPE coefficients are real for all $J$ in the small coupling regime.

\subsubsection{Discontinuity at $d=2$}
\label{sec:discontinuity at d=2}
\begin{figure}[htb]
	\begin{center}
		\scalebox{0.65}{\includegraphics{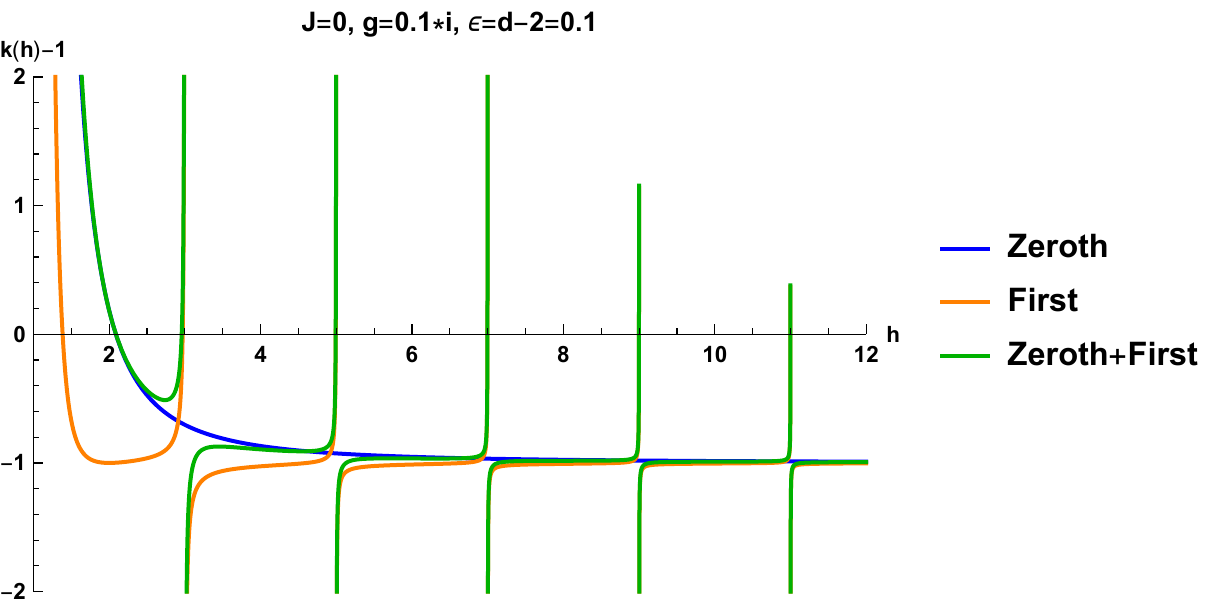}}  \qquad 
		\scalebox{0.65}{\includegraphics{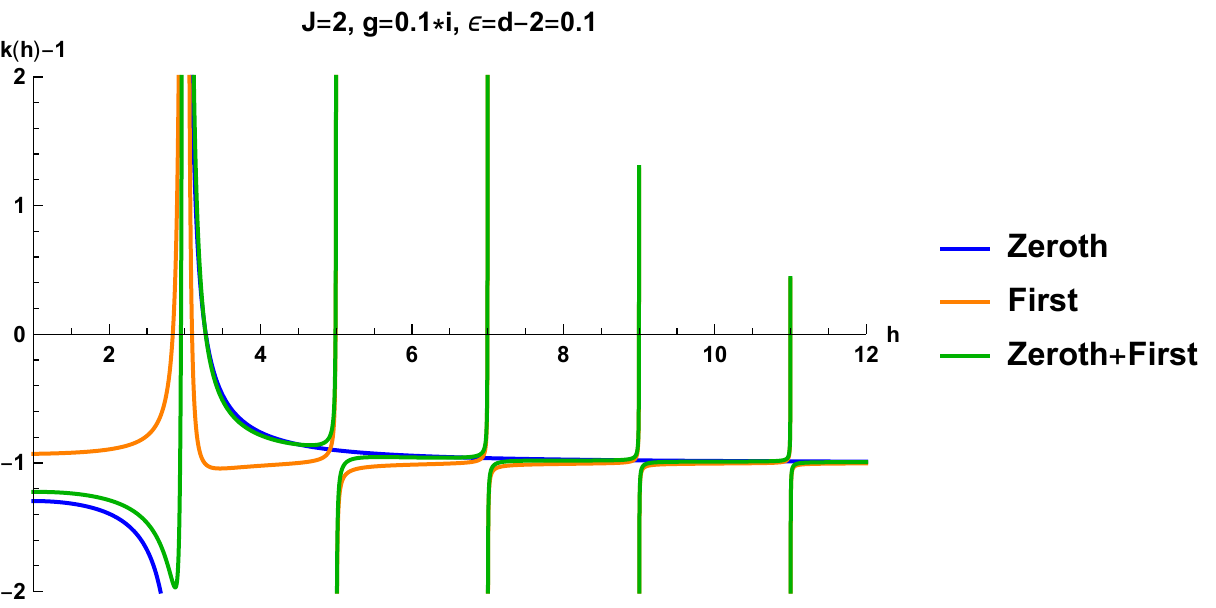}}  \\ \, \vspace{30pt}
		\scalebox{0.65}{\includegraphics{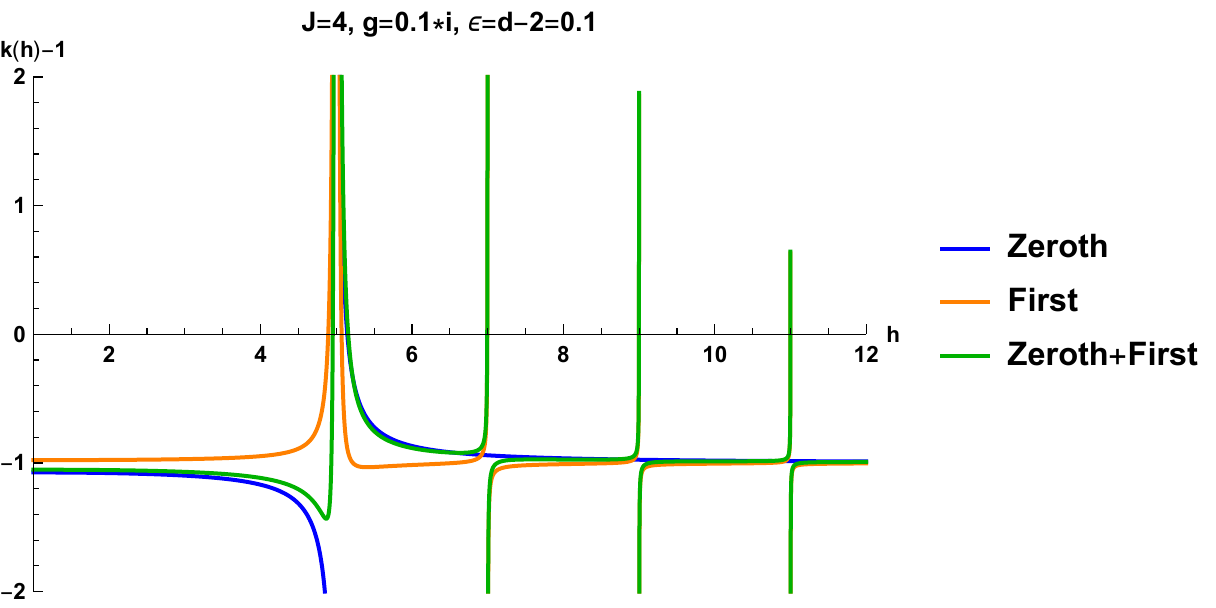}}
	\end{center}
	\caption{Plots of $k(h)-1$ around $d=2$ with $g=0.1i$, and $J=0,2,4$.
	The blue lines represent the result strictly at $d=2$ and the orange lines represent the first order correction of $\epsilon \equiv d-2$.
	For the plots, we took $\epsilon = 0.1$. The green lines are the results with the zeroth and first order corrections together.}
	\label{fig:d=2}
\end{figure}

At $d=2$, all solutions with $m>0$ in \eqref{eq:hmJSol} disappear from the spectrum.
In order to better understand this phenomenon, let us consider $d=2+\varepsilon$ and expand the eigenvalue \eqref{eq:k(h,J)} in $\varepsilon$.
This leads to 
	\begin{equation}
	\begin{split}
		k(h,J) \, = \, - \, \frac{12 \pi^2 \bar{g}^2}{(h+J-1)(h-J-1)} \bigg[& 1 \, - \, \frac{1}{2} \bigg( 2\gamma + \frac{1}{h+J-1} + \frac{3}{1-h+J} + 4\log 2 \\
		&+ \psi\left(\frac{1-h+J}{2}\right) + \psi\left(\frac{h+J-1}{2}\right) \bigg) \, \varepsilon \, + \, \mathcal{O}(\varepsilon^2) \bigg] \, ,
	\end{split}
	\end{equation}
where $\gamma$ is the Euler-Mascheroni constant.
The zeroth order in $\varepsilon$ is a monotonically decreasing function of $h$ for any $J$,
while the first digamma function appearing in the $\mathcal{O}(\varepsilon)$ order introduces an infinite number of divergences at
	\begin{equation}
		h \, = \, 1 + J + 2m \, \qquad (m=0,1,2 \cdots) \, , 
	\label{d=2+epsilon solution}
	\end{equation}
and this leads to the solutions \eqref{eq:hmJSol} with $m>0$.
Figure \ref{fig:d=2} shows this behavior of the eigenvalue.
Taking $d=2+\epsilon$ in \eqref{eq:OPE-coef-generalmJ} 
and sending $\epsilon\to 0$ we conclude that 
all the OPE coefficients $c_{m,J}$ (including those with $m>0$) have finite, non-zero limits when sending $d$ to $2$.  

In fact we will argue in section \ref{sec:character} that the correct spectrum of the free theory at $d=2$ is actually given by \eqref{d=2+epsilon solution}.
This is based on the character decomposition of the free theory partition function which we will present in detail.

\subsubsection{The $d=1$ case}
\label{sec:d=1}

The measure computed in appendix \ref{app:measure} is not defined for $d=1$. The correct measure for $d=1$ is given by \cite{Maldacena:2016hyu}:
	\begin{equation}
		\mu(h) \, =  \, \frac{2h-1}{ \pi \tan(\pi h/2)} \frac{\Gamma(h)^2}{\Gamma(2h)} \frac{\alpha_0 k(h,0)}{2} \, ,
	\label{eq:d=1 measure}
	\end{equation}
with $\alpha_0=\frac{\pi}{3 \Gamma(1/4)^4\bar{g}^2}$, and $k(h,0)$ given by \eqref{eq:k(h,J)} at $d=1$.
The on-shell value of the conformal dimensions $h_m$ are, from section \ref{sec:OPE}:
\begin{align}
h_{\pm}&=\frac{1}{2}\pm 2 \sqrt{-\frac{3\bar{g}^2}{\pi}}\,\Gamma(1/4)^2 +\mathcal{O}(\bar{g}^3) \;,\\
h_m&=\frac{1}{2}+2m-\frac{6\Gamma(1/4)^4\bar{g}^2}{m\pi} +\mathcal{O}(\bar{g}^4) \;, \;\;\; m>0 \;.
\end{align}
In $d=1$, a physical dimension needs to be greater than $1/2$. Therefore, in the weak purely imaginary coupling limit $h_{-}$ does not exist in the spectrum. For weak real coupling, $h_{\pm}$ are on the contour ${\rm Re}(h) =1/2$.
The OPE coefficients are given by:
	\begin{align}
		c_{\pm}^2 \, &= \, 2 \, \mp \, (\pi + 4\log 2) \Gamma(1/4)^2 \,  4 \sqrt{-\frac{3\bar{g}^2}{\pi}} \, + \, \mathcal{O}(\bar{g}^3) \, , \\
		c_m^2 \, &= \, \frac{4}{\pi} \frac{\Gamma(2m+1/2)^2}{ \Gamma(4m+1)} \, + \, \mathcal{O}(\bar{g}^2) \, .
	\end{align}
Therefore, the OPE coefficients are real for the free theory ($\bar{g}=0$) and for small pure imaginary coupling.

As a final comment on the $d=1$ case, if we expand the general formulae \eqref{eq:H(h)} around $d=1$, we obtain:
	\begin{equation}
		\mu_{d/4}^d(h,0) \, = \, \frac{\sqrt{\pi} \, \tan(\frac{\pi}{4}(2h+1)) \Gamma(h)}{2^{2h-2} \, \tan(\pi h/2) \, \Gamma(\frac{1}{2}+h)} \, + \, \mathcal{O}(d-1) \, .
	\end{equation}
This result does not agree with the expression given in 
\eqref{eq:d=1 measure}.
The reason for this is that the complete set of $h$ in $d=1$ is not just the principal series ($h=1/2+i r$), but also contains the discrete modes ($h=2n$) \cite{Kitaev:2017hnr}.
Therefore \eqref{eq:4ptCFT} (expressing the four-point function in terms of conformal partial waves) must be modified by adding contours around the discrete modes $h=2n$ and changing  the measure $\mu(h)$ accordingly.

\subsubsection{Reappearance of complex dimensions.}

From now on we discuss the case $\bar{g}$ (and $\bar{\lambda}$) purely imaginary. We consider only spin $J=0$ in this subsection.
We denote $\bar{g}_0 = 3^{-1/2}g_c (4\pi)^{-d/2}\Gamma(d/4)^{-2}$, with $g_c$ defined in \eqref{eq:gc}. This $\bar{g}_0$ is the maximal value of $|\bar{g}|$ for which $\bar{\lambda}(\bar{g})$ is invertible to $\bar{g}(\bar{\lambda})$ as depicted in figure~\ref{fig:tetra} (we have also taken into account the rescaling of the coupling). 

\begin{figure}[ht]
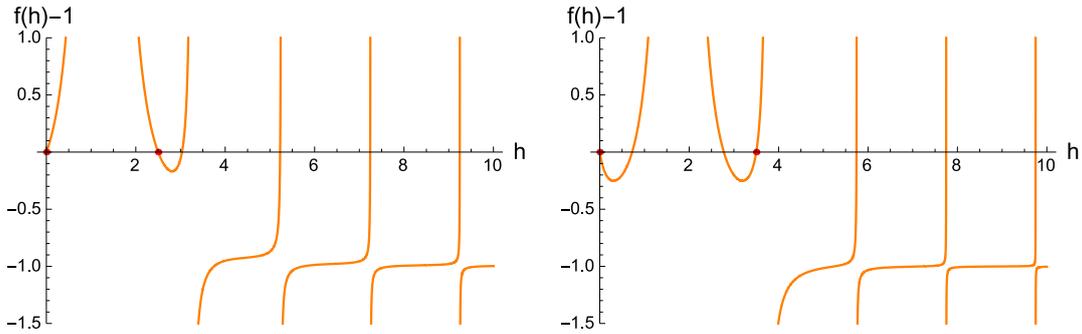

\centering
\begin{minipage}{0.4\textwidth}
           \centering 
            \includegraphics[width=\textwidth]{BSeq-2-5d-g0.pdf}
        \end{minipage}
        \hspace{0.01\textwidth}
\begin{minipage}{0.4\textwidth}
            \centering
            \includegraphics[width=\textwidth]{BSeq-3-5d-g0.pdf}
        \end{minipage}
 \caption{Plots of $k(h)-1$ for $d=2.5$ (left) and $d=3.5$ (right) at $g=g_0(d)$. The zeros on the left panel correspond (from left to right) to $h_{0-}$, $h_{0+}$, and $h_1$ to $h_4$, while on the right panel this applies only to $h>0$, as $h=0$ is a new solution. The zeros at $h=0$ and $h=d$ are marked by a red dot.} \label{fig:BSeq-g0}
\end{figure}

For $\bar{g}=\im \bar{g}_0$, we have two exact solutions at $h=0$ and $h=d$, for any $d$.\footnote{The solution $h=0$ violates the unitarity bound $h>d/2-1$ for $d>2$. Therefore, for $d>2$ we expect to find an upper value of $|\bar{g}|$ beyond which the UV CFT with bilinear operator of dimension $h_{0-}$ is necessarily non-unitary.} From a numerical check, we find that for $d\lesssim 2.9728$ these correspond to $h_{0\pm}$, while for  $d\gtrsim 2.9728$ $h=0$ comes from the continuation of a negative solution, and $h=d$ corresponds to $h_1$.
This is depicted in figure~\ref{fig:BSeq-g0}.

Interestingly, we have a neat interpretation from the point of view of AdS/CFT. According to the standard dictionary \cite{Gubser:1998bc,Witten:1998qj} we should have $h_\pm=\frac{d}{2} \pm \sqrt{\frac{d^2}{4}+m^2}$, with $m$ being the mass of a field in AdS${}_{d+1}$. The plus and minus signs correspond to different boundary conditions \cite{Klebanov:1999tb}, the plus being always allowed, and the minus only for $-\frac{d^2}{4}<m^2<-\frac{d^2}{4}+1$. For $d\lesssim 2.9728$ we have 
the following situation. At $\bar{g}=0$ we get $h_{0\pm}=d/2$ which can be interpreted
as $h_{\pm}$ with the mass saturating the Breitenlohner-Freedman bound $m^2\geq -\frac{d^2}{4}$ \cite{Breitenlohner:1982jf}. We dial up $\bar{g}$ and when we reach $\bar{g}=\im g_0$ the mass become zero; in this case $h_-$ is only allowed in the bulk for $d\leq 2$.
For $d\gtrsim 2.9728$ it is $h_1$ (instead of $h_{+}$) which starts at $\bar{g}=0$ from a positive mass $m^2=4-\frac{d^2}{4}$ and reaches $m^2=0$ at $\bar{g}=\im \bar{g}_0$.

\begin{figure}[ht]
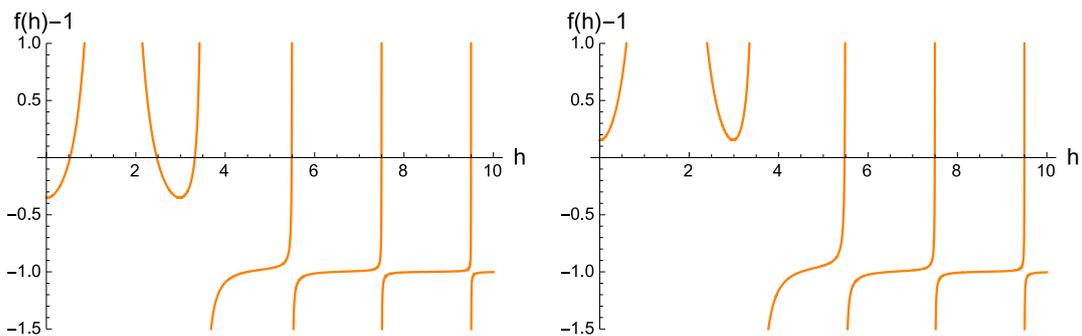

\centering
\begin{minipage}{0.4\textwidth}
           \centering 
            \includegraphics[width=\textwidth]{BSeq-3d-g015.pdf}
        \end{minipage}
        \hspace{0.01\textwidth}
\begin{minipage}{0.4\textwidth}
            \centering
            \includegraphics[width=\textwidth]{BSeq-3d-g02.pdf}
        \end{minipage}
 \caption{Plots of $k(h)-1$ for $d=3$ at $\bar{g}=0.15$ (left) and  $\bar{g}=0.19$ (right). The zeros on the left panel correspond (from left to right) to $h_{0-}$, $h_{0+}$, and $h_1$ to $h_4$. On the right panel $h_2$ to $h_4$ are the only remaining real roots. In this case $\bar{g}_0\simeq 0.186135$ and $\bar{g}_*-\bar{g}_0\simeq 5.7\times 10^{-5}$.} \label{fig:BSeq-3d}
\end{figure}

As already stated, for $\bar{g}$ purely imaginary and small enough, we obtain a real spectrum. However, it is plausible that there exists a value $\bar{g}_*$ such that for $\vert \bar{g} \vert >  \bar{g}_* $, some dimensions become complex again. 
With a numerical  study, we find that for $d>2$ there exists a $\bar{g}_*\geq \bar{g}_0$ at which $h_{0+}$ and $h_1$ merge and beyond which they become complex. The transition can be understood from the graphical solution of the equation, as in figure~\ref{fig:BSeq-3d}. 
Only at $d \simeq 2.9728$ we find that $\bar{g}_*= \bar{g}_0$, while $\bar{g}_*\to+\infty$  for $d\leq 2$. 

It has been conjectured in \cite{Kim:2019upg} that the appearance of a complex scaling dimension $h$ with ${\rm Re}(h)=d/2$ is associated to a non-zero vacuum expectation value of the associated operator, hence to a spontaneous breaking of conformal invariance. The intuitive reason is that in the AdS/CFT picture such operators correspond to fields with mass below the Breitenlohner-Freedman bound. In our case, such a phenomenon seems to take place at the transition from $g^2<0$ to $g^2>0$. It is plausible that the instability in the AdS side of the correspondence translates into a spontaneous breaking of conformal invariance of our model at real coupling.
On the other hand, the complex dimensions at $\bar{g}^2<-\bar{g}_*^2<0$ have ${\rm Re}(h)>d/2$, so they seem to correspond to fields with complex mass (with both real and imaginary parts being non-zero). 
However, we stress that from a renormalization group point of view the model makes sense only for $|\bar{g}|\leq \bar{g}_0$. Since $\bar{g}_*\geq \bar{g}_0$ for any $d$, the appearance of such complex solutions is probably not relevant to our model.

\subsection{A group-theoretic derivation of the spectrum of bilinear operators in the  free theory}
\label{sec:character}

In the limit of vanishing coupling, our model \eqref{eq:action} reduces effectively to a vector model with $O(N^3)$ symmetry. The two models are still distinguishable if one chooses to impose a singlet constraint based on one or the other group but, as we are here studying only bilinear operators, the two models should be indistinguishable in the free limit. Since our spectrum is continuous for $\bar{g}\to 0$ (see appendix \ref{app:free} for the computation directly at $\bar{g}=0$), this raises the question of why it is parametrized by $J$ and $m$ rather than by just $J$, as in the usual vector model (see for example \cite{Giombi:2016ejx} for a review).

Furthermore, we would like to understand the apparent discontinuity at $d=2$, which is present both at finite $\bar{g}$ (previous sections) and at $\bar{g}=0$ (appendix \ref{app:free}).

In this section, we reconstruct the spectrum of the free theory by a different method, as a way to cross-check our results, and in particular shed some light on the two questions above. We follow the set of ideas which have been developed in a number of papers, in connection to the Hagedorn transition in gauge theories \cite{Sundborg:1999ue,Aharony:2003sx} and the AdS/CFT duality between vector models and higher-spin theory \cite{Shenker:2011zf,Jevicki:2014mfa,Giombi:2014yra}. Similar methods have also been applied to tensor models in \cite{Beccaria:2017aqc,Bulycheva:2017ilt,Choudhury:2017tax}. Since we are here interested in the free theory, and its spectrum of bilinear operators, the $O(N)^3$ symmetry of our model will play no role, and we can actually treat its free limit as a $O(N^3)$ vector model. The main difference to the usual vector model, which we wish to highlight, is the effect of the non-canonical kinetic operator of our model on the spectrum of bilinear operators.

The spectrum of operators of a CFT on $\mathbb{R}^d$, or equivalently of CFT states on $\mathbb{R}\times S^{d-1}$, can be encoded in the grand canonical partition function with singlet constraint on $S^1 \times S^{d-1}$, where the $S^1$ is understood as Euclidean thermal circle with period $\beta$. First, we introduce the canonical or single-particle partition function:
\be \label{eq:singleZ}
Z(q,\mu) \, = \, \Tr[q^\Delta \mu^{j_3}] \, ,
\ee
where $q=e^{-\beta}$ and $\mu=e^{-\Omega}$, with $\Omega$ being the chemical potential conjugate to the eigenvalues of Cartan elements of $S^{d-1}$, denoted by $j_3$, and the trace is over all the states  built out of the elementary field $\phi$, which transforms in a real representation $R$ of the symmetry group $\cG$ (we will consider either $U(N)$ or $O(N)$). In this section, $\Delta$ denotes the conformal dimension, i.e.\ the eigenvalue of the dilation operator, which as usual plays the role of Hamiltonian in the radial quantization picture.

The grand canonical or multi-particle free energy, without singlet constraint, is related to the single-particle partition function by:
\be
F \, = \, - \ln \cZ(q,\mu) \, = \, -\Tr[\ln (1-q^\Delta \mu^{j_3})^{-1}] \, = \, - \sum_{m=1}^{\infty} \frac{1}{m} Z(q^m, \mu^m) \,.
\ee
Following \cite{Sundborg:1999ue,Aharony:2003sx,Shenker:2011zf,Giombi:2014yra}, imposing the singlet constraint amounts to writing the multi-particle partition function with an integral over the symmetry group $\cG$:
\be \label{eq:Z-singlet}
\begin{split}
\cZ_\cG(q,\mu) &= \int_\cG [dU] \exp \left\{ \sum_i \sum_{m=1}^{\infty}  \frac{1}{m} q^{m \Delta_i} \mu^{m j_{3,i}} \chi^{\cG}_{R} (U^m) \right\} \\
&= \int_\cG [dU] \exp \left\{  \sum_{m=1}^{\infty}  \frac{1}{m} Z(q^m, \mu^m) \chi^{\cG}_{R} (U^m) \right\} \\
&\equiv \exp \left\{ \sum_{m=1}^{\infty}  \frac{1}{m} Z_\cG(q^m, \mu^m) \right\}\,,
\end{split}
\ee
where $[dU]$ is the normalized Haar measure, and $\chi^{\cG}_{R} (U)$ is the character of the group element $U\in\cG$ in the representation $R$.
The two cases which are relevant for us are \cite{Beccaria:2017aqc}:
\begin{align}
\chi^{U(N)}_{N\oplus \bar{N}} (U) & = \tr(U) + \tr(U^{-1})\,, 
\\
\chi^{O(N)}_{N} (U) & = \tr(U) \,. 
\end{align}

The integral over the group can be done explicitly, and the result can be expressed in terms of characters of the conformal group, which are in fact the single-particle partition function without singlet constraint \eqref{eq:singleZ}.  
We denote the character of the $(\Delta_{\phi},J)$ representation of the $SO(2,d)$ conformal group by $\chi^{(d)}_{(\Delta_{\phi},J)}(q,\mu)$.
For the $U(N)$  gauge symmetry case one finds \cite{Shenker:2011zf}:
\be \label{eq:Z-UN}
Z^{(d)}_{U(N)}(q,\mu) \, = \, \left( \chi^{(d)}_{(\Delta_\phi,0)}(q,\mu) \right)^2 \, .
\ee
For the $O(N)$ gauge symmetry case, the $O(N)$ gauge singlet condition introduces an additional term in the partition function \cite{Jevicki:2014mfa, Giombi:2014yra} as:
\begin{equation} \label{eq:Z-ON}
Z^{(d)}_{O(N)}(q,\mu) \, = \, \frac12 \left( \chi^{(d)}_{(\Delta_{\phi},0)}(q,\mu) \right)^2 \, + \, \frac12 \, \chi^{(d)}_{(\Delta_{\phi},0)}(q^2,\mu^2) \, .
\end{equation}

The derivation above is very generic, based just on representation theory (the integral over the group is the way to count the number of singlets in a product of representations), and thus it applies also to our model with a non-canonical dimension $\Delta_\phi$ for the elementary field. The appearance of $\Delta_\phi\neq \frac{d-2}{2}$ is in fact the only difference between our $Z^{(d)}_{\cG}(q,\mu)$ and those found in the literature, and we are going to explore the consequences of this difference.

Before moving on, we should point out a subtle aspect of the above discussion.
The group integral enforcing the singlet constraint is usually introduced in the partition function by gauging the global symmetry on the compact manifold $S^1 \times S^{d-1}$, in the limit of vanishing gauge coupling, or equivalently by coupling the theory to a flat connection $A_\mu=U^{-1}\partial_\mu U$ and integrating over it. The connection can be gauged away, except for the constant $A_0$ component which has a non-trivial holonomy on $S^1$. For the usual vector model one can then show \cite{Shenker:2011zf,Giombi:2014yra} that the integral over $A_0$ reduces to the group integral in \eqref{eq:Z-singlet}.
In our case, the non-integer power of the Laplacian renders such path integral derivation more perilous.
Gauging can actually be done in the standard way, simply replacing the derivatives with covariant derivatives, as best seen by expressing our kinetic operator in terms of the heat kernel by an inverse Laplace transform:
\be
\begin{split}
S_{\rm free}[\phi]   &=   \frac{\Gamma(1+\zeta)}{2} \int_{\gamma-\im \infty}^{\gamma+\im \infty} \frac{ds}{2\pi\im} s^{-1-\zeta} \int d^dx \,\sqrt{\bar{g}} \;   \phi_{\mba}(x) e^{ - s \partial^2}\phi_{\mba}(x)\\
&=   \frac{\Gamma(1+\zeta)}{2} \int_{\gamma-\im \infty}^{\gamma+\im \infty} \frac{ds}{2\pi\im}  \sum_{n\geq 0} \frac{s^{-1-\zeta+n}}{n!} \int d^dx \,\sqrt{\bar{g}} \;   \phi_{\mba}(x) (- \partial^2)^n \phi_{\mba}(x) \, .
\end{split}
\ee
The replacement $\partial_\mu \to \partial_\mu + A_\mu$ then leads to a gauge-invariant action.
Promoting our kinetic operator to a Weyl-covariant operator is instead more problematic, and we are not aware of any such generalization for non-integer powers of a Laplacian.\footnote{For integer powers, such generalization is known as the GJMS operators \cite{graham1992conformally}.}
We thus take \eqref{eq:singleZ} and \eqref{eq:Z-singlet} as our starting point, putting aside a proper path integral derivation.

\subsubsection{The $d=3$ case}
\label{app:d=3}
For $d=3$, the long representation ($\Delta>J+1$ for $J\ge 1$ and $\Delta>1/2$ for $J=0$) of the character for $SO(2,3)$ is given by:
	\begin{equation}
		\chi^{(3)}_{(\Delta,J)}(q,\mu) \, = \, \frac{q^{\Delta} \sum_{j=-J}^J \mu^j}{(1-q) (1-q \mu) (1-q \mu^{-1})} \, .
	\end{equation}
The short representations are obtained by eliminating corresponding null states:
	\begin{equation}
		\chi^{(3)}_{(\frac12,0)}(q,\mu) \, = \, \chi^{(3)}_{(\Delta,0)}(q,\mu)\Big|_{\Delta=\frac12} \, - \, \chi^{(3)}_{(\frac52,0)}(q,\mu) \, = \, \frac{q^{1/2} (1+q) }{(1-q \mu) (1-q \mu^{-1})} \, ,
	\end{equation}
for $J=0$ and:
	\begin{equation}
		\chi^{(3)}_{(J+1,J)}(q,\mu) \, = \, \chi^{(3)}_{(\Delta,J)}(q,\mu)\Big|_{\Delta=J+1} \, - \, \chi^{(3)}_{(J+2,J-1)}(q,\mu)
		\, = \, \frac{q^{J+1} \left[(q-\mu)\mu^J +(1-q\mu) \mu^{-J} \right]}{(1-\mu)(1-q) (1-q \mu) (1-q \mu^{-1})} \, ,
	\end{equation}
for $J\ge 1$.

Let us first consider the $U(N)$ gauge symmetry case.
For the canonical dimension of the fundamental scalar $\Delta_{\phi}=1/2$, we find that:
	\begin{equation}
		Z^{(d=3)}_{U(N)}(q,\mu) \, = \, \left( \chi^{(3)}_{(\frac12,0)}(q,\mu) \right)^2 \, = \, \chi^{(3)}_{(1,0)}(q,\mu) \, + \, \sum_{J=1}^{\infty} \, \chi^{(3)}_{(J+1,J)}(q,\mu) \, .
	\end{equation}
This result is the well-known Flato-Fronsdal decomposition \cite{Flato:1978qz}, which was also generalized to any dimension in \cite{Dolan:2005wy}.
Next, we consider the $\Delta_{\phi}=d/4=3/4$ case.
For this case, we have:
	\begin{equation}
		Z^{(d=3)}_{U(N)}(q,\mu) \, = \, \left( \chi^{(3)}_{(\frac34,0)}(q,\mu) \right)^2 \, = \, \sum_{J=0}^{\infty} \sum_{m=0}^{\infty} \, \chi^{(3)}_{(\frac32+J+2m,J)}(q,\mu) \, .
	\end{equation}

For the $O(N)$ gauge symmetry case, following the same computation as above we find:
	\begin{equation}
	\begin{split}
		Z^{(d=3)}_{O(N)}(q,\mu) \, &= \, \frac12 \left( \chi^{(3)}_{(\frac12,0)}(q,\mu) \right)^2 \, + \, \frac12 \, \chi^{(3)}_{(\frac12,0)}(q^2,\mu^2) \\
		&= \, \chi^{(3)}_{(1,0)}(q,\mu) \, + \, \sum_{J=1}^{\infty} \, \chi^{(3)}_{(2J+1,2J)}(q,\mu) \, ,
	\end{split}
	\end{equation}
for $\Delta_{\phi}=(d-2)/2 = 1/2$, while in the case 
$\Delta_{\phi}= d/4=3/4$ we get:
	\begin{equation}
	\begin{split}
		Z^{(d=3)}_{O(N)}(q,\mu) \, &= \, \frac12 \left( \chi^{(3)}_{(\frac34,0)}(q,\mu) \right)^2 \, + \, \frac12 \, \chi^{(3)}_{(\frac34,0)}(q^2,\mu^2) \\
		&= \, \sum_{J=0}^{\infty} \sum_{m=0}^{\infty} \, \chi^{(3)}_{(\frac32+2J+2m,2J)}(q,\mu) \, .
		\end{split}
	\end{equation}
This agrees with the results we found in section \ref{sec:OPE}.

\subsubsection{The $d=2$ case}
\label{app:d=2}
For $d=2$, the long representation of the character for $SO(2,2)$ is given by \cite{Dolan:2005wy}:
	\begin{equation}
		\chi^{(2)}_{(\Delta,J)}(q,\mu) \, = \, \frac{q^{\Delta}\mu^{J}}{(1-q\mu)(1-q/\mu)} \, , \qquad \qquad (\Delta > J)
	\end{equation}
The short representations are again obtained by eliminating corresponding null states:
	\begin{equation}
		\chi^{(2)}_{(J,J)}(q,\mu) \, = \, \chi^{(2)}_{(\Delta,J)}(q,\mu)\Big|_{\Delta=J} \, - \, \chi^{(2)}_{(J+1,J-1)}(q,\mu)
		\, = \, \frac{q^J \mu^J}{1-q \mu} \, .
	\end{equation}

For the canonical dimension of the free scalar $\Delta_{\phi}=0$, we have:
	\begin{equation}
		Z^{(d=2)}_{U(N)}(q,\mu) \, = \, \left( \chi^{(2)}_{(0,0)}(q,\mu) \right)^2 \, = \, \sum_{J=0}^{\infty} \, \chi^{(2)}_{(J,J)}(q,\mu) \, ,
	\end{equation}
and:
	\begin{equation}
		Z^{(d=2)}_{O(N)}(q,\mu) \, = \, \frac12 \left( \chi^{(2)}_{(0,0)}(q,\mu) \right)^2 \, + \, \frac12 \, \chi^{(2)}_{(0,0)}(q^2,\mu^2)
		\, = \, \sum_{J=0}^{\infty} \, \chi^{(2)}_{(2J,2J)}(q,\mu) \, .
	\end{equation}

For $\Delta_\phi=d/4=1/2$ case, the decompositions are given by: 
	\begin{equation}
		Z^{(d=2)}_{U(N)}(q,\mu) \, = \, \left( \chi^{(2)}_{(\frac12,0)}(q,\mu) \right)^2 \, = \, \sum_{n, \bar{n}=0}^{\infty} \, \chi^{(2)}_{(1+n+\bar{n}, \, n-\bar{n})}(q,\mu) \, ,
	\end{equation}
and:
	\begin{equation}
		Z^{(d=2)}_{O(N)}(q,\mu) \, = \, \frac12 \left( \chi^{(2)}_{(\frac12,0)}(q,\mu) \right)^2 \, + \, \frac12 \, \chi^{(2)}_{(\frac12,0)}(q^2,\mu^2)
		\, = \,\sum_{\substack{n, \bar{n}=0\\ n+\bar{n}={\rm even}}}^{\infty} \, \chi^{(2)}_{(1+n+\bar{n}, \, n-\bar{n})}(q,\mu) \, ,
	\end{equation}
where for the $O(N)$ case the summations over $n$ and $\bar{n}$ are taken only for the combinations such that $n+\bar{n}$ is an even integer.
This condition can be explicitly implemented by introducing an additional parameter $a=\{0,1\}$ and parametrizing $n=2n'+a$ and $\bar{n}=2\bar{n}'+a$.
Then the partition function for $O(N)$ can be written as:
	\begin{equation}
	\begin{split}
		Z^{(d=2)}_{O(N)}(q,\mu) \,& = \, \frac12 \left( \chi^{(2)}_{(\frac12,0)}(q,\mu) \right)^2 \, + \, \frac12 \, \chi^{(2)}_{(\frac12,0)}(q^2,\mu^2)
		\\
		& = \, \sum_{a=0,1} \sum_{n'=0}^{\infty} \sum_{\bar{n}'=0}^{\infty} \, \chi^{(2)}_{(1+2a+2n'+2\bar{n}', \, 2n'-2\bar{n}')}(q,\mu) \, .
		\end{split}
	\end{equation}
We note that if we introduce the conformal weight $h$ ($\bar{h}$) of the holomorphic (anti-holomorphic) sector by:
	\begin{equation}
		\Delta \, = \, h \, + \, \bar{h} \, , \qquad J \, = \, h \, - \, \bar{h} \, ,  
	\end{equation}
then the spectrum in terms of $(h, \bar{h})$ is given by:
	\begin{equation}
		h \, = \, \frac{1+2n}{2} \, , \qquad \bar{h} \, = \, \frac{1+2\bar{n}}{2} \, .
	\end{equation}	
For a free scalar field, as in this case, the symmetric spectrum between ($h, \bar{h}$) is expected.

The spectrum identified here is not the one we found in section \ref{sec:d=2} by setting $d=2$. In fact, in order to reproduce the values $h+\bar h = 1 + J + 2\bar{n}$ we need to include not only the states with $m=0$ from section \ref{sec:d=2}, but also the states with $m>0$ which appear at $d=2+\epsilon$, \eqref{d=2+epsilon solution}.

\subsubsection{The $d=1$ case}
\label{app:d=1}
For $d=1$, there is no angular momentum or spin, so the character of the $SL(2,R)$ representation with weight $\Delta$ is given by:
	\begin{equation}
		\chi^{(1)}_{(\Delta)}(q) \, = \, \frac{q^{\Delta}}{1-q} \,.
	\end{equation}
Therefore, for the $U(N)$ gauge symmetry case, we have:\footnote{The canonical dimension of the free boson in $d=1$ gives $\Delta = -1/2$ and unitary representation of $d=1$ does not exist for $\Delta < 0$ \cite{Dolan:2005wy}.
Nevertheless, if we naively use the above character formula for this canonical dimension, still the decomposition works as we show below.}
	\begin{equation}
		Z^{(d=1)}_{U(N)}(q) \, = \, \left( \chi^{(1)}_{(-1/2)}(q) \right)^2 \, = \, \sum_{m=0}^{\infty} \, \chi^{(1)}_{(m-1)}(q) \, ,
	\end{equation}
and:
	\begin{equation}
		Z^{(d=1)}_{U(N)}(q) \, = \, \left( \chi^{(1)}_{(1/4)}(q) \right)^2 \, = \, \sum_{m=0}^{\infty} \, \chi^{(1)}_{(m+\frac{1}{2})}(q) \, .
	\end{equation}
The former corresponds to a free scalar with the canonical dimension and the latter corresponds to the generalized free scalar with $\zeta=1/4$.

For the $O(N)$ gauge symmetry case, we obtain:
	\begin{equation}
		Z^{(d=1)}_{O(N)}(q) \, = \, \frac12 \left( \chi^{(1)}_{(-1/2)}(q) \right)^2 \, + \, \frac12 \, \chi^{(1)}_{(-1/2)}(q^2) \, = \, \sum_{m=0}^{\infty} \, \chi^{(1)}_{(2m-1)}(q) \, ,
	\end{equation}
and:
	\begin{equation}
		Z^{(d=1)}_{O(N)}(q) \, = \, \frac12 \left( \chi^{(1)}_{(1/4)}(q) \right)^2 \, + \, \frac12 \, \chi^{(1)}_{(1/4)}(q^2) \, = \, \sum_{m=0}^{\infty} \, \chi^{(1)}_{(2m+\frac{1}{2})}(q) \, .
	\end{equation}
This agrees with what we found in section \ref{sec:d=1}.

\section{Conclusions}
\label{sec:conclusions}

%

In this chapter, we studied a bosonic $O(N)^3$ tensor model with a long-range propagator: $(-\Delta)^{ \zeta}$ with $\zeta = d/4$, reproducing the conformal scaling in the infrared. 
Our results are the following. We showed that the renormalized mass can be tuned to zero and the wave function renormalization is a finite rescaling. This should come as no surprise: we have fixed the scaling of the covariance to the infrared scaling, hence we do not get an additional anomalous scaling from a wave function renormalization. This is a characteristic feature of long-range models. 
We then studied the four-point function and showed that in the $N\to \infty$ limit but non-perturbatively (i.e. at all orders) in the coupling constants the RG flow has four lines of fixed points parametrized by the tetrahedral invariant $g$. Contrary to the pillow and double-trace invariants, the tetrahedral invariant does not have any positivity property. Furthermore, due the melonicity of the large-$N$ limit, the beta functions depend on $g^2$. Thus, we can consider a purely imaginary tetrahedral coupling $g = \pm\im |g|$, in which case the fixed point values are real, at least for small $g$. In particular, $g_{1+}>0$ and $\beta'_{g_1}( g_{1+}) >0$, that is, $(g_{1+},g_{2+}) $ is an \emph{infrared attractive} fixed point. This choice can seem highly non-standard but is not unprecedented. Indeed, a famous example of a field theory with imaginary coupling is given by the Lee-Yang model with an $\im \lambda \phi^3$ interaction \cite{Fisher:1978pf,Cardy:1985yy}, which is a real but non-unitary conformal field theory \cite{Gorbenko:2018dtm}.
 
In the second part of the chapter, we studied the CFT at the large-$N$ fixed point. The free theory is unitary. The conformal dimensions of the bilinear primary operators with arbitrary spin are given by $h_{m,J}=d/2 + J + 2m$ with $m\ge 0$. Near the fixed points, for $d\ne 2$, the conformal dimensions are shifted from the free value by $\mathcal{O}(g^2)$ for $(m,J) \neq (0,0)$ and by $\mathcal{O}(\sqrt{-g^2})$ for $(m,J)=(0,0)$. The OPE coefficients are real and shifted by $\mathcal{O}(g^2)$ for $(m,J) \neq (0,0)$. 
For $(m,J)=(0,0)$, the OPE coefficient is shifted by $\mathcal{O}(\sqrt{-g^2})$. It stays real for imaginary tetrahedral coupling, but becomes complex for real tetrahedral coupling.
The model at $d=2$ is very special, and still unclear. While direct computation both in the free and interacting cases (section \ref{sec:d=2} and appendix \ref{app:free}) seem to suggest that all the states with $m>0$ are absent at $d=2$, a derivation of the spectrum of the free theory based on character decomposition (section \ref{sec:character}) suggest that these states are in fact present. Inspired by the character decomposition it seems more natural to regard $d=2$ as the limit $\epsilon\to 0$ of $d=2+\epsilon$.  

Finally, in the case of a purely imaginary tetrahedral coupling, all the OPE coefficients of a bilinear primary operator and two fundamental fields are real (at all orders in the coupling). Even though we have not exhausted {\it all} the primary operators in the model, our result is a strong indication that the 
large-$N$ CFT at the infrared attractive fixed point is unitarity. This claim was then reinforced in \cite{Benedetti:2020yvb} where the two- and three-point functions of $\phi^4$ and $\phi^2$ composite operators were computed. The correlations were found to have the expected conformal behavior and the OPE coefficients were also real.

   \bigskip
   
   Let us conclude this chapter with a comparison between our results and those of \cite{Giombi:2017dtl}. Indeed, the fixed points we describe here are very different from the usual Wilson-Fisher fixed point contrary to the fixed point identified in \cite{Giombi:2017dtl} in the case of the same tensor model but  with $\zeta=1 $ instead of $\zeta=d/4$. The main differences are:
   \begin{itemize}
    \item  the Wilson-Fisher-like fixed point is reliable only for small $\epsilon =4-d$, while our results apply in any $d<4$. Our control parameter is the (bare or renormalized) tetrahedral coupling itself and not $\epsilon$. 
    \item at the Wilson-Fisher-like fixed point one gets an anomalous scaling dimension of the field, while in our case the scaling dimension of the field is fixed (although non-canonical).
    \item the Wilson-Fisher-like fixed point relies on the cancellation of the mass dimension of the coupling with the radiative corrections.\footnote{From \cite{Giombi:2017dtl}, the beta function for the tetrahedral coupling in units of cutoff reads $\beta_g=-\epsilon g+2 g^3$.} This is unlike our case, as we deal with genuinely marginal couplings in any $d$.
    \item because the mechanism of the Wilson-Fisher-like fixed point 
    requires to cancel the mass dimension of the tetrahedral coupling, the fixed point value of the tetrahedral coupling is real for $\epsilon>0$ and consequently the pillow and double-trace ones are purely imaginary. This is the origin of the instability of the fixed point discussed in 
    \cite{Giombi:2017dtl}. An imaginary tetrahedral coupling, and thus real pillow and double-trace ones, can in their case be obtained for $\epsilon<0$, i.e.\ for $d>4$, but then one deals with an ultraviolet fixed point. Furthermore, the spectrum of scalar bilinear operators computed in \cite{Giombi:2017dtl} shows an upper limit $d=4.155$ beyond which complex dimensions reappear. As the tetrahedral invariant has no positivity property, contrary to \cite{Giombi:2017dtl},  we have the freedom to consider an imaginary tetrahedral coupling. In this case we find instead a real IR fixed point with real exponents for any $d<4 $ as long as $|g|< g_*$ for some critical coupling $g_*$.
\end{itemize}

\begin{subappendices}
\section{The melon integral}
\label{app:melon}

We show that:
\be
\begin{split}
M^{\Lambda}_k(p) & = \frac{\mathcal{Z}^3}{(4\pi)^d \Gamma(\zeta)^3 }    \int_{\Lambda^{-2}}^{k^{-2}}  d\alpha_1 d\alpha_2 d\alpha_3  \;  (\alpha_1\alpha_2 \alpha_3)^{\zeta-1}    \; 
   \frac{e^{ - p^2 \frac{ \alpha_1 \alpha_2 \alpha_3}{ \alpha_1 \alpha_2 + \alpha_1 \alpha_3 + \alpha_2 \alpha_3 } }}{ ( \alpha_1 \alpha_2 + \alpha_1 \alpha_3 + \alpha_2 \alpha_3)^{d/2} }  \crcr
   & =M^{\Lambda}_k(0)- \mathcal{Z}^3\frac{p^{2d-6\zeta}}{(4\pi)^d \Gamma(\zeta)^3} f\left( \frac{ k^2 }{p^2}, \frac{p^2}{\Lambda^2} \right)
\end{split}
\ee
with $f$ a function such that $ \lim_{x,y\to 0} f(x,y) = \rm{finite}$.
The plan is:
\begin{itemize}
 \item Taylor expand at order one with integral remainder in the variable $tp^2$.

 \item Consider the integral remainder and rescale $t$ by $p^2$. This yields the 
 scaling in $p$ and $p^2$ appears only in the limits of the integral.
 Introduce Hepp sectors and compute the integral when sending the cutoffs to their limits. 
 \item Compute $M^{\Lambda}_k(0)$.
\end{itemize}
Denoting $I(p^2)=\mathcal{Z}^{-3}(4\pi)^d\Gamma(\zeta)^3M_k^{\Lambda}(p)$ in order to get rid of the overall constant we have: 
\begin{align*}
 I(p^2) =&  \int_{\Lambda^{-2}}^{k^{-2}} d\alpha \;\frac{   (\alpha_1\alpha_2 \alpha_3)^{\zeta-1}   }{ ( \alpha_1 \alpha_2 + \alpha_1 \alpha_3 + \alpha_2 \alpha_3)^{d/2} } \crcr
      & -p^2 \int_0^1 dt \int_{\Lambda^{-2}}^{k^{-2}} d\alpha   \frac{   (\alpha_1\alpha_2 \alpha_3)^{\zeta}   }{ ( \alpha_1 \alpha_2 + \alpha_1 \alpha_3 + \alpha_2 \alpha_3)^{1+d/2} }
         e^{-tp^2  \frac{ \alpha_1 \alpha_2 \alpha_3}{ \alpha_1 \alpha_2 + \alpha_1 \alpha_3 + \alpha_2 \alpha_3 } } \;.
\end{align*}
Rescaling $\alpha  = \frac{ \alpha'}{p^2}$ and dropping the primes yields the remainder term:
\[
p^{2d-6\zeta} f\left( \frac{ k^2 }{p^2}, \frac{p^2}{\Lambda^2} \right) =  ( p^2)^{1 - 3 -3\zeta +  2 + d} \int_0^1 dt \int_{ \frac{ p^2}{ \Lambda^2} }^{\frac{ p^2}{ k^2}} d\alpha   \frac{   (\alpha_1\alpha_2 \alpha_3)^{\zeta}   }{ ( \alpha_1 \alpha_2 + \alpha_1 \alpha_3 + \alpha_2 \alpha_3)^{1+d/2} }
         e^{-t  \frac{ \alpha_1 \alpha_2 \alpha_3}{ \alpha_1 \alpha_2 + \alpha_1 \alpha_3 + \alpha_2 \alpha_3 } }
\]

We split the $\alpha$ integrals in a sum over 6 Hepp sectors (total orderings of the parameters $\alpha$),  and in the sector $\alpha_1 < \alpha_2 < \alpha_3$, 
we change variables to $\alpha_3 = \rho , \alpha_2 = x  \rho , \alpha_1 =  y x  \rho $. The remainder term is then:
\begin{align*}
 f\left( \frac{ k^2 }{p^2}, \frac{p^2}{\Lambda^2} \right)  = &  6 \int_0^1 dt  \int_{ \frac{p^2}{\Lambda^2} }^{\frac{p^2}{k^2}  } d\rho  \;  \rho^{  3 \zeta - d }  
 \int_{\frac{p^2}{\Lambda^2} \rho^{-1} }^1 dx \; x^{2\zeta -\frac{d}{2}   } \int_{ \frac{p^2}{\Lambda^2} \rho^{-1}x^{-1}}^1 dy  \;   \; 
  \frac{  y ^{\zeta }  }{  ( 1 +  y  +  x y )^{1 + d/2} } e^{-t   \frac{ \rho x y  }{ 1 + y + xy }} \, .
\end{align*}
The integrals are clearly convergent when sending $\Lambda \to \infty$ and we get:
\begin{align*}
 f\left( \frac{ k^2 }{p^2}, 0  \right)  = &  6 \int_0^1 dt  \int_{ 0 }^{\frac{p^2}{k^2}  } d\rho  \;  \rho^{  3 \zeta - d }  
 \int_{ 0  }^1 dx \; x^{2\zeta -\frac{d}{2}   } \int_{ 0 }^1 dy  \;   
  \frac{  y ^{\zeta }  }{  ( 1 +  y  +  x y )^{1 + d/2} }   e^{-t   \frac{ \rho x y  }{ 1 + y + xy }}\, .
\end{align*}
We will compute this in the next subsection but for now we  check that it is convergent when sending $k\to 0$. As $ 3 \ge 1 + y + xy \ge 1$ we have an upper bound:
\[
  f\left( \frac{ k^2 }{p^2}, 0  \right)  \le   6 \int_0^1 dt  \int_{ 0 }^{\frac{p^2}{k^2}  } d\rho  \;  \rho^{  3 \zeta - d }  
 \int_{ 0  }^1 dx \; x^{2\zeta -\frac{d}{2}   } \int_{ 0 }^1 dy  \;   
   y ^{\zeta }   e^{- \frac{t}{3}   \rho x y  } \;,
\]
and rescaling $ \rho = \frac{ u }{txy }$ and sending $k\to 0$ we get:
\[
 f\le 6 \int_0^1 dt  
 \int_{ 0  }^1 dx \; x^{2\zeta -\frac{d}{2}   } \int_{ 0 }^1 dy  \;   
   y ^{\zeta }  
   \int_{ 0 }^{ \infty  } d u   \;  \left( \frac{1}{txy}\right)^{1 + 3\zeta - d} u^{  3 \zeta - d }  
   e^{- \frac{u}{3}  } \;.
\]
Recalling that $\zeta  = d/4$ the bound writes
\[
 f \le 6 \int_0^1 dt \; t^{-1 + \frac{d}{4}} \int_0^1 dx\; x^{-1+ \frac{d}{4}} \int_0^1 dy \; y^{-1 + \frac{d}{2}} \int_0^{\infty} du \; u^{-\frac{d}{4}} e^{-\frac{u}{3}} \;,
\]
which is convergent for $d<4$. For $d=4$, we still need to deal with the last integral.

 \paragraph{The integral remainder term}
 
 We now compute the numerical constant $f(0,0)$ defined by:
\[
 f\left( 0, 0 \right) =  \int_0^1 dt \int_{ 0 }^{\infty} d\alpha    \;\frac{   (\alpha_1\alpha_2 \alpha_3)^{\zeta}   }{ ( \alpha_1 \alpha_2 + \alpha_1 \alpha_3 + \alpha_2 \alpha_3)^{1+d/2} }
         e^{-t  \frac{ \alpha_1 \alpha_2 \alpha_3}{ \alpha_1 \alpha_2 + \alpha_1 \alpha_3 + \alpha_2 \alpha_3 } } \;.
\]
First, we change variables to $u=\alpha_1\alpha_2,  \; v=\alpha_1\alpha_3, \; 
w=\alpha_2\alpha_3 $ to get:
\begin{equation*}
f\left( 0, 0 \right) =  \frac{1}{2}\int_0^1 dt \int_{ 0 }^{\infty} dudvdw \; \frac{\left(uvw\right)^{\zeta/2-1/2}}{\left(u+v+w \right)^{1+d/2}}e^{-t\frac{\left(uvw\right)^{1/2}}{u+v+w}} \;,
\end{equation*}
and we perform a second change of variables $u=\alpha\delta\gamma, \;
v=\alpha\delta(1-\gamma) , \; w=\alpha(1-\delta)$ to obtain:
\begin{align*}
f\left( 0, 0 \right)=& \frac{1}{2}\int_0^1 dt \int_{ 0 }^{\infty} d\alpha \int_0^1 d\delta \int_0^1 d\gamma \crcr
    & \qquad \alpha^{-1/2+3\zeta/2-d/2}\delta^{\zeta}\gamma^{\zeta/2-1/2}(1-\delta)^{\zeta/2-1/2}(1-\gamma)^{\zeta/2-1/2}e^{-\alpha^{1/2}t\delta(1-\delta)^{1/2}(1-\gamma)^{1/2}} \;.
\end{align*}
The integral over $\alpha$ is a gamma function and we use 
$
\int_0^{\infty} dx \; x^a \exp\{-bx^{1/2}\} = 2 b^{ - 2a- 2}\Gamma(2a+2)
$
to write:
\begin{equation*}
f\left( 0, 0 \right)=\Gamma(1+3\zeta-d)\int_0^1 dt t^{d-3\zeta-1} \int_0^1 d\delta \delta^{d-2\zeta-1}(1-\delta)^{d/2-\zeta-1}\int_0^1 d\gamma \gamma^{d/2-\zeta-1}(1-\gamma)^{d/2-\zeta-1} \;.
\end{equation*}
The integral on $t$ gives a factor $ (d-3\zeta)^{-1}$ while the integrals on $\delta$ and $\gamma$ are Euler beta functions:
\begin{equation*}
f\left( 0, 0 \right)=\frac{1}{d-3\zeta}\frac{\Gamma(1-d+3\zeta)}{\Gamma(3d/2-3\zeta)}\Gamma(d/2-\zeta)^3 \;,
\end{equation*} 
which  for $\zeta=\frac{d}{4}$ simplifies to:
\begin{equation}
f\left( 0, 0 \right)=\frac{4}{d}\frac{\Gamma(1-d/4)}{\Gamma(3d/4)}\Gamma(d/4)^3 \;.
\end{equation} 

\paragraph{The local part}

We now want to compute the UV divergent piece $M_k^{\Lambda}(0)$.
We denote: 
\begin{align*}
I_0=& \mathcal{Z}^{-3}(4\pi)^d\Gamma(\zeta)^3M_k^{\Lambda}(0)=\int_{\Lambda^{-2}}^{k^{-2}} d\alpha \;\frac{   (\alpha_1\alpha_2 \alpha_3)^{\zeta-1}   }{ ( \alpha_1 \alpha_2 + \alpha_1 \alpha_3 + \alpha_2 \alpha_3)^{d/2} } \crcr
 =&\Lambda^{2(d-3\zeta)}\int_{1}^{k^{-2}\Lambda^2} d\alpha \;\frac{   (\alpha_1\alpha_2 \alpha_3)^{\zeta-1}   }{ ( \alpha_1 \alpha_2 + \alpha_1 \alpha_3 + \alpha_2 \alpha_3)^{d/2} } \;.
\end{align*}
This is convergent for $k\rightarrow 0$ so we can write $I_0$ as:
\begin{equation*}
I_0=\Lambda^{2(d-3\zeta)}\int_1^{\infty}d\alpha \;\frac{   (\alpha_1\alpha_2 \alpha_3)^{\zeta-1}   }{ ( \alpha_1 \alpha_2 + \alpha_1 \alpha_3 + \alpha_2 \alpha_3)^{d/2} }-\Lambda^{2(d-3\zeta)}\int_{k^{-2}\Lambda^2}^{\infty}d\alpha \;\frac{   (\alpha_1\alpha_2 \alpha_3)^{\zeta-1}   }{ ( \alpha_1 \alpha_2 + \alpha_1 \alpha_3 + \alpha_2 \alpha_3)^{d/2} }\, .
\end{equation*}

Let us denote:

\begin{equation*}
I_1=\Lambda^{2(d-3\zeta)}\int_1^{\infty}d\alpha \;\frac{   (\alpha_1\alpha_2 \alpha_3)^{\zeta-1}   }{ ( \alpha_1 \alpha_2 + \alpha_1 \alpha_3 + \alpha_2 \alpha_3)^{d/2} }\, ,
\end{equation*}
and 
\begin{equation*}
I_2=\Lambda^{2(d-3\zeta)}\int_{k^{-2}\Lambda^2}^{\infty}d\alpha \;\frac{   (\alpha_1\alpha_2 \alpha_3)^{\zeta-1}   }{ ( \alpha_1 \alpha_2 + \alpha_1 \alpha_3 + \alpha_2 \alpha_3)^{d/2} }\, .
\end{equation*}

These integrals can be separated into six Hepp sectors. For the sector $\alpha_1<\alpha_2<\alpha_3$ we can make the change of variables:
\begin{align*}
\alpha_1&=\rho \, ,\crcr
\alpha_2&=\rho x \, ,\crcr
\alpha_3&=\rho x y \, .
\end{align*} 

Then, we get for $I_1$, for $\zeta=\frac{d}{4}$:
\begin{equation*}
I_1= 6\Lambda^{d/2}\int_{1}^{\infty} d\rho \rho^{-d/4-1} \int_1^{\infty}dx \int_1^{\infty}dy \frac{x^{-1}y^{d/4-1}}{\left(1+y+xy\right)^{d/2}} \;,
\end{equation*}
which is
\begin{align*}
I_1=6\Lambda^{d/2}\frac{4}{d}\int_1^{\infty}dx \int_1^{\infty}dy \frac{x^{-1}y^{d/4-1}}{\left(1+y+xy\right)^{d/2}} \;.
\end{align*}
We can do the same change of variables for $I_2$, we get:
\begin{equation*}
I_2= 6\Lambda^{d/2}\int_{k^{-2}\Lambda^2}^{\infty} d\rho \rho^{-d/4-1} \int_1^{\infty}dx \int_1^{\infty}dy \frac{x^{-1}y^{d/4-1}}{\left(1+y+xy\right)^{d/2}} \;,
\end{equation*}
which is:
\begin{align*}
I_2=6\Lambda^{d/2}\frac{4}{d}\left(\frac{\Lambda}{k}\right)^{-d/2}\int_1^{\infty}dx \int_1^{\infty}dy \frac{x^{-1}y^{d/4-1}}{\left(1+y+xy\right)^{d/2}} \;.
\end{align*}

Finally, we obtain:
\begin{equation}
M_k^{\Lambda}(0)=\Lambda^{d/2} 
\frac{24\mathcal{Z}^3}{d(4\pi)^d\Gamma(d/4)^3}\left(1-\left(\frac{\Lambda}{k}\right)^{-d/2}\right) \int_1^{\infty}dx \int_1^{\infty}dy \frac{x^{-1}y^{d/4-1}}{\left(1+y+xy\right)^{d/2}}\;.
\end{equation}

For $k \rightarrow 0$ we get:
\begin{equation}
\lim_{k\rightarrow 0}M_k^{\Lambda}(0)=\Lambda^{d/2} 
\frac{24\mathcal{Z}^3}{d(4\pi)^d\Gamma(d/4)^3}\int_1^{\infty}dx \int_1^{\infty}dy \frac{x^{-1}y^{d/4-1}}{\left(1+y+xy\right)^{d/2}}\;.
\end{equation}

\section{Eigenvalues of the four-point kernel}
\label{ap:eigenvalues}

As we have seen in section \ref{sec:CFT}, the conformal three-point functions are eigenfunctions of the four-point kernel. We thus have:

\begin{equation}
\int d^dx_3d^dx_4 \; K(x_1,x_2,x_3,x_4)v(x_0,x_3,x_4)=k(h,J)v(x_0,x_3,x_4) \;,
\end{equation}
with 
\begin{equation}
K(x_1,x_2,x_3,x_4)=3\lambda^2G(x_{13})G(x_{24})G(x_{34})^2 \;,
\end{equation}
and the two-point function in position space is:
\begin{equation}
G(x)=\frac{\mathcal{Z}}{(2\pi)^{d/2}x^{d/2}}\;.
\end{equation}

We can then use the following conformal integral \cite{Klebanov:2016xxf,Giombi:2017dtl} twice to compute $k(h,J)$: 

\begin{equation}
\int d^d x_0 \frac{1}{(x_{01}^2)^{\alpha_1}(x_{02}^2)^{\alpha_2}(x_{03}^2)^{\alpha_3}}= \frac{L_d(\alpha_1,\alpha_2)}{(x_{12}^2)^{\tfrac{d}{2}-\alpha_3}(x_{13}^2)^{\tfrac{d}{2}-\alpha_2}(x_{23}^2)^{\tfrac{d}{2}-\alpha_1}} \; ,
\label{eq:conf_int}
\end{equation}
with $\alpha_1+\alpha_2+\alpha_3=d$ and
\begin{equation}
L_d(\alpha_1,\alpha_2)=\pi^{\tfrac{d}{2}}\frac{\Gamma(\tfrac{d}{2}-\alpha_1)\Gamma(\tfrac{d}{2}-\alpha_2)\Gamma(\tfrac{d}{2}-\alpha_3)}{\Gamma(\alpha_1)\Gamma(\alpha_2)\Gamma(\alpha_3)} \; .
\end{equation}

We finally obtain:
\begin{equation}
k(h,J)=3g^2\Gamma(d/4)^4
 \frac{\Gamma(-\frac{d}{4}+\frac{h+J}{2})\Gamma(\frac{d}{4}-\frac{h-J}{2})}{\Gamma(\frac{3d}{4}-\frac{h-J}{2})\Gamma(\frac{d}{4}+\frac{h+J}{2})} \;.
\end{equation}

\section{Measure and residue}
\label{app:measure}
In this appendix, we give a detailed computation of the measure and residues, which are needed for the computation of the OPE coefficients in section \ref{sec:OPE}.

 \paragraph{The measure.}
 We want to compute the measure at the physical dimensions $h_{m,J}= d/2 + J + 2m + 2z_{m,J}$. In this subsection, the results are valid for $d \neq 1,2$.
 From now on we consider only even spin, as otherwise the measure 
 in \eqref{eq:measure} is zero. Taking into account that  
 $\Delta_{\phi}=d/4$ the measure simplifies to:
\begin{equation}\label{eq:H(h)}
\begin{split}
 \mu_{d/4}^d(h, J)  & =    
   \frac{  \Gamma(J+\frac{d}{2}) 
  } {  \Gamma(J+1)} 
  \;  H^d_J(h) \; , \crcr
   H^d_J(h) & = \frac{
    \Gamma(\frac{ -\frac{d}{2}  +h+J}{2}) 
    \Gamma(\frac{ \frac{d}{2} -h+J}{2})
    \Gamma(h-1)\Gamma(d-h+J)\Gamma(\frac{h +J}{2})^2
 }{ 
  \Gamma(\frac{ \frac{3d}{2} -h+J}{2})\Gamma(\frac{\frac{d}{2} +h+J}{2}) 
  \Gamma(h-\frac{d}{2})\Gamma(h+J-1)\Gamma(\frac{d- h  +J}{2})^2
 }  \; .
 \end{split}
\end{equation}
We parametrize $h = d/2+J+2m + 2z$  and $H^d_J(d/2+J+2m + 2z) $ becomes:
\begin{equation} 
   \frac{ \Gamma(J+m+z)\Gamma( -m-z) 
    \Gamma(\frac{d}{2} + J + 2m +2z -1)
    \Gamma(\frac{d}{2} -2m -2z )\Gamma(\frac{d}{4} + J + m+z )^2
 }{ 
  \Gamma( \frac{d}{2} -m-z )\Gamma( \frac{d}{2} + J + m +z  ) 
  \Gamma(J + 2m +2z)\Gamma(\frac{d}{2} + 2J + 2m +2z -1)
  \Gamma( \frac{d}{4} -m-z )^2
 }  \; . 
\end{equation}

As the anomalous dimensions $z_{m,J}$ are small at small coupling, we can compute the measure at $h_{m,J}$ as a Laurent series in $z_{m,J}$. 
Recalling that $\Gamma'(z) = \Gamma(z) \psi(z) $ with $\psi(z)$ the digamma function, we again have two cases.

\paragraph{\it The case $(m,J) = (0,0)$} The Laurent series of $ H^d_0 (d/2 +  2z) $ at small $z$ is obtained as:    
 \begin{equation}
    \begin{split}
 &  \frac{ \Gamma( z)\Gamma(  -z) 
    \Gamma(\frac{d}{2}   -2z )\Gamma(\frac{d}{4}  +z )^2
 }{ 
  \Gamma( \frac{d}{2}  -z )\Gamma( \frac{d}{2}  +z  ) 
  \Gamma( 2z)  \Gamma( \frac{d}{4} -z )^2 }  
  = \left( -\frac{2}{z}\right) \; 
  \frac{ \Gamma(1+ z) \Gamma(1 -z) 
    \Gamma(\frac{d}{2}   -2z )\Gamma(\frac{d}{4}  +z )^2
 }{ 
  \Gamma( \frac{d}{2}  -z )\Gamma( \frac{d}{2}  +z  ) 
  \Gamma(1+ 2z) 
  \Gamma( \frac{d}{4} -z )^2
 }  \crcr
 & \qquad \qquad  =\left(  -\frac{2}{z} \right) \bigg[\frac{1}{\Gamma(d/2)}   + z  \frac{1}{\Gamma(d/2)}  \bigg( 4\psi(d/4) - 2 \psi(d/2) - 2 \psi(1)  \bigg) \bigg]
    +O(z) \;,
     \end{split}
    \end{equation}
therefore:
\begin{equation}
\begin{split}
 \mu_{d/4}^d \left( \frac{d}{2} +2 z ,  0 \right)  = -  \frac{2}{z}
+  4 \bigg[   \psi(d/2) +  \psi(1) - 2\psi(d/4)  \bigg]  + O(z) \;. 
\end{split}
\end{equation}

\paragraph{\it The case $(m,J) \neq (0,0)$} Using $\Gamma(-m-z) \Gamma(1 + m + z) = (-1)^{m+1} \Gamma(1+z) 
 \Gamma(1-z) / z $ we have:
  \begin{equation} 
 \begin{split}
& H^d_J(d/2+J+2m + 2z) = \crcr
& \qquad  =     \frac{  (-1)^{m+1}    \Gamma(J+m )
    \Gamma(\frac{d}{2} + J + 2m  -1)
    \Gamma(\frac{d}{2} -2m   )\Gamma(\frac{d}{4} + J + m  )^2
 }{ z\Gamma(m+1 )
  \Gamma( \frac{d}{2} -m  )\Gamma( \frac{d}{2} + J + m    ) 
  \Gamma(J + 2m  )\Gamma(\frac{d}{2} + 2J + 2m   -1)
  \Gamma( \frac{d}{4} -m  )^2
 }  + O(z^0) \; .
\end{split}
 \end{equation}
In order to include as much as possible explicitly positive terms, it is convenient to use:
\begin{align*}
 (-1)^{m+1} \frac{ \Gamma(\frac{d}{2} -2m   ) }{   \Gamma( \frac{d}{2} -m  )  }  =  -\frac{\Gamma\left(1 + m-\frac{d}{2} \right) }{\Gamma\left(1 + 2 m-\frac{d}{2} \right)} \;,
\end{align*}
therefore:
\begin{equation}
\begin{split}
  \mu_{d/4}^d \left( \frac{d}{2} + J + 2m + 2 z , J \right)& =   
 (-1)  \frac{  \Gamma(J+\frac{d}{2}) 
  \Gamma( 1 + m - \frac{d}{2}  )} { \Gamma(J+1)\Gamma(m+1 )
  \Gamma( 1 + 2m - \frac{d}{2}  )\Gamma( \frac{d}{4} -m  )^2} 
 \\
 &  \times
  \frac{    \Gamma(J+m )
    \Gamma(\frac{d}{2} + J + 2m  -1)
    \Gamma(\frac{d}{4} + J + m  )^2
 }{ \Gamma( \frac{d}{2} + J + m ) 
  \Gamma(J + 2m  )\Gamma(\frac{d}{2} + 2J + 2m   -1) }  \;\frac{1}{z} + O(z^0) \; .  
\end{split}
\end{equation}

\paragraph{The $k'$ term.}
Next we need to evaluate $k'$ at $h_{m,J}$. Shifting to the $z$ variables, 
\begin{equation*}
k'(d/2 + J + 2m +2z,J ) = \frac{1}{2}\frac{d}{dz}k_{(m,J)}(z)\;,
\end{equation*}
where the functions $k_{(m,J)}(z)$ are defined  in \eqref{eq:2ks}.
We have:
\begin{equation}
\begin{split}
 \frac{d}{dz} k_{(0,0)}(z)  &= k_{(0,0)}(z) \left[ -\frac{2}{z}  + \Psi(1+z) - \psi(1-z)
 + \psi\left(\frac{d}{2} -z \right)  -\psi\left(\frac{d}{2}+z \right)\right] \\
 \frac{d}{dz} k_{(m,J)}(z) & \xlongequal{ (m,J) \neq (0,0) } k_{(m,J)} (z) \bigg[ -\frac{1}{z} + \psi(J+m+z) 
  + \psi(1+z) -\psi(1-z) \crcr
  & \qquad + \psi \left(m+1- \frac{d}{2} + z \right) 
-\psi \left(\frac{d}{2} + J + m +z \right) \crcr
& \qquad -\psi(m+1+z) 
  - \psi\left( z-\frac{d}{2}\right) + \psi\left( \frac{d}{2} +1 -z \right)
  \bigg] \;.
\end{split}
\end{equation}

At the physical dimension $z_{m,J}$, $k_{(m,J)}(z_{m,J})=1$ therefore we get the Laurent series:
\begin{equation}
 \frac{1}{2}\frac{d}{dz} k_{(0,0)}(z_{0,0}) 
    = -\frac{1}{z_{0,0}} + O(z_{0,0}) \;,\qquad
\frac{1}{2} \frac{d}{dz} k_{(m,J)}(z_{m,J}) = -\frac{1}{2 z_{m,J}} 
  + O(z_{m,J}^0)  \;.
\end{equation}
Observe that the Laurent series of $k_{(0,0)}$ does not have a constant term.

\section{The free theory for $\zeta\le1$}
\label{app:free}
The four-point function in a (generalized) free CFT:
\be
  S[\phi]   =   \frac{1}{2} \int d^dx \;   \phi_{\mba}(x) (   - \partial^2)^{\zeta}\phi_{\mba}(x) \;,
\ee
with a real field of dimension $\Delta_{\phi} = d/2-\zeta$ can be written as in \eqref{eq:4ptCFT} with zero four-point kernel:
\begin{equation}
 \langle{\phi(x_1) \phi(x_3) \rangle \; \langle \phi(x_2) \phi(x_4)} \rangle 
 +  \langle{\phi(x_1) \phi(x_4) \rangle \; \langle \phi(x_2) \phi(x_3)} \rangle 
   =  \sum_J 
  \int_{\frac{d}{2}-\imath \infty}^{\frac{d}{2}+\imath\infty} \frac{dh}{2\pi \imath}
   \; \mu_{\Delta_{\phi}}^d(h,J)
     G^{\Delta_{\phi}}_{h,J}(x_i)\;.
\end{equation}
The measure is given by \eqref{eq:measure}:
\begin{equation}\label{eq:measure1}
\begin{split}
\mu_{\Delta_{\phi}}^d(h, J)  \, = & \, \left( \frac{ 1 + (-1)^J }{2} \right)
   \frac{  \Gamma(J+\frac{d}{2}) 
  }
  {  \Gamma(J+1)} 
   \\
&  \ \times    \frac{
 \Gamma( \frac{d}{2} - \Delta_{\phi})^2 
    \Gamma(\frac{ 2\Delta_{\phi} -d +h+J}{2}) 
    \Gamma(\frac{2\Delta_{\phi}-h+J}{2})
    \Gamma(h-1)\Gamma(d-h+J)\Gamma(\frac{h +J}{2})^2
 }{ \Gamma( \Delta_{\phi})^2 
  \Gamma(\frac{2d-2\Delta_{\phi}-h+J}{2})\Gamma(\frac{d-2\Delta_{\phi} +h+J}{2}) 
  \Gamma(h-\frac{d}{2})\Gamma(h+J-1)\Gamma(\frac{d- h  +J}{2})^2
 }  \;.
\end{split}
\end{equation}

As in the case of the interacting theory, we can close the contour to the right and pick up the poles of the measure with ${\rm Re}(h)\geq d/2$, from which we should exclude the ``spurious'' poles of \cite{Simmons-Duffin:2017nub}, i.e.\ the poles of the measure that cancel with the poles of the conformal blocks.
Such spurious poles are the poles of the $\Gamma(d-h+J)$ factor in the numerator of \eqref{eq:measure1}.
Since $h>d/2$, we are left with the poles of $\Gamma(\frac{2\Delta_{\phi}-h+J}{2})$, i.e.:
\be \label{eq:free-h}
h_{m,J} = 2 \Delta_{\phi} + J + 2 m \,,  \qquad m\in \mathbb{N} \;.
\ee
However, we have two gamma functions in the denominator of \eqref{eq:measure1} that can have poles, which, in the case that they coincide with any of the above poles, can lead to a zero residue (that is the absence of the corresponding pole).
The gammas in  question are $\Gamma(\frac{2d-2\Delta_{\phi}-h+J}{2})$ and $\Gamma(\frac{d- h  +J}{2})^2$. Substituting \eqref{eq:free-h} into \eqref{eq:measure} we find:
\be
\text{Res}\left[\mu_{\Delta_{\phi}}^d(h,J) \right]_{h=h_{m,J} }  \propto \frac{1}{\Gamma(d-2\Delta_{\phi}-m) \Gamma(\frac{d}{2} -  \Delta_{\phi} -m)^2} \,.
\ee

For the canonical scaling, $\Delta_{\phi}=\frac{d}{2} -1$ (that is 
$\zeta=1$), the denominator is $\Gamma(2-m) \Gamma(1-m)^2$ and all the poles with $m\geq 1$ have zero residue. This means that the genuine poles are given by \eqref{eq:free-h} with $m=0$. 
This spectrum coincides with the one of the vector model, not with the one of an interacting tensor model with standard propagator \cite{Giombi:2017dtl}.
This should not be a surprise, as the free theory is indistinguishable from a vector model with $O(N^3)$ symmetry, but as soon as interactions are turned on the symmetry is broken down to $O(N)^3$. 
Another way to understand this result is to notice that in the free theory any operator containing a factor $\partial^2\phi$ can be eliminated  by the equations of motion, regardless of the tensor rank.

For our scaling, $\Delta_{\phi}=\frac{d}{4}$ (that is $\zeta = d/4$), we find instead $\Gamma(\frac{d}{2}-m) \Gamma(\frac{d}{4}-m)^2$. Since the unitarity bounds require $d\leq 4$, we have three distinct cases: for $d=4$, we are back to the canonical case; for $d=2$, we have poles only from the first gamma function, leading to the restriction $m=0$ as in Sec.~\ref{sec:d=2}; for $0<d<4$ and $d\neq 2$, neither of the two arguments of the gamma functions are integers and all $m\geq 0$ are genuine poles.

While the discontinuity of the spectrum at $d=4$ can be understood as a consequence of the kinetic term becoming local, the discontinuity at $d=2$ remains puzzling. In fact, repeating the argument above, we would expect not to be able to remove operators with $\partial^2\phi$ factors for any $d<4$, as the Schwinger-Dyson equation:
\be
\langle \phi(x_1) \ldots \phi(x_n) (-\partial^2)^{d/4} \phi(x_0) \rangle = \sum_{i=1}^n \delta(x_0-x_i) \langle \prod_{j\neq i}^{1\ldots n} \phi(x_j)  \rangle \, ,
\ee
does not imply that we can eliminate $\partial^2 \phi$ inside correlation functions.
%

\section{The SYK model}
\label{app:original syk}
In this appendix, we review the OPE coefficients of the original SYK model and the conformal SYK model of Gross and Rosenhaus \cite{Gross:2017vhb}.

For the original SYK model, the OPE coefficients are given by \cite{Maldacena:2016hyu}:
	\begin{equation}
		c_m^2 \, = \, \alpha_0 \, \frac{(h_m-1/2)}{\pi \tan(\pi h_m / 2)} \frac{\Gamma(h_m)^2}{\Gamma(2h_m)} \, \frac{1}{k'(h_m)} \, , \qquad (m \, = \, 1, 2, \cdots) \;,
	\end{equation}
where:
	\begin{equation}
		\alpha_0 \, = \, 
		\frac{2\pi q}{(q-1)(q-2)\tan \frac{\pi}{q}} \, , 
	\end{equation}
and: 
	\begin{equation}
		k(h) \, = \, - \, (q-1) \, \frac{\Gamma(\frac{3}{2}-\frac{1}{q})\Gamma(1-\frac{1}{q})\Gamma(\frac{1}{q}+\frac{h}{2})\Gamma(\frac{1}{2}+\frac{1}{q}-\frac{h}{2})}
		{\Gamma(\frac{1}{2}+\frac{1}{q})\Gamma(\frac{1}{q})\Gamma(\frac{3}{2}-\frac{1}{q}-\frac{h}{2})\Gamma(1-\frac{1}{q}+\frac{h}{2})} \, .
	\end{equation}
The on-shell value of the conformal dimensions $h_m$ are determined by $k(h)=1$. Since $\tan(\pi h_m/2)<0$ and $k'(h_m)<0$, we have: 
	\begin{equation}
		c_m^2 \, > \, 0 \, , \qquad (m \, = \, 1, 2, \cdots) \;.
	\end{equation}

The conformal SYK model considered by Gross and Rosenhaus \cite{Gross:2017vhb} has the OPE coefficients:
	\begin{equation}
		c_m^2 \, = \, \alpha_0(q,\Delta) \, \frac{(h_m-1/2)}{\pi \tan(\pi h_m / 2)} \frac{\Gamma(h_m)^2}{\Gamma(2h_m)} \, \frac{1}{(1-2\bar{b})^2 k'(h_m)} \, , \qquad (m \, = \, 0, 1, 2, \cdots) \;,
	\end{equation}
with:
	\begin{equation}
		\alpha_0(q,\Delta) \, = \, \frac{2\pi}{(q-1)(1-2\Delta)\tan \pi\Delta} \, , 
	\end{equation}
and:
	\begin{equation}
		k(h) \, = \, - \, (q-1) \, \frac{\Gamma(\frac{3}{2}-\Delta)\Gamma(1-\Delta)\Gamma(\Delta+\frac{h}{2})\Gamma(\frac{1}{2}+\Delta-\frac{h}{2})}
		{\Gamma(\frac{1}{2}+\Delta)\Gamma(\Delta)\Gamma(\frac{3}{2}-\Delta-\frac{h}{2})\Gamma(1-\Delta+\frac{h}{2})} \, .
	\end{equation}
The on-shell value of the conformal dimensions $h_m$ are determined by $(1-2\bar{b})k(h)=1$.
The dependence of the coupling constant comes from $\bar{b}$ which is determined by:
	\begin{equation}
		\frac{\bar{b}^q}{1-2\bar{b}} \, = \, \frac{1}{2\pi J^2} \, (1-2\Delta)\tan\pi \Delta \, .
	\label{eq:bbar}
	\end{equation}

For $\Delta=1/q$, the OPE coefficients are identical to those of the original SYK model except the $(1-2\bar{b})^{-2}$ factor.
For real value of the coupling constant, this factor is always positive. Therefore in this model, for any value of real coupling constant: 
	\begin{equation}
		c_m^2 \, > \, 0 \, , \qquad (m \, = \, 0, 1, 2, \cdots) \, .
	\end{equation}

Let us now explicitly compute the OPE coefficient for small coupling $|J|\ll 1$ in this model.
To compare with our model we set $q=4$ and $\Delta=1/q=1/4$.

First from \eqref{eq:bbar}, we can explicitly solve for $\bar{b}$ as
	\begin{equation}
		\bar{b} \, = \, \frac{1}{2} \, - \, \frac{\pi}{8} J^2 \, + \, \mathcal{O}(J^4) \, .
	\end{equation}
The solution of the conformal dimensions are now given by
	\begin{equation}
		h_m \, = \, \frac{3}{2} + 2m \, + \, \frac{3J^2}{4(1+2m)} \, + \, \mathcal{O}(J^4) \, ,
	\end{equation}
and 
	\begin{equation}
		c_m^2 \, = \, \alpha_0(4,1/4) \, \frac{3\Gamma(3/2+2m)^2}{\pi^2 \Gamma(3+4m)} \, + \, \mathcal{O}(J) \, .
	\end{equation}

\end{subappendices}

\chapter{Quartic trifundamental models}
\label{chap:trif}
In chapter \ref{chap:3loops}, we studied the long-range multi-scalar model with quartic interactions. Multi-scalar models with quartic interactions describe some of the most important universality classes, such as the Ising and Heisenberg models, but other multi-scalar models, with smaller symmetry groups, are also of general interest (see for example \cite{Pelissetto:2000ek,Kleinert:2001ax} and references therein). Being able to classify or better understand all the possible universality classes appearing in such models would be of great theoretical appeal and efforts in this direction have been made (see for example in \cite{Brezin:1973jt,Michel:1983in,Michel:1985,Toledano:1985,Hatch:1985,Vicari:2006xr,Osborn:2017ucf,Rychkov:2018vya,Codello:2018nbe,Codello:2020lta,Hogervorst:2020gtc,Osborn:2020cnf}). However, as the number $\cN$ of fields increases, a full classification becomes daunting.
A natural way to broaden our understanding is then to gradually break the maximal symmetry group, i.e.\ the $O(\cN)$ group, to smaller ones, which of course can be done in many ways. We went in this directions in chapter \ref{chap:3loops}, by studying various symmetry groups: $O(N)$ vector model, cubic model and bifundamental model. 

In this chapter we go one step further in the same direction, and consider a \emph{trifundamental model}, with symmetry group $O(N_1)\times O(N_2) \times O(N_3)$, and $N_1 N_2 N_3=\cN$ with both short- and long-range propagators.
Whereas the $O(\cN)$ model has a single coupling, and the bifundamental model has two, the trifundamental model has five independent couplings (one tetrahedron and one double-trace couplings as for the $O(N)^3$ model of chapter \ref{chap:CTKT} but three different pillows), making its system of beta functions more involved. For this reason, we study its fixed points either numerically, for specific values of the $N_i$'s, or in some large-$N$ scaling limits, with either one, two, or all three  of the $N_i$'s being taken to infinity.
In the homogeneous case $N_i = N$, for $i=1,2,3$, the model reduces to the $O(N)^3$ \emph{tensor model} of chapter \ref{chap:CTKT}. However, studying it in the framework of the multi-scalar model allowed us to study the $1/N$ corrections which turned out to be surprisingly involved. 
This is due to the fact that the tetrahedron coupling receives no radiative corrections at large $N$, and therefore its beta function is either trivial (long-range case) or determined solely by the wave function renormalization (short-range case), the latter only starting with a (two-loop) cubic term.
At order $1/N$, the beta function of the tetrahedral coupling acquires  a (one-loop) quadratic term, destroying its exact marginality in the long-range model, and creating in the short-range model a delicate competition with the cubic term, the latter being leading in $1/N$ but subleading in the coupling.
In order to disentangle the effects of this quadratic term one needs to analyze scaling regimes defining a hierarchy between $1/N$ and $\epsilon$, where $\epsilon$ is either defined as the deviation from the critical dimension in the short-range case, i.e.\ $\epsilon=4-d$, or as the deviation from the critical scaling of the propagator in the long-range case, i.e.\  $C(p)=1/p^{(d+\epsilon)/2}$. This is why in this chapter, we will work with dimensional regularization as in chapter \ref{chap:3loops} and not with cutoffs as in chapter \ref{chap:CTKT}.

In section \ref{sec:short-range}, after a quick review of the short-range multi-scalar model, we compute the beta functions and fixed points of the short-range $O(N_1)\times O(N_2)\times O(N_3)$ model at two loops. First, in section \ref{sec:short-range_num}, we look for numerical solutions of the fixed point equations at finite $N_i$. Then, in sections \ref{sec:short-range_vector} and \ref{sec:short-range_matrix}, we compute the fixed points respectively in the vector-like ($N_1 \rightarrow \infty$; $N_2$ and $N_3$ fixed) and in the matrix-like ($N_2= cN_1= N \rightarrow \infty$; $c$ and $N_3$ fixed) limits. Finally, in section \ref{sec:short-range_triple}, we study the large-$N$ limit and its first subleading corrections in the case $N_1=N_2=N_3=N$, and with a single coupling for the three pillow interactions, corresponding to the $O(N)^3$ tensor model. 
Next, in section \ref{sec:long-range}, we study the long-range case. In section \ref{sec:long-range_equalN}, we directly set $N_1=N_2=N_3=N$ to study the bosonic $O(N)^3$ tensor model, with a single coupling for the three pillow interactions. We study the fixed points and critical exponents at two loops, up to and including order $1/N$. We finish with some concluding comments in \ref{sec:concltri}. In appendix \ref{app:grad_flow}, we give the beta functions of 
the trifundamental model as a gradient flow. 

\section{The short-range trifundamental model}
\label{sec:short-range}

\subsection{The short-range multi-scalar model}
\label{sec:short-range_ms}
The short-range multi-scalar model with quartic interactions in dimension $d$ is defined by the action:
\begin{equation}
		S[\phi]  \, = \, \int d^dx \, \bigg[ \frac{1}{2} \partial_{\mu} \phi_\mba(x) \partial_{\mu} \phi_{\mba}(x)
		\, + \, \frac{1}{4!} \, \lambda_{\mba \mbb \mbc \mbd} \phi_{\mba}(x) \phi_{\mbb}(x) \phi_{\mbc}(x) \phi_{\mbd}(x) \bigg] \, ,
	\end{equation}
where the indices take values from 1 to $\cN$, and a summation over repeated indices is implicit.
For the Euclidean theory in $d=4-\epsilon$ dimension the beta function up to two loops in the minimal subtraction scheme \cite{ZinnJustin:2002ru} is:
\begin{align} \label{eq:beta-general}
		\beta_{\mba \mbb \mbc \mbd} \, &= \, - \, \epsilon \gt_{\mba \mbb \mbc \mbd}
		\, + \, \left(\gt_{\mba \mbb \mbe \mbf}\gt_{\mbe \mbf \mbc \mbd} + 2 \textrm{ terms} \right) 
		\, - \, \left(\gt_{\mba \mbb \mbe \mbf}\gt_{\mbe \mbg \mbh \mbc}\gt_{\mbf \mbg \mbh \mbd}+ 5 \textrm{ terms} \right) \crcr
		&\quad + \, \frac{1}{12} \left(\gt_{\mba \mbb \mbc \mbe}\gt_{\mbe \mbf \mbg \mbh}\gt_{\mbf \mbg \mbh \mbd} + 3 \textrm{ terms} \right) +\mathcal{O}(\gt^4) \,,
	\end{align}
where we rescaled the renormalized coupling to $\gt_{\mba \mbb \mbc \mbd} = g_{\mba \mbb \mbc \mbd} (4\pi)^{-d/2}/\Gamma(d/2)$.

By imposing various symmetry restrictions on the interaction one obtains different models which have been extensively studied (see for example \cite{Pelissetto:2000ek,Kleinert:2001ax,Vicari:2006xr,Osborn:2017ucf,Rychkov:2018vya} and references therein). We study here the case with $O(N_1)\times O(N_2) \times O(N_3)$ invariance, which is relatively new.

\subsection{The short-range  trifundamental model}
The fields in the trifundamental model are rank-3 tensor fields transforming in the trifundamental representation of $O(N_1)\times O(N_2) \times O(N_3)$. This is made manifest by writing the index $\mba$ as a triplet $\mba=(a_1,a_2,a_3)$, where the first, second and third index correspond to the $O(N_1)$, $O(N_2)$ and $O(N_3)$ group respectively:
\be
\phi_{a_1 a_2 a_3} \to \sum_{b_1 b_2 b_3}^{1\ldots N} R^{(1)}_{a_1 b_1}R^{(2)}_{a_2 b_2}R^{(3)}_{a_3 b_3}\phi_{b_1 b_2 b_3}\,, \;\;\;\; R^{(i)}\in O(N_i)\,.
\ee
Notice that since each orthogonal group contains a $\mathbb{Z}_2$ subgroup, in order to have a faithful action of the symmetry group, we should quotient $O(N_1)\times O(N_2) \times O(N_3)$ by a $\mathbb{Z}_2^3$, which acts trivially. As this is irrelevant to our study, we will stick to the  unquotiented version of the symmetry group.

Under such symmetry transformation, the most general invariant tensor structure for the coupling is:\footnote{The normalization has be chosen so that the couplings are normalized by $1/4$ and not $1/4!$, as usually done in tensor models.}
	\begin{equation} \label{eq:coupling-trifund}
		\gt_{\mba \mbb \mbc \mbd} \, = \, \gt \left(\delta^t_{\mba \mbb \mbc \mbd} + 5 \textrm{ terms} \right)
		\, + \, \sum_{i=1,2,3} \gt_{p,i} \left(\delta^{p,i}_{\mba \mbb ;\mbc \mbd} + 5 \textrm{ terms} \right)
		\, + \, 2\gt_d\left(\delta^d_{\mba \mbb \mbc \mbd} + 2 \textrm{terms} \right) \, ,
	\end{equation}
with $\delta^t_{\mba \mbb \mbc \mbd}$ and $\delta^d_{\mba \mbb \mbc \mbd}$ defined in \eqref{eq:delta_invariants} and :
\begin{equation}
\delta^{p,i}_{\mba \mbb ;\mbc \mbd} =\delta_{a_ic_i}\delta{b_id_i}\prod_{j\neq i}\delta_{a_jb_j}\delta_{c_jd_j}
\end{equation}
They correspond to the quartic invariants represented graphically in figure~\ref{fig:trace_invariants} but we now have a different coupling for each pillow invariant.

Substituting \eqref{eq:coupling-trifund} in \eqref{eq:beta-general} and truncating at one loop, we obtain the beta functions:
	\begin{align}
		\beta_t=&-\epsilon\tilde{g}+4\Big[6\tilde{g}\tilde{g}_d+2(\tilde{g}_{p,1}\tilde{g}_{p,2}+\tilde{g}_{p,1}\tilde{g}_{p,3}+\tilde{g}_{p,2}\tilde{g}_{p,3})\crcr
		& +\tilde{g}((1+N_1)\tilde{g}_{p,1}+(1+N_2)\tilde{g}_{p,2}+(1+N_3)\tilde{g}_{p,3})\Big] \, , \crcr
		\beta_{p,i}=&-\epsilon\tilde{g}_{p,i}+2\Big[12\tilde{g}_d\tilde{g}_{p,i}+4\tilde{g}(\tilde{g}_{p,i+1}+\tilde{g}_{p,i+2})
		+4\tilde{g}_{p,i+1}\tilde{g}_{p,i+2}+(2+N_i)\tilde{g}^2 \crcr
		& +2\tilde{g}_{p,i}((1+N_{i+1})\tilde{g}_{p,i+1}+(1+N_{i+2})\tilde{g}_{p,i+2}+(N_{i+1}+N_{i+2})\tilde{g})+\tilde{g}_{p,i}^2(4+N_i+N_{i+1}N_{i+2})\Big]\, , \crcr
		\beta_d=&-\epsilon\tilde{g}_d+2\Big[\tilde{g}_d^2(8+N_1N_2N_3)+3(\tilde{g}_{p,1}^2+\tilde{g}_{p,2}^2+\tilde{g}_{p,3}^2)
		+2\tilde{g}(\tilde{g}_{p,1}+\tilde{g}_{p,2}+\tilde{g}_{p,3})\crcr
		& +2 (N_1\tilde{g}_{p,2}\tilde{g}_{p,3}+N_2\tilde{g}_{p,1}\tilde{g}_{p,3}+N_3\tilde{g}_{p,1}\tilde{g}_{p,2}) +2\tilde{g}_d\tilde{g}(N_1+N_2+N_3) \crcr
		&+2\tilde{g}_d((1+N_1+N_2N_3)\tilde{g}_{p,1}+(1+N_2+N_1N_3)\tilde{g}_{p,2}+(1+N_3+N_1N_2)\tilde{g}_{p,3}) \Big] \, , \label{eq:beta^(4)}
	\end{align}
where $i\in\{1,2,3\}\,\text{mod}\, 3$, i.e.\ $g_{p,4}=g_{p,1}$ and $g_{p,5}=g_{p,2}$.
The two-loop terms can be obtained by computer algebra but they are too long to write here and we will only use them in section~\ref{sec:short-range_triple}.
In appendix~\ref{app:grad_flow} we write the system \eqref{eq:beta^(4)} as a gradient flow.

Notice that the model with only double-trace interaction has an enhanced symmetry, being invariant under field transformations in the fundamental representation of $O(N_1 N_2 N_3)$; that is, it is the usual $O(\cN)$ model in disguise.
Similarly, if all the couplings except the double-trace and one pillow, e.g.\ $\gt_{p,1}$, are zero, then the model has the symmetry group  $O(N_1)\times O(N_2 N_3)$, and it is a bifundamental model in disguise. 
Such symmetry enhancements are reflected in the fact that the couplings set to zero are not turned on by the renormalization group flow.
Keeping instead at least two pillows, or just the tetrahedron, will break the symmetry back to $O(N_1)\times O(N_2) \times O(N_3)$, and the flow will generate the remaining couplings.
The tetrahedron is in fact the single coupling which is most characteristic of the full symmetry group, being capable alone to generate all the others by RG flow. Moreover, it is the coupling that in the $O(N)^3$ model leads to a melonic dominance at large $N$.
Therefore, in most of the following we will only be interested in fixed points with non-vanishing tetrahedron coupling, $\tilde{g}\ne 0$.

\paragraph{Remark.}

Let us consider the case of real coupling constants.
The tetrahedron interaction is not positive definite, thus the most general potential can be unstable. However, for particular choices of (real) couplings, the pillow and double-trace can dominate over the negative direction of the tetrahedron, thus making the whole interaction positive. In particular, as shown in  \cite{Michel:1983in} (see also \cite{ZinnJustin:2007zz}), based on the gradient flow representation \cite{Wallace:1974dy}, at order $\epsilon$ any non-trivial fixed point corresponds to a positive interaction. Moreover, we also know that an unstable interaction cannot flow to a stable one \cite{Rychkov:2018vya}.
As a consequence, any non-trivial fixed point with $\tilde{g}\neq 0$ must also have at least some of the other couplings non-zero, and so must also any initial condition that flows to such a fixed point.

\subsection{Numerical solutions for small $N_i$}
\label{sec:short-range_num}

Even at the one-loop level it is hard to solve the beta functions \eqref{eq:beta^(4)} for generic values of $N_i$.
In this subsection we solve for fixed points at low $N_i$.
In the following, we only search for fixed points at one-loop with $\tilde{g}\ne 0$ and real critical couplings.

We define the critical coupling vector by
	\begin{equation}
		\vec{g}_\star \, \equiv \, \big( \tilde{g}^\star, \, \tilde{g}_{p,1}^\star, \, \tilde{g}_{p,2}^\star, \, \tilde{g}_{p,3}^\star, \, \tilde{g}_d^\star \big) \, .
	\end{equation}
The stability matrix of the fixed point is given by 
	\begin{equation}
		M_{ab} \, = \, \frac{\partial \beta_a(\vec{g})}{\partial g_b} \bigg|_{\vec{g}_\star} \, ,
	\end{equation}
where $a, b=\{t, (p,1) , (p,2) , (p,3) , d\}$.
We arrange the eigenvalues of this matrix in a vector denoted by $\vec{\omega}$.
If the eigenvalue $\omega_a$ is positive, then the fixed point is stable in the corresponding eigendirection.

\medskip

We have checked that there is \emph{no} real fixed point with $\tilde{g}\ne 0$ that is stable in all five directions in the range $2 \le N_i \le 50$.\footnote{This was done using a computer algebraic tool, more precisely the \textit{Solve} algorithm of Mathematica.}
We explicitly show some examples below.

\paragraph{Case $N_1=2$, $N_2=2$}
By fixing $N_1=N_2=2$, there is no fixed point in the range of $2 \le N_3 \le21$.
At $N_3 = 22$, we find four fixed points:   
	\begin{align}
		\vec{g}_\star \, &= \, \left( \frac{\epsilon}{1200}, \, -\frac{\epsilon}{600}, \, -\frac{\epsilon}{600}, \, \frac{\epsilon}{400}, \, \frac{\epsilon}{800} \right) \, , \qquad
		\vec{\omega} \, = \, \big( -6.72\epsilon, \, 6\epsilon, \, -4.8\epsilon, \, 0.48\epsilon, \, -0.48\epsilon \big) \, , \\
		\vec{g}_\star \, &= \, \left( \frac{7\epsilon}{8304}, \, -\frac{7\epsilon}{4152}, \, -\frac{7\epsilon}{4152}, \, \frac{7\epsilon}{2768}, \, \frac{59\epsilon}{49824} \right) \, ,\qquad
		\vec{\omega} \, = \, \big( -6.79\epsilon, \, 6\epsilon, \, -4.85\epsilon, \, 0.485\epsilon, \, -0.485\epsilon \big) \, , \\
		\vec{g}_\star \, &= \, \left( \frac{7\epsilon}{6912}, \, -\frac{35\epsilon}{20736}, \, -\frac{35\epsilon}{20736}, \, \frac{49\epsilon}{20736}, \, \frac{157\epsilon}{124416} \right) \, ,\qquad
		\vec{\omega} \, = \, \big( -6.70\epsilon, \, 6\epsilon, \, -4.76\epsilon, \, 0.486\epsilon, \, -0.486\epsilon \big) \, , \\
		\vec{g}_\star \, &= \, \left( \frac{\epsilon}{976}, \, -\frac{5\epsilon}{2928}, \, -\frac{5\epsilon}{2928}, \, \frac{7\epsilon}{2928}, \, \frac{7\epsilon}{5856} \right) \, , \qquad
		\vec{\omega} \, = \, \big( -6.78\epsilon, \, 6\epsilon, \, -4.82\epsilon, \, -0.491\epsilon, \, -0.491\epsilon \big) \, .
	\end{align}
There is no fixed point which is stable in all five directions.
In the range of $23 \le N_3 \le33$, we find similar four types of fixed points.
At $N_3 = 34$, we find six fixed points:
	\begin{align}
		\vec{g}_\star \, &= \, \left( \frac{\epsilon}{1008}, \, -\frac{\epsilon}{1008}, \, 0, \, \frac{\epsilon}{756}, \, \frac{\epsilon}{6048} \right) \, , \qquad
		\vec{\omega} \, = \, \big( 6\epsilon, \, 5.58\epsilon, \, -4\epsilon, \, -3.86\epsilon, \, 0 \big) \, , \\
		\vec{g}_\star \, &= \, \left( \frac{11(3445\mp203\sqrt{97})\epsilon}{62966016}, \, -\frac{11(-6479\pm\sqrt{97})\epsilon}{62966016}, \, -\frac{11(-6479\pm\sqrt{97})\epsilon}{62966016}, \,
		\frac{11(3171\pm67\sqrt{97})\epsilon}{20988672}, \right. \nonumber\\
		& \left. \qquad \, \frac{(314123\mp6325\sqrt{97})\epsilon}{377796096} \right) \, ,\nonumber\\
		&\quad \vec{\omega} \, = \, \big( -6.69\epsilon, \, 6\epsilon, \, -5.39\epsilon, \, 3.21\epsilon, \, -3.21\epsilon \big) \qquad ({\rm for\ upper}) \, , \\
		&\quad \vec{\omega} \, = \, \big( -6.30\epsilon, \, 6\epsilon, \, -5.00\epsilon, \, 3.21\epsilon, \, -3.21\epsilon \big) \qquad ({\rm for\ lower}) \, , \\
		\vec{g}_\star \, &= \, \left( \frac{(161-10\sqrt{97})\epsilon}{259536}, \, \frac{(-589+3\sqrt{97})\epsilon}{519072}, \, \frac{(-589+3\sqrt{97})\epsilon}{519072}, \,
		\frac{(428+7\sqrt{97})\epsilon}{259536}, \, \frac{(428+7\sqrt{97})\epsilon}{519072} \right) \, ,\nonumber\\
		&\quad \vec{\omega} \, = \, \big( -6.37\epsilon, \, 6\epsilon, \, -5.13\epsilon, \, 3.06\epsilon, \, -3.06\epsilon \big) \qquad ({\rm for\ upper}) \, , \\
		&\quad \vec{\omega} \, = \, \big( -6.63\epsilon, \, 6\epsilon, \, -5.25\epsilon, \, 3.38\epsilon, \, -3.38\epsilon \big) \qquad ({\rm for\ lower}) \, .
	\end{align}
The other solution is given by the first solution with exchanging $\bar{g}_{p,1}^\star$ and $\bar{g}_{p,2}^\star$ with the same eigenvalue vector.
There is no fixed point which is stable in all five directions.

\paragraph{Case $N_1=2$, $N_2=3$}
By fixing $N_1=2$ and $N_2=3$, there is no fixed point in the range of $2 \le N_3 \le45$.
At $N_3 = 46$, we find two fixed points:
	\begin{align}
		\vec{g}_\star \, &= \, \left( \frac{17(126955\pm3\sqrt{1345})\epsilon}{2844275280}, \, 0, \, - \frac{17(126955\pm3\sqrt{1345})\epsilon}{2844275280}, \,
		\frac{(126955\pm3\sqrt{1345})\epsilon}{129285240}, \right. \nonumber\\
		& \left. \qquad \, \frac{(243160\mp461\sqrt{1345})\epsilon}{2844275280} \right) \, ,\nonumber\\
		&\quad \vec{\omega} \, = \, \big( 6\epsilon, \, 5.67\epsilon, \, -4.27\epsilon, \, -4.17\epsilon, \, -0.24\epsilon \big) \qquad ({\rm for\ upper}) \, , \\
		&\quad \vec{\omega} \, = \, \big( 6\epsilon, \, 5.66\epsilon, \, -4.26\epsilon, \, -4.16\epsilon, \, 0.24\epsilon \big) \qquad ({\rm for\ lower}) \, .
	\end{align}
There is no fixed point which is stable in all five directions.
At least up to $N_3=1000$, we find the same type of two fixed points, and they are unstable in some of the directions.

\subsection{Vector-like limit}
\label{sec:short-range_vector}

We now consider the limit $N_1 \rightarrow \infty$ while keeping $N_2$ and $N_3$ fixed. 
We define the new couplings:
\begin{equation}
\tilde{g}_S=\tilde{g}+\tilde{g}_{p,1} \,, \;\;\; \tilde{g}_D=\tilde{g}-\tilde{g}_{p,1} \,, \;\;\; \tilde{g}_2=\tilde{g}_d+\frac{\tilde{g}_{p,2}}{N_2}+\frac{\tilde{g}_{p,3}}{N_3} \,,
\end{equation}
which correspond to orthogonal operators at large $N_1$, and thus their beta functions will decouple. We furthermore rescale the couplings in order to obtain a large-$N_1$ expansion:
\begin{equation}
\tilde{g}_S=\frac{\bar{g}_S}{N_1} \,, \;\;\; \tilde{g}_D=\frac{\bar{g}_D}{N_1} \,, \;\;\;\tilde{g}_{p,i}=\frac{\bar{g}_{p,i}}{N_1} \,, \;\;\; \tilde{g}_2=\frac{\bar{g}_2}{N_1} \,.
\end{equation}
The three-loop terms are suppressed in $1/N_1$ and at leading order we obtain the following beta functions:
\begin{align}
\beta_S&=-\epsilon\bar{g}_S+2\bar{g}_S^2 \, ,\crcr
\beta_{D}&=-\epsilon\bar{g}_D-2\bar{g}_D^2 \, , \crcr
\beta_{p,2}&=-\epsilon\bar{g}_{p,2}+4\bar{g}_S\bar{g}_{p,2}+2N_3\bar{g}_{p,2}^2 \, , \crcr
\beta_{p,3}&=-\epsilon\bar{g}_{p,3}+4\bar{g}_S\bar{g}_{p,3}+2N_2\bar{g}_{p,3}^2 \, , \crcr
\beta_2&=-\epsilon\bar{g}_2 +4\bar{g}_S\bar{g}_2+2N_2N_3\bar{g}_2^2\, . 
\label{eq:beta_largeN1}
\end{align}

We can then solve for fixed points. We obtain the following 32 fixed points:
\begin{align}
\bar{g}_S^{\star}&=\{0,\frac{\epsilon}{2}\} \, , \;\; \bar{g}_D^{\star}=\{0,-\frac{\epsilon}{2}\}\, , \crcr
\bar{g}_{p,2}&=\{0,\pm \frac{\epsilon}{2N_3}\} \, , \;\; \bar{g}_{p,3}=\{0,\pm \frac{\epsilon}{2N_2}\} \, , \;\; \bar{g}_{2}=\{0,\pm \frac{\epsilon}{2N_2N_3}\} \, ,
\end{align}
where the sign in $(\bar{g}_{p,2}^{\star},\bar{g}_{p,3}^{\star},\bar{g}_{2}^{\star})$ is the upper one when $\bar{g}_S^{\star}=0$ and the lower one when $\bar{g}_S^{\star}=\epsilon/2$. 

The stability matrix is triangular at large $N_1$ and the critical exponents are given by the diagonal elements:
\begin{align}
\partial \beta_{S,D}(\bar{g}^{\star})&=\left\{\begin{array}{l}
-\epsilon \quad \text{ if } \bar{g}_{S,D}^{\star}=0 \, ,\\
 \epsilon \quad \text{ else,}
\end{array}\right. \crcr
\partial \beta_{p,i}=&\left\{ \begin{array}{l}
 -\epsilon \quad \text{ if } (\bar{g}_S^{\star},\bar{g}_{p,i}^{\star})=(0,0) \text{ or } (\frac{\epsilon}{2},-\frac{\epsilon}{2N_j}) \text{ with } (i,j)=\{(2,3),(3,2)\}\, , \\
 \epsilon \quad \text{ else,}
\end{array} \right. \crcr
\partial \beta_{2}=&\left\{ \begin{array}{l}
 -\epsilon \quad \text{ if } (\bar{g}_S^{\star},\bar{g}_{2}^{\star})=(0,0) \text{ or } (\frac{\epsilon}{2},-\frac{\epsilon}{2N_2N_3}) \, , \\
 \epsilon \quad \text{ else.} 
\end{array} 
\right.
\end{align}

The only stable fixed point in all five directions is: 
$(\bar{g}_S^{\star},\bar{g}_D^{\star},\bar{g}_{p,2}^{\star},\bar{g}_{p,3}^{\star},\bar{g}_2^{\star})=(\frac{\epsilon}{2},-\frac{\epsilon}{2},0,0,0).$
This corresponds to $\bar{g}_{p,1}=\frac{\epsilon}{2}$ and $\bar{g}^{\star}=\bar{g}_{p,2}^{\star}=\bar{g}_{p,3}^{\star}=\bar{g}_2^{\star}=0$. It is a chiral fixed point with symmetry $O(N_1)\times O(N_2N_3)$, similar to those found in bifundamental models $O(N)\times O(M)$.

\medskip

In summary, we find \emph{no} real stable fixed point with non-zero tetrahedral coupling in the vector-like limit.

\subsection{Matrix-like limit}
\label{sec:short-range_matrix}

We now consider the matrix-like double-scaling large-$N$ limit:
\begin{equation}
N_1=c N \,, \; N_2=  N \,, \; N \rightarrow \infty \,,
\end{equation}
with $N_3$ fixed and $c\ge 1$ fixed and of order one. We redefine the double-trace coupling, combining it with the third pillow coupling: 
\begin{equation}
\tilde{g}_{dp}=\tilde{g}_d+\frac{\tilde{g}_{p,3}}{N_3} \,,
\end{equation}
and we rescale all the couplings with $N$ as:
\begin{equation}
\tilde{g}=\frac{\bar{g}}{N} \,, \; \tilde{g}_{p,1}=\frac{\bar{g}_{p,1}}{N} \,, \; \tilde{g}_{p,2}=\frac{\bar{g}_{p,2}}{N} \,, \; \tilde{g}_{p,3}=\frac{\bar{g}_{p,3}}{N^2} \,, \; \tilde{g}_{dp}=\frac{\bar{g}_{dp}}{N^2} \,. 
\end{equation}

The barred couplings are 't Hooft couplings, fixed in the large-$N$ limit, and we recognize the standard scaling of quartic matrix invariants with a single trace (the tetrahedron and first two pillows) or a double-trace (the third pillow and the double-trace). The third pillow behaves effectively as a double-trace because the vertical line in figure~\ref{fig:trace_invariants} corresponds in this case to the index whose range remains finite (i.e.\ $N_3$).

The one-loop beta functions at leading order in $1/N$ are:
\begin{align}
\beta_t&= -\epsilon \bar{g} +4\bar{g}\left(c\bar{g}_{p,1}+\bar{g}_{p,2}\right) \,,\crcr
\beta_{p,1}&= -\epsilon \bar{g}_{p,1}+2c\left(\bar{g}_{p,1}^2+\bar{g}^2\right)+4\bar{g}_{p,1}\left(\bar{g}+\bar{g}_{p,2}\right)+2N_3\bar{g}_{p,1}^2 \,, \crcr
\beta_{p,2}&= -\epsilon \bar{g}_{p,2}+2\left(\bar{g}_{p,2}^2+\bar{g}^2\right)+4c\bar{g}_{p,2}\left(\bar{g}+\bar{g}_{p,1}\right)+2cN_3\bar{g}_{p,2}^2 \,, \crcr
\beta_{p,3}&= -\epsilon \bar{g}_{p,3} +8\bar{g}_{p,1}\bar{g}_{p,2}+4
\bar{g}_{p,3}\left(c\bar{g}_{p,1}+\bar{g}_{p,2}+(1+c)\bar{g}\right)+2c\bar{g}_{p,3}^2+8\bar{g}\left(\bar{g}_{p,1}+\bar{g}_{p,2}\right)+2(N_3+2)\bar{g}^2 \,,\crcr
\beta_{dp}&=-\epsilon \bar{g}_{dp}+4\bar{g}_{dp}(1+c)\bar{g}+2cN_3\bar{g}_{dp}^2+\frac{2(N_3+2)}{N_3}\bar{g}^2+6\left(\bar{g}_{p,1}^2+\bar{g}_{p,2}^2\right)+\frac{4(2+N_3^2)}{N_3}\bar{g}_{p,1}\bar{g}_{p,2}\crcr
& +4\bar{g}_{dp}\left((c+N_3)\bar{g}_{p,1}+(1+N_3c)\bar{g}_{p,2}\right)+\frac{4(2+N_3)}{N_3}\bar{g}\left(\bar{g}_{p,1}+\bar{g}_{p,2}\right)\,.
\end{align}

We find 32 fixed points. For $\bar{g}^{\star},\bar{g}_{p,1}^{\star},\bar{g}_{p,2}^{\star}$, we find either:
\begin{align}
\bar{g}^{\star}&=0  \,,\crcr
(\bar{g}_{p,1}^{\star},\bar{g}_{p,2}^{\star})&= \{ (0,0),(0,\frac{\epsilon}{2(1+cN_3)}),(\frac{\epsilon}{2(c+N_3)},0) \,,\crcr
&(-\frac{\epsilon(1-cN_3)}{2\left(c^2N_3+c(N_3^2-3)+N_3\right)},-\frac{\epsilon(c-N_3)}{2\left(c^2N_3+c(N_3^2-3)+N_3\right)}) \} \,,
\label{eq:fpmatrix0}
\end{align}
or
\begin{align}
\bar{g}^{\star}&=\frac{\epsilon\left(c(1-N_3)\pm(1-c)N_3\sqrt{c^2+1-cN_3}\right)}{4(1+c)(c+N_3(c-1)^2)}  \,,\crcr
\bar{g}_{p,1}^{\star}&=\frac{\epsilon\left(cN_3(c-1)+1 \mp \sqrt{c^2+1-cN_3}\right)}{4(1+c)(c+N_3(c-1)^2)} \,,\crcr
\bar{g}_{p,2}^{\star}&=\frac{\epsilon\left(N_3(1-c)+c^2 \pm c\sqrt{c^2+1-cN_3}\right)}{4(1+c)(c+N_3(c-1)^2)} \,,
\label{eq:fpmatrix1}
\end{align}
or
\begin{align}
\bar{g}^{\star}&=\frac{\epsilon\left(c(1+c)(1-N_3)\pm (N_3(c^2+1)-2c)\sqrt{c^2+1-cN_3}\right)}{4N_3(c^2+1)^2-4c(3c^2-2c+3)}  \,,\crcr
\bar{g}_{p,1}^{\star}&=\frac{\epsilon\left(c^3N_3-2c^2+c(N_3+1)-1 \pm (c-1)\sqrt{c^2+1-cN_3}\right)}{4N_3(c^2+1)^2-4c(3c^2-2c+3)} \,, \crcr
\bar{g}_{p,2}^{\star}&=\frac{\epsilon\left(N_3-2c+c^2(N_3+1)-c^3 \mp c(c-1)\sqrt{c^2+1-cN_3}\right)}{4N_3(c^2+1)^2-4c(3c^2-2c+3)} \,,
\label{eq:fpmatrix2}
\end{align}
where the signs are taken to be simultaneously either the upper or the lower ones. 

For the last two couplings, we find the following fixed points in terms of $\bar{g}^{\star},\bar{g}_{p,1}^{\star},\bar{g}_{p,2}^{\star}$:
\begin{align}
\bar{g}_{p,3}^{\star}&=\frac{1}{4c}\Bigg[ \epsilon-4c\bar{g}_{p,1}^{\star}-4\bar{g}_{p,2}^{\star}-4(1+c)\bar{g}^{\star} \crcr
& \pm \sqrt{\left(4c(\bar{g}_{p,1}^{\star}+\bar{g}^{\star})+4(\bar{g}_{p,2}^{\star}+\bar{g}^{\star})-\epsilon\right)^2-16c\left(4(\bar{g}^{\star}+\bar{g}_{p,1}^{\star})(\bar{g}^{\star}+\bar{g}_{p,2}^{\star})+\bar{g}^{\star}{}^2(N_3-2)\right)}\Bigg] \crcr
\bar{g}_{dp}^{\star}&=\frac{1}{4cN_3}\Bigg[\epsilon-4c(\bar{g}_{p,1}^{\star}+\bar{g}^{\star}+N_3\bar{g}_{p,2}^{\star})-4(\bar{g}_{p,2}^{\star}+\bar{g}^{\star}+N_3\bar{g}_{p,1}^{\star})  \,,\crcr
& \pm \Big(16c^2(\bar{g}_{p,1}^{\star}+\bar{g}^{\star}+N_3\bar{g}_{p,2}^{\star})^2+(4(\bar{g}_{p,2}^{\star}+\bar{g}^{\star}+N_3\bar{g}_{p,1}^{\star})-\epsilon)^2-32c\left(\bar{g}_{p,1}^{\star}\bar{g}_{p,2}^{\star}+\bar{g}^{\star}(\bar{g}_{p,1}^{\star}+\bar{g}_{p,2}^{\star})\right)\crcr
& \qquad -16cN_3(\bar{g}_{p,1}^{\star}{}^2+\bar{g}_{p,2}^{\star}{}^2+\bar{g}^{\star}{}^2)-8\epsilon c(\bar{g}_{p,1}^{\star}+\bar{g}^{\star}+N_3\bar{g}_{p,2}^{\star})\Big)^{1/2}\Bigg] \,,
\label{eq:fpmatrixp3d}
\end{align}
where the signs are chosen independently. Notice that since $\bar{g}^{\star},\bar{g}_{p,1}^{\star},\bar{g}_{p,2}^{\star}$ are of order $\epsilon$, so are also $\bar{g}_{p,3}^{\star},\bar{g}_{dp}^{\star}$.

We can now compute the critical exponents. The stability matrix is a block triangular matrix with a first block corresponding to the couplings $\bar{g}^{\star},\bar{g}_{p,1}^{\star},\bar{g}_{p,2}^{\star}$ and a second diagonal block for the couplings $\bar{g}_{p,3}^{\star},\bar{g}_{dp}^{\star}$. 

We first compute the critical exponents for the couplings $\bar{g}^{\star},\bar{g}_{p,1}^{\star},\bar{g}_{p,2}^{\star}$. For the fixed points of \eqref{eq:fpmatrix0}, we find the following critical exponents:
\begin{align}
(\omega_t,\omega_1,\omega_2)&=\{ (-\epsilon,-\epsilon,-\epsilon),(\epsilon,\frac{(1-cN_3)\epsilon}{1+cN_3},\frac{(1-cN_3)\epsilon}{1+cN_3}),(\epsilon,\frac{(c-N_3)\epsilon}{c+N_3},\frac{(c-N_3)\epsilon}{c+N_3}),\crcr
&(\epsilon,-\frac{(1-cN_3)(c-N_3)\epsilon}{c^2N_3+N_3+c(N_3^2-3)},\frac{(1-cN_3)(c-N_3)\epsilon}{c^2N_3+N_3+c(N_3^2-3)}) \} \,.
\end{align}
For the fixed points of \eqref{eq:fpmatrix1} we find:
\begin{align}
\omega_t&=\epsilon\crcr
\omega_1&=-\omega_2=\frac{\epsilon\sqrt{c^2+1-cN_3}}{(1+c)(c+(c-1)^2N_3)}\Bigg(c^2(1-2N_3)-c(c-1)^2N_3^3+N_3^2(c^2-c+1)^2\crcr
& \qquad \qquad \pm 2N_3(N_3-1)c(c-1)\sqrt{c^2+1-cN_3}\Bigg)^{1/2}\,.
\end{align}
For the fixed points of \eqref{eq:fpmatrix2}, we find:
\begin{align}
\omega_t&=\epsilon \crcr
\omega_1&=\omega_2=\epsilon\frac{c(N_3^2+2)(c^2+1)-N_3(c^4+4c^2+1) \pm c(N_3-1)(1+c)\sqrt{c^2+1-cN_3} }{(c^2+1)^2N_3-c(3c^2-2c+3)}\,.
\label{eq:critmatrix3}
\end{align}

We are interested in finding stable fixed points with non-zero tetrahedron coupling.
The fixed points in \eqref{eq:fpmatrix0} have zero tetrahedral coupling and the ones in equation \eqref{eq:fpmatrix1} have $\omega_1=-\omega_2$ hence cannot be stable in all five directions.

We are left with the fixed points of equation \eqref{eq:fpmatrix2}. Because of the square root, for these fixed points $\omega_{1,2}$ are real only for $N_3\leq \frac{c^2+1}{c}$. For these values of $N_3$, the branch with a minus sign always has negative or zero critical exponents. The solutions with a plus sign is positive for $c\le N_3\leq \frac{c^2+1}{c}$. If $N_3 > \frac{c^2+1}{c}$, both solutions are complex with a positive real part.

We thus have to look at the critical exponents for the last two couplings in order to conclude. As the second block of the stability matrix is diagonal, the critical exponents for the last two couplings are just the diagonal elements:

\begin{align}
\partial \beta_{p,3}(\bar{g}^{\star})=\pm \frac{\epsilon  \sqrt{R_1\pm R_2\sqrt{c^2+1-cN_3}}}{|(c^2+1)^2N_3-c(3c^2-2c+3)|} \,, \\
\partial \beta_{dp}(\bar{g}^{\star})=\pm \frac{\epsilon \sqrt{R_3 \pm R_4\sqrt{c^2+1-cN_3}}}{|(c^2+1)^2N_3-c(3c^2-2c+3)|} \,,
\end{align}
with:
\begin{align}
R_1=& c^2(c^2+1)^2N_3^4-c\left(2c^6+11c^4+2c^3+11c^2+2\right)N_3^3\crcr
& +\left(c^8+9c^6+20c^5+24c^4+20c^3+9c^2+1\right)N_3^2\crcr
& +c\left(4c^6-22c^5-5c^4-54c^3-5c^2-22c+4\right)N_3\crcr
& -c^2\left(3c^4-30c^3+14c^2-30c+3\right) \,, \crcr
R_2=& 2c\left(c+1\right)\Big(c(c^2+1)N_3^3-\left(3c^4+c^3+8c^2+c+3\right)N_3^2\crcr
& \qquad +\left(5c^4+16c^2+5\right)N_3-6c(c^2+1)\Big) \,, \crcr
R_3=&3c^2(c^2+1)^2N_3^4-3c\left(2c^6+11c^4+2c^3+11c^2+2\right)N_3^3\crcr
& +\left(c^8+31c^6+12c^5+84c^4+12c^3+31c^2+1\right)N_3^2\crcr
& -c^2\left(14c^4+39c^3+34c^2+39c+14\right)N_3\crcr
& -c^2\left(3c^4-30c^3+14c^2-30c+3\right) \,,\crcr
R_4=& 6c^2\left(c^2+1\right)\left(c+1\right)\left(N_3-1\right)\left(N_3-\frac{2c}{c^2+1}\right)\left(N_3-\frac{c^2+1}{c}\right) \,.
\end{align}
and the signs in front of  $\partial \beta_{p,3}(\bar{g}^{\star})$ and $\partial \beta_{dp}(\bar{g}^{\star})$ are the same as in $\bar{g}_{p,3}^{\star}$ and $\bar{g}_{dp}^{\star}$ while the signs inside the square roots are taken to be simultaneously the same as in \eqref{eq:fpmatrix2}.

For $c \le N_3\leq \frac{c^2+1}{c}$, $\partial \beta_{p,3}(\bar{g}^{\star})$ can be real and positive, but in this case $\partial \beta_{dp}(\bar{g}^{\star})$ is purely imaginary.
However, for $N_3 > \frac{c^2+1}{c}$, $\partial \beta_{p,3}(\bar{g}^{\star})$ and $\partial \beta_{dp}(\bar{g}^{\star})$ are both complex and we can choose the sign in front so that the real part is positive. 

\medskip

Summarizing the findings of this subsection: we find \emph{no} real stable fixed point in the matrix-like large-$N$ limit; however, for $N_3>\frac{c^2+1}{c}$ we do find a complex infrared fixed point  stable in all five directions.

\subsection{Tensor-like limit}
\label{sec:short-range_triple}
We finally consider the large-$N$  limit with:
	\begin{equation}
		N_1 \, = \, N_2  = \, N_3  =  \, N \, , \quad {\rm and} \quad N \, \to \, \infty \, .
	\end{equation}
While we could, like in the previous sections, consider an inhomogeneous scaling with ratios different from one (e.g.\ $N_1/N_2=c$) this leads to very bulky formulas, with not much qualitative gain. We will thus stick to the homogeneous case. 

The resulting $O(N)^3$ model has been studied at leading order in $1/N$ \cite{Giombi:2017dtl}. Here we analyze the fate of its fixed points at subleading orders in $1/N$. We combine the three pillow couplings into one coupling $\tilde{g}_p/3=\tilde{g}_{p,1}=\tilde{g}_{p,2}=\tilde{g}_{p,1}$, thus endowing the model with a discrete color permutation symmetry. 
After scaling the couplings as:
	\begin{equation} \label{eq:bar_g}
		\gt \, = \, \frac{\bar{g}}{N^{3/2}} \, , \qquad \gt_p \, = \, \frac{\bar{g}_p}{N^2} \, , \qquad \gt_d \, = \, \frac{\bar{g}_d}{N^3} \, , 
	\end{equation}
the two-loop beta functions up to order $\cO(N^{-3/2})$ are:
	\begin{align} \label{eq:beta_t-SR}
		\bar{\beta}_t \, &= \, - \, \epsilon\bar{g} \, + \, 2\bar{g}^3 \, +\frac{2\bar{g}\bar{g}_p}{3N}\Big[6-\bar{g}_p\Big] \, + \, \mathcal{O}(N^{-3/2}) \, , \crcr
		\bar{\beta}_{p} \, &= \, - \, \epsilon\bar{g}_{p} \, + \, 6\bar{g}^2 +  \, \frac{2\bar{g}_{p}^2}{3}
		\, - \, 2\bar{g}^2 \bar{g}_{p}   + \, \frac{8\bar{g}}{N^{1/2}} \Big[  \bar{g}_{p} - 3\bar{g}^2  \Big]  \crcr
		& \quad + \frac{2}{9N} \Big[5\bar{g}_p^2\left(3-\bar{g}_p\right)+54\bar{g}^2\left(1-2\bar{g}_p\right)\Big]\, + \, \mathcal{O}(N^{-3/2}) \, , \crcr
		\bar{\beta}_d \, &= \, - \, \epsilon\bar{g}_d \, + \,\frac{2}{3} \left(  3\bar{g}_d^2 + 6 \, \bar{g}_{p} \bar{g}_d + 2 \bar{g}_{p}^2 \right)  - 2\bar{g}^2 \left( 5\bar{g}_d + 4 \bar{g}_{p} \right) \crcr
	& \quad \, + \frac{4\bar{g}}{N^{1/2}} \Big[ \bar{g}_{p} +  3\bar{g}_d -3\bar{g}^2  \Big] \, \crcr
	& \quad +\frac{2\bar{g}_p}{9N}\Big[18\bar{g}_d+9\bar{g}_p-15\bar{g}_d\bar{g}_p-14\bar{g}_p^2-36\bar{g}^2 \Big] \, + \, \mathcal{O}(N^{-3/2}) \, .
	\end{align}
At leading order we reproduce the results obtained in \cite{Giombi:2017dtl}.

We remark that, while at leading order the tetrahedral beta function has no quadratic term, such a term appears at the first non-zero subleading order, $N^{-1}$. In order to better understand the implications of this, consider the fictitious single-coupling beta function $-\epsilon g + g^3 +\frac{2 a}{N} g^2$, with $a$ some real constant; its fixed points are:
\begin{equation} \label{eq:ex-FP}
g_{\star,\pm} = -\frac{a}{N} \pm \sqrt{\epsilon+\frac{a^2}{N^2}} \,.
\end{equation}
If we expand at large $N$, we find:
\be
g_{\star,\pm} = \pm \sqrt{\epsilon} -\frac{a}{N} \pm \frac{a^2}{2\sqrt{\epsilon} N^2} + \cO(N^{-3})\,,
\ee
and naively we seem to have a problem: the subleading orders are non-perturbative and even blow up for $\epsilon\to 0$. We would thus conclude that the $\sqrt{\epsilon}$ fixed point is spurious.
However, a more careful look reveals that the behavior of the fixed points \eqref{eq:ex-FP} is actually governed by the combination $\epsilon N^2$.
For $\epsilon N^2 \ll 1$ (towards the finite-$N$ range), the usual one-loop-driven Wilson-Fisher fixed point is obtained, $g_{\star,+}\sim N\epsilon/2a$. For $\epsilon N^2 \gg 1$, one gets instead the two-loop-driven fixed points typical of the $O(N)^3$ model in the melonic limit \cite{Giombi:2017dtl}, $\bar{g}_{\star,\pm}\sim \pm \sqrt{\epsilon}$. As we wish to study the $1/N$ corrections to the melonic leading order, we need to assume $\epsilon N^2 \gg 1$. To this end, we set:
\be
N=\tilde{N}/\sqrt{\epsilon}\,,
\ee
and we expand beta functions, fixed points and critical exponents in $1/\tilde{N}$ first, and only afterwards in $\epsilon$. 
The beta functions become:
\begin{align}
		\bar{\beta}_t \, &= \, - \, \epsilon\bar{g} \, + \, 2\bar{g}^3 \, +\frac{2\sqrt{\epsilon}\bar{g}\bar{g}_p}{3\tilde{N}}\Big[6-\bar{g}_p\Big] \, + \, \mathcal{O}(\tilde{N}^{-3/2}) \, , \crcr
		\bar{\beta}_{p} \, &= \, - \, \epsilon\bar{g}_{p} \, + \, 6\bar{g}^2 +  \, \frac{2\bar{g}_{p}^2}{3}
		\, - \, 2\bar{g}^2 \bar{g}_{p}   + \, \frac{8\epsilon^{1/4}\bar{g}}{\tilde{N}^{1/2}} \Big[  \bar{g}_{p} - 3\bar{g}^2  \Big]  \crcr
		& \quad + \frac{2\sqrt{\epsilon}}{9\tilde{N}} \Big[5\bar{g}_p^2\left(3-\bar{g}_p\right)+54\bar{g}^2\left(1-2\bar{g}_p\right)\Big]\, + \, \mathcal{O}(\tilde{N}^{-3/2}) \, , \crcr
		\bar{\beta}_d \, &= \, - \, \epsilon\bar{g}_d \, + \,\frac{2}{3} \left(  3\bar{g}_d^2 + 6 \, \bar{g}_{p} \bar{g}_d + 2 \bar{g}_{p}^2 \right)  - 2\bar{g}^2 \left( 5\bar{g}_d + 4 \bar{g}_{p} \right) \crcr
	& \quad \, + \frac{4\epsilon^{1/4}\bar{g}}{\tilde{N}^{1/2}} \Big[ \bar{g}_{p} +  3\bar{g}_d -3\bar{g}^2  \Big] \,  \crcr
	& \quad +\frac{2\sqrt{\epsilon}\bar{g}_p}{9\tilde{N}}\Big[18\bar{g}_d+9\bar{g}_p-15\bar{g}_d\bar{g}_p-14\bar{g}_p^2+36\bar{g}^2 \Big] \, + \, \mathcal{O}(\tilde{N}^{-3/2}) \, . 
	\end{align}
	
We parametrize the critical couplings as:
\begin{align}
		\bar{g}^\star \,& = \, \bar{g}^\star_{(0)} \, + \, \tilde{N}^{-\frac{1}{2}} \, \bar{g}^\star_{(1)} \, +\tilde{N}^{-1}\bar{g}^{\star}_{(2)}\, + \, \mathcal{O}(\tilde{N}^{-3/2}) \, , \crcr
\bar{g}^\star_p \, &= \, \bar{g}^\star_{p,(0)} \, + \, \tilde{N}^{-\frac{1}{2}} \, \bar{g}^\star_{p,(1)} \, +\tilde{N}^{-1}\bar{g}^{\star}_{p,(2)}\, + \, \mathcal{O}(\tilde{N}^{-3/2}) \, , \crcr
\bar{g}^\star_d \, &= \, \bar{g}^\star_{d,(0)} \, + \, \tilde{N}^{-\frac{1}{2}} \, \bar{g}^\star_{d,(1)} \, +\tilde{N}^{-1}\bar{g}^{\star}_{d,(2)}\, + \, \mathcal{O}(\tilde{N}^{-3/2}) \, .
\label{eq:param_couplings}
	\end{align}
Solving for the zeros of the beta functions at leading order we find the following complex solutions:\footnote{There are also additional four solutions with zero tetrahedral coupling:
\[
(\bar{g}_{p,(0)}^{\star},\bar{g}_{d,(0)})=\{(0,0),(0,\frac{\epsilon}{2}),(\frac{3\epsilon}{2},-\frac{3\epsilon}{2}),(\frac{3\epsilon}{2},-\epsilon)\} \,.
\]
We do not study them further as we are interested in fixed points with non-zero tetrahedral coupling.}
	\begin{equation} \label{eq:LO-FP-SR}
		\bar{g}^\star_{(0)} \, = \, \pm \, \sqrt{ \frac{\epsilon}{2}} \, , \qquad 
		\bar{g}_{p,(0)}^\star \, = \, \pm \, 3 i \sqrt{ \frac{\epsilon}{2}} \, +\frac{3\epsilon}{2} +\mathcal{O}(\epsilon^{3/2})\, , \qquad 
		\bar{g}_{d,(0)}^\star \, = \, \mp \,  i \sqrt{ \frac{\epsilon}{2}}(3\pm \sqrt{3}) +\mathcal{O}(\epsilon^{3/2})\, .
	\end{equation}
There are eight solutions by the combination of the signs in $\bar{g}^\star_{(0)}$, $\bar{g}_{p,(0)}^\star$ and the relative sign of $3$ and $\sqrt{3}$ in $\bar{g}_{d,(0)}^\star$.
The overall sign of $\bar{g}_{d,(0)}^\star$ is associated with the sign of $\bar{g}_{p,(0)}^\star$.

Next we compute the subleading $\mathcal{O}(\tilde{N}^{-1/2})$ corrections to the fixed points by substituting \eqref{eq:param_couplings}, with the leading order given by \eqref{eq:LO-FP-SR}, into the beta functions and solving for the $\mathcal{O}(\tilde{N}^{-1/2})$ order. Since $\bar{\beta}_t$ does not have a $\mathcal{O}(\tilde{N}^{-1/2})$ contribution, the equation for $\bar{g}^\star_{(1)}$ comes only from the leading order part, evaluated at linear order in the coupling correction, leading to:
	\begin{equation}
		\bar{g}^\star_{(1)} \, = \, 0 \, , \qquad ({\rm for} \ \epsilon \, \ne \, 0) \,.
	\end{equation}
For the pillow and double-trace we find instead non-trivial corrections: 
\begin{align}
		\bar{g}^\star_{p,(1)} \, = \, \mp \, 3 \sqrt{2}\epsilon^{3/4} \, , \qquad 
		\bar{g}^\star_{d,(1)} \, = \, \pm \, 3 \frac{\epsilon^{3/4}}{\sqrt{2}} \, .
	\end{align}
The choice of upper or lower sign for $\bar{g}^\star_{p,(1)}$ and $\bar{g}^\star_{d,(1)}$ is synchronized with that for $\bar{g}_{(0)}^\star$.
At next order, $\tilde{N}^{-1}$, we find:\footnote{We checked the next two orders, and we have found that the order $\tilde{N}^{-3/2}$ starts again at order $\epsilon^{3/4}$, and order $\tilde{N}^{-2}$ at order $\sqrt{\epsilon}$. We do not know whether this pattern repeats to all orders.}
\begin{align}
		\bar{g}^{\star}_{(2)} \, &= \, \pm 3i\sqrt{\epsilon}  \mp \frac{9\epsilon}{2\sqrt{2}}+ \mathcal{O}(\epsilon^{3/2})\, , \crcr
		\bar{g}^\star_{p,(2)} \, &= \,  9\sqrt{\epsilon} \mp \frac{33i \epsilon}{\sqrt{2}}+ \mathcal{O}(\epsilon^{3/2}) \, , \crcr
		\bar{g}^\star_{d,(2)} \, &= \, - \, 3 \sqrt{3\epsilon}\left(\sqrt{3}\pm 1\right) \pm \frac{3\sqrt{2}i\epsilon}{2}\left(5\pm 2\sqrt{3}\right)+ \mathcal{O}(\epsilon^{3/2}) \, .
\end{align}
where the first sign in $\bar{g}^{\star}_{(2)}$ is the upper one when the choice of sign was the same for $\bar{g}^{\star}_{(0)}$ and $\bar{g}^{\star}_{p,(0)}$, the second sign in $\bar{g}^{\star}_{(2)}$ is synchronized with the sign of $\bar{g}^{\star}_{(0)}$, the signs in $\bar{g}^{\star}_{p,(2)}$ and in front of the $\epsilon$ term in $\bar{g}^{\star}_{d,(2)}$ are synchronized with the sign of $\bar{g}^{\star}_{p,(0)}$ and the signs in the parenthesis in $\bar{g}^\star_{d,(2)}$ are synchronized with the sign in $\bar{g}^\star_{d,(0)}$.

We then compute the critical exponents up to order $\tilde{N}^{-1}$ to find:
\begin{align}
\omega_t&=2\epsilon \mp \frac{6 i \sqrt{2}\epsilon}{\tilde{N}}+\mathcal{O}(\epsilon^{3/2},\tilde{N}^{-3/2})  \,,\crcr
\omega_p&=\pm 2 i \sqrt{2\epsilon} + 12\frac{\sqrt{\epsilon}\mp i \sqrt{2}\epsilon}{\tilde{N}} +\mathcal{O}(\epsilon^{3/2},\tilde{N}^{-3/2}) \,, \crcr
\omega_d&= \pm 2i \sqrt{6\epsilon} \mp 12\sqrt{3} \frac{\sqrt{\epsilon}\mp i \sqrt{2}\epsilon}{\tilde{N}} +\mathcal{O}(\epsilon^{3/2},\tilde{N}^{-3/2})  \,.
\end{align}

The sign of the leading order in $\omega_p$ is the same one as in $\bar{g}_{p,(0)}^{\star}$. The sign of the leading order in $\omega_d$ is the upper one if the choices of signs in $\bar{g}_{p,(0)}^{\star}$ and $\bar{g}_{d,(0)}^{\star}$ are different and the lower one if they are the same. The sign in front of the $\tilde{N}^{-1}$ term of $\omega_d$ is synchronized with $\bar{g}_{d,(0)}^{\star}$. The other signs for the orders $\tilde N^{-1}$ are synchronized with $\bar{g}_{p,(0)}^{\star}$. In particular, for the fixed points with the lower choice of sign in $\bar{g}_{d,(0)}^{\star}$, the real parts of all three critical exponents are positive. 

\medskip

In conclusion, the complex fixed point of the short-range $O(N)^3$ model found in \cite{Giombi:2017dtl} persists at subleading orders in $1/N$. Importantly, the order $\tilde{N}^{-1/2}$ corrections to the critical exponents are zero, but the order $\tilde{N}^{-1}$ endow them with a real part, meaning that the fixed point is infrared stable.

\section{The long-range trifundamental model}
\label{sec:long-range}

\subsection{The long-range multi-scalar model}
\label{sec:long-range_ms}
We now consider the long-range multi-scalar model with quartic interactions introduced in chapter \ref{chap:3loops} with action \eqref{eq:actionms}. For simplicity, we will consider here the beta functions only up to two loops. We then have:
\begin{equation}
\beta_{\mba \mbb \mbc \mbd} = -\epsilon \gt_{\mba \mbb \mbc \mbd} + \alpha_{D}\left(\gt_{\mba \mbb \mbe \mbf}\gt_{\mbe \mbf \mbc \mbd} + 2 \textrm{ terms} \right)  + \alpha_{S}\left(\gt_{\mba \mbb \mbe \mbf}\gt_{\mbe \mbg \mbh \mbc}\gt_{\mbf \mbg \mbh \mbd}+ 5 \textrm{ terms}\right) 
\label{eq:beta_abcd_alpha}\,,
\end{equation}
\begin{equation}
\beta^{(2)}_{\mbc\mbd}
= - (d-2\Delta_{\phi} ) \rt_{\mbc \mbd} +\alpha_{D}  \big( \rt_{ \mbe \mbf}\gt_{\mbe \mbf \mbc \mbd} \big) + \alpha_{S} \big(\rt_{\mbe \mbf} \gt_{\mbe \mbg \mbh \mbc}\gt_{\mbf \mbg \mbh \mbd} \big) \,. 
\label{eq:beta2_abcd_alpha}
\end{equation}
The alpha coefficients were defined in \eqref{eq:alphas3}.

\subsection{Large-$N$ expansion of the long-range $O(N)^3$ tensor model}
\label{sec:long-range_equalN}
We now set $N_1=N_2=N_3=N$ and study the fixed points of the long-range $O(N)^3$ model at next-to-leading order in $1/N$. We use:
	\begin{equation}
		\gt_{\mba \mbb \mbc \mbd} \, = \, \gt \left(\delta^t_{\mba \mbb \mbc \mbd} + 5 \textrm{ terms} \right)
		\, + \,  \gt_{p} \left(\delta^{p}_{\mba \mbb; \mbc \mbd} + 5 \textrm{ terms} \right)
		\, + \, 2\gt_d \left(\delta^d_{\mba \mbb \mbc \mbd} + 2 \textrm{ terms} \right) \, ,
	\label{eq:couplingtot}
	\end{equation}
where like before each $\mba$ is a triplet of indices $\mba=(a_1,a_2,a_3)$.
$\delta^t_{\mba \mbb \mbc \mbd}$ and $\delta^d_{\mba \mbb \mbc \mbd}$ are defined as before, and 
\begin{equation}
\delta^p_{\mba \mbb ;\mbc \mbd}=\frac{1}{3}\sum_{i=1}^3 \delta^{p,i}_{\mba \mbb ; \mbc \mbd} \, .
\end{equation}

The beta functions up to two-loops are then:
\begin{align}
\beta_t=&-\epsilon \tilde{g}+\frac{4\alpha_D}{3}\Big[2\tilde{g}_p^2+18\tilde{g}\tilde{g}_d+3(N+1)\tilde{g}\tilde{g}_p\Big]\crcr
& +\frac{4\alpha_S}{9}\Big[27(3N+2)\tilde{g}^3 + 54 \left(N^3+14\right) \tilde{g}_d^2 \tilde{g} +3\left(N^3+9 N^2+51 N+53\right) \tilde{g}_p^2 \tilde{g} \crcr
& \qquad \qquad + 2\left(2 N^2+13 N+24\right)\tilde{g}_p^3 +18\left(2 N^2+5 N+14\right) \tilde{g}_p \tilde{g}^2  \crcr
& \qquad \qquad  +36\tilde{g}_d\left(4\tilde{g}_p^2+9N\tilde{g}^2+3(N^2+3N+3)\tilde{g}\tilde{g}_p\right)\Big] \, , \crcr
\beta_p=& -\epsilon \tilde{g}_p + \frac{2\alpha_D}{3}\Big[36\tilde{g}_p\tilde{g}_d+3(N+2)\left(3\tilde{g}+4\tilde{g}_p \right)\tilde{g}+(N^2+5N+12)\tilde{g}_p^2\Big] \crcr
& +\frac{4\alpha_S}{9}\Big[54(N^2+N+4)\tilde{g}^3 +54 \left(N^3+14\right) \tilde{g}_d^2 \tilde{g}_p +18\left(5 N^2+19 N+30\right) \tilde{g}_p^2 \tilde{g} \crcr
& \qquad \qquad  + \left(4 N^3+27 N^2+135 N+179\right)\tilde{g}_p^3 +9  \left(N^3+6 N^2+51 N+50\right)\tilde{g}_p \tilde{g}^2 \crcr
& \qquad \qquad +36 \tilde{g}_d \left( \left(4 N^2+8 N+15\right)\tilde{g}_p^2+3 (7 N+8) \tilde{g}_p \tilde{g} +9(N+2) \tilde{g}^2 \right)  \Big] \, , \crcr
\beta_d=& -\epsilon \tilde{g}_d+\frac{2\alpha_D}{3}\Big[ 3\left(N^3+8\right) \tilde{g}_d^2 +6\left(N^2+N+1\right) \tilde{g}_d \tilde{g}_p +18N \tilde{g}_d \tilde{g} + (2N+3)\tilde{g}_p^2+ 6 \tilde{g}\tilde{g}_p \Big] \crcr
& +\frac{4\alpha_S}{9}\Big[ 27N \tilde{g}^3+ 216 \tilde{g}_d^2 \left( \left(N^2+N+1\right)\tilde{g}_p+3N \tilde{g} \right)+18  \left(5 N^3+22\right)\tilde{g}_d^3 \crcr
& \qquad \qquad +9 \tilde{g}_d \left(\left(N^3+3 N^2+17 N+17\right)\tilde{g}_p^2 +12  \left(N^2+N+3\right)\tilde{g}_p \tilde{g}+3  \left(N^3+3 N+2\right)\tilde{g}^2\right)\crcr
& \qquad \qquad +72(N+1) \tilde{g}_p^2 \tilde{g}+7  \left(N^2+3 N+5\right)\tilde{g}_p^3+18 \left(N^2+N+4\right) \tilde{g}_p \tilde{g}^2\Big] \, , \crcr
\beta^{(2)}=& -(d-2\Delta_{\phi})\tilde{r}+2\alpha_D\tilde{r}\Big[3N\tilde{g}+(N^2+N+1)\tilde{g}_p+(N^3+2)\tilde{g}_d\Big] \crcr
& +2\alpha_S\tilde{r}\Big[36N\tilde{g}\tilde{g}_d+12(N^2+N+1)\tilde{g}_p\left(\tilde{g}_d+\tilde{g}\right)+6(N^3+2)\tilde{g}_d^2+3(N^3+3N+2)\tilde{g}^2\crcr
& \qquad \qquad +(N^3+3N^2+9N+5)\tilde{g}_p^2\Big] \, . 
\end{align}

\paragraph{Fixed points}

We rescale the couplings as:
\begin{equation}
\tilde{g}=\frac{\bar{g}}{N^{3/2}} \,, \; \tilde{g_{p}}=\frac{\bar{g}_{p}}{N^2} \,, \; \tilde{g_d}=\frac{\bar{g}_d}{N^3} \,,
\end{equation}
and first consider the large-$N$ limit. The couplings $\bar{g},\bar{g}_p$ and $\bar{g}_d$ are then the same as in chapter \ref{chap:CTKT}. 

In chapter \ref{chap:CTKT}, we showed that at $\epsilon=0$ the tetrahedron coupling $\bar{g}$ is exactly marginal in the large-$N$ limit, 
and it parametrizes a line of fixed points for the remaining two couplings. The exact marginality is due to the fact that at large $N$ the tetrahedron receives no radiative corrections, and moreover in the long-range case there is no wave function renormalization. The latter is responsible for the $2\bar{g}^3$ term in \eqref{eq:beta_t-SR}, which is absent in the long-range case.
However, at order $N^{-1}$ the tetrahedron beta function is non-zero also in the long-range model, and excluding uncontrolled non-perturbative fixed points, the line of fixed points collapses to the trivial fixed point at vanishing couplings.
Turning on $\epsilon$ does not help, as it contributes a term $-\epsilon \bar{g}$, that being the only term of order $N^0$, leads to $\bar{g}^{\star}=0$ already at leading order. As we did before, it is instructive to consider again a fictitious single-coupling beta function to guide our understanding; the situation we have in the long-range model, at $\epsilon\neq 0$ and at next-to-leading order in $1/N$, is captured by a beta function of the form $-\epsilon \bar{g} + \bar{g}^2/N$. Its fixed points are the trivial one, and $\bar{g}^{\star}= N\epsilon$, which goes to infinity if we take $N\to\infty$ at fixed $\epsilon$.
Similarly to what we have seen in the short-range case, the problem is resolved by specifying how small should $\epsilon$ be in comparison to  $1/N$. In particular, it is clear that we now need $ N \epsilon \ll 1$. In other words, we should move the $-\epsilon \bar{g}$ term to the first non-trivial order in $1/N$, by setting
\be
\epsilon=\frac{\tilde{\epsilon}}{N}\,,
\ee
and expanding as before in $1/N$ first, and then in $\tilde{\epsilon}$.
Notice that the condition $ N \epsilon \ll 1$ is compatible with the $N \sqrt{\epsilon}\gg 1$ condition which we had in the short-range case. Of course the meaning of $\epsilon$ is different in the two cases, but in practice their role is similar.
We also note that a similar tuning of $\epsilon$ and $N$ was considered in \cite{Fleming:2020qqx} in order to find a finite-$N$ precursor of the large-$N$ line of fixed points in the $O(N)$ model with $(\phi^2)^3$ interaction.

To simplify the computations we use the same two independent couplings as in chapter \ref{chap:CTKT}:

\begin{equation} \label{eq:g_1,2}
\bar{g}_1=\frac{\bar{g}_p}{3} \,, \; \bar{g}_2=\bar{g}_d+\bar{g}_p\,.
\end{equation}

Parametrizing the coefficients of the $\epsilon$ expansion  of the one- and two-loop constants $\alpha$ as: 
\begin{align}
		\alpha_D \, &= \, 1\, + \, \alpha_{D,1} \, \epsilon \, + \, \alpha_{D,2} \, \epsilon^2 \, + \, \mathcal{O}(\epsilon^3) \, ,\crcr
		\alpha_S \, &= \, \alpha_{S,0} \, + \, \alpha_{S,1} \, \epsilon \,  + \, \mathcal{O}(\epsilon^2) \, ,
	\label{eq:alpha param}
	\end{align}
the beta functions at two loops up to order $N^{-1}$ are:

\begin{align}
\beta_t \, = & \, \frac{\bar{g}}{N}\Big[12\bar{g_1}\left(1+\alpha_{S,0} \bar{g_1} \right) -\tilde{\epsilon}\Big] +\mathcal{O}(N^{-3/2})\,, \crcr
\beta_1=& 2\left(\bar{g}_1^2+\bar{g}^2\right)+4\alpha_{S,0}\bar{g}_1\bar{g}^2 +\frac{8\bar{g}}{N^{1/2}} \Big[ \bar{g}_1 + \alpha_{S,0} \bar{g}^2 \Big]  \crcr 
&+\frac{1}{N}\bigg[  10\bar{g}_1^2+4\bar{g}^2+8\alpha_{S,0}\bar{g}_1\left(2\bar{g}_1^2+3\bar{g}^2\right) +\tilde{\epsilon}\left(2\alpha_{D,1}(\bar{g}_1^2+\bar{g}^2)+\bar{g}_1(4\alpha_{S,1}\bar{g}^2-1)\right)\bigg] +\mathcal{O}(N^{-3/2})\,, \crcr
\beta_2=& 2\left(\bar{g_2}^2+3\bar{g}^2\right)+12\alpha_{S,0}\bar{g_2}\bar{g}^2 +\frac{12\bar{g}}{N^{1/2}} \Big[ \bar{g_2} + 3\alpha_{S,0} \bar{g}^2 \Big] \crcr 
&+\frac{1}{N} \bigg[ 12\left(\bar{g_1}^2+\bar{g}^2+\bar{g_1}\bar{g_2}\right)+12\alpha_{S,0}\bar{g_1}\left(2\bar{g_1}^2+3\bar{g_2}\bar{g_1}+8\bar{g}^2\right)\crcr
& +\tilde{\epsilon}\left(2\alpha_{D,1}(\bar{g}_2^2+3\bar{g}^2)+\bar{g}_2(12\alpha_{S,1}\bar{g}^2-1)\right)
 \bigg] +\mathcal{O}(N^{-3/2})\,, \crcr
\beta^{(2)}=&-\frac{d}{2}\tilde{r}+2\left(2\bar{g}_2+3\alpha_{S,0}\bar{g}^2\right)\tilde{r} +\frac{6\bar{g}\tilde{r}}{N^{1/2}}\crcr
& + \frac{\tilde{r}}{N}\bigg[6\bar{g}_1\left(1+3\alpha_{S,0}\bar{g}_1\right)+\frac{\tilde{\epsilon}}{2}\left(4\alpha_{D,1}\bar{g}_2+12\alpha_{S,1}\bar{g}^2-1\right)\bigg]+\mathcal{O}(N^{-3/2})\,.
\label{eq:beta-LR}
\end{align}

We then parametrize the critical couplings as:
\begin{align}
		\bar{g}^{\star} \, &= \,  \, \bar{g}^{\star}_{(0)} \, + \, \bar{g}^{\star}_{(1)} N^{-1/2}  + \, \mathcal{O}(N^{-1}) \, , 
		\crcr
		\bar{g}_{1}^{\star} \, &= \, \bar{g}^{\star}_{1,(0)} \, + \,  \, \bar{g}^{\star}_{1,(1)} N^{-1/2}  + \, \mathcal{O}(N^{-1}) \, , \crcr
		\bar{g}_2^{\star} \, &= \, \bar{g}^{\star}_{2,(0)} \, + \,  \, \bar{g}^{\star}_{2,(1)} N^{-1/2} \, + \, \mathcal{O}(N^{-1}) \, . 
		\label{eq:g-parametrization} 
\end{align}

\paragraph{Leading-order.}

As we already discussed, at leading order (i.e.\ $N^0$), the tetrahedron beta function is identically zero, hence $\bar{g}^{\star}_{(0)}$ is a free parameter.
For the other two couplings, the leading-order fixed points, expanded to second order in $\bar{g}^{\star}_{(0)}$, are:
\begin{align}
\bar{g}^{\star}_{1,(0)}&=\pm \sqrt{-\bar{g}^{\star}_{(0)}{}^2}-\bar{g}^{\star}_{(0)}{}^2\alpha_{S,0}+ \mathcal{O}(\bar{g}^{\star}_{(0)}{}^3)\,, \crcr
\bar{g}^{\star}_{2,(0)}&=\pm \sqrt{3}\sqrt{-\bar{g}^{\star}_{(0)}{}^2} -3\bar{g}^{\star}_{(0)}{}^2\alpha_{S,0}+ \mathcal{O}(\bar{g}^{\star}_{(0)}{}^3)\,.
\label{eq:LO_pillow_dt}
\end{align}
They correspond to the lines of fixed points found at large $N$ in chapter \ref{chap:CTKT}. For small coupling $|\bar{g}^{\star}_{(0)}|$, $\bar{g}^{\star}_{1,(0)}$ and $\bar{g}^{\star}_{2,(0)}$ are complex for real $\bar{g}^{\star}_{(0)}$ and real for purely imaginary $\bar{g}^{\star}_{(0)}$.

\paragraph{Next-to-leading order.}

Substituting \eqref{eq:g-parametrization} and \eqref{eq:LO_pillow_dt} into the beta functions \eqref{eq:beta-LR} and solving for fixed points at order $N^{-1/2}$ we find $\bar{g}^{\star}_{1,(1)}$ and $\bar{g}^{\star}_{2,(1)}$ in terms of $\bar{g}^{\star}_{(0)}$ and $\bar{g}^{\star}_{(1)}$:

\begin{align}
\bar{g}^{\star}_{1,(1)}&=-2\bar{g}^{\star}_{(0)}-2\bar{g}^{\star}_{(0)}\bar{g}^{\star}_{(1)}\alpha_{S,0}\mp \frac{\bar{g}^{\star}_{(0)}\bar{g}^{\star}_{(1)}}{\sqrt{-\bar{g}^{\star}_{(0)}{}^2}} + \mathcal{O}(\bar{g}^{\star}_{(0)}{}^3)\,, \crcr
\bar{g}^{\star}_{2,(1)}&=-3\bar{g}^{\star}_{(0)}-6\bar{g}^{\star}_{(0)}\bar{g}^{\star}_{(1)}\alpha_{S,0}\mp\frac{\sqrt{3}\bar{g}^{\star}_{(0)}\bar{g}^{\star}_{(1)}}{\sqrt{-\bar{g}^{\star}_{(0)}{}^2}} + \mathcal{O}(\bar{g}^{\star}_{(0)}{}^3) \,.
\end{align}
The signs in the two sets $\{\bar{g}^{\star}_{1,(0)},\bar{g}^{\star}_{1,(1)}\}$ and $\{\bar{g}^{\star}_{2,(0)},\bar{g}^{\star}_{2,(1)}\}$ are taken to be simultaneously either the upper or lower ones so that we still have four choices of sign. 

\paragraph{Fixing the tetrahedron coupling.}

Since the beta function of the tetrahedron is still zero at order $N^{-1/2}$, it would seem that our lines of fixed points have become surfaces (that is parametrized by two free parameters $\bar{g}^{\star}_{ (0)},\bar{g}^{\star}_{ (1)}$). On the other hand, if we homogeneously truncate all the beta functions at this order, there is no real justification for the expansion of $\bar{g}^{\star} $ in \eqref{eq:g-parametrization}; this is only justified at higher orders, as all the orders $N^{-n/2}$ with $n\geq 2$ in the tetrahedron beta function are non-trivial.
In the spirit of a $1/N$ expansion, as opposed to a strict $N\to\infty$ limit, it is more consistent to keep the same number of non-trivial orders for each beta function regardless of their different scaling in $N$. By doing so, we will be able to fix $\bar{g}^{\star}_{(0)}$ and $\bar{g}^{\star}_{(1)}$.

Substituting \eqref{eq:LO_pillow_dt} into the order $N^{-1}$ of the tetrahedron beta function,  we fix $\bar{g}^{\star}_{(0)}$. Besides the trivial solution, we find:
\begin{equation}
\bar{g}^{\star}_{(0)}=\pm\frac{1}{2\alpha_{S,0}}\sqrt{2 \pm \frac{ 6+\tilde{\epsilon}\alpha_{S,0}}{\sqrt{3(3+\tilde{\epsilon}\alpha_{S,0})}}}\,.
\label{eq:solg0s}
\end{equation}
The choice of signs is independent of the choices for the previous solutions. 

We are interested in purely imaginary solutions, as at leading order this gives real critical exponents (see section \ref{sec:line}), and a real spectrum of bilinear operators, with real OPE coefficients (see section \ref{sec:unitarity}). The solutions with a plus sign inside the square root have a non-zero real part for all values of $\tilde{\epsilon}$, in particular remain finite for $\tilde{\epsilon}\to 0$, and thus they are not to be trusted in our perturbative expansion. The solutions with a minus sign instead are purely imaginary for $\tilde{\epsilon}<-3/\alpha_{S,0}$ (notice this bound is positive as $\alpha_{S,0}$ is negative), and they vanish for $\tilde{\epsilon}\to 0$. 

In this case, we can  expand $\bar{g}^{\star}_{(0)} $ for small $\tilde{\epsilon}$, finding:
\begin{equation}
\bar{g}^{\star}_{(0)}=\pm \frac{i }{12}\left(\tilde{\epsilon}-\frac{\alpha_{S,0}}{6}\tilde{\epsilon}^2\right) +\mathcal{O}(\tilde{\epsilon}^3) \,.
\label{eq:gt0_epsilon}
\end{equation}
We can also expand $\bar{g}_{1,(0)}$ and $\bar{g}_{2,(0)}$ in $\tilde{\epsilon}$:
\begin{align}
\bar{g}_{1,(0)}&= \pm \frac{1}{12}\left(\tilde{\epsilon}-\frac{\alpha_{S,0}}{12}(2\mp1)\tilde{\epsilon}^2\right) +\mathcal{O}(\tilde{\epsilon}^3)\,, \crcr
\bar{g}_{2,(0)}&=\pm \frac{1}{4\sqrt{3}}\left(\tilde{\epsilon}-\frac{\alpha_{S,0}}{12}(2\mp\sqrt{3})\tilde{\epsilon}^2\right) +\mathcal{O}(\tilde{\epsilon}^3) \,,
\label{eq:g1-g2_epsilon}
\end{align}
where the global sign and the one inside the brackets are taken to be simultaneously either the upper or lower ones. 

The $N^{-1/2}$ correction $\bar{g}^{\star}_{(1)}$ is still a free parameter at this order. In order to fix it we need to consider the $N^{-3/2}$ contribution to $\beta_t$, which we have not displayed in \eqref{eq:beta-LR}.
This is easily obtained from the general multi-scalar results of chapter \ref{chap:3loops}, from which we find:
\begin{equation}
\beta_t \, =  \, \frac{\bar{g}}{N}\Big[12\bar{g}_1\left(1+\alpha_{S,0} \bar{g_1} \right) -\tilde{\epsilon}\Big]+\frac{48}{N^{3/2}}\alpha_{S,0}\bar{g}_1\bar{g}^2 +\mathcal{O}(N^{-2}) \,.
\end{equation}
Substituting the coupling $1/N$ expansions from \eqref{eq:g-parametrization}, the order $N^{-3/2}$ of $\beta_t $ is:
\begin{equation}
-6\tilde{\epsilon}\bar{g}_{(1)}+72\bar{g}_{1,(1)}\left(\bar{g}_{(0)}+2\bar{g}_{1,(0)}\bar{g}_{(0)}\alpha_{S,0}\right)+72\bar{g}_{1,(0)}\left(\bar{g}_{(1)}+4\bar{g}_{(0)}^2\alpha_{S,0}+\bar{g}_{(1)}\bar{g}_{1,(1)}\alpha_{S,0}\right) \, ,
\end{equation}
and substituting the values of $\bar{g}_{1,(0)}^{\star}$ and $\bar{g}_{1,(1)}^{\star}$, solving for $\bar{g}_{(1)}^{\star}$ in terms of $\bar{g}_{(0)}^{\star}$ we obtain:
\begin{equation}
\bar{g}_{(1)}^{\star}=-\frac{24\bar{g}_{(0)}^{\star}{}^2}{\tilde{\epsilon}\pm\frac{24\bar{g}_{(0)}^{\star}{}^2}{\sqrt{-\bar{g}_{(0)}^{\star}{}^2}}+72\bar{g}_{(0)}^{\star}{}^2\alpha_{S,0}} \,,
\label{eq:gt1_comp}
\end{equation}
where the choice of sign is the same as in $\bar{g}_{1,(0)}^{\star}$.
This expression is real for purely imaginary $\bar{g}_{(0)}^{\star}$. 

The expression \eqref{eq:gt1_comp} comes from a two-loop truncation and thus it should be trusted only up to  order $\tilde{\epsilon}^2$.
Therefore, we first substitute \eqref{eq:gt0_epsilon} in \eqref{eq:gt1_comp} and then expand at order two in $\tilde{\epsilon}$:
\begin{align}
\bar{g}_{(1)}^{\star}= \begin{cases} \frac{1}{6}\left(-\tilde{\epsilon}+\frac{\alpha_{S,0}}{2}\tilde{\epsilon}^2\right) + \mathcal{O}(\tilde{\epsilon}^3) \;\;\; \text{ for the upper choice of sign,} \crcr
\frac{1}{18}\left(\tilde{\epsilon}-\frac{\alpha_{S,0}}{18}\tilde{\epsilon}^2\right) + \mathcal{O}(\tilde{\epsilon}^3) \;\;\; \text{ for the lower choice of sign.} 
\end{cases}
\end{align}
We can now also give the $\tilde{\epsilon}$ expansion of $\bar{g}_{1,(1)}^{\star}$ and $\bar{g}_{2,(1)}^{\star}$.
\begin{align*}
\bar{g}_{1,(1)}^{\star} &= \begin{cases} \mp \frac{i\alpha_{S,0}}{36}\tilde{\epsilon^2} +\mathcal{O}(\tilde{\epsilon}^3) \;\;\;  \text{ for the upper choice of sign in } \bar{g}_{1,(0)}^{\star}\,, \crcr
  \mp \frac{i}{9}\left(\tilde{\epsilon}-\frac{5\alpha_{S,0}}{36}\tilde{\epsilon}^2\right)+\mathcal{O}(\tilde{\epsilon}^3)\;\;\; \text{ for the lower choice of sign in } \bar{g}_{1,(0)}^{\star}\,, \end{cases} \crcr
\bar{g}_{2,(1)}^{\star}&= \begin{cases} \pm \frac{i}{12}\left((-3\pm 2\sqrt{3})\tilde{\epsilon}+\frac{\alpha_{S,0}}{2}(3\mp 2\sqrt{3})\tilde{\epsilon}^2\right)+\mathcal{O}(\tilde{\epsilon}^3) \;\;\;  \text{ for the upper choice of sign in } \bar{g}_{1,(0)}^{\star}\,, \crcr
\pm \frac{i}{36}\left((-9 \mp 2\sqrt{3})\tilde{\epsilon}+\frac{\alpha_{S,0}}{18}(9\pm 2\sqrt{3})\tilde{\epsilon}^2\right)+\mathcal{O}(\tilde{\epsilon}^3) \;\;\;  \text{ for the lower choice of sign in } \bar{g}_{1,(0)}^{\star}\,, \end{cases}
\end{align*}
where the choice of sign in front is the same as for $\bar{g}_{(0)}^{\star}$ and the choice of sign in the parenthesis for $\bar{g}_{2,(1)}^{\star}$ is the same as for $\bar{g}_{2,(0)}^{\star}$.

\paragraph{Critical exponents}

We will now compute the critical exponents. For the quadratic coupling we obtain:
\begin{equation}
\partial  \beta^{(2)}(g^{\star})=-\nu^{-1}=-\frac{d}{2} \pm 2\sqrt{-3\bar{g}^{\star}_{(0)}{}^2} \mp \frac{1}{N^{1/2}}\frac{6\bar{g}^{\star}_{(0)}\bar{g}^{\star}_{(1)}}{\sqrt{-3\bar{g}^{\star}_{(0)}{}^2}} +\mathcal{O}(N^{-1},\bar{g}^{\star}_{(0)}{}^3) \,,
\end{equation}
where the signs are taken to be simultaneously either the upper or lower ones and are the same as for $\bar{g}_{2,(0)}$.

The critical exponents for the quartic couplings are given by:\footnote{They correspond to the diagonal elements as the stability matrix is triangular at order $\mathcal{O}(N^{-1/2})$.}
\begin{align}
\partial \beta_{1}(\bar{g}^{\star})&=\pm \Bigg[ 4\sqrt{-\bar{g}^{\star}_{(0)}{}^2}-\frac{1}{N^{1/2}}\frac{4\bar{g}^{\star}_{(0)}\bar{g}^{\star}_{(1)}}{\sqrt{-\bar{g}^{\star}_{(0)}{}^2}}\Bigg] +\mathcal{O}(N^{-1},\bar{g}^{\star}_{(0)}{}^3)\,, \crcr
\partial \beta_{2}(\bar{g}^{\star})&=\pm \Bigg[ 4\sqrt{-3\bar{g}^{\star}_{(0)}{}^2}-\frac{1}{N^{1/2}}\frac{12\bar{g}^{\star}_{(0)}\bar{g}^{\star}_{(1)}}{\sqrt{-3\bar{g}^{\star}_{(0)}{}^2}}\Bigg] +\mathcal{O}(N^{-1},\bar{g}^{\star}_{(0)}{}^3) \,,
\end{align}
where the signs are taken to be simultaneously either the upper or lower ones in the two sets $\{\bar{g}_{1,(0)},\partial \beta_{1}\}$ and $\{\bar{g}_{2,(0)},\partial \beta_{2}\}$.
At leading order, the stable fixed points are those with the choice of the upper sign in $\bar{g}_{1,(0)}$ and $\bar{g}_{2,(0)}$. There are two such fixed points depending on the choice of sign in $\bar{g}_{(0)}^{\star}$:
\begin{align}
\bar{g}^{\star}&=\pm \frac{i}{12}\left(\tilde{\epsilon}-\frac{\alpha_{S,0}}{6}\tilde{\epsilon}^2\right)+\frac{1}{6N^{1/2}}\left(\tilde{\epsilon}-\frac{\alpha_{S,0}}{3}\tilde{\epsilon}^2\right)+\mathcal{O}(\tilde{\epsilon}^3,N^{-1}) \,, \crcr 
\bar{g}_{1}^{\star}&=\frac{1}{12}\left(\tilde{\epsilon}-\frac{\alpha_{S,0}}{12}\tilde{\epsilon}^2\right)\mp \frac{i\alpha_{S,0}}{36N^{1/2}}\tilde{\epsilon}^2 +\mathcal{O}(\tilde{\epsilon}^3,N^{-1}) \,, \crcr
\bar{g}_{2}^{\star}&=\frac{1}{4\sqrt{3}}\left(\tilde{\epsilon}-\frac{\alpha_{S,0}}{12}(2-\sqrt{3})\tilde{\epsilon}^2\right)\pm \frac{i(-3+2\sqrt{3})}{12N^{1/2}}\left(\tilde{\epsilon}-\frac{\alpha_{S,0}}{2}\tilde{\epsilon}^2\right)+\mathcal{O}(\tilde{\epsilon}^3,N^{-1}) \,,
\label{eq:stablefpLO}
\end{align}
where the signs in all three couplings are taken to be simultaneously either the upper or lower ones. 
For these two fixed points, the $\tilde{\epsilon}$ expansions of the critical couplings are then: 
\begin{align}
\partial  \beta^{(2)}(g^{\star})&=-\nu^{-1}=-\frac{d}{2}+\frac{1}{2\sqrt{3}}\left(\tilde{\epsilon}-\frac{\alpha_{S,0}}{6}\tilde{\epsilon}^2\right)\pm \frac{i}{\sqrt{3}N^{1/2}}\left(\tilde{\epsilon}-\frac{\alpha_{S,0}}{2}\tilde{\epsilon}^2\right)+\mathcal{O}(\tilde{\epsilon}^3,N^{-1}) \,, \crcr
\partial \beta_{1}(g^{\star})&=\frac{1}{3}\left(\tilde{\epsilon}-\frac{\alpha_{S,0}}{6}\tilde{\epsilon}^2\right) \pm \frac{2 i }{3N^{1/2}}\left(\tilde{\epsilon}-\frac{\alpha_{S,0}}{2}\tilde{\epsilon}^2\right) +\mathcal{O}(\tilde{\epsilon}^3,N^{-1}) \,, \crcr
\partial \beta_{2}(g^{\star})&= \frac{1}{\sqrt{3}}\left(\tilde{\epsilon}-\frac{\alpha_{S,0}}{6}\tilde{\epsilon}^2\right) \pm \frac{2 i }{\sqrt{3} N^{1/2}}\left(\tilde{\epsilon}-\frac{\alpha_{S,0}}{2}\tilde{\epsilon}^2\right) +\mathcal{O}(\tilde{\epsilon}^3,N^{-1}) \,,
\end{align}
where the choice of sign is the same as in $\bar{g}_{(0)}^{\star}$.

In order to compute the critical exponent of the tetrahedron coupling, we need to compute the eigenvalues of the stability matrix as it is not triangular beyond order
$N^{-1/2}$. However, up to order $N^{-3/2}$, it depends only on the values of the critical couplings at leading and next-to-leading order. For the fixed point in \eqref{eq:stablefpLO}, we have at second order in $\tilde{\epsilon}$:
\begin{equation}
\omega_{t}=\frac{\tilde{\epsilon}}{N}\left(1+\frac{\alpha_{S,0}}{6}\tilde{\epsilon}\right)+\frac{2 i \alpha_{S,0}\tilde{\epsilon}^2}{3N^{3/2}}+\mathcal{O}(\tilde{\epsilon}^3, N^{-2}) \,.
\end{equation}

\medskip

In summary, while at leading order an imaginary tetrahedron coupling leads to four stable fixed lines of real pillow and double-trace couplings, going up to next-to-leading non-trivial order for all the beta functions fixes all the couplings to eight isolated fixed points, having the same reality properties as before at leading order, but the opposite one at subleading order (i.e.\ real tetrahedron and purely imaginary pillow and double-trace corrections). As with the fixed point values, we have for the critical exponents that what was real at leading order gets an imaginary part at subleading order. 

\section{Conclusions}
\label{sec:concltri}

We have studied a trifundamental model, that is, a multi-scalar model invariant under $O(N_1)\times O(N_2) \times O(N_3)$ transformations, of which the scalar fields form a trifundamental representation. We have considered versions of the model with either short- or long-range Gaussian part, and we have studied the renormalization group beta functions at finite or large $N_i$, in various scaling limits.
Our main conclusion is that in general we find \emph{no} stable real fixed points with non-zero tetrahedral coupling.

In order to find genuine infrared-stable fixed points with non-zero tetrahedral coupling we have to consider complex fixed points. This immediately raises the prospect that the fixed point theories are not unitary; however, complex CFTs have been considered in statistical physics and in the description of walking behavior in high-energy physics (see for example \cite{Gorbenko:2018ncu,Gorbenko:2018dtm} and references therein). Complex, stable (in all directions) infrared fixed points are obtained in the homogeneous (i.e.\ $N_i=N$ for $i=1,2,3$) large-$N$ limit
of the long-range model. In this case, the tetrahedral coupling is exactly marginal, and when taken to be purely imaginary all the CFT data available to us indicates that the leading large-$N$ CFT is real and within unitarity bounds (see section \ref{sec:unitarity} and \cite{Benedetti:2020yvb}). In this chapter we have shown that this does not survive at subleading order in $1/N$: the line of fixed point reduces to an isolated point, and unitarity is broken by the $1/N$ corrections which bring imaginary parts to the critical exponents.
A similar complex CFT, providing subleading corrections to that of \cite{Giombi:2017dtl}, is found also for the short-range model, but in that case it is the real part of the critical exponents which is suppressed in $1/N$, rather than the imaginary part; therefore, while the two models have probably qualitatively similar behavior at finite $N$, it is only in the long-range case that a real and unitary CFT arises in the strict large-$N$ limit.

A subtle aspect of our analysis of subleading corrections in $1/N$ to the fixed points of the $O(N)^3$ model is the identification of an appropriate hierarchy between the two small parameters at play, i.e.\ $1/N$ and $\epsilon$, the latter being defined as the deviation from the critical dimension in the short-range case, i.e.\ $\epsilon=4-d$, or as the deviation from the critical scaling of the propagator in the long-range case, i.e.\  $C(p)=1/p^{(d+\epsilon)/2}$.
In the former case it turns out that we need $\epsilon N^2 \gg 1$, while in the latter we need $\epsilon N \ll 1$.
The reason for that is the form of the tetrahedron beta functions, which we can roughly understand in the following way. Slightly simplifying things (in reality we have a coupled system of equations), at two-loop order the tetrahedron beta function has the form $\b_{SR}(g)=-\epsilon g + b g^3 +\frac{a}{N} g^2 +\cO(N^{-3/2})$ in the short-range case, and $\b_{LR}(g)= -\epsilon g + \frac{a}{N} g^2 +\cO(N^{-3/2})$ in the long-range case, for some constants $a$ and $b$ of order one.
The conditions on $\epsilon$ and $N$ then arise from demanding that the fixed point from the leading order in $1/N$ remains dominant in the beta function. As a perturbative solution of $\b_{SR}(g^\star)=0$ at leading order implies $g^\star\sim \sqrt{\epsilon}$, we see that the first two terms in $\b_{SR}(g)$ are of order $\epsilon^{3/2}$, while the third is of order $\epsilon/N$, hence we must have $\sqrt{\epsilon}\gg 1/N$.
For the long-range case, a non-trivial perturbative solution of $\b_{LR}(g^\star)=0$ at leading order is instead not possible for $\epsilon>0$, and we must require $\epsilon\ll 1/N$, so that the first two terms in $\b_{LR}(g)$ lead to a Wilson-Fisher type solution, with $\epsilon N$ being the effective small parameter.
A similar tuning of $\epsilon$ and $N$ as in our long-range model was also considered in \cite{Fleming:2020qqx} in order to find a finite-$N$ precursor of the line of fixed points that appears in the short-range $O(N)$ model with $(\phi^2)^3$ interaction at large $N$, for $\epsilon=0$.

It would be interesting to understand if the non-existence of stable real fixed points with non-vanishing tetrahedral coupling could be proved in general terms, for example by using group-theoretical arguments, as in \cite{Michel:1985,Toledano:1985}, or by exploiting the gradient flow representation of the renormalization group equations, along the lines of other proofs, for example as in  \cite{Michel:1983in,ZinnJustin:2007zz,Rychkov:2018vya}. We have tried the second route, but failed so far in this task; nonetheless, we report in appendix \ref{app:grad_flow} some relevant formulas for the gradient flow of the trifundamental model, hoping that they could serve as reference or inspiration for a future proof.

More generally, it would also be interesting to understand whether any stable real fixed points exist with rank-$p$ tensor symmetry, such as $O(N)^p$, for higher $p$.
We will see in chapter \ref{chap:sextic} that real fixed points have been found in short-range models with $p=3$, but with sextic interactions, for small $\epsilon=3-d$\cite{Giombi:2018qgp}. We will study the subleading orders in section \ref{sec:sexticNLO} to understand if they also become complex, or whether sextic interactions have some fundamental difference with respect to quartic ones.
\begin{subappendices}
\section{Gradient flow} 
\label{app:grad_flow}
We wish to write the beta functions \eqref{eq:beta^(4)} as a gradient flow:
\begin{equation}
\beta_{a}=T_{ab}\frac{\partial U}{\partial g_b}\, ,
\label{eq:gradient_flow}
\end{equation}
with $U$ a potential, and $T_{ab}$ a non-trivial symmetric matrix, where the indices $a,b$ run over the five couplings $t$, $(p,i)$, $d$. For the general system \eqref{eq:beta-general}, the one loop potential is:
\begin{equation}
U_{MS}=-\frac{\epsilon}{2}\tilde{g}_{ijkl}\tilde{g}_{ijkl}+\tilde{g}_{ijkl}\tilde{g}_{klmn}\tilde{g}_{mnij} \,,
\end{equation}
and we recall that the metric in the general case is trivial at this order (in fact it is trivial up to two loops \cite{Wallace:1974dy}). Substituting \eqref{eq:coupling-trifund} we find the one loop potential for the short-range tri-fundamental model:
\begin{align*}
U&=-3\epsilon N_1N_2N_3\Bigg[\left(N_1N_2N_3+N_1+N_2+N_3+2\right)\tilde{g}^2+2\left(2+N_1N_2N_3\right)\tilde{g}_d^2 +4(N_1+N_2+N_3)\tilde{g}\tilde{g}_d \crcr
&\quad +\sum_{i=1}^3\Bigg(\left((1+N_i)N_{i+1}N_{i+2}+N_i+3\right)\tilde{g}_{p,i}^2+2\tilde{g}\left(2+(1+N_i)(N_{i+1}+N_{i+2})\right)\tilde{g}_{p,i}\crcr
& \quad \qquad \quad +2\left((1+N_i)(1+N_{i+1})+2N_{i+2}\right)\tilde{g}_{p,i}\tilde{g}_{p,i+1}+4\tilde{g}_d\left(1+N_i+N_{i+1}N_{i+2}\right)\tilde{g}_{p,i}\Bigg) \Bigg] \crcr
& +4N_1N_2N_3\Bigg[\Big(12+\sum_{i=1}^3\left(6N_i+6N_iN_{i+1}+N_i^2(N_{i+1}+N_{i+2})\Big)\right)\tilde{g}^3\crcr
& \quad +2\left(N_1N_2N_3+8\right)\left(N_1N_2N_3+2\right)\tilde{g}_d^3\crcr
& \quad +6\Big(3N_1N_2N_3+\sum_{i=1}^3\left(N_i^2+2N_iN_{i+1}+3N_i\right)+6\Big)\left(\tilde{g}^2\tilde{g}_d+2\tilde{g}_{p,1}\tilde{g}_{p,2}\tilde{g}_{p,3}\right)\crcr
& \quad +6\left(N_1+N_2+N_3\right)\left(N_1N_2N_3+8\right)\tilde{g}_d^2\tilde{g} \crcr
& \quad +\sum_{i=1}^3 3\tilde{g}_{p,i}\Bigg(\left(N_{i+1}^2N_{i+2}^2(1+N_i)+N_{i+1}N_{i+2}(N_i^2+6N_i+13)+N_i^2+13N_i+18\right)\frac{\tilde{g}_{p,i}^2}{3}\crcr
& \qquad  +\left((N_{i+1}^2+N_{i+2}^2)(1+N_i)+N_{i+1}N_{i+2}(N_i^2+3N_i+6)\right. \crcr
& \hspace{2cm} \left. +(N_{i+1}+N_{i+2})(3N_i+7)+N_i^2+9N_i+10\right)\tilde{g}^2 \crcr
& \qquad  + 2\left(8+N_1N_2N_3\right)\left(1+N_i+N_{i+1}N_{i+2}\right)\tilde{g}_d^2 \crcr
& \qquad  +\left(2N_{i+1}N_{i+2}+(N_i+1)(N_{i+1}N_{i+2}(N_{i+1}+N_{i+2})+8) \right. \crcr
& \qquad \qquad \left. +(N_{i+1}+N_{i+2})(N_i^2+5N_i+10)\right)\tilde{g}\tilde{g}_{p,i}\crcr
& \qquad +2\left(N_i^2+5N_i(1+N_{i+1}N_{i+2})+N_{i+1}N_{i+2}(5+N_{i+1}N_{i+2})+10\right)\tilde{g}_d\tilde{g}_{p,i} \crcr
&\qquad  +\sum_{j,k=1 ; j\neq k \neq i}^3\left(2N_jN_k^2+(1+N_j)(N_i^2+5N_i+10)+N_k(1+N_i)(N_j^2+N_j+8)\right)\tilde{g}_{p,j}\tilde{g}_{p,i} \crcr
& \qquad  + 2\left(2N_{i+2}^2+2N_{i+2}(2N_iN_{i+1}+3(N_i+N_{i+1})+2)+N_i^2+N_{i+1}^2 \right. \crcr
& \hspace{2cm} \left. +(N_i+N_{i+1})(N_iN_{i+1}+7)+2N_{i}N_{i+1}+12\right)\tilde{g}\tilde{g}_{p,i+1} \crcr
& \qquad  +4\left(N_{i+2}^2N_iN_{i+1}+N_{i+2}(N_i^2+N_{i+1}^2+N_i+N_{i+1}+6)+4(1+N_i)(1+N_{i+1})\right)\tilde{g}_d\tilde{g}_{p,i+1} \crcr
& \qquad  +4\left(N_i^2+N_i(N_{i+1}N_{i+2}+4(N_{i+1}+N_{i+2})+1)+(N_{i+1}+N_{i+2})(N_{i+1}N_{i+2}+4)+6\right)\tilde{g}\tilde{g}_d  \Bigg) \Bigg] \,,
\end{align*}
where $i \in \{1,2,3\}$ and $N_4=N_1$, $N_5=N_2$.

The matrix $T$ can now be found following \cite{Wallace:1974dy}, by using the following expression for its inverse:
\be
(T^{-1})_{ab} = \f{\p \gt_{\mba \mbb \mbc \mbd} }{\p g_a} \f{\p \gt_{\mba \mbb \mbc \mbd} }{\p g_b} \,.
\ee
Defining $(6 N_1 N_2 N_3)\, \eta_{ab}=(T^{-1})_{ab}$, we have:
\begin{align}
\eta_{tt}&= 2+N_1+N_2+N_3+ N_1 N_2 N_3 \,,\crcr
\eta_{p_i p_i}&=2+(1+N_i)(1+N_{i+1}N_{i+2}) \,,\crcr
\eta_{d d}&=4+2 N_1N_2N_3 \,,\crcr
\eta_{t p_i}&=\eta_{p_i t}=2+(1+N_i)(N_{i+1} + N_{i+2}) \,,\crcr
\eta_{t d}&=\eta_{d t}=2(N_1+N_2+N_3) \,,\crcr
\eta_{p_i p_j}&=1+N_i+N_j+N_iN_j+2N_k  \,, \;\; \text{ with } i\neq j \neq k \in \{1,2,3\} \,, \crcr
\eta_{p_id}&=\eta_{d p_i}=2(1+N_i+N_{i+1}N_{i+2}) \,.
\end{align}

The long-range beta functions \eqref{eq:beta_abcd_alpha}, or \eqref{eq:beta-LR} differ from the short rang ones only by the presence of the $\a_D$ and $\a_S$ coefficients and the absence of the terms coming from the wave-function renormalization. In particular they are equal at one loop.

In the homogeneous large-$N$ limit $N_1=N_2=N_3=N$ we switch to rescaled variables, as in \eqref{eq:bar_g}. Accordingly, the system becomes
\be \label{eq:gradient_flow_bar}
 6 N^3 \bar{\eta}_{ab} \bar{\beta}_b = \frac{\partial U}{\partial \bar{g}_a} \,,
\qquad 
\bar{\eta}_{ab} = \frac{\p g_c}{\p \bar{g}_a} \eta_{cd} \frac{\p g_d}{\p \bar{g}_b} \, .
\ee
In order to obtain a finite limit, it turns out one needs to first multiply both sides of \eqref{eq:gradient_flow_bar} by
the diagonal matrix $\rho={\rm diag}(N^{-3},N^{-2},N^{-2},N^{-2},1)$ to obtain:
\be
\lim_{N\to\infty} N^3\, \rho \, \bar{\eta} = 
\begin{pmatrix}    
1 & 0 & 0 & 0 & 0 \\
0 & 1 & 0 & 0 & 0 \\
0 & 0 & 1 & 0 & 0 \\
0 & 0 & 0 & 1 & 0 \\
0 & 2 & 2 & 2 & 2 
\end{pmatrix} \,.
\ee
The mixing elements between pillows and double-trace are explained by the diagonalization of the system at large $N$ in \eqref{eq:g_1,2}.

Notice that $U$ by itself does not have a finite limit for $N\to\infty$ even when written in terms of $\bar{g}_a$ couplings, it is only $\rho_{ab} \partial U/\partial \bar{g}_b$ that does. However, there is no need to rescale by $\rho$ if we write the system as in \eqref{eq:gradient_flow}.

\end{subappendices}

\chapter{The F-theorem in the melonic limit}
\label{chap:Ftheorem}
Among the most intriguing features of quantum field theory in various dimensions are the so called $c$-, $a$- and $F$-theorems \cite{Zamolodchikov:1986gt,Komargodski:2011vj,Jafferis:2011zi,Klebanov:2011gs}. These lettered theorems state that under the RG flow between various fixed points some quantities (aptly denoted $c$, $a$ or $F$) always decrease. Intuitively, these quantities must in some way count the degrees of freedom in the theory, as the RG flow decimates the degrees of freedom when going from one fixed point to another.

The most well known of the lettered theorems, the $c$-theorem in dimension 2 was first proven by Zamolodchikov \cite{Zamolodchikov:1986gt}. The quantity $c$ in this case was defined using the two-point functions of the stress-energy tensor. Interestingly, the obtained $c$-function coincides at the RG fixed point with the Weyl anomaly coefficient $c$, that is the central charge. 

In $d=4$ dimensions, the $a$-theorem was first conjectured by Cardy \cite{Cardy:1988cwa}. In this case, there are two universal Weyl anomaly coefficients, usually denoted $a$ and $c$. Cardy conjectured that the quantity that should decrease along the RG flow is the $a$-coefficient, multiplying the Euler density. In practice, this coefficient can be computed from the expectation value of the trace of the stress-energy tensor in the Euclidean theory on $S^4$. After a long time, the $a$-theorem was finally proven in \cite{Komargodski:2011vj,Komargodski:2011xv}.

The latest addition to this list of monotonicity theorems is the $F$-theorem, concerning field theories in $d=3$.
In this case, $F$ has a relatively straightforward definition as the free energy of the CFT on the sphere. 
The compactness of the sphere regulates the infrared divergences, however ultraviolet divergences persist and need to be regularized properly: $F$ is defined as the finite part of the free energy.
This choice was first proposed in \cite{Jafferis:2010un,Jafferis:2011zi}, where various checks were performed on supersymmetric theories, then extended to non-supersymmetric ones in \cite{Klebanov:2011gs} (see also \cite{Pufu:2016zxm} for a review).
Shortly after, the $F$-theorem was proven in \cite{Casini:2012ei}, using the relation between the free energy and the entanglement entropy across a circle \cite{Casini:2011kv}; so far, this is the only method that works for all three theorems \cite{Casini:2004bw,Casini:2017vbe}.

One common feature of the proofs of these theorems is that unitarity plays a crucial role: it is underlying the use of positivity of two-point functions in \cite{Zamolodchikov:1986gt}, of the optical theorem in \cite{Komargodski:2011vj}, and of the strong subadditive inequality in \cite{Casini:2012ei,Casini:2004bw,Casini:2017vbe}.
At present, it remains unclear to what extent unitarity is a necessary ingredient and if the assumptions of these theorems could be relaxed to include at least some class of non-unitary models.
A non-physical counterexample to the necessity of unitarity is provided by the generalized $F$-theorem tests in non-integer dimensions \cite{Giombi:2014xxa,Fei:2015oha,Giombi:2015haa}, where it was shown to hold, despite the fact that CFTs in non-integer dimensions have been shown to be non-unitary \cite{Hogervorst:2015akt}, at least in the case of the Wilson-Fisher fixed point. 
On the other hand, as we will see below, a trivial counterexample to a generic $F$-theorem without unitarity is provided by a non-unitary generalized free field theory flow.\footnote{It should be remarked that generalized free fields, as the long-range model we will consider here, evade also another hypothesis of the standard proofs, that is, locality. Long-range models in particular do not have a local energy-momentum tensor, which plays a crucial role in the standard proofs of the $c$- and $a$-theorems. However, the embedding of such models in a larger space \cite{Paulos:2015jfa} could perhaps provide a workaround for such proofs. In fact, for the special case of an integer number of extra dimensions, boundary versions of the $F$-theorem have indeed been proposed \cite{Gaiotto:2014gha} (see also \cite{Kobayashi:2018lil} for more information on monotonicity theorems in boundary or defect CFTs).}

In most applications, one knows from the start whether the theory of interest is unitary or not, or at least one has a good degree of confidence in that, and therefore testing the $F$-theorem in a theory satisfying the hypotheses of \cite{Casini:2011kv} is at most an interesting exercise.
However, in some cases ascertaining the unitarity, or lack thereof, of a theory can be challenging; for instance, in the Wilson-Fisher fixed point at non-integer dimensions \cite{Hogervorst:2015akt}, or in the $O(N)$ model at non-integer $N$ \cite{Maldacena:2011jn,Binder:2019zqc}, non-unitarity is a non-trivial result, manifesting itself only in operators of large dimension.

The main subject of this chapter will be another non-trivial example, going in the opposite direction: the long-range $O(N)^3$ model (introduced in chapter \ref{chap:CTKT}), a manifestly non-unitary model, which however in the large-$N$ limit has so far passed all the unitarity tests.
As a warm up, in section~\ref{sec:gaussian_CFT}, we consider two Gaussian CFTs and examine the flow between them. Next, in section~\ref{sec:ON_model}, we revisit the vector $O(N)$ model and rederive its sphere free energy, previously obtained in  \cite{Klebanov:2011gs}. In section~\ref{sec:ON3-model}, we finally study the long-range $O(N)^3$ model on the sphere and confirm that the $F$-theorem holds in this case. We give abundant details on notations and computations in several appendices. Appendices \ref{app:useful} and \ref{app:sphereCFT} give formulas and definitions for CFTs on the sphere. In appendices \ref{app:F_GFFT} and \ref{app:C_dim_reg} we give details for the computations of the free energy for generalized free field theories (GFFT) with short- and long-range covariances. In appendix~\ref{app:CPW}, we review the basics of conformal partial wave expansion.
In appendix~\ref{app:monster} we prove that the exceptional diagram of the $O(N)^3$ model at NNLO has a non-vanishing finite part (i.e.\ it contributes to $F$), although it does not contribute in the assessment of the $F$-theorem.
Lastly, in appendices~\ref{app:I_eps} and \ref{app:NumericsLargeJ}, we give details on intermediate results in the computation of the sphere free energy for the $O(N)^3$ model. 

\section{Flow between Gaussian CFTs}
\label{sec:gaussian_CFT}

As a warm-up, and for later reference, let us consider the following quadratic action:
\be \label{eq:GaussianS}
S_{\rm Gauss}[\phi] = \f12 \int \dd x  \, \phi(x) (-\p^2)^{\z} \phi(x) + \f{\l}{2} \int \dd x  \, \phi(x) (-\p^2) \phi(x) \,,
\ee
with $0<\z<1$. The non-integer power of the Laplacian is defined in momentum space simply as $p^{2\z}$, or in position space by a convolution with a non-local kernel, see \eqref{eq:fracLapl_flat_phys}-\eqref{eq:flatCinv}.
The second term in \eqref{eq:GaussianS} is the standard short-range free action of scalar fields, while the first is a generalized free field theory (GFFT), constituting the free part of interacting long-range scalar models.
The models we will consider in the next two sections are short-range and long-range, respectively, hence this simple example will also allow us to introduce some useful results for later on.

The coupling $\l$ has mass dimension $2\z-2<0$, hence it is an irrelevant coupling for the GFFT. Since the theory is Gaussian, the RG flow is rather trivial: the two-point function in momentum space is $(p^{2\z}+\l p^2)^{-1}$ and goes to the GFFT propagator $1/p^{2\z}$ for $p\to 0$, while for $p\to\infty$ it goes to the canonical free theory propagator $1/p^2$ (up to normalization).
Therefore, we have a flow between two Gaussian CFTs.

The flow is rather standard from the GFFT side, as the operator $ \phi \p^2 \phi$ is a primary in the OPE spectrum of $\phi\times \phi$ in the GFFT (e.g.\ \cite{Benedetti:2019ikb}), and it has scaling dimension greater than $d$.
On the other hand, the flow is somewhat unusual from the canonical free CFT side, as the non-local operator $ \phi \p^{2\z} \phi$ is not in the CFT spectrum.
One possible way to write the perturbation in the UV in the framework of conformal perturbation theory is to introduce an additional field, following the idea proposed in \cite{Behan:2017emf} for the short-range/long-range Ising transition. We thus rewrite the action with a second field $\chi$:
\be \label{eq:GaussianS-chi}
S_{\rm Gauss-2}[\phi,\chi] = \f12 \int \dd x  \, \phi(x) (-\p^2) \phi(x)  + \f12 \int \dd x  \, \chi(x) (-\p^2)^{-\z} \chi(x) + \f{\im}{\sqrt{\l} } \int \dd x  \, \chi(x) \phi(x) \,.
\ee
Integrating out the field $\chi$ we recover \eqref{eq:GaussianS}, up to a rescaling $\phi\to\phi/\sqrt{\l}$.\footnote{The two formulations of the theory are only equivalent if we restrict to the set of correlators of operators built only with fields $\phi$, plus the mixing operator $\chi(x) \phi(x)$. In particular it would make no sense to integrate out $\phi$ in \eqref{eq:GaussianS-chi}.}
This formulation is non-standard in other ways, namely the need of an imaginary coupling (that could however be absorbed into the field $\chi$), and the negative power of the Laplacian, the latter leading to unusual features about the thermodynamic limit, such as inequivalence of statistical ensembles \cite{Campa:2009rev}.
An important difference between \eqref{eq:GaussianS} and \eqref{eq:GaussianS-chi} is that, in the second case, the UV theory has an additional, albeit decoupled, degree of freedom, the field $\chi$. 

We want to test the $F$-theorem on this flow between Gaussian CFTs. To that end, we place the fixed-point  theories on a spherical background, which is done by the standard procedure recalled in appendix~\ref{app:sphereCFT}. 
In practice, the local Laplacian is replaced by the Weyl covariant version (i.e.\ the operator in \eqref{eq:prop-def} with the choice \eqref{eq:b_W}), while the non-local one by the more complicated operator \eqref{eq:D_z} (with kernel \eqref{eq:sphereCinv}).

As we are interested in the difference between the free energies at the two limiting theories, the overall normalization of the functional integral over $\phi$ is not important and will be omitted. However, in the formulation with the action \eqref{eq:GaussianS-chi} the auxiliary field $\chi$ should better have a unit normalized Gaussian functional measure:
\be
\int [{\rm d} \chi] \, e^{-S_{\rm Gauss-2}[0,\chi]} = 1 \,,
\ee
so that integrating it out leads to the functional integral for \eqref{eq:GaussianS}, with the same normalization. Equivalently, this is demanded by imposing that the UV theories obtained from \eqref{eq:GaussianS-chi} or from \eqref{eq:GaussianS} have the same free energy.

For the weak form of the $F$-theorem we only need to compare the fixed-point theories. These are GFFTs with different values of $\z$, hence it is straightforward to compute $F$.
The free energy parametrized by $\z$ (including the standard $\z=1$ case) is given by:
\be \label{eq:F_GFFT}
F = \f12 \Tr[\ln C^{-1}] = \f12 \sum_{n\geq 0} D_n \, \ln\big(\om^{(\z)}_n \big) \,,
\ee
where $D_n$ is the multiplicity of the eigenvalues, see \eqref{eq:multi}.
The sum is clearly divergent, hence we need a regularization.

\begin{figure}[htb!]
	\centering
	\includegraphics[width=0.5\linewidth]{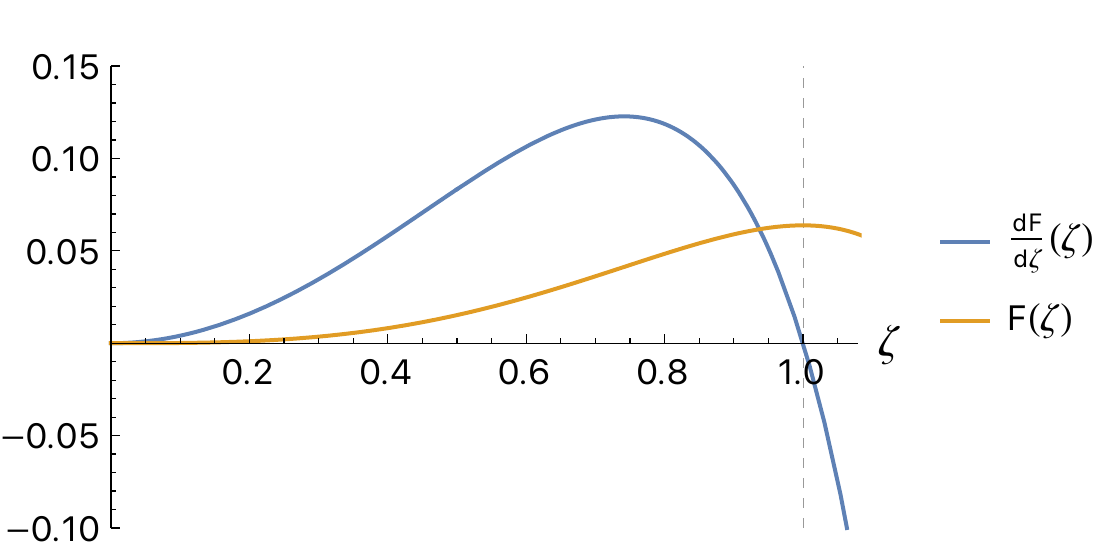}
	\caption{The free energy is $0$ when $\zeta=0$, grows with $\zeta$, and reaches its maximum at $\zeta=1$. We plot the $d=3$ case. The blue curve is the derivative of the free energy with respect to $\zeta$ and the orange curve is the free energy itself.}
	\label{fig:freeenergyfreetheory}
\end{figure}

The same kind of sum was encountered in \cite{Klebanov:2011gs} as the IR limit of a CFT perturbed by a double-trace operator.
Using dimensional regularization as in \cite{Diaz:2007an}, we find:
\begin{equation}
\frac{dF}{d\zeta}=-\zeta\frac{\sin(\pi \zeta)}{\sin(\pi d/2)}\frac{\Gamma(d/2-\zeta)\Gamma(d/2+\zeta)}{\Gamma(1+d)} \label{dF/dz} \,,
\end{equation}
which only has poles for $d$ even. See appendix~\ref{app:F_GFFT} for the detailed computation. When $d=3$, \eqref{dF/dz} simplifies to

\begin{equation}
\frac{d F}{d \zeta}=\frac{1}{24} \pi  \zeta  \left(1-4 \zeta ^2\right) \tan (\pi  \zeta ) \,,
\end{equation}
that is positive for $0 \le \zeta \le1 $. An immediate consequence is that the free energy $F$ grows with $\zeta$ (see figure~\ref{fig:freeenergyfreetheory}), showing that an RG trajectory flowing from a short-range free Gaussian model ($\z=1$) to a long-range Gaussian model ($\z<1$) satisfies the $F$-theorem.

It is interesting to consider the model \eqref{eq:GaussianS} with $\z>1$, for which the GFFT is non-unitary.\footnote{As can be seen from the K\"all\'en-Lehmann spectral representation of the propagator \cite{Benedetti:2020seh}, or from the fact that the unitarity bound $\D>d/2-1$ is violated.}
We also restrict to $\z<d/2$, to keep $\D>0$ and avoid the logarithmic two-point function at $\D=0$.
In this case, the role of UV and IR limiting theory is exchanged, with the $\z>1$ GFFT flowing in the IR to the standard free theory with $\z=1$. And since $\z=1$ is a maximum for  $F$, we find that the free energy increases in the IR. We thus have a trivial counterexample to the $F$-theorem for non-unitary theories.

\section{The $O(N)$ model revisited}
\label{sec:ON_model}

In this section, we look at the three-dimensional $O(N)$ model in the short-range case, i.e.\ $\zeta=1$ and $d=3$,
\be	\label{eq:ON-action}
\begin{split}
S[\phi] &=  \frac{1}{2} \int \dd x \, \phi_{a} (  - \partial^2)\phi_{a} + 
	\frac{ m^{2}}{2} \int \dd x \, \phi_{a}  \phi_{a} + \frac{\l}{4 N} \int \dd x \,  (\phi_{a} \phi_{a})^2 \, , 
\end{split}
\ee
where repeated indices are summed over the range $a = 1, \cdots, N$.
Although this case has been studied before in \cite{Klebanov:2011gs} and we only reproduce here the known result, we will do this by a different method. This helps us prepare the ground for the next section. The new elements of our analysis are the following.
First, we will frame the discussion within the 2PI effective action formalism (for which we follow \cite{Benedetti:2018goh,Berges:2004yj}). 
Second, we will show how the result of \cite{Klebanov:2011gs} is reproduced by means of a conformal partial wave expansion.

\subsection{The sphere free energy at leading order in the large-$N$ expansion}

At large $N$, the leading-order (LO) 2PI effective action is of order $N$, and reads
\be \label{eq:ON_Gamma_2PI}
\begin{split}
\mathbf{\G}_{\rm LO}[G] &=  N\left(\f12 \int_{x,y} C_1^{-1}(x,y) G(y,x)  + \f12 \int_{x,y} \ln (G^{-1})(x,y) + \f{m^{2}}{2} \int_x G(x,x) + \f{\l}{4} \int_x G(x,x)^2 \right) \\
& =  \f{N}{2} \Tr\left[ C_1^{-1} G +  \ln (G^{-1}) + m^{2} G + \f{\l}{2}  \cB \right] \,,
\end{split}
\ee
where $\int_x = \int \dd x \sqrt{g(x)}$, and $C_1^{-1}=-\partial^2$ is written as the inverse of the free propagator, which is defined in \eqref{eq:C_1}.
In the second line we introduced a compact trace notation, as well as the double-propagator operator
\be \label{eq:bubble-kernel}
\cB(x,y) = G(x,y)^2\,.
\ee
The logarithmic term $\ln(G^{-1})$ should be understood in terms of its eigenvalues (in particular when $G$ coincides or is proportional to the free propagator $C_1$), or by a formal expansion around the free propagator.

\begin{figure}[htb]
\begin{center}
\begin{minipage}{0.3\textwidth}
\includegraphics[scale=0.75]{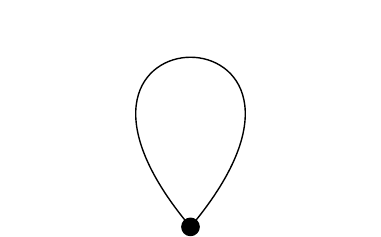}
\end{minipage}
\begin{minipage}{0.3\textwidth}
\includegraphics[scale=0.75]{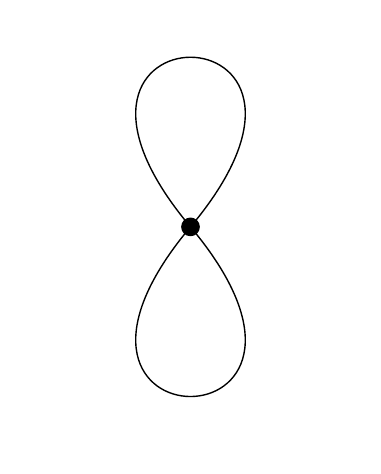}
\end{minipage}
 \caption{The only two vacuum 2PI diagrams occurring in the $O(N)$ model at large $N$. The tadpole on the left has a two-valent mass vertex, while the figure-eight on the right has a $\l$ vertex.} 
 \label{fig:vacuum-2PI-ON}
 \end{center}
\end{figure}

The first two terms in \eqref{eq:ON_Gamma_2PI} are very generic, it is the one-loop part of the effective action. The rest should in general be given by a sum over all the vacuum 2PI diagrams, built from the vertices of the theory, but with a generic propagator $G(x,y)$, to be determined self-consistently at a second stage.
The sum of diagrams becomes manageable in the large-$N$ expansion, and at the leading order written in \eqref{eq:ON_Gamma_2PI} only two diagrams survive, the mass tadpole, and the interaction double-tadpole, or figure eight (see figure~\ref{fig:vacuum-2PI-ON}).

The true full two-point function of the model is found by solving the Schwinger-Dyson equations, which are obtained as the field equations of the 2PI effective action:
\be
\f{\d \G}{\d G} = 0 \,.
\ee
From \eqref{eq:ON_Gamma_2PI} we find the following form of the Schwinger-Dyson equations:
\be
G^{-1}(x,y) = C_1^{-1}(x,y) + \left(m^2+ \l G(x,x) \right) \f{\d(x-y)}{\sqrt{g(x)}} \,,
\ee
where we used
\be
\f{ \d G(u,w)}{\d G(x,y)} = \f12 \f{\d(x-u)\d(y-w)+\d(x-w)\d(y-u)}{\sqrt{g(u)}\sqrt{g(w)}} \,.
\ee
Clearly, it is enough to tune the bare mass:
\be \label{eq:mass-tadpole}
m^2 = -\l C_1(x,x) \,,
\ee
in order to cancel the on-shell tadpole and obtain trivially the solution $G=C_1$.

The free energy $F$ is obtained by evaluating the 2PI effective action on shell, i.e.\ by substituting the solution of the Schwinger-Dyson equations into \eqref{eq:ON_Gamma_2PI}.
Evaluating it on the sphere resolves the IR problem arising from the fact that $F$ is proportional to the volume of the $d$-dimensional background space. However, there are still UV divergences originating from the functional traces, and from the evaluation of the propagator at coincident points.
As we will now review, after appropriate regularization, the result is that the tadpole terms will drop out and thus the leading-order free energy is the same as that of the free theory.

When replacing $G=C_1$ in \eqref{eq:ON_Gamma_2PI}, the first two terms should reproduce ($N$ times) the free energy of the GFFT \eqref{eq:F_GFFT} at $\z=1$.
We have already discussed the regularization and evaluation of $\Tr[\ln C_1^{-1}]$ in the previous section, but what about the first term in \eqref{eq:ON_Gamma_2PI}? Clearly, it corresponds to a divergent contribution $\Tr[\mathbf{1}]\sim \d(0)$. A similar term (with the opposite sign) is however discarded in the typical derivation of the 2PI effective action \cite{Benedetti:2018goh}, for which the first term would otherwise read $\Tr[ (C_1^{-1}-G^{-1}) G]$, and therefore the two cancel out. In any case, when regularized by analytical continuation in the dimension, such terms vanish identically, as shown in \eqref{eq:sum_multiplicity}.

The bare mass was chosen in \eqref{eq:mass-tadpole} to cancel the tadpole in the SD equation, but the cancellation does not occur in the free energy.
However, as we show in appendix~\ref{app:tadpole-SR}, analytic continuation in the dimension gives $C_1(x,x)=0$, hence the contribution of the tadpole terms  to the free energy vanishes.
Therefore, at leading order the free energy of the $O(N)$ critical theory is the same as for the free theory, which in $d=3$ reads: 
\be
F_{\rm LO} = \mathbf{\G}_{\rm LO}[C_1] =  \f12 \Tr[\ln C_1^{-1}] = \f{N}{16} \left( \ln 4 - \f{3\,\z(3)}{\pi^2} \right)\,.
\ee
In order to find a non-trivial result one thus need to consider the subleading corrections to $F$.

\subsection{The next-to-leading order of the large-$N$ expansion}
\label{sec:ON-model-NLO}

\begin{figure}[htbp]
\centering
\includegraphics[width=0.5\textwidth]{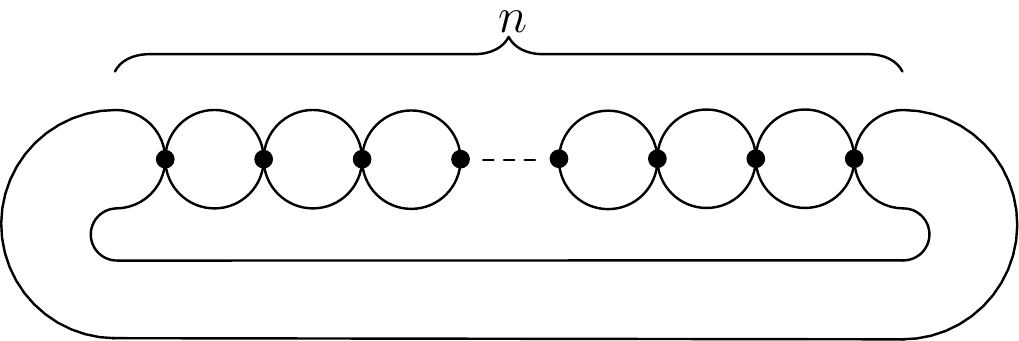}
\caption{Chain of bubbles with $n\geq 1$ vertices. The two-point kernel $\cB$ corresponds to one of the bubbles in the chain.}
\label{fig:bubble_chain}
\end{figure}

The next-to-leading order (NLO) contribution to the 2PI effective action is of order $N^0$, and reads \cite{Benedetti:2018goh}
\be \label{eq:ON_Gamma_2PI_NLO}
\mathbf{\G}_{\rm NLO}[G] =  \f12 \Tr[\ln(\mathbf{1}+ \l \cB)] \,,
\ee
where $\cB$ is the two-point kernel \eqref{eq:bubble-kernel}, corresponding to a single bubble in the chain of bubbles represented in figure~\ref{fig:bubble_chain}.
This is the same contribution found in  \cite{Klebanov:2011gs} by introducing an intermediate field, and their result will thus be reproduced: on shell we replace $G(x,y)\to G_{\rm LO}(x,y)=C_1(x,y)$,\footnote{Notice that the NLO part of the on-shell $G(x,y)$ only contributes at NNLO and beyond. This is obvious by expanding $\mathbf{\G}_{\rm NLO}[G_{\rm LO}+G_{\rm NLO}/N]$, and it is true also for $\mathbf{\G}_{\rm LO}[G_{\rm LO}+G_{\rm NLO}/N]$ because by construction $\f{\d \mathbf{\G}_{\rm LO}}{\d G}[G_{\rm LO}]=0$.} and thus the two-point kernel $\cB(x,y)$  is a two-point function of an operator of dimension $\D=d-2$, which on the sphere is diagonalized by the spherical harmonics.
After going to eigenvalues, and taking the bare coupling to infinity in order to tune to the fixed point,\footnote{\label{foot:FPcoupling}The bare coupling in \eqref{eq:ON_Gamma_2PI_NLO} should be expressed as a function of the renormalized coupling, at the leading order of the large-$N$ expansions, as the NLO part would only contribute to NNLO in the effective action. At LO, we have $\l=g/(1-b(d) g/\mu^{4-d})$, where $b(d)$ is a positive constant with a pole at $d=4$, $\mu$ is the renormalization scale, and $g$ is the renormalized coupling. The fixed point of the beta function for the dimensionless coupling $\tilde{g}=g/\mu^{4-d}$ is at $\tilde{g}_\star=1/b(d)$, hence we have $\l\to\infty$.} one is left with the same type of computation as in \eqref{eq:F_GFFT}. In $d=3$, the result is: 
\be
F_{\rm NLO} = \mathbf{\G}_{\rm NLO}[C_1] = - \f{\z(3)}{8\pi^2} \,,
\ee
and thus the free energy of the IR fixed point is smaller than the one of the UV free theory.

However, the fact that the NLO contribution to the 2PI effective action could be expressed in terms of a two-point kernel is a very peculiar feature of the $O(N)$ model at large $N$.
In particular, it does not generalize to other models, such as the melonic ones that we will discuss in the following,
for which the NNLO (the NLO is vanishing) contribution is expressed in terms of a four-point kernel.
Therefore, as a warm-up to our next computation, we would like to recover the $O(N)$ model result of   \cite{Klebanov:2011gs} by expressing the NLO part of the effective action as:
\be \label{eq:ON_Gamma_2PI_NLO_2}
\mathbf{\G}_{\rm NLO}[G] =  \f12 \Tr[\ln(\mathbb{I}-K)] \,,
\ee
where we introduced the Bethe-Salpeter four-point kernel
\be \label{eq:K-doubletrace}
K(x_1,x_2,x_3,x_4) = - \l G(x_1,x_3) G(x_2,x_4) \d(x_3-x_4) \,,
\ee
and by a slight abuse of notation we use the same symbol for the trace, which in this case refers to a trace of a bilocal to bilocal operator, that is a four-point kernel that acts by integration on two arguments, e.g.\ $\Tr[K]= \int_{x_1,x_2} K(x_1,x_2,x_1,x_2)$.

Under quite general assumptions the Bethe-Salpeter kernel of a CFT is diagonalized by the basis of three-point functions (see appendix~\ref{app:CPW}), hence once we go on shell ($G(x,y)\to C_1(x,y)$) and tune to the fixed point, we should obtain a representation of the kernel (or functions of it) by applying it on the resolution of the identity \eqref{eq:res-id-symm-2}. 
This is not so straightforward for the kernel \eqref{eq:K-doubletrace}. As explained in footnote~\ref{foot:FPcoupling}, the fixed point is at $\l\to\infty$, which requires us to work at finite $\l$ and then take the limit. However, for finite $\l$, and this being a dimensionful coupling for $d<4$, the kernel \eqref{eq:K-doubletrace} does not have the right conformal properties to ensure that its convolution with a three-point function \eqref{eq:3pt} transforms as the three-point function itself.  Thus the  kernel \eqref{eq:K-doubletrace} applied on a three-point function cannot be proportional to it. Unless of course the proportionality constant is zero or infinite. 

On the principal series, we have ${\rm Re}(2\D-h)<0$, hence when the delta function in \eqref{eq:K-doubletrace} acts on \eqref{eq:3pt} the result is zero: the three-point functions with $h\in\cP_+$ are indeed eigenfunctions of \eqref{eq:K-doubletrace} with \emph{vanishing eigenvalue}.
On the other hand, for the isolated contribution in \eqref{eq:res-id-symm-2} we have $2\D-h=0$, hence the delta function gives a finite result, which is not\footnote{At $h=2\D$ and $J=0$, the three-point function \eqref{eq:3pt} is proportional to $C(x_3,x_0)C(x_4,x_0)$, and the action of $K$ on it generates a bubble. The result can be evaluated most easily in Fourier space, and it is found to be proportional to $\l b(d) p_1^{-2} p_2^{-2}(p_1+p_2)^{d-4}$, with the same constant $b(d)$ that appeared in footnote~\ref{foot:FPcoupling}. Due to the factor $(p_1+p_2)^{d-4}$, the result is not proportional to the Fourier transform of $C(x_3,x_0)C(x_4,x_0)$.} proportional to $ \la \phi(x_1) \phi(x_2) \cO_{2\D}(x_0) \ra_{\rm cs}$. 
However, the conformal theory is reached at $\l\to\infty$, hence the three-point function with $h=2\D$ and $J=0$  is formally an eigenfunction with \emph{infinite eigenvalue}.

This has the interesting consequence that for the right-amputated four-point function, which is a geometric series in $K$ (see \eqref{eq:F_s-K}) and hence the coupling appears in the denominator of the resummed series, we obtain  (see appendix~\ref{app:CPW} for the notation):
\be
\begin{split}
\cF_s(x_1,x_2,x_3,x_4)  &= \int \dd y_1 \dd y_2 (\mathbb{I}-K)^{-1}(x_1,x_2,y_1,y_2) C_1(y_1,x_3) C_1(y_2,x_4) \\
& =   \sum_{J\in \mathbb{N}_0^{\text{even}}}  \int_{\f{d}{2}}^{\f{d}{2}+\im\infty}  \f{{\rm d}h}{2\pi\im} \r(h,J) \, \cN^{\D}_{h,J}  \cN^{\D}_{\htilde,J} \, \Psi_{h,J}^{\D,\D,\D,\D}(x_1,x_2,x_3,x_4)  \, .
\end{split}
\ee
Therefore, the isolated contribution that is present in the free theory (see \eqref{eq:cF-GFFT-extra} with $\D=d/2-1$) is suppressed in the critical theory, and the conformal partial wave expansion is the same as in \eqref{eq:cF-GFFT}, even tough $\D=d/2-1<d/4$.
We thus recover the well-known result that the spectrum of the critical $O(N)$ model (i.e.\ the poles of the integrand) at large $N$ is the same as in the free theory for $J>0$, while for $J=0$ it has the $\phi^2$ operator replaced by its shadow.

Applying the above formalism to the free energy on the sphere is however trickier than for the four-point function.
Because of the conformal nature of the basis of three-point functions, we expect the eigenbasis of $K$ on the sphere to be obtained by Weyl mapping:\footnote{Applying the same mapping in \eqref{eq:CPW}, the $\Om(z)$ factors cancel, and we obtain an overall factor $\Om(x_3)^{-d}\Om(x_4)^{-d}$, which reconstructs the correct factors of $1/\sqrt{g}$ for the delta functions in \eqref{eq:res-id-symm-2}.}
\be
\la \phi_{\D}(x_3)  \phi_{\D}(x_4) \cO_h^{\m_1 \cdots \m_J}(x_0) \ra_{\rm cs} \to \Om(x_3)^{-\D}\,\Om(x_4)^{-\D}\,\Om(x_0)^{-h}\,\la \phi_{\D}(x_3) \phi_{\D}(x_4) \cO_h^{\m_1 \cdots \m_J}(x_0) \ra_{\rm cs} \,.
\ee
In fact, for a Bethe-Salpeter kernel with the good conformal transformation properties, the Weyl mapping would give
\be
K(x_1,x_2,x_3,x_4)  \to \Om(x_1)^{-\D}\,\Om(x_2)^{-\D}\, \Om(x_3)^{\D-d}\,\Om(x_4)^{\D-d}\, K(x_1,x_2,x_3,x_4) \,,
\ee
and thus all the $\Om(x_3)$ and $\Om(x_4)$ factors would drop out in the convolution, leading to a consistent eigenvalue equation.
As explained above, the kernel in \eqref{eq:K-doubletrace} does not transform in such a nice way at finite $\l$, and we have instead:\footnote{The two transformations agree for $\D=d/4$, i.e.\ when $\l$ is dimensionless, but for the short-range $O(N)$ model this only happens at $d=4$, where there is no interacting fixed point.}
\be
K(x_1,x_2,x_3,x_4)  \to \Om(x_1)^{-\D}\,\Om(x_2)^{-\D}\, \Om(x_3)^{-\D-d/2}\,\Om(x_4)^{-\D-d/2}\, K(x_1,x_2,x_3,x_4) \,.
\ee
However, the three-point functions with operator in the principal series are still zero modes of $K$.
Therefore, inserting the resolution of the identity inside the trace in \eqref{eq:ON_Gamma_2PI_NLO_2}, we see that a non-vanishing contribution can only come from the isolated term at $h=2\D$.\footnote{One might worry that although the three-point functions with operator in the principal series have a vanishing eigenvalue, the trace of the conformal partial wave is divergent, and thus the product is undefined. However, as we will see in the following section, there is a clean way to regularize the trace of the conformal partial wave, which does not affect the eigenvalues, and hence the product is indeed zero in the present case.}
It turns out that at finite $\l$ this reproduces the same series as \eqref{eq:ON_Gamma_2PI_NLO}, and hence we can from here on follow again the same steps as in  \cite{Klebanov:2011gs} and obtain the same result. In fact, it is easy to check that the convolution of $K$ with \eqref{eq:extra} equals $K$.
While this is tautological (as we have inserted the identity and the contribution of the principal series is zero, the isolated contribution must reproduce the full identity), one can check directly that indeed the isolated contribution in \eqref{eq:cF-GFFT-extra} acts alone as the identity operator on $K$, and we are back to having to evaluate the chains of bubbles, for which we can follow the steps of \cite{Klebanov:2011gs}. 
Although from a practical point of view we have not gained anything by applying the conformal partial wave expansion to this problem, nevertheless this small detour taught us about the importance of the isolated contributions in such formalism, and it serves as a test of the method, in view of the application in the next section.


\section{The long-range $O(N)^3$ model}
\label{sec:ON3-model}

In this section, we study the long-range $O(N)^3$ tensor model with $0<\zeta<1$ and $d<4$ on the sphere. We first discuss the Schwinger-Dyson equation and then the free energy at next-to-next-to-leading order in this context.

We recall the action of the model on flat space:
\be	\label{eq:ON3-action}
\begin{split}
S[\phi] &=  \frac{1}{2} \int \dd x \, \phi_{\mba} (  - \partial^2)^{\zeta}\phi_{\mba} + 
	\frac{ m^{2\zeta}}{2} \int \dd x \, \phi_{\mba}  \phi_{\mba} \\
	&\qquad + \frac{1}{4} \int \dd x \, \left[ \im \lambda \hat{\delta}^t_{\mba \mbb\mbc\mbd} + \lambda_1 \hat{P}^{(1)}_{\mba\mbb; \mbc\mbd}
	+ \lambda_2 \hat{P}^{(2)}_{\mba\mbb; \mbc\mbd } \right] \phi_{\mba} \phi_{\mbb} \phi_{\mbc} \phi_{\mbd} \, , 
\end{split}
\ee
where we have explicitly taken out a factor $\im$ in front of the tetrahedral coupling as we will be interested in fixed points with purely imaginary tetrahedral coupling. In the present notation the large-$N$ OPE spectrum is then real for real $\l$. $\hat{\delta}^t_{\mba \mbb\mbc\mbd}$ was defined in section \ref{sec:4point} and $\hat{P}^{(1)}_{\mba\mbb; \mbc\mbd} \, , \, \hat{P}^{(2)}_{\mba\mbb; \mbc\mbd }$ were defined in \eqref{eq:proj_ortho}.

\subsection{Schwinger-Dyson equation for the two-point function}
\label{sec:SDE_tensor}

We start here with a brief review of the solution of the Schwinger-Dyson equation, with the slight modification of working on the spherical background.

\begin{figure}[htb]
\begin{center}
\begin{minipage}{0.3\textwidth}
\includegraphics[scale=0.75]{vacuum_mass.pdf}
\end{minipage}
\begin{minipage}{0.3\textwidth}
\includegraphics[scale=0.75]{vacuum-2PI-8.pdf}
\end{minipage}
\begin{minipage}{0.3\textwidth}
\includegraphics[scale=0.75]{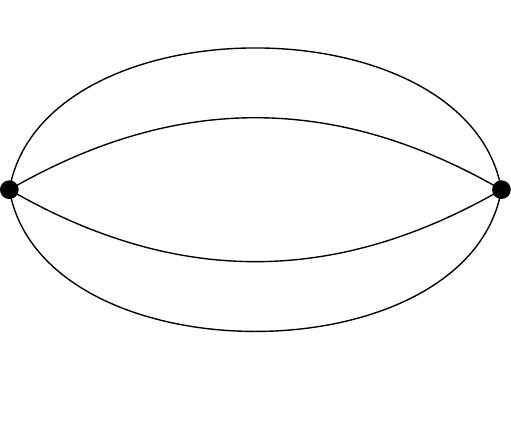}
\end{minipage}
 \caption{The only three types of vacuum 2PI diagrams occurring at large $N$. The tadpole on the left has a two-valent mass vertex. The figure-eight in the middle has a $\l_2$ vertex, and it is of the same type as in the $O(N)$ model. The melon diagram on the right has two $\l$ vertices, and it is characteristic of tensor models.} \label{fig:vacuum-2PI}
 \end{center}
\end{figure}

For large $N$, the dominant 2PI diagrams are depicted in figure~\ref{fig:vacuum-2PI}.
The resulting leading-order 2PI effective action is of order $N^3$, and reads
\be \label{eq:ON3_Gamma_2PI}
\begin{split}
\mathbf{\G}[G]  &= N^3\left(\f12 \int_{x,y} C^{-1}(x,y) G(y,x)  + \f12 \int_{x,y} \ln (G^{-1})(x,y) + \f{m^{2\zeta}}{2} \int_x G(x,x) \right.\\
&\qquad\quad \left.+ \f{\l_2}{4} \int_x G(x,x)^2 +\f{\l^2}{8}\int_{x,y} G(x,y)^4\right)\\
& = \f{N^3}{2} \Tr\left[  C^{-1} G +  \ln (G^{-1}) + m^{2\z} G + \f{\l_2}{2}  \cB +\f{\l^2}{4} \cB^2\right] \,,
\end{split}
\ee
where we used the notation introduced in \eqref{eq:ON_Gamma_2PI}, and we used the symmetry of $G(x,y)$ to write the melon integral as a trace of a convolution of two $\cB(x,y)$. The latter is of course an arbitrary choice, and we could as well write it as the trace of the convolution of $G(x,y)$ with $G(y,x)^3$, which will be a useful point of view later on.

The Schwinger-Dyson equations obtained from \eqref{eq:ON3_Gamma_2PI}:
\be \label{eq:SDeq}
G^{-1}(x,y) =  C^{-1}(x,y)+ \left(m^{d/2} +\l_2  G(x,x) \right) \f{\d(x-y)}{\sqrt{g(x)}} + \l^2     G(x,y)^3 \, ,
\ee
should be understood in the sense of distributions, that is, denoting by $\cS(x,y)$ the right-hand side of \eqref{eq:SDeq}, we should demand that in the limit in which any regularization is removed we obtain:
\be
\int \dd y \sqrt{g(y)} \,  \cS(x,y) \int \dd z \sqrt{g(z)} \, G(y,z) \phi(z) = \phi(x) \,.
\ee

Tuning the bare mass to cancel the tadpole and the divergent part of the melonic integral,
and taking the ansatz  
\begin{equation}
G_{\star}(x,y)=\cZ C(x,y) = \cZ \frac{c(\Delta)}{s(x,y)^{2\Delta}} \,,
\label{eq:ansatzF}
\end{equation}
where the coefficient $c(\D)$ is defined in \eqref{eq:freeC-flat}, we obtain
\begin{align}  \label{eq:SDeq2}
\l^2 \cZ^4 \int \dd z \sqrt{g(z)}\, C(x,z)\left( C(z,y)^3  -\f{\delta(z-y)}{\sqrt{g(z)}} B \right) =(1-\cZ)\f{\delta(x-y)}{ \sqrt{g(x)}} \,.
\end{align}
The constant $B$ comes from the mass counterterm
$m^{d/2} = - \l_2  C(x,x) - B$
and should be chosen so as to cancel the divergence of the convolution of $C$ with $C^3$ and yield a delta function as a result.
Since for $\D=d/4$ we have $C(x,y)^3 = c(\D)^3/s(x,y)^{2(d-\D)}$, by comparison to footnote~\ref{foot:Dz-subtr} we have: 
\be
B = c(\D)^3  \int_{s(u,y)>r} \dd u \sqrt{g(u)}\,\f{1}{s(u,y)^{2(d-\Delta)}} - \f{c(\D)^3}{c(d-\D)}\f{\G(d-\D)}{\G(\D)} \,.
\ee
In the spirit of defining (power-) divergent quantities by analytic continuation, we might instead simply set $B=0$.
Either way we obtain for $\cZ$ the equation:
\be
\l^2 \cZ^4 \f{c(\D)^3}{c(d-\D)} = 1-\cZ \,,
\ee
or equivalently:
\be \label{eq:Catalan-Z}
\cZ=1+\l^2 \cZ^4\frac{4\Gamma(1-d/4)}{d(4\pi)^d\Gamma(3d/4)} \,,
\ee
which is the same equation as \eqref{eq:wave} in flat space.\footnote{with $\lambda \rightarrow \im \lambda$} The solution of this equation is the generating function of 4-Catalan (or Fuss-Catalan) numbers:
\be \label{eq:4Catalan}
\cZ(\l) = \sum_{n=0}^{+\infty} \f{1}{4n+1} \binom{4n+1}{n} \left( - \l^2 \f{c(\D)^3}{c(d-\D)}\right)^n \, ,
\ee
and can also be written in an explicit closed form (see\ \cite{Bonzom:2011zz}). Here we just point out that it has a square root singularity at
\be \label{eq:lambda_c}
\l^2_c =  - \f{3^3}{2^8}  \f{c(d-\D)}{c(\D)^3} =    \f{3^3}{2^8} \frac{d(4\pi)^d\Gamma(3d/4)}{4\Gamma(1-d/4)}  \,,
\ee
at which $\cZ(\l_c)=4/3$. 
The series \eqref{eq:4Catalan} resums all the melonic insertions in the two-point function and the critical point corresponds to the radius of convergence of this series, thus determining the maximal value of the coupling for which the model is defined, which we defined as $g_c$ in chapter \ref{chap:CTKT}.

In conclusion, at large $N$ the two-point function of the long-range $O(N)^3$ model is proportional to the free propagator, with proportionality constant satisfying \eqref{eq:Catalan-Z}.
This holds both on flat and spherical background, and the two-point function on $S^d$ is obtained from the two-point function in $\mathbb{R}^d$ by simply replacing the distance in $\mathbb{R}^d$ by the chordal distance in $S^d$, as expected.

\subsection{The sphere free energy at leading order in the large-$N$ expansion}
\label{sec:ON3-model-LO}

We now want to evaluate the on-shell 2PI effective action, i.e.\ the free energy. We then start from \eqref{eq:ON3_Gamma_2PI}, and replace $G$ by the solution \eqref{eq:ansatzF}: 
\be \label{eq:Gamma_2PI_on_shell}
\begin{split}
\mathbf{\G}[G] = & N^3\left(\f12 \cZ \Tr[C^{-1} C]  + \f12 \Tr[\ln (\cZ^{-1} C^{-1})] + \f{m^{2\zeta}}{2} \cZ \int_x C(x,x) \right.\\
&\qquad\quad \left.+ \f{\l_2 \cZ^2}{4} \int_x C(x,x)^2 +\f{\l^2 \cZ^4}{8}\int_{x,y} C(x,y)^4\right)\,.
\end{split}
\ee
We have five terms to evaluate, all of which are UV divergent. The first four are similar to those in the $O(N)$ model, except for some $\cZ $ coefficients and for the long-range exponent $\z=d/4<1$.
The latter plays no role in the first term, which is proportional to $\Tr[\mathbf{1}]$, and it is set to zero by analytic continuation in $d$, as before. Similarly, the $\ln(\cZ^{-1})\Tr[\mathbf{1}]$ coming from the second term can be dropped. Therefore, the first two terms reproduce \eqref{eq:F_GFFT}.

In order to compute the next two terms, we need first to compute $C(x,x)$ for long-range scaling:
\begin{equation}
C(x,x)=\frac{(d-1)!}{a^{d}\Gamma(d/2)(4\pi)^{d/2}}\sum_{n=0}^{\infty}\frac{D_n}{\omega_n^{(\zeta)}} \,.
\end{equation}
Using dimensional regularization, we find again $C(x,x)=0$, see appendix~\ref{app:tadpole-LR}. 

Finally, we need to evaluate the melon integral, which at leading order is the only qualitative difference with respect to the $O(N)$ model. We call it $M$:
\begin{equation} \label{eq:melon-int}
M= \int_{x,y}  \, C(x,y)^4 \,.
\end{equation}

In order to regulate the UV divergences we set $\Delta=\frac{d-\epsilon}{4}$, and obtain the $\epsilon$ regularized melon integral:
\begin{equation} \label{eq:Mepsilon}
M_\eps 
=c(\Delta)^4 \int_{x,y}  \,  \frac{1}{s(x,y)^{2(d-\epsilon)}} = c(\Delta)^4 (2 a)^{2\eps}\int \dd  x \int \dd  y \, \f{1}{(1+x^2)^\eps (1+y^2)^\eps |x-y|^{2(d-\eps)}}\,.
\end{equation}
Alternatively, as we already mentioned, $M$ can be thought of as the trace of the convolution $G(x,y)$ with $G(y,x)^3$. On shell and for $\Delta = d/4$ we have $G_{\star}(y,x)^3 \propto C(x,y)^3 \propto 1/s(x,y)^{3d/2} = 1/s(x,y)^{2(d-\D)}$ which is also the two-point function of the shadow field. The regularization can then be viewed as a shift by $\epsilon$ of the dimension of the shadow, that is the replacement $\wtD=d-\D\to\wtD-\eps$ at fixed $\D=d/4$:
\be  \label{eq:Mepsilon-2}
M_\eps = c(\Delta)^4 \int_{x,y}  \,  \la \phi_\D(x) \phi_\D(y) \ra_{\rm cs} \la \phi_{\wtD-\eps}(x) \phi_{\wtD-\eps}(y) \ra_{\rm cs} \,.
\ee
This point of view will be useful in the next subsection.

The  regularized melon integral \eqref{eq:Mepsilon} can be reduced to an integral over a single point by exploiting the homogeneity of the sphere and factoring out a volume of the d-sphere  $V_d=\int \dd z\, \Omega(z)^d$, given in \eqref{eq:Vol-Sd}.
The remaining integral, which has appeared for example in \cite{Cardy:1988cwa,Klebanov:2011gs},  can be computed for $\eps>d/2>0$ and then analytically continued to small $\eps$, leading to: 
\begin{equation}\label{eq:vanishmelo}
M_\eps=\frac{a^{2\epsilon}\, \Gamma(\frac{d+\epsilon}{4})^4 \, \Gamma(-\frac{d}{2}+\epsilon)}{2^{3d-1}\,\pi^{d-1/2}\,\Gamma(\frac{d-\epsilon}{4})^4\,\Gamma(\frac{d+1}{2})\, \Gamma(\epsilon)} \,,
\end{equation}
which vanishes in the limit $\eps\to 0$, whenever $d$ is not an even number.
Given that $C(x,y)^3$ is proportional to $C^{-1}(x,y)$ (see \eqref{eq:SDeq2}), we can interpret this result as another instance of $\Tr[\mathbf{1}]$ being set to zero by analytic regularization.

In conclusion, at leading order the sphere free energy of the interacting long-range $O(N)^3$ model reduces to that of the free model, i.e.\ $N^3$ times the GFFT free energy \eqref{eq:F_GFFT}.
This is an interesting feature, shared with the $O(N)$ model. However, we do not know if there is any deeper reason behind it, or if it is just an accident of these specific large-$N$ limits dominated by tadpole or melonic diagrams.

\subsection{The next-to-next-to-leading order of the large-$N$ expansion}
\label{sec:ON3-model-NLO}

The free energy of the $O(N)^3$ model has a series expansion in $1/\sqrt{N}$ \cite{Carrozza:2015adg}.
At NLO the only 2PI diagram is a figure eight with one tetrahedron vertex \cite{Benedetti:2018goh}, and hence its contribution vanishes like the similar LO contributions from the $\l_2$ coupling.

\begin{figure}[htbp]
\centering
\includegraphics[width=0.4\textwidth]{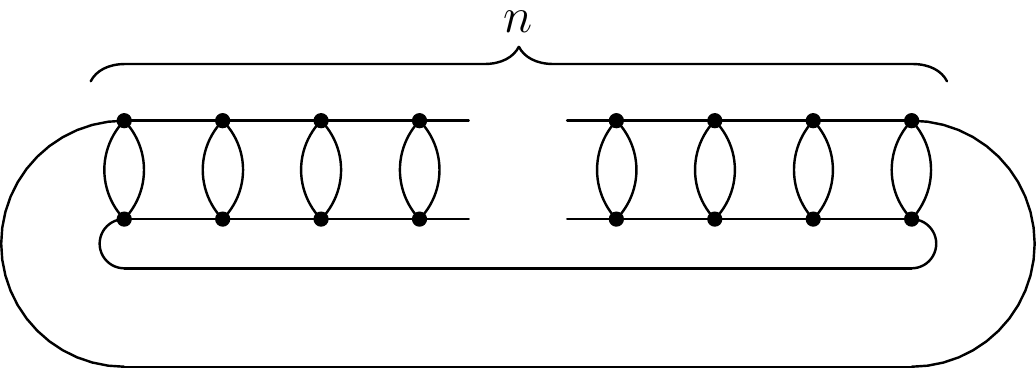}
\caption{A generic NNLO vacuum 2PI diagram having the form of a closed ladder with $n\geq 2$ rungs, and vertices corresponding to the tetrahedron interaction. Similar diagrams but  with a twist in the rails, thus forming a M\"obius strip, appear only at lower order in $N$.
}
\label{fig:laddersF}
\end{figure}

\begin{figure}[htbp]
	\centering
	\includegraphics[width=0.20\linewidth]{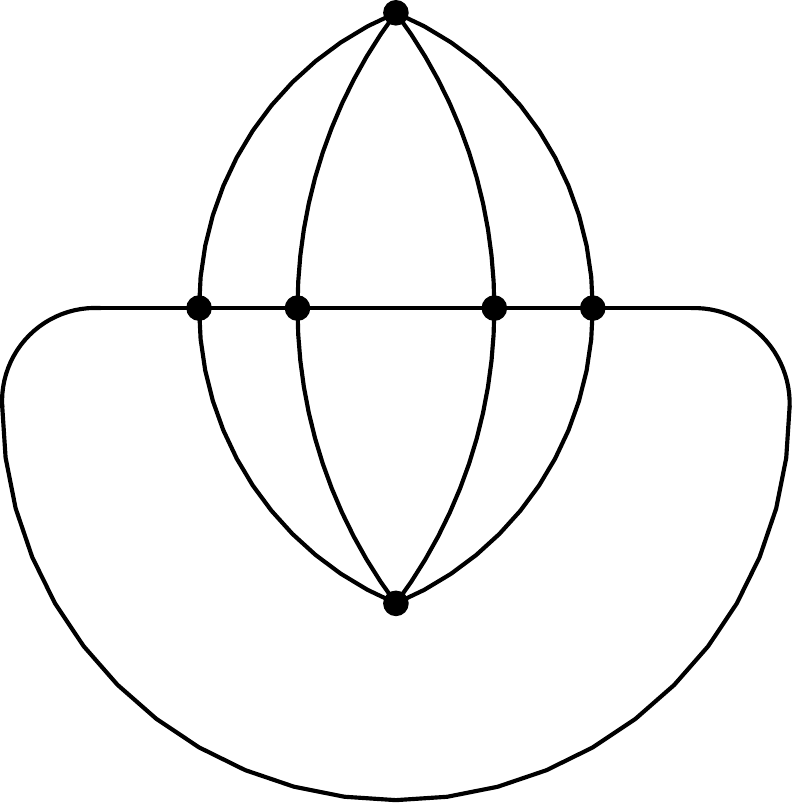}
	\caption{The unique NNLO vacuum 2PI diagram besides the ladders. All six vertices are tetrahedral.}
	\label{fig:monstru}
\end{figure}

At NNLO the combinatorics of the $O(N)^3$ model with only the tetrahedron interaction has been carefully studied in  \cite{Bonzom:2019yik}.
Restricting to 2PI diagrams, it turns out that there is an infinite family of ladder-like diagrams, closed in a planar way as shown in figure~\ref{fig:laddersF}, plus one special diagram, shown in figure~\ref{fig:monstru}. It is straightforward to complement the analysis of  \cite{Bonzom:2019yik} by adding the effect of the pillow and double-trace interactions.
It turns out that we only need to add to those, diagrams obtained from the ladder diagrams by replacing one or more rungs (each made by two $\l$ vertices) with one or more local $\l_1$ vertices, as in the chain diagrams of the $O(N)$ model of figure~\ref{fig:bubble_chain}.\footnote{Similar diagrams with $\l_2$ instead of $\l_1$, appear only farther in the $1/\sqrt{N}$ expansion. In fact, the $\l_2$ coupling is associated to a double-trace interaction, which is the same as the $O(N)$ model interaction, with the replacement $N\to N^3$. Therefore, chains of bubbles with $\l_2$ vertices will only contribute at order $N^0$, as in the $O(N)$ model. }


As we will show in appendix~\ref{app:monster}, the graph of figure~\ref{fig:monstru} gives a finite contribution to the sphere free energy. However, since it only depends on the tetrahedron coupling, it takes the same value at all the fixed points, and thus it does not play a role in checking the $F$-theorem for this model. We will omit it in the rest of this section.

\begin{figure}[htbp]
\begin{center}
$$K_1 = -\l^2\, \vcenter{\hbox{\includegraphics[width=1.8cm]{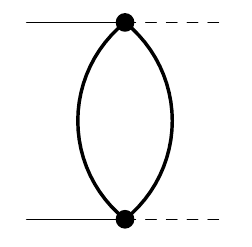}}} - \l_1 \, \vcenter{\hbox{\includegraphics[width=1.8cm]{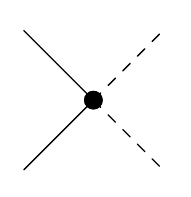}}}$$
 \caption{Graphical representation of the kernel $K_1$ \eqref{eq:ON3-K}. Solid lines represent full two-point functions, while dashed lines represent amputated external legs. For obvious reasons we call the first term ``ladder kernel" and the second ``local kernel".} 
 \label{fig:kernel}
 \end{center}
\end{figure}

We thus have an infinite series of diagrams that are a mixture of ladders and chains. Formally, they can be easily resummed in terms of a kernel that is the sum of a ladder kernel and a local kernel:\footnote{We have subtracted a $\Tr[K_1]$ from the expansion of the logarithm because its ladder contribution does not correspond to a 2PI diagram. As for its local contribution, it is another figure eight diagram, which evaluates to zero, and hence we can add or subtract it at will.}
%
\be \label{eq:ON3_Gamma_2PI_NNLO}
\mathbf{\G}_{\rm NNLO}[G] =  \f{N^2}{2} \left( \Tr[\ln(\mathbb{I}-K_1)] +  \Tr[K_1] \right)\,,
\ee
where $K_1$ is the familiar Bethe-Salpeter kernel of melonic theories, namely
\be \label{eq:ON3-K}
K_1(x_1,x_2,x_3,x_4) = - G(x_1,x_3) G(x_2,x_4) \Big(\l^2 \,  G(x_3,x_4)^2 +\lambda_1 \delta(x_3,x_4)\Big)     \,,
\ee
represented in figure~\ref{fig:kernel}.
%
It should be noted that this kernel displays two differences with respect to the usual ladder kernel of melonic theories found in the literature (e.g.\ \cite{Klebanov:2016xxf,Giombi:2017dtl}). First, we have a minus sign in the ladder kernel, due to the imaginary unit multiplying the tetrahedron interaction in \eqref{eq:ON3-action}. Second, a factor $3$ is missing from the ladder kernel. The reason for the latter can be understood from the full Bethe-Salpeter kernel obtained from the leading-order 2PI effective action, namely \eqref{eq:newker}:
\be
\begin{split}
\label{eq:full-K}
\hat{K}_{\mba\mbb; \mbc\mbd}(x_1,x_2,x_3,x_4) & = -G(x_1,x_3) G(x_2,x_4) \Big(\l^2 \,  G(x_3,x_4)^2 +\lambda_1 \delta(x_3,x_4)\Big) \hat{P}^{(1)}_{\mba\mbb; \mbc\mbd} \\
& \quad - G(x_1,x_3) G(x_2,x_4)  \Big(3\l^2 \,  G(x_3,x_4)^2 +\lambda_2 \delta(x_3,x_4)\Big) \hat{P}^{(2)}_{\mba\mbb; \mbc\mbd} \\
& \equiv  K_1(x_1,x_2,x_3,x_4) \hat{P}^{(1)}_{\mba\mbb; \mbc\mbd} + K_2(x_1,x_2,x_3,x_4) \hat{P}^{(2)}_{\mba\mbb; \mbc\mbd}\,.
\end{split}
\ee
In most of the previous literature on these models the focus was on the spectrum of singlet operators, which can be obtained by the $s$-channel OPE of the four-point function $\la \phi_{\mba}(x_1)\phi_{\mba}(x_2) \phi_{\mbb}(x_3) \phi_{\mbb}(x_4)\ra$. The latter requires inverting $(\mathbb{I}-\hat{K})_{\mba\mbb; \mbc\mbd}$ and taking a trace on the tensor indices by contracting with $\delta_{\mba\mbb}\delta_{\mbc\mbd}$. In such case, the first term is zero because $\hat{P}^{(1)}$ is traceless in this channel, and thus we have a factor $3\l^2$ in the ladder part of the surviving kernel contribution. However, the index trace leading to \eqref{eq:ON3_Gamma_2PI_NNLO} is taken by contracting with $\delta_{\mba\mbc}\delta_{\mbb\mbd}$.  In this case, the $\hat{P}^{(1)}$ term gives the leading contribution of order $N^2$, while the $\hat{P}^{(2)}$ term only contributes at order $N^0$, thus explaining the absence of the factor $3$ in \eqref{eq:ON3-K}. 
Notice that the same part of the kernel is the relevant one for the four-point function $\hat{P}^{(1)}_{\mba\mbb; \mba'\mbb'} \hat{P}^{(1)}_{\mbc\mbd; \mbc'\mbd'} \la \phi_{\mba'}(x_1)\phi_{\mbb'}(x_2) \phi_{\mbc'}(x_3) \phi_{\mbd'}(x_4)\ra$, whose  $s$-channel OPE provides the spectrum of bilinear operators that are in a symmetric-traceless matrix representation of one of the $O(N)$'s, and the singlet one of the other two.

As before,  in order to evaluate the NNLO free energy we substitute \eqref{eq:ansatzF} into \eqref{eq:ON3_Gamma_2PI_NNLO}, and we use the conformal partial wave formalism reviewed in appendix~\ref{app:CPW}.
Inserting the resolution of the identity \eqref{eq:res-id-nonsymm} inside the trace in \eqref{eq:ON3_Gamma_2PI_NNLO}, we find the following formal expression:
\be 
\begin{split}
F_{\rm NNLO} &= \f{N^2}{2} \sum_{J\in \mathbb{N}_0}  \int_{\f{d}{2}}^{\f{d}{2}+\im\infty}  \f{{\rm d}h}{2\pi\im} \r(h,J) \,\big(\ln(1-k(h,J))+k(h,J)\big)\, \cN^{\D}_{h,J}  \cN^{\wtD}_{\htilde,J} \, \Tr[ \Psi_{h,J}^{\D,\D,\wtD,\wtD} ]  \,,
\end{split}
\label{eq:F_NNLO}
\ee
with notation defined in appendix~\ref{app:CPW}.
In addition, we here have $\D=d/4$, and the kernel eigenvalue 
\begin{equation}
k(h,J)=-\f{g^2}{(4\pi)^d}
 \frac{\Gamma(-\frac{d}{4}+\frac{h+J}{2})\Gamma(\frac{d}{4}-\frac{h-J}{2})}{\Gamma(\frac{3d}{4}-\frac{h-J}{2})\Gamma(\frac{d}{4}+\frac{h+J}{2})} \,,
\label{eq:ON3-k}
\end{equation}
where $g$ is the effective tetrahedral coupling $g= \l\, \cZ(\l)^2 $, which resums all the two-point melonic insertions. The latter are absent by construction in the 2PI effective action, but reappear when going on shell, i.e.\ when replacing the generic $G$ by the solution of the SD equations $G_\star(x,y) = {\cal Z}(\l) C(x,y)$. By writing all quantities in terms of $g$ we can keep ignoring the melonic insertions and use the free propagator, but we should restrict its range to $|g|<g_c\equiv \l_c \, \cZ(\l_c)^2 $, because of the square root singularity at the critical coupling \eqref{eq:lambda_c}.

It will actually be convenient to consider the derivative of the free energy in order to get rid of the logarithm:
\be \label{eq:F-derviative}
- g\f{\p}{\p g}F_{\rm NNLO}  = N^2 \sum_{J\in \mathbb{N}_0}  \int_{\f{d}{2}}^{\f{d}{2}+\im\infty}  \f{{\rm d}h}{2\pi\im} \r(h,J) \,\f{k(h,J)^2}{1-k(h,J)}\, \cN^{\D}_{h,J}  \cN^{\wtD}_{\htilde,J} \, \Tr[ \Psi_{h,J}^{\D,\D,\wtD,\wtD} ]   \,.
\ee

A striking feature of \eqref{eq:F_NNLO}, or \eqref{eq:F-derviative}, is that the kernel eigenvalue is only sensitive to the ladder part of the kernel, because the local part has vanishing eigenvalue on the principal series.\footnote{This is straightforward for ${\rm Re}(h)>d/2$, and it is extended by analytic continuation to the principal series and beyond.} 
This fact can be puzzling, as the diagrams having $\l_1$ vertices are necessary at the perturbative level: expressing $\l_1$ as a series in the renormalized coupling $g_1$ and in $g$, they have to cancel the UV divergences of the ladder diagrams, as seen in chapter \ref{chap:CTKT}.
Nevertheless, the result of the resummed series of diagrams, evaluated at the fixed point, where $g_1$ takes a specific $g$-dependent value, turns out to be expressible in terms of only ladder diagrams.
 This is a familiar situation in the four-point function of these models,\footnote{As well as in the fishnet model \cite{Grabner:2017pgm,Kazakov:2018qbr}.} and it is due to the fact that in the conformal limit, the local kernel has zero eigenvalues.  
The resummed series captures the contribution of the chain diagrams in a subtle manner. When evaluating \eqref{eq:F-derviative} using conformal partial waves, only the ladder kernel contributes, and thus one needs to integrate over the principal series an analytic function of $g^2$. However, the result of the integration (for $J=0$) is a non-analytic function with a $\sqrt{g^2}$ branch cut. In the perturbative expansion such a branch cut can only come from the $\lambda_1$ diagrams, due to the branch cut in the $g_{1\pm}$ fixed points \eqref{eq:FP-ON3}. 
Therefore, the non-perturbative resummation of the ladder diagrams automatically includes the contribution of the chain diagrams as well, which is a very non-trivial fact.
We will provide a cross check of this statement below.

The problem with the expression \eqref{eq:F-derviative} is that the trace of the conformal partial wave is divergent. From \eqref{eq:CPW} we have:
\be \label{eq:TrCPW}
\Tr[ \Psi_{h,J}^{\D,\D,\wtD,\wtD} ] = \int_{x_1,x_2,z}  \,  \la \phi_{\D}(x_1) \phi_{\D}(x_2) \cO_h^{\m_1\cdots \m_J}(z) \ra_{\rm cs}\la \phi_{\wtD}(x_1) \phi_{\wtD}(x_2) {\cO}_{\htilde}^{\m_1\cdots \m_J}(z)\ra_{\rm cs} \,.
\ee
Formally this integral is conformally invariant, but as a consequence it is also divergent because of the infinite volume of the conformal group. Notice that the same type of integral appears as a natural pairing (or inner product) of $n$-point functions \cite{Karateev:2018oml}; however, in that case one divides by the volume of $SO(d+1,1)$ (or in other words, one considers a gauge-fixed version of the integral) in order to define a finite pairing. 
In our case, we do not have the freedom to divide the free energy by a diverging quantity: the idea of the $F$-theorem is  that instead we should look at the finite part of the free energy. This might be hiding behind some divergence, which we have to regulate and subtract.
The melon integral in \eqref{eq:melon-int} is an example of the same kind: for $\D=d/4$ it is proportional to a pairing of two-point structures $\int \dd x_1 \dd x_2   \, \la \phi_{\D}(x_1) \phi_{\D}(x_2)  \ra_{\rm cs}\la  \phi_{\wtD}(x_1) \phi_{\wtD}(x_2)\ra_{\rm cs}$, and it is divergent for the same reason as above. We have regularized the melon integral by analytic continuation in \eqref{eq:Mepsilon-2}, which is equivalent to subtracting the divergent part, and we have found a vanishing finite part.
We are now going to employ a similar analytic continuation in order to extract the finite part of \eqref{eq:TrCPW}.

In the case of \eqref{eq:TrCPW}, setting $\Delta=\frac{d-\epsilon}{4}$ everywhere would not help, as the dependence of the integrand on $\D$ drops out. It is thus useful to take the second point of view we presented on the regularization of the melon integral and shift only the dimensions of the shadow operators. That is, we define:
\be \label{Iepsilon}
\cI_\eps(J) =  \int_{x_1,x_2,z}  \,  \la \phi_\D(x_1) \phi_\D(x_2) \cO_h^{\m_1\cdots \m_J}(z) \ra_{\rm cs}\la \phi_{\wtD-\eps}(x_1)\phi_{\wtD-\eps}(x_2) {\cO}_{\htilde-\eps}^{\m_1\cdots \m_J}(z)\ra_{\rm cs} \,.
\ee
This analytic regularization breaks the conformal invariance of the integral, but not its translation invariance. Therefore, on flat space there is still a space-time volume divergence, which is instead regularized on the sphere. UV divergences (at coincident points) are still there, but will be cured in an appropriate range of $\eps$, and then we will use analytic continuation to take the limit $\eps\to 0$.

It can be shown (see appendix~\ref{app:I_eps}) that all the $J$-dependence in \eqref{Iepsilon} can be factored out of the integral
\begin{equation}
\cI_\eps (J)= \frac{\Gamma(d-2+J)\Gamma(\tfrac{d-2}{2})}{2^J\Gamma(d-2)\Gamma(\tfrac{d-2}{2}+J)} \cI_\eps (0)\,,
\end{equation}
with 
\begin{equation}\label{eq:I0e}
\cI_\eps (0)=\int \dd x_1 \dd x_2   \dd  z 
\frac{\left( \Omega(x_1)\Omega(x_2) \Omega(z)\right)^d}{s(x_1,x_2)^{d-\eps}s(x_1,z)^{d-\eps} s(x_2,z)^{d-\eps}}\,.
\end{equation}
Because of the homogeneity of the sphere we are free to set $z=0$, and factor out  the volume of the $d$-sphere  $V_d=\int \dd z\, \Omega(z)^d$, given in \eqref{eq:Vol-Sd}. The integral $\frac{\cI_\eps(0)}{V_d}$ has already been computed  in \cite{Cardy:1988cwa} and the results is\footnote{In appendix~\ref{app:I_eps} we provide a detailed derivation.}
\begin{equation} \label{eq:I_eps}
	\frac{\cI_\eps(0)}{V_d}=(2a)^{3\epsilon-d}\frac{\pi ^d \Gamma \left(\frac{\epsilon }{2}\right)^3 \Gamma \left(\frac{3 \epsilon }{2}-\frac{d}{2}\right)}{\Gamma \left(\frac{d}{2}\right) \Gamma \left(\epsilon\right)^3} \,,
\end{equation}
which has a finite limit for $\epsilon \rightarrow 0$ as long as $d$ is not an even number:
\begin{equation}
\lim_{\eps \to 0}	\frac{\cI_\eps(0)}{V_d}=\frac{8 \left(\pi ^d \Gamma \left(-\frac{d}{2}\right)\right)}{(2a)^d\Gamma \left(\frac{d}{2}\right)} \,.
\end{equation}
Reinserting the spin and volume factors we have:
\begin{equation}
\cI_\eps (J)= (2a)^{3\epsilon}\frac{\pi^{3d/2}\Gamma(\frac{\eps}{2})^3\Gamma(\frac{3\eps}{2}-\tfrac{d}{2})\Gamma(d-2+J)\Gamma(\tfrac{d-2}{2})}{2^J\Gamma(\eps)^3\Gamma(d-2)\Gamma(\tfrac{d-2}{2}+J)\Gamma(d)}\,.
\end{equation}
and after removing the regulator $\epsilon$ we get:
\begin{equation}
\cI_0 (J)= \frac{8\pi^{3d/2}\Gamma(-\tfrac{d}{2})\Gamma(d-2+J)\Gamma(\tfrac{d-2}{2})}{2^J\Gamma(d-2)\Gamma(\tfrac{d-2}{2}+J)\Gamma(d)} \,.
\end{equation}

The $\epsilon$ regularization thus provides a finite result for the trace of the conformal partial wave.
However, it turns out that it is important to consistently shift by $\epsilon$ also the normalization factor of the three-point function of shadow operators, as otherwise the resulting series in $J$ would diverge.
The product of normalization factors \eqref{eq:prod-cN} is then replaced, at large $J$, by:
\be
\cN^{\D}_{h,J}  \cN^{\wtD-\eps}_{\htilde-\eps,J} \sim \f{2^{3(d+\eps)/2+J}}{(2\pi)^{d}} \, J^{-3\eps} \left( 1 + \mathcal{O}(1/J) \right) \,.
\ee
This $J^{-3\eps} $ factor renders the series over $J$, whose coefficients otherwise behaves asymptotically as $1/J$, convergent.

We can now perform the integral on $h$ and sum over $J$ in \eqref{eq:F-derviative}.
As explained in appendix~\ref{app:NumericsLargeJ}, at large $J$, the integral behaves as $\frac{f(\epsilon)}{J^{1+3\epsilon}}$ with $f(\epsilon)$ an analytic function at $\epsilon=0$. We then write:
\be\label{eq:ladderssum}
\begin{split}
- g\f{\p}{\p g}F_{\rm NNLO}^{\eps}  =  &\, N^2 \int_{\f{d}{2}}^{\f{d}{2}+\im\infty}  \f{{\rm d}h}{2\pi\im} \r(h,0) \,\f{k(h,0)^2}{1-k(h,0)}\, \cN^{\D}_{h,0}  \cN^{\wtD-\eps}_{\htilde-\eps,0} \, \cI_\eps (0) \\
&+ N^2 \Bigg[ \sum_{J\in \mathbb{N}_+} \Bigg( \int_{\f{d}{2}}^{\f{d}{2}+\im\infty}  \f{{\rm d}h}{2\pi\im} \r(h,J) \,\f{k(h,J)^2}{1-k(h,J)}\, \cN^{\D}_{h,J}  \cN^{\wtD-\eps}_{\htilde-\eps,J} \, \cI_\eps (J) -\frac{f(\epsilon)}{J^{1+3\epsilon}}\Bigg) \\
 & \quad + \sum_{J\in \mathbb{N}_+} \frac{f(\epsilon)}{J^{1+3\epsilon}} \Bigg] \,.
\end{split}
\ee
The first sum is now convergent for $\epsilon=0$, and thus can be computed numerically, while the second sum gives and explicit pole in $\eps$.
We thus define the renormalized sphere free energy, or rather its derivative, as:
\be
- g\f{\p}{\p g}F_{\rm NNLO} = \lim_{\eps\to 0} \left(- g\f{\p}{\p g}F_{\rm NNLO}^{\eps} - \frac{N^2 f(0)}{3\epsilon} \right) \,,
\ee
which for example at $d=3$, $g=1$ and $a=1$ gives:
\be
- g\f{\p}{\p g}F_{\rm NNLO}   = 7.57 \times 10^{-4} \, N^2 \, .
\ee

\begin{figure}[htbp]
\centering
\includegraphics[scale=0.5]{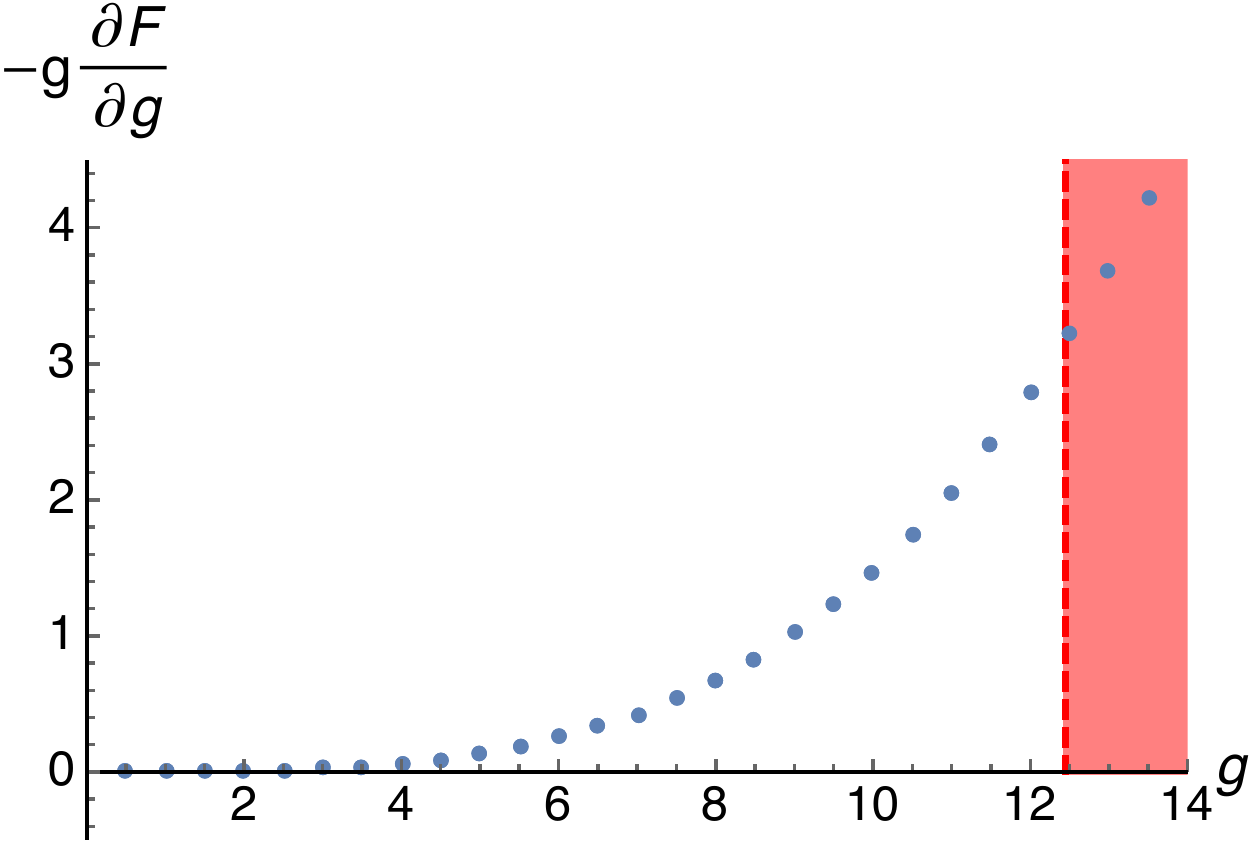}
\caption{Derivative of the free energy at $d=3$ and $a=1$. The red area corresponds to $g>g_c$, where nothing seems to happen, but in fact there is no $\l$ giving such values of $g$.}
\label{fig:plotDF}
\end{figure}

We computed this value at $d=3$, for $a=1$ and different values of $g$ up to $g_c\equiv \l_c \, \cZ(\l_c)^2$ (with $\l_c$ given in \eqref{eq:lambda_c}). The result is a positive convex function, vanishing at the origin, as shown in figure~\ref{fig:plotDF}.

\paragraph{The non-normalizable contribution.}
In the context of the $F$-theorem we are interested in the difference between the free energy at the UV fixed point and at the IR fixed point. The value of $g$ being the same at the two fixed points, and the above result being seemingly independent of $g_1$ or $g_2$, it would naively seem that the free energy is the same at the two fixed points. 

Things are however more subtle than this. As explained in appendix~\ref{app:CPW}, the resolution of the identity \eqref{eq:res-id-nonsymm} is valid in a functional space with appropriate integrability conditions, and the latter are violated by four-point functions whose $s$-channel OPE contains operators of dimension smaller than $d/2$.
It turns out that this is precisely what happens in the ultraviolet CFT, due to a primary operator in the OPE of $\hat{P}^{(1)}_{\mba\mbb; \mbc\mbd}(\phi_{\mbc} \times\phi_{\mbd} )$ whose dimension descends below $d/2$. 
In fact, at $J=0$, the equation $k(h,0)=1$ has two solutions $h_{\pm}$ lying respectively on the right and on the left of the integration contour $\cP_+$:
\begin{equation}  \label{eq:h_pm}
h_{\pm}=\frac{d}{2} \pm \frac{2\sqrt{g^2}}{\Gamma(d/2)(4\pi)^{d/2}} + \mathcal{O}(|g|^3) \,,
\end{equation} 
and in the UV, the physical dimension is actually the one on the left of the contour. 
Therefore, in evaluating the free energy of the UV theory by the CPW method we need to subtract these contributions from the operator being traced before applying it on the resolution of the identity \eqref{eq:res-id-nonsymm}, and then add them back.
This amounts to including, besides  the principal series integral, an isolated non-normalizable contribution, as in \eqref{eq:cF-extra}.
That is, in the UV version of \eqref{eq:F-derviative} we have to add minus the residue of the integrand at $h=h_-$.

\begin{figure}[htbp]
\centering
\includegraphics[scale=0.5]{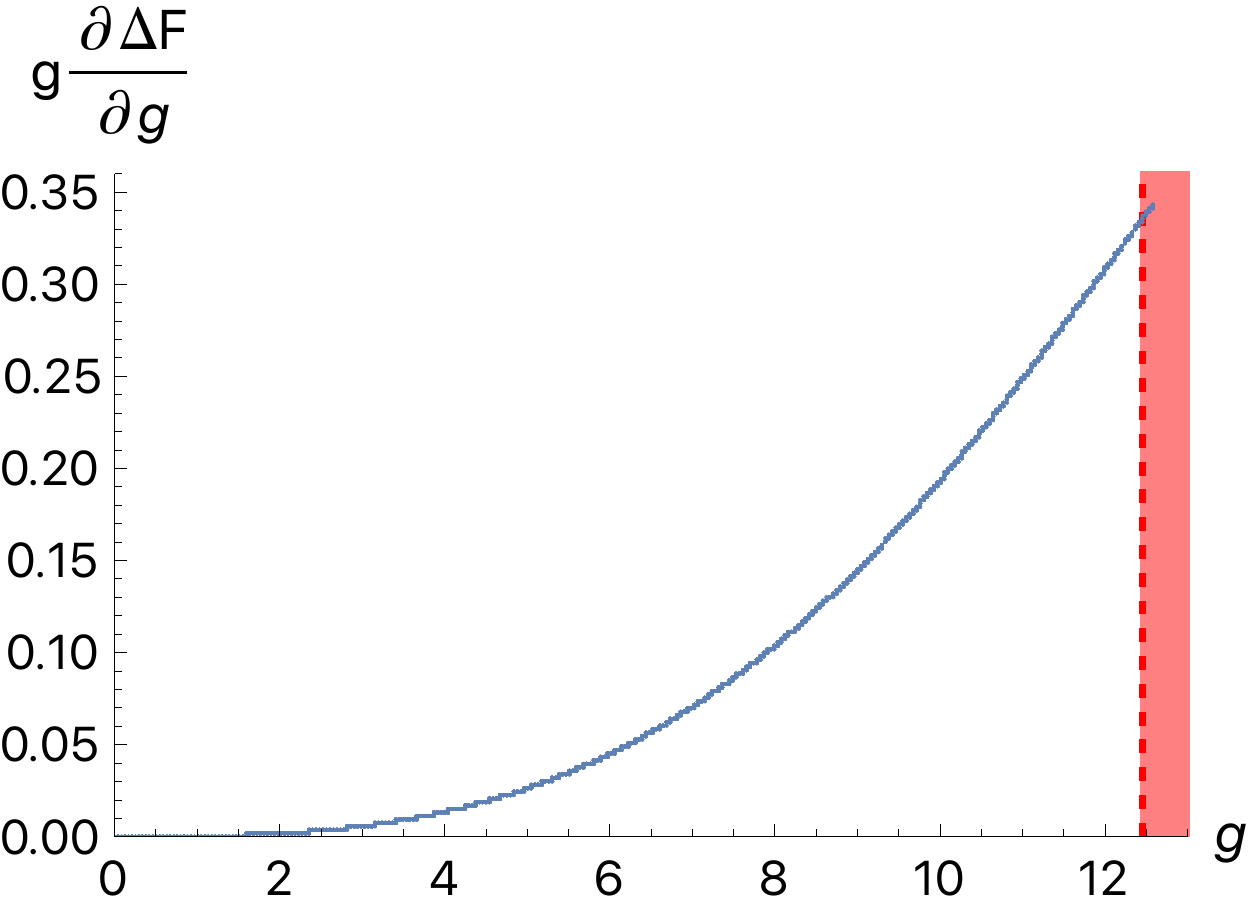}
\caption{Difference between the free energy in the UV and the free energy in the IR at $d=3$.  The red area corresponds to $g>g_c$.}
\label{fig:residue}
\end{figure}

With this in mind, it is clear that the difference between the free energy of the UV theory and the one of the IR theory is given precisely by the isolated non-normalizable contribution of the former.
Going again through the same regularization procedure as in the IR case, we thus find:
\begin{equation} \label{eq:dF_tm}
\begin{split}
g\f{\p}{\p g}\left(F_{\rm NNLO}^{UV}-F_{\rm NNLO}^{IR}\right)
&=  N^2\, {\rm Res}\left[ \r(h,0) \,\f{k(h,0)^2}{1-k(h,0)}\, \cN^{\D}_{h,0}  \cN^{\wtD}_{\htilde,0} \, \cI_0 (0) \right]_{h= h_{-}} \\
& =16\frac{\Gamma(-d/2) |g|^3}{2^{3d}\pi^{3d/2}\Gamma(d)} N^2 + \mathcal{O}(|g|^5)\,,
\end{split}
\end{equation}
which is positive for $2<d<4$. By numerical evaluation at finite $g$, it can be checked that the positivity remains valid also at all values of $g$, within the radius of convergence of the melonic series (see figure~\ref{fig:residue}).

\begin{figure}[htbp]
\centering
\includegraphics[scale=1]{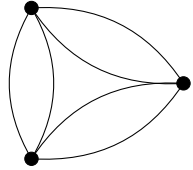}
\caption{Feynman graph contributing to the free energy at order $3$ in the coupling constant. The vertices are either two tetrahedron and one $g_1$ or three $g_1$.}
\label{fig:triangle}
\end{figure}

This result can also be checked perturbatively. At the fixed points the critical coupling $g_1$ chapters \ref{chap:CTKT} and \ref{chap:trif}:
\begin{equation}
g_{1\pm}=\pm \sqrt{g^2}\big( 1+ \mathcal{O}(g^2)\big) + g^2 \big(\psi(1)+\psi(d/2)-2\psi(d/4)  + \mathcal{O}(g^2)\big)   \,,
\end{equation}
where  $\psi(z)$ is the digamma function. When flowing from the UV to the IR, the fixed point value goes from $g_{1-}\simeq -\sqrt{g^2}$ to $g_{1+}\simeq \sqrt{g^2}$  (except for the vertical trajectories in  figure~\ref{fig:trajectory}, which we will discuss below): therefore, at leading order in $g$, graphs with an even number of vertices have the same amplitude in the UV as in the IR, while graphs with an odd number of vertices have opposite signs. The difference between the free energy in the UV and the free energy in the IR is thus expanded in odd powers of $|g|$. Up to order $|g|^3$, only the graph of figure~\ref{fig:triangle} contributes, where the vertices are either two tetrahedron and one $g_1$ or three $g_1$.

Using the expression of the kernel given in \eqref{eq:full-K} to fix the combinatorial factors, we have:
\begin{equation}
 g\f{\p}{\p g}(F_{\rm NNLO}^{UV}- F_{\rm NNLO}^{IR}) =2 |g|^3 N^2 \mathcal{A}  + \cO(|g|^5)\;,
\end{equation}
where we denoted $\mathcal{A}$ the amplitude of the graph in figure~\ref{fig:triangle}:
\begin{equation}
\mathcal{A}=c(\tfrac{d}{4})^6 \int d^dx d^dy d^dz \frac{\left( \Omega(x)\Omega(y) \Omega(z)\right)^d}{s(x,y)^{4\Delta}s(x,z)^{4\Delta} s(y,z)^{4\Delta}} \;.
\end{equation}

Using again dimensional regularization with $\Delta=\frac{d-\epsilon}{4}$, we notice that this amplitude is proportional to $\cI_\eps (0)$ in \eqref{eq:I0e}. We then obtain:
\begin{equation}
\mathcal{A}=8\frac{\Gamma(-d/2) }{2^{3d}\pi^{3d/2}\Gamma(d)} \quad \Rightarrow \quad
g\f{\p}{\p g}\left(F_{\rm NNLO}^{UV}-F_{\rm NNLO}^{IR}\right)=16\frac{\Gamma(-d/2) |g|^3}{2^{3d}\pi^{3d/2}\Gamma(d)} N^2+ \cO(|g|^5)\;,
\end{equation} 
in agreement with \eqref{eq:dF_tm}.

In conclusion, the difference between the sphere free energy at the fixed points $(g_{1-},g_{2-})$ or $(g_{1-},g_{2+})$ and the one at the fixed points $(g_{1+},g_{2-})$ or $(g_{1+},g_{2+})$ (see figure~\ref{fig:trajectory})\footnote{Notice that $(g_{1+},g_{2-})$ can be reached only through one trajectory emanating from $(g_{1-},g_{2-})$, while $(g_{1-},g_{2+})$ has only one trajectory to flow to an IR fixed point, i.e.\ $(g_{1+},g_{2+})$. Between $(g_{1-},g_{2-})$ and $(g_{1+},g_{2+})$ there is instead an infinite set of trajectories.} grows with growing $|g|$. Since the difference vanishes at $g=0$, we conclude that for the fixed points at fixed $g\neq 0$ the sphere free energy satisfies $F_{\rm NNLO}^{UV}>F_{\rm NNLO}^{IR}$, in accordance with the $F$-theorem.

\paragraph{Trajectories at fixed $g_1$.}
The reader will note in figure~\ref{fig:trajectory} the presence of two vertical lines: these are two trajectories at fixed $g_1$ connecting two different pairs of fixed points. As neither the tetrahedral coupling nor $g_1$ change along these trajectories, the above computation implies that at NNLO the free energy at the two ends of such trajectories is the same. We expect that in order to see a change of the free energy along these trajectories one would need to push the evaluation to lower orders in $1/N$.

The first non-trivial contribution involving the double trace coupling $g_2$, which does vary along the vertical trajectories, comes from ladder graphs generated by the double trace part $K_2 \hat P^{(2)}$ of the Bethe-Salpeter kernel \eqref{eq:full-K}. They essentially behave the same as the ladders generated by $K_1 \hat P^{(1)}$, up to replacing $g^2$ by $3g^2$, but the isolated contribution to the conformal partial wave expansion in the UV would in this case depend on the fixed-point value of $g_2$. 

However, it should be noted that, as such ladders appear only at order $N^0$, one should include the effect of the $1/N$ corrections to the on-shell two-point function as well as to the fixed-point values of the couplings. We know from chapter \ref{chap:trif} that such corrections have a drastic effect on the fixed-point theory: in order to find a non-trivial precursor of the large-$N$ theory, one needs to keep a finite $\eps$, with $\eps N\ll 1$; then, the lines of fixed points collapse to isolated points and the scaling dimensions acquire an imaginary part.
The whole analysis becomes much more involved, and since in this case the theory is manifestly non-unitary we do not expect the $F$-theorem to necessarily hold.

\section{Conclusions}
\label{sec:concl}

Our main result is the confirmation that the $F$-theorem holds for the long-range $O(N)^3$ model. We summarize here the content of the chapter and the main steps leading to this conclusion. 

\paragraph{Flow between Gaussian CFTs.} As a warm up, we first considered two Gaussian CFTs with action 
$\frac{1}{2} \int \dd x\phi(x)(-\partial^2)^{\zeta} \phi(x)$, one being the standard short-range action $\zeta =1$ and the second one having $\zeta \neq 1$ (also known as generalized free field theory). We examined the flow between two such CFTs and found that, on the one hand, the RG always flows in the infrared towards lower $\zeta$, while on the other hand the sphere free energy is concave with a maximum at $\zeta=1$. An RG trajectory flowing from a short-range model $\zeta=1$ to a long-range one $0<\zeta<1$ thus satisfies the $F$-theorem. However a trajectory starting in the ``strong short-range'' regime $\zeta >1$ is such that the free energy increases in the IR. This gives a trivial counter-example of the $F$-theorem for non-unitary (and non-local) theories. 

\paragraph{Revisiting the $O(N)$ model: conformal partial wave expansion.} Next, we revisited the vector $O(N)$ model and rederived its sphere free energy, previously obtained in \cite{Klebanov:2011gs}. This was not a new result but this computation allowed us to introduce the set of techniques relevant for the rest of the chapter. 

We first briefly recalled the formalism of the two-particle irreducible (2PI) effective action, which is particularly natural for discussing the free energy of large-$N$ models. At large $N$, the free energy of the interacting fixed point is the same as in the free theory, hence to test the $F$-theorem one needs to go beyond the leading order.
The diagrams contributing to the free energy at next-to-leading order (NLO) in $1/N$ are resummed to $F_{\rm NLO} = \frac{1}{2} \Tr\ln( 1-K )$, where $K$ is the Bethe-Salpeter kernel, which at the relevant order is a local operator.

We then analyzed $ {\cal F}_s $ and $F_{\rm NLO}$ using the conformal partial waves (CPW) expansion. The use of this technique for the sphere free energy is new, and it has the advantage that it can be generalized to models with different kernels $K$. 
In the $O(N)$ model, the CPW expansion of the bilocal to bilocal identity operator includes the principal series and an additional non-normalizable state, as the dimension of the field is smaller than $d/4$. When evaluating the four-point function ${\cal F}_s$ in the critical model, one finds that the Bethe-Salpeter kernel is zero on the principal series, and infinite on the non-normalizable state, thus 
${\cal F}_s$ reduces to the free contribution ${\cal F}^{\rm free}_s$ restricted to the principal series. This implies the well-known result that the spectrum of the critical $O(N)$ model at large $N$ is the same as in the free theory, except that the $\phi^2$ operator is replaced by its shadow \cite{Gubser:2002vv}.

Using the same CPW expansion of the identity for $F_{\rm NLO}$, we obtained a zero contribution from the principal series and a non-trivial one from the isolated non-normalizable state. This isolated contribution gives the only non-trivial part and reproduces by itself the value of $F_{\rm NLO}$ computed in \cite{Klebanov:2011gs}.

\paragraph{The long-range $O(N)^3$ model on the sphere.}
Finally, we studied the long-range $O(N)^3$ model on the sphere.
The situation is similar to the one in the $O(N)$ model: 
 at the first non-trivial order in $1/N$, i.e.\ at next-to-next-to-leading order (NNLO), the diagrams contributing  to the free energy  (up to an exceptional diagram which is finite and equal at all the fixed points) are resummed to $F_{\rm NNLO} = \frac{1}{2} \Tr\ln( 1-K )$. However, in stark contrast to the $O(N)$ model, $K$ is now a bilocal operator, and the CPW expansion is the only available non-perturbative method of evaluation.

Two different regimes are encountered. In the infrared CFT, $F_{\rm NNLO}$ has a CPW expansion restricted to the principal series only. Such expansion of the free energy can be evaluated non-perturbatively (numerically) showing in particular how the CPW expansion in CFT can be used to resum infinite classes of vacuum diagrams.

In contrast, for the ultraviolet CFT one needs to add a non-normalizable state besides the principal series, because the dimension of one of the primary operators in the $\phi \phi\sim \sum c_{\phi\phi O} \; O$ OPE descends below $d/2$. Although the inclusion of a non-normalizable state is reminiscent of the $O(N)$ model, the situation was conceptually and practically different. 
Due to the inclusion of this non-normalizable state, the free energy decreases between the ultraviolet and the infrared CFT, satisfying the $F$-theorem. In perturbation theory, this jump can be seen as the reversal of the sign of an infinite class of diagrams due to the flow of a coupling constant.

\begin{subappendices}

\section{Useful formulas on $S^d$}
\label{app:useful}

%
%

We collect here some useful formulas about $S^d$ and the spectrum of the Laplace-Beltrami operator on it.

The $d$-dimensional round sphere $S^d$ can be defined by the equation
\be
\sum_{\bar{\m}=1}^{d+1} (\bar{X}^{\bar{\m}})^2 \equiv \sum_{\m=1}^{d} (X^\m)^2 + Z^2= a^2\, ,
\ee
where $\bar{X}^{\bar{\m}}=\{X^\m,Z\}$ are the Cartesian coordinates in the embedding space $\mathbb{R}^{d+1}$, and $a$ is the radius of the sphere.
In the northern and southern hemispheres, the equation can be solved as $Z_\pm(X)= \pm \sqrt{a^2-X^2}$, respectively, with $X^2=\sum_{\m=1}^{d} (X^\m)^2$.

It is convenient to describe the metric on $S^d$ through the (equatorial) stereographic projection to the $d$-dimensional flat space $\mathbb{R}^d$.
The stereographic projection from the North pole to the equatorial plane is obtained by the change of variables
\be \label{eq:x-stereo}
x^{\m}_\pm(X) = \f{X^\m}{1- Z_\pm(X)/a} \,.
\ee
Introducing polar coordinates on the equatorial plane, $\{r=\sqrt{x^2},\theta_1,\ldots,\theta_{d-1}\}$, and denoting $\theta_d$ the additional angular coordinate on $S^d$ (i.e.\ the geodesic distance from the North pole), the stereographic mapping reads simply
\be
r(\theta_d) = a \cot(\theta_d/2)\,.
\ee

In stereographic coordinates, the line element takes the following form:
\begin{equation}
ds^2=\frac{4a^2}{\left(1+x^2\right)^2}\sum_{i=1}^d \, dx_i{}^2\; , \qquad x^2\equiv \sum_{i=1}^d x_i{}^2 \,.
\end{equation}
The metric is conformally flat, i.e.
\be \label{eq:Weyl_g}
g_{\m\n}(x) = \Om(x)^2 \d_{\m\n} \,,
\ee
with
\be \label{eq:Omega}
\Om(x) = \frac{2a}{(1+x^2)} \,.
\ee
The square root of the determinant of the metric is then given by $\sqrt{g}=\Om(x)^d$, 
and the Ricci scalar is $R=d(d-1)/a^2$.
The volume of the $d$-sphere is
\be \label{eq:Vol-Sd}
V_d = \int_{S^d} \dd x\, \sqrt{g(x)} =  \f{ 2 \pi^{\f{d+1}{2}}}{\G\left(\f{d+1}{2}\right)} a^d\, .
\ee

\

The eigenmodes of the scalar Laplacian on the sphere are the spherical harmonics (e.g. \cite{Rubin:1984,Samko-book})
\be \label{eq:spherHarmonics}
\Psi_{n,j} (\bar{X}) = \r^{-n}\, T^{(j)}_{\bar{\m}_1 \cdots \bar{\m}_n} \bar{X}^{\bar{\m}_1} \cdots \bar{X}^{\bar{\m}_n} \,,
\ee
where $n=0,1,2,...+\infty$, $\r=(\bar{X}^{\bar{\m}}\bar{X}_{\bar{\m}})^{1/2}$ and $T^{(j)}_{\bar{\m}_1 \cdots \bar{\m}_n}$ form a basis of constant traceless-symmetric tensors on $\mathbb{R}^{d+1}$, each basis element being labeled by a different value of $j$. Therefore, we take $j=1,2,...D_n$, with 
\be \label{eq:multi}
D_n=\frac{(n+d-2)!\, (2n+d-1)}{n!(d-1)!}\,.
\ee
The corresponding eigenvalues are:
\be
\om_n=n(n+d-1)/a^2 \,.
\ee
They are independent of $j$, hence they have multiplicity $D_n$.

The addition theorem of spherical harmonics states that:
\be \label{eq:additionTh}
\sum_{j=1}^{D_n}  \Psi_{n,j}(x) \Psi^*_{n,j}(y) = \f{D_n}{V_d}    P_n(\bar{X}\cdot \bar{Y}) \,,
\ee
where
\be
P_n(z) = \begin{cases} \f{n! (d-2)!}{(n+d-2)!} C_n^{(d-1)/2}(z)\,, &\; \text{if } d> 2\,,\\
T_n(z)\,, &\; \text{if } d= 2\,, \end{cases}
\ee
and $C_n^{\a}(z)$ and $T_n(z)$ are the Gegenbauer and Chebyshev polynomials, respectively.

\section{CFTs on $S^d$}
\label{app:sphereCFT}

Given a CFT on $\mathbb{R}^d$, and assuming that conformal invariance can be promoted to Weyl invariance\footnote{See \cite{Farnsworth:2017tbz} and references therein for the relation between Weyl and conformal invariance.} (possibly up to an anomaly), we can then define a corresponding CFT on $S^d$ by performing the Weyl transformation \eqref{eq:Weyl_g}, 
together with the transformation of primary fields
\be \label{eq:Weyl_Op}
\cO(x) \to \Om(x)^{-\D_{\cO}} \cO(x) \,,
\ee
such that $n$-point functions are obtained as
\be
\la \cO_1(x_1) \cdots \cO_n(x_n) \ra_{S^d} = \Om(x)^{-\D_1} \cdots \Om(x)^{-\D_n} \la \cO_1(x_1) \cdots \cO_n(x_n) \ra_{\mathbb{R}^d} \,.
\ee
In practice, the latter amounts to replacing the flat-space distances $|x-y|$ appearing in the conformal correlators with the chordal distance
\begin{equation}
s(x,y)=2a \frac{|x-y|}{(1+x^2)^{1/2}(1+y^2)^{1/2}} = |x-y| \Om(x)^{1/2}  \Om(y)^{1/2} \,.
\end{equation}
Notice that this is not the geodesic distance $\s(x,y)$ on $S^d$, but the Euclidean distance in the embedding space $\mathbb{R}^{d+1}$, i.e.\ $s(x(p),x(p'))=|X(p)-X(p')|$ where $X(p)$ is the embedding map $X: S^d\to \mathbb{R}^{d+1}$ while $x(p)$ is the stereographic map $x: S^d\to \mathbb{R}^{d}$. Of course, the two are trivially related by a trigonometric relation: $s(x,y)=2 a \sin(\s(x,y)/2a)$.

We can check that for a usual free scalar, the propagator on $S^d$ matches the flat one with the flat distance replaced by the chordal distance, if we  appropriately tune the non-minimal coupling with the curved background of the sphere.
The covariance, or propagator, $C_1(x,y;b)$ is the solution of the equation
\be \label{eq:prop-def}
(-\nabla_x^2 + b) C_1(x,y;b) = \f{\d(x-y)}{\sqrt{g}} \, ,
\ee
where $b$ is a constant (of squared-mass dimension),  $\nabla^2= \nabla^\m \nabla_\m$ is the covariant Laplacian on the $d$-sphere, and we specified by a subscript the coordinate on which derivatives as in this case there could be an ambiguity.
Due to homogeneity of space the propagator depends only on the geodesic distance $\s(x,y)$, hence we will also write $C_1(\s;b)$ for the propagator.
The reason for the subscript 1 in the latter is that this corresponds to the case $\z=1$ of the more general propagator we will consider in the following subsection.

The propagator on $S^d$ has been computed in \cite{Allen:1985wd} directly solving \eqref{eq:prop-def}, or from an explicit mode sum in \cite{Dowker:1975tf}. Defining
\be
\g_\pm = \f{d-1}{2} \pm \sqrt{\f{(d-1)^2}{4}-a^2 b}\, ,
\ee
the propagator is given by
\be
C_1(\s;b) = a^{2-d} \f{ \G(\g_+) \G(\g_-)}{\G(d/2)\, 2^d\, \pi^{d/2}}\, {}_{2}F_1(\g_+,\g_-;d/2; \cos(\s/2a)^2) \, ,
\ee
where  ${}_{2}F_1(\a,\b;\g;z)$ is the hypergeometric function.

The case of a Weyl invariant free scalar field is obtained with the choice 
\be \label{eq:b_W}
b= b_W\equiv  \f{d-2}{4(d-1)} R = \f{d(d-2)}{4a^2}\,,
\ee 
for which $\g_+=d/2$ and $\g_-=(d-2)/2$. In this case, the hypergeometric function reduces to a simple power and we find:
\be \label{eq:C_1}
C_1(x,y;b_W) = a^{2-d} \f{  \G(d/2-1)}{ 2^d\, \pi^{d/2}}\, (\sin(\s(x,y)/2a) )^{2-d} =  \f{  \G(d/2-1)}{ 4\, \pi^{d/2}}\, s(x,y)^{2-d}\, ,
\ee
which is precisely the massless free scalar propagator of flat space with the replacement $|x-y|\to s(x,y)$.

\subsection{Generalized free field theory}
\label{app:GFFT}

We now consider the case of a conformal generalized free field  theory (GFFT), i.e. a long-range massless Gaussian theory, sometimes called a mean-field theory. It is worth discussing it in some detail because it is the simplest case of (non-local) CFT, and because typical long-range models can be defined as perturbations of a GFFT.
By definition this is a CFT whose only non-vanishing connected $n$-point function is the two-point function, which however has a scaling exponent $\D\neq d/2-1$, and which moreover we take to be in the range $\D\in(0,d/2)$.
On flat space, the two-point function is 
\begin{equation} \label{eq:freeC-flat}
C_{\rm flat}(x,y)=\frac{c(\Delta)}{|x-y|^{2\Delta}} \,, \qquad c(\Delta)=\frac{\Gamma(\Delta)}{2^{d-2\Delta}\pi^{d/2}\Gamma(\frac{d}{2}-\Delta)} \,.
\end{equation}
Writing $\D=d/2-\z$, such GFFT can be obtained from a functional integral with the action\footnote{We typically assume $0<\z<1$. The restriction to $\z<1$ is imposed to preserve reflection positivity (unitarity in Lorentzian signature), but also because $\z>1$ would correspond to a strong short-range rather than long-range action, and moreover the operator with $\z=1$ would in that case be a relevant perturbation. The restriction to $\z>0$ is instead chosen to avoid a strong long-range action, with its associated unusual thermodynamic features  \cite{Campa:2009rev}.}
\be \label{eq:flatGFFT}
S_{\rm GFFT}[\phi] = \f12 \int \dd x  \, \phi(x) (-\p^2)^{\z} \phi(x) \,,
\ee
where the fractional power of the Laplacian can be defined in many equivalent ways \cite{Kwasnicki:2017}, among which in particular as the inverse of the ``Riesz potential'' \eqref{eq:freeC-flat}.
The easiest definition is of course in Fourier space, where $(-\p^2)^{\z}$ is defined as the multiplication operator $p^{2\z}$, which is the inverse of the Fourier transform of \eqref{eq:freeC-flat}. 
Going to position space one finds instead a representation as a hypersingular integral operator:\footnote{This can be derived by first writing
\be \nn
p^{2\z} = \f{1}{\G(-\z)} \int_0^{+\infty} {\rm d} t \f{e^{-t p^2}-1}{t^{1+\z}} \,,
\ee
whose validity is trivially checked by rescaling $t\to t/p^2$ and recognizing that the integral reduces to $p^{2\z}$ times the Cauchy-Saalsch\"utz representation of $\G(-\z)$ for $0<\z<1$.
The singular integral representation is then found by going back to position space and exchanging the order of integration  \cite{Stinga:2009}.
}
\be \label{eq:fracLapl_flat}
(-\p^2)^{\z} \phi(x) = \lim_{r\to 0} \,  c(d-\Delta) \int_{|x-y|>r} \dd y \,\f{\phi(y)-\phi(x)}{|x-y|^{2(d-\Delta)}} 
\,.
\ee
In the physics literature such representation is often expressed as
\be  \label{eq:fracLapl_flat_phys}
(-\p^2)^{\z} \phi(x) = \int \dd y \, C^{-1}_{\rm flat}(x,y) \phi(y) \,,
\ee
with  convolution kernel
\be \label{eq:flatCinv}
C^{-1}_{\rm flat}(x,y) = \frac{c(d-\Delta)}{|x-y|^{2(d-\Delta)}} \,,
\ee
without any subtraction term. For $\z>0$ the convolution is a formal divergent expression (a ``hypersingular'' integral), which is to be interpreted through analytic continuation from $\z<0$. For simplicity we will stick to this point of view.

In order to place the GFFT on the $d$-sphere, we can apply again the Weyl mapping to \eqref{eq:freeC-flat}, and thus write
\begin{equation} \label{eq:freeC}
C(x,y)= \Om(x)^{-\D}\Om(y)^{-\D} C_{\rm flat}(x,y) =\frac{c(\Delta)}{s(x,y)^{2\Delta}} \,. 
\end{equation}
Constructing an action associated to such propagator requires as usual identifying the inverse propagator, and from this the type of non-minimal coupling to the background geometry that is needed in order to obtain a conformal theory.

The covariance $C(x,y)$ in \eqref{eq:freeC} is also known as the Riesz potential, and it can be written
\be \label{C-harmonics}
C(x,y) = \sum_{n\geq 0} \sum_{j=1}^{D_n} \f{1}{\om^{(\z)}_n}  \Psi_{n,j}(x) \Psi^*_{n,j}(y) = \f{1}{V_d}  \sum_{n\geq 0} \f{D_n}{\om^{(\z)}_n} P_n(\bar{X}\cdot \bar{Y})\,,
\ee
where we used the addition theorem \eqref{eq:additionTh}.

The inverse of \eqref{eq:freeC} is defined by the equation
\be
\int \dd z \sqrt{g(z)} \, C^{-1}(x,z) \int \dd y \sqrt{g(y)} \, C(z,y) \phi(y) = \phi(x)\,,
\ee
or
\be \label{eq:defCinv}
\int \dd z \sqrt{g(z)} \, C^{-1}(x,z) C(z,y) = \f{1}{\sqrt{g}} \d(x-y) \, .
\ee
Given that on flat space \eqref{eq:flatCinv} is the inverse of \eqref{eq:freeC-flat}, it is easily seen that the above equations are solved by
\be \label{eq:sphereCinv}
\begin{split}
C^{-1}(x,y) &=  \Om(x)^{\D-d} \, \Om(y)^{\D-d} \, C_{\rm flat}^{-1}(x,y) 
= \frac{c(d-\Delta)}{s(x,y)^{2(d-\Delta)}}\,,
\end{split}
\ee
whose convolution should again be interpreted by analytic continuation.
This is of the expected form we would obtain by the Weyl mapping applied to $C_{\rm flat}^{-1}(x,y)$, formally viewed as  the two-point function of the shadow operators \cite{Ferrara:1972uq} of dimension $\widetilde{\D}=d-\D$.
It also means that defining, for  $\z=d/2-\D$, the operator whose kernel is \eqref{eq:sphereCinv} as\footnote{\label{foot:Dz-subtr}As a subtracted hypersingular integral, a rigorous covariant expression is given by (see \cite{Samko:2003,Samko-book}):
\be \nn
\cD_\z \phi (x) = \lim_{r\to 0} c(d-\Delta) \int_{s(x,y)>r} \dd y \sqrt{g(y)}\,\f{\phi(y)-\phi(x)}{s(x,y)^{2(d-\Delta)}} +\f{\G(d-\D)}{\G(\D)} \phi(x) \,.
\ee
}
\be
\cD_\z \phi (x) =  \int \dd y \, \sqrt{g(y)}\, C^{-1}(x,y) \phi(y) \,,
\ee
we find that 
\be
\cD_\z \phi (x) = \Om(x)^{\D-d} (-\p^2)^{\z} \left( \Om(x)^{\D} \phi(x) \right) \,.
\ee 
Given the Weyl transformations \eqref{eq:Weyl_g}, \eqref{eq:Weyl_Op} relating the flat space to the sphere, one recognizes in such a relation the definition of conformally covariant operator of order $\z=d/2-\D$, or conformal biweight $(\D,d-\D)$ \cite{Branson:1997,gonzalez2016recent}.

Therefore, the action replacing \eqref{eq:flatGFFT} on the sphere is
\be
S_{\rm GFFT}[\phi] = \f12 \int \dd x  \, \sqrt{g(x)}\\, \phi(x) \cD_\z \phi(x) \,.
\ee
However, calling $\cD_\z$ a conformal ``fractional Laplacian'' would be deceiving, as it turns out that the operator $\cD_\z$ is not of the form $(-\nabla^2 +b)^{\z}$:
the conformal Laplacian of biweight $(\D,d-\D)$ on the $d$-sphere can be related to the Laplace-Beltrami operator by the expression  \cite{Branson:1995}
\be \label{eq:D_z}
\cD_\z = a^{-2\z} \,\f{ \G(a \cD_{1/2} +\f12 +\z) }{ \G(a \cD_{1/2} +\f12 -\z) } \,, \qquad \cD_{1/2} 
= a \, \sqrt{ -\nabla^2 + \left( \f{d-1}{2a}\right)^2 } \,,
\ee
which should of course be interpreted in terms of the eigenvalues
\be
\begin{split}
\om^{(\z)}_n = a^{-2\z} \,\f{ \G(n +\f{d}{2} +\z) }{ \G(n +\f{d}{2}-\z) } =  a^{-2\z} \,\f{ \G(a\, \om^{(1/2)}_n +\f12 +\z) }{ \G(a\, \om^{(1/2)}_n +\f12 -\z) } \,,\\
 \om^{(1/2)}_n = a^{-1}\, \left(n+\f{d-1}{2}\right) =   \sqrt{ \om_n + \left( \f{d-1}{2a}\right)^2 }
=  \sqrt{ \om^{(1)}_n +  \f{1}{4a^2} } \,.
\end{split}
\ee
In the last equality we introduced $\om_n = n(n+d-1)/a^2$, the eigenvalues of Laplace-Beltrami operator on the sphere. 
Notice that for $\z=1$ we have $\cD_1=-\nabla^2+b_W$, as expected, and that for $n\to+\infty$ the eigenvalues of $\cD_\z$ do asymptotically approach $n^{2\z}$, as for a Laplacian to  the power $\z$.
The eigenvalues $\om^{(\z)}_n$ are of course the inverse of the eigenvalues of the Riesz potential \eqref{eq:freeC}, which were known since long to mathematicians (e.g. \cite{PavSam84}), and have been later rederived also in the physics literature \cite{Gubser:2002vv}.

\

Notice that, denoting $c_n\equiv n +\f{d}{2}-\z$, we can write
\be
\f{1}{a^{2\z} \om^{(\z)}_n} = \f{\G(c_n)}{\G(c_n+2\z)} = \f{1}{\G(2\z)} B(c_n,2\z) \,,
\ee
where $B(x,y)$ is the Euler beta function. Therefore, we have various useful representations, among which in particular the following integral representation:
\be \label{eq:EulerBeta}
\f{1}{a^{2\z} \om^{(\z)}_n} = \f{1}{\G(2\z)} \int_0^1 dt\, t^{c_n-1} (1-t)^{2\z-1} = \f{1}{\G(2\z)} \int_0^{+\infty} ds\, e^{-s\, c_n}  (1-e^{-s})^{2\z-1}\,.
\ee
One way to introduce a UV cutoff in the theory is then to replace the beta function with the incomplete beta function, i.e.\ truncating the upper end of $t$-integration at $1-e^{-s_0}$, or the lower end of the $s$-integration at $s_0>0$. From the latter one can see that such cutoff is roughly proportional to an exponential $e^{-s_0 c_n}$.
This should be compared to the flat space representation 
\be \label{eq:EulerGamma}
\f{1}{p^{2\z}} = \f{1}{\G(2\z)} \int_0^{+\infty} ds\, e^{-s\, p}  s^{2\z-1} \,,
\ee
which again can be regularized by replacing the integral representation of the gamma function with that of the incomplete gamma function.
This is in the same spirit of what was done in chapter \ref{chap:CTKT}, where however the representation of $\G(z)$ was used instead of $\G(2\z)$, i.e. the exponential cutoff was with respect to $p^2$ rather than $p$.
Notice that as expected the two $s$-integral representations in \eqref{eq:EulerBeta} and \eqref{eq:EulerGamma} coincide in the deep UV (small $s$) but differ in the IR (large $s$).

\section{Computation of the free energy for GFFT}
\label{app:F_GFFT}

In this appendix we present a detailed computation of the following sum:
\begin{equation}
F=\frac{1}{2}\sum_{n=0}^{\infty} D_n \ln \left(a^{-2\zeta}\frac{\Gamma(n+d/2+\zeta)}{\Gamma(n+d/2-\zeta)}\right)
\label{eq:free_GFFT}.
\end{equation}
This computation was done in \cite{Diaz:2007an}, but we reproduce it here with more details.

For $d>0$, this sum is divergent. We will compute it in the regime $2\zeta-2<d<0$ and then perform an analytic continuation to deduce the result for $d>0$. This computation can also be done for $\zeta>1$. For $k-1<\zeta <k \; ,  k\geq 2$, the sum \eqref{eq:free_GFFT} would have to be computed in the range $2\zeta-2k<d<0$. However, the computation is very similar than the one for $0<\zeta<1$ and leads to the same result so we will detail here only the computation for $0<\zeta<1$. 

Let us first show that the sum of multiplicity is zero in this regularization. 

\begin{equation} \label{eq:sum_multiplicity}
	\begin{split}
\sum_{n=0}^{\infty} D_n&= \sum_{n=0}^{\infty} \frac{(n+d-2)!(2n+d-1)}{n!(d-1)!}=\sum_{n=0}^{\infty} \frac{(n+d-1)!}{n!(d-1)!}\frac{2n+d-1}{n+d-1} \\
& =\sum_{n=0}^{\infty}  \frac{(n+d-1)!}{n!(d-1)!}\frac{n}{n+d-1} +  \sum_{n=0}^{\infty} \frac{(n+d-1)!}{n!(d-1)!} \\
& =\sum_{n=1}^{\infty}  \frac{(n+d-2)!}{(n-1)!(d-1)!} +(1-1)^{-d} = \sum_{n=0}^{\infty}  \frac{(n+d-1)!}{(n)!(d-1)!}+0 = (1-1)^{-d}=0 \; . 
 	\end{split}
\end{equation}

The term $\ln(a^{-2\zeta})$ can thus be dropped from the expression of $F$. Taking the derivative with respect to $\zeta$ of the remaining expression, we obtain:

\begin{equation}
\frac{dF}{d\zeta}=\frac{1}{2}\sum_{n=0}^{\infty} D_n\left( \psi(n+d/2+\zeta)+\psi(n+d/2-\zeta)\right) \, .
\end{equation}

We will now use the following integral representation of the digamma function:
\begin{equation}
\psi(z)=\int_0^{\infty} dt \left(\frac{e^{-t}}{t}-\frac{e^{-tz}}{1-e^{-t}}\right) \, ,
\end{equation}
which is valid for $z>0$. 

Leaving out the $n=0$ term,\footnote{For the case $1<\zeta<2$, we would also need to leave out the term $n=1$.} we then get:
\begin{equation}
	\begin{split}
\frac{dF}{d\zeta}&=\frac{1}{2}\left(\psi(d/2+\zeta)+\psi(d/2-\zeta)\right)+\frac{1}{2}\sum_{n=1}^{\infty} D_n\int_0^{\infty} dt \left(\frac{2e^{-t}}{t}-\frac{e^{-t(n+d/2)}}{1-e^{-t}}\left(e^{-t\zeta}+e^{t\zeta}\right)\right) \\
&=\frac{1}{2}\left(\psi(d/2+\zeta+1)+\psi(d/2-\zeta+1)-\frac{1}{d/2+\zeta}-\frac{1}{d/2-\zeta}\right)\\
& \qquad +\frac{1}{2}\sum_{n=1}^{\infty} D_n\int_0^{\infty} dt \left(\frac{2e^{-t}}{t}-\frac{e^{-t(n+d/2)}}{1-e^{-t}}\left(e^{-t\zeta}+e^{t\zeta}\right)\right) \, ,
	\end{split}
\end{equation}
where we have used $\psi(z)=\psi(1+z)-\frac{1}{z}$. 

We can now use the integral representation of the digamma function for $\psi(n+d/2 \pm \zeta +1)$. Rearranging the terms and exchanging sum and integral, we obtain:
\begin{equation}
\begin{split}
\frac{dF}{d\zeta}=& -\frac{1}{2}\left(\frac{1}{d/2+\zeta}+\frac{1}{d/2-\zeta}\right)+\int_0^{\infty} dt \frac{e^{-t}}{t}\sum_{n=0}^{\infty} D_n \crcr
& \qquad -\frac{1}{2} \int_0^{\infty} \frac{e^{-t\zeta}+e^{t\zeta}}{1-e^{-t}}e^{-td/2}\left(e^{-t}+\sum_{n=1}^{\infty}D_n e^{-tn}\right) \ .
\end{split}
\end{equation}
Again, as the sum of the multiplicities is zero, the second term vanishes. The remaining sum is:
\begin{align}
	\begin{split}
\sum_{n=1}^{\infty}D_n e^{-tn}&= \sum_{n=1}^{\infty} \frac{(n+d-2)!(2n+d-1)}{n!(d-1)!}e^{-tn} \\
&= 2 \sum_{n=1}^{\infty} n\frac{(n+d-2)!}{n!(d-1)!}e^{-tn} + \sum_{n=1}^{\infty} \frac{(n+d-2)!}{n!(d-2)!}e^{-tn} \\
&= 2e^{-t}  \sum_{n=0}^{\infty} \frac{(n+d-1)!}{n!(d-1)!}e^{-tn} +(1-e^{-t})^{-(d-1)}-1 \\
&= 2e^{-t} (1-e^{-t})^{-d}+(1-e^{-t})^{-(d-1)}-1  = (1-e^{-t})^{-d}(1+e^{-t})-1 \, . 
	\end{split}
\end{align}
Substituting this result into $\frac{dF}{d\zeta}$ and changing variables $u=e^{-t}$, we obtain:
\begin{equation}
\frac{dF}{d\zeta}=-\frac{1}{2}\left(\frac{1}{d/2+\zeta}+\frac{1}{d/2-\zeta}\right)-\frac{1}{2}\int_0^1 du\, u^{d/2-1}(u^{\zeta}+u^{-\zeta})\left((1-u)^{-d-1}(1+u)-1\right) \ .
\end{equation}

We can now compute the last integral using the regular as well as the subtracted integral representations of the beta function\footnote{In the case $1<\zeta<2$, we also need the following subtracted integral representation of the beta function:
\begin{equation*}
B(a,b)-B(a,1)+(b-1)B(a+1,1)=\int_0^1 dt \, t^{a-1}\left((1-t)^{b-1}+(b-1)t-1\right) \; , \; a>-2 \, , \; b>0 \, .
\end{equation*}}
\begin{equation}
	\begin{split}
B(a,b)&= \int_0^1 dt\, t^{a-1}(1-t)^{b-1} \; , \; a,b>0 \\
B(a,b)-B(a,c)&= \int_0^1 dt \, (1-t)^{a-1}\left(t^{b-1}-t^{c-1}\right) \; , \; a>-1 \, , \; b,c>0 \, .
	\end{split} 
\label{eq:Beta_int}
\end{equation}

Indeed, we have:
\begin{equation}
	\begin{split}
&\int_0^1 du\, u^{d/2+\zeta-1}\left((1-u)^{-d-1}(1+u)-1\right) \\
&=\int_0^1 du \left(u^{d/2+\zeta}(1-u)^{-d-1} +u^{d/2+\zeta-1}\left((1-u)^{-d-1}-(1-u)^{1-1}\right) \right) \\
&= B(d/2+\zeta+1,-d)+B(d/2+\zeta,-d)-B(d/2+\zeta,1) = 2\zeta\frac{\Gamma(d/2+\zeta)\Gamma(-d)}{\Gamma(\zeta-d/2+1)}-\frac{1}{\zeta+d/2} \,,
	\end{split}
\end{equation}
and similarly:
\begin{align}
\int_0^1 du\, u^{d/2-\zeta-1}\left((1-u)^{-d-1}(1+u)-1\right)&= -2\zeta\frac{\Gamma(d/2-\zeta)\Gamma(-d)}{\Gamma(-\zeta-d/2+1)}-\frac{1}{-\zeta+d/2} \, .
\end{align}

Thus we obtain:
\begin{equation}
\frac{dF}{d\zeta}=\zeta \Gamma(-d)\left(\frac{\Gamma(d/2-\zeta)}{\Gamma(1-\zeta-d/2)}-\frac{\Gamma(d/2+\zeta)}{\Gamma(1+\zeta-d/2)}\right) =-\zeta\frac{\sin(\pi \zeta)}{\sin(\pi d/2)}\frac{\Gamma(d/2-\zeta)\Gamma(d/2+\zeta)}{\Gamma(1+d)}\, , 
\end{equation}
which can be analytically continued to $d>0$ not even, and is valid for any value of $\zeta$, except at the poles at $\z=d/2+k$.

\section{Computation of $C(x,x)$ in dimensional regularization}
\label{app:C_dim_reg}

From the expansion of $C(x,y)$ in spherical harmonics in \eqref{C-harmonics}, we find that at coinciding points we have
\be
C(x,x) = \f{1}{V_d} \, \sum_{n\geq 0} \f{D_n}{\om^{(\z)}_n}   \,,
\label{eq:covxx}
\ee
which of course is divergent and needs regularization. We employ here analytic continuation in the dimension $d$, treating separately the two cases $\z=1$ and $\z<1$.

\subsection{$\z=1$ case}
\label{app:tadpole-SR}

With $\zeta=1$, the expression of the covariance simplifies to:
\begin{equation}
C_1(x,x)=\frac{a^{2-d}}{\Gamma(d/2)(4\pi)^{d/2}}\sum_{n=0}^{\infty} \frac{4\,\Gamma(n+d-1)(2n+d-1)}{n!(2n+d)(2n+d-2)} \,.
\end{equation}

The Weyl invariant coupling  $b= b_W\equiv  \f{d(d-2)}{4a^2}$ in \eqref{eq:prop-def} provides an IR regularization by removing the zero mode, for $d\neq 2$.
However, the sum is divergent for $d>2$ and needs a regularization.
A convenient approach is to compute it for $0<d<2$, where it converges, and where we find $C(x,x)=0$, thanks to a cancellation between the $n=0$ contribution (negative because $b_W<0$ in this range of dimensions), and the rest of the series. 
We then analytically continue the result to $d>2$. 

Let us begin by rewriting the sum as
\begin{equation}
	\begin{split}
\sum_{n=0}^{\infty}&  \frac{\Gamma(n+d-1)(2n+d-1)}{n!(n+d/2)(n+d/2-1)}= \sum_{n=0}^{\infty} \frac{\Gamma(n+d-1)}{n!(n+d/2-1)}+\sum_{n=0}^{\infty}\frac{\Gamma(n+d-1)}{n!(n+d/2)} \\
&=\frac{\Gamma(d-2)}{d/2-1}\sum_{n=0}^{\infty}\frac{1}{n!}\frac{(d-1)_n(d/2-1)_n}{(d/2)_n}+\frac{2\Gamma(d-2)}{d}\sum_{n=0}^{\infty}\frac{1}{n!}\frac{(d-1)_n(d/2)_n}{(d/2+1)_n} \,,
	\end{split}
\label{eq:sumd2}
\end{equation}
where $(b)_n=b(b+1)\dots (b+n-1)=\frac{\Gamma(b+n)}{\Gamma(b)}$ and we have used $\frac{(b)_n}{(b+1)_n}=\frac{b}{b+n}$. 

We then recognize the hypergeometric function of argument $1$:
\begin{equation}
{}_2F_1(a,b,c,1)=\sum_{n=0}^1 \frac{(a)_n(b)_n}{n!(c)_n}=\frac{\Gamma(c)\Gamma(c-b-a)}{\Gamma(c-b)\Gamma(c-a)} \,,
\end{equation} 
which is valid for ${\rm Re}(b),{\rm Re}(c)>0$ and ${\rm Re}(c-a-b)>0$. 

In order to apply this formula, we thus need $d>0$ and $d<2$: in the first sum of \eqref{eq:sumd2} we have $a=d/2-1,b=d-1,c=d/2$ and in the second sum we have $a=d/2,b=d-1,c=d/2+1$. 

We thus get:
\begin{equation}
	\begin{split}
C_1(x,x)&\propto 2\Gamma(d-2)\left(\frac{\Gamma(d/2)\Gamma(2-d)}{(d-2)\Gamma(1-d/2)}+\frac{\Gamma(d/2+1)\Gamma(2-d)}{d\Gamma(2-d/2)}\right) \\
&=\frac{2\Gamma(d-2)\Gamma(d/2)\Gamma(2-d)}{\Gamma(1-d/2)}\left(\frac{1}{d-2}+\frac{d}{2d(1-d/2)}\right)=0 \; .
	\end{split}
\end{equation}

\subsection{$\z<1$ case}
\label{app:tadpole-LR}

We want to compute the following sum:
\begin{equation}
\frac{a^{2\zeta-d}(d-1)!}{\Gamma(d/2)(4\pi)^{d/2}}\sum_{n=0}^{\infty}\frac{D_n}{a^{2\zeta}\omega_n^{(\zeta)}}\,.
\end{equation}

We use the integral representation of \eqref{eq:EulerBeta}, which in the range $2\zeta-2<d<0$ is valid for $n>0$. Taking out the $n=0$ term, and exchanging the sum and the integral, we obtain:
\begin{equation}
C(x,x)=\frac{a^{2\zeta-d}(d-1)!}{\Gamma(d/2)(4\pi)^{d/2}}\Bigg[\frac{\Gamma(d/2-\zeta)}{\Gamma(d/2+\zeta)}+\frac{1}{\Gamma(2\zeta)}\int_0^{\infty}ds (1-e^{-s})^{2\zeta-1}e^{-s(\frac{d}{2}-\zeta)}\sum_{n=1}^{\infty}D_n e^{-s n}\Bigg]\,.
\end{equation}

The remaining sum was already computed in the previous appendix and doing the change of variable $u=e^{-s}$ we obtain:
\begin{align}
 C(x,x)&=\frac{a^{2\zeta-d}(d-1)!}{\Gamma(\frac{d}{2})(4\pi)^{d/2}}\Bigg[\frac{\Gamma(\frac{d}{2}-\zeta)}{\Gamma(\frac{d}{2} +\zeta)}+ \frac{1}{\Gamma(2\zeta)}\int_0^1 du\, u^{d/2-\zeta-1}(1-u)^{2\zeta-1}\left((1-u)^{-d}(1+u)-1\right)\Bigg] \crcr
&=\frac{a^{2\zeta-d}(d-1)!}{\Gamma(d/2)(4\pi)^{d/2}}\Bigg[\frac{\Gamma(d/2-\zeta)}{\Gamma(d/2+\zeta)}\crcr
&  + \frac{1}{\Gamma(2\zeta)}\int_0^1 du\, \left(u^{d/2-\zeta}(1-u)^{2\zeta-d-1}+u^{d/2-\zeta-1}\left((1-u)^{2\zeta-d-1}-(1-u)^{2\zeta-1}\right)\right)\Bigg]\,.
\end{align}

We perform the integral using the regular and subtracted representations of the Beta function \eqref{eq:Beta_int} and obtain:
\begin{align}
C(x,x)&=\frac{a^{2\zeta-d}(d-1)!}{\Gamma(\frac{d}{2})(4\pi)^{d/2}}\Bigg[\frac{\Gamma(\frac{d}{2}-\zeta)}{\Gamma(\frac{d}{2}+\zeta)}+\frac{\Gamma(\frac{d}{2}-\zeta+1)\Gamma(2\zeta-d)}{\Gamma(-\frac{d}{2}+\zeta+1)\Gamma(2\zeta)}+\frac{\Gamma(\frac{d}{2}-\zeta)\Gamma(2\zeta-d)}{\Gamma(\zeta-\frac{d}{2})\Gamma(2\zeta)}-\frac{\Gamma(\frac{d}{2}-\zeta)}{\Gamma(\zeta+\frac{d}{2})}\Bigg] \crcr
&= \frac{a^{2\zeta-d}(d-1)!\Gamma(d/2-\zeta)\Gamma(2\zeta-d)}{\Gamma(d/2)(4\pi)^{d/2}\Gamma(2\zeta)\Gamma(\zeta-d/2)}\Bigg[\frac{d/2-\zeta}{\zeta-d/2}+1\Bigg] =0\,.
\end{align}

\section{Basics of conformal partial wave expansion}
\label{app:CPW}

We provide here some important formulas and background on the conformal partial wave expansion used in this chapter. Some formulas have already been introduced in section \ref{sec:CFT} but we give here more details. The main results of this appendix have been derived by Dobrev et al.\ in \cite{Dobrev:1976vr,Dobrev:1975ru,Dobrev:1977qv}, and largely revived in recent years \cite{Caron-Huot:2017vep,Simmons-Duffin:2017nub,Liu:2018jhs,Karateev:2018oml}.\footnote{These methods have been at the heart of a very active field in recent years, see for example their use with Mellin amplitudes  \cite{Mack:2009mi,Costa:2012cb}, their application to the Sachdev-Ye-Kitaev model \cite{Maldacena:2016hyu,Murugan:2017eto}, to the bootstrap crossing equations \cite{Gadde:2017sjg,Hogervorst:2017sfd,Sleight:2018ryu,Sleight:2018epi}, and to the construction of an AdS/CFT map \cite{deMelloKoch:2018ivk,Aharony:2020omh}.} 
Here we mostly follow the notation of \cite{Benedetti:2021qyk}, where a more detailed review can be found.

We work on flat space and comment at the end on the straightforward extension to the sphere. In the conformal limit, the Bethe-Salpeter kernel $K(x_1,x_2,x_3,x_4)$ of scalar fields of dimension $\D$ is diagonalized by functions with the structure of a conformal  three-point function as given in \eqref{eq:3ptCFT}\footnote{where we again use the notation $x_{ij}=x_i-x_j$}
\be \label{eq:3pt}
\la \phi_{\D}(x_3) \phi_{\D}(x_4) \cO_h^{\m_1 \cdots \m_J}(x_0) \ra_{\rm cs} = 
\f{ Z^{\m_1} \cdots Z^{\m_J} - \text{``traces"} }{ |x_{34}|^{2\D-h} |x_{30}|^{h} |x_{40}|^{h} } \,,   \;\;\;\; Z^{\m} = \f{|x_{30}||x_{40}|}{|x_{34}|} \left(\f{x_{30}^{\m}}{|x_{30}|^2} - \f{x_{40}^{\m}}{|x_{40}|^2}\right) \,,
\ee
such that
\be \label{eq:eigK}
\int_{x_3x_4} \, K(x_1,x_2,x_3,x_4) \, \la \phi_{\D}(x_3) \phi_{\D}(x_4) \cO_h^{\m_1 \cdots \m_J}(x_0) \ra_{\rm cs}
= k(h,J) \, \la \phi_{\D}(x_1) \phi_{\D}(x_2) \cO_h^{\m_1 \cdots \m_J}(x_0) \ra_{\rm cs} \,.
\ee
We recall that the subscript  ``cs" stands for conformal structure, meaning that the three-point function is just a notation for the structure on right-hand side. In particular, there is no structure constant, and the operator $\cO_h^{\m_1 \cdots \m_J}(z)$, of conformal dimension $h$ and in the spin-$J$ symmetric-traceless representation of the rotation group, is in general not part of the spectrum of the CFT. We denote $\phi_{\D}(x)$ a generic scalar primary of dimension $\D$, without introducing any flavor/color index structure, which we assume to be already diagonalized, as for example in \eqref{eq:full-K}.

The precise form of the kernel eigenvalue depends on the specific model. In the case of the long-range $O(N)^3$ model studied in section~\ref{sec:ON3-model}, we have:
\begin{equation}
k(h,J)=-\f{g^2}{(4\pi)^d}
 \frac{\Gamma(-\frac{d}{4}+\frac{h+J}{2})\Gamma(\frac{d}{4}-\frac{h-J}{2})}{\Gamma(\frac{3d}{4}-\frac{h-J}{2})\Gamma(\frac{d}{4}+\frac{h+J}{2})} \,.
\end{equation}

Multiplied by the following normalization factor ($\htilde=d-h$ is the dimension of the shadow operator \cite{Ferrara:1972uq})
\be \label{eq:norm3pt}
\begin{split}
\cN^{\D}_{h,J} = & \f{2^{(2\D+h+J)/2}}{(2\pi)^{d/2}} \left(  \frac{\G(\tfrac{\htilde+J+2\D-d}{2}) \G(\tfrac{h+J+2\D-d}{2})  }{ \G(\tfrac{\htilde+J-2\D+d}{2}) \G(\tfrac{h+J-2\D+d}{2}) }   \right)^{1/2}\frac{\G(\tfrac{h+J }{2})}{ \G(\tfrac{\htilde+J}{2})  } \,,
\end{split}
\ee
 the three-point functions \eqref{eq:3pt}  with fixed ${\rm Re}(\D)\in (d/4,3d/4)$  form a complete and orthonormal basis in an appropriate space of bilocal functions \cite{Dobrev:1976vr,Dobrev:1977qv}, the basis elements being labeled by the spin $J\in \mathbb{N}_0$, the position $x_0\in \mathbb{R}^d$, and the scaling dimension $h\in \cP_+$,
where
\be \label{eq:principal-series}
\cP_+ = \left\{ h \Bigm| h=\f{d}{2}+\im r,\, r\in\mathbb{R}_+ \right\}\,,
\ee
labels the principal series representations of the conformal group. More precisely, the space of bilocal functions $\cV_\D$ can be defined as the space of smooth functions $f(x_1,x_2)$ that are square integrable with respect to the scalar product 
\be \label{eq:scalar-prod}
\begin{split}
(f_1 , f_2) =  \int_{x_1\ldots x_4 } \overline{f_1(x_1,x_2)}  C^{-1}(x_1,x_3) C^{-1}(x_2,x_4) f_2(x_3,x_4) \,,
\end{split}
\ee
i.e.\ $(f,f)<\infty$, and satisfy the asymptotic boundary condition $f(x_1,x_2) \sim |x_1|^{-2\D}$ for $|x_1|\to\infty$ and similar for $|x_2|\to\infty$. Here, we have assumed that the bilocal functions have no symmetry under permutation of their two arguments, and we denoted\footnote{In this appendix we use $C(x,y)$ to denote the full-two-point function $\la\phi(x)\phi(y)\ra$ of the CFT, for which we are free to choose the same normalization as the one we used for the GFFT, even if the theory we have in mind is in general interacting.}
 $C(x_1,x_3) =c(\D) / |x_1-x_3|^{2\D}$.
Similarly, we can introduce the shadow space $\cV_{\wtD}$ with its basis of three-point functions defined as above but with $\D$ replaced by its shadow $\wtD=d-\D$. Since the two-point function of $\phi_\D$ and that of $\phi_{\wtD}$ are the inverse of each other (see \eqref{eq:flatCinv}), we can write the analogue of the scalar product \eqref{eq:scalar-prod} for $\cV_{\wtD}$ by replacing $C^{-1}$ with $C$, the two-point function of  $\phi_\D$.

The relation between $\cV_{\D}$ and $\cV_{\wtD}$ can better be understood in terms of raising and lowering of indices by the metric associated to the scalar product on them.
Let us denote $f^{x_1x_2}$, with contravariant indices $x_1,x_2$, the elements of $\cV_\D $, signaling that $f$ has dimension $\Delta$ on each of its arguments.
The factor $g_{x_1x_2;x_3x_4 }=C^{-1}(x_1,x_3)C^{-1}(x_2,x_4)$ in the scalar product in \eqref{eq:scalar-prod}  is a metric on $\cV_\D $ with covariant indices, that is with dimension $\tilde \D = d-\D$ on each of its arguments. The inverse metric is $g^{x_1x_2;x_3x_4} =  C(x_1,x_3)C(x_2,x_4) $ and the contraction on an index (integral over the position) has dimension $-d$. The metric and its inverse allow one to lower respectively raise indices, i.e. map $\cV_{\D}$ to its dual $\cV_{\tilde \D}$. 
The mapping holds also for the basis elements:\footnote{The three-point functions are not in $\mathcal{V}$, as they are not integrable, but they form a basis in the continuous sense, just like the Fourier basis does for ${\rm L}^2(\mathbb{R}^d)$.}
\be \label{eq:shadowTrans}
\begin{split}
\int  & \dd x_3 \dd x_4 \, C^{-1}(x_1,x_3)C^{-1}(x_2,x_4) \, \la \phi_{\D}(x_3) \phi_{\D}(x_4) \cO_h^{\m_1 \cdots \m_J}(x_0) \ra_{\rm cs} \,  \cN^{\D}_{h,J} \\
&=  \la \phi_{\wtD}(x_1) \phi_{\wtD}(x_2) \cO_h^{\m_1 \cdots \m_J}(x_0) \ra_{\rm cs} \, \cN^{\wtD}_{h,J} \,.
\end{split}
\ee
The completeness relation, or resolution of the identity, reads
\be \label{eq:res-id-nonsymm}
\begin{split}
\mathbb{I}(x_1,x_2,x_3,x_4) &\equiv  \d(x_1-x_3)\d(x_2-x_4)  \\
&=  \sum_{J\in \mathbb{N}_0}  \int_{\f{d}{2}}^{\f{d}{2}+\im\infty}  \f{{\rm d}h}{2\pi\im} \r(h,J) \, \cN^{\D}_{h,J}  \cN^{\wtD}_{\htilde,J} \, \Psi_{h,J}^{\D,\D,\wtD,\wtD}(x_1,x_2,x_3,x_4)  \,,
\end{split}
\ee
where the equality holds in a distributional sense when acting to the left (integration over $x_3$ and $x_4$) on $\cV_{\D}$, or to the right (integration over $x_1$ and $x_2$) on $\cV_{\tilde \D}$.
We have introduced  the Plancherel weight 
\be
\begin{split}
\r(h,J) &= \f{\G(\tfrac{d}{2}+J)}{2 (2\pi)^{d/2} J!} \f{\G(\htilde-1)\G(h-1)}{\G(\f{d}{2}-h)\G(\f{d}{2}-\htilde)} (h+J-1) (\htilde+J-1)
\,,
\end{split}
\ee
and we recall the definition of the conformal partial wave,
\be \label{eq:CPW}
 \Psi_{h,J}^{\D,\D,\wtD,\wtD}(x_1,x_2,x_3,x_4) = \int \dd  z \, \la \phi_{\D}(x_1) \phi_{\D}(x_2) \cO_h^{\m_1\cdots \m_J}(z) \ra_{\rm cs}\la \phi_{\wtD}(x_3) \phi_{\wtD}(x_4) {\cO}_{\htilde}^{\m_1\cdots \m_J}(z)\ra_{\rm cs} \,.
\ee
We notice that the product of normalization factors of the basis simplifies to
\be \label{eq:prod-cN}
\cN^{\D}_{h,J}  \cN^{\wtD}_{\htilde,J} = \f{2^{3d/2+J}}{(2\pi)^{d}} \,.
\ee

Any endomorphism $\cE:\cV_\D \to \cV_\D$ associated to a conformal kernel can be diagonalized by convoluting the kernel with the appropriate resolution of the identity, e.g.
\be
\begin{split}
\cE(x_1,x_2,x_3,x_4) & = \int \dd y_1 \dd y_2 \, \cE(x_1,x_2,y_1,y_2) \, \mathbb{I}(y_1,y_2,x_3,x_4) \\
&=  \sum_{J\in \mathbb{N}_0}  \int_{\f{d}{2}}^{\f{d}{2}+\im\infty}  \f{{\rm d}h}{2\pi\im} \r(h,J) \, \L_{\cE}(h,J) \, \cN^{\D}_{h,J}  \cN^{\wtD}_{\htilde,J} \, \Psi_{h,J}^{\D,\D,\wtD,\wtD}(x_1,x_2,x_3,x_4) \,,
\end{split}
\ee
where $\L_{\cE}(h,J)$ is the eigenvalue of $\cE$, satisfying an equation similar to \eqref{eq:eigK}.
Using the following relation between the conformal partial waves and the conformal blocks $\cG_{h,J}$ \cite{Dolan:2011dv,Simmons-Duffin:2017nub},
\be \label{eq:CPW-block} 
\Psi_{h,J}^{\D,\D,\wtD,\wtD}(x_1,x_2,x_3,x_4) =  \left(-\tfrac{1}{2}\right)^J  \left(S_{\tilde{h},J}\, \cG_{h,J}^{\D,\D,\wtD,\wtD}(x_1,x_2,x_3,x_4)  +    S_{h,J}\, \cG_{\tilde{h},J}^{\D,\D,\wtD,\wtD}(x_1,x_2,x_3,x_4) \right) \,,
\ee
with
\be
S_{h,J} =\f{\pi^{d/2} \G(h-\f{d}{2}) \G(h+J-1) \G(\f{\tilde{h}+J}{2})^2}{\G(h-1) \G(d-h+J) \G(\f{h+J}{2})^2} \,,
\ee
one can then write
\be
\begin{split}
\cE(x_1,x_2,x_3,x_4) &=  
\sum_{J\in \mathbb{N}_0}  \left(-\tfrac{1}{2}\right)^J \int_{\f{d}{2}-\im\infty}^{\f{d}{2}+\im\infty}  \f{{\rm d}h}{2\pi\im} \r(h,J)\, \L_{\cE}(h,J) \, \cN^{\D}_{h,J}  \cN^{\wtD}_{\htilde,J} \, S_{\tilde{h},J}\, \cG_{h,J}^{\D,\D,\wtD,\wtD}(x_1,x_2,x_3,x_4) \,,
\end{split}
\ee
where we used the symmetry of the measure factor $\r(h,J) \, \cN^{\D}_{h,J}  \cN^{\wtD}_{\htilde,J} $ under shadow reflection $h\to\htilde$  to extend the integration to negative imaginary parts and keep only one conformal block term.

Acting by convolution on the last two arguments of $\cE(x_1,x_2,x_3,x_4)$ with the inverse metric, we obtain an operator mapping $\cV_{\wtD}$ to $\cV_{\D}$, with a similar conformal partial wave expansion, except that the $\wtD$ arguments are replaced by $\D$:
\be \label{eq:4ptCPW}
\begin{split}
\int \dd y_3 \dd y_4 \,& \cE(x_1,x_2,y_3,y_4) C(y_3,x_3) C(y_4,x_4)\\
&=  \sum_{J\in \mathbb{N}_0}  \int_{\f{d}{2}}^{\f{d}{2}+\im\infty}  \f{{\rm d}h}{2\pi\im} \r(h,J) \, \L_{\cE}(h,J) \, \cN^{\D}_{h,J}  \cN^{\D}_{\htilde,J} \, \Psi_{h,J}^{\D,\D,\D,\D}(x_1,x_2,x_3,x_4) \\
&=\sum_{J\in \mathbb{N}_0}  \left(-\tfrac{1}{2}\right)^J \int_{\f{d}{2}-\im\infty}^{\f{d}{2}+\im\infty}  \f{{\rm d}h}{2\pi\im} \r(h,J)\, \L_{\cE}(h,J) \, \cN^{\D}_{h,J}  \cN^{\D}_{\htilde,J} \, S_{\tilde{h},J}\, \cG_{h,J}^{\D,\D,\D,\D}(x_1,x_2,x_3,x_4)\,.
\end{split}
\ee
In this case, the product of normalization factors has a ratio of gamma functions, with its own poles:\footnote{These poles do not cross the principal series as long as $\D>d/4$. Similarly, the poles of $\cN^{\wtD}_{h,J}  \cN^{\wtD}_{\htilde,J}$ stay away from $h=d/2$ for $\D<3d/4$. Together, these two conditions explain the condition on $\D$ mentioned below \eqref{eq:norm3pt}.}
\be \label{eq:prod-cN-2}
\cN^{\D}_{h,J} \cN^{\D}_{\htilde,J}=  \f{2^{(2\D+d/2+J)}}{(2\pi)^{d}}  \frac{\G(\tfrac{\htilde+J+2\D-d}{2}) \G(\tfrac{h+J+2\D-d}{2})  }{ \G(\tfrac{\htilde+J-2\D+d}{2}) \G(\tfrac{h+J-2\D+d}{2}) }  \,.
\ee
For $\cE=\mathbb{I}$, \eqref{eq:4ptCPW}, with $\L_{\mathbb{I}}(h,J)=1$, gives a conformal partial wave expansion of the inverse metric.
A similar expansion is obtained for the metric itself, replacing $C$ with $C^{-1}$ in the first line, and $\D$ with $\wtD$ in the expansion.

In the case of symmetric bilocal functions,\footnote{The corresponding space $\cV_\D^{\rm symm}$  is defined as before, except that the metric needs also symmetrization: $g^{\rm symm}_{x_1x_2;x_3x_4 }=\f12 (C^{-1}(x_1,x_3)C^{-1}(x_2,x_4)+C^{-1}(x_1,x_4)C^{-1}(x_2,x_3))$.} the completeness relation is of the same type, but with contribution only from even spin:
\be \label{eq:res-id-symm}
\begin{split}
\mathbb{I}_{\rm symm}(x_1,x_2,x_3,x_4) &\equiv  \f12\left( \d(x_1-x_3)\d(x_2-x_4) + \d(x_1-x_4)\d(x_2-x_3) \right) \\
&=  \sum_{J\in \mathbb{N}_0^{\text{even}}}  \int_{\f{d}{2}}^{\f{d}{2}+\im\infty}  \f{{\rm d}h}{2\pi\im} \r(h,J) \, \cN^{\D}_{h,J}  \cN^{\wtD}_{\htilde,J} \, \Psi_{h,J}^{\D,\D,\wtD,\wtD}(x_1,x_2,x_3,x_4)  \,. 
\end{split}
\ee

\paragraph{Non-normalizable contributions.} 
In practical applications, such as those we encounter in the bulk of this chapter, some of the hypotheses behind what we just reviewed can be violated.
Typically, we have two possible situations:
\begin{enumerate}
\item A four-point kernel $\cE(x_1,x_2,x_3,x_4)$ with right conformal transformation might nevertheless not be an endomorphism on $\cV_{\D}$ (or a map $\cV_{\wtD}\to \cV_{\D}$) because its action on an element $f\in\cV_{\D}$ (or  $\tilde{f}\in\cV_{\wtD}$) leads to a function not satisfying the integrability condition associated to the scalar product \eqref{eq:scalar-prod}.
\item The scalar field dimension might lie outside the range $(d/4,3d/4)$. This is in particular the case of the standard free theory (or the critical $O(N)$ model at large $N$) with $\D=d/2-1<d/4$, for $d<4$.
\end{enumerate}

In both cases, we can still use the conformal partial wave machinery, as long as we take care of deforming the contour of integration over $h$, or isolating the non-normalizable contributions from the four-point kernel \cite{Simmons-Duffin:2017nub}.

The typical example of a four-point kernel which is not an endomorphism is a physical four-point function of one scalar field $\phi$ whose
$s$-channel OPE contains operators of dimension smaller than $d/2$.
The identity operator is one such operator and it is always present, hence we always need to subtract the contribution that is disconnected in the  $s$-channel, $C(x_1,x_2)C(x_3,x_4)$, before applying the expansion \eqref{eq:4ptCPW}.\footnote{Convoluting $C(x_1,x_2)C(x_3,x_4)$ with $f(x_3,x_4)\in\cV_{\D}$, we obtain a new function proportional to $C(x_1,x_2)$. Regardless of whether the proportionality constant is finite or not, $C(x_1,x_2)$ is not square integrable with respect to the scalar product in \eqref{eq:scalar-prod}, and therefore it is not in $\cV_{\D}$.} Similarly, if the field $\phi$ has $\D<d/2$ and the three-point function with itself is non-vanishing, then we need to subtract the contribution that is one-particle reducible in the  $s$-channel (the $s$-channel skeleton tree diagram). It is then useful to define
\be \label{eq:F_s}
\begin{split}
\cF_s(x_1,x_2,x_3,x_4)   \equiv &\, \langle{\phi(x_1) \phi(x_2) \phi(x_3) \phi(x_4)} \rangle 
 -   C(x_1,x_2)  C(x_3,x_4) 
\\
 &-\int \dd y_1 \dd y_2 \langle \phi(x_1) \phi(x_2) \ph(y_1) \rangle C^{-1}(y_1,y_2) \langle \phi(y_2) \phi(x_3) \phi(x_4) \rangle \,,
\end{split}
\ee
which is obtained in terms of the Bethe-Salpeter kernel $K$ as (e.g.\ \cite{Benedetti:2021qyk})
\be \label{eq:F_s-K}
\cF_s(x_1,x_2,x_3,x_4)  = \int \dd y_1 \dd y_2 (\mathbb{I}-K)^{-1}(x_1,x_2,y_1,y_2) C(y_1,x_3) C(y_2,x_4)\,.
\ee
Applying to $\cF_s$ the expansion in the last line of \eqref{eq:4ptCPW}, and pushing the integration contour to the right, the integral is reduced to a sum over the residues at the poles of the integrand (the poles of $\L_{(\mathbb{I}-K)^{-1}}(h,J)$ being now the solutions of $k(h,J)=1$). Together with the sum over $J$, this reproduces the operator product expansion of the four-point function in the $s$ channel, if no other physical operators have dimension smaller than $d/2$. If instead other primaries have dimension smaller than $d/2$, then on the right of the principal series we pick their shadow pole; 
this must be corrected by deforming the contour in the conformal block representation to keep only the physical poles on the right, or equivalently (because of \eqref{eq:CPW-block}), by adding to the expansion the appropriate $\Psi$ contributions:
\be \label{eq:cF-extra}
\begin{split}
\cF_s(x_1,x_2,x_3,x_4)  
&=  \sum_{J\in \mathbb{N}_0}  \int_{\f{d}{2}}^{\f{d}{2}+\im\infty}  \f{{\rm d}h}{2\pi\im} \f{\r(h,J)}{1-k(h,J)}  \, \cN^{\D}_{h,J}  \cN^{\D}_{\htilde,J} \, \Psi_{h,J}^{\D,\D,\D,\D}(x_1,x_2,x_3,x_4) \\
&\quad - \sum_{i,J} {\rm Res}\left[  \f{\r(h,J) }{1-k(h,J)}  \, \cN^{\D}_{h,J}  \cN^{\D}_{\htilde,J} \, \Psi_{h,J}^{\D,\D,\D,\D}(x_1,x_2,x_3,x_4) \right]_{h=h_i(J)<d/2}\,,
\end{split}
\ee
where $h_i(J)$ are the physical solutions of $k(h,J)=1$ on the left of the principal series.
These isolated contributions are exactly analogue to the contributions we subtracted from the four-point function to define $\cF_s$. If such operators are present, we first subtract them from $\cF_s$, then we use the resolution of the identity to decompose the subtracted $\cF_s$, and finally we add them back as in \eqref{eq:cF-extra} to give the expansion of $\cF_s$ itself.

An example of the second situation in which \eqref{eq:4ptCPW} fails, namely when the dimension of the scalar field is outside the range $(d/4,3d/4)$, can be obtained from a generalized free field theory. Consider $\cF_s$ for a GFFT:
\be
\begin{split}
\cF_s^{GFFT}(x_1,x_2,x_3,x_4)  
&= C(x_1,x_3) C(x_2,x_4) + C(x_1,x_4) C(x_2,x_3) \,,
\end{split}
\ee
which is also twice the inverse metric of $\cV_{\D}^{\rm symm}$.
Using \eqref{eq:4ptCPW} with $\cE=\mathbb{I}$, but restricted to even spins to provide the symmetrization, we find
\begin{align}
\label{eq:cF-GFFT}
&\cF_s^{GFFT}(x_1,x_2,x_3,x_4)  
= \crcr
&2 \sum_{J\in \mathbb{N}_0^{\text{even}}}  \left(-\tfrac{1}{2}\right)^J \int_{\f{d}{2}-\im\infty}^{\f{d}{2}+\im\infty}  \f{{\rm d}h}{2\pi\im} \r(h,J)\, \cN^{\D}_{h,J}  \cN^{\D}_{\htilde,J} \, S_{\tilde{h},J}\, \cG_{h,J}^{\D,\D,\D,\D}(x_1,x_2,x_3,x_4)\,,
\end{align}
and from \eqref{eq:prod-cN-2} we know the normalization factors at  $J=0$ have a pole at $h=2\D$, that is the dimension of $\phi^2$. This pole crosses to the left of the principal series when $\D$ moves below $d/4$.
In that case, in order to recover the correct OPE, we must deform the integration contour in such a way that the pole at $h=2\D$ stays to its right, and the shadow pole at $h=d-2\D$ stays to its left.
Exploiting \eqref{eq:CPW-block}, it can be verified that the same result is obtained by writing, for $\D<d/4$:
\be \label{eq:cF-GFFT-extra}
\begin{split}
\f12 \cF_s^{GFFT}(x_1,x_2,x_3,x_4)  
&=  \sum_{J\in \mathbb{N}_0^{\text{even}}}  \int_{\f{d}{2}}^{\f{d}{2}+\im\infty}  \f{{\rm d}h}{2\pi\im} \r(h,J)  \, \cN^{\D}_{h,J}  \cN^{\D}_{\htilde,J} \, \Psi_{h,J}^{\D,\D,\D,\D}(x_1,x_2,x_3,x_4) \\
&\quad - {\rm Res}\left[  \r(h,0)  \, \cN^{\D}_{h,0}  \cN^{\D}_{\htilde,0} \, \Psi_{h,0}^{\D,\D,\D,\D}(x_1,x_2,x_3,x_4) \right]_{h=2\D}\,.
\end{split}
\ee

This has the exact same form as \eqref{eq:cF-extra}, but is quite different in nature. Contrary to the previous case, it is now the functional space itself that is deviating from the original definition. 
In particular, convoluting \eqref{eq:cF-GFFT-extra} with the  metric of $\cV_{\D}^{\rm symm}$, we obtain a modified resolution of the identity:
\be \label{eq:res-id-symm-2}
\begin{split}
\mathbb{I}_{\rm symm}(x_1,x_2,x_3,x_4) &\equiv  \f12\left( \d(x_1-x_3)\d(x_2-x_4) + \d(x_1-x_4)\d(x_2-x_3) \right) \\
&=  \sum_{J\in \mathbb{N}_0^{\text{even}}}  \int_{\f{d}{2}}^{\f{d}{2}+\im\infty}  \f{{\rm d}h}{2\pi\im} \r(h,J) \, \cN^{\D}_{h,J}  \cN^{\wtD}_{\htilde,J} \, \Psi_{h,J}^{\D,\D,\wtD,\wtD}(x_1,x_2,x_3,x_4)  \\
&\qquad -   {\rm Res}\left[ \r(h,0) \, \cN^{\D}_{h,0}  \cN^{\wtD}_{\htilde,0} \, \Psi_{h,0}^{\D,\D,\wtD,\wtD}(x_1,x_2,x_3,x_4) \right]_{h=2\D} \,.
\end{split}
\ee

Since \eqref{eq:prod-cN} has no poles, it would seem that the isolated contribution here is trivial. However, $\Psi_{2\D,0}^{\D,\D,\wtD,\wtD}(x_1,x_2,x_3,x_4)$ is a singular distribution proportional to $1/|x_{34}|^d\sim \Gamma(0) \d(x_{34})$; 
writing explicitly the limit involved in the definition of the residue, we find that  such singularity leads to a non-trivial isolated contribution to the resolution of the identity:
\be \label{eq:extra}
\begin{split}
-{\rm Res} &\left[  \r(h,0) \, \cN^{\D}_{h,0}  \cN^{\wtD}_{\htilde,0} \, \Psi_{h,0}^{\D,\D,\wtD,\wtD}(x_1,x_2,x_3,x_4) \right]_{h=2\D}\\
&= \lim_{\eps\to 0} \,  2\eps \,  \r(2\D-2\eps,0) \,  \f{2^{d/2}}{\pi^{d}} \, \Psi_{2\D-2\eps,0}^{\D,\D,\wtD,\wtD}(x_1,x_2,x_3,x_4)\\
&= \lim_{\eps\to 0} \, c(d/2-\epsilon) \, c(2\D-2\eps) \, c (d-2\D+2\eps) \, \Psi_{2\D-2\eps,0}^{\D,\D,\wtD,\wtD}(x_1,x_2,x_3,x_4)\\
&= c(2\D) \, c (d-2\D) \, \int \dd x_0 \f{1}{|x_{10}|^{2\D} |x_{20}|^{2\D}}  \f{1}{|x_{30}|^{d-2\D} |x_{40}|^{d-2\D}}  \d(x_{34}) \,,
\end{split}
\ee
where we used the distributional identity $\lim_{\epsilon\to 0} c(d/2-\epsilon) / |x|^{d-2\epsilon} = \delta(x)$.

\paragraph{Conformal partial waves on the sphere.}
Everything we discussed in this appendix transcribes mutatis mutandis for CFTs on the sphere via the Weyl mapping (i.e. replacing all the distances by chordal distance and adding the adequate volume factors). By conformality of all the integrals involved (e.g.\ in \eqref{eq:shadowTrans}, \eqref{eq:CPW}, and so on), the analytic properties of the conformal partial wave expansion are unchanged, and the effect of being on the sphere is only visible in the external points being multiplied by the appropriate $\Om(x)$ factors.

However,  in the evaluation of the sphere free energy we encounter traces of endomorphisms on $\cV_\D$,
\be
\Tr(\cE) =  
\int \dd x_1 \dd x_2 \, \cE( x_1,x_2 , x_1,x_2 ) \,,
\ee
which are divergent due to their conformal invariance. The regularization of such traces necessarily breaks conformal invariance, and as a result $\Om(x)$ factors survive, which play a crucial role in leading to finite results.
The precise choice of regularization and its effects are described in the bulk of the chapter.

\section{The special graph at NNLO}
\label{app:monster}

In this section we show that the contribution of the graph in
figure~\ref{fig:monstru} to the sphere free energy is finite. As it is not especially informative, we will not compute this contribution exactly.

The amplitude of any vacuum graph can be written as the convolution of a free covariance (representing any of its edges) and a two-point kernel (representing the amplitude of the remaining two-point amputated graph $\mathcal{G}$):
\be
 I = \int_{x,y} C(x,y) A^{\mathcal{G}}(x,y) \;.
\ee
In the case of the melon in \eqref{eq:melon-int} for instance $A^{\mathcal{G}}(x,y) = C(x,y)^3$. Using dimensional regularization and $\Delta = \frac{d-\epsilon}{4}$, dimensional analysis leads to:
\be
 I = \int_{x,y} \frac{c(\Delta)}{s(x,y)^{ \frac{d- \epsilon}{2}} } \; \frac{A^{\mathcal{G}}(\epsilon)}{s(x,y)^{\frac{3}{2}d - b \epsilon}} \;.
\ee
The precise scaling $b$ depends on the particular graph one considers (e.g. $b=3/2$ for the melon). Following the same steps leading to \eqref{eq:vanishmelo} we conclude that:
\be
 I \sim \frac{1}{\Gamma(\frac{1+2b}{4} \epsilon)} A^{\mathcal{G}}(\epsilon) \;.
\ee

\begin{figure}[htbp]
	\centering
	\includegraphics[width=0.16\linewidth]{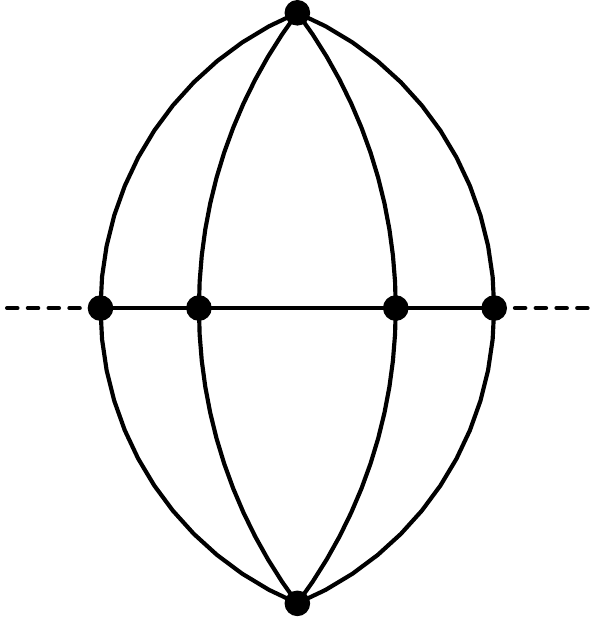}
	\caption{Amputated two-point diagram obtained by opening any edge in the graph of figure~\ref{fig:monstru}.}
	\label{fig:monstruopen}
\end{figure}

For the melon $A^{\mathcal{G}}(\epsilon)\sim \epsilon^0$, that is the melon does not contribute to the sphere free energy in the $\epsilon\to 0$ limit. We show below that for the graph in figure~\ref{fig:monstru} opened on any of its edges, as depicted in figure~\ref{fig:monstruopen},
there exist two constants $A_1$ and $A_2$ such that: 
\be
\label{appeq:amplibound} 
\frac{A_1}{\epsilon} < A^{\mathcal{G}}(\epsilon) < \frac{A_2}{\epsilon} \;,
\ee 
hence this graph brings a finite contribution to the sphere free energy. This contribution cancels between the different fixed points of interest as they all have the same tetrahedral coupling.

The remainder of this appendix is a rigorous proof of \eqref{appeq:amplibound}. It is quite lengthy and uses a set of techniques that, while standard, are quite apart from the ones used in the rest of the chapter.

One can of course apply the techniques we discuss here to the ladder diagrams. Graph by graph the ladder diagrams are divergent, and present increasingly singular poles in $1/\epsilon$. Accounting for the diagrams with $\lambda_1$ vertices allows one to subtract the poles and one can in principle resum the finite parts to reproduce the finite part of \eqref{eq:ladderssum}. This graph by graph computation is exceedingly difficult. The main gain of the conformal partial waves techniques we introduced in the main body of this chapter is to bypass this analysis entirely and directly give us the end result.

\paragraph{Going to flat space.}
We are interested in identifying the leading singular behavior in the $\epsilon\to 0$ limit of the amplitude $A^{\mathcal{G}}(x,y)$ of amputated two-point graphs $\mathcal{G}$. As already mentioned, dimensional analysis implies that, up to local terms:
$ A^{\mathcal{G}}(x,y) =  A^{\mathcal{G}}(\epsilon) / s(x,y)^{\frac{3}{2}d - b \epsilon}$
where $A^{\mathcal{G}}(\epsilon)$ might display poles in $1/\epsilon$. As this is an ultraviolet divergence it is insensitive to the details of the infrared regularization hence the leading divergence is the same on the sphere and on flat space. From now on we work on flat space.

\paragraph{Why it is non-trivial.} In the long-range model, two-point graphs are primitively power divergent. Once this local power divergence is dealt with (either set to zero in dimensional regularization or subtracted by a mass counterterm in other schemes) the only remaining primitively divergent graphs are the four-point ones. The latter bring $1/\epsilon$ poles that pile up if they come from subgraphs fully included in larger subgraphs (we review how this occurs below).

For two-point amputated graphs this naive power counting yields $A^{\mathcal{G}}(\epsilon) \sim \epsilon^{-1}$ for the two-point melon (as it has four-point subgraphs) and $A^{\mathcal{G}}(\epsilon)\sim \epsilon^{-2}$ for the two-point graph in figure~\ref{fig:monstruopen}, as it has a four-point subgraph that is a subgraph of another four-point subgraph. The non-trivial result of this section is that a detailed analysis of the sub divergences of these two graphs improves on the naive expectation, that is $A^{\mathcal{G}}(\epsilon)\sim \epsilon^0$ for the melon and $A^{\mathcal{G}}(\epsilon) \sim \epsilon^{-1}$ for the graph in figure~\ref{fig:monstruopen}

\paragraph{Structure of the amplitudes.}
On flat space the divergences and their subtractions are captured by the Bogoliubov--Parasiuk--Hepp--Zimmermann theorem \cite{Hepp:1966eg,Zimmermann:1969jj,rivasseau2014perturbative}.
For long-range models only two and four-point subgraphs
are primitively divergent and they are subtracted by applying local Taylor operators. 

The momentum and direct space formulation of the BPHZ theorem
are reviewed for instance in \cite{rivasseau2014perturbative} and
the parametric space version is discussed in 
\cite{Bergere:1974zh,Bergere:1980sm}. In flat space, using Schwinger parameters, the amplitude of an amputated graph in the long-range model with external momenta $p_i$, $V$ vertices and $E$ edges is $ \bar A_{\mu}^{\mathcal{G}}(p_i)  = (2\pi)^d \delta(\sum_ip_i)    A_{\mu}^{\mathcal{G}}(p_i)  $ with:
\be
\begin{split}
   A^{\mathcal{G}}(p_i) & = 
   \frac{ 1 }{\Gamma(\zeta)^E (4\pi)^{\frac{d}{2}(E-V+1) }  }
 \int_0^{k^{-2} }  \left( \prod_{e\in \mathcal{G}} d\alpha_e \; \alpha_e^{\zeta-1}  \right) \; 
 \frac{ e^{-\frac{ V(\mathcal{G})  } { U(\mathcal{G}) } } } { U(\mathcal{G})^{d/2}  }   \;,\crcr
 U(\mathcal{G}) & = \sum_{T \subset \mathcal{G} } \prod_{e\notin T} \alpha_e  \;,\;\;
  V(\mathcal{G}) =  \sum_{T_1,T_2 \subset \mathcal{G} } (\sum_{i \in T_2} p_i)^2\;  \prod_{e\notin T_1\cup T_2} \alpha_e \; ,
\end{split}
\ee
where $T$ runs over the spanning trees in $\mathcal{G}$, respectively $T_1,T_2$ run over the pairs of trees, and $i\in T_2$ runs over the external vertices that belong to the tree $T_2$. 
The integral is cutoffed by an infrared cutoff $k$ on the integration interval, but any other infrared cutoff will do.
The overall factor $ (2\pi)^d \delta(\sum_ip_i) $ reproduces the bare vertex. 
We work in dimensional regularization and we set $\zeta = \frac{d+\epsilon}{4}$. 
The amplitude at zero external momenta of $\mathcal{G}$ is, up to prefactors:
\be
 A^{\mathcal{G}} \sim  \int_0^{k^{-2}}  \left( \prod_{e\in \mathcal{G}} d\alpha_e \; \alpha_e^{\zeta-1} \right) \; 
 \frac{ 1} { U(\mathcal{G})^{d/2}  }  = k^{-2\zeta E + d [E -V +1 ]}
 \int_0^{1}  \left( \prod_{e\in \mathcal{G}} d\alpha_e \; \alpha_e^{\zeta-1} \right) \; 
 \frac{ 1} { U(\mathcal{G})^{d/2}  } \; .
\ee

If $\mathcal{G}$ is a two-point graph its amplitude in momentum space writes:
\be\label{appeq:ampli2point}
 A^{\mathcal{G}}(p) = p^{ \frac{d + \epsilon }{2} - V \epsilon} A^{\mathcal{G}}(\epsilon) + A^{\mathcal{G}} \; ,
\ee
where we denote somewhat abusively by $A^{\mathcal{G}}(\epsilon)$ the coefficient of the scaling behavior in $p$. Note that for two-point graphs $A^{\mathcal{G}}$ is zero in dimensional regularization, as it is power divergent. This is of course not the case for four-point graphs. In principle $A^{\mathcal{G}}(\epsilon)$ is a function of $p/k$ and in the case of the melon we showed in chapter \ref{chap:CTKT} that it has a finite limit for $k\to 0$. 

\paragraph{Taylor operators and renormalized amplitudes.}
The Taylor operators acting on amplitudes are localization operators. For any $\gamma$ subgraph of $\mathcal{G}$ we define:
\be
\tau_{\gamma} A^{\mathcal{G}}(p_i) \equiv  A^{\gamma} \; A^{\mathcal{G}/\gamma}(p_i) \;,
\ee
where $A^{\gamma}$ is the amplitude of $\gamma$ at zero momentum and $\mathcal{G}/\gamma$ denotes the graph obtained from $\mathcal{G}$ by contracting $\gamma$ to a point. As we deal with the long-range case the subtraction of local parts suffices in order to render the amplitudes ultraviolet finite.

We consider amputated subgraphs $\gamma$ of $\mathcal{G}$. A subgraph $\gamma$ contains all the vertices hooked to its edges.
An inclusion forest \cite{Zimmermann:1969jj} of subgraphs of $\mathcal{G}$ is a family ${\cal F}$ of subgraphs such that any two $\gamma_1 ,\gamma_2 \in {\cal F}$ are either totally disjoint (i.e. they do not share neither edges nor vertices) or one of them is fully contained in the other:
\be
 {\cal F} = \big\{\gamma \in \mathcal{G} \big{|}\forall \gamma_1,\gamma_2  \text{ either } \gamma_1 \subset \gamma_2 \; \text{, or } \gamma_2 \subset \gamma_1 \; \text{, or }  \gamma_1 \cap \gamma_2 = \emptyset \big\} \;.
\ee

The renormalization operator  \cite{Hepp:1966eg,Zimmermann:1969jj,rivasseau2014perturbative} $R_{\mathcal{G}}$ associated to the graph $\mathcal{G}$ is a sum over the inclusion forests ${\cal F} \subset \mathcal{G}$ 
of primitively divergent subgraphs (i.e. two and four-point subgraphs in our case) of a product of Taylor operators associated to the graphs in the forest:
\be
R_{\mathcal{G}} =\sum_{\cal F \subset \mathcal{G} } \prod_{\gamma \in \cal F} (-\tau_{\gamma} ) \;.
\ee
the forests include the empty forest, which contributes a $1$ to this formula. 

\begin{theorem}(BPHZ \cite{Bergere:1974zh,rivasseau2014perturbative}). 
The renormalized amplitude $R_{\mathcal{G}}A^{\mathcal{G}}(p_i)$ of any graph $\mathcal{G}$ is ultraviolet convergent, that is $\lim_{\epsilon \to 0} R_{\mathcal{G}}A^{\mathcal{G}}(p_i)$ is finite.
\end{theorem}

\paragraph{How we will use the BPHZ theorem.}
We are interested in identifying the $1/\epsilon$ behavior of the bare amplitude of a two-point graph $\mathcal{G}$. Separating the empty forest in the renormalization operator we have:
\be
R_{\mathcal{G}} A^{\mathcal{G}}(p)  = A^{\mathcal{G}}(p) + \sum_{\cal F \subset \mathcal{G}}^{ {\cal F} \neq \emptyset} \prod_{\gamma \in \cal F} (-\tau_{\gamma} ) A^{\mathcal{G}}(p) \; ,
\ee
and the BPHZ theorem ensures that this expression is convergent in the $\epsilon\to 0$ limit.
It follows that the singular part of the coefficient $A^{\mathcal{G}}(\epsilon)$ in \eqref{appeq:ampli2point} \emph{must be entirely canceled by the subtractions}, hence it equals the divergent part of the counterterms.

\paragraph{The melon.} Let us first consider the melon $\mathcal{G}$. As primitively divergent subgraphs it has itself $\gamma = \mathcal{G}$, and three subgraphs $\gamma_i,i=1,2,3$ made of two edges. It follows that:
\be\label{appeq:melofin}
 R_{\mathcal{G}}A^{\mathcal{G}}(p) = A^{\mathcal{G}}(p) - \tau_\gamma A^{\mathcal{G}}(p) - \sum_i \tau_{\gamma_i}A^{\mathcal{G}}(p)  + \sum_i \tau_{\gamma}\tau_{\gamma_i}A^{\mathcal{G}}(p) =
 A^{\mathcal{G}}(p) - A^{\mathcal{G}}  \; .
\ee
In this equation, $ \tau_\gamma A^{\mathcal{G}}(p)$ is just the local part $A^{\mathcal{G}}$ of the melon (and in particular it is zero in dimensional regularization). Moreover, $ \tau_{\gamma}\tau_{\gamma_i}A^{\mathcal{G}}(p) = \tau_{\gamma_i} A^{\mathcal{G}}(p)$ because $\mathcal{G}/\gamma_i$ is a tadpole graph, therefore the action of $\tau_{\gamma}$ is trivial.
It follows that the two terms summed over $i$ cancel exactly, which, together with \eqref{appeq:ampli2point}, yields:
\be
p^{ \frac{d + \epsilon }{2} - V \epsilon} A^{\mathcal{G}}(\epsilon) = 
A(p) - A^{\mathcal{G}}  =  R_{\mathcal{G}}A^{\mathcal{G}}(p) \; .
\ee
As $R_{\mathcal{G}}A^{\mathcal{G}}(p)$ has a finite limit for $\epsilon \to 0$ we conclude that $ A^{\mathcal{G}}(\epsilon) $ has no poles in $1/\epsilon$.

\paragraph{The graph in figure~\ref{fig:monstruopen}.} This graph has many four-point subgraphs.
To identify them we label $x,y$ the two external vertices, $v_1,v_2$ the top and bottom vertices and $z_1,z_2$ the two vertices on the horizontal ($x$ and $z_1$ are connected by an edge). The list of two and four-point subgraphs of the graph in figure~\ref{fig:monstruopen} comprises: 
\begin{itemize}
 \item two kite graphs, see figure~\ref{fig:monstruopensubgraph2}, $\gamma_x$ (and $\gamma_y$) obtained by cutting the edges $(x,v_1)$, $(x,z_1)$ and $(x,v_2)$ (resp. 
  $(y,v_1)$, $(y,z_2)$ and $(y,v_2)$).
  \begin{figure}[htbp]
  	\centering
  	\includegraphics[width=0.18\linewidth]{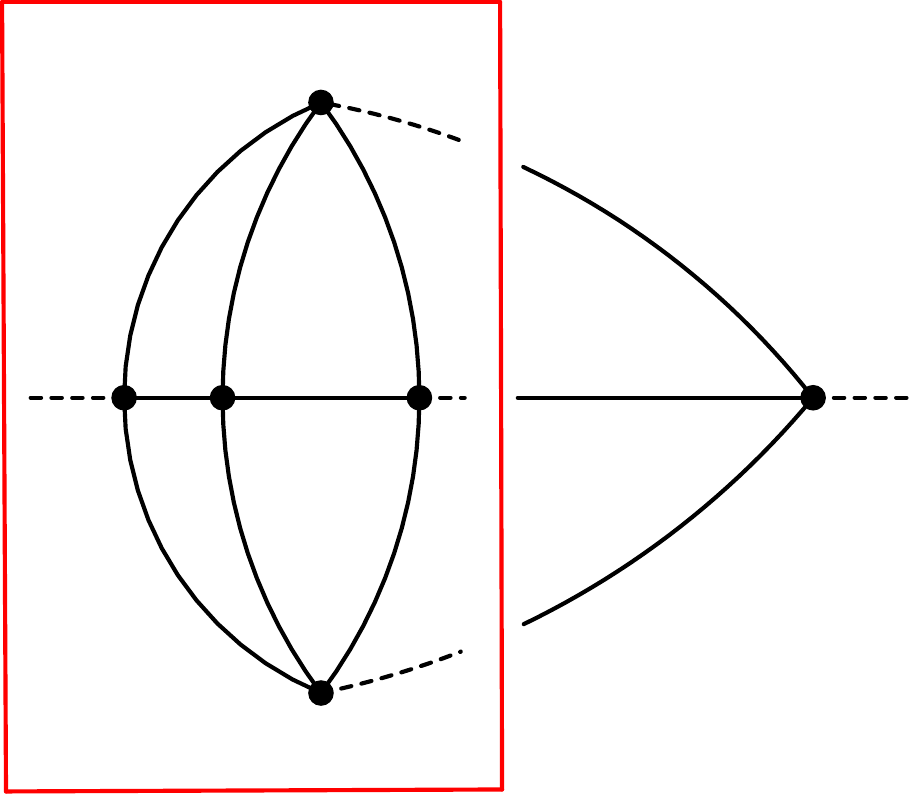}
  	\caption{Kite graph obtained by cutting $(y,v_1)$, $(y,z_2)$ and $(y,v_2)$.}
  	\label{fig:monstruopensubgraph2}
  \end{figure}
 \item 11 graphs $\gamma^{ab}$ (see figure~\ref{fig:monstruopensubgraph1}) obtained by cutting any of the internal edges $(a,b)$ in the graph.
 \begin{figure}[htbp]
 	\centering
 	\includegraphics[width=0.14\linewidth]{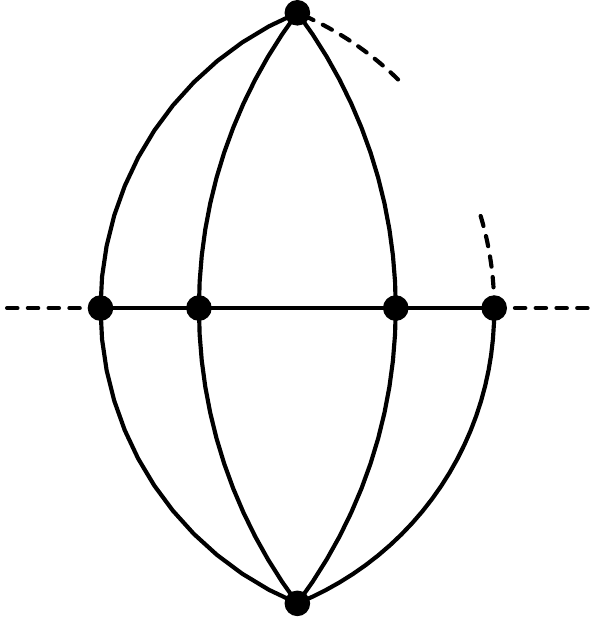}
 	\caption{Example of one $\gamma^{ab}$ graph, obtained by cutting one of the internal edges.}
 	\label{fig:monstruopensubgraph1}
 \end{figure}
 
 \item itself, that is $\gamma=G$
\end{itemize}

These graphs are organized in several inclusion forests:
\begin{itemize}
 \item the empty forest
 \item the one-graph forests. They consist in either
    \begin{itemize}
     \item the graph $\gamma$
     \item one of the kite graphs $\gamma_x$ or $\gamma_y$
     \item any one of the remaining 11 graphs $\gamma^{ab}$
    \end{itemize}
  \item the two-graph forests. They are
    \begin{itemize}
    \item 11 of the form $\{\gamma, \gamma^{ab}\}$ for some  $ab$.
    \item three of the form $\{ \gamma^{xa},\gamma_x \} $ for the $a$s connected to $x$ by some edge
    respectively three of the form $\{\gamma^{ya} ,\gamma_y \} $ for the $a$s connected to $y$ by some edge in $\mathcal{G}$
    \item two special ones  $\{ \gamma, \gamma_x\}$ and
    $\{ \gamma, \gamma_y\}$
    \end{itemize}

  \item the three-graph forests. There are six of these ones
     \begin{itemize}
      \item three of the form $\{\gamma,\gamma^{xa}, \gamma_x\} $ for some $a$ and three of the form  $\{\gamma,\gamma^{ya}, \gamma_y\} $ for some $a$.
     \end{itemize}
\end{itemize}
Plenty of the contributions to the renormalized amplitude cancel
as for any $\gamma^{ab}$ , $\mathcal{G}/\gamma^{ab}$ is contracted to a tadpole hence $\tau_{\gamma^{ab}}A^{\mathcal{G}} = \tau_{\gamma}\tau_{\gamma^{ab}}A^{\mathcal{G}}$ and $\tau_{\gamma^{ab}}\tau_{\gamma_i}A^{\mathcal{G}} = \tau_{\gamma}\tau_{\gamma^{ab}}\tau_{\gamma_i}A^{\mathcal{G}}$. The renormalized amplitude is then: 
\be
 R_{\mathcal{G}}A^{\mathcal{G}}(p) = A^{\mathcal{G}}(p) - A^{\mathcal{G}} - 2 A^{\gamma_x}  \big[ A^{\mathcal{G}/\gamma_x}(p) - A^{\mathcal{G}/\gamma_x}  \big] \;.
\ee
Observe that $\mathcal{G}/\gamma_x$ is the two-point melon graph, hence incidentally $A^{\mathcal{G}/\gamma_x}(p) - A^{\mathcal{G}/\gamma_x}$ is nothing but the subtracted melon $R_{\mathcal{G}/\gamma_x} A^{\mathcal{G}/\gamma_x}(p) $. Combining this with \eqref{appeq:ampli2point} we find that:
\be
p^{ \frac{d + \epsilon }{2} - V \epsilon} A^{\mathcal{G}}(\epsilon) = 
A(p) - A^{\mathcal{G}}  =  R_{\mathcal{G}}A^{\mathcal{G}}(p) + 2 A^{\gamma_x}  R_{\mathcal{G}/\gamma_x} A^{\mathcal{G}/\gamma_x}(p) \;,
\ee
and as the renormalized amplitudes on the right-hand side above have finite limits at $\epsilon \to 0$, the leading divergence of $A^{\mathcal{G}}(\epsilon)$ is the same as the local part of the kite graph $A^{\gamma_x}$.

\paragraph{The amplitude of the kite graph.} We will now conclude this section by proving that the amplitude of the kite graph has a pole of order $1$ in $\epsilon$. The amplitude of any four-point graph $\mathcal{G}$  with no two-point subgraphs at zero external momenta:
\be
 A^{\mathcal{G}} =  \int_0^{k^{-2}}  \left( \prod_{e\in \mathcal{G}} d\alpha_e \; \alpha_e^{\zeta-1} \right) \; 
 \frac{ 1} { U(\mathcal{G})^{d/2}  }  = k^{-2\zeta E + d [E -V +1 ]}
 \int_0^{1}  \left( \prod_{e\in \mathcal{G}} d\alpha_e \; \alpha_e^{\zeta-1} \right) \; 
 \frac{ 1} { U(\mathcal{G})^{d/2}  } \; ,
\ee
is a convergent integral over $\alpha$ for $\zeta = \frac{d+\epsilon}{4}$ but exhibits poles in $1/\epsilon$. 

For a graph with $E$ edges, we divide the integration interval into Hepp sectors, which we denote $\sigma$, that is total orderings of the $\alpha$ parameters $\alpha_{e_{\sigma(1)} } \le \alpha_{e_{\sigma(2) } } \le \dots  \le \alpha_{e_{\sigma(E)} }$. The edge $e_{\sigma(1)} $ is the most ultraviolet (lowest $\alpha$), $e_{\sigma(2)}$ is the next most ultraviolet and so on up to $e_{\sigma(E)}$, which is the most infrared. There are $E!$ sectors, as many as there are permutations. Up to the global scaling the amplitude is:
\be
A^{\mathcal{G}} =  k^{ \epsilon- V \epsilon } \sum_{\sigma} A^{\mathcal{G}}_{\sigma} \;,\quad
 A^{\mathcal{G}}_{\sigma} =  \int_{0\le \alpha_{\sigma(1)} \le \dots \le \alpha_{\sigma(E)}  \le 1}  \left( \prod_{e\in \mathcal{G}} d\alpha_e \; \alpha_e^{\zeta-1} \right) \;  \frac{ 1} { U(\mathcal{G})^{d/2}  }  \;.
\ee

The polynomial $U(\mathcal{G})$ is a sum of positive terms. It is thus bounded from below by any of its terms, and from above by the number of terms times the largest of them. The number of trees in a graph is bounded by the number of subsets of edges, that is $2^E$.

In each Hepp sector there is exactly one leading monomial corresponding to the tree $T_{\sigma}$ built by proceeding from $1$ to $E$ and at each step adding the edge $e_{\sigma(q)}$ if it does not form a loop. We thus ensure that the edges with the lowest possible $\alpha$ parameters are in the tree, hence the complement of $T_{\sigma}$ has the highest possible $\alpha$s, that is in the sector $\sigma$:
\be
0\le \alpha_{\sigma(1)} \le \dots \le \alpha_{\sigma(E)}  \le 1  \Rightarrow 2^E \prod_{e \notin T_{\sigma}} \alpha_e \ge  U(\mathcal{G}) \ge
 \prod_{e \notin T_{\sigma}} \alpha_e \; .
\ee

Associated to a Hepp sector we have the set of ``high graphs'' $\gamma_q =\{e_{\sigma(1)} ,\dots e_{\sigma(q)}\}$ formed by the $q$ most ultraviolet edges. The subgraphs are considered amputated, that is they contain all the edges $\gamma_q =\{e_{\sigma(1)} ,\dots e_{\sigma(q)}\}$ and the vertices hooked to them, but not the external edges (which can be either genuine external edges of $\mathcal{G}$, or some of the ``lower'' edges $e_{\sigma(q+1)}\dots e_{\sigma(E)}$). We remark that $\gamma_E = \mathcal{G}$ and that $\gamma_q$ is not necessarily connected. By construction the leading tree $T_{\sigma}$ is a tree in every connected component of $\gamma_{q}$, and it is the unique tree with this property.

At fixed Hepp sector $\sigma$ we perform the diagonal change of variables:
\be
 \alpha_{\sigma(i)} = \prod_{j=i}^E t_j^2 \;,\qquad \alpha_{\sigma(E) } = t_E^2 \;, \; \; \alpha_{\sigma(E-1)} = t_{E-1}^2t_E^2  \;, \;\; \dots \;\; 
  \alpha_{ \sigma(1) } = t_1^2 t_2^2 \dots t_E^2 \;,
\ee
and we note that all the edges of $\gamma_q$ have a factor $t_q^2$. Now $T_{\sigma}$ is a tree in every connected component of $\gamma_q$, hence the number of edges of $\gamma_q$ not in $T_{\sigma}$ is $E(\gamma_q) - V(\gamma_q) + C(\gamma_q)$ with 
$ V(\gamma_q), E(\gamma_q)$ and $C(\gamma_q)$ the numbers of vertices, edges and connected components of $\gamma_q$.
Every other tree $T$ in $\mathcal{G}$ will be a forest (a collection of trees) in some of the connected components of some $\gamma_q$s and the number of edges of such a $\gamma_q$ not belonging to $T$ is strictly larger that $E(\gamma_q) - V(\gamma_q) + C(\gamma_q)$, hence: 
\be
 U(\mathcal{G})\bigg{|}_{ \alpha_{\sigma(i)} = \prod_{j=i}^E t_j^2 } =  \prod_{i=1}^E  t_i^{2[E(\gamma_i) - V(\gamma_i)  + C(\gamma_i) ]  } \big[ 1 + O(t) \big] \;.
\ee
As the change of variables is diagonal we have:
\be
 A^{\mathcal{G}}_{\sigma} = 2^{E} \int_0^1 \left(  \prod_{i=1}^E dt_i \right)
   \prod_{i=1}^E t_i^{ - 1 + 2 i \zeta - d [E(\gamma_i) - V(\gamma_i)  + C(\gamma_i) ] }   \;\frac{1}{ [1 + O(t)]^{d/2} } \;.
\ee
An upper/lower bound is obtained by using  $2^E > 1 + O(t) >1$. Denoting the convergence degree of $\gamma_i$ by 
$ \omega(\gamma_i) = 2 i\zeta - d [E(\gamma_i) - V(\gamma_i)  + C(\gamma_i) ]$, if all the convergence degrees are positive an upper/lower bound is:
\be
  \prod_{i=1}^E \frac{1}{\omega(\gamma_i)}  \le A^{\mathcal{G}}_{\sigma} \le  2^E \; \prod_{i=1}^E \frac{1}{\omega(\gamma_i)}\;.
\ee

In order to conclude it is enough to examine the possible convergence degrees for all the subgraphs of $\mathcal{G}$. The number of edges of $\gamma_i$ is exactly $i$ and as we deal with quartic vertices $2i = 4V(\gamma_i) - n(\gamma_i)$ with $n(\gamma_i)$ the number of external half-edges of $\gamma_i$. 
Denoting $\gamma_i^{\rho}$ the connected components of $\gamma_i$, with $\rho = 1,\dots C(\gamma_i)$, and observing that the edges, vertices and external legs are distributed among the connected components, the convergence degree of $\gamma_i$ is:
\be
 \omega(\gamma_i) = 2 i\zeta - d [E(\gamma_i) - V(\gamma_i)  + C(\gamma_i) ]
  = i(2\zeta - \frac{d}{2}) + d \sum_{\rho =1}^{C(\gamma_i )}  \frac{n(\gamma_i^{\rho}) -4 }{4}  = \frac{\epsilon}{2} i + 
  d \sum_{\rho =1}^{C(\gamma_i )}  \frac{n(\gamma_i^{\rho}) -4 }{4}   \;,
\ee
where we used $\zeta  = \frac{d+\epsilon}{4}$. So far we have been quite general: we only assumed that there are no subgraphs with zero or negative convergence degree, that is no two-point subgraphs. For the kite graph:
\begin{itemize}
 \item every strict subgraph (i.e. subgraph different from itself) has at least six external half-edges but less that 20 half-edges hence $8\epsilon + 4 d \ge \omega(\gamma_i) \ge \frac{d}{2}$.
 \item the wheel itself (corresponding to $t_E$) has $4$ external half-edges and 8 edges $\omega(\gamma_E) = 4\epsilon$.
\end{itemize}
Thus (recalling that there are $E!$ sectors) upper and lower bounds are:
\[
\frac{1}{\epsilon} \left( \frac{1}{ 4 (4d+ 8\epsilon)^E }  \right) \le A^{\mathcal{G}}_{\sigma} \le \frac{1}{\epsilon}  \left( \frac{4^{E-1}}{ d^E} \right) \;\;\Rightarrow
\; \;\frac{ k^{ \epsilon- 5 \epsilon } }{\epsilon} \left( \frac{8!}{ 4 (4d+ 8\epsilon)^8 }  \right) \le A^{\mathcal{G}} \le \frac{  k^{ \epsilon- 5 \epsilon } }{\epsilon}  \left( \frac{8! \; 4^{7}}{ d^8}  \right) \; .
\]

\section{Regularized trace of conformal partial waves} 
\label{app:I_eps}
In this appendix we show how to compute $\cI_\eps(J)$ \eqref{Iepsilon}. By homogeneity of the sphere, we can set $z=0$, and factor out  the volume of the $d$-sphere  $V_d=\int \dd z\, \Omega(z)^d$, given in \eqref{eq:Vol-Sd}. The most generic form of a three-point function $\la \phi \phi O \ra$ is fixed by conformal symmetry as in \eqref{eq:3pt} and we obtain:
\begin{equation}
	\cI_\eps(J)=V_d\int \dd x_1 \dd x_2 \frac{\left(Z^{\m_1} \cdots Z^{\m_J} - \text{``traces"}\right)\left(Z^{\m_1} \cdots Z^{\m_J} - \text{``traces"}\right)}{(1+x_1^2)^{\eps} (1+x_2^2)^{\eps}|x_1|^{d-\eps}|x_2|^{d-\eps}|x_1-x_2|^{d-\eps}} \, ,
\end{equation}
where $Z^{\m} = \f{|x_{1}||x_{2}|}{|x_1-x_2|} \left(\f{x_{1}^{\m}}{|x_{1}|^2} - \f{x_{2}^{\m}}{|x_{2}|^2}\right)$ has unit norm.
In order to take care of the spin structure we use the following identity (e.g.\ \cite{Dolan:2011dv,Costa:2016hju}), which follows from the addition theorem \eqref{eq:additionTh}:
\begin{equation}
\left(Z^{\m_1} \cdots Z^{\m_J} - \text{``traces"}\right)\left(Z^{\m_1} \cdots Z^{\m_J} - \text{``traces"}\right)=\frac{(d-2)_J}{2^l (\frac{d-2}{2})_J}=\frac{\Gamma(d-2+J)\Gamma(\tfrac{d-2}{2})}{2^J\Gamma(d-2)\Gamma(\tfrac{d-2}{2}+J)} \,.
\end{equation}
Then the relation $\mathcal{I}_\eps(J)=\frac{\Gamma(d-2+J)\Gamma(\tfrac{d-2}{2})}{2^J\Gamma(d-2)\Gamma(\tfrac{d-2}{2}+J)} \mathcal{I}_\eps(0)$ holds and it is sufficient to compute the integral at $J=0$.
Now we are ready to deal with the integral 
\begin{equation}
	\frac{\mathcal{I}_\eps(0)}{V_d}=(2a)^{3\epsilon-d}\int \dd x_1 \dd x_2 \frac{1}{(1+x_1^2)^{\eps} (1+x_2^2)^{\eps}|x_1|^{d-\eps}|x_2|^{d-\eps}|x_1-x_2|^{d-\eps}} \, .
\end{equation}
It is convenient to perform the change of variable: $x^\mu=\frac{x_1^\mu}{x_1^2}$ and $y^\mu=\frac{x_2^\mu}{x_2^2}$. The Jacobian determinant of the transformation is $x^{-2d}y^{-2d}$ and the integral simplifies to:
\begin{equation}
\frac{\mathcal{I}_\eps(0)}{V_d}=(2a)^{3\epsilon-d}\int \dd x \dd y \frac{1}{(1+x^2)^{\eps}(1+y^2)^{\eps}(|x-y|^2+\mu^2)^{\frac{d}{2}-\frac{\eps}{2}}} \,,
\end{equation}
where we have introduced another regulator $\mu^2$ that is needed to make the integral convergent.
Assuming $0<\eps<d$, we introduce three Schwinger parameters:
\begin{equation}
	\begin{split}
\frac{\mathcal{I}_\eps(0)}{V_d}=(2a)^{3\epsilon-d}\int \dd x \dd y \frac{1}{\Gamma(\eps)^2\Gamma(\tfrac{d}{2}-\tfrac{\eps}{2})}\int d \alpha_1 d \alpha_2 d \alpha_3  (\alpha_1 \alpha_2)^{\eps-1}\alpha_3^{\tfrac{d}{2}-1-\tfrac{\eps}{2}} \\
 \times e^{-\alpha_1(1+x^2)-\alpha_2(1+y^2)-\alpha_3(x^2+y^2-2xy)-\alpha_3 \mu^2} \, .
	\end{split}
\end{equation}
We can now perform the Gaussian integrals in x and y:
\begin{equation}
\frac{\mathcal{I}_\eps(0)}{V_d}=(2a)^{3\epsilon-d}\frac{\pi^d}{\Gamma(\eps)^2\Gamma(\frac{d}{2}-\frac{\eps}{2})}\int d \alpha_1 d  \alpha_2 d \alpha_3 \frac{(\alpha_1 \alpha_2)^{\eps-1}\alpha_3^{\frac{d}{2}-1-\frac{\eps}{2}}}{(\alpha_1\alpha_2+\alpha_3(\alpha_1+\alpha_2))^{d/2}}e^{-\alpha_1-\alpha_2-\mu^2\alpha_3}\,.
\end{equation}
Using the Mellin-Barnes representation introduced in appendix \ref{sec:MB} of chapter \ref{chap:3loops} we find:
\begin{equation}
	\begin{split}
\frac{\mathcal{I}_\eps(0)}{V_d}=\frac{(2a)^{3\epsilon-d}\pi^d}{\Gamma(\eps)^2\Gamma(\frac{d}{2}-\frac{\eps}{2})}\int_{x_0- i \infty}^{x_0+i\infty} \left( \frac{dz}{2\pi i} \right) \int d\alpha_1d \alpha_2 d\alpha_3 e^{-\alpha_1-\alpha_2-\mu^2\alpha_3}\\ \times
\frac{\Gamma(-z)\Gamma(z+\frac{d}{2})(\alpha_1\alpha_2)^{\epsilon+z-1} \alpha_3^{-z-\frac{\eps}{2}-1}}{\Gamma(\tfrac{d}{2})(\alpha_1+\alpha_2)^{z+d/2}}  \,.
	\end{split}
\end{equation}
Next, for  ${\rm Re}(z)<-\eps/2$ we can use the integral representation of the Gamma function to integrate $\alpha_3$.
And if we require also ${\rm Re} (z)>-\eps$ and ${\rm Re} (z)+2\eps>d/2$ the integration over $\alpha_1$ and $\alpha_2$ can be found in appendix \ref{sec:MB}. We finally obtain:
\begin{equation}
\frac{\mathcal{I}_\eps(0)}{V_d}=\frac{(2a)^{3\epsilon-d} \pi^d}{\Gamma(\epsilon)^2\Gamma(\frac{d}{2}-\frac{\eps}{2})} \int_{x_0- i \infty}^{x_0+i\infty} \left( \frac{dz}{2\pi i} \right) \mu^{2z+\epsilon}  \frac{\Gamma (-z) \Gamma \left(\frac{d+2z}{2}\right) \Gamma \left(-\frac{2z+\epsilon }{2}\right) \Gamma \left(z+\epsilon\right)^2 \Gamma \left(\frac{-d+2z+4\epsilon}{2}\right)}{\Gamma \left(\frac{d}{2}\right) \Gamma \left(2 z+2\epsilon\right)} \,.
\end{equation}

Since we have the two conditions ${\rm Re}(z)<-\eps/2$ and ${\rm Re} (z)>-\eps$, we are free to choose $x_0 \in (-\eps,-\eps/2)$. 
Moreover we must take $\eps>d/2$ in order to avoid poles of the last gamma function in this range, but of course we will then analytically continue the result to small $\eps$. Closing the  contour to the right we pick only the poles at $z=n$ and $z=-\eps/2+n$ with $n\in\mathbb{N}_0$.  
Computing the residues at the poles we find the result:
\begin{equation}
	\begin{split}
\frac{\mathcal{I}_\eps(0)}{V_d}= (2a)^{3\epsilon-d}\left(\frac{\pi ^d \Gamma \left(\frac{\epsilon }{2}\right)^3 \Gamma \left(\frac{3 \epsilon }{2}-\frac{d}{2}\right)}{\Gamma \left(\frac{d}{2}\right) \Gamma \left(\epsilon\right)^3}\right)\left(1+ \mathcal{O}(\mu^2)\right) \\
+ \mu^{\eps} (2a)^{3\epsilon-d}\left( \frac{\pi ^d \Gamma \left(-\frac{\epsilon }{2}\right) \Gamma \left(\frac{4\epsilon -d}{2}\right)}{\Gamma \left(2\epsilon\right) \Gamma \left(\frac{d}{2}-\frac{\epsilon }{2}\right)}\right)\left(1+ \mathcal{O}(\mu^2)\right) \, .
	\end{split}
\end{equation}
When we remove the $\mu$ regulator, while keeping $\eps$ finite, only the first term coming from $z=-\eps/2$ survives and we find the final result \eqref{eq:I_eps}.

\section{Large-$J$ expansion} 
\label{app:NumericsLargeJ}

In this appendix we detail how the finite part from \eqref{eq:F-derviative} is extracted. As explained in the main text, in order to regularize the sum over $J$ it is important to consistently shift $\wtD\rightarrow\wtD-\eps$ and $\htilde\rightarrow\htilde-\eps$ everywhere. In particular, it is crucial to introduce $\eps$ in the product of normalization factors \eqref{eq:prod-cN}. At large $J$ this product reduces to:
\be
\cN^{\D}_{h,J}  \cN^{\wtD-\eps}_{\htilde-\eps,J} \sim \f{2^{3(d+\eps)/2+J}}{(2\pi)^{d}} \, J^{-3\eps} \left( 1 + \mathcal{O}(1/J) \right) \,,
\ee
and the factor $J^{-3\eps}$ turns out to suffice to render the sum convergent.
After setting $\Delta=\frac{d}{4}$,  the regularized version of \eqref{eq:F-derviative} reads:
\be 
- g\f{\p}{\p g}F_{\rm NNLO}^{\eps}  =N^2 \sum_{J\in \mathbb{N}_0}  \int_{\f{d}{2}}^{\f{d}{2}+\im\infty}  \f{{\rm d}h}{2\pi\im} \r(h,J) \,\f{k(h,J)^2}{1-k(h,J)}\, \cN^{\frac{d}{4}}_{h,J}  \cN^{\frac{3d}{4}-\eps}_{\htilde-\eps,J} \, \mathcal{I}_\eps(J)  \,,
\ee
where $\rho(h,J)$, $k(h,J)$, $\cN^{\D}_{h,J}$ and $\mathcal{I}_\eps(J)$ are all ratios of gamma functions.  

Integrating numerically each term at fixed $J$, and using standard convergence tests, one finds that the resulting series is divergent at $\eps=0$. 
In order to isolate and subtract the divergence, we need to identify the asymptotic behavior at large $J$. A naive expansion of the integrand in $1/J$ leads however to a divergent integral over $h$, due to the exchange of limit and integral, and a more careful analysis is needed.
It turns out to be convenient to make the change of variable $h=d/2 + \im \a J$ in the integral, after which we can use Stirling formula on the gamma functions with $J$ in the argument. 
With this procedure we find the following asymptotic behavior in $J$: 
\begin{equation} \label{eq:J-asympt}
	- \f{1}{N^2} g\f{\p}{\p g}F_{\rm NNLO}^{\eps} \to \sum_{J\in \mathbb{N}_+} \frac{1}{J^{1+3\eps}} \int_{0}^{+\infty} {\rm d}\a\, F(\a,\eps)=\zeta(1+\eps)f(\eps) \,,
\end{equation}
hence the series has a simple pole at $\eps=0$. 
The precise expressions of $F(\a,\eps)$ and $f(\eps)$ are:
\begin{align}
	&F(\a,\eps)=\frac{\pi ^{-\frac{d}{2}} \, g^4 \, 2^{-2 d+\frac{9 \epsilon }{2}-1} \,a^{3 \epsilon} \, \Gamma \left(\frac{d-2}{2}\right) \Gamma \left(\frac{\epsilon }{2}\right)^3   \Gamma \left(\frac{3 \epsilon }{2}-\frac{d}{2}\right)}{\Gamma (d-2) \Gamma \left(d\right) \Gamma (\epsilon )^3} \, \a^{d-2} \left(\a^2+1\right)^{1-d-\frac{3 \epsilon }{2}}\,,\\
	&f(\eps)=\frac{\pi ^{2-\frac{d}{2}} \, g^4 \, 2^{-3 d+\frac{3 \epsilon }{2}+4} \, a^{3 \epsilon}\, \Gamma \left(\frac{3 \epsilon }{2}-\frac{d}{2}\right) \Gamma \left(\frac{1}{2} (d+3 \epsilon -1)\right)}{\Gamma \left(d\right) \Gamma \left(\frac{\epsilon +1}{2}\right)^3 \Gamma \left(d+\frac{3 \epsilon }{2}-1\right)} \,,
\end{align}
and they are both analytic functions at $\eps=0$ (for $2<d<4$).
For numerical computation, it is convenient to add and subtract the asymptotic contribution \eqref{eq:J-asympt} to the original series, and write:
\be
\begin{split} \label{RegularizedSum}
	- g\f{\p}{\p g}F_{\rm NNLO}^{\eps}  =  &\,  N^2 \int_{\f{d}{2}}^{\f{d}{2}+\im\infty}  \f{{\rm d}h}{2\pi\im} \r(h,0) \,\f{k(h,0)^2}{1-k(h,0)}\, \cN^{\D}_{h,0}  \cN^{\wtD-\eps}_{\htilde-\eps,0} \, \cI_\eps (0) \\
	&+ N^2  \sum_{J\in \mathbb{N}_+} \Bigg( \int_{\f{d}{2}}^{\f{d}{2}+\im\infty}  \f{{\rm d}h}{2\pi\im}  \r(h,J) \,\f{k(h,J)^2}{1-k(h,J)}\, \cN^{\D}_{h,J}  \cN^{\wtD-\eps}_{\htilde-\eps,J} \, \cI_\eps (J) -\frac{f(\eps)}{J^{1+3\epsilon}}\Bigg) \\
	&  +N^2 \zeta(1+\eps)f(\eps)  \,,
\end{split}
\ee
where in the first term we isolated the $J=0$ contribution. The sum in the second line is now convergent for $\eps=0$, hence it can be computed numerically.
The last term expands in $\epsilon$ as:
\begin{equation}
	N^2f(\epsilon)\zeta(1+3\epsilon)=N^2\Big(\frac{f(0)}{3\epsilon}+\f13 f'(0)+\gamma f(0) + \mathcal{O}(\epsilon) \Big) \, ,
\end{equation}
and after subtracting the pure pole part, it yields an additional finite contribution. Overall we get the derivative of the sphere free energy:
\be
- g\f{\p}{\p g}F_{\rm NNLO} = \lim_{\eps\to 0} \left(- g\f{\p}{\p g}F_{\rm NNLO}^{\eps} - \frac{N^2 f(0)}{3\epsilon} \right) \, ,
\ee
which can be evaluated numerically. 
For example, for $d=3$, $g=1$ and $a=1$, we find:
\be
- g\f{\p}{\p g}F_{\rm NNLO}   = 7.57 \times 10^{-4} \, N^2 \, .
\ee

\end{subappendices}

\chapter{Sextic tensor models}
\label{chap:sextic}

In this chapter, we study models with sextic interactions in rank 3 and 5, and with either short- or long-range propagators. 
Short-range sextic models have been considered before, but either without actually studying the existence of fixed points \cite{Giombi:2017dtl} (and only for rank 5), or for a different scaling in $N$ of the couplings than the optimal one \cite{Giombi:2018qgp}.
The chapter is organized as follows. In section~\ref{sec:models}, we start by giving the definitions of our models in rank 3 and 5, short- and long-range, and a description of the leading-order diagrams. We continue in sections~\ref{sec:SDeq} and \ref{sec:kernels} by computing the two- and four-point functions. Section~\ref{sec:betas} contains a detailed derivation of the $\beta$-functions of our sextic couplings, as well as their fixed points. We compute in section~\ref{sec:BSeq} the spectrum of bilinears (including spin dependence) through the eigenvalue equation. In section~\ref{sec:sexticNLO}, we study a generic sextic multi-scalar model in order to compute $1/N$ corrections to the fixed points in rank $3$. We first study the short-range case and then the long-range one. We finish with concluding remarks in section~\ref{sec:concl_sextic}. Finally, we give more details in several appendices. In appendix~\ref{ap:conventions}, we give our explicit conventions for the interaction terms. Appendices~\ref{ap:melon} and \ref{ap:betafun4} contain detailed computations of the melon integral and the Feynman amplitudes contributing to the beta functions. Lastly, in appendix~\ref{ap:O(N)}, we compare our next-to-leading order beta functions with the result of \cite{hager2002six} in the case of the sextic $O(N)$ model.

\section{The models}
\label{sec:models}

Both models we are going to consider can be viewed as symmetry-breaking perturbations of a free $O(\cN)$-invariant action for $\cN$ scalar fields $\phi_{\bf a}(x)$, with ${\bf a}= 1,\ldots,\cN$, $x\in \mathbf{R}^d$:
\be
S_{\rm free}[\phi,\phib] = \int d^d x  \, \phib_{\bf a}(x) (   - \p_\m\p^\m)^{\zeta} \phi_{\bf a}(x) \, .
\label{eq:freemulti}
\ee
The scalar fields will be either complex or real (in the latter case $\phib_{\bf a}=\phi_{\bf a}$ and we multiply the action by a factor $1/2$).
As before, $\zeta$ is a free parameter, which must be positive in order to have a well-defined thermodynamic limit, and it must be bounded above by one in order to satisfy reflection positivity.
We will later fix it to be either $\zeta=1$, as in \cite{Klebanov:2016xxf,Giombi:2017dtl,Giombi:2018qgp}, or $\zeta=d/3$, as in chapter~\ref{chap:CTKT}.

We recall the expression of the propagator of a generalized free field theory:
\be
C(p) = \f{1}{p^{2\zeta}}\,, \;\;\;\; C(x,y) = \f{\G\left(\Delta_{\phi}\right)}{2^{2\z}\pi^{d/2}\G(\z)} \f{1}{|x-y|^{2\D_{\phi}}}\,,
\ee
with $\Delta_{\phi}= \f{d-2\zeta}{2}$.

Perturbing the free action above by a quartic $O(\cN)$-invariant potential leads to the usual short-range ($\zeta=1$, e.g.\ \cite{Moshe:2003xn}) or long-range ($\zeta<1$, e.g.\ \cite{brezin2014crossover,Defenu:2014bea}) $O(\cN)$ model.

The general type of tensor field theories we have in mind will have $\cN=N^r$, and a potential explicitly breaking the $O(\cN)$ symmetry group down to $G^r$, with either $G=O(N)$ (for real fields) or $G=U(N)$ (for complex fields).
We will explicitly consider two models with sextic interactions, for $r=3$ and $r=5$. For $r=4$ we could write a model qualitatively very similar to $r=5$, but we would not learn much more, so we will not present it.

\subsection{Rank $3$}

\paragraph{Action.}

We first consider a rank-3 bosonic tensor model in $d\leq 3$ dimensions, with $U(N)^3$ symmetry and sextic interactions.
The bare action is
\begin{align} \label{eq:action}
S[\phi,\phib] &= \int d^d x  \, \phib_{abc} (   - \p_\m\p^\m)^{\zeta} \phi_{abc} + S_{\rm int}[\phi,\phib] \, ,\\
 \label{eq:int-action}
S_{\rm int}[\phi,\phib] &= \int d^d x  \sum_{b=1}^5 \f{\l_b}{6 N^{3+\r(I_b)}} I_b \,.
\end{align}
The $U(N)^3$ invariants $I_b$ are all those that can be constructed with six fields, and their respective parameter $\r(I_b)$ will be chosen according to the optimal scaling defined in \cite{Carrozza:2015adg}:
\be
\r(I_b)=\frac{F(I_b)-3}{2} \,,
\ee
with $F(I_b)$ counting the total number of cycles of alternating colors $i$ and $j$ with $i,j ~\in \lbrace 1,2,3\rbrace$.
We represent the tensor invariants, or bubbles, as bipartite $3$-colored graphs (as defined in section~\ref{sec:colored_graphs}).
With the aid of such representation we can write the interacting part of the action as:
\be
\begin{split} \label{eq:int-action-graph}
S_{\rm int}[\phi,\phib] = & \int d^d x  
\left( \f{\l_1}{6 N^{3}} \vcenter{\hbox{\includegraphics[scale=0.5]{wheel.pdf}}}
+ \f{\l_2}{6 N^{4}} \vcenter{\hbox{
\includegraphics[scale=0.5]{long_pillow_rank3.pdf}}}
+ \f{\l_3}{6 N^{4}} \vcenter{\hbox{\includegraphics[scale=0.5]{circle.pdf}}}
 \right. \\
 & \left.
 + \f{\l_4}{6 N^{5}} \vcenter{\hbox{\includegraphics[scale=0.5]{pillow_trace_rank3.pdf}}}
  + \f{\l_5}{6 N^{6}} \vcenter{\hbox{\includegraphics[scale=0.5]{triple_trace_rank3.pdf}}}
  \right)\,,
\end{split}
\ee
where a (normalized) sum over color permutations should be understood, whenever it is non-trivial (see appendix~\ref{ap:conventions} for more details on our conventions). Bubbles which are composed of one, two, or three connected components are referred to as single-trace, double-trace, or triple-trace, respectively, for analogy with the matrix case, and bubbles $I_b$ for which $\rho(I_b)=0$ are called \emph{maximally single trace}  (MST), as each of their 2-colored subgraphs are single trace.
The $I_1$ invariant is the only MST bubble in our action.

\begin{figure}[htbp]
\centering
\includegraphics[scale=0.5]{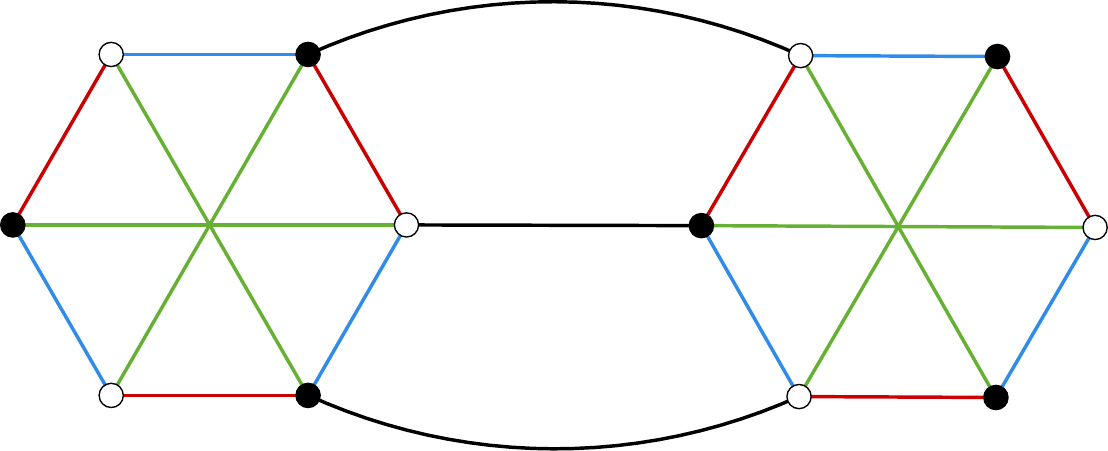}
\caption{$4$-colored graph corresponding to a two-loop Feynman diagram with external tensor contractions equivalent to $I_2$.}
\label{fig:4colored}
\end{figure}

\paragraph{Colored graphs and Feynman diagrams.} 
As we did before for quartic models, we represent Feynman graphs as $4$-colored graphs with the color black (or label $0$) for propagators. 
We give an example of the resulting $4$-colored graphs in figure~\ref{fig:4colored}.

\begin{figure}[htbp]
\centering
\includegraphics[scale=0.4]{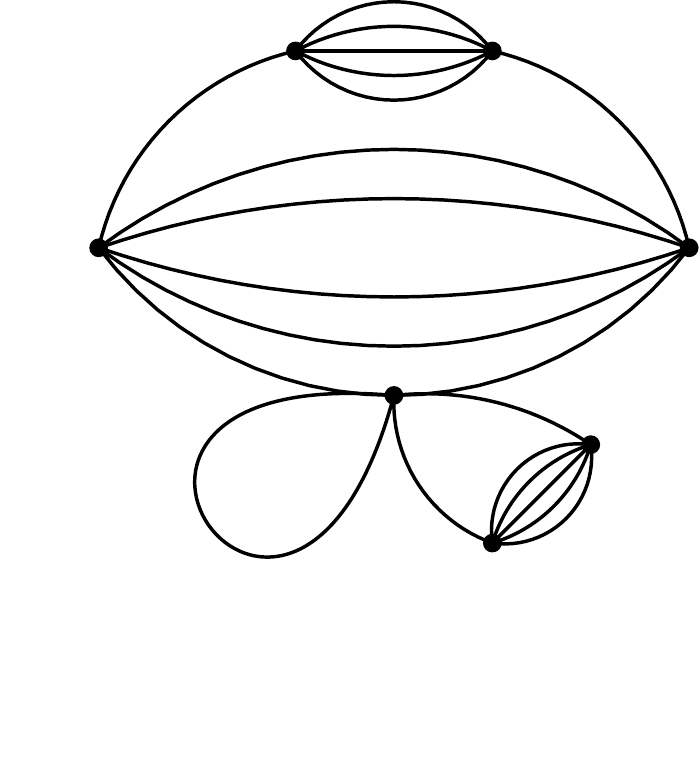}
\caption{An example of melon-tadpole Feynman diagram. Double tadpoles are based on the $I_b$ ($b \in [1,5]$) vertices and melons are based on $I_1$ vertices.}
\label{fig:melon_dtadpole}
\end{figure}

As before, ordinary Feynman diagrams are obtained by shrinking each interaction bubble to a point, which we will call an interaction vertex, or just vertex.
We give an example of such a Feynman diagram in figure~\ref{fig:melon_dtadpole}. While Feynman diagrams are sufficient for representing Feynman integrals, the $4$-colored graphs are necessary in order to identify the scaling in $N$ as each face of color $(0,i)$ gives rise to a factor $N$. 
The amplitude of a Feynman diagram $\mathcal{G}$ thus scales as $A(\mathcal{G})\sim N^{F-3n_{1}-4n_2-4n_3-5n_4-6n_5}$, with $F$ the total number of faces in the associated $4$-colored graph and $n_i$ the number of bubbles of the interaction $i$. 
The existence of the large-$N$ limit relies on the fact that the power of $N$ is bounded from above \cite{RTM,Carrozza:2015adg}.

\paragraph{Melonic graphs and melonic diagrams.} 


An important result in rank-$r$ tensor models is that if one only allows for interaction bubbles which are melonic $r$-valent graphs, then in the perturbative expansion the leading order vacuum graphs at large $N$ are melonic $(r+1)$-valent graphs \cite{Bonzom:2012hw}.
However, it is important to notice that melonic $(r+1)$-valent graphs do not correspond to melonic Feynman diagrams, i.e.\  they do not remain melonic after shrinking the colors from 1 to $r$.
From the point of view of the Feynman diagrams, melonic $(r+1)$-valent graphs reduce to the same type of cactus diagrams appearing in the large-$N$ limit of vector models, and therefore field theories based on such interaction are not expected to lead to very different results than vector models.\footnote{They can nevertheless lead to new phases with patterns of spontaneous symmetry breaking which are impossible in the vector case \cite{Benedetti:2018ghn}.} 

Adding non-melonic bubbles, things get more complicated, and possibly more interesting. In particular, it was found in \cite{Carrozza:2015adg} that non-melonic interaction bubbles can be scaled in such a way that they also contribute at leading order in the $1/N$ expansion, and that for some interactions (in that specific example, the quartic tetrahedron interaction) their leading-order Feynman diagrams are melonic. This has lead to melon-tadpoles graphs in the quartic case (see chapter~\ref{chap:CTKT}).

\begin{figure}[htbp]
\centering
\captionsetup[subfigure]{labelformat=empty}
\subfloat[]{\includegraphics[scale=1]{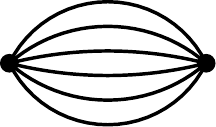}}
\hspace{1cm}
\subfloat[]{\includegraphics[scale=1]{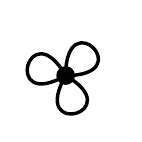}}
\caption{The two Feynman diagrams (the fundamental melon on the left, and the triple-tadpole, or trefoil, on the right) starting from which all the vacuum melon-tadpole diagrams can be built. The melon is based on the wheel vertices and the triple-tadpole is based on any of the interactions $I_i$ (for explicit examples of the corresponding colored graphs in rank 5, see figure~\ref{fig:fund_vacuum_rank5}).}
\label{fig:fund_vacuum}
\end{figure}

\begin{figure}[htbp]
\centering
\captionsetup[subfigure]{labelformat=empty}
\subfloat[]{\includegraphics[scale=1]{melonSDE.pdf}}
\hspace{1cm}
\subfloat[]{\includegraphics[scale=1]{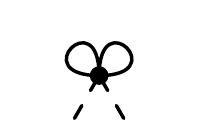}}
\caption{The two minimal two-point function Feynman diagrams used in the iterative construction of melon-tadpole diagrams. The melon is based on wheel vertices and the double tadpole is based on any of the interactions $I_i$.}
\label{fig:SDE}
\end{figure}

\paragraph{The large-$N$ limit.}  The $I_1$ invariant in \eqref{eq:int-action} (i.e.\ the first bubble in \eqref{eq:int-action-graph}, which we call the {\it wheel} graph, and which is also known as the complete bipartite graph $K_{3,3}$) stands out as the only non-melonic bubble in our action, and as a consequence, as the only interaction that does not lead only to tadpole corrections to the propagator at large $N$. It leads instead to Feynman diagrams which are of melon-tadpole type \cite{Lionni:2017xvn,Lionni:2017yvi,Prakash:2019zia} (see figure~\ref{fig:melon_dtadpole}), i.e. diagrams obtained by repeated insertions of either melon or tadpole two-point functions (figure~\ref{fig:SDE}) on the propagators of either one of the two fundamental vacuum graphs in figure~\ref{fig:fund_vacuum}.
The 4-colored graph corresponding to the fundamental melon is built from two mirror wheel graphs (i.e.\ completing in a straightforward way figure~\ref{fig:4colored}), while the triple-tadpole is built on any of the interactions.\footnote{Notice that the leading 4-colored graph of the trefoil is unique for the melonic bubbles (essentially tadpoles like to be based on multilines), while there are three leading-order trefoils that can be built on the wheel. \label{foot:trefoil}}

As tadpole corrections just renormalize the mass, the effect of $I_2$ to $I_5$, and of the $I_1$ tadpoles, will be ignored in the discussion of the Schwinger-Dyson equations for  the two-point function, assuming that we are tuning the bare mass to exactly set the effective mass to zero.
Along the same line of thoughts, we have not included quartic interactions in our action, assuming that they can be tuned to zero. 
In fact, we will be using dimensional regularization, which for massless theories results in the tadpoles (and other power-divergent integrals) being regularized to zero (e.g.\  \cite{ZinnJustin:2002ru}); thus we will actually need no non-trivial tuning of bare parameters, and we will be able to keep mass and quartic couplings identically zero.

\subsection{Rank $5$}
\label{sec:rank5}
\paragraph{Action.}

The second sextic model we will consider in this chapter is a $O(N)^5$ bosonic tensor model in $d$ dimensions. 
We consider a real tensor field of rank $5$, $\phi_{abcde}$ transforming under $O(N)^5$ with indices distinguished by their position. The action of the model is:\footnote{The optimal scaling is now defined as $\r(J_b)=\frac{F(J_b)-10}{4}$ with a straightforward generalization of \cite{Carrozza:2015adg}.}
\begin{align}
S[\phi] & = \f12 \int d^d x  \, \phi_{abcde} (   - \p_\m\p^\m)^{\zeta} \phi_{abcde} + S_{\rm int}[\phi] \, ,\\
 \label{eq:int-action-rank5}
S_{\rm int}[\phi] &= \int d^d x  \sum_{b=1}^6 \f{\k_b}{6 N^{5+\r(J_b)}} J_b \,.
\end{align}

The interaction part of the action can be written with the same graphical representation as for the previous model. However, because we are now considering a rank-5 model, the graphs representing the interactions will be $5$-colored graphs, and because we have real fields with $O(N)^5$ symmetry, the graphs will not be bipartite and the nodes will all have the same color (black). An action containing all the $O(N)^5$ invariants would be rather long,\footnote{Using the code provided in \cite{Avohou:2019qrl} (built on a generalization of the methods of \cite{BenGeloun:2013lim,BenGeloun:2017vwn}), we can count the total number of sextic invariants, with their different coloring choices, to be 1439.} and difficult to handle. We will restrict the potential by exploiting the large-$N$ limit: we start from the interaction whose bubble is a complete graph (i.e.\ in which for every pair of nodes there is an edge connecting them), and then include only the other interactions which are generated as radiative corrections, until  we obtain a renormalizable model, at large $N$. A set of interactions of this type has been introduced in \cite{Ferrari:2017jgw} with the name of \emph{melo-complete family}.
As we will explain further below, it turns out that besides the complete graph we need to include only the melonic bubbles (a straightforward generalization of the melonic bubbles of rank 3) and one new non-bipartite bubble:\footnote{Following the same logic for  rank-3, starting with the wheel interaction, we would have obtained the same action as in \eqref{eq:int-action-graph}. As in that case the set of interactions exhausts the sextic $U(N)^3$ invariants, we have chosen a different perspective in its presentation.}
\be
\begin{split} \label{eq:int-action-graph-rank5}
S_{\rm int}[\phi] = & \int d^d x  
\left( \f{\kappa_1}{6 N^{5}} \vcenter{\hbox{\includegraphics[scale=0.5]{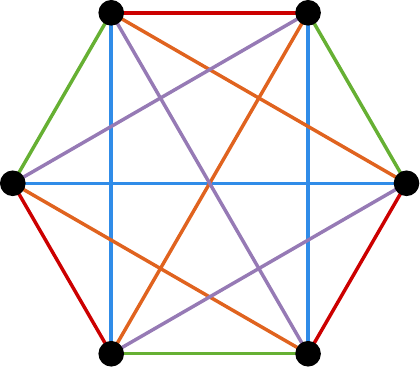}}}
+ \f{\kappa_2}{6 N^{8}} \vcenter{\hbox{\includegraphics[scale=0.5]{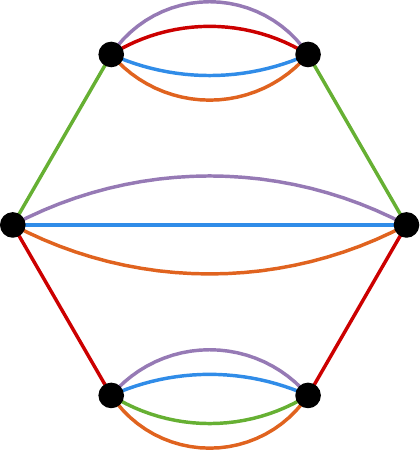}}}
+ \f{\kappa_3}{6 N^{8}} \vcenter{\hbox{\includegraphics[scale=0.5]{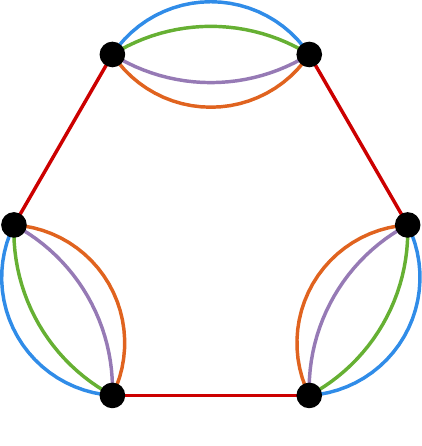}}}
 \right. \\
 & \left.
 + \f{\kappa_4}{6 N^{9}} \vcenter{\hbox{\includegraphics[scale=0.5]{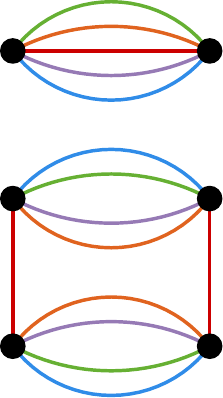}}}
 + \f{\kappa_5}{6 N^{10}} \vcenter{\hbox{\includegraphics[scale=0.5]{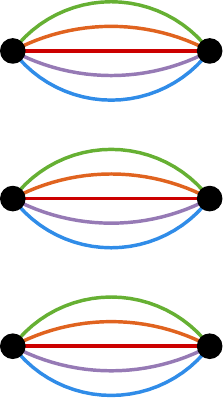}}}
  + \f{\kappa_6}{6 N^{7}} \vcenter{\hbox{\includegraphics[scale=0.5]{triangle.pdf}}}
  \right)\,,
\end{split}
\ee
where a sum over color permutations should be understood. The conventions are detailed in appendix~\ref{ap:conventions}.

\paragraph{Colored graphs and Feynman diagrams.}

\begin{figure}[htbp]
\centering
\captionsetup[subfigure]{labelformat=empty}
\subfloat[]{\includegraphics[scale=0.5]{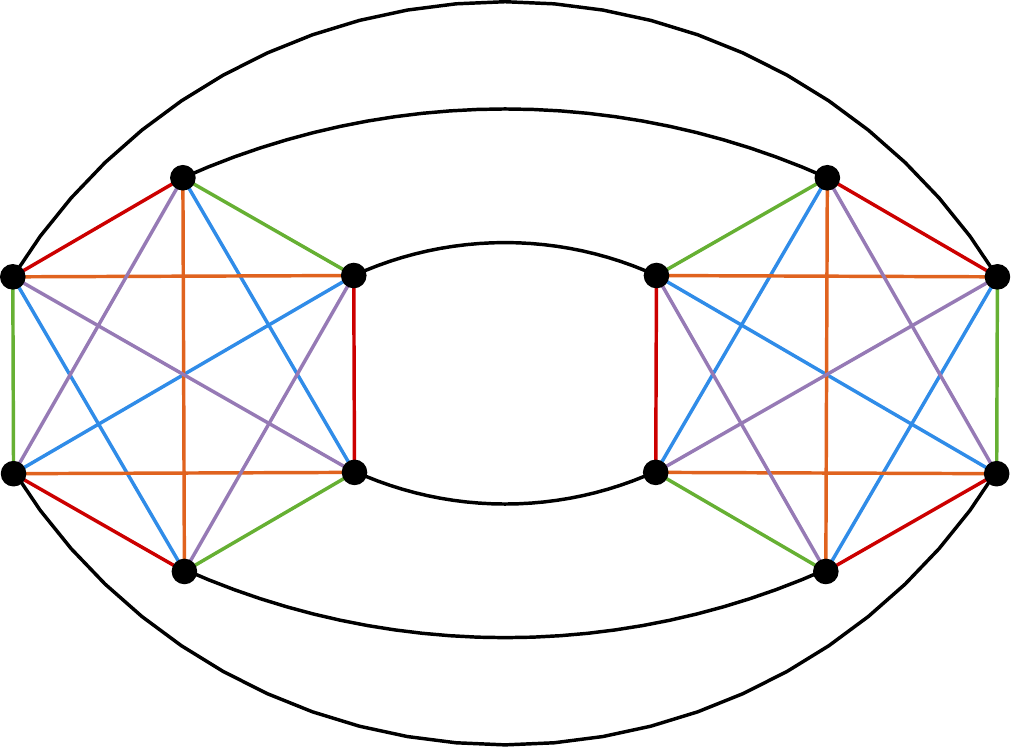}}
\hspace{1cm}
\subfloat[]{\includegraphics[scale=0.5]{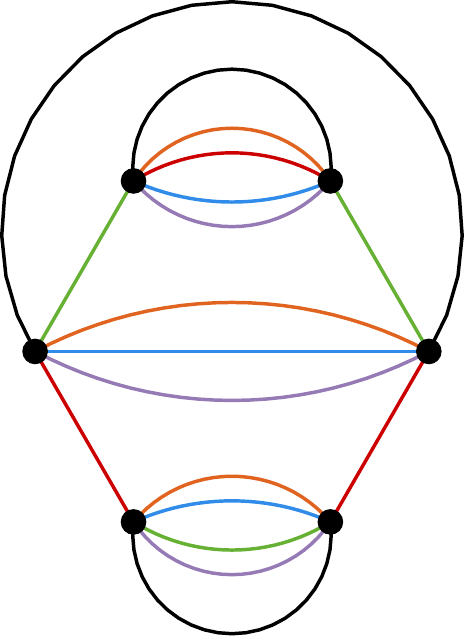}}
\caption{Two examples of vacuum $6$-colored graphs.}
\label{fig:fund_vacuum_rank5}
\end{figure}

\begin{figure}[htbp]
\centering
\includegraphics[scale=0.5]{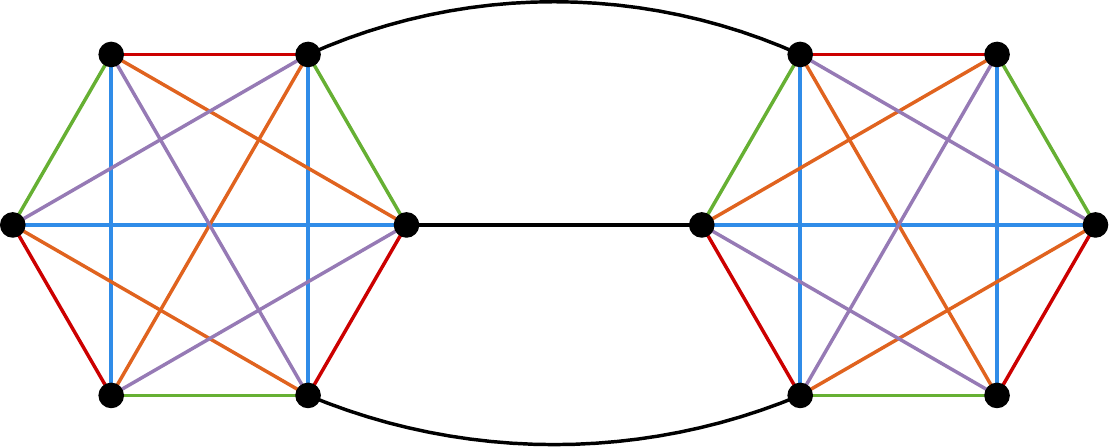}
\caption{$6$-colored graph corresponding to a two-loop correction to the six-point function, with external tensor contractions equivalent to $J_6$ (the prism).}
\label{fig:6colored}
\end{figure}

The expansion into Feynman diagrams is done similarly as for the previous model. Again, the propagators are represented by black edges. We give some examples of resulting $6$-colored graphs in figure~\ref{fig:fund_vacuum_rank5} and \ref{fig:6colored}.

\begin{figure}[htbp]
\centering
\includegraphics[scale=0.5]{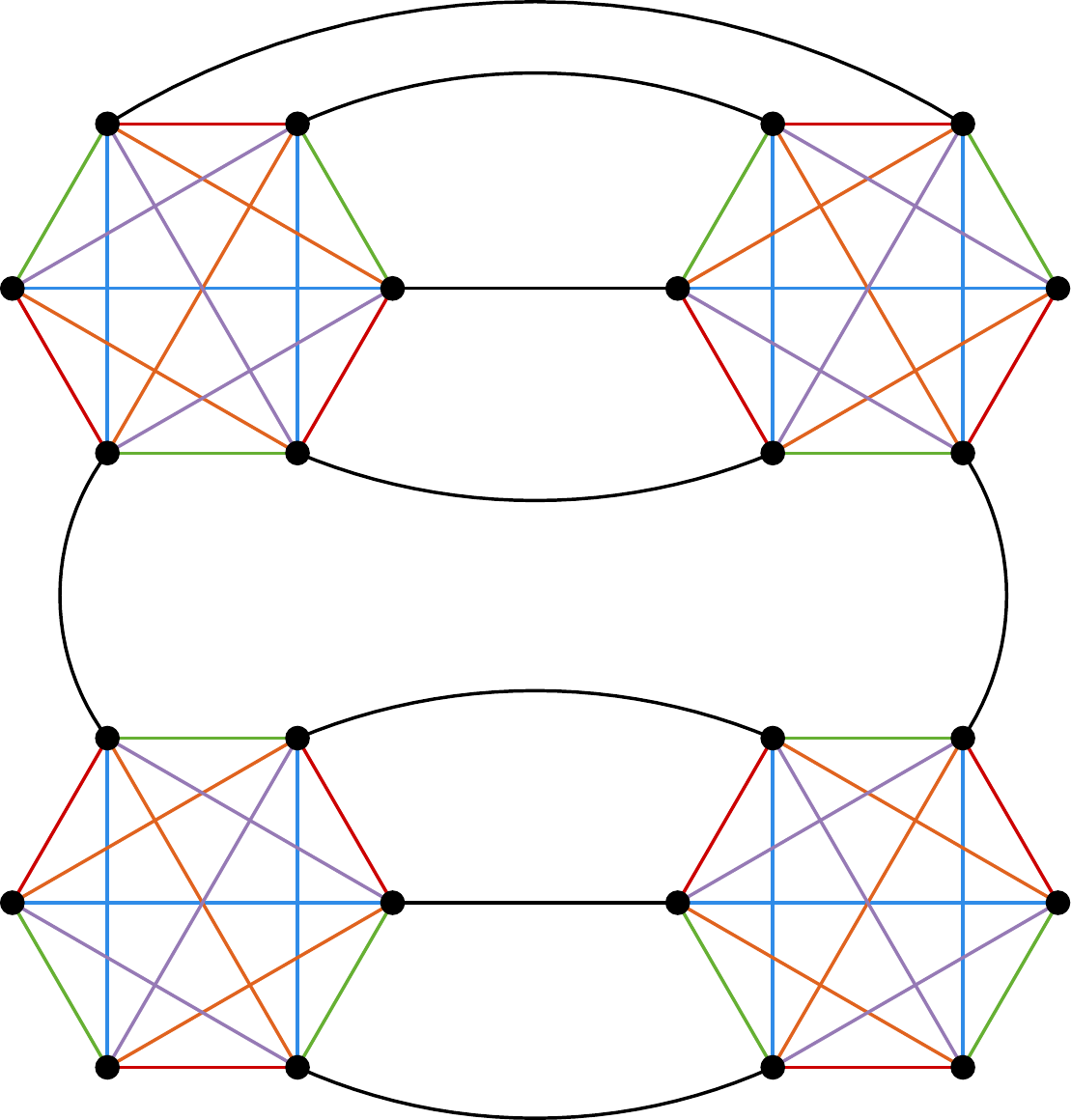}
\caption{Feynman diagram consisting in a ladder with two rungs and complete vertices. Its external tensor contractions are equivalent to $J_2$ (the long-pillow).}
\label{fig:2rungs}
\end{figure}

\paragraph{The large-$N$ expansion.} Like other tensor models, this model has also a $1/N$ expansion. The derivation is similar to the one of the $O(N)^3$ model of chapter~\ref{chap:CTKT}. First, we observe that every sextic interactions can be obtained as radiative corrections from the first interaction term $J_1$ (we call it the complete vertex, as its bubble is the complete graph on six vertices, also known as $K_6$). For example, the interaction $J_6$ (or the prism) is a rung with $3$ edges between two complete vertices (see figure~\ref{fig:6colored}), $J_2$ (or the long-pillow) and $J_4$ (or the pillow-dipole, our only double-trace interaction) are ladders made of two such rungs with different permutations of the colors between the rungs (see figure~\ref{fig:2rungs}). $J_5$ (or the triple-dipole, our only triple-trace interaction) is a ladder made of three rungs and $J_3$ a ladder made of four rungs. 

Then in any graph $\mathcal{G}$, we replace every interaction by their minimal representations in terms of complete vertices. This way, we obtain a new graph $\hat{\mathcal{G}}$ with only complete vertices. Since the rank of our model is a prime number, and the complete graph is the unique maximally single trace  (MST) invariant, we can use the result of \cite{Ferrari:2017jgw} (see also \cite{Klebanov:2019jup}), where it has been proved that in this case, the leading order vacuum Feynman diagrams are the melons constructed with two mirrored complete MST interactions (see the diagram on the left in figure~\ref{fig:fund_vacuum_rank5}), and the usual iterative insertions of melonic two-point functions. Notice that unlike for the rank-3 wheel (which is MST, but  not a complete graph), the leading order diagrams include no tadpoles. This means that the leading order diagrams of our rank-5 model are melonic \textit{after} substituting every sextic interactions by their minimal representations in terms of the complete vertex. In terms of the original interactions, the leading order diagrams are again melon-tadpole diagrams (see figure~\ref{fig:melon_dtadpole}, with tadpoles now associated to  $J_b$ with $b \in [2,6]$). The double tadpoles are based on the interactions $J_b$ vertices ($b \in [2,6]$) and the end vertices of melons are complete vertices. 
Therefore, the diagrammatics is somewhat similar to that of the quartic model, where the tetrahedron is a complete graph and it is associated to melonic diagrams, while the melonic bubbles (pillow and double-trace) are associated to tadpoles.

Again, as explained for the previous model, we will ignore the effects of the tadpoles formed by $J_2$ to $J_6$, as tadpole corrections just renormalize the mass. We will also not include quartic interactions, assuming that they can be tuned to zero.

\begin{figure}[htbp]
\centering
\includegraphics[scale=0.5]{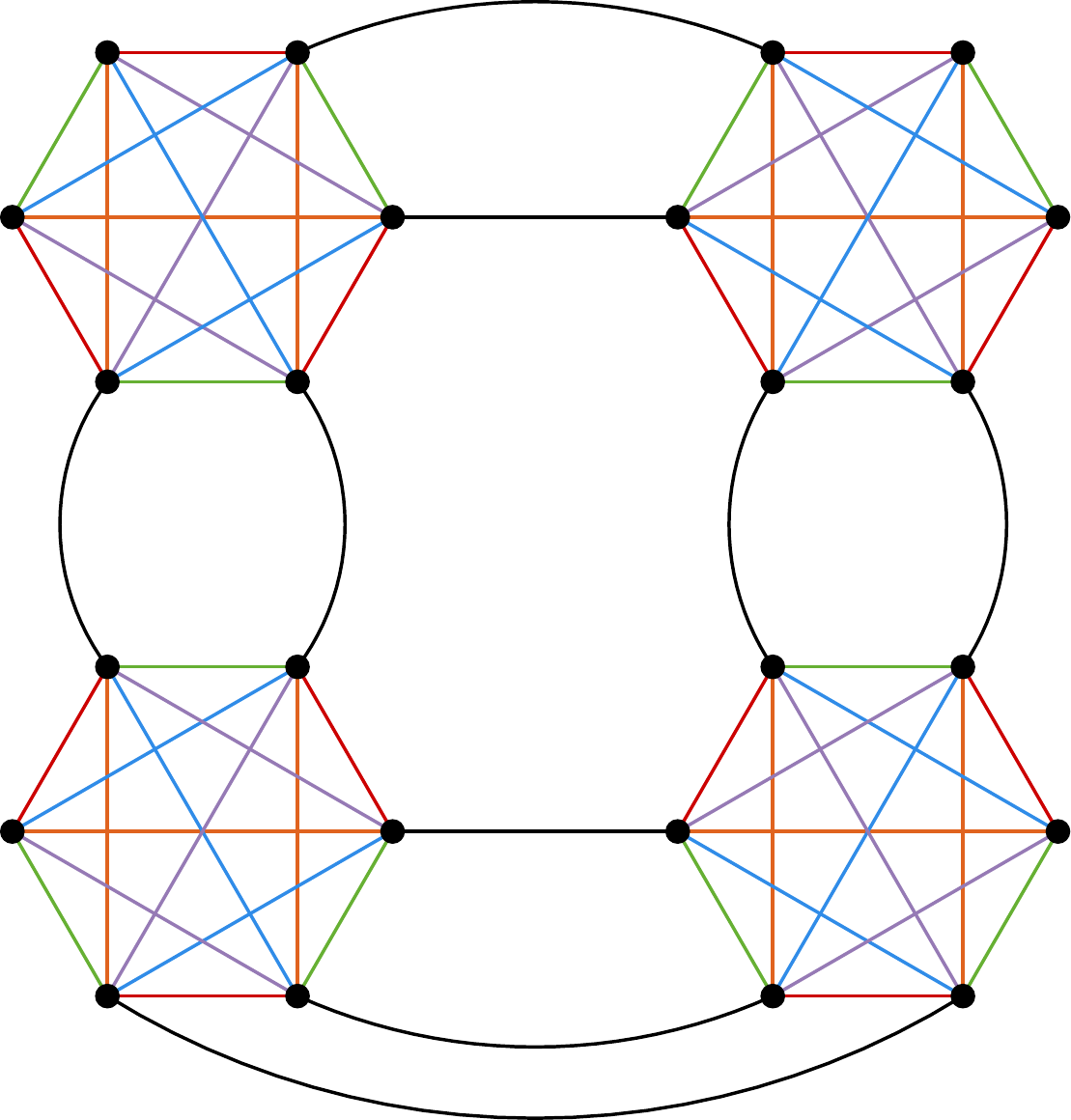}
\caption{6-colored graph corresponding to a non-melonic Feynman diagram whose exterior tensor contractions are equivalent to $J_6$.}
\label{fig:non-melonicI6}
\end{figure}

\paragraph{Radiative corrections to the prism interaction.} 
A comment is in order regarding the non-melonic interaction $J_6$. We presented in figure~\ref{fig:6colored} a melonic contraction of two $J_1$ interactions that has $J_6$ as a boundary graph. It turns out that it is the only melonic diagram built with $J_1$ vertices that produces it. Indeed, we notice in $J_6$ the presence of two mirrored triangles (with edges red-green-blue in \eqref{eq:int-action-graph-rank5}) and each can result from 1, 2 or more complete graphs. The first case corresponds to figure~\ref{fig:6colored}, but we see that the second case already requires non-melonic diagrams as in figure~\ref{fig:non-melonicI6}. In order to construct such a triangle from more that two $J_1$ vertices, we need at least two propagators (for the two colored edges that leave the nodes of the triangle) between each vertex, which in addition to at least two other propagators required to connect the mirror symmetric nodes of the two triangles, make the diagram non-melonic. 

\paragraph{Incomplete set of invariants.} As we said, the set of invariants we considered in the action is incomplete: there are more $O(N)^5$ invariants. However, it is closed. Indeed, we just showed that a $O(N)^5$ model with the complete interaction is dominated in the large-$N$ limit by melonic graphs. Therefore, it is enough to consider only the $O(N)^5$ invariants that can be generated from a melonic graph constructed with complete vertices. Those invariants are exactly $J_2$ to $J_6$ in the action of the model. The other $O(N)^5$ invariants will never be generated by a leading order six-point graph as they cannot be obtained from a melonic graph with complete vertices. Thus, at leading order in $N$, the set of invariants we consider is closed.

\paragraph{Rank-4 model.} 
Lastly, we notice that, as in rank 5, also in rank 4 there is a unique MST interaction \cite{Prakash:2019zia}. It turns out that the set of interactions it generates as radiative corrections are exactly of the same form as $J_2$ to $J_6$ in \eqref{eq:int-action-graph-rank5}, except that each multi-edge has one edge less than in rank 5 (for example, they can be obtained by removing the purple color in \eqref{eq:int-action-graph-rank5}).
Therefore, besides some different combinatorial factors, we do not expect important qualitative differences with respect to rank 5, and we chose to work with rank 5 as it contains a complete bubble, making the analogy to the quartic model of chapter~\ref{chap:CTKT} more evident.

\subsection{Renormalization: power counting}
\label{sec:powercount}

We consider $\mathcal{G}$ a connected amputated Feynman diagram with $V_6(\mathcal{G})$ $6$-valent vertices, $V_4(\mathcal{G})$ $4$-valent vertices, $E(\mathcal{G})$ edges and $n(\mathcal{G})$ external points. Computing the amplitude of the diagram $\mathcal{G}$ in momentum space, we get an independent integral $d^d p$ for every loop and a propagator $p^{-2\zeta}$ for every edge. Then, under a global rescaling of all the momenta by $t$, the amplitude is rescaled by:
\begin{align}
t^{d(E(\mathcal{G})-V_6(\mathcal{G})-V_4(\mathcal{G})+1)-2\zeta E(\mathcal{G})}&=t^{d(V_4(\mathcal{G})+2V_6(\mathcal{G})+1-\frac{n(\mathcal{G})}{2})-\zeta\left(4V_4(\mathcal{G})+6V_6(\mathcal{G})-n(\mathcal{G})\right)}\crcr
&=t^{d-\frac{n(\mathcal{G})}{2}(d-2\zeta)+V_6(\mathcal{G})(2d-6\zeta)+V_4(\mathcal{G})(d-4\zeta)}
\end{align}
where we have used $2E(\mathcal{G})=6V_6(\mathcal{G})+4V_4(\mathcal{G})-n(\mathcal{G})$.

The UV degree of divergence of the graph is then:
\begin{equation}
\text{deg}(\mathcal{G})=d-\frac{n(\mathcal{G})}{2}(d-2\zeta)+V_6(\mathcal{G})(2d-6\zeta)+V_4(\mathcal{G})(d-4\zeta) \, .
\end{equation}

\paragraph{Short-range propagator.}

For $d=3$ and $\zeta=1$, the UV degree of divergence is:
\begin{equation}
\text{deg}(\mathcal{G})=3\left(1-\frac{n(\mathcal{G})}{6}\right)\,.
\end{equation}

Thus, in $d=3$, the sextic interactions are marginal (the power counting does not depend on the number of internal vertices). The two-point and four-point diagrams are power divergent and the six-point diagrams are logarithmically divergent in the UV. Diagrams with more than eight external points are UV convergent. 

Therefore, in the following, we will use dimensional regularization, setting $d=3-\epsilon$. We will be interested in Wilson-Fisher type of fixed points, hence we will also consider $\epsilon$ finite, but small. This allows UV divergences to be regularized. Moreover, we choose the BPHZ zero momentum subtraction scheme. One could wonder why we chose this prescription and not the usual Gell-Mann and Low subtraction at non-zero momentum. The reason is that, with this prescription, we were not able to obtain analytic results for the amplitudes of four-loop diagrams in the long-range case. For consistency, we chose the zero momentum subtraction scheme for both short and long-range models.

However, as we work with a massless propagator, we also need an IR regulator. Following chapters~\ref{chap:3loops} and \ref{chap:trif}, we introduce an IR regulator by modifying the covariance as:

\begin{equation}
C_{\mu}(p)=\frac{1}{p^2+\mu^2}=\int_0^{\infty} da \,e^{-ap^2-a\mu^2}\; ,
\end{equation}
for some mass parameter $\mu>0$.

\paragraph{Long-range propagator.}

For $d<3$ and $\zeta=\frac{d}{3}$, the UV degree is:

\begin{equation}
\text{deg}(\mathcal{G})=d\left(1-\frac{n(\mathcal{G})}{6}\right)\,.
\end{equation}

Again, the sextic interactions are marginal. The two-point and four-point diagrams are power divergent and the six-point diagrams are logarithmically divergent in the UV. Graphs with more than eight external point are UV convergent. 

We will again use dimensional regularization but in this case we will keep $d<3$ fixed and set $\zeta=\frac{d+\epsilon}{3}$. We will also use a BPHZ subtraction scheme at zero momentum with modified covariance.
In this case, we will be interested in fixed points that arise at $\epsilon=0$, as in chapter~\ref{chap:CTKT}, by a different mechanism than in Wilson-Fisher.

\section{Two-point function}
\label{sec:SDeq}

\subsection{Rank $3$}

The standard Schwinger-Dyson equation (SDE) for the two-point function is, in momentum space:
\be \label{eq:SDE}
G(p)^{-1}=C(p)^{-1}-\Sigma(p) \,,
\ee
where $G(p)$ is the Fourier transform of the full two-point function $N^{-3}\la\phib_{abc}(x)\phi_{abc}(0)\ra$, and $\Sigma(p)$ is the self-energy, i.e.\ the sum of non-trivial one-particle irreducible two-point diagrams. 

In a theory which is dominated by melon-tadpole diagrams, the self-energy at leading order in $1/N$ is obtained by summing up all the Feynman diagrams which can be obtained from those in figure~\ref{fig:SDE} by repeated insertions of either one of the two diagrams on internal lines. 
The resummation of all such diagrams can be represented by the same diagrams as in figure~\ref{fig:SDE}, but with the
edges decorated by the full two-point function. Therefore, it can be expressed in momentum space as:\footnote{See footnote~\ref{foot:trefoil} for the factor 3 in the double-tadpole contribution of the wheel.}
\begin{align} \label{eq:Sigma-rank3}
\Sigma(p)&= 
\frac{\lambda_1^2}{4}\int_{q_1,q_2,q_3,q_4}G(q_1)G(q_2)G(q_3)G(q_4)G(p+q_1+q_2+q_3+q_4)\crcr
& \qquad -\frac{1}{2}(3\lambda_1+\lambda_2+\lambda_3+\lambda_4+\lambda_5)\left(\int_q G(q)\right)^2 \,.
\end{align}

\subsubsection{Short-range propagator}

When $\zeta=1$, using a power counting argument, we see that the solution admits two regimes for  $d<3$ (similarly as what happened for the quartic case). First, in the ultraviolet, there is a free scaling regime $G(p)^{-1}\sim p^2$: the free propagator dominates over the self energy. Second, in the infrared, there is an anomalous scaling regime $G(p)^{-1} \sim p^{2\Delta}$ with $\Delta=\frac{d}{3}$: the self energy dominates over the free propagator. Indeed, if we rescale the $q_i$ by $|p|$, the melon integral gives a global factor of $|p|^{4d-10\Delta}$ which must scale as $|p|^{2\Delta}$. This gives indeed $\Delta=\frac{d}{3}$.

We thus choose the following ansatz for the IR two-point function:
\begin{equation}
G(p)=\frac{\mathcal{Z}}{p^{2d/3}}\,.
\label{eq:ansatz}
\end{equation}

Neglecting the free propagator in the IR, the SDE reduces to:
\begin{equation}
\frac{p^{2d/3}}{\mathcal{Z}}=-\frac{\lambda_1^2}{4}\mathcal{Z}^5 M_{d/3}(p)\,,
\end{equation}
with $M_{d/3}(p)$ the first integral in \eqref{eq:Sigma-rank3}. This melon integral is computed in appendix~\ref{ap:melon}, giving
\begin{equation}
M_{d/3}(p)=-\frac{p^{2d/3}}{(4\pi)^{2d}}\frac{3}{d}\frac{\Gamma(1-\frac{d}{3})\Gamma(\frac{d}{6})^5}{\Gamma(\frac{d}{3})^5\Gamma(\frac{5d}{6})}~.
\end{equation}

We thus obtain:
\begin{equation}
\mathcal{Z}=\left(\frac{\lambda_1^2}{4(4\pi)^{2d}}\frac{3}{d}\frac{\Gamma(1-\frac{d}{3})\Gamma(\frac{d}{6})^5}{\Gamma(\frac{d}{3})^5\Gamma(\frac{5d}{6})}\right)^{-1/6} \,.
\end{equation}

\paragraph{Wave function renormalization.}

We introduce the wave function renormalization as $\phi=\phi_R\sqrt{Z}$ with $\phi$ the bare field and $\phi_R$ the renormalized field.
Notice that $Z$ is distinguished from $\mathcal{Z}$, as the latter is the full coefficient of the non-perturbative solution in the IR limit, while $Z$ is the usual perturbative wave function renormalization, to be fixed by a renormalization condition, as we will specify below.

After renormalization of the mass terms to zero, we have for the expansion of the renormalized two-point function at lowest order:
\begin{equation}
\Gamma^{(2)}_R(p)\equiv G_R(p)^{-1} =Zp^2-\frac{\lambda_1^2}{4} Z^{-5} M_1(p) \,,
\label{eq:gamma2}
\end{equation}
where $M_1(p)$ is the melon integral with free propagators on the edges. It is again computed in appendix~\ref{ap:melon}. At leading order in $\epsilon$, we have:
\begin{equation}
M_1(p)=-p^{2-4\epsilon}\frac{2\pi^2}{3\epsilon(4\pi)^{6}} + \mathcal{O}(1)\,.
\label{eq:melon1}
\end{equation}
At last, we fix $Z$ such that
\be
\lim_{\epsilon\to 0}\frac{d\Gamma^{(2)}_R(p)}{dp^2}|_{p^2=\mu^2}=1\,,
\ee
with $\mu$ the renormalization scale. 
At quadratic order in $\lambda_1$, we obtain:
\begin{equation}
Z=1+\frac{\lambda_1^2}{4}\tilde{M}_1(\mu) =1-\mu^{-4\epsilon}\frac{\lambda_1^2\pi^2}{6\epsilon(4\pi)^{6}}\,,
\label{eq:wavef3}
\end{equation}
with $\tilde{M}_1(\mu)=\frac{d}{dp^2}M_1(p)|_{p^2=\mu^2}$.

\subsubsection{Long-range propagator}

The value of $\zeta$ in this case is chosen to match the infrared scaling of the two-point function. We now have only one regime and the full SDE is solved by the ansatz:
\begin{equation}
G(p)=\frac{\mathcal{Z}}{p^{2d/3}} \,.
\label{eq:ansatz2}
\end{equation}
For the vertex renormalization in section~\ref{sec:betas} we will use analytic regularization, keeping $d<3$ fixed and setting $\zeta=\frac{d+\epsilon}{3}$, but since the two-point function is finite,  as we will now see, we can here set $\epsilon=0$.

The computations are the same as in the IR limit of the previous section, but we do not neglect the free propagator. Thus, we obtain:
\begin{equation}
\frac{1}{\mathcal{Z}^6}-\frac{1}{\mathcal{Z}^5}=\frac{\lambda_1^2}{4(4\pi)^{2d}}\frac{3\Gamma(1-\frac{d}{3})\Gamma(\frac{d}{6})^5}{d\Gamma(\frac{d}{3})^5\Gamma(\frac{5d}{6})} \,.
\label{Z-norm-d/3}
\end{equation}
At the first non-trivial order in the coupling constant, this gives:
\begin{equation}  \label{eq:wavef3-LR}
\mathcal{Z}=1-\frac{\lambda_1^2}{4(4\pi)^{2d}}\frac{3\Gamma(1-\frac{d}{3})\Gamma(\frac{d}{6})^5}{d\Gamma(\frac{d}{3})^5\Gamma(\frac{5d}{6})}+\mathcal{O}(\lambda_1^4).
\end{equation}
This expression is finite for $d<3$. Moreover, as we did not neglect the free propagator, $\mathcal{Z}$ has an expansion in $\lambda_1$, as the perturbative wave function renormalization, with which it can be identified in our non-minimal subtraction scheme. Therefore, in the case $\zeta=d/3$, the wave function renormalization is finite. This is the same mechanism we saw for the long-range quartic model in chapter~\ref{chap:CTKT}.

\subsection{Rank $5$}

For the $O(N)^5$ model, the Schwinger-Dyson equation in the large-$N$ limit is:
\be
G(p)^{-1}=C(p)^{-1}-\Sigma(p) \,,
\ee
with
\begin{align} \label{eq:Sigma-rank5}
\Sigma(p)&= 
\frac{\kappa_1^2}{6}\int_{q_1,q_2,q_3,q_4}G(q_1)G(q_2)G(q_3)G(q_4)G(p+q_1+q_2+q_3+q_4)\crcr
& \qquad -(\kappa_2+\kappa_3+\kappa_4+\kappa_5+\kappa_6)\left(\int_q G(q)\right)^2  \,.
\end{align}
The only differences with the rank-3 model are the combinatorial factors in front of the melon and tadpole integrals. Thus, we can use the results of the previous section.

\subsubsection{Short-range propagator}

In the IR limit, the SDE is solved again by $G(p)=\frac{\mathcal{Z}}{p^{2d/3}}$ with:
\begin{equation}
\mathcal{Z}=\left(\frac{\kappa_1^2}{6(4\pi)^{2d}}\frac{3}{d}\frac{\Gamma(1-\frac{d}{3})\Gamma(\frac{d}{6})^5}{\Gamma(\frac{d}{3})^5\Gamma(\frac{5d}{6})}\right)^{-1/6} \,.
\end{equation}
The wave function renormalization is given by:
\begin{equation} \label{eq:wavef5}
Z=1+\frac{\kappa_1^2}{6}\tilde{M}_1(\mu) =1-\mu^{-4\epsilon}\frac{\lambda_1^2\pi^2}{9\epsilon (4\pi)^{6}}\,.
\end{equation}

\subsubsection{Long-range propagator}

The full SDE is solved again by $G(p)=\frac{\mathcal{Z}}{p^{2\zeta}}$ with:
\begin{equation} \label{eq:wavef5-LR}
\mathcal{Z}=1-\frac{\kappa_1^2}{2(4\pi)^{2d}}\frac{\Gamma(1-\frac{d}{3})\Gamma(\frac{d}{6})^5}{d\Gamma(\frac{d}{3})^5\Gamma(\frac{5d}{6})}+ \mathcal{O}(\kappa_1^3) \,.
\end{equation}
This is directly the wave function renormalization which is thus finite.

\section{2PI effective action and four-point kernels}
\label{sec:kernels}

In this section, we compute the four-point kernels of both models using the 2PI formalism.
We will make use of them first for showing that indeed there is no need of counterterms with quartic interactions, and then, in the next section, for the discussion of the all-orders beta functions for the sextic couplings.

\subsection{Rank 3}

In rank 3 and at leading order in $1/N$, the 2PI effective action is given by:\footnote{We use again the short-hand notation $\mba =(a_1,a_2,a_3)$ as in chapter~\ref{chap:CTKT}.} 
\begin{align}
\label{eq:2PIrank3}
-\G^{2PI}[\mbG] = &-\f{1}{6}\left(\f{3 \l_1}{N^3}\d^{(1)}_{\mba\mbd\mbb\mbc\mbe\mbf}+\sum_{i=2}^5 \f{\l_i}{N^{3+\rho(I_i)}}\d^{(i)}_{\mba\mbb;\mbc\mbd;\mbe\mbf}\right) \int\dd x ~\mbG_{(\mba,x)(\mbb,x)}\mbG_{(\mbc,x)(\mbd,x)}\mbG_{(\mbe,x)(\mbf,x)} \crcr
&+\f{1}{2}\left(\f{\l_1}{6N^3}\right)^2 3\, \d^{(1)}_{\mba\mbb\mbc\mbd\mbe\mbf} \d^{(1)}_{\mbg\mbh\mbj\mbk\mbm\mbn} \times\crcr
&\quad \int\dd x\dd y ~\mbG_{(\mba,x)(\mbg,y)}\mbG_{(\mbb,x)(\mbh,y)} \mbG_{(\mbc,x)(\mbj,y)}\mbG_{(\mbd,x)(\mbk,y)}\mbG_{(\mbe,x)(\mbm,y)}\mbG_{(\mbf,x)(\mbn,y)} \,.
\end{align}
This is obtained by summing the contributions of the leading-order vacuum diagrams which are also two-particle irreducible (2PI) and with arbitrary propagator $\mbG$ on each line. As we already know, all the leading-order vacuum diagrams are obtained from the diagrams in figure~\ref{fig:fund_vacuum} by repeated insertions of the two-point diagrams in figure~\ref{fig:SDE}, but since all such insertions lead to two-particle reducible diagrams, we are left with just the two fundamental diagrams of figure~\ref{fig:fund_vacuum}, whose evaluation leads to \eqref{eq:2PIrank3}.

One recovers the self-energy from (using the further condensed notation $A=(\mba,x)$):
\begin{equation} \label{eq:Sigma2PI-complex}
\Sigma[\mbG]_{AB} = -\f{\d\G^{2PI}[\mbG]}{\d\mbG_{AB}} \,,
\end{equation}
which can be seen to reproduce \eqref{eq:Sigma-rank3} in momentum space.

\begin{figure}[htbp]
\centering
\captionsetup[subfigure]{labelformat=empty}
\subfloat[]{\includegraphics[scale=0.5]{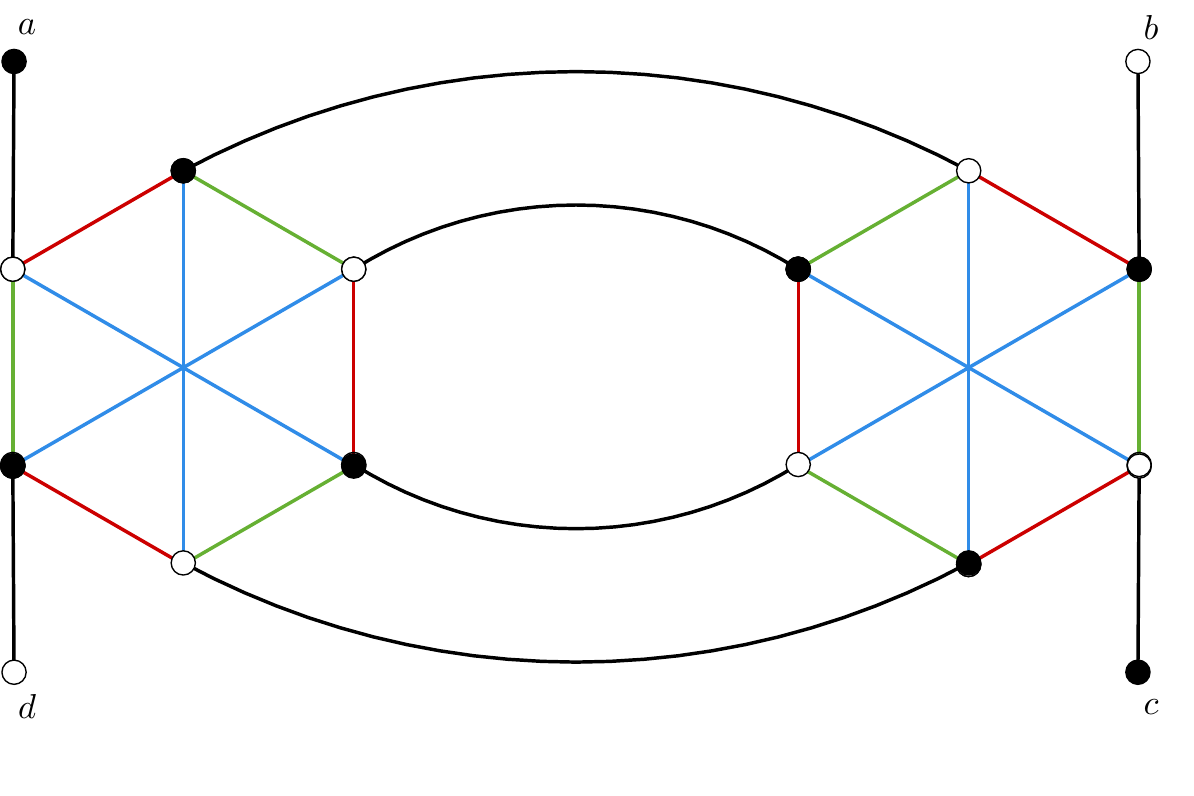}}
\hspace{1cm}
\subfloat[]{\includegraphics[scale=0.5]{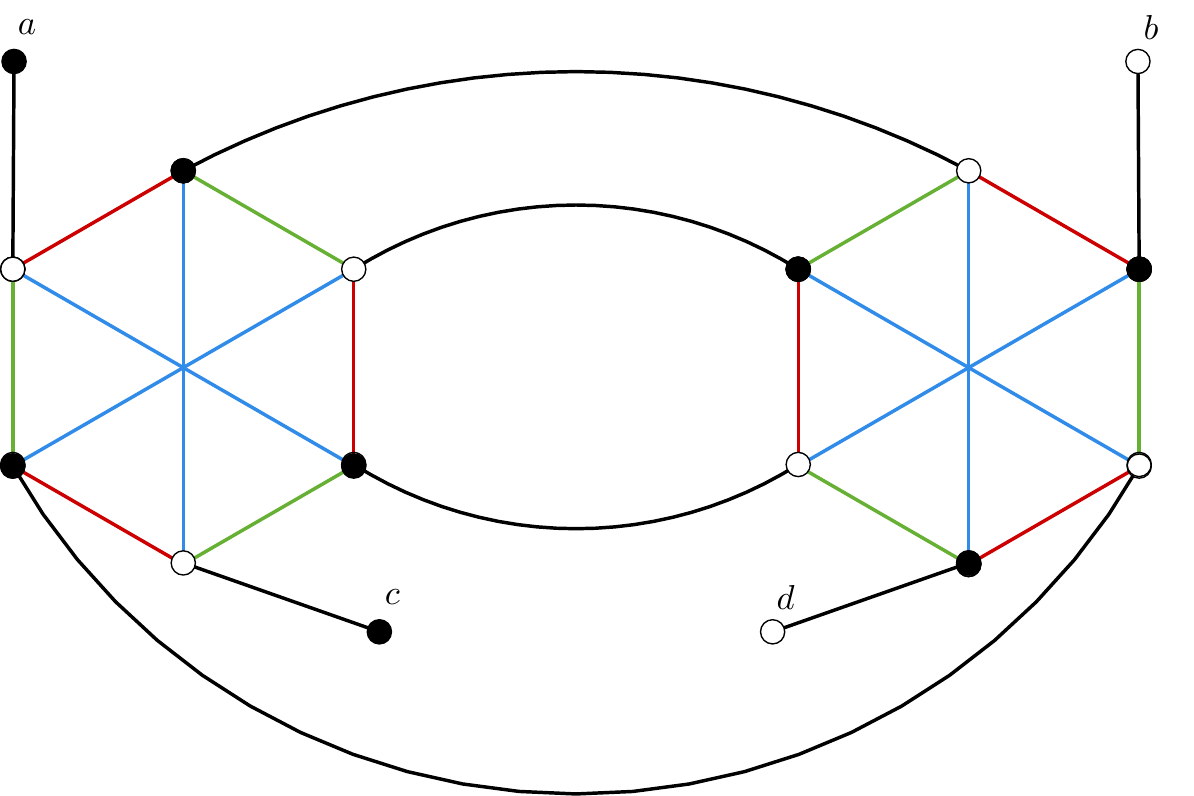}}
\caption{The contractions corresponding to a 4-point rung in the ladder making the 4-point function (a pillow on the left and a double trace on the right).}
\label{fig:kernel_wheel}
\end{figure}

The right-amputated four-point kernel on-shell is obtained by taking two derivatives of $\G^{2PI}[\mbG]$ with respect to $\mbG$  and then multiplying by two full propagators on the left:
\be \label{eq:defK}
K[\mbG]_{AB,CD} = \mbG_{AA'}\mbG_{BB'}\f{\d\Sigma[\mbG]_{CD}}{\d\mbG_{A'B'}}\,.
\ee
Applying such definition to \eqref{eq:2PIrank3} we obtain:
\begin{align}
K_{(\mba,x)(\mbb,y)(\mbc,z)(\mbd,w)}=&\int dx'dy' \, G_{xx'}G_{yy'}\left[ 
 -\f{1}{3}(9\l_1+ 2\l_2 + 3\l_3 + \l_4)\delta_{x'y'}\delta_{x'z}\delta_{x'w}G_{x'x'}\hat{\delta}^p_{\mba \mbb;\mbc\mbd}\right.\crcr
&\quad -\f{1}{3}(\l_2 + 2 \l_4+3\l_5)\delta_{x'y'}\delta_{x'z}\delta_{x'w}G_{x'x'}\hat{\delta}^d_{\mba \mbb;\mbc\mbd} \crcr
&\quad \left.+\frac{\l_1^2}{4}G_{x'y'}^4(3\hat{\delta}^p_{\mba \mbb;\mbc\mbd} \delta_{x'w}\delta_{y'z}+ 2\hat{\delta}^d_{\mba \mbb;\mbc\mbd}\delta_{x'z}\delta_{y'w})\right] \,,
\end{align}
where we defined the rescaled pillow and double-trace contraction operators $\hat{\delta}^p_{\mba \mbb;\mbc\mbd}=\frac{1}{N^2}\delta^p_{\mba \mbb;\mbc\mbd}$ and $\hat{\delta}^d_{\mba \mbb;\mbc\mbd}=\frac{1}{N^3}\delta^d_{\mba \mbb;\mbc\mbd}$ (see appendix~\ref{ap:conventions}).
The colored graphs corresponding to the last line are depicted in figure~\ref{fig:kernel_wheel}.

In momentum space this four-point kernel becomes:
\begin{align}
&K_{(\mba,p_1)(\mbb,p_2)(\mbc,p_3)(\mbd,p_4)}= \crcr
& \quad (2\pi)^d\delta(p_1+p_2+p_3+p_4)G(p_1)G(p_2)\left[ -\f{1}{3}(9\l_1+ 2\l_2 + 3\l_3 + \l_4)\int_q G(q)\hat{\delta}^p_{\mba \mbb;\mbc\mbd} \right.\crcr
&\qquad \quad -\f{1}{3}(\l_2 + 2 \l_4+3\l_5)\int_q G(q)\hat{\delta}^d_{\mba \mbb;\mbc\mbd} \crcr
&\qquad \quad +\frac{\l_1^2}{4}\left(3\hat{\delta}^p_{\mba \mbb;\mbc\mbd} \int_{q_1,q_2,q_3}G(q_1)G(q_2)G(q_3)G(-p_1-p_4-q_1-q_2-q_3)\right.\crcr
&\qquad \qquad \left.\left.  + 2\hat{\delta}^d_{\mba \mbb;\mbc\mbd}\int_{q_1,q_2,q_3}G(q_1)G(q_2)G(q_3)G(-p_1-p_3-q_1-q_2-q_3)\right) \right] \,.
\end{align}
For convenience, we introduce also a reduced kernel, with the tadpoles set to zero, i.e.:
\begin{equation}
\hat{K}_{(\mba,x)(\mbb,y)(\mbc,z)(\mbd,w)}=\frac{\l_1^2}{4}G_{zw}^4(3\hat{\delta}^p_{\mba \mbb;\mbc\mbd} G_{xw}G_{yz} + 2\hat{\delta}^d_{\mba \mbb;\mbc\mbd} G_{xz}G_{yw}) \,.
\label{kernel-rank3}
\end{equation}
In fact, since the propagator is massless, the tadpoles are zero in dimensional regularization, hence the reduced kernel is all we need.

\begin{figure}[htbp]
\centering
\vspace{.4cm}
\includegraphics[scale=1.2]{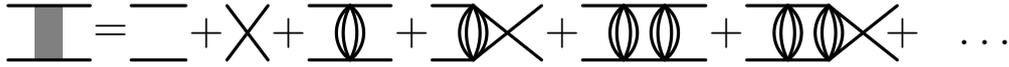}
\caption{The full forward four-point function as a series of ladders. The crossings should be omitted for our rank-3 model, because it is built on complex fields, with bipartite graphs. However, they contribute for the rank-5 model, which has real fields.
}
\label{fig:ladders}
\end{figure}

The full four-point function  at leading order in $1/N$ is obtained by summing ladders of arbitrary lengths with the (reduced) four-point kernel acting as rung (see \cite{Benedetti:2019eyl,Gurau:2019qag}). 
More precisely, defining the forward four-point function as
\be \label{eq:fw4pt}
\cF_{(\mba,x)(\mbb,y)(\mbc,z)(\mbd,w)} \equiv  \langle{\phi_{\mba}(x) \bar{\phi}_{\mbb}(y) \phi_{\mbc}(z) \bar{\phi}_{\mbd}(w)} \rangle - G(x-y) G(z-w) \d_{\mba\mbb} \d_{\mbc\mbd} \,,
\ee
one finds that at leading order in $1/N$ it is given by a geometric series on the (reduced) kernel:
\be \label{eq:fw4pt-ladders}
\cF_{(\mba,x)(\mbb,y)(\mbc,z)(\mbd,w)} = \int dz'dw' \,  (\mathbf{1} - \hat{K})^{-1}_{(\mba,x)(\mbb,y)(\mbc,z')(\mbd,w')} 
 \, G_{w'w}G_{z'z}\,.
\ee
We represent the series of ladder diagrams in figure~\ref{fig:ladders}, where the crossings do not contribute here because we consider a bipartite model with $U(N)^3$ symmetry.

For dimensional reasons, the propagators being massless and by the use of dimensional regularization, we do not expect the four-point function to require a renormalization of the quartic couplings, which are dimensionful. We verify this explicitly at lowest order in perturbation theory, that is by considering the fully amputated four-point kernel, with $G(q)$ replaced by the bare propagator $C(q)$.
Therefore, the reduced kernel writes:
\begin{equation}
\frac{\lambda_1^2}{4}\mathcal{Z}^4 (3\hat{\delta}^p_{\mba \mbb;\mbc\mbd}U_{\zeta}(p_1+p_4) + 2\hat{\delta}^d_{\mba \mbb;\mbc\mbd}U_{\zeta}(p_1+p_3))\,,
\end{equation}
with 
\begin{equation}
U_{\zeta}(p_1+p_4)=\int_{q_1,q_2,q_3}\frac{1}{q_1^{2\zeta}q_2^{2\zeta}q_3^{2\zeta}(p_1+p_4+q_1+q_2+q_3)^{2\zeta}}\,.
\end{equation}
Using \eqref{eq:intG}, we find:
\begin{equation}
U_{\zeta}(p_1+p_4)=\frac{|p_1+p_4|^{3d-8\zeta}}{(4\pi)^{3d/2}}\frac{\Gamma(d/2-\zeta)^4\Gamma(4\zeta-3d/2)}{\Gamma(\zeta)^4\Gamma(2d-4\zeta)}\,.
\end{equation}

\paragraph{Short-range propagator.} For $\zeta=1$ and $d=3-\epsilon$, this is finite (no poles in $\epsilon$):
\begin{equation}
U_1(p_1+p_4)=-\frac{|p_1+p_4| \pi}{4} \,.
\end{equation}

\paragraph{Long-range propagator.} For $\zeta=d/3$ and $d<3$, this is also finite:
\begin{equation}
U_{d/3}(p_1+p_4)=\frac{|p_1+p_4|^{d/3}}{(4\pi)^{3d/2}}\frac{\Gamma(d/6)^4\Gamma(-d/6)}{\Gamma(d/3)^4\Gamma(2d/3)} \,.
\end{equation}

In both cases, there are no divergences in the four-point kernel. We thus do not need to renormalize the four-point couplings and we can take them to be zero from the beginning.

\subsection{Rank 5}

In rank $5$ and at leading order in $1/N$, the 2PI effective action is given by:\footnote{We now have $\mba=(a_1,a_2,a_3,a_4,a_5)$ and so on.} 
\begin{align}
\label{eq:2PIrank5}
-\G^{2PI}[\mbG] = &-\f{1}{6}\left(\sum_{i=2}^6 \f{\kappa_i}{N^{5+\rho(J_i)}}\d^{(i)}_{\mba\mbb;\mbc\mbd;\mbe\mbf}\right) \int\dd x ~\mbG_{(\mba,x)(\mbb,x)}\mbG_{(\mbc,x)(\mbd,x)}\mbG_{(\mbe,x)(\mbf,x)} \crcr
&+\f{1}{2}\left(\f{\kappa_1}{6N^5}\right)^2 \d^{(1)}_{\mba\mbb\mbc\mbd\mbe\mbf} \d^{(1)}_{\mbg\mbh\mbj\mbk\mbm\mbn}\times\crcr
&\qquad \int\dd x\dd y ~\mbG_{(\mba,x)(\mbg,y)}\mbG_{(\mbb,x)(\mbh,y)} \mbG_{(\mbc,x)(\mbj,y)}\mbG_{(\mbd,x)(\mbk,y)}\mbG_{(\mbe,x)(\mbm,y)}\mbG_{(\mbf,x)(\mbn,y)} \,.
\end{align}

One recovers the self-energy from:
\begin{equation}
\Sigma[\mbG] = -2\f{\d\G^{2PI}[\mbG]}{\d\mbG} \,,
\end{equation}
where the extra factor  2 with respect to \eqref{eq:Sigma2PI-complex} is due to the difference between real and complex fields.
The amputated four-point kernel is still obtained by derivating the self-energy with respect to $\mbG$.

The right-amputated four-point kernel on-shell is then:
\begin{align}
K_{(\mba,x)(\mbb,y)(\mbc,z)(\mbd,w)}=&G_{xx'}G_{yy'}\left[ -2\left(\kappa_{6}+\kappa_{3}+\frac{2\kappa_{2}}{3}+\frac{\kappa_{4}}{3}\right)\delta_{x'y'}\delta_{x'z}\delta_{x'w}G_{x'x'}\hat{\delta}^p_{\mba\mbb;\mbc\mbd} \right.\crcr
&-2\left(\frac{\kappa_{2}}{3}+\frac{2\kappa_{4}}{3}+\kappa_{5}\right)\delta_{x'y'}\delta_{x'z}\delta_{x'w}G_{x'x'}\hat{\delta}^d_{\mba\mbb;\mbc\mbd} \crcr
&\left.+\frac{5\kappa_1^2}{6}\delta_{x'w}\delta_{y'z}G_{x'y'}^4\hat{\delta}^p_{\mba\mbb;\mbc\mbd}\right] \,. 
\end{align}

The structure is the same as for the rank-$3$ model: the only difference comes from the combinatorial factors. Then, the Feynman amplitudes are the same as before and there are still no divergences. We can thus again take the four-point couplings to be zero from the beginning. Eliminating also the tadpoles, the four-point kernel is reduced to:
\begin{equation}
\hat{K}_{(\mba,x)(\mbb,y)(\mbc,z)(\mbd,w)}=G_{xw}G_{yz}\frac{5\kappa_1^2}{6}G_{zw}^4\hat{\delta}^p_{\mba\mbb;\mbc\mbd} \,.
\end{equation}

\section{Beta functions}
\label{sec:betas}

We have seen in section~\ref{sec:SDeq} that the Schwinger-Dyson equations for the two-point functions admit a conformal IR limit for $\zeta=1$, and a conformal solution valid at all scales for $\zeta=d/3$. However, to claim that we found a non-trivial conformal field theory, we should identify an interacting fixed point of the renormalization group.\footnote{In principle  a fixed point provides us only with a scale invariant theory, full conformal invariance having to be proved case by case or on the basis of the available theorems in dimensions two and four. See for example \cite{Nakayama:2013is} for a review.}
Therefore, in this section we will study the beta functions for the full actions \eqref{eq:int-action-graph} and \eqref{eq:int-action-graph-rank5}, and their relative fixed points.

We will explain the general structure of the beta functions in the rank-3 case. As we will see, the rank-5 case is very similar, except for the presence of an additional type of interaction, $J_6$ (the prism), a difference which however turns out to be crucial.

\subsection{Rank $3$}

The amputated connected six-point function can be decomposed into the different interaction terms:
\begin{equation}
\Gamma^{(6)}(0,\dots, 0)=\sum_{i=1}^5\Gamma^{(6,i)}(0,\dots, 0)\hat{\delta}^{i}\,.
\end{equation}

The renormalized sextic couplings $g_i$ are defined in terms of the  bare expansion of the six-point functions by the renormalization condition:
\begin{equation}
\label{eq:renorcouplings}
g_i=\mu^{-2\epsilon}Z^3\Gamma^{(6,i)}(0,\dots, 0) \, ,
\end{equation}
where the power of the renormalization scale $\mu$ is  fixed by demanding that $g_i$ are dimensionless, and it is the same both for $\zeta=1$ in $d=3-\epsilon$ dimensions  and for $\zeta=\frac{d+\epsilon}{3}$ in general $d<3$.

%


At leading order in $1/N$ the wheel vertex does not receive any radiative corrections, i.e.:
\be
g_1=\mu^{-2\epsilon}Z^3\l_1\,.
\ee
Since $Z=1+\cO(\l_1^2)$, the inverse $\l_1(g_1)$ is guaranteed to exist in the perturbative expansion, at least.\footnote{For the long-range model, it is actually easier to write the inverse relation, because at $\epsilon=0$ we can solve the exact equation \eqref{Z-norm-d/3} by multiplying it by $\mathcal{Z}^6$ and using $\mathcal{Z}^6\l_1^2=g_1^2$:
\begin{equation*}
\mathcal{Z} = 1- \f{g_1^2}{g_c^2}\,, \;\;\;\; g_c^{-2} = \frac{1}{4(4\pi)^{2d}}\frac{3\Gamma(1-\frac{d}{3})\Gamma(\frac{d}{6})^5}{d\Gamma(\frac{d}{3})^5\Gamma(\frac{5d}{6})} \,.
\end{equation*}
Therefore, $\l_1=g_1/\mathcal{Z}^3$ exists for $g_1<g_c$.
\label{foot:g_c}
}
Its beta function will then be
\be
\beta_1 \equiv \m \partial_\m g_1 = (-2\epsilon+3 \eta)  g_1\,,
\ee
where we defined the anomalous dimension $\eta= (\m\partial_\m \ln Z)|_{\l_1(g_1)}$. Clearly, if $\epsilon=0$ and $Z$ is finite, as in the long-range case, then the beta function is identically zero, and we have a chance of finding a one-parameter family of fixed points, as in chapter~\ref{chap:CTKT}. On the other hand, for $\epsilon>0$, in order to find a non-trivial fixed point we have to rely on a Wilson-Fisher type of cancellation between the tree-level term and the quantum corrections, hence we need $\eta\neq 0$, that is, we need a short-range propagator.

\begin{figure}[htbp]
\centering
\captionsetup[subfigure]{labelformat=empty}
\subfloat[]{\includegraphics[scale=1]{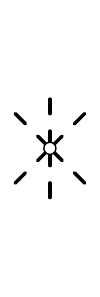}}
\hspace{1cm}
\subfloat[]{\includegraphics[scale=1]{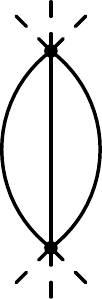}}
\hspace{1cm}
\subfloat[]{\includegraphics[scale=1]{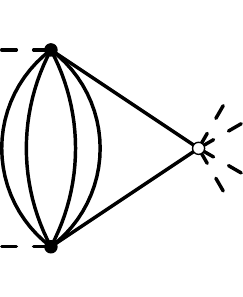}}
\caption{The three diagrams that contribute to the bare expansion of the six-point couplings up to order three. The black circles represent wheel vertices and the white circles can be any of the $I_i$ interactions (including the wheel itself).}
\label{fig:bare3}
\end{figure}

We now compute the bare expansion of the other couplings. The expansion starts of course at tree level, with a bare vertex with any $I_i$ interaction.
At order two in the coupling constants, there is only one diagram which contributes: two wheel vertices connected by three internal edges (we call this Feynman diagram the \textit{candy}). At order three, we have one more diagram: two wheel vertices connected to each other by four internal edges and each of them connected by another internal edge to a vertex with any $I_i$ interaction (including the wheel itself). These diagrams are the only tadpole-free six-point diagrams that can be obtained by cutting edges of vacuum melon-tadpole diagrams, at this order in the couplings, and they are pictured in figure~\ref{fig:bare3}.

We can actually construct the leading order $6$-point graphs at all orders using the forward four-point function introduced in \eqref{eq:fw4pt-ladders}. 
Indeed, the amputated connected six-point functions can be obtained by deleting  three different lines in the vacuum diagrams, without disconnecting the diagrams. On the other hand, vacuum diagrams are given in figure~\ref{fig:fund_vacuum}, with lines decorated by melonic and tadpole insertions, but we should not leave any closed tadpoles otherwise the diagram will evaluate to zero in dimensional regularization. Therefore, we can have at most one tadpole vertex; this fact does not limit the number of wheel vertices, as they can appear in melonic insertions as well, but it has the important consequence that the couplings $\l_2$ to $\l_5$ appear at most linearly in $\Gamma^{(6)}$.
Equivalently, we can just consider the two diagrams in figure~\ref{fig:fund_vacuum} with only melonic insertions. Furthermore, for the trefoil on the right of figure~\ref{fig:fund_vacuum}, we should cut an internal line on each of the three (decorated) leaves. At last, we should notice that each time we delete a line in a melonic two-point function, we generate a ladder diagram. In fact, starting from the SDE equation $G=(C^{-1}-\Sigma[G])^{-1}$, and using \eqref{eq:defK}, we obtain
\be
\f{\d G_{AB}}{\d C_{EF}} = (1-K)^{-1}_{AB,A'B'} G_{A'E'} C^{-1}_{E'E} G_{B'F'} C^{-1}_{F'F} + (E\leftrightarrow F)\,.
\ee
When using this formula on vacuum diagrams, we should then strip off the factors $G\cdot C^{-1}$ in order to obtain amputated $n$-point functions. We thus obtain the right-amputated version of \eqref{eq:fw4pt-ladders}.

\begin{figure}[htbp]
\centering
\captionsetup[subfigure]{labelformat=empty}
\subfloat[]{\includegraphics[scale=0.5]{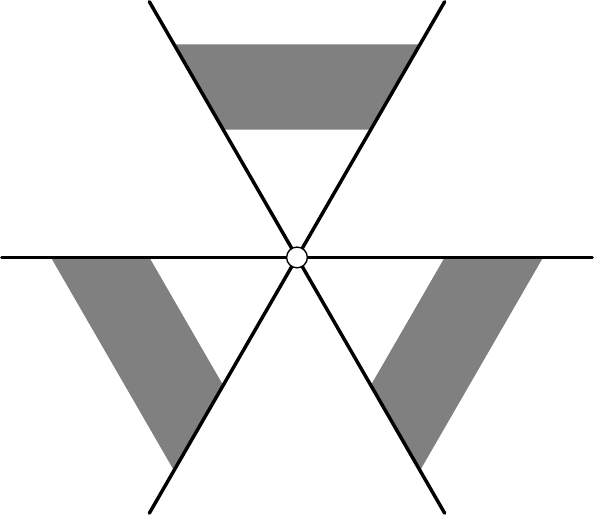}}
\hspace{1cm}
\subfloat[]{\includegraphics[scale=0.5]{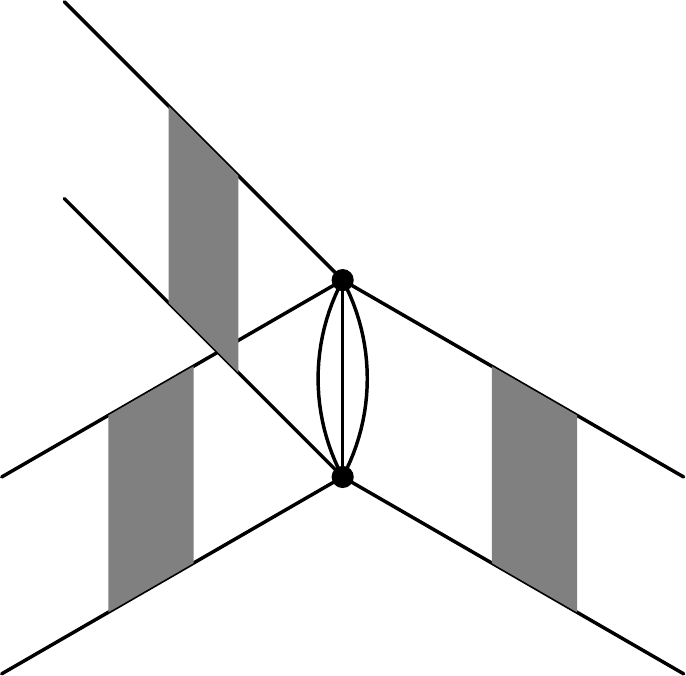}}
\hspace{1cm}
\subfloat[]{\includegraphics[scale=0.5]{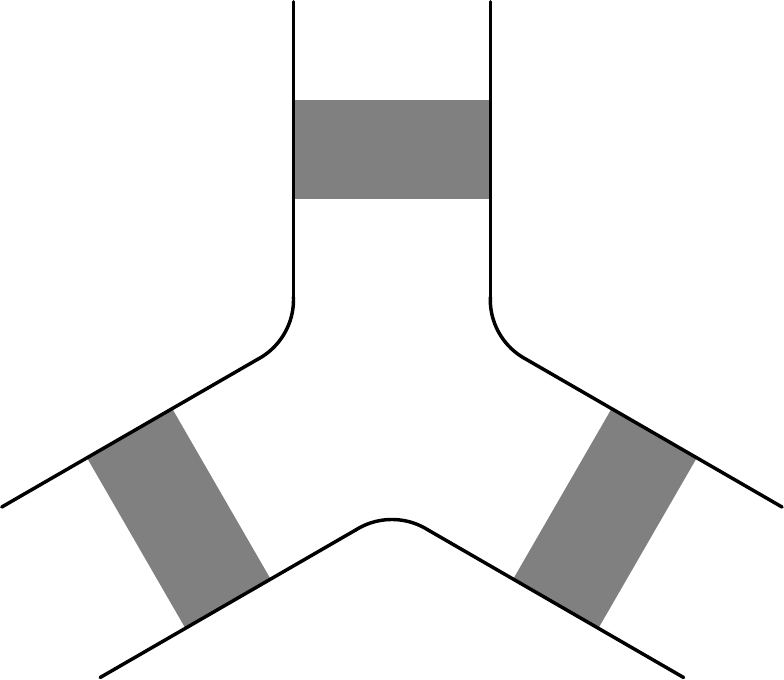}}
\caption{The three types of diagrams contributing to the bare expansion of the six-point couplings in the large-$N$ limit at all order in the coupling constants. The black circles represent wheel vertices and the white circles can be any of the $I_i$ interactions (including the wheel). The grey squares represent the full forward four-point function.}
\label{fig:allorders}
\end{figure}

In conclusion, we then have three different types of leading-order $6$-point graphs. First, we can connect three full forward four-point functions on every pairs of external legs of the bare vertex of figure~\ref{fig:bare3}, thus obtaining the graph on the left of figure~\ref{fig:allorders}. We can also do the same with the candy and obtain the graph in the middle of figure~\ref{fig:allorders}.  Finally, we can also connect three full forward four-point functions and obtain the graph on the right of figure~\ref{fig:allorders}. The last two have been encountered for example in  \cite{Gross:2017hcz,Gross:2017aos}, where they have been called \emph{contact} and \emph{planar} diagrams, respectively.

This implies that renormalized couplings $g_i$, with $i>1$, have a bare expansion of the form:
\be
g_i = \m^{-2\epsilon} Z^3 \left( \l_i +  A_i(\l_1^2 )+ \sum_{j=1\ldots 5}  B_{ij}(\l_1^2) \l_j \right)\,,
\ee
with $A_i(x)$ and $B_{ij}(x)$ starting at linear order in $x$. The term $ \l_i + \sum_{j=1\ldots 5}  B_{ij}(\l_1^2) \l_j$ is a resummation of contributions from the graph on the left of figure~\ref{fig:allorders}, while $A_i(\l_1^2 )$ resums the other two. Although we could at least write the relative Feynman integral expressions in terms of forward four-point functions and six-point kernels, as we will not need them, and the combinatorics is different for different bubbles, we will not be more precise than that.

For $i>1$ the relation between bare and renormalized couplings is linear and thus it can be easily inverted:
\be
\l_i = (\mathbf{1}+B)^{-1}_{ij} \left( \f{\m^{2\epsilon} g_j}{Z^3} - A_j \right)\big|_{\l_1=\l_1(g_1)}\,,
\ee
where $\mathbf{1}_{ij}=\delta_{ij}$.

Using the fact that the flow of $g_1$ is independent of the others, then one arrives at the conclusion that the beta functions of the other couplings are linear combinations of the couplings, with coefficients that are functions of $g_1^2$:
\be
\beta_i = (-2\epsilon+3 \eta)  g_i + \tilde{A}_i(g_1^2) + \sum_j  \tilde{B}_{ij}(g_1^2)  g_j \,,
\ee
where 
\begin{align}
\tilde{A}_i(g_1^2)  &=  \m^{-2\epsilon} Z^3 \left(  \m\p_\m A_i - \sum_{j,k}  (\m\p_\m B_{ij}) (\mathbf{1}+B)^{-1}_{jk} A_k \right)\big|_{\l_1=\l_1(g_1)} \,,\crcr
\tilde{B}_{ij}(g_1^2) &=  \sum_{j}  (\m\p_\m B_{ij}) (\mathbf{1}+B)^{-1}_{jk} \,.
\end{align}

As we saw above, the combination $-2\epsilon+3 \eta$ is either identically zero, or it is zero at the fixed point of $g_1$.
In order to find the fixed points we are left with a linear problem.

In the following we will compute explicitly the beta functions at lowest order in the perturbative expansion, i.e.\ we will only include the contribution of the diagrams in figure~\ref{fig:bare3}.

\subsubsection{Short-range propagator}
\label{sec:betas1}

We will now compute the beta function only up to order $3$ in the coupling constants for $\zeta=1$. 

Expanding the six-point functions of \eqref{eq:renorcouplings} to order three, we have the bare expansions:
\begin{align}
g_{2}&=\mu^{-2\epsilon}Z^3\left(\l_{2}-\mu^{-2\epsilon}\frac{9}{2}D_1\l_1^2+\mu^{-4\epsilon}S_1\l_1^2\left(\frac{9}{2}\l_1 + \frac{1}{2}\l_2\right)\right)\,,\crcr
g_{3}&=\mu^{-2\epsilon}Z^3\left(\l_{3}+ \mu^{-4\epsilon}S_1\frac{3}{4}\l_1^2\l_{3}\right)\,,\crcr
g_{4}&=\mu^{-2\epsilon}Z^3\left(\l_{4}+ \mu^{-4\epsilon}S_1\l_1^2\left(\f{27}{4}\l_1 +\frac{5}{2} \l_2+3\l_3+\f{7}{4}\l_4\right)\right)\,,\crcr
g_{5}&=\mu^{-2\epsilon}Z^3\left(\l_{5}-\mu^{-2\epsilon}D_1\f{1}{2}\l_1^2+ \mu^{-4\epsilon}S_1\l_1^2\left(\f{3}{4}\l_2+2\l_4+\f{15}{4}\l_5\right)\right)\,,
\end{align}
where $Z$ is given in \eqref{eq:wavef3}, $D_1$ and $S_1$ are the amplitudes of the candy and the Feynman diagram on the right of figure~\ref{fig:bare3} respectively. The integrals are both computed in appendix~\ref{ap:betafun4} using the Schwinger parametrization given in \eqref{eq:amp_final} and made explicit for sextic models in \eqref{eq:amp_final_sextic}.
It is convenient to rescale the couplings as $\bar{g}_i=g_i/(4\pi)^d$. 
%
Then, using $\mu\partial_{\mu} \l_i=0$, the beta functions $\b_i\equiv \mu\partial_{\mu} g_i$ are:
\begin{align}
\b_1 &= -2 \bar{g}_1\left(\eps-\bar{g}_1^2\pi^2\right)\,, \crcr
\b_2 &= -2 \bar{g}_2\left(\eps-\bar{g}_1^2\pi^2\right) + 4\bar{g}_1^2\pi^2\left(\f{9}{2\pi}+9\bar{g}_1+\bar{g}_2\right)\,,\crcr
\b_3 &= -2 \bar{g}_3\left(\eps-4\, \bar{g}_1^2\pi^2\right)\,,\crcr
\b_4 &= -2 \bar{g}_4\left(\eps-\bar{g}_1^2\pi^2\right) + \bar{g}_1^2\pi^2\left(54\bar{g}_1 +20\bar{g}_2 + 24\bar{g}_3 + 14\bar{g}_4\right)\,,\crcr
\b_5 &= -2 \bar{g}_5\left(\eps-\bar{g}_1^2\pi^2\right)+\bar{g}_1^2\pi^2\left(\f{2}{\pi} +6\bar{g}_2+16\bar{g}_4+30\bar{g}_5\right)\,.
\end{align}
First, we notice that if $\bar{g}_1=0$, then all the other couplings have a trivial running, dictated by their canonical dimension ($2\epsilon$).
In such case, for $\epsilon>0$ we have only the trivial fixed point,  $\bar{g}^*_i=0~\forall i$.
For $\epsilon=0$ instead, we have a 4-dimensional manifold of fixed points. This is a generalization of the vector model case, where the $(\phi_i \phi_i)^3$ interaction is exactly marginal at large $N$, and which in fact corresponds to the case in which we retain only the triple-trace term $I_5$ in our action. In that case it is known that at some critical coupling non-perturbative effects lead to vacuum instability with a consequent breaking of conformal invariance  \cite{Bardeen:1983rv,Amit:1984ri}. It would be interesting to study the vacuum stability of our model with $\bar{g}_1=0$ in order to explore the possibility of a similar phenomenon, but we leave this for future work.

We are here interested in melonic fixed points, with $\bar{g}_1\neq 0$, for which we need  $\epsilon>0$.
In that case, we obtain two interacting fixed points:
\begin{gather}
\bar{g}^*_1=\pm\f{\sqrt{\eps}}{\pi} \, ; \qquad \bar{g}^*_2=\f{9}{2\pi}\left(-1\mp 2\sqrt{\eps}\right) \, ; \qquad \bar{g}^*_3=0 \, ;
\crcr
\bar{g}^*_4=\f{9}{7\pi}\left(5 \pm 7\sqrt{\eps}\right) \, ;\qquad \bar{g}^*_5 = \f{-109\mp 126\sqrt{\eps}}{42\pi} \, .
\end{gather}
The standard linear stability analysis of the system of beta functions consists in diagonalizing the stability matrix $\cB_{ij} \equiv \p\b_i/\p \bar{g}_j|_{\bar{g}^*}$,
thus identifying the scaling operators and their scaling dimensions, from its right-eigenvectors and eigenvalues, respectively. 
In the present case, we find the slightly unusual situation of having a non-diagonalizable stability matrix.
In fact, we find that both melonic fixed points have the same eigenvalues (critical exponents):
\begin{equation}
(4 \eps;\; 4 \eps;\; 6 \eps;\; 14 \eps;\; 30 \eps)\,,
\end{equation}
with the $4 \eps$ eigenvalue having algebraic multiplicity two, but geometric multiplicity one; hence the stability matrix is not diagonalizable.
In terms of the couplings $\{\bar{g}_1,\bar{g}_2,\bar{g}_3,\bar{g}_4,\bar{g}_5\}$, the associated eigendirections are, respectively:
\begin{gather}
 \{0,1,0,-2,1\} \, ;\quad \{\mp\f{1}{18\sqrt{\eps}}, \f{2(- 632 \pm 4095\sqrt{\eps})}{12285\eps}, 0, \f{1124\mp 4095\sqrt{\eps}}{24570 \eps} ,0 \} \, ; \crcr
 \{0,0,\frac{1}{2},-\frac{3}{2},1\} \, ; \quad
 \{0,0,0,-1,1\} \, ;\quad \{0,0,0,0,1\}\, ,
\end{gather} 
with the first two forming a Jordan chain of length two.
Each (generalized) eigendirection, by its scalar product with the vector of renormalized operators arranged in the same order as the corresponding couplings, defines a scaling operator $\cO_i$ of dimension $\Delta_i=d+\theta_i$, with the $\theta_i$ being the critical exponent associated to that eigendirection.
As our critical exponents are all positive, all the scaling operators are irrelevant at the fixed points, and therefore the latter are infrared stable.
The fact that the stability matrix is not diagonalizable implies that the fixed point theory is a logarithmic conformal field theory (see for example \cite{Hogervorst:2016itc}). Therefore, although we find real exponents, as opposed to the complex ones of the quartic model \cite{Giombi:2017dtl}, the fixed-point theory is still non-unitary.

\subsubsection{Long-range propagator}
Using the results of appendix~\ref{ap:betafun4}, along with the fact that there is no diverging wave-function renormalization in this case, the bare expansion gives:
\begin{align}
g_{1}&=\mu^{-2\epsilon}\cZ^3\l_{1}\,,\crcr
g_{2}&=\mu^{-2\epsilon}\cZ^3\left(\l_{2}-\mu^{-2\epsilon}\frac{9}{2}\cZ^3D_{d/3}\l_1^2+\mu^{-4\epsilon}\cZ^6S_{d/3}\l_1^2\left(\frac{9}{2}\l_1 + \frac{1}{2}\l_2\right)\right)\,,\crcr
g_{3}&=\mu^{-2\epsilon}\cZ^3\left(\l_{3}+ \mu^{-4\epsilon}\cZ^6S_{d/3}\frac{3}{4}\l_1^2\l_{3}\right)\,,\crcr
g_{4}&=\mu^{-2\epsilon}\cZ^3\left(\l_{4}+\mu^{-4\epsilon}\cZ^6S_{d/3}\l_1^2\left(\f{27}{4}\l_1 + \frac{5}{2}\l_2+3\l_3+\f{7}{4}\l_4\right)\right)\,,\crcr
g_{5}&=\mu^{-2\epsilon}\cZ^3\left(\l_{5}-\mu^{-2\epsilon}\cZ^3D_{d/3}\f{1}{2}\l_1^2+ \mu^{-4\epsilon}\cZ^6S_{d/3}\l_1^2\left(\f{3}{4}\l_2+2\l_4+\f{15}{4}\l_5\right)\right)\,,
\end{align}
with $\cZ$ given in \eqref{eq:wavef3-LR}.
After rescaling of the coupling constants by $(4\pi)^d$, the beta functions  at $\epsilon=0$ read:
\begin{align}
\b_1 &= 0\,, \crcr
\b_2 &= \bar{g}_1^2\f{\Gamma(d/6)^3}{\Gamma(d/3)^3\Gamma(d/2)}\left(-\f{\Gamma(-d/6)\Gamma(d/6)}{\Gamma(d/3)\Gamma(2d/3)}\left(9\bar{g}_1+\bar{g}_2\right)+9\right)\,,\crcr
\b_3 &= - 3\bar{g}_1^2 \bar{g}_3\frac{\Gamma(-d/6) \Gamma(d/6)^4}{2 \Gamma(d/3)^4 \Gamma(d/2)\Gamma(2d/3)}\,,\crcr
\b_4 &= -\bar{g}_1^2 \frac{\Gamma(-d/6) \Gamma(d/6)^4}{\Gamma(d/3)^4 \Gamma(d/2)\Gamma(2d/3)}\left(\frac{27}{2}\bar{g}_1+5\bar{g}_2+6\bar{g}_3+\frac{7}{2}\bar{g}_4\right)\,,\crcr
\b_5 &= \bar{g}_1^2\f{\Gamma(d/6)^3}{\Gamma(d/3)^3\Gamma(d/2)}\left(-\frac{2\Gamma(-d/6)\Gamma(d/6)}{\Gamma(d/3)\Gamma(2d/3)}\left(\frac{3}{4} \bar{g}_2+2\bar{g}_4+\frac{15}{4}\bar{g}_5\right)+1\right)\,.
\end{align}
This time, in addition to a 4-dimensional manifold of fixed points (set by $\bar{g}_1^*=0$, and thus analogue to what we discussed for the case $\zeta=1$ at  $\epsilon=0$), we also find a line of fixed points parametrized by the exactly marginal coupling $\bar{g}_1$: 
\begin{gather}
\bar{g}_2^*=-9\bar{g}_1+\f{9\Gamma(d/3)\Gamma(2d/3)}{\Gamma(-d/6)\Gamma(d/6)} \, ;\qquad \bar{g}_3^*=0 \, ;\crcr
\bar{g}_4^*=9\bar{g}_1 - \frac{90}{7}\f{2\Gamma(d/3)\Gamma(2d/3)}{\Gamma(-d/6)\Gamma(d/6)} \, ;\crcr
\bar{g}_5^*=-3\bar{g}_1 + \f{109\Gamma(d/3)\Gamma(2d/3)}{21\Gamma(-d/6)\Gamma(d/6)} \, .
\end{gather}
The critical exponents are: 
\begin{equation}
 \left( \f{15 \bar{g}_1^2\a}{ 2};\;
   \f{7 \bar{g}_1^2\a}{ 2 };\; 
   \f{3 \bar{g}_1^2\a}{2 };\;
    \bar{g}_1^2\a\right)\,,
\end{equation}
with
\be
\a= - \f{\Gamma(-d/6) \Gamma(d/6)^4}{
  \Gamma(d/3)^4 \Gamma(d/2) \Gamma(2 d/3)} >0\,, \;\;\; \text{for } d<3\,.
\ee
The respective eigendirections in terms of $\{\bar{g}_2,\bar{g}_3,\bar{g}_4,\bar{g}_5\}$ are: 
\begin{equation}
\{0,0,0,1\} \, ;\quad \{0,0,-1,1\} \, ;\quad\{0,1,-3,2\} \, ;\quad\{1,0,-2,1\} \, .
\end{equation}
Since the critical exponents are positive, the eigendirections correspond again to irrelevant perturbations.
In this case, the stability matrix is diagonalizable, with real exponents, hence we have so far no signal of non-unitarity.

\subsection{Rank $5$}

The diagrams contributing to the six-point function at large $N$ are again the ones of figure~\ref{fig:bare3} (or figure~\ref{fig:allorders} at all orders). However, now the black vertices represent the complete interaction and the white vertices represent only the other interactions $J_b$ for $b>1$. This will slightly change the bare expansion of the couplings and their beta functions. 

\subsubsection{Short-range propagator}

There is no radiative corrections for the coupling of the complete interaction, the renormalized coupling is just rescaled by the wave function renormalization of \eqref{eq:wavef5}:
\begin{equation}
g_1=\mu^{-2\epsilon}Z^3\kappa_1 \,.
\end{equation}

Then, we obtain the following bare expansions up to order three in the coupling constants:
\begin{align}
g_{2}&=\mu^{-2\epsilon}Z^3\left(\kappa_{2}+\mu^{-4\epsilon}\kappa_1^2S_1\left(2\kappa_{6}+\frac{2}{3}\kappa_{2}\right)\right)\,,\crcr
g_{3}&=\mu^{-2\epsilon}Z^3\left(\kappa_{3}+\mu^{-4\epsilon}\kappa_1^2\kappa_{3}S_1\right)\,,\crcr
g_{4}&=\mu^{-2\epsilon}Z^3\left(\kappa_{4}+\mu^{-4\epsilon}\kappa_1^2S_1\left(3\kappa_{6}+4\kappa_{3}+\frac{10}{3}\kappa_{2}+\frac{7}{3}\kappa_{4}\right)\right)\,,\crcr
g_{5}&=\mu^{-2\epsilon}Z^3\left(\kappa_{5}+\mu^{-4\epsilon}\kappa_1^2S_1\left(\kappa_{2}+\frac{8}{3}\kappa_{4}+5\kappa_{5}\right)\right)\,,\crcr
g_{6}&=\mu^{-2\epsilon}Z^3\left(\kappa_{6}-\mu^{-2\epsilon}\frac{10}{3}\kappa_1^2 D_1\right) \,,
\end{align}

with $D_{1}$ and $S_{1}$ defined in the previous section.

Let us rescale all the coupling constants as $\bar{\kappa_1}=\frac{\kappa_1}{(4\pi)^d}$ and so on.
Then the beta functions are:
\begin{align}
\beta_{1}&=-2\epsilon \bar{g}_1+\frac{4}{3}\pi^2\bar{g}_1^3 \,,\crcr  
\beta_{2}&=-2\epsilon \bar{g}_{2}+\frac{4}{3}\pi^2\bar{g}_1^2\bar{g}_{2}+\frac{8\pi^2}{3}\bar{g}_1^2\left(6\bar{g}_{6}+2\bar{g}_{2}\right)\,, \crcr
\beta_{3}&=-2\epsilon \bar{g}_{3}+\frac{4}{3}\pi^2\bar{g}_1^2\bar{g}_{3}+8\pi^2\bar{g}_1^2\bar{g}_{3} \,, \crcr
\beta_{4}&=-2\epsilon \bar{g}_{4}+\frac{4}{3}\pi^2\bar{g}_1^2\bar{g}_{4}+\frac{8\pi^2}{3}\bar{g}_1^2\left(9\bar{g}_{6}+12\bar{g}_{3}+10\bar{g}_{2}+7\bar{g}_{4}\right)\,,\crcr
\beta_{5}&=-2\epsilon \bar{g}_{5}+\frac{4}{3}\pi^2\bar{g}_1^2\bar{g}_{5}+\frac{8\pi^2}{3}\bar{g}_1^2\left(3\bar{g}_{2}+8\bar{g}_{4}+15\bar{g}_{5}\right)\,, \crcr
\beta_{6}&=-2\epsilon \bar{g}_{6}+\frac{4}{3}\pi^2\bar{g}_1^2\bar{g}_{6}+\frac{40\pi}{3}\bar{g}_1^2 \,.
\end{align}

The only fixed point when $\epsilon \neq 0$ is the trivial one: $\bar{g}_i^*=0,~\forall i$. We do not find any Wilson-Fisher like fixed point. 
This is due to the beta function of the prism. The non-zero fixed point of $\beta_{1}$ is $\bar{g}_1^*=\frac{\sqrt{\epsilon}}{2\pi}$. If we put it in the beta function of the prism, we obtain an expression independent of $\bar{g}_6$ and proportional to $\bar{g}_1^2$. This would imply $\bar{g}_1=0$ which is incompatible with $\bar{g}_1^*=\frac{\sqrt{\epsilon}}{2\pi}$ when $\epsilon \neq 0$. This solution is not a fixed point of the whole system.  

\subsubsection{Long-range propagator}

When $\zeta=d/3$, the wave function renormalization is finite and equal to $\mathcal{Z}$, given in \eqref{eq:wavef5-LR}. In this case the bare expansion is:
\begin{align}
g_1&=\mu^{-2\epsilon}\mathcal{Z}^3\kappa_1  \,,\crcr
g_{2}&=\mu^{-2\epsilon}\left(\mathcal{Z}^3\kappa_{2}+\mu^{-4\epsilon}\kappa_1^2\mathcal{Z}^9S_{d/3}\left(2\kappa_{6}+\frac{2}{3}\kappa_{2}\right)\right)\,,\crcr
g_{3}&=\mu^{-2\epsilon}\left(\mathcal{Z}^3\kappa_{3}+\mu^{-4\epsilon}\kappa_1^2\kappa_{3}\mathcal{Z}^9S_{d/3}\right)\,,\crcr
g_{4}&=\mu^{-2\epsilon}\left(\mathcal{Z}^3\kappa_{4}+\mu^{-4\epsilon}\kappa_1^2\mathcal{Z}^9S_{d/3}\left(3\kappa_{6}+4\kappa_{3}+\frac{10}{3}\kappa_{2}+\frac{7}{3}\kappa_{4}\right)\right)\,,\crcr
g_{5}&=\mu^{-2\epsilon}\left(\mathcal{Z}^3\kappa_{5}+\mu^{-4\epsilon}\kappa_1^2\mathcal{Z}^9S_{d/3}\left(\kappa_{2}+\frac{8}{3}\kappa_{4}+5\kappa_{5}\right)\right)\,,\crcr
g_{6}&=\mu^{-2\epsilon}\left(\mathcal{Z}^3\kappa_{6}-\mu^{-2\epsilon}\frac{10}{3}\mathcal{Z}^6\kappa_1^2 D_{d/3}\right)\,.
\end{align}

Then, the beta function of the complete interaction is again exactly zero. The other beta functions are, after rescaling of the coupling constants by $(4\pi)^d$:
\begin{align}
\beta_{2}&=-2\bar{g}_1^2\frac{\Gamma(d/6)^4 \Gamma(-d/6)}{\Gamma(d/3)^4\Gamma(d/2)\Gamma(2d/3)}\left(2\bar{g}_6+\frac{2}{3}\bar{g}_2\right) \,,\crcr
\beta_{3}&=-2\bar{g}_1^2\bar{g}_3\frac{\Gamma(d/6)^4 \Gamma(-d/6)}{\Gamma(d/3)^4\Gamma(d/2)\Gamma(2d/3)} \,,\crcr
\beta_{4}&=-2\bar{g}_1^2\frac{\Gamma(d/6)^4 \Gamma(-d/6)}{\Gamma(d/3)^4\Gamma(d/2)\Gamma(2d/3)}\left(3\bar{g}_6+4\bar{g}_3+\frac{10}{3}\bar{g}_2+\frac{7}{3}\bar{g}_4 \right)\,,\crcr
\beta_{5}&=-2\bar{g}_1^2\frac{\Gamma(d/6)^4 \Gamma(-d/6)}{\Gamma(d/3)^4\Gamma(d/2)\Gamma(2d/3)}\left(\bar{g}_2+\frac{8}{3}\bar{g}_4+5\bar{g}_5 \right) \,,\crcr
\beta_{6}&=\frac{20}{3}\frac{\Gamma(d/6)^3}{\Gamma(d/3)^3\Gamma(d/2)}\bar{g}_1^2 \,.
\end{align}

The beta function for $g_6$ admits a unique fixed point with $\bar{g}_1^*=0$. The other beta functions are then exactly zero. 
Starting from non-zero couplings, we find that the flow is driven by $\bar{g}_6$ flowing to minus infinity in the IR, and the other couplings flow towards: 
\begin{equation}
\bar{g}_2^*=-3\bar{g}_6 \, ; \quad  \bar{g}_3^* = 0 \, ; \quad \bar{g}_4^*=3\bar{g}_6 \, ; \quad \bar{g}_5^* = -3\bar{g}_6 \, .
\end{equation}

\section{Spectrum of operators}
\label{sec:BSeq}

For the rank-3 case we found IR fixed points with non-zero wheel coupling, both in the short-range and long-range versions of the model.
In order to better understand the conformal field theory at such IR fixed points,\footnote{We assume here that our fixed points correspond to conformal field theories.} we wish to compute the spectrum of operators that appear in the operator product expansion of $\phi_{abc}(x)\bar{\phi}_{abc}(0)$. Schematically, these are expected to be the bilinear operators $\phi_{abc}(\partial^2)^n\bar{\phi}^{abc}$, and their spectrum can  be obtained using the conformal Bethe-Salpeter equation \cite{Klebanov:2016xxf,Giombi:2017dtl}, or equivalently, the spectral decomposition of the four-point function \cite{Liu:2018jhs,Gurau:2019qag}.

As we saw in section~\ref{sec:CFT}, the dimensions of bilinear operators can be found by solving the equation $k_{\zeta}(h,J)=1$ with $k_{\zeta}(h,J)$ the eigenvalues of the two-particle irreducible four-point kernel. Moreover, as eigenfunctions of the kernel are three-point functions of two fundamental scalars with an operator, we need to find the eigenvalues $k_{\zeta}(h,J)$ of the kernel from the equation:
\begin{equation}
    k_{\zeta}(h,J) v_J(x_0,x_1,x_2) = \int\,d^dx_3\,d^dx_4\, K(x_1,x_2;x_3,x_4)v_J(x_0,x_3,x_4),
    \label{BS}
\end{equation}
where the form of the kernel is obtained from \eqref{kernel-rank3} to be
\begin{equation}
    K(x_1,x_2;x_3,x_4) = \f{\l_1^2}{4} \left[3G(x_{14})G(x_{23})+2G(x_{13})G(x_{24})\right]G(x_{34})^4\,,
    \label{kernel}
\end{equation}
and since we integrate over $x_3$ and $x_4$, both terms can be combined into one.

\subsection{Short-range propagator}

Since the corresponding integrals are simpler to solve in position space, we wish to set up the eigenvalue equation in position space. For that, we need the two-point function in position space, which for the case $\zeta=1$ is as follows:
\begin{align}
 G(x) &= \int\,\f{d^d p }{(2\pi)^d} e^{-i p\cdot x}\,G(p)=\cZ\int\,\f{d^d p }{(2\pi)^d} \f{e^{-i p\cdot x}}{p^{2d/3}}\crcr
 &= \cZ\f{2^{d/3}}{(4\pi)^{d/2}}\f{\Gamma(\f{d}6)}{\Gamma(\f{d}3)}\f1{(x^2)^{d/6}}= F_1 \f1{(x^2)^{d/6}} \,,
 \label{2-pt-x}
\end{align}
where $F_1 =\cZ\f{2^{d/3}}{(4\pi)^{d/2}}\f{\Gamma(\f{d}6)}{\Gamma(\f{d}3)} $.
To perform the integrals at $J=0$, we use the conformal identity \eqref{eq:conf_int}. 
Using it a first time to perform the integral over $x_3$, we obtain
\begin{equation}
     \int\,d^d x_3 \f1{(x_{03}^2)^{h/2}(x_{23}^2)^{d/6}(x_{34}^2)^{5d/6-h/2}}=\f{L_d\big(\f{h}2,\f{d}6\big)}{(x_{02}^2)^{-d/3+h/2}(x_{04}^2)^{d/3}(x_{24}^2)^{d/2-h/2}} \, ,
     \label{int-1}
\end{equation}
with
\begin{equation}
   L_d\bigg(\f{h}2,\f{d}6\bigg) =\pi^{d/2}\f{\Gamma(\f{d}3)\Gamma(-\f{d}3+\f{h}2)\Gamma(\f{d}2-\f{h}2)}{\Gamma(\f{d}6)\Gamma(\f{5d}6-\f{h}2)\Gamma(\f{h}2)}\,.
   \label{part-1}
\end{equation}
Now, performing the remaining integral over $x_4$, we get
\begin{equation}
     \int\,d^d x_4 \f1{(x_{04}^2)^{d/3+h/2}(x_{24}^2)^{d/2-h/2}(x_{14}^2)^{d/6}}=\f{L_d(\f{d}3+\f{h}2,\f{d}2-\f{h}2)}{(x_{02}^2)^{d/3}(x_{01}^2)^{h/2}(x_{12}^2)^{d/6-h/2}},
     \label{int-2}
\end{equation}
with
\begin{equation}
     L_d\bigg(\f{d}3+\f{h}2,\f{d}2-\f{h}2\bigg) =\pi^{d/2}\f{\Gamma(\f{d}3)\Gamma(\f{h}2)\Gamma(\f{d}6-\f{h}2)}{\Gamma(\f{d}3+\f{h}2)\Gamma(\f{d}2-\f{h}2)\Gamma(\f{d}6)}.
   \label{part-2}
\end{equation}
Collecting the terms from the first and second integrals, and combining their coefficients from \eqref{2-pt-x}, \eqref{part-1} and \eqref{part-2}, we get the $J=0$ eigenvalues of the kernel to be
\begin{align}
    k_1(h,0) &= \f54\, \l_1^2\,F_1^6\, \pi^d\,\f{\Gamma(\f{d}3)^2\Gamma(-\f{d}3+\f{h}2)\Gamma(\f{d}6-\f{h}2)}{\Gamma(\f{d}6)^2\Gamma(\f{5d}6-\f{h}2)\Gamma(\f{d}3 + \f{h}2)}\crcr
    &=\f54 \l_1^2\bigg(\f1{\pi^{d}}\f{4}{\lambda_1^2}\f{d}{3}\f{\Gamma(\frac{d}{6})\Gamma(\frac{5d}{6})}{\Gamma(1-\frac{d}{3})\Gamma(\frac{d}{3})}\bigg)\bigg(\pi^d\f{\Gamma(\f{d}3)^2\Gamma(-\f{d}3+\f{h}2)\Gamma(\f{d}6-\f{h}2)}{\Gamma(\f{d}6)^2\Gamma(\f{5d}6-\f{h}2)\Gamma(\f{d}3 + \f{h}2)}\bigg)
    \crcr
    &= -5\times \f{\Gamma(\f{5d}6)\Gamma(\f{d}3)\Gamma(-\f{d}3+\f{h}2)\Gamma(\f{d}6-\f{h}2)}{\Gamma(-\f{d}3)\Gamma(\f{d}6)\Gamma(\f{5d}6-\f{h}2)\Gamma(\f{d}3 + \f{h}2)}.
    \label{eq:eigenvalue}
\end{align}
To find the spectrum of the bilinears, we must solve the above equation for $k_1(h,0)=1$, with $d=3-\epsilon$. We use the method of section \ref{sec:unitarity},  setting $h=1+2n+2z$, and treating $z$ as a perturbation of the classical dimension, which is justified for small $\epsilon$. 

For $n=0$ and $n=1$, we find the following solutions:
\begin{align}
h_{0}&=1+\frac{29}{3}\epsilon + \mathcal{O}(\epsilon^2), \\
h_1&=3+3\epsilon + \mathcal{O}(\epsilon^2).
\end{align}
We also find a solution corresponding to the mixing with a quartic operator (as we deduce from the dimension at $\epsilon=0$):
\begin{equation}
h_{q}=2-\frac{32}{3}\epsilon + \mathcal{O}(\epsilon^2).
\end{equation}
Lastly, for $n>1$ we find:
\begin{equation}
h_n=1+2n-\frac{\epsilon}{3}+\frac{20}{3n(n-1)(4 n^2-1)}\epsilon^2+\mathcal{O}(\epsilon^3)~~, ~~ n>1.
\end{equation}
The solutions we just found are exactly the ones found in \cite{Giombi:2017dtl}, which is not surprising, as their equation 4.6 for $q=6$, giving the eigenvalues of the kernel, is the same as \eqref{eq:eigenvalue}. However, it was assumed to hold for rank $5$, but as we have seen, it turns out that in that case there is no Wilson-Fisher fixed point, hence no interacting CFT to which these equations might apply. On the other hand, we have shown here that we still recover the same spectrum for the model in rank $3$, which admits a melonic Wilson-Fisher fixed point.

\begin{figure}[htbp]
\centering
\captionsetup[subfigure]{labelformat=empty}
\subfloat[]{\includegraphics[scale=0.45]{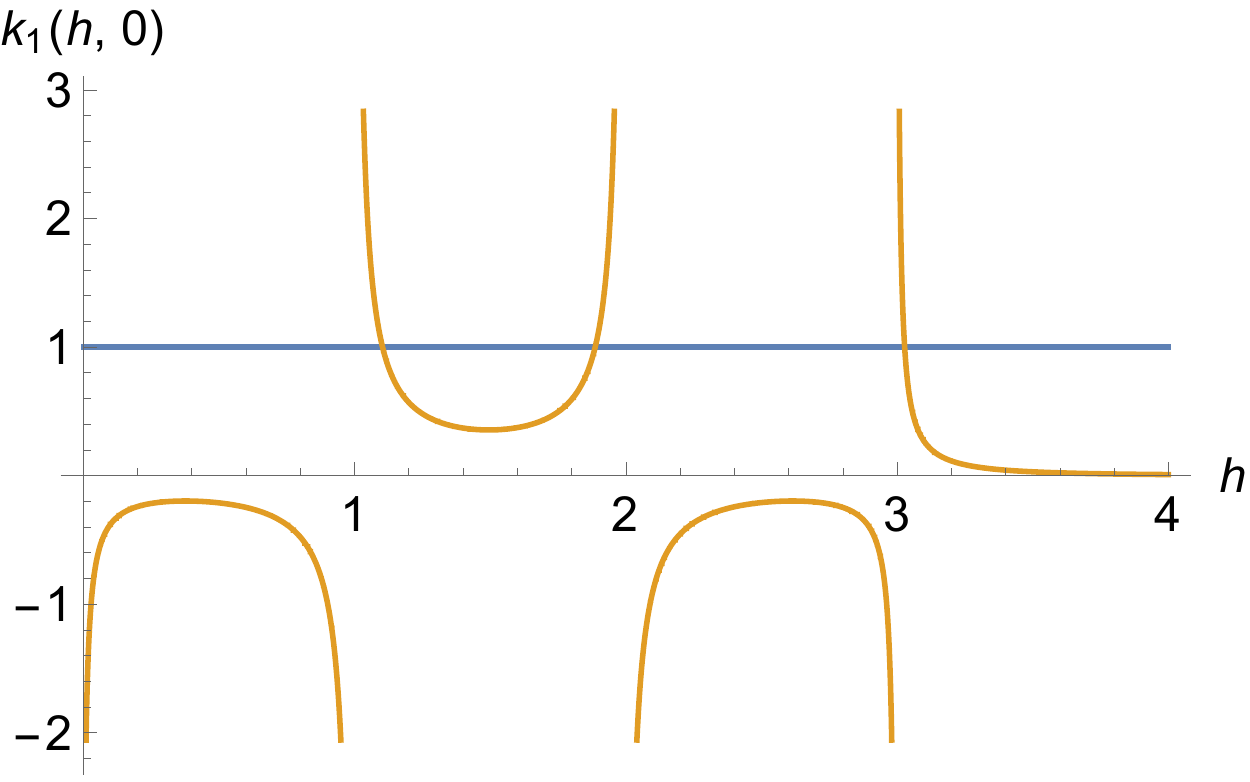}}
\subfloat[]{\includegraphics[scale=0.45]{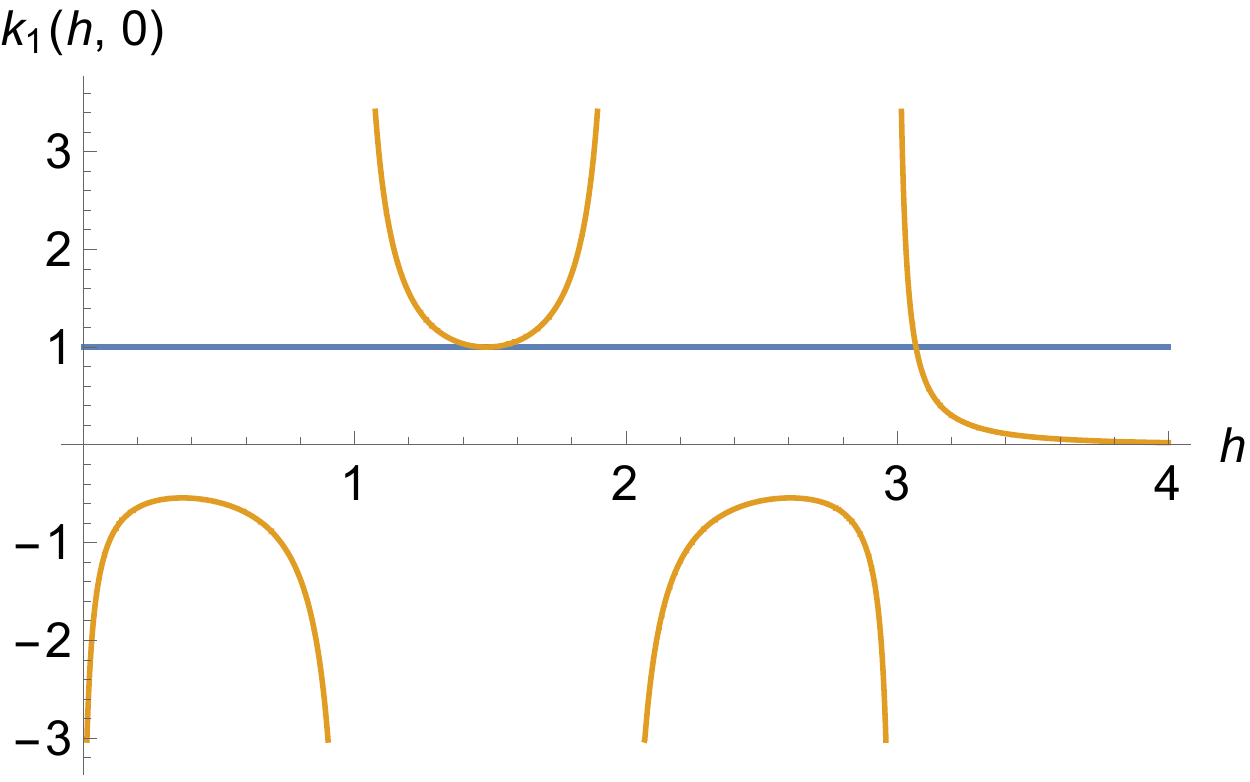}}
\subfloat[]{\includegraphics[scale=0.45]{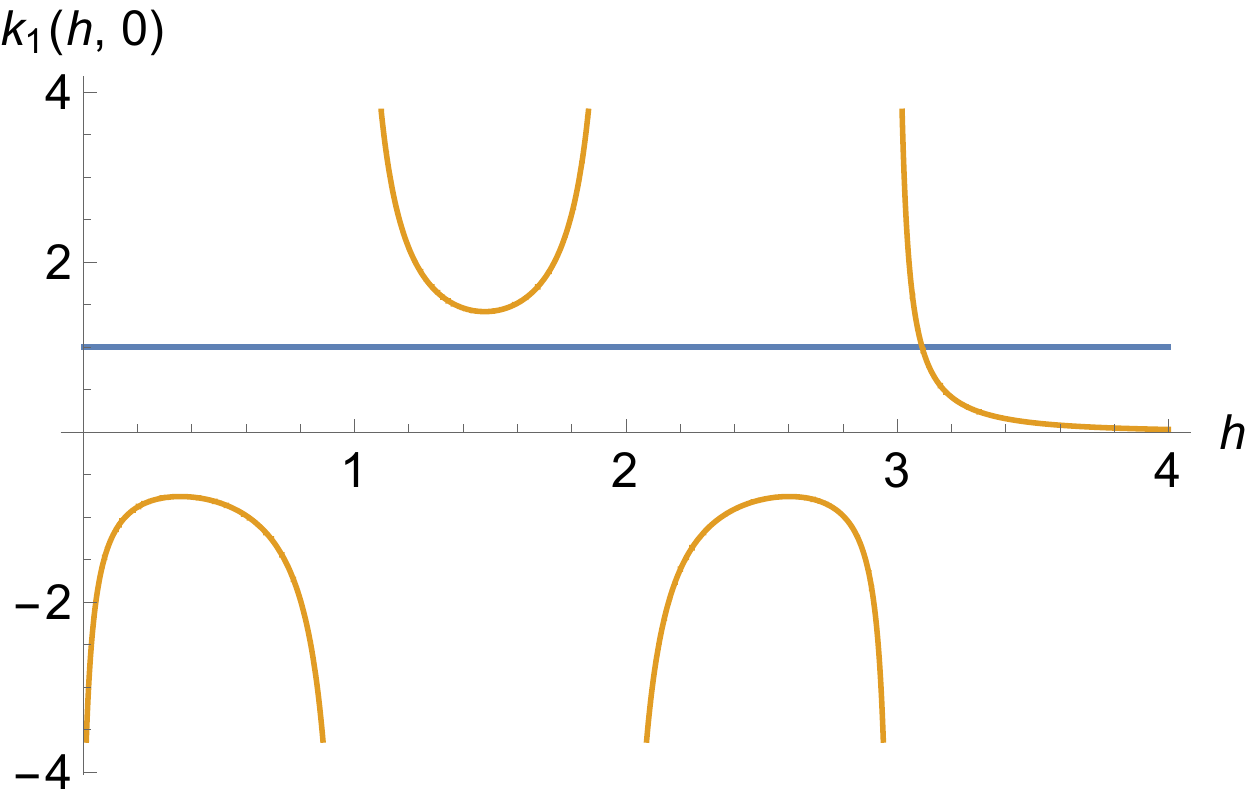}}
\caption{Plots of $k_1(h,0)$ at $d=3-\epsilon$ for, from left to right, $\epsilon=0.01$, $\epsilon=0.02819$, and $\epsilon=0.04$. On the left panel, the intersections with the blue line  correspond to $h_0$, $h_q$ and $h_1$. On the middle panel, $h_0$ and $h_q$ have merged, and on the right panel only $h_1$ remains.}
\label{fig:complexsol}
\end{figure}

As $\epsilon>0$, all the solutions we found are real. If we send $\epsilon$ to zero, we recover the classical dimensions $h^{classical}_n=1+2n$ of the bilinear operators $\phi_{abc}(\partial^2)^n\phi^{abc}$, except for $h_{q}$ corresponding to a quartic operator. 
However, this is only true for $\epsilon$ small enough. As $\epsilon$ increases, the two solutions $h_{0}$ and $h_q$ merge and become complex, see figure~\ref{fig:complexsol}. This happens around $\epsilon=0.02819$. Again, the same phenomenon appeared in \cite{Giombi:2017dtl}.

\paragraph{Higher spins.}
We can also compute the spectrum of bilinears at higher spin. Using \cite{Gurau:2019qag}, $k_1$ becomes:
\begin{equation} \label{eq:f1J-zeta1}
k_1(h,J)=-5\times \f{\Gamma(\f{5d}6)\Gamma(\f{d}3)\Gamma(-\f{d}3+\f{h}2+J/2)\Gamma(\f{d}6-\f{h}2+J/2)}{\Gamma(-\f{d}3)\Gamma(\f{d}6)\Gamma(\f{5d}6-\f{h}2+J/2)\Gamma(\f{d}3 + \f{h}2+J/2)}.
\end{equation}
We find the following solutions for $k_1(h,J)=1$:
\begin{align}
\label{eq:h0J-zeta1}
h_{0,J}&=1+J-\frac{4J^2+29}{3(4J^2-1)}\epsilon + \mathcal{O}(\epsilon^2), \\
h_{1,J}&=3+J+\frac{-4J^2-8J+27}{3(2J+3)(2J+1)}\epsilon + \mathcal{O}(\epsilon^2), \\
h_{n,J}&=1+2n+J-\frac{\epsilon}{3}+\frac{5\epsilon^2}{3n(n-1)(n+1/2+J)(n-1/2+J)} + \mathcal{O}(\epsilon^3)~~, \quad n>1.
\end{align}
Notice that these can all be written in the form $h_{n,J}=d-2+2n+J+2z_{n,J}$, with $d=3-\epsilon$.

For $J=0$, we recover the solutions we found in the beginning of this section, except for $h_q$. This is due to the fact that the factor $\Gamma(-\f{d}3+\f{h}2+J/2)$ in \eqref{eq:f1J-zeta1} only leads to a singularity for $h>0$ if $J=0$. Therefore, for $J> 0$, we only have dimensions corresponding to bilinear operators and no longer have a dimension corresponding to a quartic operator. 

One can check at leading order in $\epsilon$ from \eqref{eq:h0J-zeta1}, or to all orders directly from \eqref{eq:f1J-zeta1}, that the spin-2 operator with $n=0$ has the classical dimension $h_{0,2}=3-\epsilon=d$, as expected from a conserved energy-momentum tensor.

\subsection{Long-range propagator}

The computation of the spectrum of bilinears of the long-range model with the modified propagator goes exactly along the same lines as the one with the normal propagator. 
The only difference lies in the structure of the two-point function. The position space expression for the renormalized propagator (or two-point function) is:
\begin{equation}
	G(x) = \f{F_{d/3}}{(x^2)^{d/6}},\hspace{.2cm}F_{d/3} =\cZ\f{2^{d/3}}{(4\pi)^{d/2}}\f{\Gamma(\f{d}6)}{\Gamma(\f{d}3)},
\end{equation}
where $\cZ$ is the solution of \eqref{Z-norm-d/3}. \\
Once again we solve the same eigenvalue \eqref{BS} using the same kernel \eqref{kernel}. The resulting eigenvalue, for $J=0$, is:
\begin{align}
	k_{d/3}(h,0) &= \f54\, \l_1^2\,F_{d/3}^6\, \pi^d\,\f{\Gamma(\f{d}3)^2\Gamma(-\f{d}3+\f{h}2)\Gamma(\f{d}6-\f{h}2)}{\Gamma(\f{d}6)^2\Gamma(\f{5d}6-\f{h}2)\Gamma(\f{d}3 + \f{h}2)}\crcr
	&=\f54\, \l_1^2\,\cZ^6\f1{(4\pi)^{2d}}\bigg(\f{\Gamma(\f{d}6)}{\Gamma(\f{d}3)}\bigg)^4 \,\f{\Gamma(-\f{d}3+\f{h}2)\Gamma(\f{d}6-\f{h}2)}{\Gamma(\f{5d}6-\f{h}2)\Gamma(\f{d}3 + \f{h}2)}\crcr
	&= \f54 \bar{g}_1^2\bigg(\f{\Gamma(\f{d}6)}{\Gamma(\f{d}3)}\bigg)^4 \,\f{\Gamma(-\f{d}3+\f{h}2)\Gamma(\f{d}6-\f{h}2)}{\Gamma(\f{5d}6-\f{h}2)\Gamma(\f{d}3 + \f{h}2)} \,,
	\label{eigenvalue-2}
\end{align}
where in the last line we used the renormalized coupling defined in section~\ref{sec:betas1}, namely  $\bar{g}_1 = \f1{(4\pi)^d}\l_1\,\cZ^3$.

In order to find the OPE spectrum we have to solve for $k_{d/3}(h,0)=1$. The main difference with respect to the previous case is that the spectrum will now depend on the value of the exactly marginal coupling, which will replace $\epsilon$ in the role of small parameter. 

Again we use the method of section~\ref{sec:unitarity} to solve $k_{d/3}(h,0)=1$, and we find the following solutions:
\begin{align}
h_0&=\frac{d}{3}+\frac{15\Gamma(1-d/6)}{d\Gamma(2d/3)\Gamma(d/2)}\left(\frac{\Gamma(d/6)}{\Gamma(d/3)}\right)^4\bar{g}_1^2 + \mathcal{O}(\bar{g}_1^4)\,, \\
h_n&=\frac{d}{3}+2n+\frac{(-1)^{n+1}}{n!}\frac{5\Gamma(n-d/6)}{2\Gamma(2d/3-n)\Gamma(d/2+n)}\left(\frac{\Gamma(d/6)}{\Gamma(d/3)}\right)^4\bar{g}_1^2+ \mathcal{O}(\bar{g}_1^4)\,.
\end{align}
Notice that at $\bar{g}_1=0$, we recover the classical dimensions $h^{classical}_n=d/3+2n$. At $\bar{g}_1 \neq 0$, all dimensions are real, and they are greater than $d/3$ for $\bar{g}_1^2>0$ and small. 

As before, there is also a solution corresponding to a quartic operator:
\begin{equation}
h_{q}=\frac{2d}{3}-\frac{15\Gamma(1-d/6)}{d\Gamma(2d/3)\Gamma(d/2)}\left(\frac{\Gamma(d/6)}{\Gamma(d/3)}\right)^4\bar{g}_1^2 + \mathcal{O}(\bar{g}_1^4)\,.
\end{equation}

\begin{figure}[htbp]
\centering
\includegraphics[scale = 0.5]{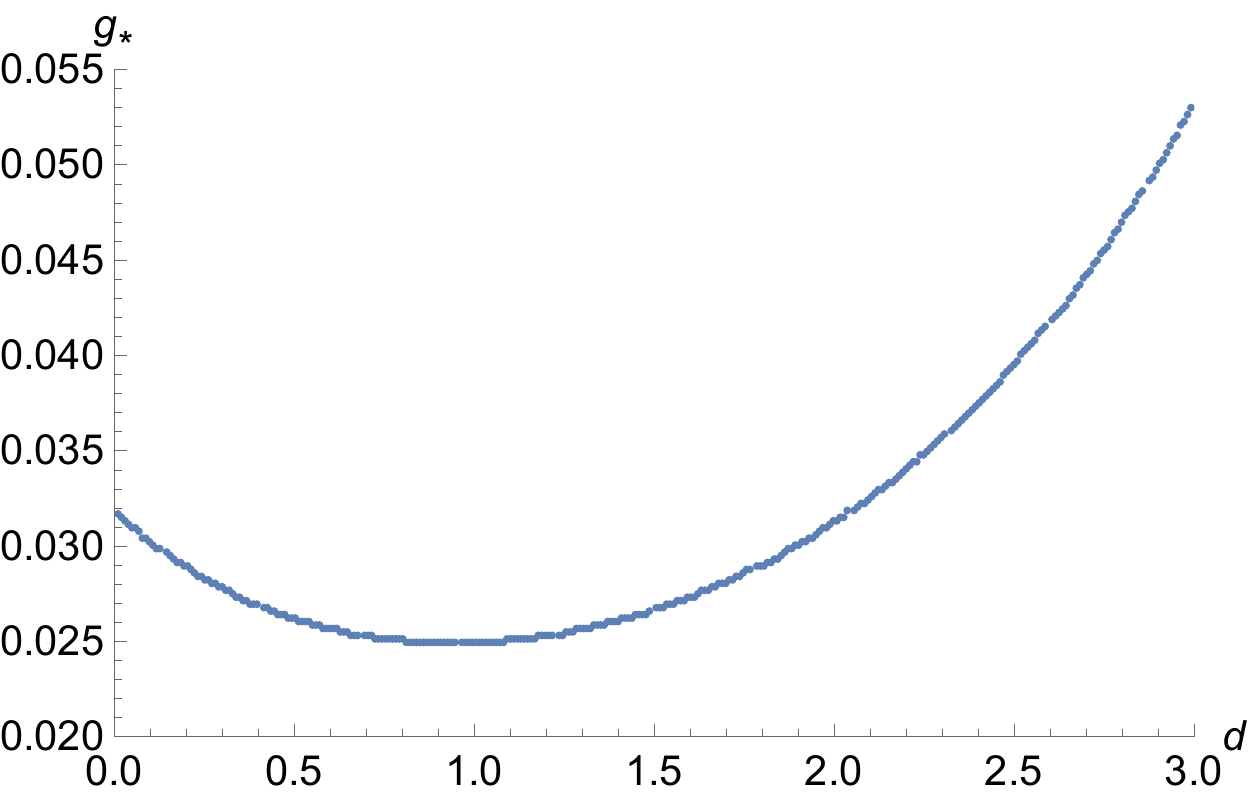}
\hspace{1cm}
\includegraphics[scale = 0.5]{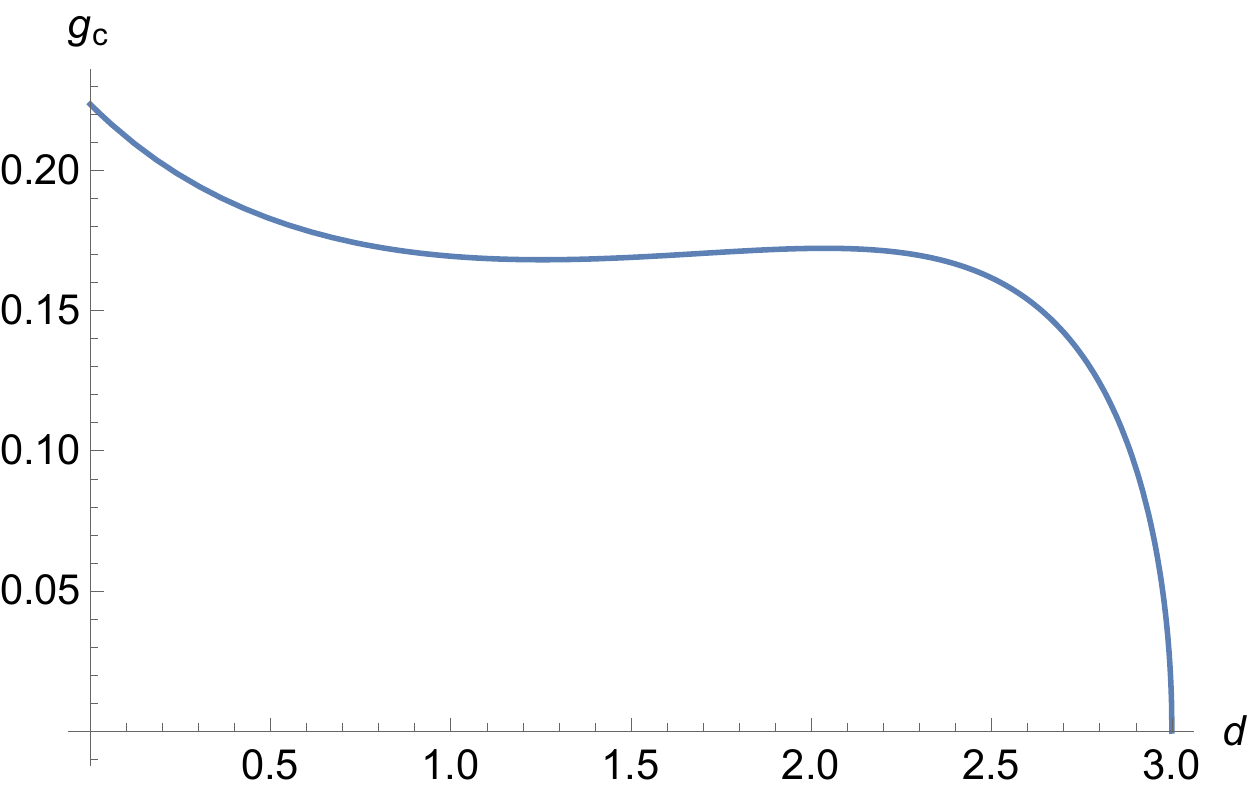}
\caption{Plots of $g_\star$ and $g_c$ as functions of $d$. The two curves cross at $d\simeq 2.97$.}
\label{fig:g_star}
\end{figure}

The plots of $k_{d/3}(h,0)$ are qualitatively similar to those in figure~\ref{fig:complexsol}, and we find the appearance of a pair of complex solutions for $g_1> g_\star >0$. For $d=2$, we have $g_\star \simeq 0.0313$, which is smaller than the value $g_c$ defined in footnote~\ref{foot:g_c}, at which the relation between bare $\l_1$ and renormalized $g_1$ becomes non-invertible, and which for $d=2$ is $g_c\simeq 0.1722$. A similar situation is found for any $d\lesssim 2.97$, while for $d\gtrsim 2.97$ we find $g_c<g_\star$. Comparative plots of $g_\star$ and $g_c$ as functions of $d$ are shown in figure~\ref{fig:g_star}. Therefore, for $d\lesssim 2.97$, the scenario differs from the one encountered in chapter \ref{chap:CTKT}, as in the present case the complex transition lies within the regime of validity of the fixed point solution. Furthermore, at the transition where the first two solutions of $k_{d/3}(h,0)=1$ merge (and then become complex) their value is $d/2$, within numerical precision. Such transition thus seems to be compatible with the scenario advanced in \cite{Kim:2019upg}, where the appearance of complex dimensions of the form $d/2+\im f$ for a given operator has been conjectured to be a signal that such operator acquires a non-zero vacuum expectation value.

\paragraph{Higher spins.}

Again we can compute the spectrum of bilinears for spin $J>0$. The eigenvalue becomes:
\begin{equation}
k_{d/3}(h,J) = \f54 \bar{g}_1^2\bigg(\f{\Gamma(\f{d}6)}{\Gamma(\f{d}3)}\bigg)^4 \,\f{\Gamma(-\f{d}3+\f{h}2+J/2)\Gamma(\f{d}6-\f{h}2+J/2)}{\Gamma(\f{5d}6-\f{h}2+J/2)\Gamma(\f{d}3 + \f{h}2+J/2)}\,.
\end{equation}
We find the following solutions for $k_{d/3}(h,J)=1$:
\begin{align}
h_{0,J}&=\frac{d}{3}+J-\frac{5}{2}\frac{\Gamma(-d/6+J)}{\Gamma(2d/3)\Gamma(d/2+J)}\left(\frac{\Gamma(d/6)}{\Gamma(d/3)}\right)^4\bar{g}_1^2 + \mathcal{O}(\bar{g}_1^4)\,, \\
h_{n,J}&=\frac{d}{3}+2n+J+\frac{(-1)^{n+1}}{n!}\frac{5\Gamma(n-d/6+J)}{2\Gamma(2d/3-n)\Gamma(d/2+n+J)}\left(\frac{\Gamma(d/6)}{\Gamma(d/3)}\right)^4\bar{g}_1^2+ \mathcal{O}(\bar{g}_1^4)\,.
\end{align}

Again, when $J=0$ we recover the dimensions we computed in the beginning of this section, except for the one corresponding to a quartic operator.

However, differently from the $\zeta=1$ case, and as in section \ref{sec:unitarity}, we find no spin-two operator of dimension $d$. This is due to the fact that the energy momentum tensor is not a local operator.

\section{Next-to-leading order in rank $3$}
\label{sec:sexticNLO}
In order to compute $1/N$ corrections to the fixed points of the sextic model in rank $3$, we go back to the $O(\mathcal{N})$ invariant action of \eqref{eq:freemulti} and consider a generic sextic multi-scalar models both in short and long-range. This will be very similar to the computations we did in the quartic case in chapters \ref{chap:3loops} and \ref{chap:trif}.

\subsection{The short-range sextic multi-scalar model}
\label{sec:sexticMS}

\subsubsection{Action}

The short-range multi-scalar model with sextic interactions and complex fields in dimension $d$ is defined by the action:

\begin{equation}
		S[\phi]  \, = \, \int d^dx \, \bigg[ \frac{1}{2} \partial_{\mu} \bar{\phi}_\mba(x) \partial_{\mu} \phi_{\mba}(x)
		\, + \, \frac{1}{(3!)^2} \, \lambda_{\mba \mbb \mbc ; \mbd \mbe \mbf } \phi_{\mba}(x)\phi_{\mbb}(x)\phi_{\mbc}(x) \bar{\phi}_{\mbd}(x)  \bar{\phi}_{\mbe}(x)  \bar{\phi}_{\mbf}(x) \bigg] \, ,
	\end{equation}
where the indices take values from 1 to $\cN$, and a summation over repeated indices is implicit. The coupling $\lambda_{\mba \mbb \mbc ; \mbd \mbe \mbf }$ is fully symmetric into the first three indices (corresponding to fields $\phi$) and the last three indices (corresponding to fields $\bar{\phi}$). It thus amounts in general to $2\binom{\cN+2}{3}=\frac{\cN(\cN+1)(\cN+2)}{3}$ couplings.

We work with the same renormalization scheme as before. As we showed in the first part, we can renormalize the mass terms to zero and there are no divergences in the four-point kernel. We can thus take the four-point couplings to be zero from the beginning.  

\subsubsection{Wave function renormalization}

We introduce the wave function renormalization by rescaling the bare field $\phi_{\mba}$ as $\phi_{\mba}=\left(\sqrt{Z}\right)_{\mba \mbb}\phi^R_{\mbb}$ with $\phi^R_{\mbb}$ the renormalized field. The computation is the same as in section \ref{sec:SDeq}. We just have to keep track of the indices. 

%
%

At quadratic order in the coupling constant, we obtain:
\begin{equation}
Z_{\mba \mbb}=\delta_{\mba \mbb}+\frac{\lambda_{\mba \mbc \mbd ;\mbe \mbf \mbg}\lambda_{\mbe \mbf \mbg ; \mbc \mbd \mbb}}{24}\tilde{M}_1(\mu) =\delta_{\mba \mbb}-\mu^{-4\epsilon}\frac{\lambda_{\mba \mbc \mbd ;\mbe \mbf \mbg}\lambda_{\mbe \mbf \mbg ;\mbc \mbd \mbb}\pi^2}{36\epsilon(4\pi)^{6}}\,,
\label{eq:wavef3NLO}
\end{equation}
with $\tilde{M}_1(\mu)=\frac{d}{dp^2}M_1(p)|_{p^2=\mu^2}$ and $M_1$ given in \eqref{eq:melon1}.

\subsubsection{Beta functions}

We define the renormalized sextic coupling $g_{\mba \mbb \mbc ; \mbd \mbe \mbf}$ in terms of the bare expansion of the six-point function by the following renormalization condition:

\begin{equation}
g_{\mba \mbb \mbc ;\mbd \mbe \mbf}=\mu^{-2\epsilon}\Gamma^{(6)}_{\mbg \mbh \mbj ;\mbk \mbp \mbq}(0,\dots,0)\left(\sqrt{Z}\right)_{\mbg \mba}\left(\sqrt{Z}\right)_{\mbh \mbb}\left(\sqrt{Z}\right)_{\mbj \mbc}\left(\sqrt{Z}\right)_{\mbk \mbd} \left(\sqrt{Z}\right)_{\mbp \mbe}\left(\sqrt{Z}\right)_{\mbq \mbf} \, ,
\end{equation}
where $\Gamma^{(6)}_{\mbg \mbh \mbj ;\mbk \mbp \mbq}(0,\dots,0)$ is the one-particle irreducible six-point function at zero external momentum. We compute it up to order three in the coupling constant (four-loops) using the bare expansion in terms of connected amputated one particle irreducible Feynman graphs. Similarly as what we did for quartic models, we write the amplitude of the latter in Schwinger parametrization as:

\be\label{eq:amp_final_sextic}
\mathcal{A}(\mathcal{G} ) =  \mu^{ (2d-6\zeta)(V-1)} \; \mathcal{\hat{A}}(\mathcal{G}) \,, \quad
\mathcal{\hat{A}}(\mathcal{G}) =
 \frac{1} { 
  \big[ (4\pi)^{d/2} \Gamma(\zeta)^3 \big]^{V-1} }
\int_0^{\infty}
\prod_{e \in \mathcal{G}} d a_e
\;\;
\frac{\prod_{e \in \mathcal{G}} a_e^{\zeta-1} \; e^{-\sum_{e \in \mathcal{G}} a_e}}
{\big(\sum_{\cT \in \mathcal{G}  } \prod_{e \notin \cT } a_e\big)^{d/2} } \,,
\ee
where $V$ denote the numbers of vertices of $\mathcal{G}$, $e \in \mathcal{G}$ runs over the edges of $\mathcal{G}$, and $\cT$ denotes the spanning trees in $\mathcal{G}$ (e.g.\ \cite{Rivasseau:1991ub}). Note that we used the fact that we only deal with six-point graphs with sextic vertices, as these are sufficient to describe the divergent graphs described above.

\begin{figure}[htbp]
\centering
\captionsetup[subfigure]{labelformat=empty}
\subfloat[$D_{\zeta}$]{\includegraphics[scale=1]{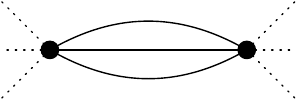}}
\hspace{1cm}
\subfloat[$D_{\zeta}^2$]{\includegraphics[scale=1]{D2.pdf}}
\\
\subfloat[$S_{\zeta}$]{\includegraphics[scale=1]{S.pdf}}
\hspace{1cm}
\subfloat[$I_{\zeta}$]{\includegraphics[scale=1]{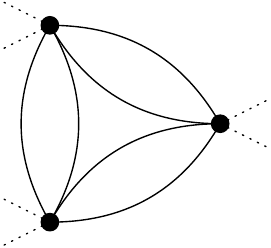}}
\hspace{1cm}
\subfloat[$J_{\zeta}$]{\includegraphics[scale=1]{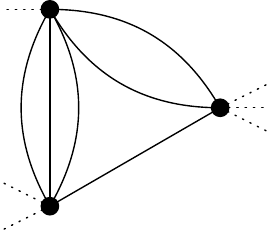}}
\caption{The five graphs contributing to the bare expansion up to order $3$ in the coupling constant. For the short-range case, $\zeta=1$.}
\label{fig:graphs}
\end{figure}

There are five graphs contributing to the bare expansion up to order $3$ in the coupling constant. They are represented in figure~\ref{fig:graphs} and we call $D_{\zeta},S_{\zeta},I_{\zeta},J_{\zeta}$ the amplitudes $\mathcal{\hat{A}}(\mathcal{G})$ of these diagrams. The two graphs, $D_{\zeta}$ and $S_{\zeta}$ were already contributing at leading order while the other two graphs are new. For the second diagram in figure~\ref{fig:graphs}, we use the fact that the amplitude of a one-vertex reducible diagram factors into the product of the amplitudes of its one-vertex irreducible parts.

We should also be careful to conserve the permutation symmetry of the six-point function in its indices. Therefore, we should completely symmetrize over external white and black vertices. However, due to specific symmetries of the diagrams under relabeling, certain terms are equal. Grouping them together and setting $\zeta=1$, we get:

\begin{align}
&\Gamma^{(6)}_{\mba \mbb \mbc ;\mbd \mbe \mbf}(p_1,\dots,p_6)=\lambda_{\mba \mbb \mbc ;\mbd \mbe \mbf}-\frac{\mu^{-2\epsilon}}{6}\left[ 3\left(\lambda_{\mba \mbb \mbg ;\mbh \mbj \mbd}\lambda_{\mbc \mbh \mbj ;\mbe \mbf \mbg}+ 8\text{ terms}\right) +\lambda_{\mba \mbb \mbc ;\mbg \mbh \mbj}\lambda_{\mbg \mbh \mbj ;\mbd \mbe \mbf}\right]D_1 \crcr
& +\frac{\mu^{-4\epsilon}}{12}\left[3\left(\lambda_{\mba \mbg \mbh ; \mbj \mbk \mbl}\lambda_{\mbb \mbj \mbk ;\mbg \mbh \mbm}\lambda_{\mbl \mbm \mbc ;\mbd \mbe \mbf} + 2\text{ terms}\right)+ 3\left(\lambda_{ \mbj \mbk \mbl ; \mbd \mbg \mbh}\lambda_{ \mbg \mbh \mbm ;\mbe \mbj \mbk}\lambda_{ \mba \mbb \mbc ;\mbf \mbl \mbm} + 2\text{ terms}\right) \right.\crcr
&\left.  \quad + 3\left(\lambda_{\mba \mbg \mbh ;\mbj \mbk \mbl}\lambda_{ \mbj \mbk \mbm ;\mbg \mbh \mbd}\lambda_{\mbl \mbb \mbc ;\mbe \mbf \mbm} + 8\text{ terms}\right)+2\left(\lambda_{\mba \mbg \mbh ;\mbj \mbk \mbl }\lambda_{ \mbj \mbk \mbl ;\mbh \mbm \mbd}\lambda_{\mbm \mbb \mbc ;\mbe \mbf \mbg} + 8\text{ terms}\right)\right]S_1 \crcr
& +\frac{\mu^{-4\epsilon}}{36}\left[ 9\left(\lambda_{\mba \mbg \mbh ;\mbj \mbe \mbf}\lambda_{\mbj \mbk \mbl ;\mbm \mbg \mbh}\lambda_{\mbm \mbb \mbc ;\mbk \mbl \mbd}+ 8\text{ terms}\right) +\lambda_{\mba \mbb \mbc ;\mbg \mbh \mbj}\lambda_{\mbg \mbh \mbj ;\mbk \mbl \mbm}\lambda_{\mbk \mbl \mbm ;\mbd \mbe \mbf}\right]D_1^2 \crcr
& +\frac{\mu^{-4\epsilon}}{4}\left[\left(\lambda_{\mba \mbg \mbh ; \mbd \mbj \mbk}\lambda_{\mbb \mbc \mbl ;\mbg  \mbh \mbm}\lambda_{\mbj  \mbk \mbm ; \mbl \mbe \mbf} + 8 \text{ terms}\right)+ 2\left(\lambda_{\mba \mbg  \mbh ;\mbd  \mbj \mbk }\lambda_{\mbb \mbc \mbj; \mbg \mbl \mbm}\lambda_{\mbk  \mbl \mbm ;\mbh \mbe \mbf} + 8 \text{ terms}\right) \right. \crcr
& \left. \quad +\left(\lambda_{\mba \mbg \mbh ;\mbd  \mbj \mbk}\lambda_{\mbb \mbl \mbm ;\mbe \mbg \mbh}\lambda_{\mbc  \mbj \mbk ;\mbf \mbl \mbm} + 5 \text{ terms}\right)+4\left(\lambda_{\mba \mbg \mbh ; \mbd  \mbj \mbk}\lambda_{\mbb \mbj \mbl ;\mbe \mbg \mbm}\lambda_{\mbc  \mbk \mbm ;\mbf \mbh \mbl} + 5 \text{ terms}\right)\right]I_1 \crcr
& +\frac{\mu^{-4\epsilon}}{12}\left[\left(\lambda_{\mba \mbg \mbh ;\mbj \mbe \mbf }\lambda_{\mbb \mbc \mbj ;\mbk \mbl \mbm }\lambda_{\mbk \mbl \mbm ; \mbg \mbh   \mbd} +  \text{ 8 terms}\right) + 6\left(\lambda_{\mba \mbg \mbh ;\mbj \mbe \mbf }\lambda_{\mbb \mbc \mbk ;\mbg \mbl \mbm }\lambda_{\mbl \mbm\mbj ;\mbh \mbk  \mbd} +  \text{ 8 terms}\right)\right. \crcr
& \quad+ \left(\lambda_{\mbj \mbb \mbc ;\mbd \mbg \mbh }\lambda_{\mbk \mbl \mbm ;\mbe \mbf \mbj  }\lambda_{\mba \mbg \mbh ;\mbk \mbl \mbm} +  \text{ 8 terms}\right) + 6\left(\lambda_{\mbj \mbb \mbc ;\mbd  \mbg \mbh }\lambda_{\mbg \mbl \mbm ;\mbk \mbe \mbf }\lambda_{\mba \mbh \mbk ;\mbl  \mbm \mbj} +  \text{ 8 terms}\right)\crcr
& \quad +3\left(\lambda_{\mba \mbk \mbl ;\mbm \mbe \mbd}\lambda_{\mbb \mbj \mbm  ;\mbf \mbg  \mbh }\lambda_{\mbc \mbg \mbh ; \mbj  \mbk \mbl} +  \text{ 17 terms}\right)+6\left(\lambda_{\mba \mbl \mbm ;\mbh \mbd \mbe }\lambda_{\mbb  \mbj \mbk ;\mbg\mbm  \mbf }\lambda_{\mbc \mbg \mbh ;\mbj  \mbk \mbl} +  \text{ 17 terms}\right)\crcr
& \quad +3\left(\lambda_{ \mbm \mba \mbb ;\mbd \mbk \mbl}\lambda_{\mbc   \mbg \mbh ;\mbe \mbj  \mbm}\lambda_{\mbj \mbk \mbl  ;\mbg \mbh   \mbf} +  \text{ 17 terms}\right)+6\left(\lambda_{ \mba \mbb  \mbh ;\mbd \mbl\mbm}\lambda_{\mbc   \mbg \mbm ;\mbe \mbj   \mbk}\lambda_{\mbj \mbk \mbl ;\mbg \mbh   \mbf} +  \text{ 17 terms}\right) \crcr
& \quad \left. +3\left(\lambda_{\mba \mbb \mbc ;\mbg \mbh \mbj}\lambda_{\mbj \mbl \mbm ;\mbk \mbd \mbe}\lambda_{\mbg \mbh \mbk ;\mbl \mbm \mbf} + \text{ 2 terms}\right)+ 3\left(\lambda_{\mbg \mbh \mbj ;\mbd \mbe \mbf}\lambda_{ \mba \mbb \mbk ;\mbj \mbl \mbm}\lambda_{ \mbl \mbm \mbc ;\mbg \mbh \mbk }+ \text{ 2 terms}\right)\right]J_1 \, ,
\label{eq:bare_series}
\end{align}
where the notation $+ \dots$ terms corresponds to a sum over terms obtained by permuting external indices into non-equivalent ways. For example, the nine terms in the first line correspond to the choice of the white index $\mba, \mbb$ or $\mbc$ on the second coupling and  the choice of the black index $\mbd,\mbe$ or $\mbf$ on the first coupling.
The integrals $D_{\zeta},S_{\zeta},I_{\zeta},J_{\zeta}$ are computed in appendix~\ref{ap:betafun4} both for the short-range case $\zeta=1$ and the long-range case $0<\zeta<1$.


We rescale the couplings by $\tilde{g}_{\mba \mbb \mbc ;\mbd \mbe \mbf}=(4\pi)^{-d}g_{\mba \mbb \mbc ;\mbd \mbe \mbf}$ and we define:
\begin{align}
\alpha_{D_1}&=\epsilon(4\pi)^d\frac{D_1}{3}=\frac{2\pi}{3} \; ,  \; \alpha_{S_1}=-\epsilon(4\pi)^{2d} \frac{S_1}{3}= \frac{2\pi^2}{3} \; ,  \; \alpha_{I_1}=-\epsilon(4\pi)^{2d}I_1= -\pi^4 \crcr
\alpha_{J_1}&=\epsilon(4\pi)^{2d}\frac{(D_1^2-2J_1)}{6}= -\frac{4\pi^2}{3}  \; , \; \alpha_{M_1}=-\epsilon(4\pi)^{2d}\frac{\tilde{M}_1}{12}=\frac{\pi^2}{18} \, .
\end{align}

We finally obtain the following beta function:
\begin{align}
&\beta_{\mba \mbb \mbc ;\mbd \mbe \mbf}=-2\epsilon \tilde{g}_{\mba \mbb \mbc ;\mbd \mbe \mbf}+\alpha_{D_1}\left[ 3\left(\tilde{g}_{\mba \mbb \mbg ;\mbh \mbj \mbd}\tilde{g}_{\mbc \mbh \mbj ;\mbe \mbf \mbg}+ 8\text{ terms}\right) +\tilde{g}_{\mba \mbb \mbc ;\mbg \mbh \mbj}\tilde{g}_{\mbg \mbh \mbj ;\mbd \mbe \mbf}\right] \crcr
&\quad +\alpha_{S_1}\left[3\left(\tilde{g}_{\mba \mbg \mbh ; \mbj \mbk \mbl}\tilde{g}_{\mbb \mbj \mbk ;\mbg \mbh \mbm}\tilde{g}_{\mbl \mbm \mbc ;\mbd \mbe \mbf} + 2\text{ terms}\right)+ 3\left(\tilde{g}_{ \mbj \mbk \mbl ; \mbd \mbg \mbh}\tilde{g}_{ \mbg \mbh \mbm ;\mbe \mbj \mbk}\tilde{g}_{ \mba \mbb \mbc ;\mbf \mbl \mbm} + 2\text{ terms}\right) \right.\crcr
&\quad\left.  \qquad + 3\left(\tilde{g}_{\mba \mbg \mbh ;\mbj \mbk \mbl}\tilde{g}_{ \mbj \mbk \mbm ;\mbg \mbh \mbd}\tilde{g}_{\mbl \mbb \mbc ;\mbe \mbf \mbm} + 8\text{ terms}\right)+2\left(\tilde{g}_{\mba \mbg \mbh ;\mbj \mbk \mbl }\tilde{g}_{ \mbj \mbk \mbl ;\mbh \mbm \mbd}\tilde{g}_{\mbm \mbb \mbc ;\mbe \mbf \mbg} + 8\text{ terms}\right)\right] \crcr
&\quad +\alpha_{I_1}\left[\left(\tilde{g}_{\mba \mbg \mbh ; \mbd \mbj \mbk}\tilde{g}_{\mbb \mbc \mbl ;\mbg  \mbh \mbm}\tilde{g}_{\mbj  \mbk \mbm ; \mbl \mbe \mbf} + 8 \text{ terms}\right)+ 2\left(\tilde{g}_{\mba \mbg  \mbh ;\mbd  \mbj \mbk }\tilde{g}_{\mbb \mbc \mbj; \mbg \mbl \mbm}\tilde{g}_{\mbk  \mbl \mbm ;\mbh \mbe \mbf} + 8 \text{ terms}\right) \right. \crcr
&\quad \left. \qquad +\left(\tilde{g}_{\mba \mbg \mbh ;\mbd  \mbj \mbk}\tilde{g}_{\mbb \mbl \mbm ;\mbe \mbg \mbh}\tilde{g}_{\mbc  \mbj \mbk ;\mbf \mbl \mbm} + 5 \text{ terms}\right)+4\left(\tilde{g}_{\mba \mbg \mbh ; \mbd  \mbj \mbk}\tilde{g}_{\mbb \mbj \mbl ;\mbe \mbg \mbm}\tilde{g}_{\mbc  \mbk \mbm ;\mbf \mbh \mbl} + 5 \text{ terms}\right)\right]\crcr
&\quad +\alpha_{J_1}\left[\left(\tilde{g}_{\mba \mbg \mbh ;\mbj \mbe \mbf }\tilde{g}_{\mbb \mbc \mbj ;\mbk \mbl \mbm }\tilde{g}_{\mbk \mbl \mbm ; \mbg \mbh   \mbd} +  \text{ 8 terms}\right) + 6\left(\tilde{g}_{\mba \mbg \mbh ;\mbj \mbe \mbf }\tilde{g}_{\mbb \mbc \mbk ;\mbg \mbl \mbm }\tilde{g}_{\mbl \mbm\mbj ;\mbh \mbk  \mbd} +  \text{ 8 terms}\right)\right. \crcr
&\quad \qquad +\left(\tilde{g}_{\mbj \mbb \mbc ;\mbd \mbg \mbh }\tilde{g}_{\mbk \mbl \mbm ;\mbe \mbf \mbj  }\tilde{g}_{\mba \mbg \mbh ;\mbk \mbl \mbm} +  \text{ 8 terms}\right) + 6\left(\tilde{g}_{\mbj \mbb \mbc ;\mbd  \mbg \mbh }\tilde{g}_{\mbg \mbl \mbm ;\mbk \mbe \mbf }\tilde{g}_{\mba \mbh \mbk ;\mbl  \mbm \mbj} +  \text{ 8 terms}\right)\crcr
& \quad \qquad +3\left(\tilde{g}_{\mba \mbk \mbl ;\mbm \mbe \mbd}\tilde{g}_{\mbb \mbj \mbm  ;\mbf \mbg  \mbh }\tilde{g}_{\mbc \mbg \mbh ; \mbj  \mbk \mbl} +  \text{ 17 terms}\right)+6\left(\tilde{g}_{\mba \mbl \mbm ;\mbh \mbd \mbe }\tilde{g}_{\mbb  \mbj \mbk ;\mbg\mbm  \mbf }\tilde{g}_{\mbc \mbg \mbh ;\mbj  \mbk \mbl} +  \text{ 17 terms}\right)\crcr
&\quad \qquad +3\left(\tilde{g}_{ \mbm \mba \mbb ;\mbd \mbk \mbl}\tilde{g}_{\mbc   \mbg \mbh ;\mbe \mbj  \mbm}\tilde{g}_{\mbj \mbk \mbl  ;\mbg \mbh   \mbf} +  \text{ 17 terms}\right)+6\left(\tilde{g}_{ \mba \mbb  \mbh ;\mbd \mbl\mbm}\tilde{g}_{\mbc   \mbg \mbm ;\mbe \mbj   \mbk}\tilde{g}_{\mbj \mbk \mbl ;\mbg \mbh   \mbf} +  \text{ 17 terms}\right) \crcr
&\quad \qquad \left. +3\left(\tilde{g}_{\mba \mbb \mbc ;\mbg \mbh \mbj}\tilde{g}_{\mbj \mbl \mbm ;\mbk \mbd \mbe}\tilde{g}_{\mbg \mbh \mbk ;\mbl \mbm \mbf} + \text{ 2 terms}\right)+ 3\left(\tilde{g}_{\mbg \mbh \mbj ;\mbd \mbe \mbf}\tilde{g}_{ \mba \mbb \mbk ;\mbj \mbl \mbm}\tilde{g}_{ \mbl \mbm \mbc ;\mbg \mbh \mbk }+ \text{ 2 terms}\right)\right]\crcr
&\quad +\alpha_{M_1}\left(\tilde{g}_{\mba \mbg \mbh ;\mbj \mbk \mbl}\tilde{g}_{\mbj \mbk \mbl ;\mbg \mbh \mbm}\tilde{g}_{\mbm \mbb \mbc; \mbd \mbe \mbf} + 5 \text{ terms}\right)\, .
\label{eq:beta_SR}
\end{align}

We also compute the field critical exponent defined by:
\begin{equation}
\eta_{\mba \mbb}=2\beta_{\mbk \mbc \mbd ;\mbe \mbf \mbg}\left(\frac{\partial Z^{1/2}}{\partial \tilde{g}_{\mbk \mbc \mbd ;\mbe \mbf \mbg}}Z^{-1/2}\right)_{\mba \mbb} \, .
\end{equation}
Using \eqref{eq:wavef3NLO} and \eqref{eq:beta_SR}, we obtain:
\begin{equation}
\eta_{\mba \mbb}=\frac{\pi^2}{9}\tilde{g}_{\mba \mbc \mbd ;\mbe \mbf \mbg}\tilde{g}_{\mbe \mbf \mbg; \mbc \mbd \mbb} \, .
\end{equation}

By imposing various symmetry restrictions on the interaction, one obtains different models. We study here the case with $U(N)^3$ invariance in order to obtain subleading corrections to the fixed point of the previous sections.

\subsubsection{Application: $U(N)^3$ symmetry}
\label{sec:tensor_SR}

In this subsection, we specify the symmetry to $U(N)^3$ with $\mathcal{N}=N^3$. Each index is now a triplet of indices going from $1$ to $N$. There are five different invariants, thus five couplings. We set:\footnote{The normalization was chosen so that the couplings are normalized by $1/6$ as usually done in sextic tensor models.}

\begin{align}
\tilde{g}_{\mba \mbb  \mbc ;\mbd \mbe\mbf}=&\tilde{g}_1 \left(\delta^{(1)}_{\mba \mbb  \mbc ;\mbd \mbe\mbf}+ 5 \text{ terms}\right)+\frac{1}{2}\tilde{g}_2\left( \delta^{(2)}_{\mba \mbd ; \mbb \mbe; \mbc\mbf} + 11 \text{ terms}\right) +\frac{1}{2}\tilde{g}_3 \left(\delta^{(3)}_{\mba \mbd ; \mbb \mbe; \mbc\mbf}+11 \text{ terms}\right)\crcr
& +\tilde{g}_4 \left(\delta^{(4)}_{\mba \mbd ; \mbb \mbe; \mbc\mbf}+ 5 \text{ terms}\right)+\tilde{g}_5 \left(\delta^{(5)}_{\mba \mbd ; \mbb \mbe; \mbc\mbf}+ 5 \text{ terms}\right)\, ,
\label{eq:invariants}
\end{align}
where the contractions are specified in appendix~\ref{ap:conventions}. The corresponding invariants are represented in \eqref{eq:int-action-graph}. 


We rescale the couplings as:

\begin{equation}
\tilde{g}_1=\frac{\bar{g}_1}{N^3} \; ,\; \tilde{g}_2=\frac{\bar{g}_2}{N^4}  \; , \; \tilde{g}_3=\frac{\bar{g}_3}{N^4}  \; , \; \tilde{g}_4=\frac{\bar{g}_4}{N^5} \; , \; \tilde{g}_5=\frac{\bar{g}_5}{N^6} \, .
\end{equation}

This rescaling ensures that the model admits a well-behaved large-$N$ expansion and the barred couplings are the same as in the first part of the chapter.

We then obtain the following beta functions up to order $N^{-1}$:

\begin{align}
\beta_1&=-2 \bar{g}_1\left(\epsilon-\bar{g}_1^2\pi^2\right)-\frac{24\pi^4}{N}\bar{g}_1^3 +\mathcal{O}(N^{-2}) \, ,\crcr
\beta_2&=-2 \bar{g}_2\left(\epsilon-\bar{g}_1^2\pi^2\right)+ 4\pi^2\bar{g}_1^2\left(\frac{9}{2\pi}+9\bar{g}_1+\bar{g}_2\right) \crcr
& +\frac{4}{N}\Bigg[\frac{\pi}{9}\left(81\bar{g}_1^2+36\bar{g}_1\bar{g}_2+\bar{g}_2^2+6\bar{g}_3(9\bar{g}_1+\bar{g}_2)\right)-2\pi^4\bar{g}_1^2\bar{g}_2 -\pi^2\bar{g}_1^2(63\bar{g}_1-2\bar{g}_2+9\bar{g}_3)\Bigg] +\mathcal{O}(N^{-2})\, , \crcr
\beta_3&= -2 \bar{g}_3\left(\epsilon-4\bar{g}_1^2\pi^2\right)+\frac{2}{N}\Bigg[\frac{\pi}{9}\left(36\bar{g}_1\bar{g}_2+2\bar{g}_2^2+9\bar{g}_3^2\right)  +4\pi^2\bar{g}_1^2\left(\bar{g}_2-18\bar{g}_1\right)\Bigg] +\mathcal{O}(N^{-2}) \, ,\crcr
\beta_4&= -2 \bar{g}_4\left(\epsilon-\bar{g}_1^2\pi^2\right)+2\pi^2\bar{g}_1^2\left(27\bar{g}_1+10\bar{g}_2+12\bar{g}_3+7\bar{g}_4\right) \crcr
& + \frac{1}{N}\Bigg[\frac{2\pi}{9}\left(2\bar{g}_2(5\bar{g}_2+12\bar{g}_3+4\bar{g}_4)+36\bar{g}_1(\bar{g}_2+3\bar{g_3}+2\bar{g}_4)+3\bar{g}_3(9\bar{g}_3+4\bar{g}_4)\right) \crcr
& \qquad -\frac{\pi^4}{81}\left(162\bar{g}_1^2(54\bar{g}_1+5\bar{g}_2)+3\bar{g}_3(648\bar{g}_1^2+2\bar{g}_2^2+12\bar{g}_2\bar{g}_3+9\bar{g}_3^2)+2\bar{g}_4(324\bar{g}_1^2+\bar{g}_2^2+18\bar{g}_3^2) \right)\crcr
& \qquad  - 4\pi^2\bar{g}_1^2(36\bar{g}_1+25\bar{g}_2+6\bar{g}_3+6\bar{g}_4) \Bigg] +\mathcal{O}(N^{-2})\, , \crcr
\beta_5&=-2 \bar{g}_5\left(\epsilon-\bar{g}_1^2\pi^2\right)+\pi^2\bar{g}_1^2\left(\frac{2}{\pi}+6\bar{g}_2+16\bar{g}_4+30\bar{g}_5\right) \crcr
& + \frac{1}{N}\Bigg[ -\frac{2\pi^4}{243}\left(3(31\bar{g}_3+7\bar{g}_4)\bar{g}_2^2+10\bar{g}_2^3 +9\bar{g}_3\bar{g}_4(3\bar{g}_3+2\bar{g}_4)+2\bar{g_4}^3+243\bar{g}_1^2(\bar{g}_2+3\bar{g}_3+\bar{g}_4) \right.\crcr
& \qquad \qquad \qquad \left. +6\bar{g}_2(9\bar{g}_3^2+12\bar{g}_3\bar{g}_4+2\bar{g}_4^2)\right) \crcr
& \qquad +\frac{4\pi}{9}\left(\bar{g}_2(\bar{g}_2+6\bar{g}_3+4\bar{g}_4)+2\bar{g}_4(3\bar{g}_3+\bar{g}_4)\right)-4\pi^2\bar{g}_1^2(3\bar{g}_2+3\bar{g}_3+4\bar{g}_4) \Bigg] +\mathcal{O}(N^{-2}) \, .
\end{align}

Similarly, we find for the field critical exponent:
\begin{equation}
\eta=\frac{2\pi^2\bar{g}_1^2}{3}+\mathcal{O}(N^{-2}) \, .
\end{equation}

If we try to solve naively these beta functions, we find non-perturbative fixed points that blow up when we send $\epsilon \rightarrow 0$. For example, $g_2^{\star}$ has the following form:
\begin{equation}
g_2^{\star}=a+b\sqrt{\epsilon}+\frac{c}{\epsilon N}+\mathcal{O}(N^{-2}) \, .
\end{equation}
with $a,b,c$ constants of order $1$. 

It is because here the behavior of the fixed point is governed by the combination $\epsilon N$. Indeed, the fixed points of the typical melonic large-$N$ limit are obtained for $1 \gg \frac{1}{\epsilon N}$ or $\epsilon N \gg 1$. As we wish to study the $1/N$ corrections to these fixed points we set:
\begin{equation}
\tilde{N}=\epsilon N \, ,
\end{equation}
and we expand first in $1/\tilde{N}$ and then in $\epsilon$. This is very similar to what happened in chapter \ref{chap:trif}. 

We parametrize the critical couplings as $\bar{g}_i=\bar{g}_{i,0}+\frac{\bar{g}_{i,1}}{\tilde{N}} +\mathcal{O}(\tilde{N}^{-2})$ for $i=1,\dots ,5$. Solving for the zeros of the beta functions at leading order, apart from the Gaussian fixed point, we find the following solutions:
\begin{gather}
\bar{g}^*_{1,0}=\pm\f{\sqrt{\eps}}{\pi}\, ; \qquad \bar{g}^*_{2,0}=\f{9}{2\pi}\left(-1\mp 2\sqrt{\eps}\right)\, ; \qquad \bar{g}^*_{3,0}=0\, ;
\crcr
\bar{g}^*_{4,0}=\f{9}{7\pi}\left(5 \pm 7\sqrt{\eps}\right)\, ;\qquad \bar{g}^*_{5,0} = \f{-109\mp 126\sqrt{\eps}}{42\pi} \, .
\label{eq:FP_LO}
\end{gather}

The signs for all five couplings are taken to be simultaneously either the upper or lower ones. We thus recover the two interacting large-$N$ fixed points of section~\ref{sec:betas}. 
%

Substituting \eqref{eq:FP_LO} into the order $\tilde{N}^{-1}$ of the beta functions, we find the following corrections to the fixed points:

\begin{gather}
\bar{g}^*_{1,1}= \pm 6 \pi \epsilon^{3/2} , \qquad \bar{g}^*_{2,1}= \frac{9}{4\pi}\left(-1\pm4\sqrt{\epsilon}\right)+\mathcal{O}(\epsilon) , \qquad \bar{g}^*_{3,1}=- \frac{3}{2\pi} +\mathcal{O}(\epsilon), 
\crcr
\bar{g}^*_{4,1}= \frac{9(68-\pi^2)}{98\pi}\mp \frac{9(392+15\pi^2)\sqrt{\epsilon}}{196\pi}+\mathcal{O}(\epsilon) , \\
\bar{g}^*_{5,1}= \frac{-18459+566\pi^2}{6860 \pi} \pm \frac{3(2940+193\pi^2)\sqrt{\epsilon}}{980 \pi}+\mathcal{O}(\epsilon) \, . \nonumber
\end{gather}
where the choice of sign is the same as for the leading order so that we still have two fixed points.

We then compute the critical exponents up to order $\tilde{N}^{-1}$. They are the eigenvalues of the stability matrix $\frac{\partial \beta_i}{\partial \tilde{g}_j}$ evaluated at the fixed points. We find for both fixed points:

\begin{equation}
\left(4\epsilon  \, , \, 4\epsilon-\frac{4\epsilon}{\tilde{N}}\pm\frac{8\epsilon^{3/2}}{\tilde{N}} \, , \,  6\epsilon  \, , \, 14\epsilon-\frac{(16+\pi^2)\epsilon}{2\tilde{N}}\mp\frac{2\pi^2}{\tilde{N}}\epsilon^{3/2} \, , \, 30 \epsilon   \right) \, ,
\end{equation}
where the choice of sign is the same as in \eqref{eq:FP_LO}. 

All critical exponents are real positive. Therefore, both fixed points are infrared stable. Moreover, contrary to the large-$N$ case, we now have five different eigenvalues: the stability matrix is diagonalizable at order $\tilde{N}^{-1}$. 

We can finally compute the field critical exponent at the fixed points. We find for both fixed points:

\begin{equation}
\eta(\bar{g}^{\star})=\frac{2\pi^2}{3}\left(\frac{\epsilon}{\pi^2}+\frac{12\epsilon^2}{N}\right)+\mathcal{O}(N^{-2}) \, .
\end{equation}

\subsection{The long-range sextic multi-scalar model}
\label{sec:MS_LR}

\subsubsection{Action}

The long-range multi-scalar model with sextic interactions and complex fields in dimension $d$ is defined by the action:

\begin{equation}
		S[\phi]  \, = \, \int d^dx \, \bigg[ \frac{1}{2}  \bar{\phi}_\mba(x)\left(- \partial^2\right)^{\zeta} \phi_{\mba}(x)
		\, + \, \frac{1}{(3!)^2} \, \lambda_{\mba \mbb \mbc; \mbd \mbe \mbf } \phi_{\mba}(x)\phi_{\mbb}(x)\phi_{\mbc}(x) \bar{\phi}_{\mbd}(x)  \bar{\phi}_{\mbe}(x)  \bar{\phi}_{\mbf}(x) \bigg] \, .
	\end{equation}
	
For this entire section, the dimension is now fixed to be smaller than three (but not necessarily close to three) and we use the same renormalization scheme as for the large-$N$ tensor field theory. Again, the key difference with the short-range case is that we now don't have any wave function renormalization. 

%
%


\subsubsection{Beta functions}

%
The running coupling is defined by:

\begin{equation}
g_{\mba \mbb \mbc; \mbd \mbe \mbf}=\mu^{-2\epsilon}\Gamma^{(6)}_{\mba \mbb \mbc ;\mbd \mbe \mbf}(0,\dots,0) \, ,
\end{equation}
with the following bare expansion:

\begin{align}
&\Gamma^{(6)}_{\mba \mbb \mbc; \mbd \mbe \mbf}(p_1,\dots,p_6)=\lambda_{\mba \mbb \mbc ;\mbd \mbe \mbf}-\frac{\mu^{-2\epsilon}}{6}\left[ 3\left(\lambda_{\mba \mbb \mbg ;\mbh \mbj \mbd}\lambda_{\mbc \mbh \mbj ;\mbe \mbf \mbg}+ 8\text{ terms}\right) +\lambda_{\mba \mbb \mbc ;\mbg \mbh \mbj}\lambda_{\mbg \mbh \mbj ;\mbd \mbe \mbf}\right]D_{d/3} \crcr
& +\frac{\mu^{-4\epsilon}}{12}\left[3\left(\lambda_{\mba \mbg \mbh ; \mbj \mbk \mbl}\lambda_{\mbb \mbj \mbk ;\mbg \mbh \mbm}\lambda_{\mbl \mbm \mbc ;\mbd \mbe \mbf} + 2\text{ terms}\right)+ 3\left(\lambda_{ \mbj \mbk \mbl ; \mbd \mbg \mbh}\lambda_{ \mbg \mbh \mbm ;\mbe \mbj \mbk}\lambda_{ \mba \mbb \mbc ;\mbf \mbl \mbm} + 2\text{ terms}\right) \right.\crcr
&\left.  \quad + 3\left(\lambda_{\mba \mbg \mbh ;\mbj \mbk \mbl}\lambda_{ \mbj \mbk \mbm ;\mbg \mbh \mbd}\lambda_{\mbl \mbb \mbc ;\mbe \mbf \mbm} + 8\text{ terms}\right)+2\left(\lambda_{\mba \mbg \mbh ;\mbj \mbk \mbl }\lambda_{ \mbj \mbk \mbl ;\mbh \mbm \mbd}\lambda_{\mbm \mbb \mbc ;\mbe \mbf \mbg} + 8\text{ terms}\right)\right]S_{d/3} \crcr
& +\frac{\mu^{-4\epsilon}}{36}\left[ 9\left(\lambda_{\mba \mbg \mbh ;\mbj \mbe \mbf}\lambda_{\mbj \mbk \mbl ;\mbm \mbg \mbh}\lambda_{\mbm \mbb \mbc ;\mbk \mbl \mbd}+ 8\text{ terms}\right) +\lambda_{\mba \mbb \mbc ;\mbg \mbh \mbj}\lambda_{\mbg \mbh \mbj ;\mbk \mbl \mbm}\lambda_{\mbk \mbl \mbm ;\mbd \mbe \mbf}\right]D_{d/3}^2 \crcr
& +\frac{\mu^{-4\epsilon}}{4}\left[\left(\lambda_{\mba \mbg \mbh ; \mbd \mbj \mbk}\lambda_{\mbb \mbc \mbl ;\mbg  \mbh \mbm}\lambda_{\mbj  \mbk \mbm ; \mbl \mbe \mbf} + 8 \text{ terms}\right)+ 2\left(\lambda_{\mba \mbg  \mbh ;\mbd  \mbj \mbk }\lambda_{\mbb \mbc \mbj; \mbg \mbl \mbm}\lambda_{\mbk  \mbl \mbm ;\mbh \mbe \mbf} + 8 \text{ terms}\right) \right. \crcr
& \left. \quad +\left(\lambda_{\mba \mbg \mbh ;\mbd  \mbj \mbk}\lambda_{\mbb \mbl \mbm ;\mbe \mbg \mbh}\lambda_{\mbc  \mbj \mbk ;\mbf \mbl \mbm} + 5 \text{ terms}\right)+4\left(\lambda_{\mba \mbg \mbh ; \mbd  \mbj \mbk}\lambda_{\mbb \mbj \mbl ;\mbe \mbg \mbm}\lambda_{\mbc  \mbk \mbm ;\mbf \mbh \mbl} + 5 \text{ terms}\right)\right]I_{d/3} \crcr
& +\frac{\mu^{-4\epsilon}}{12}\left[\left(\lambda_{\mba \mbg \mbh ;\mbj \mbe \mbf }\lambda_{\mbb \mbc \mbj ;\mbk \mbl \mbm }\lambda_{\mbk \mbl \mbm ; \mbg \mbh   \mbd} +  \text{ 8 terms}\right) + 6\left(\lambda_{\mba \mbg \mbh ;\mbj \mbe \mbf }\lambda_{\mbb \mbc \mbk ;\mbg \mbl \mbm }\lambda_{\mbl \mbm\mbj ;\mbh \mbk  \mbd} +  \text{ 8 terms}\right)\right. \crcr
& \quad + \left(\lambda_{\mbj \mbb \mbc ;\mbd \mbg \mbh }\lambda_{\mbk \mbl \mbm ;\mbe \mbf \mbj  }\lambda_{\mba \mbg \mbh ;\mbk \mbl \mbm} +  \text{ 8 terms}\right) + 6\left(\lambda_{\mbj \mbb \mbc ;\mbd  \mbg \mbh }\lambda_{\mbg \mbl \mbm ;\mbk \mbe \mbf }\lambda_{\mba \mbh \mbk ;\mbl  \mbm \mbj} +  \text{ 8 terms}\right)\crcr
& \quad +3\left(\lambda_{\mba \mbk \mbl ;\mbm \mbe \mbd}\lambda_{\mbb \mbj \mbm  ;\mbf \mbg  \mbh }\lambda_{\mbc \mbg \mbh ; \mbj  \mbk \mbl} +  \text{ 17 terms}\right)+6\left(\lambda_{\mba \mbl \mbm ;\mbh \mbd \mbe }\lambda_{\mbb  \mbj \mbk ;\mbg\mbm  \mbf }\lambda_{\mbc \mbg \mbh ;\mbj  \mbk \mbl} +  \text{ 17 terms}\right)\crcr
& \quad +3\left(\lambda_{ \mbm \mba \mbb ;\mbd \mbk \mbl}\lambda_{\mbc   \mbg \mbh ;\mbe \mbj  \mbm}\lambda_{\mbj \mbk \mbl  ;\mbg \mbh   \mbf} +  \text{ 17 terms}\right)+6\left(\lambda_{ \mba \mbb  \mbh ;\mbd \mbl\mbm}\lambda_{\mbc   \mbg \mbm ;\mbe \mbj   \mbk}\lambda_{\mbj \mbk \mbl ;\mbg \mbh   \mbf} +  \text{ 17 terms}\right) \crcr
& \quad \left. +3\left(\lambda_{\mba \mbb \mbc ;\mbg \mbh \mbj}\lambda_{\mbj \mbl \mbm ;\mbk \mbd \mbe}\lambda_{\mbg \mbh \mbk ;\mbl \mbm \mbf} + \text{ 2 terms}\right)+ 3\left(\lambda_{\mbg \mbh \mbj ;\mbd \mbe \mbf}\lambda_{ \mba \mbb \mbk ;\mbj \mbl \mbm}\lambda_{ \mbl \mbm \mbc ;\mbg \mbh \mbk }+ \text{ 2 terms}\right)\right]J_{d/3} .
\end{align}

We rescale the couplings by $\tilde{g}_{\mba \mbb \mbc ;\mbd \mbe \mbf}=(4\pi)^{-d}\Gamma(d/3)^{-3}g_{\mba \mbb \mbc ; \mbd \mbe \mbf}$ and we define:

\begin{align}
\alpha_{D_{d/3}}&=\epsilon(4\pi)^d\Gamma(\tfrac{d}{3})^{3}\frac{D_{d/3}}{3}=\frac{\Gamma(\tfrac{d}{6})^3}{3\Gamma(\tfrac{d}{2})} \; , \; \alpha_{S_{d/3}}=-\epsilon(4\pi)^{2d} \Gamma(\tfrac{d}{3})^{6}\frac{S_{d/3}}{3}= -\frac{\Gamma(\tfrac{d}{6})^4\Gamma(-\tfrac{d}{6})\Gamma(\tfrac{d}{3})^2}{6\Gamma(\tfrac{d}{2})\Gamma(\tfrac{2d}{3})} \, ,\crcr
\alpha_{I_{d/3}}&=-\epsilon(4\pi)^{2d}\Gamma(\tfrac{d}{3})^{6}I_{d/3}= -\frac{\Gamma(\tfrac{d}{6})^9}{2\Gamma(\tfrac{d}{3})^3\Gamma(\tfrac{d}{2})}\; ,  \crcr
\alpha_{J_{d/3}}&=\epsilon(4\pi)^{2d}\Gamma(\tfrac{d}{3})^{6}\frac{(D_{d/3}^2-2J_{d/3})}{6}= \frac{\Gamma(\tfrac{d}{6})^6}{6\Gamma(\tfrac{d}{2})^2\Gamma(\tfrac{d}{3})^6}\Bigg[\psi(\tfrac{d}{6})-\psi(1)+\psi(\tfrac{d}{3})-\psi(\tfrac{d}{2})\Bigg] \, .
\end{align}

We finally obtain the following beta functions:
\begin{align}
&\beta_{\mba \mbb \mbc ;\mbd \mbe \mbf}=-2\epsilon \tilde{g}_{\mba \mbb \mbc ;\mbd \mbe \mbf}+\alpha_{D_{d/3}}\left[ 3\left(\tilde{g}_{\mba \mbb \mbg ;\mbh \mbj \mbd}\tilde{g}_{\mbc \mbh \mbj ;\mbe \mbf \mbg}+ 8\text{ terms}\right) +\tilde{g}_{\mba \mbb \mbc ;\mbg \mbh \mbj}\tilde{g}_{\mbg \mbh \mbj ;\mbd \mbe \mbf}\right] \crcr
&  +\alpha_{S_{d/3}}\left[3\left(\tilde{g}_{\mba \mbg \mbh ; \mbj \mbk \mbl}\tilde{g}_{\mbb \mbj \mbk ;\mbg \mbh \mbm}\tilde{g}_{\mbl \mbm \mbc ;\mbd \mbe \mbf} + 2\text{ terms}\right)+ 3\left(\tilde{g}_{ \mbj \mbk \mbl ; \mbd \mbg \mbh}\tilde{g}_{ \mbg \mbh \mbm ;\mbe \mbj \mbk}\tilde{g}_{ \mba \mbb \mbc ;\mbf \mbl \mbm} + 2\text{ terms}\right) \right.\crcr
&\left.  \quad + 3\left(\tilde{g}_{\mba \mbg \mbh ;\mbj \mbk \mbl}\tilde{g}_{ \mbj \mbk \mbm ;\mbg \mbh \mbd}\tilde{g}_{\mbl \mbb \mbc ;\mbe \mbf \mbm} + 8\text{ terms}\right)+2\left(\tilde{g}_{\mba \mbg \mbh ;\mbj \mbk \mbl }\tilde{g}_{ \mbj \mbk \mbl ;\mbh \mbm \mbd}\tilde{g}_{\mbm \mbb \mbc ;\mbe \mbf \mbg} + 8\text{ terms}\right)\right] \crcr
& +\alpha_{I_{d/3}}\left[\left(\tilde{g}_{\mba \mbg \mbh ; \mbd \mbj \mbk}\tilde{g}_{\mbb \mbc \mbl ;\mbg  \mbh \mbm}\tilde{g}_{\mbj  \mbk \mbm ; \mbl \mbe \mbf} + 8 \text{ terms}\right)+ 2\left(\tilde{g}_{\mba \mbg  \mbh ;\mbd  \mbj \mbk }\tilde{g}_{\mbb \mbc \mbj; \mbg \mbl \mbm}\tilde{g}_{\mbk  \mbl \mbm ;\mbh \mbe \mbf} + 8 \text{ terms}\right) \right. \crcr
& \left. \quad +\left(\tilde{g}_{\mba \mbg \mbh ;\mbd  \mbj \mbk}\tilde{g}_{\mbb \mbl \mbm ;\mbe \mbg \mbh}\tilde{g}_{\mbc  \mbj \mbk ;\mbf \mbl \mbm} + 5 \text{ terms}\right)+4\left(\tilde{g}_{\mba \mbg \mbh ; \mbd  \mbj \mbk}\tilde{g}_{\mbb \mbj \mbl ;\mbe \mbg \mbm}\tilde{g}_{\mbc  \mbk \mbm ;\mbf \mbh \mbl} + 5 \text{ terms}\right)\right]\crcr
& +\alpha_{J_{d/3}}\left[\left(\tilde{g}_{\mba \mbg \mbh ;\mbj \mbe \mbf }\tilde{g}_{\mbb \mbc \mbj ;\mbk \mbl \mbm }\tilde{g}_{\mbk \mbl \mbm ; \mbg \mbh   \mbd} +  \text{ 8 terms}\right) + 6\left(\tilde{g}_{\mba \mbg \mbh ;\mbj \mbe \mbf }\tilde{g}_{\mbb \mbc \mbk ;\mbg \mbl \mbm }\tilde{g}_{\mbl \mbm\mbj ;\mbh \mbk  \mbd} +  \text{ 8 terms}\right)\right. \crcr
& \quad +\left(\tilde{g}_{\mbj \mbb \mbc ;\mbd \mbg \mbh }\tilde{g}_{\mbk \mbl \mbm ;\mbe \mbf \mbj  }\tilde{g}_{\mba \mbg \mbh ;\mbk \mbl \mbm} +  \text{ 8 terms}\right) + 6\left(\tilde{g}_{\mbj \mbb \mbc ;\mbd  \mbg \mbh }\tilde{g}_{\mbg \mbl \mbm ;\mbk \mbe \mbf }\tilde{g}_{\mba \mbh \mbk ;\mbl  \mbm \mbj} +  \text{ 8 terms}\right)\crcr
& \quad +3\left(\tilde{g}_{\mba \mbk \mbl ;\mbm \mbe \mbd}\tilde{g}_{\mbb \mbj \mbm  ;\mbf \mbg  \mbh }\tilde{g}_{\mbc \mbg \mbh ; \mbj  \mbk \mbl} +  \text{ 17 terms}\right)+6\left(\tilde{g}_{\mba \mbl \mbm ;\mbh \mbd \mbe }\tilde{g}_{\mbb  \mbj \mbk ;\mbg\mbm  \mbf }\tilde{g}_{\mbc \mbg \mbh ;\mbj  \mbk \mbl} +  \text{ 17 terms}\right)\crcr
& \quad +3\left(\tilde{g}_{ \mbm \mba \mbb ;\mbd \mbk \mbl}\tilde{g}_{\mbc   \mbg \mbh ;\mbe \mbj  \mbm}\tilde{g}_{\mbj \mbk \mbl  ;\mbg \mbh   \mbf} +  \text{ 17 terms}\right)+6\left(\tilde{g}_{ \mba \mbb  \mbh ;\mbd \mbl\mbm}\tilde{g}_{\mbc   \mbg \mbm ;\mbe \mbj   \mbk}\tilde{g}_{\mbj \mbk \mbl ;\mbg \mbh   \mbf} +  \text{ 17 terms}\right) \crcr
& \quad \left. +3\left(\tilde{g}_{\mba \mbb \mbc ;\mbg \mbh \mbj}\tilde{g}_{\mbj \mbl \mbm ;\mbk \mbd \mbe}\tilde{g}_{\mbg \mbh \mbk ;\mbl \mbm \mbf} + \text{ 2 terms}\right)+ 3\left(\tilde{g}_{\mbg \mbh \mbj ;\mbd \mbe \mbf}\tilde{g}_{ \mba \mbb \mbk ;\mbj \mbl \mbm}\tilde{g}_{ \mbl \mbm \mbc ;\mbg \mbh \mbk }+ \text{ 2 terms}\right)\right].
\label{eq:beta_LR}
\end{align}

\subsubsection{Application: $U(N)^3$ symmetry}
\label{sec:tensor_LR}

We specify again the symmetry to $U(N)^3$ as in section~\ref{sec:tensor_SR}. Now, for the long-range case, setting $\zeta=\frac{d+\epsilon}{3}$, we obtain the following beta functions up to order $N^{-1}$:
\begin{align}
\beta_1&=-2\epsilon \bar{g}_1+\frac{48}{N}\alpha_{I_{d/3}}\bar{g}_1^3 +\mathcal{O}(N^{-2})\, ,\crcr
\beta_2&=-2\epsilon\bar{g}_2+3\bar{g}_1^2\left(9\alpha_{D_{d/3}}+2\alpha_{S_{d/3}}(9\bar{g}_1+\bar{g}_2)\right) +\frac{2}{N}\Bigg[\frac{\alpha_{D_{d/3}}}{3}\left(81\bar{g}_1^2+36\bar{g}_1\bar{g}_2+\bar{
g}_2^2+6\bar{g}_3(9\bar{g}_1+\bar{g}_2)\right)\crcr
& \qquad +2\bar{g}_1^2\left(9(4\alpha_{J_{d/3}}+\alpha_{S_{d/3}})(3\bar{g}_1+\bar{g}_3)+2(4\alpha_{I_{d/3}}+3\alpha_{S_{d/3}})\bar{g}_2\right)\Bigg]+\mathcal{O}(N^{-2})\, , \crcr
\beta_3&=-2\epsilon\bar{g}_3+9\alpha_{S_{d/3}}\bar{g}_1^2\bar{g}_3 \crcr
& \quad +\frac{1}{N}\Bigg[3\alpha_{D_{d/3}}\bar{g}_3^2+\frac{2\alpha_{D_{d/3}}}{3}\bar{g}_2\left(18\bar{g}_1+\bar{g}_2\right)+108\alpha_{J_{d/3}}\bar{g}_1^3+12\alpha_{S_{d/3}}\bar{g}_1^2\bar{g}_2\Bigg]+\mathcal{O}(N^{-2})\, ,\crcr
\beta_4&=-2\epsilon\bar{g}_4+3\alpha_{S_{d/3}}\bar{g}_1^2\left(27\bar{g}_1+10\bar{g}_2+12\bar{g}_3+7\bar{g}_4\right)\crcr
& \quad +\frac{1}{27N}\Bigg[9\alpha_{D_{d/3}}\left(36\bar{g}_1\left(\bar{g}_2+3\bar{g}_3+2\bar{g}_4\right)+2\bar{g}_2\left(5\bar{g}_2+12\bar{g}_3+4\bar{g}_4\right)+3\bar{g}_3\left(9\bar{g}_3+4\bar{g}_4\right)\right) \crcr
& \qquad +\frac{2}{3}\alpha_{I_{d/3}}\big(162\bar{g_1}^2\left(54\bar{g}_1+5\bar{g}_2+12\bar{g}_3+4\bar{g}_4\right)+2\bar{g}_2^2\left(2\bar{g}_2+\bar{g}_4\right)\crcr
& \qquad \qquad \qquad +3\bar{g}_3\left(2\bar{g}_2^2+12\bar{g}_2\bar{g}_3+9\bar{g}_3^2+6\bar{g}_3\bar{g}_4\right)  \big)\crcr
& \qquad +324\alpha_{S_{d/3}}\bar{g}_1^2\left(4\bar{g}_2+6\bar{g}_3+\bar{g}_4\right)+81\alpha_{J_{d/3}}\bar{g}_1^2\left(36\bar{g}_1+33\bar{g}_2+18\bar{g}_3+8\bar{g}_4\right) \Bigg] +\mathcal{O}(N^{-2})\, ,\crcr
\beta_5&=-2\epsilon\bar{g}_5+3\bar{g}_1^2\left(\alpha_{D_{d/3}}+\alpha_{S_{d/3}}(3\bar{g}_2+8\bar{g}_4+15\bar{g}_5)\right) \crcr
& \quad +\frac{1}{N}\Bigg[\frac{2}{3}\alpha_{D_{d/3}}\left(\bar{g}_2(\bar{g}_2+6\bar{g}_3+4\bar{g}_4)+2\bar{g}_4(3\bar{g}_3+\bar{g}_4)\right)  +6\alpha_{S_{d/3}}\bar{g}_1^2\left(3\bar{g}_3+4\bar{g}_4\right) \crcr
& \qquad +\frac{4}{243}\alpha_{I_{d/3}}\big(243\bar{g_1}^2\left(\bar{g}_2+3\bar{g}_3+\bar{g}_4\right)+10\bar{g}_2^3 +9\bar{g}_3\left(7\bar{g}_2^2+6\bar{g}_2\bar{g}_3\right) \crcr
& \qquad \qquad \qquad +\bar{g}_4\left(3\bar{g}_2\left(7\bar{g}_2+24\bar{g}_3+4\bar{g}_4\right)+9\bar{g}_3\left(3\bar{g}_3+2\bar{g}_4\right)+2\bar{g}_4\right)\big) \crcr
&  \qquad    +3\alpha_{J_{d/3}}\bar{g}_1^2\left(3\bar{g}_2+6\bar{g}_3+8\bar{g}_4\right) \Bigg]+\mathcal{O}(N^{-2})\, .
\end{align}

In the long-range case, at $\epsilon=0$, the wheel coupling $\bar{g}_1$ is exactly marginal. However, at order $N^{-1}$ the wheel beta function is non-zero and the line of fixed points found in section~\ref{sec:betas} collapses to the trivial fixed point. 
Turning on $\epsilon$ does not solve the problem as it contributes a term $-2\epsilon\bar{g}_1$ which already gives $\bar{g}_1^{\star}=0$ at leading order. As for the short-range case, we should also consider how small $\epsilon$ is compared to $N$. At next-to-leading order the wheel beta function has the following form $-2\epsilon \bar{g}_1+\bar{g}_1^3/N$. Its fixed points are the trivial one and $\bar{g}_1^{\star}=\sqrt{N\epsilon}$. The latter goes to infinity for $N \rightarrow \infty$ at fixed $\epsilon$. This is solved by imposing $N\epsilon \ll 1$. We thus set:
\begin{equation}
\epsilon=\frac{\tilde{\epsilon}}{N} \, ,
\end{equation}
and as before we expand first in $1/N$ and then in $\tilde{\epsilon}$. This is once more similar to what happens in chapter~\ref{chap:trif}.

We parametrize again the critical couplings as $\bar{g}_i=\bar{g}_{i,0}+\frac{\bar{g}_{i,1}}{N} +\mathcal{O}(N^{-2})$ for $i=1,\dots , 5$. Solving for the zeros of the beta functions at leading order we find the following solutions:\footnote{There is also a solution with zero wheel coupling leading to a $4$-dimensional manifold of fixed points. We do not study this solution further as we are only interested in solutions with non-zero wheel coupling in order to have a melonic fixed point.}

\begin{gather}
\bar{g}_{2,0}^*=-9g_{1,0}+\f{9\Gamma(2d/3)}{\Gamma(-d/6)\Gamma(d/6)\Gamma(d/3)^2}\, ,\qquad \bar{g}_{3,0}^*=0 \, ,\crcr
\bar{g}_{4,0}^*=9g_{1,0} - \frac{90}{7}\f{\Gamma(2d/3)}{\Gamma(-d/6)\Gamma(d/6)\Gamma(d/3)^2}\, ,\crcr
\bar{g}_{5,0}^*=-3g_{1,0} + \f{109\Gamma(2d/3)}{21\Gamma(-d/6)\Gamma(d/6)\Gamma(d/3)^2}\, .
\end{gather}

We thus again recover the line of fixed points found in section~\ref{sec:betas}.
Solving the wheel beta functions at order $N^{-1}$, we find:
\begin{equation}
\bar{g}_{1,0}=\pm \frac{ \sqrt{\tilde{\epsilon}}}{2\sqrt{6\alpha_{I_{d/3}}}} \;.
\end{equation}

$I_{d/3}$ being negative, $\bar{g}_{1,0}$ is thus purely imaginary. This implies that the other four couplings are also complex at leading order. 
However, substituting these results into the order $N^{-1}$ of the beta functions, we find non-perturbative corrections to the fixed points which blow up when sending $\tilde{\epsilon} \rightarrow 0$. This cannot be fixed by rescaling $\tilde{\epsilon}$ or $N$. 
This is due to the form of the beta functions. Indeed, as we saw in the short-range model, all couplings except the wheel start at order $0$ in $\epsilon$. When solving at next-to-leading order, this will lead to non-perturbative results. To cure this, we would have to rescale $N$ as we did for the short-range model. However, doing so instead of rescaling $\epsilon$, the beta function of the wheel would give $g_{1}^{\star}=0$ and the only fixed point would then be the trivial one. 
We therefore conclude that there is no precursor at next-to-leading order of the large-$N$ fixed point.

\section{Conclusions}
\label{sec:concl_sextic}

In this chapter, we presented an analysis of the melonic large-$N$ limit in various versions of bosonic tensor models with sextic interactions.
We considered explicitly tensors of rank 3 and 5, but we expect rank 4 to behave similarly to rank 5. We chose as free propagator either the standard short-range propagator, or a critical long-range propagator. We discussed in detail some standard properties of melonic theories, as the closed Schwinger-Dyson equation for the two-point function, and the Bethe-Salpeter equation for the spectrum of bilinear operators. However, as we emphasized, the conformal solution of these equations are only justified if the quantum field theory actually admits a fixed point of the renormalization group. In this respect, we found a striking difference between the rank-3 and rank-5 models, as only the former (both in the short-range and long-range versions) admits a non-trivial (and real) fixed point for $d<3$, with an interaction leading to melonic dominance. 
The rank-5 model instead has only one trivial (i.e.\ non-interacting) fixed point. It would be interesting to check whether such conclusion would remain valid after including in the action \eqref{eq:int-action-graph-rank5} the other possible sextic interactions that we have omitted by restricting to a melo-complete family.

In the second part of the chapter, we studied the next-to-leading order of the rank-$3$ model both in short- and long-range. In both cases, we studied the renormalization group and computed fixed points of the beta functions at next-to-leading order. However, the results are radically different in the two cases. 
In the short-range case, the theory admits a non-trivial real stable IR fixed point with non-zero wheel coupling, thus leading to melonic dominance.  
In the long-range case, the corrections to the large-$N$ fixed points are not perturbative in $\tilde{\epsilon}$ and even blow up when $\tilde{\epsilon}$ goes to zero. This indicates that the large-$N$ fixed point found in the first part of the chapter has no precursor at next-to-leading order.

As for the computation of $1/N$ corrections in quartic models of chapter \ref{chap:trif}, a subtle part of our analysis is the identification of a proper hierarchy between our two small parameters $1/N$ and $\epsilon$. Indeed, in the short-range case we need $\epsilon N \gg 1$ while in the long-range case we need $\epsilon N \ll 1$. These conditions are found by demanding that the large-$N$ fixed points remain dominant in the beta functions. However, for the long-range case, contrary to quartic models, this condition is not enough to ensure a perturbative solution of the beta functions at next-to-leading order. This is due to the presence of a term of order $\mathcal{O}(\epsilon^0)$ in the large-$N$ fixed point.

Nevertheless, this is an interesting new feature of our fixed point that also appears in the short-range case. Indeed, the wheel coupling at large $N$ is of order $\sqrt{\epsilon}$ while the other couplings start at order $\epsilon^0$. This is very different from usual Wilson-Fisher-like fixed points and from the quartic model fixed points of chapter \ref{chap:trif}. It is due to the fact that the graph $D_1$ contributes to leading order in $N$ only with wheel vertices whereas in quartic models the one-loop Feynman graph contributes to leading order with all three quartic interactions on the vertices. This model thus leads to a new kind of melonic fixed point. 

Comparing our findings for the short range model with those of the sextic model in \cite{Giombi:2017dtl}, we observe similar results for two-point function and spectrum  of operators. However, we do so for the rank-3 model, where such analysis is justified by the existence of a melonic fixed point, whereas their analysis was formally based on a rank-5 model, which we showed is inconsistent. The fact that we find the same result is not a coincidence: our kernel eigenvalue \eqref{eq:eigenvalue} coincides with the $q=6$ case of the eigenvalue computed in \cite{Giombi:2017dtl} for a general $q$-valent melonic theory. Such eigenvalue depends only on the assumption that a $q$-valent interaction leads to melonic dominance. The latter can for example be obtained with a rank-$(q-1)$ model with a complete interaction, as assumed in \cite{Giombi:2017dtl}. However, as argued in \cite{Prakash:2019zia}, and as we saw also here, rank $q-1$ is not necessary: a $q$-valent interaction can lead to a melonic limit even in a tensor model of rank $r<q-1$ (in which case the model was called \emph{subchromatic} in \cite{Prakash:2019zia}); this is the case of our rank-3 model with wheel interaction.

Comparing instead the long-range model of this chapter to the quartic long-range model of chapter \ref{chap:CTKT}, we see some similarity but also an important difference: on one hand, both models admit a line of fixed points, parametrized by the interaction that leads to melonic dominance; on the other, in the quartic case, the fixed point and conformal dimensions are real only for purely imaginary tetrahedral coupling, while in the sextic model, we have a real fixed point and real spectrum for a real wheel coupling.
Furthermore, unlike in chapter \ref{chap:CTKT}, in the present case the appearance of complex dimensions at some critical value of the marginal coupling seems to be compatible with the scenario conjectured in \cite{Kim:2019upg}, according to which it is a signal of an instability of the vacuum.

We have also encountered some of the recurring aspects of melonic theories (for rank 3, at least): for the short-range version, reality of the CFT constrains $\epsilon$ to stay very small; in the long-range version, we have instead the freedom to reach an integer dimension ($d=2$ in this case), by keeping the marginal coupling small, but at the price of loosing the energy-momentum tensor (as usual in long-range models \cite{Paulos:2015jfa}).
It would be interesting to get a better understanding of how general these features are.

One new feature that we found is that the fixed point of the short-range model has a non-diagonalizable stability matrix, even in the range of $\epsilon$ for which the exponents are real. This is an indication that the fixed-point theory is a logarithmic CFT, and thus it is non-unitary. At next-to-leading order, the stability matrix becomes diagonalizable.  However, the theory at the fixed point still cannot be unitary at next-to-leading order. Indeed, corrections in $1/N$ cannot cancel logarithmic terms in correlation functions and thus we cannot recover unitarity at next-to-leading order.
\begin{subappendices}
\section{Conventions for the interaction terms}
\label{ap:conventions}

We write here in an explicit form the interactions appearing in \eqref{eq:int-action} and \eqref{eq:int-action-rank5}, as well as the quartic invariants, in terms of contraction operators built as linear combinations of products of Kronecker delta functions.

\subsection{Rank 3}
Using the compact notation $\mathbf{a}=(a_1a_2a_3)$,
the $U(N)^3$ quartic invariants, also known as pillow and double-trace invariants, respectively, are:
\begin{align}
I_p &=  \delta^p_{\mba\mbb; \mbc\mbd}\phi_{\mba}(x) \phib_{\mbb}(x)  \phi_{\mbc}(x) \phib_{\mbd }(x)\,,\\
I_d &= \delta^d_{\mba\mbb; \mbc\mbd }  \phi_{\mba}(x) \phib_{\mbb}(x)  \phi_{\mbc}(x) \phib_{\mbd }(x)\,,
\end{align}
with: 
\be
    \delta^p_{\mba\mbb; \mbc\mbd }=\frac{1}{3} \sum_{i=1}^3  \delta_{a_id_i} \delta_{b_ic_i} \prod_{j\neq i}  \delta_{a_jb_j}\delta_{c_j d_j} \; , \quad\quad  \delta^d_{\mba\mbb; \mbc\mbd }  = \delta_{\mba \mbb}  \delta_{\mbc \mbd}\,,
\ee
and $\delta_{\mba \mbb}  = \prod_{i=1}^3 \delta_{a_i b_i} $.

The sextic invariants depicted in \eqref{eq:int-action} are instead:
\begin{align}
I_1 &=  \delta^{(1)}_{\mba\mbb \mbc\mbd\mbe\mbf} \phi_{\mba}(x) \phib_{\mbb}(x)  \phi_{\mbc}(x) \phib_{\mbd }(x)\phi_{\mbe }(x)\phib_{\mbf }(x)\,,\\
I_b &=\delta^{(b)}_{\mba\mbb; \mbc\mbd; \mbe\mbf }  \phi_{\mba}(x) \phib_{\mbb}(x)  \phi_{\mbc}(x) \phib_{\mbd }(x)\phi_{\mbe }(x)\phib_{\mbf }(x)\,,   \qquad b=2,\ldots,5\,,
\end{align}
with
\begin{align*}
   & \delta^{(1)}_{\mba\mbb\mbc\mbd \mbe\mbf }= \d_{a_1b_1}\d_{a_2f_2}\d_{a_3d_3}\d_{c_1d_1}\d_{c_2b_2}\d_{c_3f_3}\d_{e_1f_1}\d_{e_2d_2}\d_{e_3b_3}\, ,\crcr
  & \delta^{(2)}_{\mba\mbb; \mbc\mbd; \mbe\mbf }= \frac{1}{9}\left( \sum_{i=1}^3 \sum_{j \neq i}  \delta_{a_if_i} \delta_{b_ic_i} \delta_{c_j d_j}\delta_{e_jf_j}\left(\prod_{k\neq i}  \delta_{a_kb_k} \right)\left(\prod_{l\neq j}  \delta_{e_l d_l} \right) \left( \prod_{m \neq i,j} \delta_{c_m f_m}\right) \right. \crcr
  & \qquad \qquad \qquad + \mbc\mbd \leftrightarrow \mbe\mbf + \mbc\mbd \leftrightarrow \mba\mbb \bigg) \, ,\crcr
  & \delta^{(3)}_{\mba\mbb; \mbc\mbd; \mbe\mbf }= \frac{1}{3} \sum_{i=1}^3 \delta_{a_if_i} \delta_{b_ic_i} \delta_{d_i e_i}\prod_{j\neq i}  \delta_{a_j b_j}\delta_{c_j d_j}\delta_{e_j f_j} \, ,\crcr
  & \delta^{(4)}_{\mba\mbb; \mbc\mbd; \mbe\mbf }=\frac{1}{3}\left(\delta_{\mba\mbb}\delta^p_{\mbc\mbd;\mbe\mbf} + \delta_{\mbc\mbd}\delta^p_{\mba\mbb;\mbe\mbf}+ \delta_{\mbe\mbf}\delta^p_{\mba\mbb;\mbc\mbd}\right)\, ,\crcr
  & \delta^{(5)}_{\mba\mbb; \mbc\mbd; \mbe\mbf }=\delta_{\mba\mbb}\delta_{\mbc\mbd}\delta_{\mbe\mbf}\, .
\end{align*}
Besides the color symmetrization, to simplify the computation of the beta functions, we have included a symmetrization with respect to exchange of pairs of black and white vertices.

\subsection{Rank 5}

Using the compact notation $\mathbf{a}=(a_1a_2a_3a_4a_5)$, the $O(N)^3$ melonic quartic invariants are:
\begin{align}
I_p &=  \delta^p_{\mba\mbb; \mbc\mbd}\phi_{\mba}(x) \phi_{\mbb}(x)  \phi_{\mbc}(x) \phi_{\mbd }(x)\,,\\
I_d &= \delta^d_{\mba\mbb; \mbc\mbd }  \phi_{\mba}(x) \phi_{\mbb}(x)  \phi_{\mbc}(x) \phi_{\mbd }(x)\,,
\end{align}
with: 
\be
   \delta^p_{\mba\mbb; \mbc\mbd }=\frac{1}{5} \sum_{i=1}^5  \delta_{a_id_i} \delta_{b_ic_i} \prod_{j\neq i}  \delta_{a_jb_j}\delta_{c_j d_j} \; , \quad\quad  \delta^d_{\mba\mbb; \mbc\mbd }  = \delta_{\mba \mbb}  \delta_{\mbc \mbd} \,,
\ee
and $\delta_{\mba \mbb}  = \prod_{i=1}^5 \delta_{a_i b_i} $.

The sextic invariants depicted in \eqref{eq:int-action-rank5} are instead:

\begin{align}
J_1 &=  \delta^{(1)}_{\mba\mbb \mbc\mbd\mbe\mbf} \phi_{\mba}(x) \phi_{\mbb}(x)  \phi_{\mbc}(x) \phi_{\mbd }(x)\phi_{\mbe }(x)\phi_{\mbf }(x)\,,\\
J_b &=\delta^{(b)}_{\mba\mbb; \mbc\mbd; \mbe\mbf } \phi_{\mba}(x) \phi_{\mbb}(x)  \phi_{\mbc}(x) \phi_{\mbd }(x)\phi_{\mbe }(x)\phi_{\mbf }(x)\,,   \qquad b=2,\ldots,6\,,
\end{align}
with
\begin{align*}
   & \delta^{(1)}_{\mba\mbb\mbc\mbd\mbe\mbf}  = \delta_{a_1 b_1}  \delta_{a_2f_2} \delta_{a_3 e_3}  \delta_{a_4 d_4 } \delta_{a_5 c_5}   \delta_{b_2 c_2}\delta_{b_3 d_3}\delta_{b_4 f_4} \delta_{b_5 e_5} \delta_{c_3 f_3}\delta_{c_4 e_4}\delta_{c_1 d_1}\delta_{e_1 f_1}\delta_{e_2 d_2}\delta_{d_5 f_5} \,  , \crcr
   & \delta^{(2)}_{\mba\mbb; \mbc\mbd; \mbe\mbf }= \frac{1}{60}\left( \sum_{i=1}^5 \sum_{j \neq i}  \delta_{a_ic_i} \delta_{b_id_i} \delta_{c_j e_j}\delta_{d_jf_j}\left(\prod_{k\neq i}  \delta_{a_kb_k} \right)\left(\prod_{l\neq j}  \delta_{e_l f_l} \right) \left( \prod_{m \neq i,j} \delta_{c_m d_m}\right) \right. \crcr
   & \qquad \qquad \qquad + \mbc\mbd \leftrightarrow \mbe\mbf + \mbc\mbd \leftrightarrow \mba\mbb \bigg) \, ,\crcr
   & \delta^{(3)}_{\mba\mbb; \mbc\mbd; \mbe\mbf }= \frac{1}{5} \sum_{i=1}^5 \delta_{a_if_i} \delta_{b_ic_i} \delta_{d_i e_i}\prod_{j\neq i}  \delta_{a_j b_j}\delta_{c_j d_j}\delta_{e_j f_j} \, ,\crcr
  & \delta^{(4)}_{\mba\mbb; \mbc\mbd; \mbe\mbf }=\frac{1}{3}\left(\delta_{\mba\mbb}\delta^p_{\mbc\mbd;\mbe\mbf} + \delta_{\mbc\mbd}\delta^p_{\mba\mbb;\mbe\mbf}+ \delta_{\mbe\mbf}\delta^p_{\mba\mbb;\mbc\mbd}\right)\, ,\crcr
  & \delta^{(5)}_{\mba\mbb; \mbc\mbd; \mbe\mbf }=\delta_{\mba\mbb}\delta_{\mbc\mbd}\delta_{\mbe\mbf}\, , \crcr
  & \delta^{(6)}_{\mba\mbb; \mbc\mbd; \mbe\mbf }= \frac{1}{60} \sum_{i=1}^5 \sum_{j\neq i}\sum_{k\neq i,j}  \delta_{a_ic_i} \delta_{b_id_i} \delta_{c_j e_j}\delta_{d_jf_j}\delta_{a_k e_k}\delta_{b_k f_k}\left(\prod_{l\neq i,k}  \delta_{a_l b_l} \right)\left(\prod_{m\neq j,k}  \delta_{e_m f_m} \right) \left( \prod_{n \neq i,j} \delta_{c_n d_n}\right) \, . 
\end{align*}

\section{The melon integral}
\label{ap:melon}

In this section we compute the melon integral contributing to the wave function renormalization in the short-range case and to the coefficient of the full two-point function in both the short- and long-range case. 

We want to compute:
\begin{equation}
M_{\zeta}(p)=\int_{q_1,q_2,q_3,q_4}G_0(q_1)G_0(q_2)G_0(q_3)G_0(q_4)G_0(p+q_1+q_2+q_3+q_4) \,,
\end{equation}
with $G_0(p)=\frac{1}{p^{2\zeta}}$.

We will use the following formula to compute $M(p)$:
\begin{equation}
\int\f{\dd[d] k}{(2\pi)^d} \f{1}{k^{2\a}(k+p)^{2\b}}  = \f{1}{(4\pi)^{d/2}}\f{\G(d/2 - \a)\G(d/2 - \b)\G(\a +\b -d/2)}{\G(\a)\G(\b) \G(d - \a - \b)}\f{1}{|p|^{2(\a+\b - d/2)}} \,.
\label{eq:intG}
\end{equation}
We obtain:
\begin{equation}
M_{\zeta}(p)=\frac{p^{4d-10\zeta}}{(4\pi)^{2d}}\frac{\Gamma(d/2-\zeta)^5\Gamma(5\zeta-2d)}{\Gamma(\zeta)^5\Gamma(5d/2-5\zeta)} \,.
\end{equation}
For $\zeta=\frac{d}{3}$, this simplifies to:
\begin{equation} \label{eq:M-d/3}
M_{d/3}(p)=-\frac{p^{2d/3}}{(4\pi)^{2d}}\frac{3}{d}\frac{\Gamma(1-\frac{d}{3})\Gamma(\frac{d}{6})^5}{\Gamma(\frac{d}{3})^5\Gamma(\frac{5d}{6})} \,.
\end{equation}
We will also need the melon integral for $d=3-\epsilon$ and $\Delta=1$:
\begin{equation}  \label{eq:M-1}
M_1(p)=\frac{p^{2-4\epsilon}}{(4\pi)^{6-2\epsilon}}\frac{\Gamma(2\epsilon-1)\Gamma(\frac{1-\epsilon}{2})^5}{\Gamma(\frac{5}{2}(1-\epsilon)}\,.
\end{equation}
At first order in $\epsilon$, this gives:
\begin{equation}
M_1(p)=- \frac{p^{2-4\epsilon}}{(4\pi)^{6}}\frac{2\pi^2}{3\epsilon} + \mathcal{O}(1)\,.
\end{equation}

\section{Beta functions details}
\label{ap:betafun4}

\subsection{2-loop amplitude}
\label{app:D}

We want to compute the two-loop amputated Feynman integral $D_{\zeta}$ represented in figure~\ref{fig:graphs}.
Applying \eqref{eq:amp_final_sextic}, this can be written as:
\begin{equation}
D_{\zeta}=\frac{1}{\Gamma(\zeta)^2(4\pi)^d}\int da_1da_2da_3 \frac{(a_1a_2a_3)^{\zeta-1}}{\left(a_1a_2+a_1a_3+a_2a_3\right)^{\tfrac{d}{2}}}e^{-(a_1+a_2+a_3)}\,.
\end{equation}

In the following we will rely on the Mellin-Barnes representation introduced in appendix \ref{sec:MB}. Using \eqref{eq:MB_rep} we can rewrite the denominator of $D_{\zeta}$ as:

\begin{equation}
\frac{1}{\left(a_1a_2+a_1a_3+a_2a_3\right)^{\tfrac{d}{2}}}=\int_{0^-} [dz] \frac{\Gamma(z)\Gamma(z+\tfrac{d}{2})}{\Gamma(\tfrac{d}{2})}\frac{(a_2a_3)^z}{(a_1(a_2+a_3))^{z+\tfrac{d}{2}}} \, .
\end{equation}

We can then integrate the Schwinger parameters using the following formula:

\begin{equation}
\int da_1 da_2 \frac{(a_1a_2)^{u-1}}{(a_1+a_2)^{\gamma}}=\frac{\Gamma(u)^2\Gamma(2u-\gamma)}{\Gamma(2u)}\,,
\label{eq:int_class}
\end{equation}
which is valid for $Re(u)>0$ and $Re(2u)>Re(\gamma)$.

\paragraph{Short-range: $\zeta=1$, $d=3-\epsilon$.}

We then obtain:
\begin{equation}
D_1=\f{1}{(4\pi)^3\Gamma(\tfrac{3}{2})}\int_{-\tfrac{1}{2}^-} [dz] \Gamma(-z)\Gamma(z+\tfrac{3-\epsilon}{2})\frac{\Gamma(1+z)^2\Gamma(\tfrac{1+\epsilon}{2}+z)}{\Gamma(2+2z)}\Gamma(-\tfrac{1}{2}+\tfrac{\epsilon}{2}-z)\, ,
\end{equation}
where we have moved the integration contour such that all gamma functions have positive arguments.

There is only one pole giving a singularity in $\epsilon$ located at $z=-\frac{1}{2}+\frac{\epsilon}{2}$. 
We thus obtain:

\begin{align}
D_1=&\frac{1}{(4\pi)^3\Gamma(\tfrac{3-\epsilon}{2})}\bigg[\frac{\Gamma(\tfrac{1-\epsilon}{2})\Gamma(\tfrac{1+\epsilon}{2})^2\Gamma(\epsilon)}{\Gamma(1+\epsilon)} \crcr
& \qquad \qquad \qquad  +\int_{0^-} [dz] \Gamma(-z)\Gamma(z+\tfrac{3}{2})\frac{\Gamma(1+z)^2\Gamma(\tfrac{1}{2}+z)}{\Gamma(2+2z)}\Gamma(-\tfrac{1}{2}-z)\bigg]+ \mathcal{O}(\epsilon)\, .
\end{align}
The remaining integral has two types of poles, situated at $z=n_1$ for $n_1\geq 0$ and $z=-\frac{1}{2}+n_2$ for $n_2\geq 1$. We have:

\begin{align}
&\int_{0^-} [dz] \Gamma(-z)\Gamma(z+\tfrac{3}{2})\frac{\Gamma(1+z)^2\Gamma(\tfrac{1}{2}+z)}{\Gamma(2+2z)}\Gamma(-\tfrac{1}{2}-z) \crcr
&=\sum_{n=0}^{\infty} \frac{(-1)^n}{n!}\frac{\Gamma(n+\tfrac{3}{2})\Gamma(n+1)^2\Gamma(n+\tfrac{1}{2})\Gamma(-\tfrac{1}{2}-n)}{\Gamma(2+2n)} \crcr
& \quad + \sum_{n=1}^{\infty} \frac{(-1)^n}{n!}\frac{\Gamma(\tfrac{1}{2}-n)\Gamma(n+1)\Gamma(n+\tfrac{1}{2})^2\Gamma(n)}{\Gamma(2n+1)} \, .
\end{align}

Both sums are convergent and can be computed to obtain:
\begin{equation}
\int_{0^-} [dz] \Gamma(-z)\Gamma(z+\tfrac{3}{2})\frac{\Gamma(1+z)^2\Gamma(\tfrac{1}{2}+z)}{\Gamma(2+2z)}\Gamma(-\tfrac{1}{2}-z)=\pi^{\tfrac{3}{2}}\ln(\tfrac{4}{9}) \, .
\label{eq:part_int}
\end{equation}

Finally, we obtain:
\begin{equation}
D_1=\frac{\pi}{(4\pi)^3}\left[\frac{2}{\epsilon}+\psi(\tfrac{1}{2})+\psi(\tfrac{3}{2})+4\ln(\tfrac{2}{3})\right] +\mathcal{O}(\epsilon) \, .
\end{equation}

\paragraph{Long-range: $\zeta=\frac{d+\epsilon}{3}$, $d<3$.}

Using the same method as for the short-range case, we obtain:

\begin{align}
D_{d/3}=&\frac{\Gamma(\tfrac{d}{6})^3}{(4\pi)^d \Gamma(\tfrac{d}{3})^3\Gamma(\tfrac{d}{2})}\left[ \frac{1}{\epsilon}+\psi(1)+\psi(\tfrac{d}{6})-2\psi(\tfrac{d}{3}) \right. \crcr
& \qquad +\frac{\Gamma(-\tfrac{d}{6})\Gamma(\tfrac{d}{3})^3\Gamma(\tfrac{d}{2})}{\Gamma(\tfrac{d}{6})^2\Gamma(\tfrac{2d}{3})}{}_3F_2\left(\tfrac{d}{6},\tfrac{d}{3},\tfrac{d}{2};1+\tfrac{d}{6},\tfrac{1}{2}+\tfrac{d}{3};\tfrac{1}{4}\right) \crcr
& \qquad \left. +\frac{d^2}{2(d+3)(6-d)}{}_4F_3\left(1,1,1+\tfrac{d}{6},1+\tfrac{d}{3};2,2-\tfrac{d}{6},\tfrac{3}{2}+\tfrac{d}{6};\tfrac{1}{4}\right) \right]+\mathcal{O}(\epsilon)\, . 
\end{align}
where ${}_p F_q(a_1 \dots a_p ; b_1 \dots b_q ; z)$ are the generalized hypergeometric functions.

\subsection{4-loop amplitude}

\subsubsection{$S_{\zeta}$ integral}

The $S_{\zeta}$ integral with Schwinger parameters can be written as:

\begin{align}
S_{\zeta}=&\frac{1}{\Gamma(\zeta)^6(4\pi)^{2d}}\int da_1da_2db_1db_2db_3db_4 \, (a_1a_2b_1b_2b_3b_4)^{\zeta-1}e^{-(a_1+a_2+b_1+b_2+b_3+b_4)}\crcr
& \qquad \qquad \times \frac{1}{\left[(a_1+a_2)(b_1b_2(b_3+b_4)+b_3b_4(b_1+b_2))+b_1b_2b_3b_4\right]^{\tfrac{d}{2}}}\, .
\end{align}

Doing the change of variables $a_1=\alpha \beta$ and $a_2=\alpha(1-\beta)$, we can integrate $\beta$ to obtain:

\begin{equation}
S_{\zeta}=\frac{1}{\Gamma(\zeta)^4\Gamma(2\zeta)(4\pi)^{2d}}\int d\alpha db_1db_2db_3db_4 \frac{\alpha^{2\zeta-1}(b_1b_2b_3b_4)^{\zeta-1}e^{-(\alpha+b_1+b_2+b_3+b_4)}}{\left[\alpha(b_1b_2(b_3+b_4)+b_3b_4(b_1+b_2))+b_1b_2b_3b_4\right]^{\tfrac{d}{2}}}\,.
\end{equation}

However, one needs to take into account the subtraction of the local part of the four-point insertion. Using a Taylor expansion with integral rest, the subtracted $S_{\zeta}$ can then be written as:

\begin{align}
S_{\zeta}=\frac{-\tfrac{d}{2}}{\Gamma(\zeta)^4\Gamma(2\zeta)(4\pi)^{2d}}&\int_0^1 dt\int d\alpha db_1db_2db_3db_4\, \alpha^{2\zeta-1}(b_1b_2b_3b_4)^{\zeta}e^{-(\alpha+b_1+b_2+b_3+b_4)}\crcr
& \qquad \times \frac{1}{\left[\alpha(b_1b_2(b_3+b_4)+b_3b_4(b_1+b_2))+tb_1b_2b_3b_4\right]^{1+\tfrac{d}{2}}}\, .
\end{align}
Using two Mellin-parameters, the denominator can be written as:
\begin{align}
&\frac{1}{\left[\alpha(b_1b_2(b_3+b_4)+b_3b_4(b_1+b_2))+tb_1b_2b_3b_4\right]^{1+\tfrac{d}{2}}}=\crcr
&\int [dz_1][dz_2] \frac{\Gamma(-z_1)\Gamma(-z_2)\Gamma(z_1+z_2+\tfrac{d}{2}+1)}{\Gamma(1+\tfrac{d}{2})}\frac{(tb_1b_2b_3b_4)^{z_1}(\alpha b_3b_4(b_1+b_2))^{z_2}}{(\alpha b_1b_2(b_3+b_4))^{z_1+z_2+\tfrac{d}{2}+1}}\, .
\end{align}
We can then integrate the Schwinger parameters using \eqref{eq:int_class} as well as perform the $t$ integral.

\paragraph{Short-range: $\zeta=1$, $d=3-\epsilon$.}

We obtain:

\begin{align}
S_1=-\frac{3-\epsilon}{2(4\pi)^6\Gamma(\tfrac{5}{2})}&\int_{-\tfrac{1}{2}^-} [dz_1] \int_{-1^-}[dz_2] \Gamma(-z_1)\Gamma(-z_2)\Gamma(z_1+z_2+\tfrac{5-\epsilon}{2})\Gamma(-\tfrac{1}{2}+\tfrac{\epsilon}{2}-z_1)\crcr
& \times \frac{\Gamma(2+z_1+z_2)^2\Gamma(\tfrac{3}{2}+\tfrac{\epsilon}{2}+z_1+z_2)}{\Gamma(4+2z_1+2z_2)(1+z_1)}\frac{\Gamma(-\tfrac{1}{2}+\tfrac{\epsilon}{2}-z_2)^2\Gamma(-1+\epsilon-z_2)}{\Gamma(-1+\epsilon-2z_2)} \, ,
\end{align}
where we have moved the contours so that all gamma functions have positive argument.

The poles in $z_1$ and $z_2$ are independent. There is only one pole giving a contribution to order $\epsilon^{-1}$, situated at $z_1=-\tfrac{1}{2}+\tfrac{\epsilon}{2}$ and $z_2=-1+\epsilon$. 
We obtain:

\begin{equation}
S_1=-\frac{2\pi^2}{\epsilon(4\pi)^6}+\mathcal{O}(\epsilon^0) \, .
\end{equation}

\paragraph{Long-range: $\zeta=\tfrac{d+\epsilon}{3}$.}

With the same method, we obtain in the long-range case:

\begin{equation}
S_{d/3}=\frac{1}{2\epsilon(4\pi)^{2d}}\frac{\Gamma(-\tfrac{d}{6})\Gamma(\tfrac{d}{6})^4}{\Gamma(\tfrac{d}{3})^4\Gamma(\tfrac{2d}{3})\Gamma(\tfrac{d}{2})} +\mathcal{O}(\epsilon^0) \, .
\end{equation}

\subsubsection{$I_{\zeta}$ integral}

We now compute the $I$ integral:

\begin{align}
I_{\zeta}=&\frac{1}{\Gamma(\zeta)^6(4\pi)^{2d}}\int da_1da_2db_1db_2 dc_1dc_2 (a_1a_2b_1b_2c_1c_2)^{\zeta-1}e^{-(a_1+a_2+b_1+b_2+c_1+c_2)}\crcr
&\quad \times \frac{1}{\left[c_1c_2(a_1+a_2)(b_1+b_2)+b_1b_2(a_1+a_2)(c_1+c_2)+a_1a_2(b_1+b_2)(c_1+c_2)\right]^{\tfrac{d}{2}}} \, .
\end{align}

The denominator can be written as:

\begin{align}
&\frac{1}{\left[c_1c_2(a_1+a_2)(b_1+b_2)+b_1b_2(a_1+a_2)(c_1+c_2)+a_1a_2(b_1+b_2)(c_1+c_2)\right]^{\tfrac{d}{2}}}= \crcr
&\int \int [dz_1][dz_2] \frac{\Gamma(-z_1)\Gamma(-z_2)\Gamma(z_1+z_2+\tfrac{d}{2})}{\Gamma(\tfrac{d}{2})}\crcr
& \qquad \times \frac{(a_1a_2(b_1+b_2)(c_1+c_2))^{z_1}(b_1b_2(a_1+a_2)(c_1+c_2))^{z_2}}{(c_1c_2(a_1+a_2)(b_1+b_2))^{z_1+z_2+\tfrac{d}{2}}} \, .
\end{align}

\paragraph{Short-range: $\zeta=1$, $d=3-\epsilon$.}

Integrating the Schwinger parameters, we find in the short-range case:

\begin{align}
&I_1=\frac{1}{\Gamma(\tfrac{3}{2})(4\pi)^6}\int_{-\tfrac{1}{2}^-} [dz_1] \int_{-\tfrac{1}{2}^-} [dz_2] \Gamma(-z_1)\Gamma(-z_2)\Gamma(z_1+z_2+\tfrac{3-\epsilon}{2})\crcr 
& \times \frac{\Gamma(\tfrac{\epsilon-1}{2}-z_1-z_2)^2\Gamma(-1+\epsilon-z_1-z_2)}{\Gamma(-1+\epsilon-2z_1-2z_2)}\frac{\Gamma(1+z_2)^2\Gamma(\tfrac{1+\epsilon}{2}+z_2)}{\Gamma(2+2z_2)}\frac{\Gamma(1+z_1)^2\Gamma(\tfrac{1+\epsilon}{2}+z_1)}{\Gamma(2+2z_1)} \, .
\end{align}

The poles in $z_1$ and $z_2$ are not independent. Between, $-\frac{1}{2}^-$ and $\frac{1}{2}^-$, there are three poles in $z_1$: $0\; ,\, -\tfrac{1}{2}+\tfrac{\epsilon}{2}-z_2 \; , \, -1+\epsilon-z_2$. We obtain:

\begin{align}
I_1=&\frac{1}{\Gamma(\tfrac{3}{2})(4\pi)^6}\int_{-\tfrac{1}{2}^-} [dz_2]\Gamma(-z_2)\frac{\Gamma(1+z_2)^2\Gamma(\tfrac{1+\epsilon}{2}+z_2)}{\Gamma(2+2z_2)} \crcr
& \qquad \times \Bigg[ \Gamma(z_2+\tfrac{3-\epsilon}{2})\Gamma(\tfrac{1+\epsilon}{2})\frac{\Gamma(-\tfrac{1}{2}+\tfrac{\epsilon}{2}-z_2)^2\Gamma(-1+\epsilon-z_2)}{\Gamma(-1+\epsilon-2z_2)}\crcr
& \qquad \quad + \Gamma(\tfrac{1}{2}-\tfrac{\epsilon}{2}+z_2)\Gamma(-\tfrac{1}{2}+\tfrac{\epsilon}{2})\frac{\Gamma(\tfrac{1}{2}+\tfrac{\epsilon}{2}-z_2)^2\Gamma(\epsilon-z_2)}{\Gamma(1+\epsilon-2z_2)}\crcr
& \qquad\quad + \Gamma(1-\epsilon+z_2)\Gamma(\tfrac{1}{2}+\tfrac{\epsilon}{2})\frac{\Gamma(\tfrac{1-\epsilon}{2})^2\Gamma(\epsilon-z_2)^2\Gamma(-\tfrac{1}{2}+\tfrac{3\epsilon}{2}-z_2)}{\Gamma(1-\epsilon)\Gamma(2\epsilon-2z_2)} \Bigg] \crcr
& + \frac{1}{\Gamma(\tfrac{3}{2})(4\pi)^6}\int_{\tfrac{1}{2}^-} [dz_1] \int_{-\tfrac{1}{2}^-} [dz_2] \Gamma(-z_1)\Gamma(-z_2)\Gamma(z_1+z_2+\tfrac{3-\epsilon}{2})\crcr 
& \qquad \times \frac{\Gamma(\tfrac{\epsilon-1}{2}-z_1-z_2)^2\Gamma(-1+\epsilon-z_1-z_2)}{\Gamma(-1+\epsilon-2z_1-2z_2)}\crcr
& \qquad \times \frac{\Gamma(1+z_2)^2\Gamma(\tfrac{1+\epsilon}{2}+z_2)}{\Gamma(2+2z_2)}\frac{\Gamma(1+z_1)^2\Gamma(\tfrac{1+\epsilon}{2}+z_1)}{\Gamma(2+2z_1)} .
\end{align}

The remaining double integral is $\mathcal{O}(\epsilon^0)$. There is one contribution of order $\mathcal{O}(\epsilon^{-1})$ from the first term from the pole $z_2=-\tfrac{1}{2}+\tfrac{\epsilon}{2}$:
\begin{equation}
\frac{2}{\Gamma(\tfrac{3}{2})(4\pi)^6}\Gamma(-\tfrac{1}{2})\Gamma(\tfrac{1}{2})^4\Gamma(\epsilon)+\mathcal{O}(\epsilon^0) \, .
\end{equation}

For the second term, there are three singular contributions from the poles at $z_2=0$, $z_2=-\tfrac{1}{2}+\tfrac{\epsilon}{2}$ and $z_2=\epsilon$:
\begin{equation}
\frac{1}{\Gamma(\tfrac{3}{2})(4\pi)^6}\Bigg[\Gamma(-\tfrac{1}{2})\Gamma(\tfrac{1}{2})^4\Gamma(\epsilon)+\Gamma(\tfrac{1}{2})^4\Gamma(-\tfrac{1}{2})\Gamma(-\epsilon)-2\Gamma(\tfrac{1}{2})^4\Gamma(-\tfrac{1}{2})\Gamma(\epsilon)\Bigg] \, .
\end{equation}

For the third term, again three poles give singular contributions, $z_2=0$, $z_2=-\tfrac{1}{2}+\tfrac{3\epsilon}{2}$ and $z_2=\epsilon$:
\begin{equation}
\frac{1}{\Gamma(\tfrac{3}{2})(4\pi)^6}\Bigg[\Gamma(-\tfrac{1}{2})\Gamma(\tfrac{1}{2})^4\frac{\Gamma(\epsilon)^2}{\Gamma(2\epsilon)}+2\Gamma(\tfrac{1}{2})^4\Gamma(-\tfrac{1}{2})\Gamma(-\epsilon)+\Gamma(\tfrac{1}{2})^9\Gamma(2\epsilon)\Bigg]\, .
\end{equation}

Putting all singular contributions together, we finally obtain:
\begin{equation}
I_1=\frac{1}{(4\pi)^6}\frac{\pi^4}{\epsilon} +\mathcal{O}(\epsilon^0) \, .
\end{equation}

\paragraph{Long-range: $\zeta=\tfrac{d+\epsilon}{3}$.}

Using the same computation method, we find in the long-range case:

\begin{equation}
I_{d/3}=\frac{1}{(4\pi)^{2d}}\frac{\Gamma(\tfrac{d}{6})^9}{2\epsilon\Gamma(\tfrac{d}{3})^9\Gamma(\tfrac{d}{2})} + \mathcal{O}(\epsilon^0) \, .
\end{equation}

\subsubsection{$J_{\zeta}$ integral}

We now compute the $J_{\zeta}$ integral:

\begin{align}
J_{\zeta}=&\frac{1}{\Gamma(\zeta)^6(4\pi)^{2d}}\int da_1da_2da_3db_1db_2dc \, (a_1a_2a_3b_1b_2c)^{\zeta-1}e^{-(a_1+a_2+a_3+b_1+b_2+c)}\crcr
& \qquad \qquad \times \frac{1}{\left[a_1a_2a_3(b_1+b_2)+(c(b_1+b_2)+b_1b_2)(a_1a_2+a_1a_3+a_2a_3)\right]^{\tfrac{d}{2}}} \, .
\end{align}

This integral has a leading divergence in $\mathcal{O}(\epsilon^{-2})$. We thus have to compute the first two singular contributions. 
We write the denominator as:

\begin{align}
&\frac{1}{\left[a_1a_2a_3(b_1+b_2)+b_1b_2(a_1a_2+a_1a_3+a_2a_3)+c(b_1+b_2)(a_1a_2+a_1a_3+a_2a_3)\right]^{\tfrac{d}{2}}} = \crcr
& \int \int \int [dz_1] [dz_2][dz_3] \frac{\Gamma(-z_1)\Gamma(-z_2)\Gamma(-z_3)\Gamma(z_1+z_2+\tfrac{d}{2})\Gamma(z_3+z_2+\tfrac{d}{2})}{\Gamma(\tfrac{d}{2})\Gamma(z_2+\tfrac{d}{2})} \crcr
& \qquad \times \frac{(b_1b_2)^{z_1}(a_1a_2a_3(b_1+b_2)^{z_2}(a_2a_3)^{z_3}}{(c(b_1+b_2))^{z_1+z_2+\tfrac{d}{2}}(a_1(a_2+a_3))^{z_2+z_3+\tfrac{d}{2}}} \, .
\end{align}

We can then integrate the Schwinger parameters using \eqref{eq:int_class}. 

\paragraph{Short-range: $\zeta=1$, $d=3-\epsilon$.}

We obtain:

\begin{align}
J_1&=\frac{1}{(4\pi)^6\Gamma(\tfrac{3-\epsilon}{2})}\int_{-\tfrac{1}{2}^-} [dz_1]\int_{0^-} [dz_2]\int_{-\tfrac{1}{2}^-} [dz_3] \frac{\Gamma(-z_1)\Gamma(-z_2)\Gamma(-z_3)}{\Gamma(z_2+\tfrac{3-\epsilon}{2})}\Gamma(z_1+z_2+\tfrac{3-\epsilon}{2}) \crcr
& \qquad \times \Gamma(z_2+z_3+\tfrac{3-\epsilon}{2})\Gamma(\tfrac{-1+\epsilon}{2}-z_1-z_2)\Gamma(\tfrac{-1+\epsilon}{2}-z_3)\crcr
&\qquad \times \frac{\Gamma(1+z_1)^2\Gamma(\tfrac{1+\epsilon}{2}+z_1)}{\Gamma(2+2z_1)}\frac{\Gamma(1+z_2+z_3)^2\Gamma(\tfrac{1+\epsilon}{2}+z_2+z_3)}{\Gamma(2+2z_2+2z_3)} \, .
\end{align}

Between $-\frac{1}{2}^-$ and $\frac{1}{2}^-$, there are two poles in $z_3$: $z_3=0$ and $z_3=-1/2+\epsilon/2$. We thus have:

\begin{align}
J_1&=\frac{1}{(4\pi)^6\Gamma(\tfrac{3-\epsilon}{2})}\int_{-\tfrac{1}{2}^-} [dz_1]\int_{0^-} [dz_2]\Gamma(-z_1)\Gamma(-z_2)\frac{\Gamma(z_1+z_2+\tfrac{3-\epsilon}{2})}{\Gamma(z_2+\tfrac{3-\epsilon}{2})}\Gamma(\tfrac{-1+\epsilon}{2}-z_1-z_2) \crcr
& \qquad \times \frac{\Gamma(1+z_1)^2\Gamma(\tfrac{1+\epsilon}{2}+z_1)}{\Gamma(2+2z_1)}\Bigg[ \Gamma(-\tfrac{1}{2}+\tfrac{\epsilon}{2})\Gamma(z_2+\tfrac{3-\epsilon}{2})\frac{\Gamma(1+z_2)^2\Gamma(\tfrac{1+\epsilon}{2}+z_2)}{\Gamma(2+2z_2)} \crcr
& \qquad \qquad + \Gamma(\tfrac{1}{2}-\tfrac{\epsilon}{2})\Gamma(1+z_2)\frac{\Gamma(\tfrac{1}{2}+\tfrac{\epsilon}{2}+z_2)^2\Gamma(z_2+\epsilon)}{\Gamma(1+\epsilon+2z_2)}\Bigg] \crcr
&+\frac{1}{(4\pi)^6\Gamma(\tfrac{3-\epsilon}{2})}\int_{-\tfrac{1}{2}^-} [dz_1]\int_{0^-} [dz_2]\int_{\tfrac{1}{2}^-} [dz_3] \frac{\Gamma(-z_1)\Gamma(-z_2)\Gamma(-z_3)}{\Gamma(z_2+\tfrac{3-\epsilon}{2})}\Gamma(z_1+z_2+\tfrac{3-\epsilon}{2}) \crcr
&\qquad  \times \Gamma(z_2+z_3+\tfrac{3-\epsilon}{2})\Gamma(\tfrac{-1+\epsilon}{2}-z_1-z_2)\Gamma(\tfrac{-1+\epsilon}{2}-z_3)\crcr
& \qquad \times\frac{\Gamma(1+z_1)^2\Gamma(\tfrac{1+\epsilon}{2}+z_1)}{\Gamma(2+2z_1)}\frac{\Gamma(1+z_2+z_3)^2\Gamma(\tfrac{1+\epsilon}{2}+z_2+z_3)}{\Gamma(2+2z_2+2z_3)} \, .
\label{eq:J_1}
\end{align}

Let us call $J_a$ the first term in \eqref{eq:J_1}. The first two poles in $z_2$ are located at $0$ and $-\frac{1}{2}+\frac{\epsilon}{2}-z_1$. We obtain:

\begin{align}
J_a=&\frac{1}{(4\pi)^6\Gamma(\tfrac{3-\epsilon}{2})}\int_{-\tfrac{1}{2}^-} [dz_1]\Gamma(-z_1)\frac{\Gamma(1+z_1)^2\Gamma(\tfrac{1+\epsilon}{2}+z_1)}{\Gamma(2+2z_1)} \crcr
& \qquad \times \Bigg[ \Gamma(z_1+\tfrac{3-\epsilon}{2})\Gamma(-\tfrac{1}{2}+\tfrac{\epsilon}{2}-z_1)\Gamma(\tfrac{1}{2}+\tfrac{\epsilon}{2}) \crcr
& \qquad \qquad + \Gamma(-\tfrac{1}{2}+\tfrac{\epsilon}{2})\Gamma(\tfrac{1}{2}-\tfrac{\epsilon}{2}+z_1)\frac{\Gamma(-\tfrac{1}{2}+\tfrac{\epsilon}{2}-z_1)\Gamma(\epsilon-z_1)}{\Gamma(1+\epsilon-2z_1)} \Bigg] \crcr
& +\frac{1}{(4\pi)^6\Gamma(\tfrac{3-\epsilon}{2})}\int_{-\tfrac{1}{2}^-} [dz_1]\int_{1^-} [dz_2]\Gamma(-z_1)\Gamma(-z_2)\frac{\Gamma(z_1+z_2+\tfrac{3-\epsilon}{2})}{\Gamma(z_2+\tfrac{3-\epsilon}{2})}\Gamma(\tfrac{-1+\epsilon}{2}-z_1-z_2) \crcr
& \qquad \times \frac{\Gamma(1+z_1)^2\Gamma(\tfrac{1+\epsilon}{2}+z_1)}{\Gamma(2+2z_1)} \Gamma(-\tfrac{1}{2}+\tfrac{\epsilon}{2})\Gamma(z_2+\tfrac{3-\epsilon}{2})\frac{\Gamma(1+z_2)^2\Gamma(\tfrac{1+\epsilon}{2}+z_2)}{\Gamma(2+2z_2)} \, .
\end{align}

The first term has only a singular contribution from the pole at $z_1=-\frac{1}{2}+\frac{\epsilon}{2}$. However, it is canceled by the pole at $z_1=-\frac{1}{2}+\epsilon$ of the second term. The second term has also two singular contributions from the poles at $0$ and $\epsilon$ but they cancel. The remaining double integral is also finite. 
Finally, $J_a=\mathcal{O}(\epsilon^0)$.

Let us now call $J_b$ the second term in \eqref{eq:J_1}. The first poles in $z_2$ are located at $z_2=0$ and $z_2=-\frac{1}{2}+\frac{\epsilon}{2}-z_1$. We obtain: 
\begin{align}
J_b=&\frac{1}{(4\pi)^6\Gamma(\tfrac{3-\epsilon}{2})}\int_{-\tfrac{1}{2}^-} [dz_1]\Gamma(-z_1)\Gamma(\tfrac{1}{2}-\tfrac{\epsilon}{2})\frac{\Gamma(1+z_1)^2\Gamma(\tfrac{1+\epsilon}{2}+z_1)}{\Gamma(2+2z_1)} \crcr
& \quad \quad \times \Bigg[ \frac{\Gamma(\tfrac{1}{2}+\tfrac{\epsilon}{2})^2\Gamma(\epsilon)}{\Gamma(1+\epsilon)}\Gamma(z_1+\tfrac{3-\epsilon}{2})\Gamma(\tfrac{\epsilon-1}{2}-z_1) +\frac{\Gamma(\tfrac{1-\epsilon}{2}+z_1)}{\Gamma(1-z_1)}\frac{\Gamma(\epsilon-z_1)^2\Gamma(\tfrac{3\epsilon-1}{2}-z_1)}{\Gamma(2(\epsilon-z_1))} \Bigg] \crcr
& + \frac{1}{(4\pi)^6\Gamma(\tfrac{3-\epsilon}{2})}\int_{-\tfrac{1}{2}^-} [dz_1]\int_{1^-} [dz_2]\Gamma(-z_1)\Gamma(-z_2)\frac{\Gamma(z_1+z_2+\tfrac{3-\epsilon}{2})}{\Gamma(z_2+\tfrac{3-\epsilon}{2})}\Gamma(\tfrac{-1+\epsilon}{2}-z_1-z_2) \crcr
& \quad \quad \times \frac{\Gamma(1+z_1)^2\Gamma(\tfrac{1+\epsilon}{2}+z_1)}{\Gamma(2+2z_1)}\Gamma(\tfrac{1}{2}-\tfrac{\epsilon}{2})\Gamma(1+z_2)\frac{\Gamma(\tfrac{1}{2}+\tfrac{\epsilon}{2}+z_2)^2\Gamma(z_2+\epsilon)}{\Gamma(1+\epsilon+2z_2)} \, .
\end{align}

The first term can be written as:
\begin{equation}
\frac{1}{(4\pi)^3\Gamma(\tfrac{3-\epsilon}{2})}\frac{\Gamma(\tfrac{1}{2}-\tfrac{\epsilon}{2})\Gamma(\tfrac{1}{2}+\tfrac{\epsilon}{2})^2}{\Gamma(1+\epsilon)}\Gamma(\epsilon)D_1 \, .
\end{equation}

The second term has four poles giving singular contributions. The two located at $0$ and $\epsilon$ cancel. The two located at $-\frac{1}{2}+\tfrac{3\epsilon}{2}$ and $-\tfrac{1}{2}+\epsilon$ give the following contribution:

\begin{align}
\frac{1}{(4\pi)^6\Gamma(\tfrac{3-\epsilon}{2})}\Bigg[\frac{\Gamma(\tfrac{1}{2}-\tfrac{3\epsilon}{2})\Gamma(\tfrac{1}{2}+\tfrac{3\epsilon}{2})^2\Gamma(\tfrac{1}{2}-\tfrac{\epsilon}{2})^3}{\Gamma(\tfrac{3}{2}-\tfrac{3\epsilon}{2})\Gamma(1+3\epsilon)}\Gamma(\epsilon)\Gamma(2\epsilon) -\frac{\Gamma(\tfrac{1}{2}-\tfrac{\epsilon}{2})^2\Gamma(\tfrac{1}{2}+\tfrac{\epsilon}{2})^4}{\Gamma(\tfrac{3}{2}-\tfrac{\epsilon}{2})\Gamma(1+\epsilon)^2}\Gamma(\epsilon)^2\Bigg] \, .
\end{align}

The last double integral is finite and we thus obtain:
\begin{align}
J_b&= \frac{1}{(4\pi)^3\Gamma(\frac{3-\epsilon}{2})}\frac{\Gamma(\tfrac{1}{2}-\tfrac{\epsilon}{2})\Gamma(\tfrac{1}{2}+\tfrac{\epsilon}{2})^2}{\Gamma(1+\epsilon)}\Gamma(\epsilon)D_1 \crcr
&+ \frac{1}{(4\pi)^6\Gamma(\tfrac{3-\epsilon}{2})}\Bigg[\frac{\Gamma(\tfrac{1}{2}-\tfrac{3\epsilon}{2})\Gamma(\tfrac{1}{2}+\tfrac{3\epsilon}{2})^2\Gamma(\tfrac{1}{2}-\tfrac{\epsilon}{2})^3}{\Gamma(\tfrac{3}{2}-\tfrac{3\epsilon}{2})\Gamma(1+3\epsilon)\Gamma(1-\epsilon)}\Gamma(\epsilon)\Gamma(2\epsilon)\crcr
& \qquad \quad + -\frac{\Gamma(\tfrac{1}{2}-\tfrac{\epsilon}{2})^2\Gamma(\tfrac{1}{2}+\tfrac{\epsilon}{2})^4}{\Gamma(\tfrac{3}{2}-\tfrac{\epsilon}{2})\Gamma(1+\epsilon)^2}\Gamma(\epsilon)^2\Bigg] + \mathcal{O}(\epsilon^0) \, .
\end{align}

Let us call $J_c$ the triple integral in \eqref{eq:J_1}. 
We now first integrate $z_2$, the first two poles are located at $0$ and $-\frac{1}{2}+\frac{\epsilon}{2}-z_1$. We have:
\begin{align}
J_c&=\frac{1}{(4\pi)^6\Gamma(\tfrac{3-\epsilon}{2})}\int_{-\tfrac{1}{2}^-} [dz_1]\int_{\tfrac{1}{2}^-} [dz_3] \Gamma(-z_1)\Gamma(-z_3)\Gamma(\tfrac{-1+\epsilon}{2}-z_3)\frac{\Gamma(1+z_1)^2\Gamma(\tfrac{1+\epsilon}{2}+z_1)}{\Gamma(2+2z_1)} \crcr
& \qquad \times \Bigg[ \Gamma(-\tfrac{1}{2}+\tfrac{\epsilon}{2}-z_1)\frac{\Gamma(z_1+\tfrac{3-\epsilon}{2})\Gamma(z_3+\tfrac{3-\epsilon}{2})}{\Gamma(\tfrac{3-\epsilon}{2})}\frac{\Gamma(1+z_3^2)\Gamma(\tfrac{1+\epsilon}{2}+z_3)}{\Gamma(2+2z_3)} \crcr
&\qquad \qquad + \Gamma(\tfrac{1-\epsilon}{2}+z_1)\Gamma(1+z_3-z_1)\frac{\Gamma(\tfrac{1+\epsilon}{2}+z_3-z_1)^2\Gamma(\epsilon+z_3-z_1)}{\Gamma(1-z_1)\Gamma(1+\epsilon+2z_3-2z_1)}\Bigg] \crcr
& +\frac{1}{(4\pi)^6\Gamma(\tfrac{3-\epsilon}{2})}\int_{-\tfrac{1}{2}^-} [dz_1]\int_{1^-} [dz_2]\int_{\tfrac{1}{2}^-} [dz_3] \frac{\Gamma(-z_1)\Gamma(-z_2)\Gamma(-z_3)}{\Gamma(z_2+\tfrac{3-\epsilon}{2})}\Gamma(z_1+z_2+\tfrac{3-\epsilon}{2})\crcr
&\qquad  \times \Gamma(z_2+z_3+\tfrac{3-\epsilon}{2}) \Gamma(\tfrac{-1+\epsilon}{2}-z_1-z_2)\Gamma(\tfrac{-1+\epsilon}{2}-z_3)\crcr
&\qquad \times\frac{\Gamma(1+z_1)^2\Gamma(\tfrac{1+\epsilon}{2}+z_1)}{\Gamma(2+2z_1)}\frac{\Gamma(1+z_2+z_3)^2\Gamma(\tfrac{1+\epsilon}{2}+z_2+z_3)}{\Gamma(2+2z_2+2z_3)} \, .
\end{align}

For the first term, the integrals on $z_1$ and $z_3$ are independent and the integral on $z_1$ is exactly $D_1$. The integral on $z_3$ is finite and was already computed in appendix \ref{app:D}. The contribution from this term is then:
\begin{equation}
\frac{D_1}{(4\pi)^3\Gamma(\tfrac{3-\epsilon}{2})}\pi^{\tfrac{3}{2}}\left(1+\log(\tfrac{4}{9})\right) \, .
\end{equation}

For the second term, the poles in $z_1$ and $z_3$ are no longer independent. We have five poles in $z_1$ located at $0,-\frac{1}{2}+\frac{\epsilon}{2},1+z_3,\frac{1}{2}+\frac{\epsilon}{2}+z_3,\epsilon+z_3$. However, only the pole in $z_1=-\frac{1}{2}+\frac{\epsilon}{2}$ leads to a singular contribution:
\begin{align}
&-\frac{\Gamma(\tfrac{1}{2}-\tfrac{\epsilon}{2})\Gamma(\tfrac{1}{2}+\tfrac{\epsilon}{2})\Gamma(\epsilon)}{\Gamma(1+\epsilon)(4\pi)^6\Gamma(\tfrac{3-\epsilon}{2})}\int_{\tfrac{1}{2}^-} [dz_3] \frac{\Gamma(-z_3)\Gamma(-\tfrac{1}{2}+z_3)\Gamma(\tfrac{1}{2}+z_3)\Gamma(1+z_3)^2\Gamma(\tfrac{3}{2}+z_3)}{\Gamma(2+2z_3)}\crcr
&=-\frac{\Gamma(\tfrac{1}{2}-\tfrac{\epsilon}{2})\Gamma(\tfrac{1}{2}+\tfrac{\epsilon}{2})^2\Gamma(\epsilon)}{\Gamma(1+\epsilon)(4\pi)^6\Gamma(\tfrac{3-\epsilon}{2})^2}\pi^{\tfrac{3}{2}}(1+\log(\tfrac{4}{9})) \, ,
\end{align}
where the integral on $z_3$ is the same as in \eqref{eq:part_int}. 

We can then show, with a long but straightforward computation, that the remaining triple integral is finite. 
Gathering all contributions, we finally obtain for $J_1$: 
\begin{align}
J_1=\frac{1}{(4\pi)^6}&\Bigg[ (4\pi)^3\frac{\Gamma(\tfrac{1}{2}-\tfrac{\epsilon}{2})\Gamma(\tfrac{1}{2}+\tfrac{\epsilon}{2})^2\Gamma(\epsilon)}{\Gamma(\tfrac{3}{2}-\tfrac{\epsilon}{2})\Gamma(1+\epsilon)}D_1-\frac{\Gamma(\tfrac{1}{2}-\tfrac{\epsilon}{2})^2\Gamma(\tfrac{1}{2}+\tfrac{\epsilon}{2})^4\Gamma(\epsilon)^2}{\Gamma(\tfrac{3}{2}-\tfrac{\epsilon}{2})^2\Gamma(1+\epsilon)^2}\crcr
& \quad + \frac{\Gamma(\tfrac{1}{2}-\tfrac{3\epsilon}{2})\Gamma(\tfrac{1}{2}-\tfrac{\epsilon}{2})^3\Gamma\tfrac{1}{2}+\tfrac{3\epsilon}{2})^2\Gamma(\epsilon)\Gamma(2\epsilon)}{\Gamma(\tfrac{3}{2}-\tfrac{\epsilon}{2})\Gamma(\tfrac{3}{2}-\tfrac{3\epsilon}{2})\Gamma(1+3\epsilon)}\crcr
& \quad + \frac{\pi^{\tfrac{3}{2}}\left(1+\log(\tfrac{4}{9})\right)}{\Gamma(\tfrac{3-\epsilon}{2})}\left((4\pi)^3D_1-\frac{\Gamma(\tfrac{1}{2}-\tfrac{\epsilon}{2})\Gamma(\tfrac{1}{2}+\tfrac{\epsilon}{2})^2\Gamma(\epsilon)}{\Gamma(1+\epsilon)\Gamma(\tfrac{3}{2}-\tfrac{\epsilon}{2})}\right) \Bigg] + \mathcal{O}(\epsilon^0)\, .
\end{align}

We then obtain for $\alpha_{J_1}$:
\begin{equation}
\alpha_{J_1}= \frac{2\pi^2}{3}\left[ \psi(\tfrac{1}{2})-\psi(\tfrac{3}{2})\right] =-\frac{4\pi^2}{3} \, .
\end{equation}

\paragraph{Long-range $\zeta=\frac{d+\epsilon}{3}$.}

Using the same method, we find for $\alpha_{J_{d/3}}$ in the long-range case:
\begin{equation}
\alpha_{J_{d/3}}=\frac{\Gamma(\tfrac{d}{6})^6}{6\Gamma(\tfrac{d}{2})^2\Gamma(\tfrac{d}{3})^6}\Bigg[\psi(\tfrac{d}{6})-\psi(1)+\psi(\tfrac{d}{3})-\psi(\tfrac{d}{2})\Bigg] \, .
\end{equation}

\section{Comparison with the sextic $O(N)$ model}
\label{ap:O(N)}

In this appendix, we compare our results with the beta functions of the sextic $O(N)$ model up to four loops. This comparison is not straight-forward. First, we specify the symmetry of the generic sextic multi-scalar model of section~\ref{sec:sexticMS} to $U(N)$. Then, we notice that for vector fields $U(N)$ symmetry is equivalent to $O(2N)$ symmetry. We thus substitute $N \rightarrow M/2$ in the $U(N)$ beta functions in order to compare with known results for the $O(N)$ beta functions.

Let us first specify the symmetry in \eqref{eq:beta_SR} to $U(N)$ by setting:
\begin{equation}
\tilde{g}_{\mba \mbb  \mbc ;\mbd \mbe\mbf}=\frac{\tilde{g}}{6}\left(\delta_{ad}\delta_{be}\delta_{cf}+\delta_{ad}\delta_{bf}\delta_{ce}+\delta_{ae}\delta_{bd}\delta_{cf}+\delta_{ae}\delta_{bf}\delta_{cd}+\delta_{af}\delta_{bd}\delta_{ce}+\delta_{af}\delta_{be}\delta_{cd}\right) \; ,
\end{equation}
where each index is now a single index going from $1$ to $N$ and $\mathcal{N}=N$.

We then obtain the following beta function up to cubic order in $\tilde{g}$:
\begin{align}
\beta_{U}=&-2\epsilon \tilde{g}+\frac{\tilde{g}^2}{48\pi^2}\left(3N+11\right)\crcr
& +\frac{\tilde{g}^3}{9216\pi^4}\left(53N^2-429N-826-\frac{\pi^2}{4}\left(N^3+17N^2+155N+340\right)\right) \, . 
\end{align}

We now substitute $N \rightarrow M/2$ and rescale the coupling by $\tilde{g}=\frac{g_O}{5\pi}$ in order to match the conventions of \cite{hager2002six}. We finally obtain:
\begin{align}
\beta_O=&-2\epsilon g_O +\frac{2g_O^2}{15}\left(3M+22\right) \crcr
& -\frac{g_O^3}{1800}\left(8\left(53M^2+858M+3304\right)+\pi^2\left(M^3+34M^2+620M+2720\right)\right) \, ,
\end{align}
which agrees with the results of \cite{hager2002six}.
\end{subappendices}

\chapter{Large-$N$ limit of irreducible random tensors in rank $5$}
\label{chap:rank5}
In this chapter we generalize in rank $5$ the proof of the existence of a large-$N$ melonic limit for irreducible tensor models in rank $3$ of \cite{Benedetti:2017qxl, Carrozza:2018ewt, Carrozza:2018psc}.
In section \ref{sec:model}, we introduce the models and our main results: theorem~\ref{theorem_princ} and theorem~\ref{theorem_LO}. In section \ref{sec:expansion}, we perform the perturbative expansion and define two different types of diagrams: Feynman maps and stranded graphs. We also introduce in more detail the problematic subgraphs that could potentially lead to violations of the maximum scaling in $N$. In section \ref{sec:tmd}, we introduce the important notion of \textit{boundary graph}, as well as various particular subgraphs that will play important roles in the rest of the proof. In section \ref{sec:subtraction}, we perform the explicit subtraction of melons and double-tadpoles. We then arrive to an equivalent theory with renormalized covariance in which melons and double-tadpoles have been subtracted from the Feynman expansion. In section \ref{sec:deletions}, we establish a number of lemmas and propositions enabling the recursive deletion of certain subgraphs, that will be instrumental to ultimately prove theorem~\ref{theorem_princ}. Finally, in section \ref{sec:LO}, we prove theorem~\ref{theorem_LO} and show that melons dominate the large-$N$ expansion. 
In appendix~\ref{ap:bounds}, we prove some useful bounds on the number of faces of stranded graphs, while appendix~\ref{app:particular} provides a proof of lemma~\ref{lemma:particular_cases} (which handles a number of particular cases). 

%
%

\section{The models and the main results}
\label{sec:model}

We consider a real tensor $T_{abcde}$, transforming in the tensor product of five fundamental representations of the orthogonal group $O(N)$ (hence, $a,b,c,d,e=1,\dots,N$).
The action of the symmetric group $\mathcal{S}_5$ and the trace operation allows to decompose $T_{abcde}$ into irreducible components, which are themselves tensors of rank $5$, $3$ or $1$.  
In this chapter, we will focus on the seven inequivalent representations of rank $5$. They are the traceless representations with index symmetry given by the following Young tableaux.
$$
\begin{Young}  \cr  \cr  \cr \cr  \cr  \end{Young}\quad \begin{Young}  & & & & \cr \end{Young}\quad 
\begin{Young}  & & &  \cr \cr \end{Young}\quad 
\begin{Young}  & &  \cr \cr \cr \end{Young} \quad \begin{Young}  &   \cr \cr \cr \cr\end{Young}\quad \begin{Young}  & &  \cr & \cr \end{Young}\quad 
\begin{Young}  &   \cr & \cr \cr \end{Young}
$$
The first two correspond to the antisymmetric and symmetric traceless representations, respectively, while the other five have mixed index permutation symmetries (that is, they carry representations of $\mathcal{S}_5$ of dimension higher than $2$). Given such a representation, a central object in our construction will be the orthogonal projector on the associated linear subspace of tensors, with respect to the canonical scalar product: $\langle T \vert T' \rangle := T_{abcde} T'_{abcde}$. The kernel of this projector will serve as a degenerate covariance, it is therefore crucial for it to be symmetric. As an illustration, let us find the orthogonal projectors on completely symmetric traceless tensors and completely antisymmetric tensors.

 \paragraph{ Symmetric traceless representation.}
 
Let us start from a completely symmetric tensor $T_{abcde}$. We can decompose $T$ into a traceless part $T^0_{abcde}$, a symmetric traceless tensor of rank tree $T^1_{abc}$, and a vector $T^2_a$.\footnote{Here we use the notation of footnote \ref{fn:sym} In particular, we have 
\begin{equation*}
T^{1}_{(cde}\delta_{ab)}=\frac{1}{10}\left(T^1_{cde}\delta_{ab}+T^1_{bde}\delta_{ac}+T^1_{bce}\delta_{ad}+T^1_{bcd}\delta_{ae}+T^1_{ade}\delta_{bc}
 +T^1_{ace}\delta_{bd}+T^1_{acd}\delta_{be}+T^1_{abe}\delta_{cd}+T^1_{abd}\delta_{ce}+T^{1}_{abc}\delta_{de}\\\right).
\end{equation*}
}:
\begin{equation*}
T_{abcde}=T^0_{abcde}+10T^1_{(cde}\delta_{ab)}
+ 30T^2_{(a}\delta_{bc}\delta_{de)}\,.
\end{equation*} 
By taking successive traces over pairs of indices, we obtain 
\begin{equation*}
T_{abcdd}=(N+6)T^1_{abc}+2(N+4)(T^2_{a}\delta_{bc}+T^2_{b}\delta_{ac}+T^2_{c}\delta_{ab})\, ,
\end{equation*}
and
\begin{equation*}
T_{abbdd}=2(N+2)(N+4)T^2_{a}\,,
\end{equation*}
which, combined, leads to the following expressions for $T^1$ and $T^2$:
\begin{align*}
T^1_{abc}&=\frac{1}{N+6}\left(T_{abcdd}-\frac{1}{(N+2)}\left(T_{addee}\delta_{bc}+T_{bddee}\delta_{ac}+T_{cddee}\delta_{ab}\right)\right) \,, \\ 
T^2_a &= \frac{1}{2(N+2)(N+4)} T_{abbdd}\,.
\end{align*}
This allows us to define a projector on symmetric traceless tensors as the projector on symmetric tensors minus the projector on traces. We find:
\begin{align}
\pmb S_{\pmb a,\pmb b}=\frac{1}{5!}&\left[\sum_{\sigma \in \mathcal{S}_5}\prod_{i=1}^5\delta_{a_ib_{\sigma(j)}}-\frac{2}{N+6}\sum_{\substack{\{i_1,i_2,i_3\} \cup\{i_4,i_5\} \\ = \llbracket 1,5 \rrbracket}}\sum_{\substack{\{j_1,j_2,j_3\}\cup\{j_4,j_5\}\\ = \llbracket 1,5 \rrbracket}}\delta_{a_{i_4}a_{i_5}}\delta_{b_{j_4}b_{j_5}}\sum_{\sigma \in \mathcal{S}_3}\prod_{k=1}^3\delta_{a_{i_k}b_{j_{\sigma(k)}}}\right. \crcr
&\left.+\frac{2}{(N+4)(N+6)}\sum_{\substack{\{i_1\}\cup\{i_2,i_3\}\cup\{i_4,i_5\} \\ =\llbracket 1,5 \rrbracket}}\sum_{\substack{\{j_1\}\cup\{j_2,j_3\}\cup\{j_4,j_5\} \\= \llbracket 1,5 \rrbracket}}\delta_{a_{i_1}b_{j_1}}\delta_{a_{i_2}a_{i_3}}\delta_{a_{i_4}a_{i_5}}\delta_{b_{j_2}b_{j_3}}\delta_{b_{j_4}b_{j_5}}\right] \,,
\label{sym}
\end{align}
where we use again the short-hand notation $\pmb a=(a_1,a_2,a_3,a_4,a_5)$, $\pmb b=(b_1,b_2,b_3,b_4,b_5)$ (and so on).
Moreover, one can readily check that $\pmb S_{\pmb a,\pmb b} = \pmb S_{\pmb a,\pmb b}$, so that $\pmb S$ is the looked-for orthogonal projector.

\paragraph{Antisymmetric representation.}

The orthogonal projector on completely antisymmetric tensors takes the form:
\begin{equation}\label{antisym}
\pmb A_{\pmb a,\pmb b}=\frac{1}{5!}\sum_{\sigma \in \mathcal{S}_5}\epsilon(\sigma)\prod_{i=1}^5\delta_{a_ib_{\sigma(j)}} \, .
\end{equation}

A covariance for the other five inequivalent irreducible representations can be obtained in a similar fashion. As it turns out, the explicit form of this projector is not necessary for our proofs to go through, so we only provide a brief sketch of the general construction. As a first step, one can construct the canonical projector associated to the target Young tableau, by first symmetrizing over indices appearing in a same row, then antisymmetrizing over indices appearing in a same column. After projecting out the trace components, one obtains a projector with the desired image. However, in contrast to what happened with the completely symmetric and symmetric traceless representations, this first projector will not in general be orthogonal. If so, one needs to orthogonalize it as a last step in the construction.  

\

The generic tensor model with $5$-simplex (or complete graph) interaction is defined by the action:
\begin{align}
&S(T)=\frac{1}{2}\sum_{a_1, \, \ldots, \, a_5}T_{a_1a_2a_3a_4a_5}T_{a_1a_2a_3a_4a_5}\crcr
& -\frac{\lambda}{6N^5}\sum_{a_1, \, \ldots, \, a_{15}}T_{a_1a_2a_3a_4a_5}T_{a_5a_6a_7a_8a_9}T_{a_{9}a_{4}a_{10}a_{11}a_{12}}T_{a_{12}a_{8}a_{3}a_{13}a_{14}}T_{a_{14}a_{11}a_{7}a_{2}a_{15}}T_{a_{15}a_{13}a_{10}a_{6}a_{1}} \, .
\end{align}
We will denote the $5$-simplex pattern of contraction by
\begin{equation}\label{eq:5-simplex}
\delta^h_{\pmb a\pmb b\pmb c\pmb d\pmb e\pmb f}=\delta_{a_1f_5}\delta_{a_2e_4}\delta_{a_3d_3}\delta_{a_4c_2}\delta_{a_5b_1}\delta_{b_2f_4}\delta_{b_3e_3}\delta_{b_4d_2}\delta_{b_5c_1}\delta_{c_3f_3}\delta_{c_4e_2}\delta_{c_5d_1}\delta_{d_4f_2}\delta_{d_5e_1}\delta_{e_5f_1}\,,
\end{equation}
such that the action can be simplified to:
\begin{equation}
S(T)=\frac{1}{2}T_{\pmb a}T_{\pmb a} -\frac{\lambda}{6N^5}\delta^h_{\pmb a\pmb b\pmb c\pmb d\pmb e\pmb f}T_{\pmb a}T_{\pmb b}T_{\pmb c}T_{\pmb d}T_{\pmb e}T_{\pmb f}\,.
\end{equation}

We also denote by $\pmb 1$ the identity operator $\pmb 1_{\pmb a, \pmb b}=\pmb 1_{a_1a_2a_3a_4a_5,b_1b_2b_3b_4b_5}=\prod_{i=1}^5\delta_{a_ib_i}=\delta_{\pmb a\pmb b}$, and by $\partial_T$ the tensor of derivative operators $(\partial_T)_{\pmb a}\equiv \frac{\partial}{\partial T_{\pmb a}}$.
With these notations at hand, we can write the partition function, the free energy and its first derivative as:
\begin{align}
Z_{\pmb 1}(\lambda)&=\int [dT]e^{-S(T)}=\left[e^{\frac{1}{2}\partial_T\pmb 1 \partial_T}e^{\frac{\lambda}{6N^5}\delta^h_{\pmb a\pmb b\pmb c\pmb d\pmb e\pmb f}T_{\pmb a}T_{\pmb b}T_{\pmb c}T_{\pmb d}T_{\pmb e}T_{\pmb f}}\right]_{T=0} ~,~\crcr
\ln Z_{\pmb 1}(\lambda)&=\ln \lbrace \int [dT]e^{-S(T)} \rbrace , ~~F_{\pmb 1}(\lambda)=\frac{6}{N^{5}}\lambda\partial_{\lambda}\ln Z_{\pmb 1}(\lambda) \,.
\end{align}
In the following, $\pmb P$ will denote any one of the seven orthogonal projectors on irreducible rank-$5$ tensor representations. We will sometimes illustrate our calculations with $\pmb P =\pmb A$ or $\pmb S$, but our main results hold for any irreducible representation. The irreducible tensor model of interest can be obtained from the generic model by disallowing the propagation of modes which are in the kernel of $\pmb P$. This in turn amounts to replacing the non-degenerate covariance $\pmb 1$ by the degenerate covariance $\pmb P$:
\begin{equation}
F_{\pmb P}(\lambda)=\frac{6}{N^{5}}\lambda\partial_{\lambda}\ln \left\lbrace \left[e^{\frac{1}{2}\partial_T \pmb P \partial_T}e^{\frac{\lambda}{6N^5}\delta^h_{\pmb a\pmb b\pmb c\pmb d\pmb e\pmb f}T_{\pmb a}T_{\pmb b}T_{\pmb c}T_{\pmb d}T_{\pmb e}T_{\pmb f}}\right]_{T=0}\right\rbrace \; . 
\label{eq:model1}
\end{equation}
Note that, in this equation, the tensor $T$ has no symmetry property under permutation of its indices. However, as only the projected modes $\pmb P T$ propagate, we can equivalently change variables to $P=\pmb P T$ as done in \cite{Benedetti:2017qxl}. We can then write: 
\begin{align}
F_{\pmb P}(\lambda)&=\frac{6}{N^{5}}\lambda\partial_{\lambda}\ln \left\lbrace \left[e^{\frac{1}{2}\partial_P \pmb P \partial_P}e^{\frac{\lambda}{6N^5}\delta^h_{\pmb a\pmb b\pmb c\pmb d\pmb e\pmb f}P_{\pmb a}P_{\pmb b}P_{\pmb c}P_{\pmb d}P_{\pmb e}P_{\pmb f}}\right]_{P=0}\right\rbrace \; , \crcr
\frac{\partial}{\partial P_{\pmb a}}P_{\pmb b} & \equiv \pmb P_{\pmb a, \pmb b} \; ,
\label{eq:model}
\end{align}
where the tensor $P$ is in the image of $\pmb P$ and thereby irreducible, and the second line is a definition. The factor $6/N^5$ is for later convenience; it will in particular ensure that $F_{{\pmb P}}$ is an order $1$ quantity in the large-$N$ limit.  

\subsection{Main theorems}

The main result of this chapter is the existence of a $1/N$ expansion for all seven irreducible rank-$5$ tensor models with complete graph interaction. It is given by the following theorem.

\begin{theorem}
We have (in the sense of perturbation series):
\begin{equation}\label{eq:largeN}
F_{\pmb P}(\lambda)=\sum_{\omega \in \mathbb{N}} N^{-\omega}F_{\pmb P}^{(\omega)}(\lambda)\,.
\end{equation}
\label{theorem_princ}
\end{theorem}
\begin{proof}
This follows from \eqref{eq:model_subtracted}, remark~\ref{rem:broken_unbroken} and proposition~\ref{prop:positive_degree}. 
\end{proof}

In section \ref{sec:LO}, we further prove that these models are dominated by melon diagrams. This is given by the following theorem.

\begin{theorem}
In equation \eqref{eq:largeN}, the leading order contribution $F_{\pmb P}^{(0)}(\lambda)$ is a sum over melonic stranded graphs. For small enough $\lambda$, it is the unique continuous solution of the polynomial equation
\begin{equation}
1- X + m_{\pmb P} \lambda^2 X^6  = 0 \, ,
\end{equation} 
such that $F_{\pmb P}^{(0)}(0)=1$, and where $m_{\pmb P}$ is a model-specific real constant. In particular, $m_{\pmb S} = m_{\pmb A} = \left( \frac{1}{5!} \right)^4$.
\label{theorem_LO}
\end{theorem}

\begin{proof}
This follows from proposition~\ref{propo:LO}, as well as \eqref{eq:m} and \eqref{eq:sde} in section~\ref{sec:subtraction}.
\end{proof}

\medskip

For completeness, we note that we could consider other $5$-simplex interactions, that differ from our choice in \eqref{eq:5-simplex} by a permutation of the strands on each half-edge. Namely, in general, we could introduce the modified kernel:
\begin{equation}\label{eq:modified_5-simplex}
\tilde{\delta}^{h}_{\pmb a\pmb b\pmb c\pmb d\pmb e\pmb f} = \delta^{h}_{(\sigma_1 \cdot \pmb a ) (\sigma_2 \cdot \pmb b) (\sigma_3 \cdot \pmb c) (\sigma_4 \cdot \pmb d) (\sigma_5 \cdot \pmb e)   (\sigma_6 \cdot\pmb f) } \, ,
\end{equation}
where $\{ \sigma_k \}$ are permutations in $\mathcal{S}_5$, and $\cdot$ denotes the natural action of $\mathcal{S}_5$ on a $5$-tuple. If $\pmb P = \pmb S$ or $\pmb A$, any two such choices differ at most by a sign, and are therefore equivalent. A priori, this is not necessarily so for other irreducible representations, since permuting two indices which are neither in a same column nor in a same row of the Young tableau involves non-trivial linear combinations of tensors. We leave the evaluation of the dimension of the space of $5$-simplex invariants for each irreducible representation to future work. However, we note that our main theorems remain valid for any such interaction, and in fact any linear combination thereof. The only reason we decide to focus exclusively on the kernel of \eqref{eq:5-simplex} is to keep the combinatorial structure of the Feynman diagrams (see the next section) as elementary as possible. Indeed, the modified $5$-simplex of \eqref{eq:modified_5-simplex} is in general not symmetric under cyclic permutation of its half-edges, and would therefore require the introduction of vertices with marked half-edges in the Feynman rules. A particularly interesting example of such a non-cyclic kernel is 
\begin{align}\label{eq:colored_5-simplex}
\tilde{\delta}^{h}_{\pmb a\pmb b\pmb c\pmb d\pmb e\pmb f} &= \delta^{h}_{\pmb a  (\sigma \cdot \pmb b) (\sigma^2 \cdot \pmb c) (\sigma^3 \cdot \pmb d) (\sigma^4 \cdot \pmb e)   (\sigma^5 \cdot\pmb f) } \crcr
&= \delta_{a_1f_1}\delta_{a_2e_2}\delta_{a_3 d_3}\delta_{a_4c_4}\delta_{a_5b_5}\delta_{b_3f_3}\delta_{b_4e_4}\delta_{b_2d_2}\delta_{b_1c_1}\delta_{c_2f_2}\delta_{c_3e_3}\delta_{c_5d_5}\delta_{d_4f_4}\delta_{d_1e_1}\delta_{e_5f_5} \,, 
\end{align}
where $\sigma = (15)(234)$. A specificity of this pattern of contractions is that every tensor index in position $k$ is contracted with another tensor index in position $k$, and is known as a \emph{colorable} interaction in the random tensor literature. Up to a permutation of the half-edges and to a global permutation of the tensor indices, the kernel of \eqref{eq:colored_5-simplex} is in fact the unique colorable $5$-simplex interaction \cite{Ferrari:2017jgw} and it corresponds to the complete interaction of the rank-$5$ model studied in chapter \ref{chap:sextic}. With this choice of interaction, it is actually possible to prove a slightly improved version of theorem~\ref{theorem_princ}, guaranteeing that $m_{\pmb P} > 0$ for \emph{any} irreducible representation $\pmb P$.\footnote{This stems from the fact that we can prove a slightly improved version of lemma~\ref{lemma:melon_tadpole}, see remark~\ref{rem:improved_lemma}.} In particular, this observation implies that the colorable interaction \eqref{eq:colored_5-simplex} is non-vanishing for \emph{any} irreducible representation, and therefore, that our results have non-trivial implications for any choice of irreducible propagator.\footnote{From our current understanding, we cannot guarantee that $m_{\pmb P}$ is necessarily non-vanishing with the cyclic vertex \eqref{eq:5-simplex} (unless $\pmb P = \pmb S$ or $\pmb A$). In particular, we cannot exclude the possibility that this vertex might identically vanish for some specific choice of representation $\pmb P$.} While straightforward, we leave the detailed treatment of non-cyclic vertices, as well as the general proof that $m_{\pmb P} > 0$ in the case of a colorable interaction to the interested reader.

\section{Perturbative expansion}
\label{sec:expansion}

\subsection{Feynman maps}\label{sec:feynman_maps}

Given the structure of the propagator $\pmb P$, which is in general not invariant under index permutations, it will be convenient to view the Feynman expansion as a weighted sum of \emph{combinatorial maps} (or \emph{embedded graphs}) rather than ordinary graphs. Even though combinatorial maps always provide a natural way of representing Wick contractions, they are often dispensed with in field theory because, in many instances, the Feynman amplitudes themselves only depend on the graph structure. In our context, this will remain true in representations such as the symmetric traceless or antisymmetric ones, but not in general
\cite{Carrozza:2018ewt}.\footnote{\label{ft:vertex_inv} Under permutation of the first and second half-edges of the vertex (and similarly for any other pair), we find that
\begin{equation*}
\delta^{h}_{\pmb b  \pmb a \pmb c \pmb d \pmb e \pmb f} = \delta^{h}_{(\gamma \cdot \pmb a ) (\gamma^{-1} \cdot \pmb b) ((12) \cdot \pmb c) ((23) \cdot \pmb d) ((34) \cdot \pmb e)   ((45) \cdot\pmb f) } \, , 
\end{equation*} 
where $\gamma = (12345)$. The product of the signatures of the permutations appearing on the right-hand side being even, the invariance of the vertex under permutation of its half-edges follows for $\pmb P \in \{ \pmb A ,\pmb S\}$.} We therefore resort to the language of combinatorial maps.

There are three steps to obtain the perturbative expansion of $F_{\pmb P}$. First, we Taylor expand in $\lambda$ and compute the Gaussian integrals. This leads to a sum over six-valent \emph{combinatorial maps}. We then take the logarithm, which results in a sum over only \emph{connected} combinatorial maps. Finally, we apply the operator $6\lambda \partial_{\lambda}$, which leads to \emph{rooted} connected combinatorial maps. We call a rooted map a map with a half-edge on a vertex marked with an incoming arrow. 

At first order in $\lambda$, $F_{\pmb P}$ corresponds to $\frac{6\times5}{2} = 15$ rooted, connected, combinatorial maps. Contrary to non-rooted maps, unlabeled rooted maps $\mathcal{M}$ come with a combinatorial weight $1$. This is why we chose to study $F_{\pmb P}$ instead of $\ln Z_{\pmb P}(\lambda)$.

\begin{figure}[htbp]
\centering
\captionsetup[subfigure]{labelformat=empty}\subfloat[]{\includegraphics[scale=1]{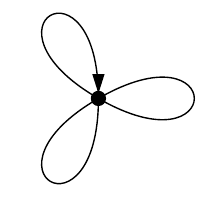}}
\hspace{1cm}
\subfloat[]{\includegraphics[scale=1]{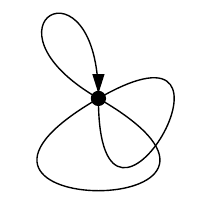}}
\hspace{1cm}
\subfloat[]{\includegraphics[scale=1]{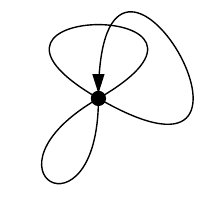}}
\caption{Three of the fifteen first order contributions to $F_{\pmb P }(\lambda)$.}
\end{figure}

We can then write $F_{\pmb P}(\lambda)$ as:
\begin{equation}
F_{\pmb P}(\lambda)=\sum_{\mathcal{M} ~ \mathrm{connected},~ \mathrm{rooted}} \lambda^{V(\mathcal{M})}\mathcal{A}(\mathcal{M}) \, ,
\end{equation}
with $V(\mathcal{M})$ the number of vertices of $\mathcal{M}$.

\subsection{Stranded graphs}\label{sec:stranded_graphs}

We will now go from this representation to a more detailed one in terms of \emph{stranded graphs} $G$. Indeed, each half-edge in a map $\mathcal{M}$ carries five indices and can therefore be represented by five strands. In turn, each term in the propagator has a specific tensor structure, which induces a particular pairing of the strands being propagated along an edge. Since there are $q=10$ half-strands to be paired along a propagator, there are $(2 q -1)!! = 945$ such tensor structures, all of which appear in the symmetric traceless propagator \eqref{sym}. A \emph{stranded graph} is a combinatorial map, together with a choice of one such tensor structure per edge. As a result, a combinatorial map $\cG$ with $E$ edges gives rise to $945^{E}$ stranded graphs, which we will sometimes call \emph{stranded configurations of $\cG$}. Note that, depending on the model, only a subset of those stranded configurations may be relevant. This is clear from the expression of $\pmb A$ in \eqref{antisym}, which only features $5!=120$ of the $945$ possible tensor structures of the propagator.

We will distinguish three types of edge configurations:
\begin{itemize}
\item In an \emph{unbroken edge}, all the strands are traversing and connecting half-strands at the two ends 
\begin{figure}[htbp]
\centering
\includegraphics[scale=1]{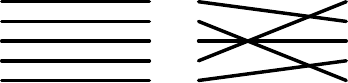}
\caption{Two examples of unbroken edges (out of $5! = 120$).}
\label{fig:unbroken}
\end{figure}
\item In a \emph{broken edge} a pair of half-strands is connected at each end of the edge, and the three other strands are traversing
\begin{figure}[htbp]
\centering
\includegraphics[scale=1]{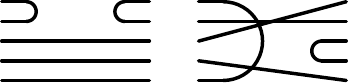}
\caption{Two examples of simply-broken edges (out of ${5 \choose 2}^2 \times 3 ! =600$).}
\label{fig:broken}
\end{figure}
\item In a \emph{doubly-broken edge}, two pairs of half-strands are connected at each end of the edge, and the fifth strand is traversing
\begin{figure}[htbp]
\centering
\includegraphics[scale=1]{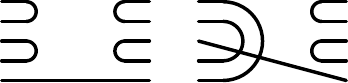}
\caption{Two examples of doubly-broken edges (out of $\left[\frac{1}{2}{5 \choose 2}{3 \choose 2}\right]^2 = 225$).}
\label{fig:doubly_broken}
\end{figure}
\end{itemize}
In particular, the $5!$ tensor structures common to $\pmb A$ and $\pmb S$ lead to unbroken edges. Furthermore, in $\pmb S$, the $600$ terms proportional to $\frac{1}{N+6}$ are associated to  broken edges, while the $225$ terms proportional to $\frac{1}{(N+4)(N+6)}$ lead to doubly-broken edges. Moreover, the large-$N$ scaling of each type of edge is universal:  for any choice of propagator $\pmb P$, unbroken tensor structures appear with a coefficient of order one, while broken (resp. doubly-broken) contributions are rescaled by factors of order $1/N$ (resp. $1/N^2$).

\begin{figure}[htbp]
\centering
\includegraphics[scale=0.5]{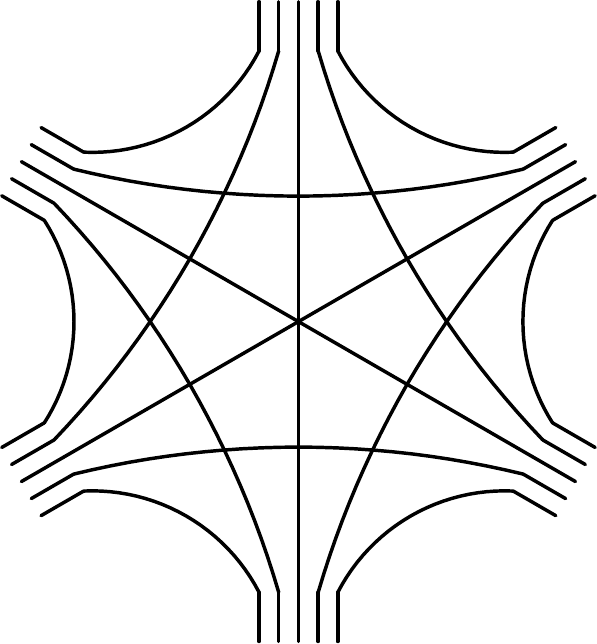}
\caption{Stranded graph representation of the interaction vertex.}\label{fig:vertex}
\end{figure}

We now turn to the stranded representation of the interaction vertex. We call each pair of indices contracted in the $5$-simplex interaction a \emph{corner}. The whole pattern of contractions is represented as a six-valent vertex with fifteen corners, as shown in figure~\ref{fig:vertex}. The vertices are then combined with the stranded edges to form a complete stranded diagram. A closed cycle of strands in such a diagram is called a \emph{face}. Finally, we will respectively denote by $F(G)$, $U(G)$, $B_1(G)$ and $B_2(G)$ the number of faces, unbroken edges, simply-broken edges and doubly-broken edges of $G$.

With these definitions in place, we can write the amplitude of a Feynman map as a sum of amplitudes of its stranded configurations and thus recast the perturbative expansion in terms of stranded graphs: 
\begin{equation*}
F_{\pmb P}(\lambda)=\sum_{G~\text{connected, rooted}} \lambda^{V(G)}\mathcal{A}(G) \,,
\end{equation*}
where $\mathcal{A}(G)$ is the amplitude of the stranded graph $G$. A key advantage is that the large-$N$ behavior of $\mathcal{A}(G)$ is explicitly encoded in the stranded structure of $G$. By inspection of the Feynman rules, each vertex contributes a scaling factor $N^{-5}$, while each broken (resp. doubly-broken) propagator is weighted by a factor $N^{-1}$ (resp. $N^{-2}$) relative to unbroken propagators. Moreover, after contracting the Kronecker delta functions entering the definition of the propagator and vertex kernels, one is left with one free sum and therefore one factor of $N$ per face. This leads to the following large-$N$ asymptotics of the amplitudes:
\begin{equation}\label{eq:ampli}
\mathcal{A}(G) = K(G) N^{-\omega(G)} \left( 1 + \mathcal{O} (1/N) \right)\,, 
\end{equation}
where $K(G)$ is a non-vanishing rational number independent from $N$, and the \emph{degree} $\omega$ of the stranded graph $G$ is:\footnote{Note that the first term in \eqref{eq:degree} reflects the (conventional) $N^{-5}$ scaling introduced in the definition of $F_{\bf P}$; see \eqref{eq:model}.}
\begin{equation}\label{eq:degree}
\omega(G)=5+5V(G)+B_1(G)+2B_2(G)-F(G)\,.
\end{equation}
For a given choice of $\pmb P$, we can work out an exact formula for $\mathcal{A}(G)$. For instance, when $\pmb P = \pmb A$ or $\pmb S$, we find: 
\begin{equation}
\mathcal{A}(G)=\left(\frac{\varepsilon(G)2^{B_1(G)}2^{B_2(G)}}{5!^{U(G)+B_1(G)+B_2(G)}\left(1+\frac{6}{N}\right)^{B_1(G)+B_2(G)}\left(1+\frac{4}{N}\right)^{B_2(G)}}\right)N^{-\omega(G)}\, ,
\label{eq:ampliS}
\end{equation}
where $\varepsilon(G)=(-1)^{B_1(G)}\prod_{e\in G\text{ unbroken}}\epsilon(\sigma^e)$, $\sigma^e$ is the permutation associated to the unbroken edge $e$, and $\epsilon = 1$ (when $\pmb P = \pmb S$) or $\sgn$ (when $\pmb P = \pmb A$).   

The degree $\omega$ is an integer quantity, which can \emph{a priori} take arbitrarily negative values. If one were able to prove it to be bounded from below, the existence of a large-$N$ expansion would immediately follow. We will see that this is not true in general: stranded graphs with arbitrarily negative degrees do exist. However, for any map $\cG$, we will prove that none of its stranded configurations $G$ with $\omega(G) < 0$ (if they exist) actually contribute to the full amplitude $\mathcal{A}(\cG)$.

The stranded graphs and combinatorial maps appearing in the rest of the chapter will always be assumed to be connected, unless specified otherwise.  

\subsection{Problematic cases}

Consider a stranded graph $G$, and let us simplify the expression of its degree. We denote by $F_p$ the number of \emph{faces of length $p$}, that is, the number of faces that have exactly $p$ corners. Each vertex contributing exactly $15$ corners to the graph, we have the relation: 
\begin{equation}
15V=\sum_{p\geq 1} pF_p\,.
\end{equation}
Together with $F=\sum_{p\geq 1} F_p$, this leads to the following expression for the degree:
\begin{equation}
\omega =5+B_1 +2B_2 +\sum_{p \geq 1} F_p \left(\frac{p}{3}-1\right)\,.
\label{eq:degree_simplified}
\end{equation}

We thus obtain the elementary but important proposition:
\begin{proposition}
Let $G$ be a stranded graph. If $F_1 (G) = F_2 (G) = 0$, then 
\begin{equation}
\omega(G) \geq 0 \,.
\end{equation}
\end{proposition} 
\begin{proof}
This immediately follows from \eqref{eq:degree_simplified}: the only terms that are not explicitly non-negative are proportional to $F_1$ and $F_2$.
\end{proof}

\begin{figure}[htbp]
\centering
\includegraphics[scale=.8]{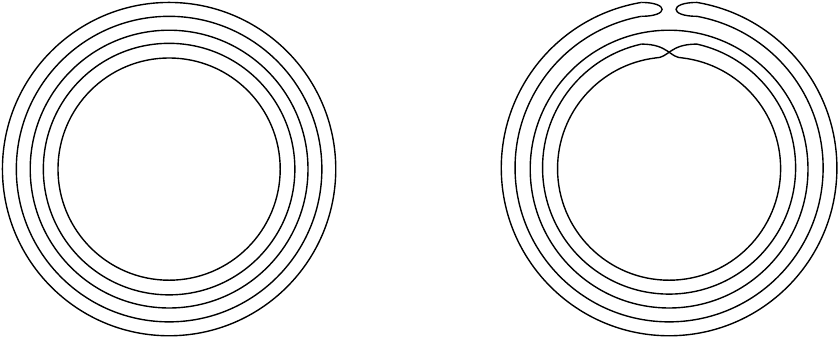}
\caption{Two stranded graphs associated to a ring map, in unbroken (left) and simply-broken (right) configurations.}
\label{fig:ring_graph}
\end{figure}

In section~\ref{sec:deletions}, we will prove that an even larger class of stranded graphs have non-negative degrees. As we will be working by induction on the number of vertices, it is convenient to introduce the notion of \emph{ring graph}, defined as a stranded graph with a single edge closed onto itself, and therefore, no vertex (see figure~\ref{fig:ring_graph}). The degree of a ring graph is then defined by equation \eqref{eq:degree_simplified}, and is clearly non-negative.  

We finally introduce some nomenclature.
\begin{definition}
A \emph{short face} is a face of length one or two.
An \emph{end graph} is either a graph with no short face, or a ring graph. End graphs have non-negative degrees.
\end{definition}
We have reduced our problem significantly. The rest of the chapter is dedicated to the analysis of graphs containing short faces and vertices. 

\section{Combinatorial structure of subgraphs with short faces}
\label{sec:tmd}

The purpose of this section is to introduce specific submap and subgraph structures which may support short faces, and will therefore require special attention. Before that, we also introduce the general notion of boundary graph, which conveniently captures the relation between external legs and external faces of a stranded subgraph.  

\subsection{Boundary graph}\label{sec:bdy_graph}

To any $n$-point stranded graph $G$, we associate a canonically constructed $5$-regular graph with $n$-vertices $G_\partial$, which we call the \emph{boundary graph of $G$} \cite{Gurau:2009tz}. It is constructed in such a way as to faithfully represent the tensorial structure of the correlator $G$ contributes to, up to permutations of the indices appearing in a same tensor. 

\begin{figure}[htbp]
\centering
\includegraphics[scale=0.3]{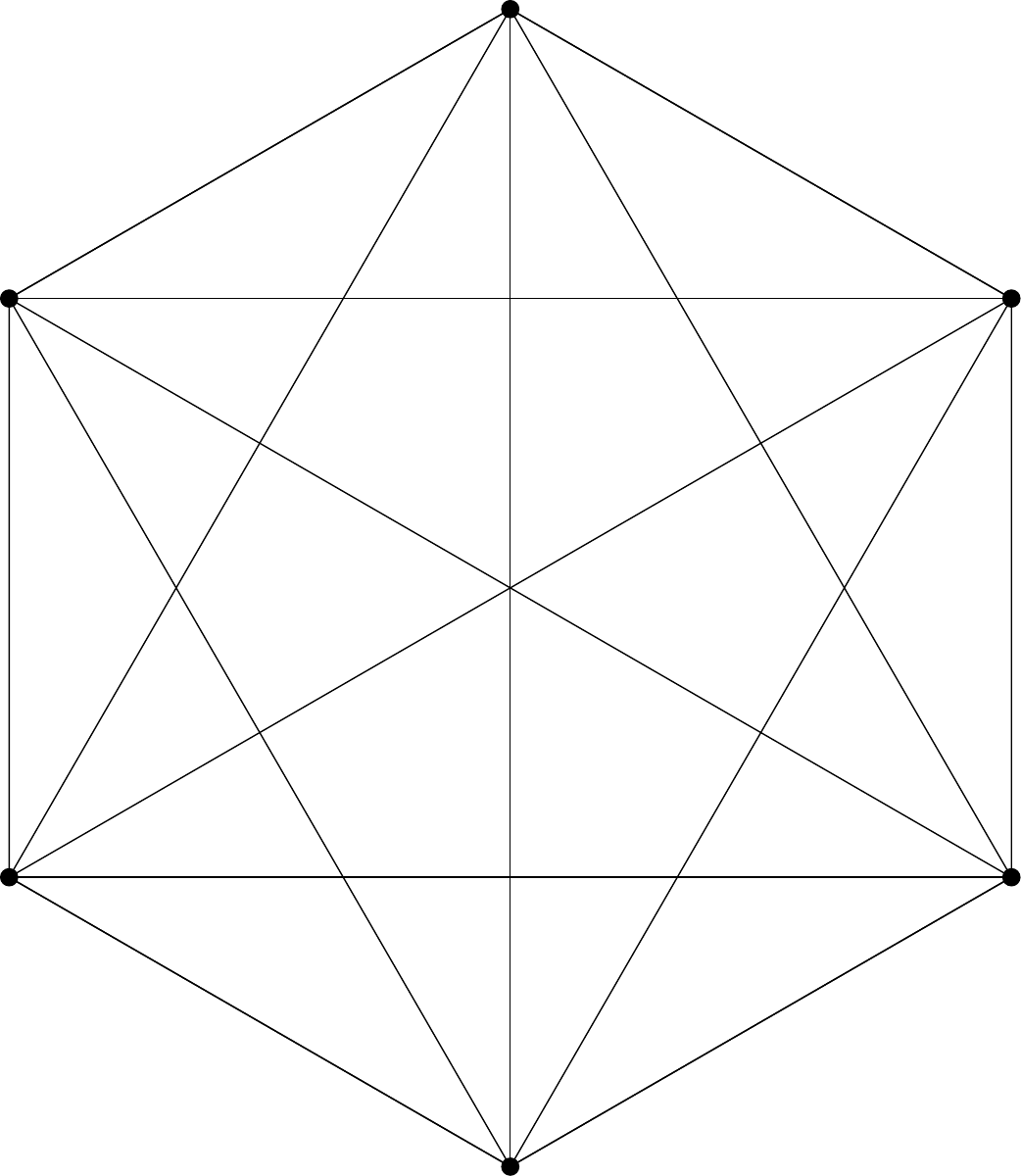}
\caption{The boundary graph of the stranded interaction vertex is the complete graph $K_6$. }
\label{fig:vertex_bndy}
\end{figure}

\begin{figure}[htbp]
\centering
\includegraphics[scale=1.5]{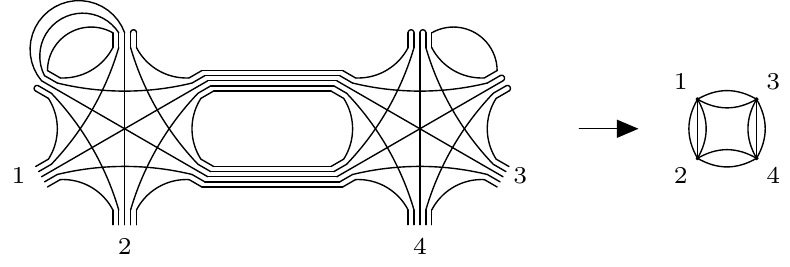}
\caption{A four-point stranded graph and its corresponding boundary graph.}
\label{fig:ex_stranded_bndy}
\end{figure}

More precisely, we define $G_\partial$ through the following procedure. First, each external leg of $G$ is represented in $G_\partial$ by a $5$-valent vertex. Then, for every external strand connecting two external legs of $G$, we draw an edge between the corresponding vertices in $G_\partial$. 
For example, the stranded six-point graph consisting in a single interaction vertex has for boundary graph the complete graph on six vertices $K_6$, as represented in figure~\ref{fig:vertex_bndy}. Graphically, one can obtain $G_\partial$ from $G$ by deleting all its internal faces, and pinching its external legs to form vertices as represented in figure~\ref{fig:ex_stranded_bndy}. Finally, insofar as the external legs of $G$ are labeled, we will consider $G_\partial$ as a labeled graph. 

\subsection{Faces of length one: tadpoles}

\begin{figure}[htbp]
\centering
\captionsetup[subfigure]{labelformat=empty}
\subfloat[]{\includegraphics[scale=1]{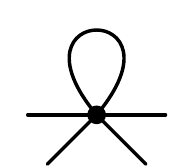}}
\hspace{1cm}
\subfloat[]{\includegraphics[scale=1]{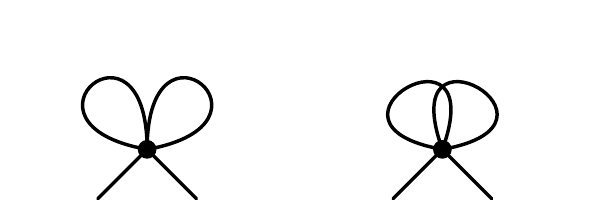}}
\caption{The unique single-tadpole map (left); and two examples of double-tadpole maps that differ through their embedding (right).}
\label{fig:simp_db_tadpole}
\end{figure}

A stranded graph can only have faces of length one if its parent map contains tadpole lines. For convenience, we will distinguish two types of elementary tadpole submaps or subgraphs.  
\begin{definition}
A \emph{single-tadpole} (or, equivalently, a \emph{tadpole}) is a four-point Feynman map or stranded graph with one vertex and one self-loop. A \emph{double-tadpole} is a two-point Feynman map or stranded graph with one vertex and two self-loops. See figure~\ref{fig:simp_db_tadpole}.
\end{definition}

By extension, a graph or map obtained from a single-tadpole (resp.  double-tadpole) by substituting internal edges with non-trivial two-point submaps or subgraphs will be called a \emph{generalized single-tadpole} (resp. \emph{generalized double-tadpole}).

\paragraph{Bad double-tadpoles.}

\begin{figure}[htbp]
\centering
\captionsetup[subfigure]{labelformat=empty}
\subfloat[]{\includegraphics[scale=1]{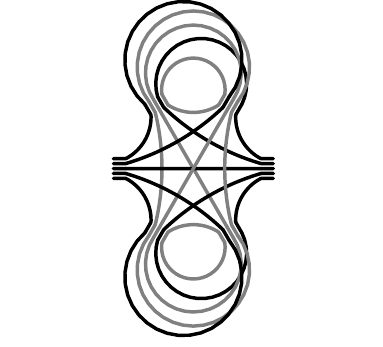}}
\hspace{1cm}
\subfloat[]{\includegraphics[scale=0.25]{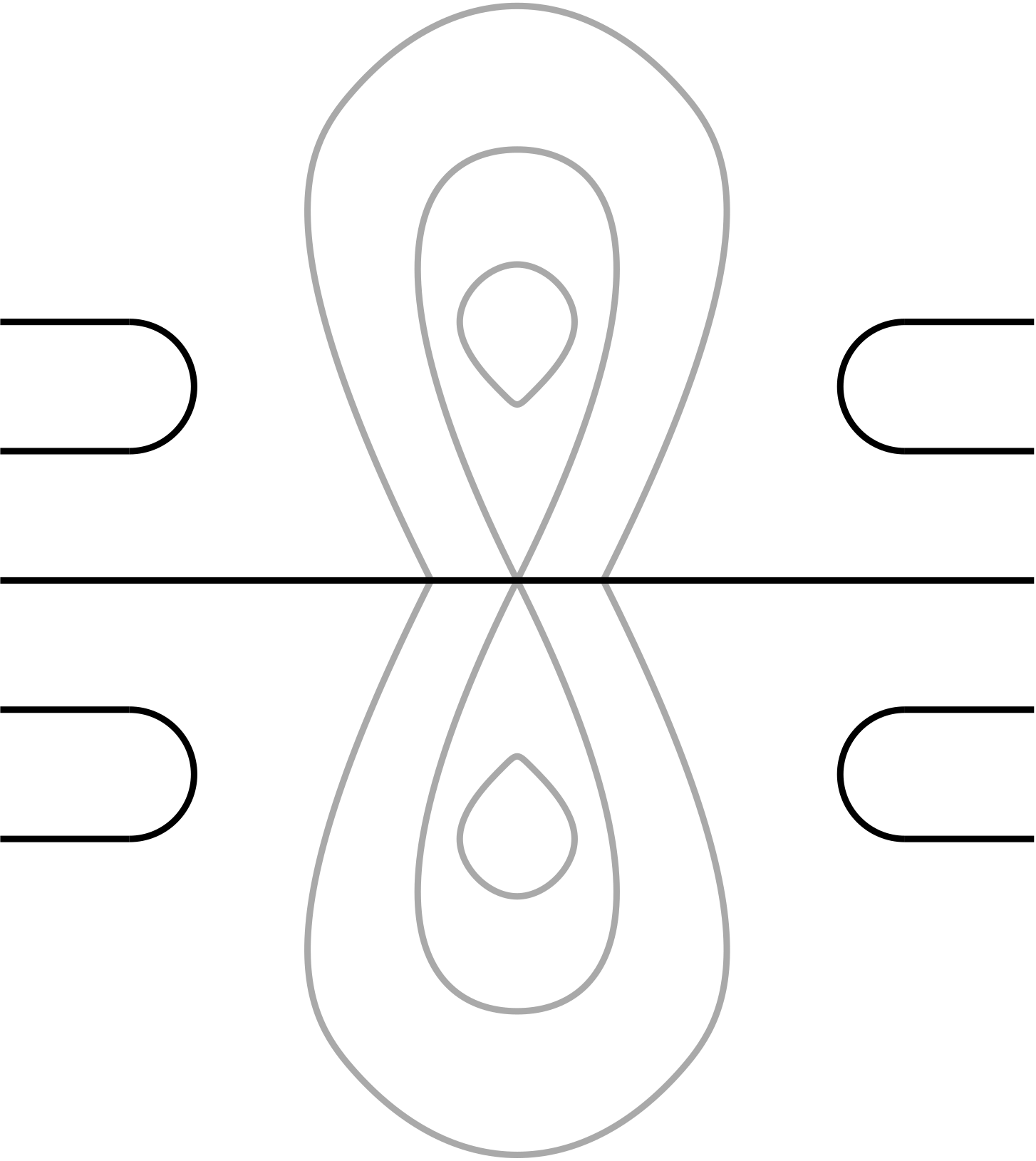}}
\hspace{1.5cm}
\subfloat[]{\includegraphics[scale=0.75]{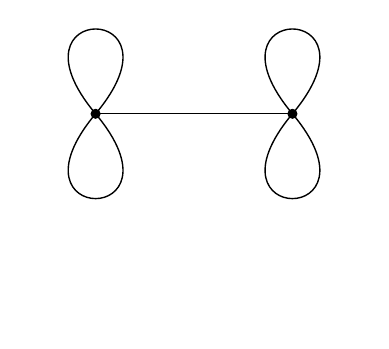}}
\caption{An example of bad double-tadpole subgraph (left panel). The equivalent representation provided in the central panel gives a clearer picture of the face structure. The associated boundary graph is represented in the rightmost panel.}
\label{fig:double_tadpole}
\end{figure}

A double-tadpole graph has at most four internal faces: two of length one, and two of length two. Taking the factor $N^{-5}$ from the vertex into account, we conclude that a double-tadpole graph scales at most like $N^{-1}$. At first sight, it therefore seems that double-tadpoles cannot lead to graphs with negative degrees. But this conclusion is not warranted, which we can illustrate by considering the stranded graph represented in figure~\ref{fig:double_tadpole}. Such a configuration supports four faces and hence saturates our scaling bound. Furthermore, its boundary graph is that of a doubly-broken edge. We call any such configuration a \emph{bad double-tadpole}. 

\begin{figure}[htbp]
\centering
\includegraphics[scale=0.75]{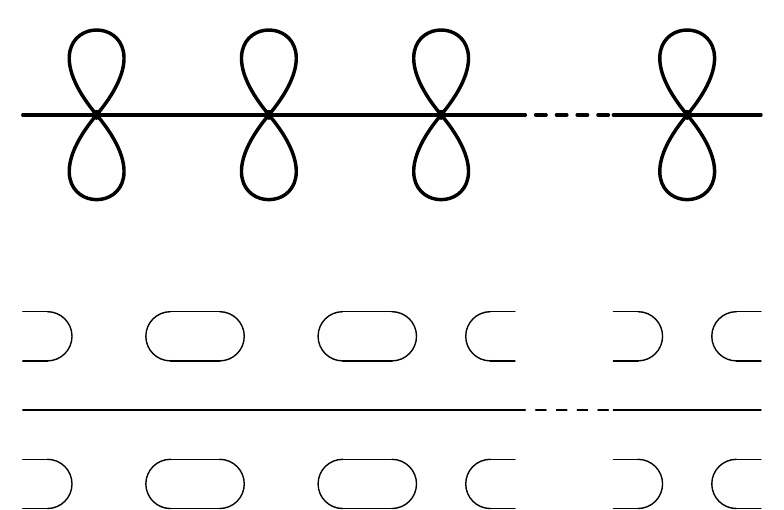}
\caption{A chain of bad double-tadpoles: two additional faces are being closed for each double-tadpole one adds to the chain.}
\label{fig:chain_bad}
\end{figure}

A serious difficulty arises from the fact that we can construct chains of bad double-tadpoles, arranged in such a way that: for every double-tadpole we add to the chain, two additional faces are being closed (see figure~\ref{fig:chain_bad}). 
With $p$ double-tadpoles, the scaling of such a chain is:
\begin{equation}
\left(\frac{1}{N}\right)^pN^{2p-1}=N^{p-1}\,,
\end{equation}
which is unbounded from above. 

As a result, we observe that the degree is unbounded from below in the class of all stranded graphs. Nevertheless, we will see that the irreducible nature of the tensor representations we are working with allows to tame the contributions of such diagrams.

\subsection{Face of length two: melons, dipoles and dipole-tadpoles}

\begin{figure}[htbp]
\centering
\includegraphics[scale=1]{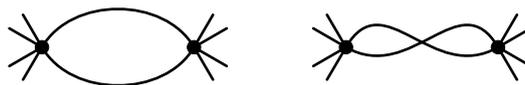}
\caption{Examples of (planar and non-planar) dipoles.}
\label{fig:dipole}
\end{figure}

We will focus on three particular submap structures that can support faces of length two. We start with the minimal one.
\begin{definition}
A \emph{dipole} is an eight-point Feynman map or stranded graph with two vertices, two edges (which we call internal edges) and no self-loop. See figure~\ref{fig:dipole}.
\end{definition}

\begin{figure}[htbp]
\centering
\includegraphics[scale=1]{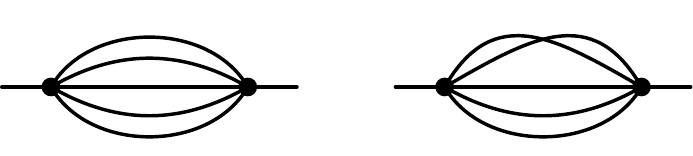}
\caption{Examples of (planar and non-planar) melon two-point maps.}
\label{fig:melon}
\end{figure}

As will become clear later on, we will have to pay extra attention to dipole subgraphs which appear in two other types of structures, which we now introduce. The first one is the familiar melon that we already introduced in previous chapters. Let us recall its definition for the model of this chapter.   
\begin{definition}
A \emph{melon} is a two-point Feynman map or stranded graph with two vertices, five edges, and no self-loop. See figure~\ref{fig:melon}.
\end{definition}

As for tadpoles and double-tadpoles, a graph or map obtained from a melon by dressing its propagator edges with non-trivial two-point functions will be called a \emph{generalized melon}.

\begin{figure}[htbp]
\captionsetup[subfigure]{labelformat=empty}
\centering
\subfloat[]{\includegraphics[scale=1]{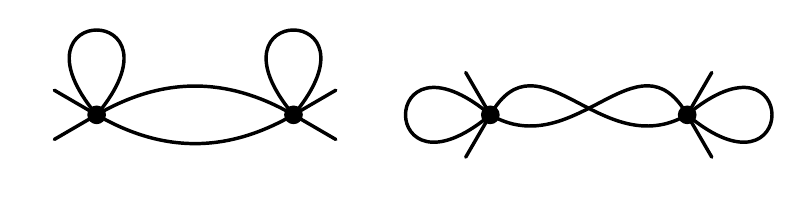}}
\hspace{1cm}
\subfloat[]{\includegraphics[scale=1]{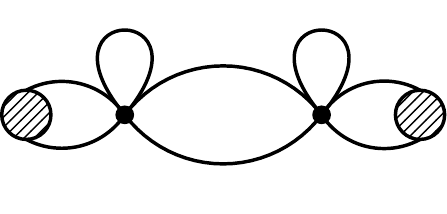}}
\caption{Examples of dipole-tadpole maps. The rightmost dipole-tadpole is separating.}
\label{fig:dipole-tadpole}
\end{figure}

We finally introduce a particular subgraph containing a dipole and two tadpoles, which we call a dipole-tadpole. 
\begin{definition}\label{def:dipole-tadpole}
A \emph{dipole-tadpole} is a four-point Feynman map or stranded graph with two vertices, four edges, and exactly one self-loop on each vertex. See figure~\ref{fig:dipole-tadpole}.
A dipole-tadpole will be called \emph{separating} if it is adjacent to a generalized double-tadpole, as represented in the right panel of figure~\ref{fig:dipole-tadpole}.
\end{definition}
We will also talk of \emph{generalized dipole-tadpole} if we allow the internal edges of a dipole-tadpole to be dressed by non-trivial two-point functions. 

\subsection{Type-$I$ and type-$II$ configurations}\label{sec:types}

\begin{figure}[H]
\centering
\captionsetup[subfigure]{labelformat=empty}
\subfloat[]{\includegraphics[scale=1]{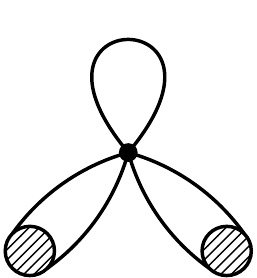}}
\hspace{1cm}
\subfloat[]{\includegraphics[scale=1]{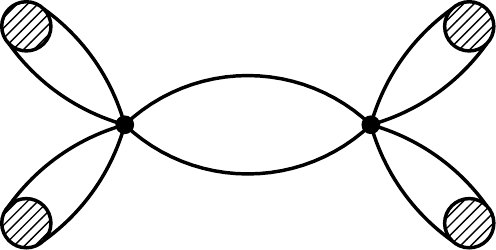}}
\hspace{1cm}
\subfloat[]{\includegraphics[scale=1]{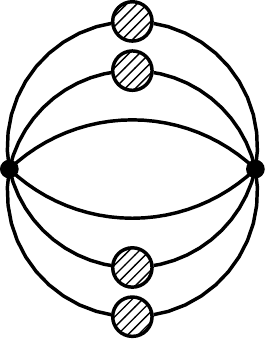}}
\caption{Type-$II$ tadpole (left) and dipoles (middle and right). By exclusion, a tadpole or dipole in any other configuration is of type $I$.}
\label{fig:not_easy}
\end{figure}

Finally, it will later prove convenient to distinguish two types of tadpoles and dipoles: those which appear as subgraphs of generalized double-tadpoles, generalized melons or generalized dipole-tadpoles, as represented in figure~\ref{fig:not_easy}; and all the others. We will label the latter as \emph{type-$I$}, the former as \emph{type-$II$}. 
 
\section{Subtraction of double-tadpoles and melons}
\label{sec:subtraction}

In this section, we show that the irreducible model with propagator $\pmb P$ in \eqref{eq:model} is equivalent to a theory with renormalized covariance, in which melons and double-tadpoles have been subtracted from the Feynman expansion. While not strictly necessary to prove the existence of the large-$N$ expansion \cite{Carrozza:2018ewt}, this reformulation is convenient. It cleanly separates Feynman maps that support stranded configurations with non-positive degrees, from those that do not. Only the latter can be accurately estimated by analyzing the combinatorial structure of their stranded configurations, a task we will turn to in section \ref{sec:deletions}.

Let us start by estimating the amplitude of melon and double-tadpole maps. 
\begin{lemma}\label{lemma:melon_tadpole}
Let $\mathcal{G}$ be a (non-amputated) two-point Feynman map. The associated amplitude $\mathcal{A}(\mathcal{G})_{\pmb a,\pmb b}$ can be written as:
\begin{equation}\label{eq:schur1}
\mathcal{A}(\mathcal{G})_{\pmb a,\pmb b}=\lambda^{V(\mathcal{G})}f_{\mathcal{G}}(N)\pmb P_{\pmb a,\pmb b}\,,
\end{equation}
where $f_{\mathcal{G}}$ is some (rational) function. 
Furthermore:
\begin{itemize} 
\item if $\cG$ is a double-tadpole, then $f_{\mathcal{G}}(N) = \mathcal{O}(1/N)$;
\item if $\cG$ is a melon, then $f_{\mathcal{G}}(N) = f_\cG^{(0)} + \mathcal{O}(1/N)$ where $f_\cG^{(0)} \in \mathbb{R}$. Moreover, when $\pmb P \in \{ \pmb S , \pmb A \}$, one necessarily has $f_\cG^{(0)} > 0$.
\end{itemize}
\end{lemma}
\begin{proof}
The functional form of \eqref{eq:schur1} is a direct consequence of the irreducibility of the representation. It  follows from Schur's lemma for any two-point graph $\cG$. 

Let us first assume that $\cG$ is a double-tadpole. It is clear that any stranded configuration of $\cG$ has at most four faces (this can be formalized with the help of e.g. the bounds of appendix~\ref{ap:bounds}). They contribute a factor of order at most $N^4$, which is compensated by the $1/N^5$ scaling of the vertex. Hence $f_{\mathcal{G}}(N) = \mathcal{O}(1/N)$.

Next, we assume $\cG$ to be a melon. Consider one of its stranded configurations $G$. As will be clear from remark~\ref{rem:broken_unbroken} below, we can assume that $G$ contains only unbroken edges. We have $(5\times4)/2=10$ internal corners at our disposal on each vertex to build up faces, so $20$ corners in total. From the structure of the melon and the unbroken character of the edges of $G$, it is also clear that any face must have length at least two. It immediately follows that $F(G)\leq 10$, leading to a contribution to the amplitude scaling like $N^{10}$ at most. Taking the two factors of $1/N^5$ coming from the vertices into account, we infer that $f_{\mathcal{G}}(N) = \mathcal{O}(1)$. Let us finally specialize to $\pmb P  \in \{ \pmb S , \pmb A\}$. Given that any unbroken edge configuration contributes to $\pmb P$, it is straightforward to show that: 
a) this bound can be saturated; b) the configurations that do so have only unbroken edges and the same boundary graph, namely, that of an unbroken edge; c) despite having identical boundary graphs, the way in which the external strands are being paired up in any two such configurations differ by a permutation. As a result, there can be no cancellation between leading order stranded configurations, which implies that $f_\cG^{(0)} \neq 0$.\footnote{This is a crucial difference with double-tadpoles. For the latter, leading order stranded configurations are of the doubly-broken type, and as a result, necessarily cancel out once resummed into the full amplitude.} Given the symmetric structure of the melon and of its leading order contributions, it is also possible to show that $f_\cG^{(0)} > 0$, irrespectively of the choice of irreducible representation. This is direct for the symmetric traceless propagator since there are no signs involved in its unbroken stranded contributions. For the antisymmetric representation, we can infer from the structure of a leading-order melon stranded graph that the product of the signatures of the permutations labeling its unbroken edges (including the external one) is necessarily even, leading to an overall positive sign. We leave the details of the proof, which follows from footnote \ref{ft:vertex_inv}, to the interested reader.  
\end{proof}

\begin{remark}\label{rem:improved_lemma}
If one works with the colorable interaction kernel \eqref{eq:colored_5-simplex} instead of \eqref{eq:5-simplex}, it is possible to prove that $f_\cG^{(0)} \geq 0$ for any melon $\cG$, and $f_\cG^{(0)} > 0$ for at least one such $\cG$. Indeed, it can be shown that, in this particular case, the unique leading-order stranded configuration of a closed melon happens to be decorated by the same unbroken edge (that is, the same permutation $\sigma \in \mathcal{S}_5$) on all six propagators. Hence, the coefficient associated to this particular unbroken edge is raised to an even power, and the overall sign of the amplitude is always positive. Moreover, it is straightforward to see that any $\sigma$ can contribute to a leading-order melon in such a way, as there is always at least one non-zero unbroken contribution in each propagator. We then conclude that at least one melon is non-vanishing at leading order.
\end{remark}

In light of the previous lemma, it is clear that double-tadpole submaps are well-behaved in the large-$N$ limit, even though some of their stranded configurations are not. To prove the existence of the large-$N$ limit, we must therefore make sure to always bound a double-tadpole Feynman map as a whole. Moreover, it is also clear from lemma~\ref{lemma:melon_tadpole} that melon two-point functions will contribute to the leading order. By dressing double-tadpole subgraphs with such two-point functions, we can generate a family of stranded graphs with arbitrarily negative degrees, but no double-tadpoles. This indicates that the whole family of two-point functions generated by double-tadpoles and melons needs to be treated with care: the existence of the large-$N$ expansion cannot be deduced from bounds on their individual stranded configurations. 

We then follow the method of \cite{Benedetti:2017qxl} and adapt it to rank $5$. We consider a modified theory with covariance $K\pmb P$ where $K$ is a real number. Let us denote by $\Sigma^{(2)}$ the contribution of melon and double-tadpole maps to the self-energy. By lemma~\ref{lemma:melon_tadpole}, we have:
\begin{equation}
\Sigma^{(2)}_{\pmb a,\pmb b}=\left(\lambda K f_1^{\pmb P}+\lambda^2K^5f_2^{\pmb P}\right) \pmb P_{\pmb a,\pmb b}\,,
\end{equation}
where $f_1^{\pmb P}(N)$ and $f_2^{\pmb P}(N)$ are series in $1/N$ verifying
\begin{equation}
f_1^{\pmb P}(N)=\mathcal{O}(1/N) \qquad \mathrm{and} \qquad f_2^{\pmb P}(N)= m_{\pmb P} + \mathcal{O}(1/N)\,.
\end{equation}
Moreover, for $\pmb P \in \{ \pmb S , \pmb A\}$, the constant $m_{\pmb P}$ is necessarily non-vanishing and positive.\footnote{Owing to remark~\ref{rem:improved_lemma}, we also have $m_{\pmb P} >0$ for \emph{any} $\pmb P$ with the alternative choice of vertex kernel \eqref{eq:colored_5-simplex}.} Up to symmetry factors, it essentially counts the number of leading order melon stranded graphs.

As an illustration, for $\pmb P = \pmb A$ or $\pmb S$, we have the exact formula:\footnote{Analogous formulas exist for mixed representations, but they are slightly more involved as in those cases the Feynman amplitudes may depend on the embedding information of the Feynman maps.}
\begin{align}
\Sigma^{(2)}_{\pmb a,\pmb b}&=15\frac{\lambda K^2}{N^5}\sum_c \bigg(\pmb P_{a_1a_2a_3a_4a_5,c_1c_2c_3c_4c_5}\pmb P_{c_5c_6c_7c_8c_9,c_9c_4c_{10}c_{11}c_{12}}\crcr
& \qquad \qquad \qquad \pmb P_{c_{10}c_6c_3c_{13}c_{14},c_{14}c_{11}c_7c_2c_{15}}\pmb P_{c_{15}c_{13}c_{12}c_8c_1,b_1b_2b_3b_4b_5}\bigg) \crcr
& +120 \frac{\lambda^2K^5}{N^{10}}\sum_{c,d}\bigg(\pmb P_{a_1a_2a_3a_4a_5,c_1c_2c_3c_4c_5}\pmb P_{c_5c_6c_7c_8c_9,d_5d_6d_7d_8d_9}\pmb P_{c_9c_4c_{10}c_{11}c_{12},d_9d_4d_{10}d_{11}d_{12}}\crcr
& \qquad \qquad \qquad  \pmb P_{c_{12}c_8c_{3}c_{13}c_{14},d_{12}d_8d_{3}d_{13}d_{14}}\pmb P_{c_{14}c_{11}c_{7}c_{2}c_{15},d_{14}d_{11}d_{7}d_{2}d_{15}} \crcr
& \qquad \qquad \qquad  \pmb P_{c_{15}c_{13}c_{16}c_{2}c_{1},d_{15}d_{13}d_{10}d_{6}d_{1}}\pmb P_{d_1d_2d_3d_4d_5,b_1b_2b_3b_4b_5} \bigg)\, .
\end{align}
Using the explicit expression of the propagators, we can find exact expressions for $f_1^{\pmb P}$ and $f_2^{\pmb P}$. For instance, we determined (by numerical methods) that:\footnote{The computation could in principle be performed for $f_2^{\pmb P}$ as well, but it is more costly.}
\begin{align}
f_1^{\pmb A}&= \frac{(N-4)^2(N^2-13N+34)}{115200 N^5} \,, \crcr
f_1^{\pmb S}&= \frac{(N+8)^2(N^5+19N^4+50N^3-356N^2+8N+672)}{960N^4(N+4)^2(N+6)^2}  \, . 
\end{align}
More interestingly for the large-$N$ limit itself, we can in fact evaluate $m_{\pmb P} = \underset{N\to \infty}{\lim} f_2^{\pmb P}(N)$ exactly, which we briefly sketch. Consider a melon map $\cG$. A leading order stranded configuration of $\cG$ with unbroken boundary graph can only have unbroken edges. Furthermore, once we fix the structure of the external strands, there is a unique choice of configuration of the five internal edges that makes the graph leading order. Since in both $\pmb S$ and $\pmb A$, an unbroken edge is weighted by the combinatorial factor $1/5!$ (up to a sign), we conclude that the contribution of $\cG$ to $m_{\pmb P}$ is $(1/5!)^5$. Given that there are $5!$ melon maps, this finally leads to: 
\begin{equation}\label{eq:m}
m_{\pmb S} = m_{\pmb A} = (1/5!)^4\,.
\end{equation}

Following \cite{Benedetti:2017qxl}, we denote $\Sigma^{(2)}=\lambda K f_1^{\pmb P} + \lambda^2 K^5 f_2^{\pmb P}$, $T^6$ the interaction of equation \eqref{eq:model1}, and define the subtracted interaction:
\begin{equation}
\frac{\lambda}{6N^5}:T^6:_K=\frac{\lambda}{6N^5}T^6-\frac{1}{2}\Sigma^{(2)}T\pmb P T\,.
\end{equation}
As is clear from the notation, this enforces a form of Wick ordering with respect to the covariance $K \pmb P$, which subtracts the double-tadpole and melon interactions. As a result, the model with covariance $K \pmb P$ and interaction $:T^6:_K$ can be expanded in terms of Feynman maps which have neither double-tadpoles nor melons subgraphs. 

\begin{figure}[htbp]
\centering
\includegraphics[scale=.8]{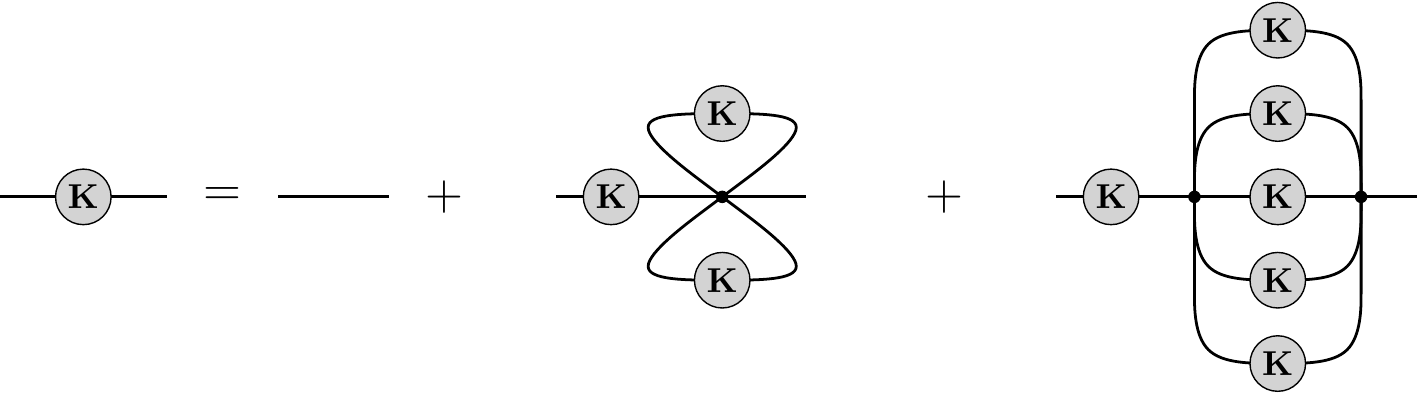}
\caption{Schematic structure of the Schwinger-Dyson equation resumming arbitrary two-point melon and double-tadpole maps (for simplicity, we are ignoring embedding information).}
\label{fig:SDE5}
\end{figure}

The last step amounts to choosing $K$ in such a way that the model with covariance $K \pmb P$ and subtracted interaction is nothing but our original model of \eqref{eq:model}:
\begin{align}
F_{\pmb P}(\lambda)&=\frac{6}{N^{5}}\lambda\partial_{\lambda}\ln \left\lbrace \left[e^{\frac{1}{2}\partial_T \pmb P \partial_T}e^{\frac{\lambda}{6N^5}T^6}\right]_{T=0}\right\rbrace \crcr
&=\frac{6}{N^{5}}\lambda\partial_{\lambda}\ln \left\lbrace \left[e^{\frac{1}{2}\partial_T \pmb P \partial_T}e^{\frac{\lambda}{6N^5}:T^6:_K+\frac{\Sigma^{(2)}}{2}T\pmb P T}\right]_{T=0}\right\rbrace \crcr
&=\frac{6}{N^{5}}\lambda\partial_{\lambda}\ln \left\lbrace \left[e^{\frac{1}{2}\frac{1}{1-\Sigma^{(2)}}\partial_T \pmb P \partial_T}e^{\frac{\lambda}{6N^5}:T^6:_K}\right]_{T=0}\right\rbrace \, .
\end{align}
From the last line, we need to ensure that $K = (1-\Sigma^{(2)})^{-1}$, which results in a polynomial equation for $K$:
\begin{equation}\label{eq:sde}
1-K+\lambda f_1^{\pmb P}K^2+\lambda^2f_2^{\pmb P} K^6=0 \,. 
\end{equation}
Equivalently, this equation can be deduced from the Schwinger-Dyson equation of melon and double-tadpole two-point functions, which we have illustrated in figure~\ref{fig:SDE5}. For $N$ large and $\lambda$ small enough this equation admits a unique solution $K(\lambda,N)$ with the following properties: it is a series in both $\lambda$ and $1/N$, it is uniformly bounded in both $N$ and $\lambda$, and
\begin{equation}
\lim_{\lambda \rightarrow 0}\left[\lim_{N\rightarrow \infty}K(\lambda,N)\right]=1 \, . 
\end{equation} 
Furthermore, $\lim_{N\rightarrow \infty}K(\lambda,N)$ is a series in $\lambda^2$ which coincides with the generating function of Fuss-Catalan numbers $A_n (6,1)$.

We can then write equation \eqref{eq:model} as:
\begin{equation}
F_{\pmb P}(\lambda)=\frac{6}{N^{5}}\lambda\partial_{\lambda}\ln \left\lbrace \left[e^{\frac{K(\lambda,N)}{2}\partial_T \pmb P \partial_T}e^{\frac{\lambda}{6N^5}:T^6:_{K(\lambda,N)}}\right]_{T=0}\right\rbrace\,,
\end{equation}
and obtain the looked-for perturbative expansion in terms of Feynman maps with no double-tadpoles or melons:
\begin{equation}
F_{\pmb P}(\lambda)=\sum_{\substack{\hat{G} \text{ connected, rooted }\\  \text{with no double-tadpoles or melons}}} \lambda^{V(\hat{G})}\left[ K(\lambda,N) \right]^{U(\hat{G})+B_1(\hat{G})+B_2(\hat{G})}\mathcal{A}(\hat{G})\,.
\label{eq:model_subtracted}
\end{equation}
In this equation, $\mathcal{A}$ designates the same amplitude map as defined in \eqref{eq:ampli}.
Given that $K(\lambda,N)$ is also a series in $1/N$, the $1/N$ expansion in theorem \ref{theorem_princ} follows from remark \ref{rem:broken_unbroken} and from proposition \ref{prop:positive_degree}.

\section{Non-negativity of the degree}\label{sec:deletions}

In this section, we prove that the degree of a stranded graph with no melon and no double-tadpole is non-negative. Because the distinction between graphs and embedded graphs does not matter for this purpose, we will ignore it. In particular, our figures should now be understood as representing equivalent classes of maps which only differ through their embedding. 

\subsection{Flip distance between boundary graphs and scaling bounds}\label{sec:flip_distance}

In the following, it will be convenient to extract large-$N$ scaling information by direct inspection of the boundary graph of a given stranded configuration. 

\begin{figure}[htbp]
\centering
\includegraphics[scale=1]{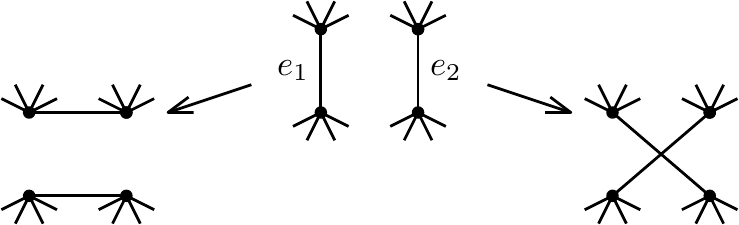}
\caption{The two possible flips of edges $e_1$ and $e_2$ in a boundary graph.}
\label{fig:flips_boundary}
\end{figure}

To this effect, we first introduce combinatorial moves acting on pairs of edges in a boundary graph, which we call \emph{flips}. Given two distinct edges $e_1$ and $e_2$ in a boundary graph $B$, a flip amounts to: 1) cutting $e_1$ and $e_2$ open; and 2) recombining the resulting four half-edges in one of two possible channels, to obtain a new boundary graph $\tilde{B}$. This is illustrated in figure~\ref{fig:flips_boundary}. It is easy to see that the set of (not necessarily connected) boundary graphs with prescribed number of vertices is stable under flips. Moreover, these moves are ergodic in this space: given two $5$-regular graphs, it is always possible to transform one into the other through a finite number of successive flips. As a result, we can introduce a notion of \emph{flip distance} between such graphs.  

\begin{definition}
Let $B_1$ and $B_2$ be boundary graphs with $n\geq 2$ vertices. We define the \emph{flip distance} between $B_1$ and $B_2$, denoted by $d(B_1 , B_2)$, as the minimal number of successive flips required to map $B_1$ to $B_2$. 
\end{definition}
It is elementary to check that $d$ defines a proper notion of distance on the space of boundary graphs with $n$ vertices. By convention, we will also postulate that $d(B_1 , B_2) = \infty$ whenever $B_1$ and $B_2$ do not have the same number of vertices. 
The relation between flip distance and scaling is captured by the following proposition.
\begin{proposition}\label{propo:general_deletion}
Consider a stranded graph $G$, and a strict subgraph $S \subset G$. Let $B$ be a boundary graph such that $d(B, S_\partial) < \infty$. Then, there exists a stranded graph $S'$ such that $S'_\partial = B$ and
\begin{equation}\label{propo:bound1}
\vert F(G') - F(G) \vert \leq \vert F(S) - F(S') \vert + d(B, S_\partial)\,,
\end{equation}
where $G'$ is the graph obtained by substitution of $S$ by $S'$ into $G$.

In particular, if $G$ and $G'$ contain only unbroken edges, $F(S') = 0$, and $G'$ remains connected, we find:
\begin{equation}\label{propo:bound2}
\omega(G) \geq \omega(G') + 5 \left( V(S) - V(S') \right) - F(S) - d(S_\partial , S'_\partial )\,.
\end{equation}
\end{proposition}
\begin{proof}
The idea is to perform a succession of cut-and-glue operations on the internal strands of $S$ (while leaving the rest of the graph unchanged), until we obtain a new subgraph with boundary $B$. Since each cut-and-glue operation is reflected by a flip at the level of boundary graphs, this can be done in at most $d(B,S_\partial)$ steps. We then insert or remove internal faces to change $F(S)$ into $F(S')$, and obtain the target stranded subgraph $S'$. This last step is responsible for the first term in the right-hand-side of \eqref{propo:bound1}. Furthermore, it is clear that each cut-and-glue operation changes the number of faces in the graph by $-1$, $0$ or $1$, which explains the second term.  
Finally, if we assume that $G$ and $G'$ contain only unbroken edges, then $B_1 = B_2 =0$ for both graphs, and together with $F(S')=0$, \eqref{propo:bound2} follows from \eqref{propo:bound1} and \eqref{eq:degree}.
\end{proof}
Equation \eqref{propo:bound2} will be particularly relevant because it will allow us to derive inductive bounds of the form $\omega(G) \geq \omega(G')$ from the local combinatorial condition:
\begin{equation}
d(S_\partial , S'_\partial ) \leq  5 \left( V(S) - V(S') \right) - F(S) \,.
\end{equation}

\begin{figure}[htbp]
\centering
\includegraphics[scale=1]{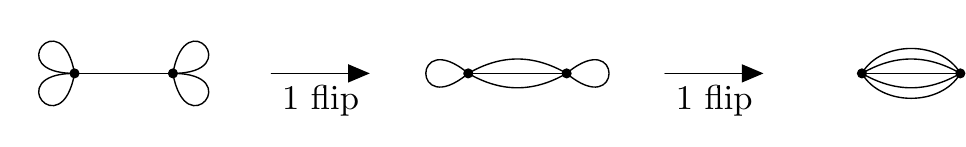}
\caption{Flip distance between the boundary graphs of the doubly-broken, broken and unbroken propagators.}
\label{fig:propagator_bndy}
\end{figure}

As a simple illustration of equation \eqref{propo:bound1}, consider the boundary graph of a single edge $e$. If $e$ is doubly-broken, it is at flip distance one from the boundary graph of a broken edge, which is itself at flip distance one from the boundary graph of an unbroken edge; see figure~\ref{fig:propagator_bndy}. 

Therefore, we can replace a broken edge by an unbroken one in such a way that the number of faces decreases at most by one. As a result, the number of broken edges decreases by one, which implies that the degree \eqref{eq:degree} can only decrease. Likewise, we can replace a doubly-broken edge by an unbroken one in such a way that the number of faces decreases at most by two, whereas the number of doubly-broken edges decreases by one. Again, the degree can only decrease. This leads to the following observation.
\begin{remark}\label{rem:broken_unbroken}
For any stranded graph $G$, there exists a stranded graph $G'$ with $B_1(G')= B_2(G') = 0$, and such that
\begin{equation}
\omega(G)\geq \omega(G')\,.
\end{equation}
\end{remark}
Hence, for the purpose of finding lower bounds on the degree, we can restrict ourselves to graphs with only unbroken edges. This property is assumed in the remainder of the present section.

We now turn to the definition of basic combinatorial moves, which we will use in combination in the proof of subsection \ref{subsec:main_proof}. A first straightforward example concerns double-tadpoles.
\begin{lemma}\label{lemma:double_tadpole_deletion}
Consider a stranded graph $G$ with a double-tadpole subgraph $S$. It is possible to replace $S$ by an unbroken edge in such a way that the resulting graph $G'$ verifies:
\begin{equation}
\omega(G) \geq \omega(G') - 1\,.
\end{equation}
\end{lemma}
\begin{proof}
We notice that: $F(S)\leq 4$ if $S_\partial$ is of the doubly-broken type; $F(S)\leq 3$ if $S_\partial$ is of the simply-broken type; and $F(S)\leq 2$ otherwise. The result then follows from \eqref{propo:bound2}.
\end{proof}
Less straightforward examples will be the focus of the next four subsections. 

\subsection{Single-tadpole deletions}

\begin{figure}[htbp]
\centering
\includegraphics[scale=1]{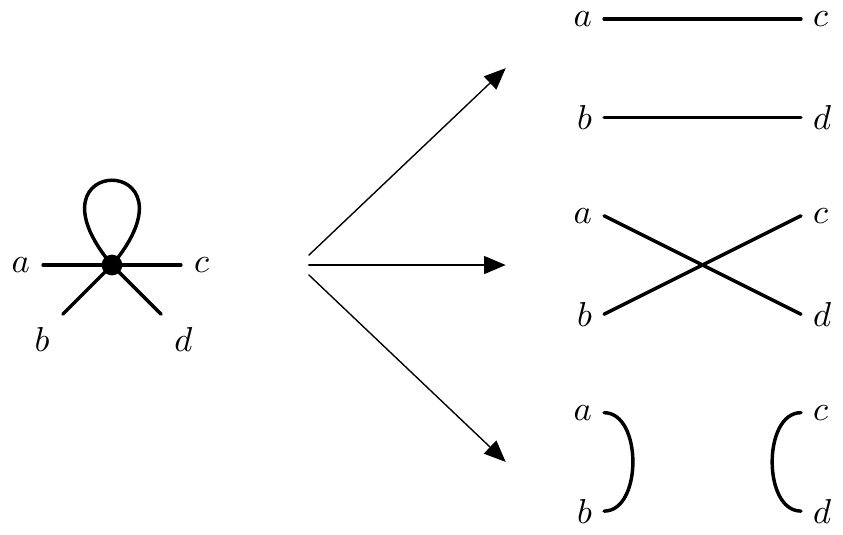}
\caption{The three deletion channels of a single-tadpole. From top to bottom: parallel, cross and orthogonal channels.}
\label{fig:channel_single_tadpole}
\end{figure}

We first look for combinatorial moves that replace a single-tadpole subgraph with two (unbroken) propagators, and delete as few faces as possible. If we ignore for the moment the permutations labeling the two edges after the deletions, there are exactly three ways of doing so, which amount to a choice of pairing of the external legs of the subgraph: we call these \textit{deletion channels}, or simply \emph{channels}. They are the \emph{parallel} (pairing $(a,c)$ and $(b,d)$), \emph{cross} (pairing $(a,d)$ and $(b,c)$) and \emph{orthogonal} (pairing $(a,b)$ and $(c,d)$) channels, as illustrated in figure~\ref{fig:channel_single_tadpole}. Note that this nomenclature is purely conventional: it depends on an arbitrary labeling of the external legs of the tadpole. In the following, we will fix a canonical labeling for each possible structure of the boundary graph. 

\begin{figure}[htbp]
\centering
\begin{tabular}{ccccc}
\subfloat[1+1+1+1]{\includegraphics[width=0.17\textwidth]{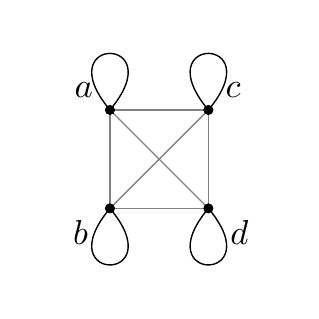}}
&
\subfloat[1+1+2]{\includegraphics[width=0.17\textwidth]{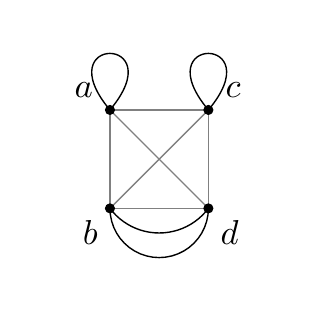}}
&
\subfloat[1+3]{\includegraphics[width=0.17\textwidth]{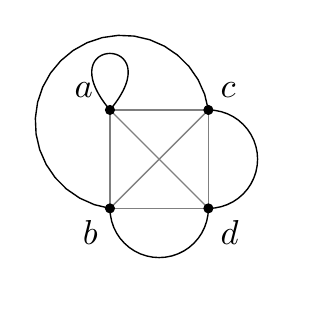}}
&
\subfloat[2+2\label{fig:tadpole_config_d}]{\includegraphics[width=0.17\textwidth]{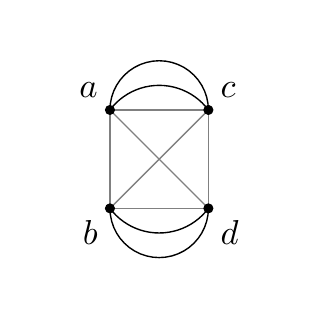}}
&
\subfloat[4\label{fig:tadpole_config_e}]{\includegraphics[width=0.17\textwidth]{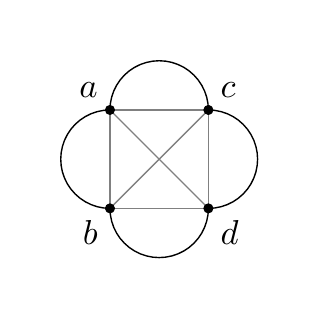}}
\end{tabular}
\caption{The five possible boundary graphs of a single-tadpole.}
\label{fig:tadpole_config}
\end{figure}

To find a suitable deletion along the lines of proposition \ref{propo:general_deletion}, we first need to determine the structure of the boundary graph $S_\partial$. Up to a relabeling of the vertices, we find the five possibilities represented in figure~\ref{fig:tadpole_config}. Indeed, first notice that the structure of the vertex imposes the presence of a $K_4$ subgraph (the complete graph on $4$ vertices). We have represented this subgraph in grey in figure~\ref{fig:tadpole_config}. We are left with a choice of pairing of eight remaining half-edges (two per vertex), to form the four edges that we have represented in black. 
The five configurations we end up with are distinguished by the lengths of the cycles formed by the black edges, and can be labeled by the partitions of $4$. Indeed, we have a budget of four edges, which can be split up into: four cycles of length one ($1+1+1+1$); two cycles of length one and one of length two ($1+1+2$); one cycle of length one and one of length three ($1+3$); two cycles of length two ($2+2$); or one cycle of length four ($4$). 

We can now write the following lemma.
\begin{lemma}\label{lemma:single_tadpole}
Let $G$ be a stranded graph, and $S$ a strict single-tadpole subgraph of $G$. Call $G'$ the graph obtained after a deletion of $S$ in the channel $c$, and assume that $G'$ remains connected. 
\begin{enumerate}
\item If $S_\partial$ is in the configuration $1+1+2$, it is possible to choose $G'$ such that:
\begin{enumerate}
\item $\omega(G) \geq \omega(G') + 1$ 
when $c$ is the parallel channel,
\item $\omega(G) \geq \omega(G') - 1$ 
when $c$ is any other channel.
\end{enumerate}
\item If $S_\partial$ is in any other configuration, it is possible to choose $G'$ such that
$\omega(G) \geq \omega(G')$.
\end{enumerate}
\end{lemma}

\begin{proof}
The single-tadpole $S$ can support at most one internal face, and exactly one vertex is lost upon deletion. $G'$ being connected, proposition~\ref{propo:general_deletion} guarantees that we can arrange the strands in such a way that:
\begin{equation}
\omega(G) \geq \omega(G') + 4 - d(S_\partial , B_c)\,,
\end{equation} 
where $B_c$ is the boundary graph characterizing the channel $c$. For instance, if $c$ is the parallel channel, $B_c$ is the four-vertex graph in which vertices $a$ and $c$ are connected by five edges, and likewise for vertices $b$ and $d$. It remains to bound the flip distance between $S_\partial$ and $B_c$. It helps to first determine the flip distance between the various possible configurations of $S_\partial$, which we have represented in figure~\ref{fig:distance_tadpole}. It is then apparent that the lemma follows from the following sufficient conditions: 
\begin{itemize}
\item if $S_\partial$ is in the configuration $1+1+1+1$, $d(S_\partial , B_c)\leq 4$ for any $c$;
\item if $S_\partial$ is in the configuration $2+2$ and $c$ is the parallel channel, then $d(S_\partial , B_c)\leq 2$;
\item if $S_\partial$ is in the configuration $4$, $d(S_\partial , B_c)\leq 3$ if $c$ is the parallel or orthogonal channel, and $d(S_\partial , B_c)\leq 4$ otherwise;
\item if $S_\partial$ is in the configuration $3+1$ and $c$ is the cross channel, then $d(S_\partial , B_c)\leq 4$.
\end{itemize}

\begin{figure}[htbp]
\centering
\includegraphics[scale = 1]{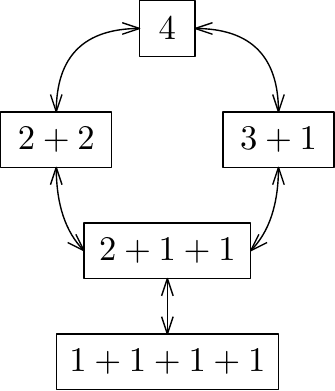}
\caption{Distance between the five single-tadpole configurations: any two partitions connected by an edge are at flip distance one from each other.}
\label{fig:distance_tadpole}
\end{figure}

If $S_\partial$ is in the configuration $1+1+1+1$, we need $2$ flips to disconnect the graph in the appropriate channel, and $2$ more flips to remove the self-loops. Hence, we have $d(S_\partial , B_c) \leq 4$
. This is illustrated in figure~\ref{fig:ex_deletion_tadpole}, for the parallel channel.

\begin{figure}[htbp]
\centering
\includegraphics[scale=1]{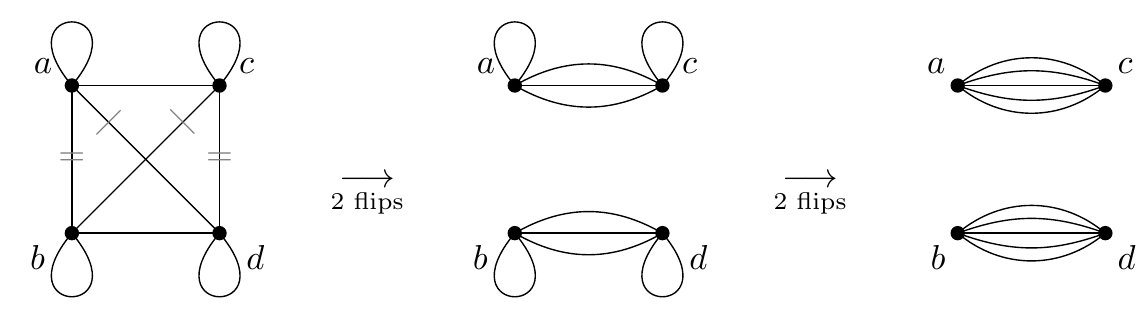}
\caption{Flip distance between a tadpole in the configuration $1+1+1+1$ and the parallel channel.}
\label{fig:ex_deletion_tadpole}
\end{figure}

If $S_\partial$ is in the configuration $2+2$ and $c$ is the parallel channel, we can infer that $d(S_\partial, B_c) \leq 2$ by first flipping the edges $(a,b)$ and $(c,d)$, then the edges $(a,d)$ and $(b,c)$.

Likewise, if $S_\partial$ is in the configuration $4$, we can show that $d(S_\partial, B_c) \leq 3$ if $c$ is the parallel or orthogonal channel. One needs an extra flip in the cross channel, because $a$ and $d$ (resp. $b$ and $c$) are initially connected by a single edge; hence, $d(S_\partial, B_c) \leq 4$  in that case.

If $S_\partial$ is in the configuration $3+1$ and $c$ is the cross channel, we need $3$ flips to disconnect. This has the effect of creating a second self-loop. We can then perform one more flip to remove the self-loops, and obtain the boundary graph $B_c$. As a result, $d(S_\partial , B_c) \leq 4$. This is illustrated in figure~\ref{fig:ex_deletion_tadpole2}.
\begin{figure}[htbp]
\centering
\includegraphics[scale=1]{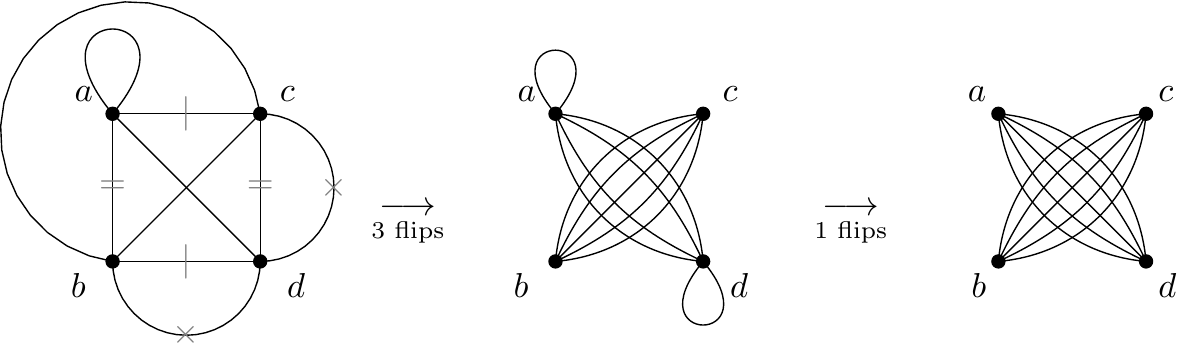}
\caption{Flip distance between a tadpole in the configuration $1+3$ and the cross channel.}
\label{fig:ex_deletion_tadpole2}
\end{figure}
\end{proof}

\subsection{Dipole deletions}

\begin{figure}[htbp]
\centering
\captionsetup[subfigure]{labelformat=empty}
\subfloat[]{\includegraphics[scale=1]{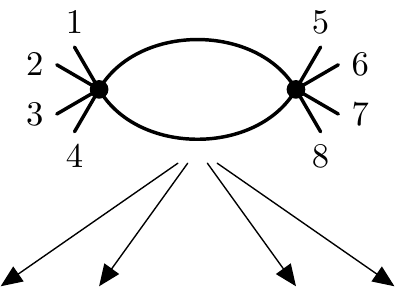}}
\\
\subfloat[(1)]{\includegraphics[scale=0.8]{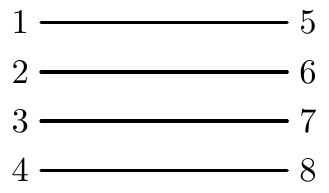}}
\hspace{1cm}
\subfloat[(2a)]{\includegraphics[scale=0.8]{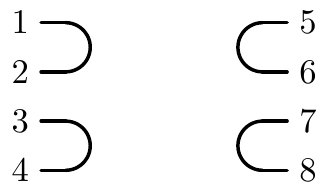}}
\hspace{1cm}
\subfloat[(2b)]{\includegraphics[scale=0.8]{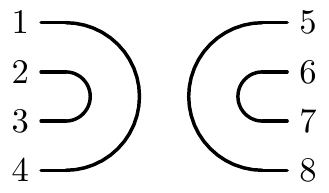}}
\hspace{1cm}
\subfloat[(2c)]{\includegraphics[scale=0.8]{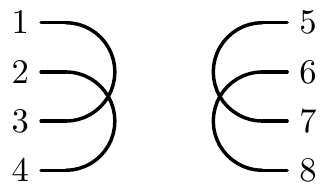}}
\caption{The four deletion channels we consider for a dipole; from left to right: channels $(1)$, $(2a)$, $(2b)$ and $(2c)$.}
\label{fig:channel_dipole}
\end{figure}

We will now look for combinatorial moves that replace a dipole subgraph with four (unbroken) propagators and delete as few faces as possible. In contrast to the single-tadpole deletions of the previous section, there are many more ways of doing so, leading to many more than three channels of deletions. However, for our purpose, it will be sufficient to consider only four of those channels. Indeed, all we need is a sufficiently rich set of deletion moves to ensure that, in all situations, at least one of them can be performed while maintaining our combinatorial constraints (connectedness, and the absence of melons or double-tadpoles). This subset of channels is presented in figure~\ref{fig:channel_dipole}. Note that, apart from the fact that the groups of half-edges $\{ 1, 2, 3, 4 \}$ and $\{ 5, 6, 7, 8 \}$ are attached to different vertices, the labeling is purely conventional at this stage. This will be taken advantage of and made more precise in the proof of lemma \ref{lemma:dipole} (see also figure~\ref{fig:dipole_config}).

\begin{figure}[htbp]
\centering
\subfloat[]{\label{fig:dipole_connected_a}\includegraphics[scale=1]{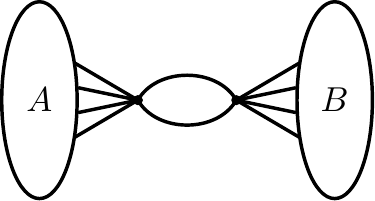}}
\hspace{1cm}
\subfloat[]{\label{fig:dipole_connected_b}\includegraphics[scale=1]{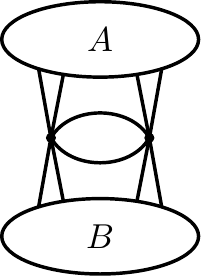}}
\hspace{1cm}
\subfloat[]{\label{fig:dipole_connected_c}\includegraphics[scale=1]{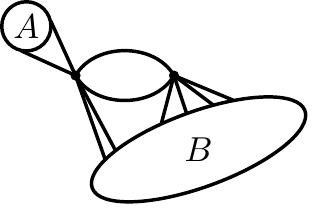}}
\hspace{1cm}
\subfloat[]{\label{fig:dipole_connected_d}\includegraphics[scale=1]{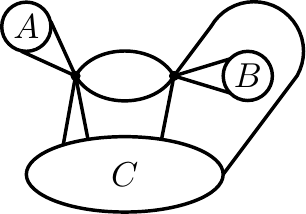}}
\caption{Possible configurations of a dipole which disconnects the graph upon deletion in channel $(2a)$.}
\label{fig:dipole_connected}
\end{figure}

The reason for choosing these four channels is that, if we assume that the dipole is of type $I$, then at least one of them does not disconnect the graph. Indeed, suppose channel $(2a)$ disconnects. Then the graph is in either one of the configurations depicted in figure~\ref{fig:dipole_connected_a}, \ref{fig:dipole_connected_b} and \ref{fig:dipole_connected_c}, with subgraphs $A$ and $B$ not necessarily connected. 
\begin{itemize}
\item If it is in the first configuration, then channels $(2b)$ and $(2c)$ also disconnect but channel $(1)$ does not. Indeed, otherwise there would be generalized double-tadpoles on both vertices of the dipole, which means that the latter would be of type $II$.
\item If it is in the second configuration, channels $(2b)$ and $(2c)$ do not disconnect. Otherwise, the subgraphs $A$ and $B$ would have to be disconnected, in a way that would either generate two generalized double-tadpoles, or a generalized melon. Both cases are excluded given that the dipole is of type $I$.
\item If it is in the third configuration, suppose that channel $(2b)$ also disconnects. Then, the subgraph $B$ must be disconnected and the dipole is in the configuration of figure~\ref{fig:dipole_connected_d} with $A$, $B$ and $C$ subgraphs not necessarily connected. Then, channel $(2c)$ does not disconnect, otherwise $C$ would have to be disconnected and there would again be two generalized double-tadpoles. 
\end{itemize}
All in all, at least one channel does not disconnect if the dipole is of type $I$.
  
The following lemma gives us tools to recursively remove dipoles from a stranded graph. 
\begin{lemma}\label{lemma:dipole}
Let $G$ be a stranded graph and $S$ a strict dipole subgraph of $G$. Call $G'$ the graph obtained after deletion of $S$ in the channel $c$, and assume $G'$ remains connected. There exists a conventional labeling of the external legs of $S$ (see figure~\ref{fig:channel_dipole}) such that: if $c$ is channel $(1),\,(2a),\,(2b)$ or $(2c)$, then it is possible to choose $G'$ such that 
$\omega(G)\geq \omega(G')$. 
\end{lemma}

\begin{proof}
We have two cases to consider: the dipole contains one internal face or none. In the latter case ($F(S)=0$), the two corners of the dipole can either be on the same external face or on two distinct ones, as represented in figure~\ref{fig:dipole_reduction}. In both those cases, this subset of strands can be reconfigured in such a way as to ensure that the dipole contains an internal face. Moreover, such a move does not affect the rest of the graph. We obtain in this way a graph $\tilde{G}$ containing a dipole subgraph $\tilde{S}$ such that: $F(\tilde{S})=1$ and $\omega(G) \geq \omega(\tilde{G}) + 1$. It is then clear that the lemma will hold in general if we can prove it for configurations like $\tilde{G}$, which we now turn too.
\begin{figure}[htbp]
\centering
\includegraphics[scale=.8]{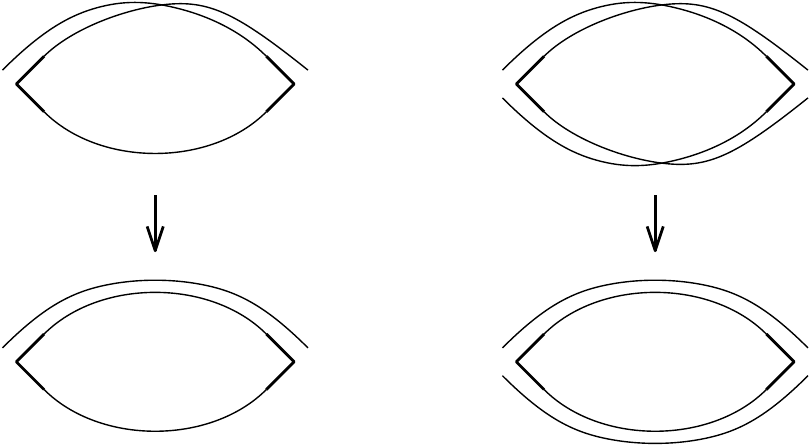}
\caption{Two types of configurations of the dipole when $F(S)=0$: the two internal corners can lie on the same external face (left), or on two distinct ones (right). In both cases, we can reconfigure this subset of strands in such a way as to ensure $F(S)=1$, without affecting the rest of the graph $G$.}
\label{fig:dipole_reduction}
\end{figure}

We can thus assume, without loss of generality, that $F(S)=1$, which makes it easier to determine the structure of the boundary graph $S_\partial$. We can proceed similarly as for single-tadpoles, and associate a $K_4$ subgraph to each of the two vertices in the dipole, which we represent in gray. We are left with a choice of pairing between eight additional half-edges, four of them attached to each $K_4$ subgraph, which we can represent in black. Given that the two internal corners of the dipole have been used to build up the internal face, any pairing of the black half-edges must connect one $K_4$ subgraph to the other. Consequently, we can again classify the allowed contractions in terms of the number and lengths of cycles with support on black edges only. The resulting boundary graphs can be labeled by partitions of 8 into even integers, yielding five possibilities: $8 = 6 + 2 = 4 + 4 = 4 + 2 + 2 = 2 + 2 + 2 + 2$. See figure~\ref{fig:dipole_config}.

\begin{figure}[htbp]
\centering
\begin{tabular}{ccc}
\subfloat[2+2+2+2]{\includegraphics[width=0.25\textwidth]{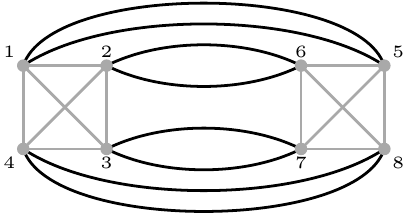}}
&
\subfloat[4+4]{\includegraphics[width=0.25\textwidth]{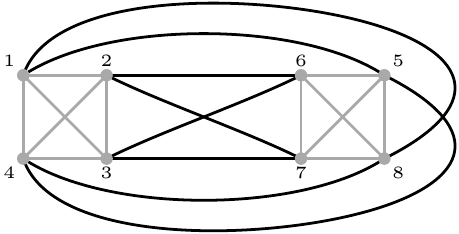}}
&
\subfloat[6+2]{\includegraphics[width=0.25\textwidth]{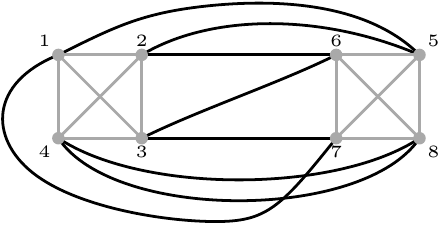}} \\
\subfloat[8]{\includegraphics[width=0.25\textwidth]{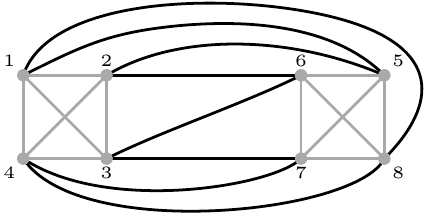}}
& &
\subfloat[4+2+2]{\includegraphics[width=0.25\textwidth]{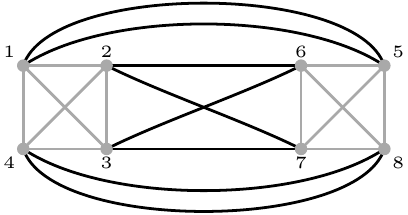}} 
\end{tabular}
\caption{The five possible boundary graphs of a dipole.}
\label{fig:dipole_config}
\end{figure}

Given that $G'$ remains connected, $F(S)=1$ and $V(S)=2$, proposition~\ref{propo:general_deletion} guarantees that we can arrange the strands in such a way that:
\begin{equation}
\omega(G) \geq \omega(G') + 9 - d(S_\partial , B_c)\,,
\end{equation} 
where $B_c$ is the boundary graph characterizing the channel $c$. Hence, the looked-for bound will follow from $d(S_\partial , B_c) \leq 9$, which we now prove. 

We can first determine the flip distance between the five boundary graphs of figure~\ref{fig:dipole_config}, which is reported in figure~\ref{fig:distance}. As a result, it is sufficient to prove that:
\begin{itemize}
\item if $c=1$ and $S$ is in the configuration $2+2+2+2$, then $d(S_\partial , B_c )\leq 6$;
\item if $c=2a$ and $S$ is in the configuration $4+2+2$ or $4+4$, then $d(S_\partial , B_c )\leq 8$;
\item if $c=2b$ and $S$ is in the configuration $8$ or $4+2+2$, then $d(S_\partial , B_c )\leq 8$;
\item if $c=2c$ and $S$ is in the configuration $2+2+2+2$ or $4+4$, then $d(S_\partial , B_c )\leq 8$; if, on the other hand, $S$ is in configuration $6+2$, then $d(S_\partial , B_c )\leq 9$.
\end{itemize}

\begin{figure}[htbp]
\centering
\includegraphics[scale = 1]{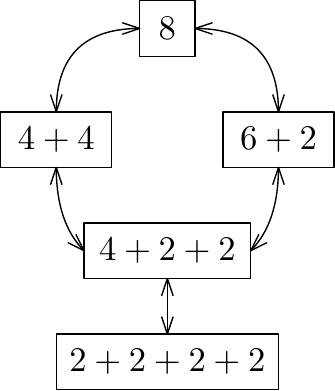}
\caption{Distance between the five dipole configurations: any two partitions connected by an edge are at flip distance one from each other.}
\label{fig:distance}
\end{figure}

\textbf{Channel $1$ (parallel channel).} To map the $2+2+2+2$ configuration to the parallel configuration, we need to cut all twelve grey edges. This can be achieved in $6$ flips. The other four dipole configurations being at distance at most $3$ from $2+2+2+2$, we always have $d(S_\partial , B_1) \leq 9$.

\textbf{Channel $(2a)$.} Let us first look at the grey edges. 
In the two configurations $4+4$ and $4+2+2$, we need to cut eight of the twelve grey strands; this requires $4$ flips. 
We then have to perform four more flips on pairs of black edges to obtain the boundary graph $B_c$. We thus have $d(S_\partial , B_c) \leq 8$ in the configurations $4+4$ and $4+2+2$. Since $8$ is at distance one from $4+4$, while $6+2$ and $2+2+2+2$ are at distance one from $4+2+2$, we conclude that $d(S_\partial , B_c) \leq 9$ in all cases.

\textbf{Channel $(2b)$.} We again need to perform $4$ flips on grey edges. In configurations $8$ and $4+2+2$, we can then obtain $B_{2b}$ after $4$ more flips on black edges. As $4+2+2$ is at distance one from $4+4$, $6+2$ and $2+2+2+2$, we always have $d(S_\partial , B_c) \leq 9$. 

\textbf{Channel $(2c)$.} As before, we have to cut eight of the twelve grey strands, which can be achieved in $4$ flips. Then, for the configurations $2+2+2+2$ and $4+4$, we can implement $4$ additional flips on black strands to obtain the boundary graph $B_{2c}$. Given that $8$ and $4+4$ are at distance one from either $2+2+2+2$ or $4+4$, we infer that $d(S_\partial , B_{2c}) \leq 9$ for these four configurations.  
We can finally check that the configuration $6+2$ can also be mapped to $B_{2c}$ in $9$ flips (4 flips on grey edges, and $5$ on black edges), as illustrated in figure~\ref{fig:moves_62}).
\begin{figure}[htbp]
\centering
\includegraphics[scale=1.2]{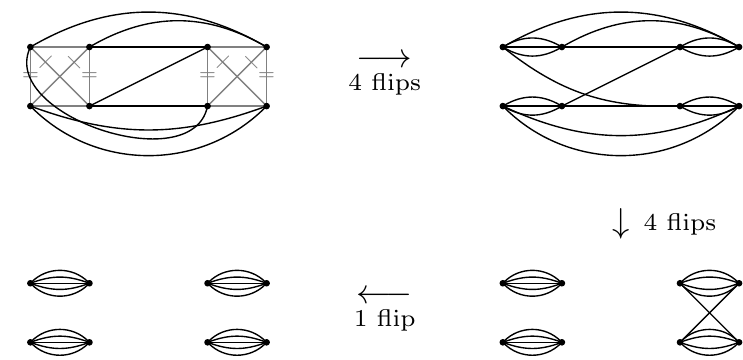}
\caption{Graphical proof that the configuration $6+2$ is at flip distance (at most) $9$ from $B_{2c}$.}
\label{fig:moves_62}
\end{figure}
\end{proof}

\subsection{Dipole-tadpole and quartic rung deletions}

\begin{figure}[htbp]
\centering
\includegraphics[scale=1]{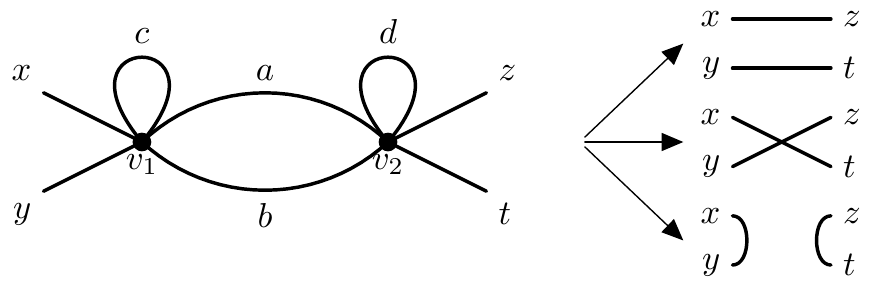}
\caption{The three channels of deletion for the dipole-tadpole four-point subgraph.}
\label{fig:dipole_tadpole_channel}
\end{figure}

We will now proceed with the deletion of the four-point \textit{dipole-tadpole} subgraphs. As for the deletion of single-tadpoles, there are three possible channels: parallel, cross and orthogonal. They are represented in figure~\ref{fig:dipole_tadpole_channel}.

In the following lemma, we prove that it is always advantageous to delete a dipole-tadpole in the orthogonal channel, and in at least one of the parallel or cross channels. 
\begin{lemma}
Let $G$ be a stranded graph and $S$ a strict dipole-tadpole subgraph of $G$. Call $G'$ the graph obtained after deletion of $S$ in the channel $c$ and assume $G'$ remains connected. 

\begin{itemize}
\item[(a)] Suppose $c$ is the orthogonal channel. Then, it is possible to choose $G'$ such that $\omega(G)\geq \omega(G')$. 

\item[(b)] Suppose $c$ is the parallel (resp. cross) channel, and call $G''$ the graph obtained after deletion of $S$ in the cross (resp. parallel) channel. Then, if it is not possible to choose $G'$ such that $\omega(G)\geq \omega(G')$, provided that $G''$ remains connected, it is possible to choose $G''$ such that $\omega(G)\geq \omega(G'')$.

\end{itemize}
\label{lemma:dipole_tadpole}
\end{lemma}

\begin{proof}

Call $v_1$ the vertex connected to $(x,y)$ and $v_2$ the other vertex. We first observe that a number of situations can be dealt with lemma \ref{lemma:single_tadpole}, through successive deletions of the tadpoles at $v_1$ and $v_2$. We distinguish three cases. 
\begin{itemize}
\item If neither $v_1$ nor $v_2$ are of type $1+1+2$, then we can perform the move in any channel (and in particular in the orthogonal channel): delete $v_1$ in the parallel channel,  then $v_2$ in the desired channel. This is illustrated in figure~\ref{fig:lemma_diptad_case1}. 

\item If one of them (say $v_1$) is of type $1+1+2$, there is a single channel $c$ in which $v_1$ can be deleted (in which case we gain a factor $1/N$). We then distinguish two subcases. If the channel $c$ pairs $(x,y)$ and $(a,b)$, we first delete $v_2$ in the parallel channel, then $v_1$ in the channel $c$. This implements the orthogonal deletion, as illustrated in figure~\ref{fig:lemma_diptad_case2a}. 

If, on the other hand, the channel $c$ is parallel/cross (say, it pairs $(x,a)$ and $(y,b)$), we delete $v_1$ in this channel, and then $v_2$ in any desired channel, as shown in figure~\ref{fig:lemma_diptad_case2b}. 

\item Assume, finally, that both $v_1$ and $v_2$ are of type $1+1+2$. If at least one of them (say $v_1$) can be deleted in the parallel or cross channel, we perform this move and then delete $v_2$ in the appropriate channel: the first deletion yields a $1/N$ suppression, while the second deletion brings at most a factor of $N$. We can therefore perform the deletion in all three channels, as illustrated in figure~\ref{fig:lemma_diptad_case3a}. 

On the other hand, if both $v_1$ and $v_2$ can only be deleted in the orthogonal channel, we are still able to implement the orthogonal channel: delete $v_1$ in the orthogonal channel (yields a $1/N$ suppression), then $v_2$ in the parallel channel (results in an additional factor of $N$). This is illustrated in figure~\ref{fig:lemma_diptad_case3b}. 

\end{itemize}

\begin{figure}[htbp]
\centering
\begin{tabular}{c}
\subfloat[Neither $v_1$ nor $v_2$ are of type $1+1+2$\label{fig:lemma_diptad_case1}]{\includegraphics[scale=.9]{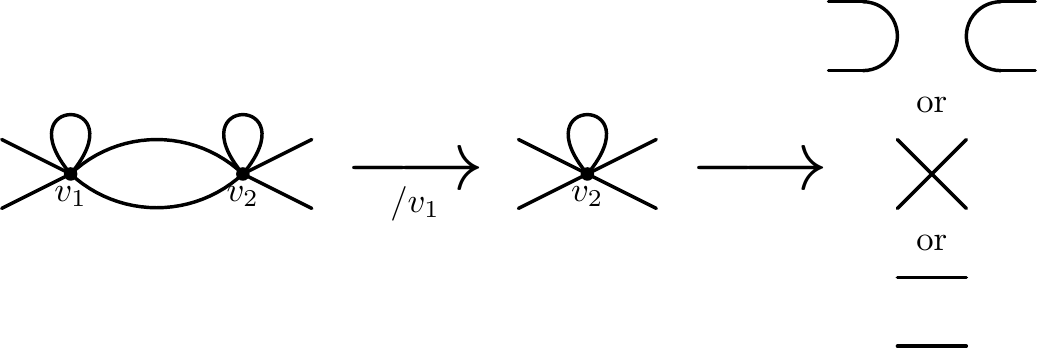}} \\
\subfloat[Only $v_1$ is of type $1+1+2$ and can be deleted in the orthogonal channel\label{fig:lemma_diptad_case2a}]{\includegraphics[scale=0.9]{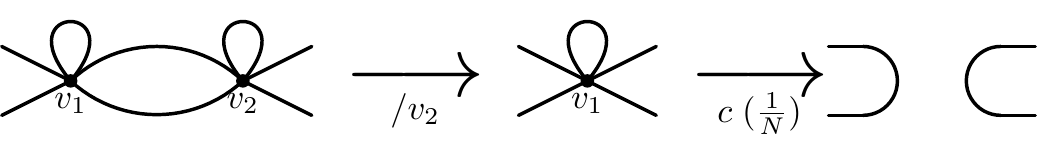}} \\
\subfloat[Only $v_1$ is of type $1+1+2$ and can be deleted in the parallel/cross channel\label{fig:lemma_diptad_case2b}]{\includegraphics[scale=0.9]{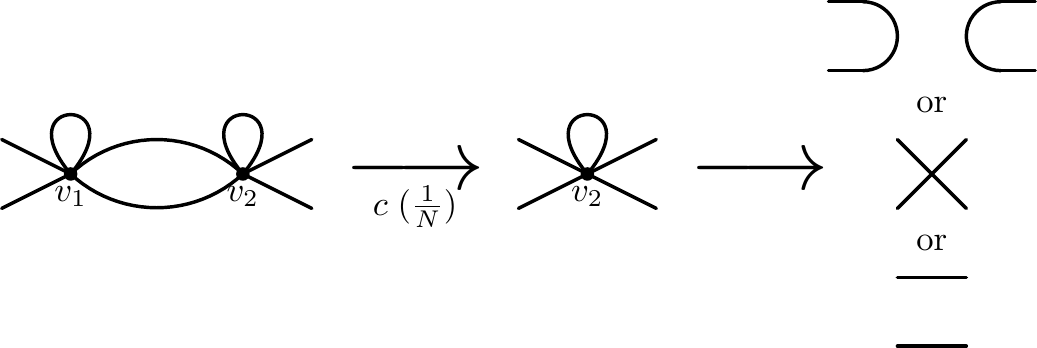}} \\
\subfloat[$v_1$ and $v_2$ are of type $1+1+2$, and $v_1$ can be deleted in the parallel/cross channel\label{fig:lemma_diptad_case3a}]{\includegraphics[scale=0.9]{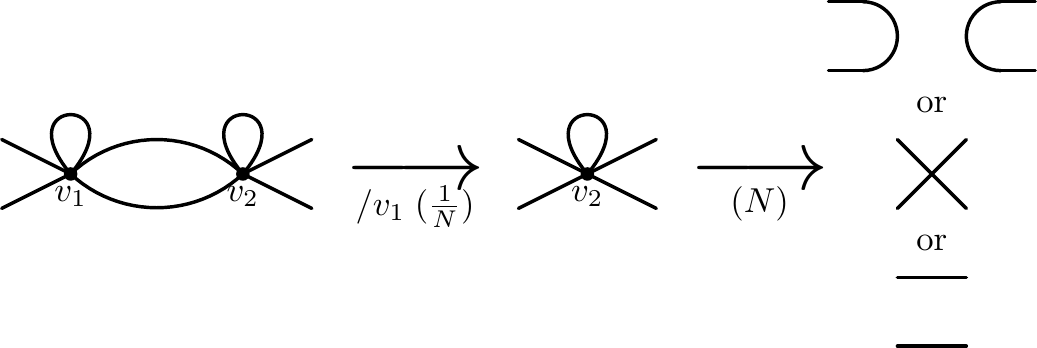}} \\
\subfloat[$v_1$ and $v_2$ are of type $1+1+2$, and both can be deleted in the cross channel\label{fig:lemma_diptad_case3b}]{\includegraphics[scale=0.9]{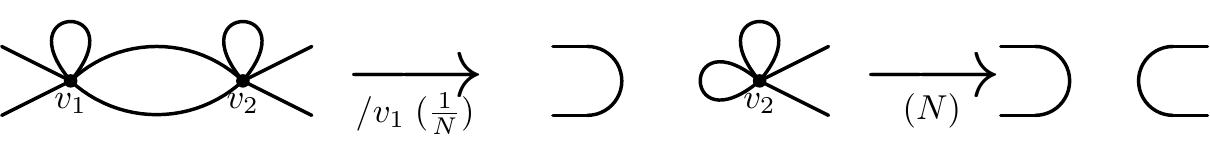}} 
\end{tabular}
\caption{Dipole-tadpole deletions from tadpole deletions (lemma \ref{lemma:dipole_tadpole}).}
\label{fig:dip_tad_1st}
\end{figure}

All in all, the only subcases left to investigate are about the parallel/cross channels, in the following situation: $v_2$ is of type $1+1+2$ and can be deleted in the orthogonal channel; furthermore, if $v_1$ is of type $1+1+2$, its easy channel is also the orthogonal one. In particular, we can now assume that $v_2$ is in one of the two configurations shown in figure~\ref{fig:config_v2}.

\begin{figure}[htbp]
\centering
\captionsetup[subfigure]{labelformat=empty}
\subfloat[]{\includegraphics[scale=.6]{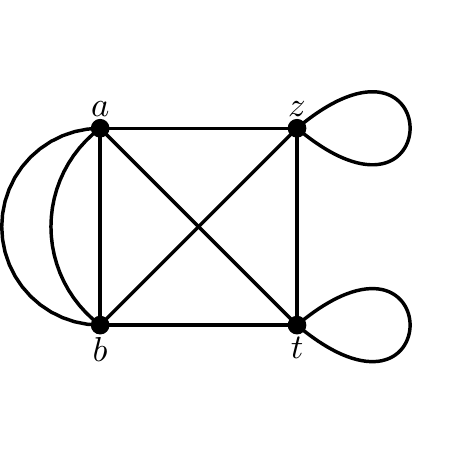}} \hspace{2cm}
\subfloat[]{\includegraphics[scale=.6]{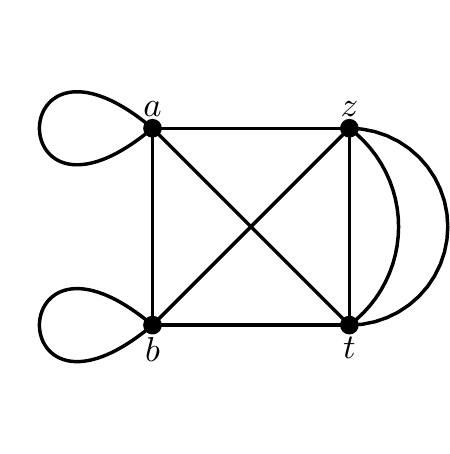}}
\caption{Special configurations of $v_2$.}\label{fig:config_v2}
\end{figure}

\medskip

We now determine the allowed boundary graphs in the remaining configurations. The lemma will follow if we can prove that $d(S_\partial,B_c) \leq 10 - F(S)$, with $c$ either the parallel or cross channel. For this purpose, it will be sufficient to consider equivalent classes of boundary graphs under exchanges of $x$ and $y$, $z$ and $t$, as well as $(x,y)$ and $(z,t)$.
As $v_2$ can only be in the configurations of figure~\ref{fig:config_v2}, the full boundary graph must have one of the two structures depicted in figure~\ref{fig:full_bndy_v2}. We distinguish three cases.

\begin{figure}[htbp]
\centering
\includegraphics[scale=1]{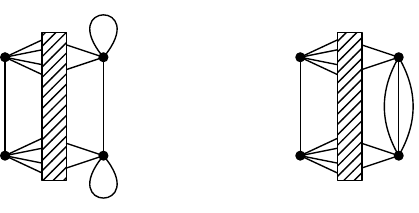}
\caption{Two special configurations for the boundary graph of a dipole-tadpole.}
\label{fig:full_bndy_v2}
\end{figure}

Assume first that $S_\partial$ has two connected components. In that case, we necessarily have $F(S)\leq 3$. Indeed, we infer from the boundary structure of tadpoles in figure~\ref{fig:tadpole_config} that at most three corners are available at $v_1$ (resp. $v_2$) to support faces of length two or higher. Furthermore, each such face will use at least one corner. But when $S_\partial$ is disconnected, two of those corners must already be occupied by strands that connect $z$ to $t$ (resp. $x$ to $y$). We have therefore at most one face of length two or higher. Remembering that each tadpole line can support an additional face, we obtain the claimed bound: $F(S) \leq 3$. The six inequivalent boundary graphs which can be realized under those conditions are represented in figure~\ref{fig:config_ortho_del_diptad}. It is straightforward to check that $d(S_\partial , B_c) \leq 7 \leq 10 - F(S)$, where $c$ is e.g. the parallel channel. 

\begin{figure}[htbp]
\centering
\includegraphics[scale=1]{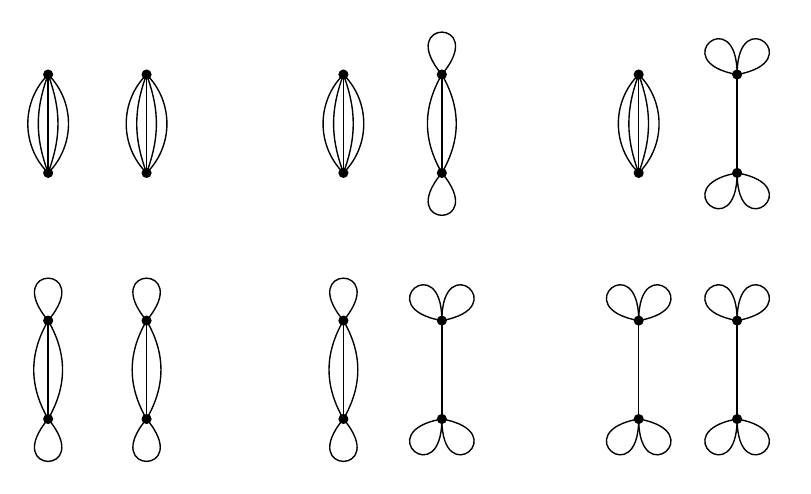}
\caption{The six inequivalent boundary graphs with two connected components.}
\label{fig:config_ortho_del_diptad}
\end{figure}

We can now assume that $S_\partial$ is connected. We note that there is at least two edges connecting the pair of vertices $(x,y)$ and $(z,t)$. Indeed, if there were only one such edge, we would have an additional seven half-edges to match from the pair $(x,y)$, which by parity is impossible. Let us first assume that there are exactly two edges connecting $(x,y)$ to $(z,t)$, and that furthermore, they are not both in the same parallel/cross channel.  For definiteness, and without loss of generality, we can suppose those edges are between $y$ and $t$, and between $y$ and $z$. There are then four possible boundary graphs, as represented in figure~\ref{fig:bndy_exactly_1para}. Compared to the previous paragraph, one additional corner is available at $v_1$ (resp. $v_2$) to build up faces of length two or higher. This leads to the weaker bound $F(S) \leq 4$. However, a straightforward inspection of the graphs of figure~\ref{fig:bndy_exactly_1para} shows that $d(S_\partial , B_c) \leq 6 \leq 10 - F(S)$ always holds (and is saturated for the last configuration), where $c$ is the parallel channel. 

\begin{figure}[htbp]
\centering
\includegraphics[scale=1]{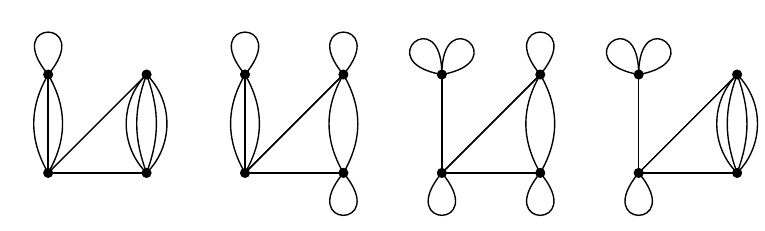}
\caption{The four inequivalent boundary graphs with exactly one edge in the cross channel, and one edge in the parallel channel.}
\label{fig:bndy_exactly_1para}
\end{figure}

Let us finally suppose that there are at least two edges in the same parallel or cross channel. Without loss of generality, we can assume it to be the parallel channel. After a straightforward (but tedious) inspection, we find another sixteen inequivalent boundary graphs, which we have depicted in figure~\ref{fig:bndy_16}. Any such configuration verifies $d(S_\partial , B_c) \leq 5 \leq 10 - F(S)$, where we have used that $F(S) \leq 5$ for any dipole-tadpole $S$. 

This concludes the proof.
\begin{figure}[htbp]
\centering
\includegraphics[scale=1]{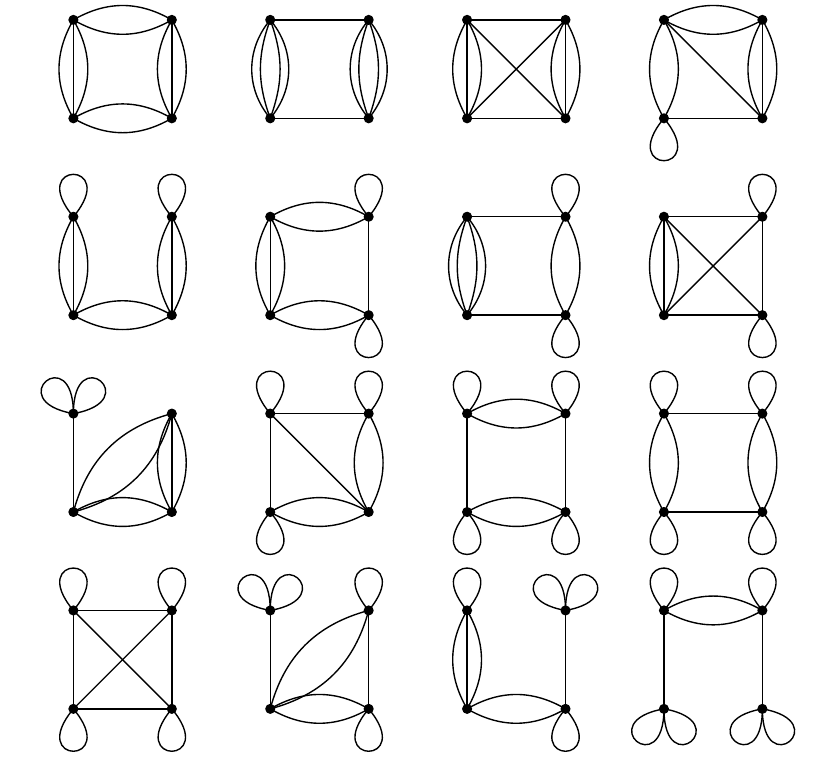}
\caption{The sixteen inequivalent boundary graphs with at least two edges in the parallel channel.}
\label{fig:bndy_16}
\end{figure}
\end{proof}

\begin{figure}[htbp]
\centering
\includegraphics[scale=1]{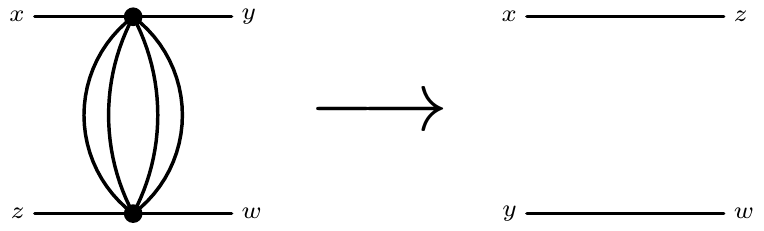}
\caption{Deletion of a quartic rung subgraph.}
\label{fig:deletion_melon_4pt}
\end{figure}

We will also need to delete a particular type of four-point subgraph represented in figure~\ref{fig:deletion_melon_4pt}. We call this type of graph \textit{quartic rung}. There are three different channels of deletion but we will only consider one (as represented in figure~\ref{fig:deletion_melon_4pt}). 

\begin{lemma}
Let $G$ be a stranded graph and $S$ a strict quartic rung subgraph of $G$. Call $G'$ the graph obtained after deletion of $S$ in the channel depicted in figure~\ref{fig:deletion_melon_4pt} and assume $G'$ remains connected. 

It is always possible to perform the deletion in such a way that $\omega(G)\geq \omega(G')$. 
\label{lemma:melon4pt}
\end{lemma}

\begin{proof}

We want to prove the following bound:
\begin{equation}
d(S_{\partial},B)\leq 10-F(S) \, ,
\end{equation}
with $B$ the boundary graph of the deletion channel depicted in figure~\ref{fig:deletion_melon_4pt} with no self-loops.

First notice that we can apply the argument of figure~\ref{fig:dipole_reduction} (from the proof of lemma~\ref{lemma:dipole}) to each of the six dipole subgraphs of $S$. This allows to assume, without loss of generality, that $F(S)= 6$. 
In such a situation, $S_\partial$ is one of the five boundary graphs represented in figure~\ref{fig:bndy_lemma5}: there is one edge from $x$ to $y$, one edge from $z$ to $w$, and each of the eight remaining edges connects a vertex of the pair $(x,y)$ to a vertex of the pair $(z,w)$. It is then clear that eight edges need to be reconfigured to map $S_\partial$ to $B$. Given that those do not include any self-loop, we conclude that $d(S_\partial, B) \leq 4 = 10 - F(S)$. 
\begin{figure}[htbp]
\centering
\includegraphics[scale=1]{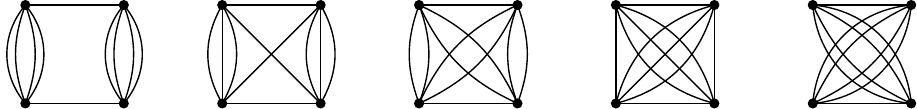}
\caption{The five possible boundary graphs of a quartic rung $S$ with $F(S)=6$.}
\label{fig:bndy_lemma5}
\end{figure}
\end{proof}

\subsection{Two-point subgraph deletions}\label{subsec:main_proof}

The following lemma will allow us to find inductive bounds on two-particle reducible graphs.
\begin{lemma}\label{lemma:2PR}
Let $G$ be a two-particle reducible stranded graph. That is, $G$ has the structure represented in figure~\ref{fig:2PR}, where $S_1$ and $S_2$ are (non-empty) two-point subgraphs. Denote by $G_1$ (resp. $G_2$) the graph obtained by closing $S_1$ (resp. $S_2$) with an unbroken edge $e_1$ (resp. $e_2$). It is possible to choose $e_1$ and $e_2$ such that:
\begin{equation}
\omega(G) \geq \omega(G_1) + \omega(G_2) - 4\,.
\end{equation}
Moreover, if $S_2$ has no tadpole and no dipole, then 
\begin{equation}
\omega(G) \geq \omega(G_1)+1 \,.
\end{equation}

\begin{figure}[htbp]
\centering
\includegraphics[scale=1]{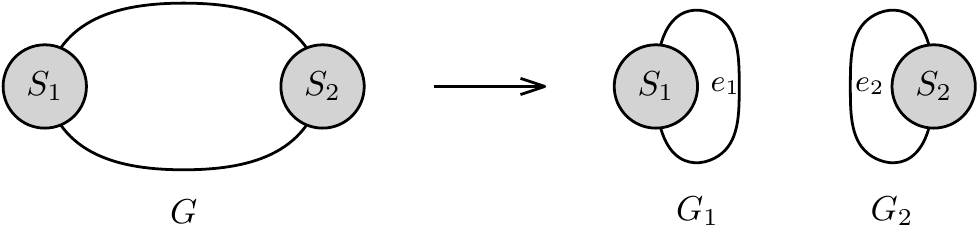}
\caption{Deletion of two-particle reducible components.}
\label{fig:2PR}
\end{figure}
\end{lemma}
\begin{proof}
The boundary graph of $S_1$ (resp. $S_2$) is one of the three configurations shown in figure~\ref{fig:propagator_bndy}. It is then apparent that we can choose $e_1$ and $e_2$ such that:\footnote{Note that we are constrained by the fact that $e_1$ and $e_2$ are required to be unbroken. Without this restriction, we could ensure that $F(G_1) = F(S_1) + 5$ and $F(G_2) = F(S_2) + 5$.}
\begin{equation}
F(G_1) \geq F(S_1) +3 \quad \mathrm{and} \quad F(G_2) \geq F(S_2) + 3\,.
\end{equation}
Furthermore, it is clear that $F(G) \leq F(S_1) + F(S_2) + 5$ (there are ten strands in $G\setminus(S_1 \cup S_2)$, and they all belong to faces of length at most two). Hence, $F(G_1)+ F(G_2) - F(G) \geq 1$, which is equivalent to $\omega(G_1)+\omega(G_2) - 4 \leq \omega(G)$.
Finally, if we assume that $S_2$ has no tadpole or dipole, it is clear that $G_2$ has at most one dipole or one tadpole. Hence, 
\begin{equation}
\omega(G_2)=5+\sum_{k \geq 1} \frac{k-3}{3}F_k (G_2) \geq 5-2/3 \, ,
\end{equation}
which implies $\omega(G_2)\geq 5$ (since $\omega \in \mathbb{N}$). As a consequence, $\omega(G) \geq \omega(G_1) +1$.
\end{proof}

Finally, to prove the main result of this section (proposition~\ref{prop:positive_degree}), we will also rely on a number of special two-point moves, which we gather in the next lemma.  
\begin{lemma}
Let $G$ be a stranded graph and $S$ a strict subgraph of $G$. Suppose $S$ is one of the two-point subgraphs of figure~\ref{fig:particular_cases}. Call $G'$ the graph obtained from $G$ by substituting $S$ with an unbroken edge $e$. 
Then, it is always possible to choose $e$ in such a way that $\omega(G)\geq \omega(G')+1$.

\begin{figure}[H]
\captionsetup[subfigure]{labelformat=empty}
\begin{center}
\begin{tabular}{cccc}
\subfloat[$H_0$]{\includegraphics[scale=0.9]{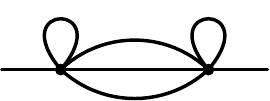}}
&
\subfloat[$H_1$]{\includegraphics[scale=0.9]{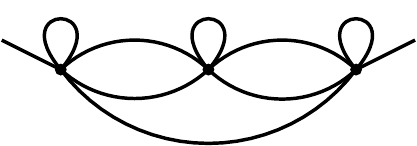}}
&
\subfloat[$H_2$]{\includegraphics[scale=0.9]{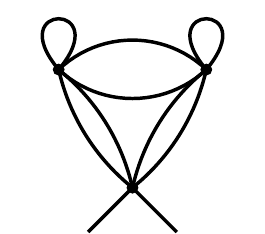}}
&
\subfloat[$H_3$]{\includegraphics[scale=0.9]{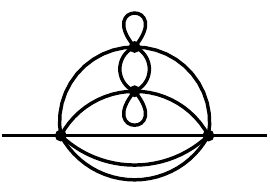}}
\\
\subfloat[$H_4$]{\includegraphics[scale=0.9]{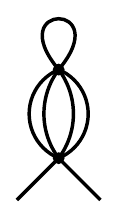}}
&
\subfloat[$H_{5}$]{\includegraphics[scale=0.9]{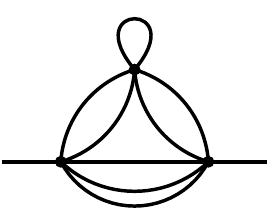}}
&
\subfloat[$H_{6}$]{\includegraphics[scale=0.9]{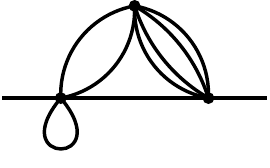}}
&
\subfloat[$H_{7}$]{\includegraphics[scale=0.9]{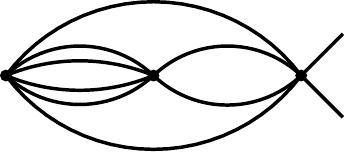}}
\\
\subfloat[$H_{8}$]{\includegraphics[scale=0.9]{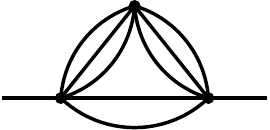}}
&
\subfloat[$H_{9}$]{\includegraphics[scale=0.9]{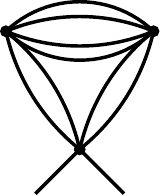}}
&
\subfloat[$H_{10}$]{\includegraphics[scale=0.9]{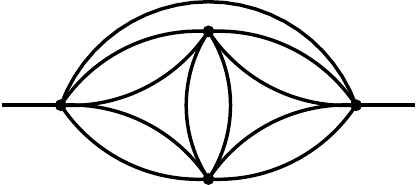}}
&
\subfloat[$H_{11}$]{\includegraphics[scale=0.9]{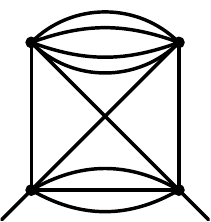}}
\\
\subfloat[$H_{12}$]{\includegraphics[scale=0.9]{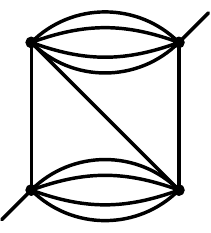}}
&
&
&
\end{tabular}
\end{center}
\caption{Particular two-point subgraphs which can always decrease the degree upon deletion.}
\label{fig:particular_cases}
\end{figure}
\label{lemma:particular_cases}
\end{lemma}
\begin{proof}
See appendix~\ref{app:particular}.
\end{proof}

\begin{lemma}\label{lemma:two-point_deletion}
Let $G$ be a stranded graph with no double-tadpole, no melon, and no \emph{separating dipole-tadpole} (as introduced in definition~\ref{def:dipole-tadpole} and figure~\ref{fig:dipole-tadpole}). Suppose there exists a proper two-point subgraph $H \subset G$, which can be deleted (i.e. replaced by an unbroken edge) in a way that strictly decreases the degree. Then, there exists a graph $G'$, with no double-tadpole and no melon, such that:
\begin{equation}
V(G') < V(G) \quad \mathrm{and} \quad \omega(G) \geq \omega(G')\,.
\end{equation}  
\end{lemma}
\begin{proof}
Let us call $G_1$ the (connected) graph obtained from $G$ by deletion of $H$, and such that $\omega(G) \geq \omega(G_1)+1$.

If $G_1$ does not contain double-tadpoles or melons, we take $G' = G_1$.

If, on the other hand, $G_1$ contains a melon, then $G$ was in the configuration of figure~\ref{fig:config_deletion_2pt}, namely: the subgraph $H$ is adjacent to a quartic rung. 
In this case, we define $G'$ as the graph obtained by deletion of the quartic rung, following lemma~\ref{lemma:melon4pt}. It is clear that this move cannot create melons or double-tadpoles, and by lemma~\ref{lemma:melon4pt}, $\omega(G) \geq \omega(G')$. 
\begin{figure}[H]
\centering
\captionsetup[subfigure]{labelformat=empty}
\includegraphics[scale=1]{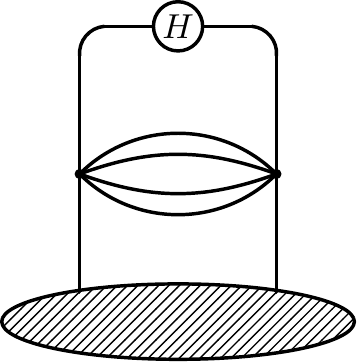}
\caption{Configuration of the graph generating a melon upon deletion of $H$.}
\label{fig:config_deletion_2pt}
\end{figure}

Finally, $G_1$ can contain a double-tadpole, in which case $G$ contains the subgraph depicted on the left side of figure~\ref{fig:deletion_2pt}. As illustrated in the same figure, we can subsequently remove the double-tadpole from $G_1$ to obtain a graph $G_2$ such that $\omega(G_1) \geq \omega(G_2) - 1$ (following lemma~\ref{lemma:double_tadpole_deletion}). As a result, $\omega(G) \geq \omega(G_2)$. $G_2$ cannot contain a double-tadpole, otherwise $G$ would contain a separating dipole-tadpole (as illustrated in figure~\ref{fig:ind_dipole_tadpole_a}), which is excluded by assumption. If $G_2$ does not contain a melon either, we define $G' = G_2$ and conclude. If, on the other hand, $G_2$  contains a melon, then $G$ contains a subgraph with a quartic rung as represented in figure~\ref{fig:config_3_dbtad}. In that case, we can again invoke lemma~\ref{lemma:melon4pt} to delete the quartic rung, and obtain a graph $G'$ with no double-tadpole or melon.
\begin{figure}[H]
\centering
\captionsetup[subfigure]{labelformat=empty}
\includegraphics[scale=1]{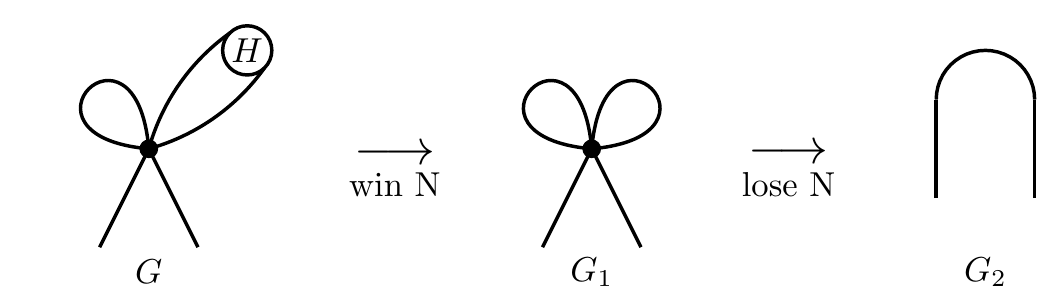}
\caption{Steps for deleting $H$ if it appears within a generalized double-tadpole. We gain a factor $N$ with the first step and lose one with the second step.}
\label{fig:deletion_2pt}
\end{figure}

\begin{figure}[H]
\centering
\captionsetup[subfigure]{labelformat=empty}
\includegraphics[scale=1]{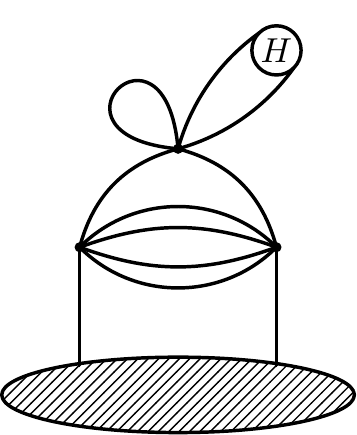}
\caption{Configuration leading to a melon when deleting the generalized double-tadpole containing $H$.}
\label{fig:config_3_dbtad}
\end{figure}
\end{proof}

\subsection{Main proposition}

\begin{proposition}
Let $G$ be a stranded graph. If $G$ has no double-tadpole and no melon, then $\omega(G) \geq 0$. 
\label{prop:positive_degree}
\end{proposition}

\begin{proof}
If $G$ has no short face or no vertex, we have already seen that $\omega(G) \geq 0$. In all other cases, we proceed by induction on the number of vertices. From now on, we assume that $V(G)\geq 1$, and that $G$ contains at least one tadpole or one dipole.  

Even if $G$ cannot have double-tadpoles or melons, generalized double-tadpoles and melons are still allowed. To avoid difficulties with such subgraphs, we will first deal with type-$I$ dipoles and tadpoles, and study the more involved case of type-$II$ configurations separately (see subsection~\ref{sec:types} and figure~\ref{fig:not_easy} for definitions).  

We now proceed with an exhaustive graph-theoretic distinction of cases. In each situation, we will look for a strict subgraph of $G$ which can be deleted without increasing the degree, and while preserving our combinatorial constraints (namely: connectedness, the absence of melons or double-tadpoles, as well as the absence of broken or doubly-broken edges). From the induction hypothesis, it will then follow that $\omega(G) \geq 0$.

\begin{figure}[htbp]
\centering
\subfloat[\label{fig:ind_dipole_tadpole_a}]{\includegraphics[scale=1.2]{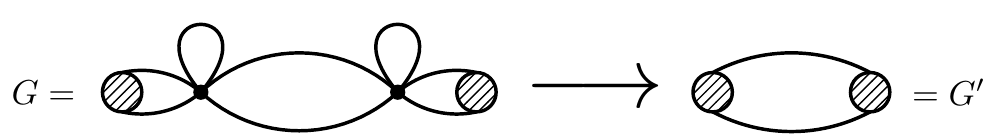}}
\\
\subfloat[\label{fig:ind_dipole_tadpole_b}]{\includegraphics[scale=1.2]{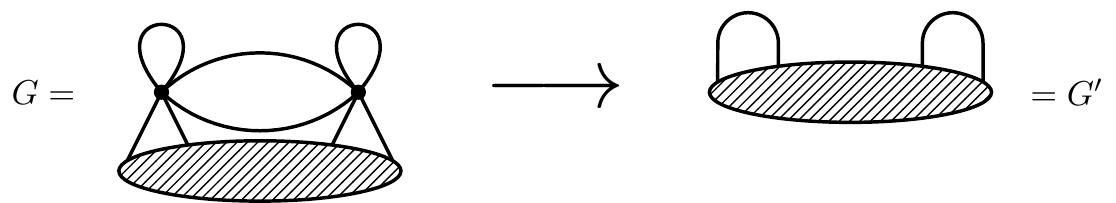}}
\caption{The two possible primary configurations of a dipole-tadpole. It can be separating (top), in which case the graph remains connected upon deletion in the parallel/cross channel; or non-separating (bottom), in which case the graph remains connected upon deletion in the orthogonal channel.}
\label{fig:ind_dipole_tadpole}
\end{figure}

\begin{itemize}
\item[\emph{Case A.}] First suppose that there exists a separating dipole-tadpole in $G$ (see figure~\ref{fig:ind_dipole_tadpole_a}). Performing a deletion of this dipole-tadpole in the parallel or cross channel, we obtain a graph $G'$ which cannot contain double-tadpoles or melons. Furthermore, by lemma~\ref{lemma:dipole_tadpole} (b), we can make sure that $\omega(G)\geq \omega(G') \geq 0$ (the last inequality follows from the induction hypothesis, because $G'$ has strictly fewer vertices than $G$).
\end{itemize}

We can now assume that there are no more separating dipole-tadpoles.

\begin{itemize}
\item[\emph{Case B.}] Suppose that there exists a non-separating dipole-tadpole in $G$ (see figure~\ref{fig:ind_dipole_tadpole_b}). We can then use lemma~\ref{lemma:dipole_tadpole} to delete the latter in the orthogonal channel, and obtain a connected graph $G'$ with $\omega(G) \geq \omega(G')$.
If $G'$ has no double-tadpole and no melon, then we are done. If, on the other hand, $G'$ has a double-tadpole, then $G$ is in either one the configuration represented in figure~\ref{fig:ind_dipole_tadpole_ortho_a} or contains the subgraph $H_2.$
If, instead, $G'$ has a melon, then $G$ either contains the two-point subgraph $H_4$ (see figure~\ref{fig:particular_cases}) or contains the four-point function depicted in figure~\ref{fig:ind_dipole_tadpole_ortho_c}.

\begin{figure}[H]
\centering
\subfloat[]{\label{fig:ind_dipole_tadpole_ortho_a}\includegraphics[scale=0.8]{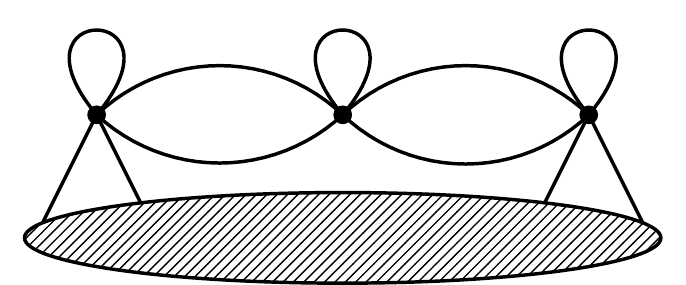}}
\hspace{1cm}
\subfloat[]{\label{fig:ind_dipole_tadpole_ortho_c}\includegraphics[scale=0.8]{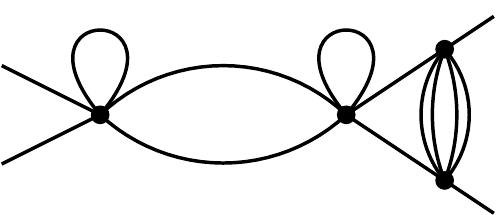}}
\caption{Two configurations of a non-separating dipole-tadpole that can create double-tadpoles or melons upon deletion in the orthogonal channel. 
}
\label{fig:ind_dipole_tadpole_ortho}
\end{figure}

In the two situations of figure~\ref{fig:ind_dipole_tadpole_ortho}, we can try to perform the deletion of the dipole-tadpole in the parallel channel, which cannot disconnect the graph. One may however create double-tadpoles or melons, in which case $G$ contains one of the subgraph $H_1$, $H_2$, $H_3$, or the two-point subgraph depicted in figure~\ref{fig:ind_dipole_tadpole_2pt}. In the latter case, we can use lemma~\ref{lemma:melon4pt} to remove the quartic rung, and reduce the problem to the situation in which $G$ contains the subgraph $H_0$. In conclusion, we have shown that $G$ always contains a subgraph $H_i$ covered by lemma~\ref{lemma:particular_cases}. We can therefore apply lemma~\ref{lemma:two-point_deletion} as a last step, which outputs a suitable graph $G''$ with $\omega(G)\geq \omega(G'')$.

\begin{figure}[H]
\centering
\includegraphics[scale=1]{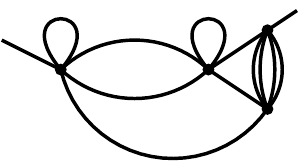}
\caption{Configuration of a dipole-tadpole subgraph which can generate a melon in the parallel or cross channel.}
\label{fig:ind_dipole_tadpole_2pt}
\end{figure}
 
\end{itemize}

We can now assume that there are no more dipole-tadpole subgraphs in $G$. 

\begin{itemize}
\item[\emph{Case C.}] Suppose that there exists a type-$I$ tadpole in $G$. Then the graph remains connected upon deletion of this tadpole in any channel. By application of lemma~\ref{lemma:single_tadpole}, we obtain a graph $G'$ with $\omega(G)\geq \omega(G')$. 
\end{itemize}
\begin{itemize}
\item If $G'$ has a double-tadpole, then $G$ either contained a dipole-tadpole or the subgraph $H_4$ depicted in figure~\ref{fig:particular_cases}. The first situation has already been excluded, and we can use lemmas~\ref{lemma:particular_cases} and \ref{lemma:two-point_deletion} in combination to deal with the second. 

\item If $G'$ has a melon, then  $G$ either contained the subgraph $H_5$ from figure~\ref{fig:particular_cases} or the four-point graph of figure~\ref{fig:ind_easy_tadpole}.
In the second case, we can use lemma~\ref{lemma:melon4pt} to delete the quartic rung. This last step cannot create a melon. If it does not create a double-tadpole either, we conclude.  If it does, then $G$ necessarily contained the subgraph $H_6$ from figure~\ref{fig:particular_cases}. We can deal with this situation, as well as with the configuration $H_5$, by application of lemmas~\ref{lemma:particular_cases} and \ref{lemma:two-point_deletion}. 
\end{itemize}

\begin{figure}[htbp]
\centering
\includegraphics[scale=1]{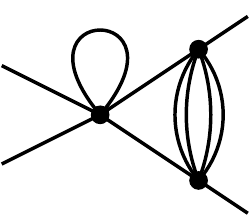}
\caption{A four-point subgraph that can lead to the creation of a melon when deleting a type-$I$ tadpole.}
\label{fig:ind_easy_tadpole}
\end{figure}

From now on, we assume that there is no type-$I$ tadpole in $G$. 

\begin{itemize}
\item[\emph{Case D.}] Suppose that $G$ contains a type-$I$ dipole. We can then attempt to delete this dipole in one of the channels covered by lemma~\ref{lemma:dipole}. 
\end{itemize}

\begin{itemize}
\item If the channels $(2a),(2b)$ and $(2c)$ all disconnect the graph, we instead perform the deletion in the parallel channel to obtain a new graph $G'$. Thanks to our assumption that the deleted dipole was of type-$I$, $G'$ is necessarily connected (see our earlier discussion around figure~\ref{fig:dipole_connected}). Moreover, $G'$ cannot have double-tadpoles or melons. By application of lemma~\ref{lemma:dipole}, we can ensure that $\omega(G) \geq \omega(G')$.

\item If one of the three channels $2$ does not disconnect the graph, for example the channel $(2a)$ (which we can assume without loss of generality), we perform the deletion in this channel. However, this can create a double-tadpole. In this case $G$ was in one of the first three configurations of figure~\ref{fig:ind_config_easy_dipole}.

\begin{figure}[htpb]
\centering
\captionsetup[subfigure]{labelformat=empty}
\subfloat[(i)]{\includegraphics[scale=1]{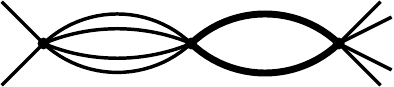}}
\hspace{1cm}
\subfloat[(ii)]{\includegraphics[scale=1]{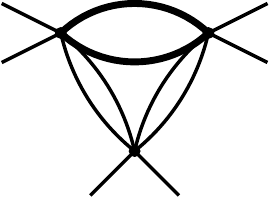}}
\hspace{1cm}
\subfloat[(iii)]{\includegraphics[scale=1]{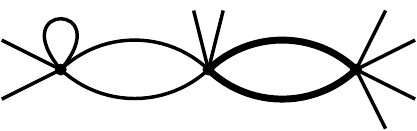}}
\\
\subfloat[(iv)]{\includegraphics[scale=1]{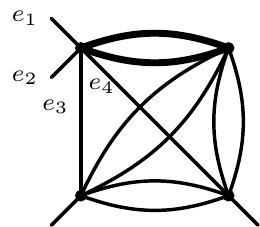}}
\hspace{1cm}
\subfloat[(v)]{\includegraphics[scale=1]{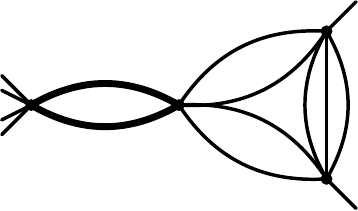}}
\hspace{1cm}
\subfloat[(vi)]{\includegraphics[scale=1]{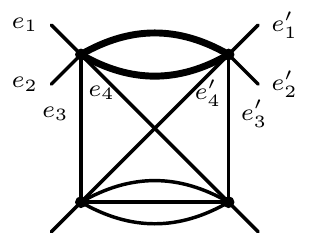}}
\hspace{1cm}
\subfloat[(vii)]{\includegraphics[scale=1]{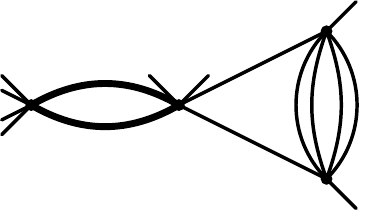}}
\caption{Seven configurations of a dipole (in bold) that lead to the creation of a melon or a double-tadpole. The first three lead to the creation of a double-tadpole while the four other lead to the creation of a melon.}
\label{fig:ind_config_easy_dipole}
\end{figure}
This can also create a melon. In this case, $G$ was either in one of the last four configurations of figure~\ref{fig:ind_config_easy_dipole} or contains the subgraph $H_{10}$ depicted in figure~\ref{fig:particular_cases}.
$H_{10}$ can be dealt with by application of lemmas~\ref{lemma:particular_cases} and \ref{lemma:two-point_deletion}. We have to look at the other configurations in more detail.

\begin{itemize}
\item[(i)] We delete the dipole in the parallel channel instead. If this creates a double-tadpole, the graph is either in the configuration of figure~\ref{fig:ind_deletion_easy_dipole_a} or contains the subgraph $H_7$ of figure~\ref{fig:particular_cases}.

\begin{figure}[htbp]
\centering
\subfloat[]{\label{fig:ind_deletion_easy_dipole_a}\includegraphics[scale=1]{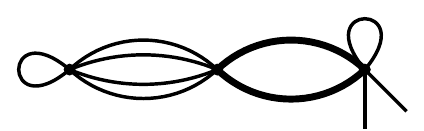}}
\hspace{0.125cm}
\subfloat[]{\label{fig:ind_deletion_easy_dipole_b}\includegraphics[scale=1]{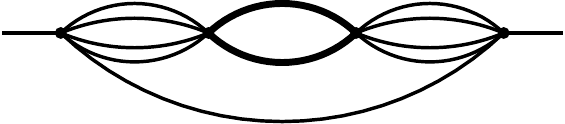}}
\hspace{0.125cm}
\subfloat[]{\label{fig:ind_deletion_easy_dipole_c}\includegraphics[scale=1]{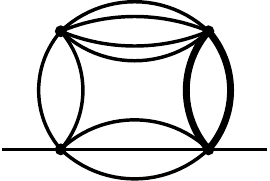}}
\\
\subfloat[]{\label{fig:ind_deletion_easy_dipole_d}\includegraphics[scale=1]{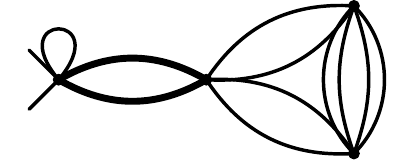}}
\hspace{1cm}
\subfloat[]{\label{fig:ind_deletion_easy_dipole_e}\includegraphics[scale=1]{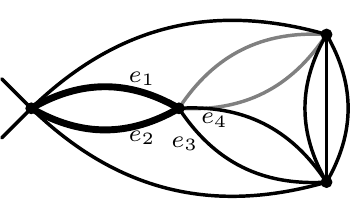}}
\\
\subfloat[]{\label{fig:ind_deletion_easy_dipole_f}\includegraphics[scale=1]{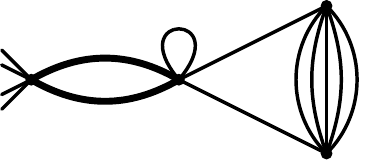}}
\hspace{1cm}
\subfloat[]{\label{fig:ind_deletion_easy_dipole_g}\includegraphics[scale=1]{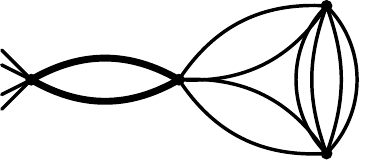}}
\caption{The interesting configurations when deleting a type-$I$ dipole. The type-$I$ dipole we want to delete is represented in bold.}
\label{fig:ind_deletion_easy_dipole}
\end{figure}

The first graph contains a type-$I$ tadpole (at its left end) so it has already been excluded.  We can deal with $H_7$ by means of lemmas~\ref{lemma:particular_cases} and \ref{lemma:two-point_deletion}.

If this creates a melon, the graph is in one of the configurations of figure~\ref{fig:ind_deletion_easy_dipole_b} and \ref{fig:ind_deletion_easy_dipole_c}.

We can first use lemma~\ref{lemma:melon4pt} to delete a quartic rung in both of these graphs. The first one reduces to $H_4$, the second one to $H_0$. In both situations, we can then conclude by invoking lemmas~\ref{lemma:particular_cases} and \ref{lemma:two-point_deletion}.

\item[(ii)] We perform the move of figure~\ref{fig:ind_move_easy_dipole}, which can be justified from lemma~\ref{lemma:dipole} as follows. We first try to delete the dipole in the (unique) channel $2$ that connects $a$ to $w$ and $b$ to $x$. If this move turns out to connect $c$ to $y$ (resp. $z$) and $d$ to $z$ (resp. $y$), we are done. If not, $c$ is mapped to $d$, which may result in a disconnected graph. In that situation, we instead perform a deletion in the (again unique) channel that connects $a$ to $x$ and $b$ to $y$. We are then guaranteed that $c$ is not mapped to $d$, and we have successfully implemented the combinatorial move of figure~\ref{fig:ind_move_easy_dipole}. 
\begin{figure}[H]
\centering
\includegraphics[scale=1]{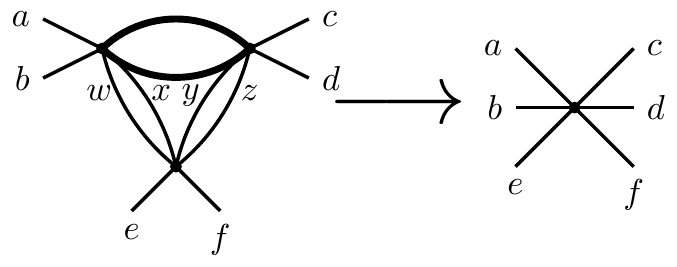}
\caption{Graphical representation of the move used to delete a type-$I$ dipole (in bold) in the configuration (ii).}
\label{fig:ind_move_easy_dipole}
\end{figure}

If this creates a double-tadpole, then $G$ must contain one of the subgraphs $H_8$ or $H_9$, as depicted in figure~\ref{fig:particular_cases}.

If this creates a melon, we are instead led to the two-point subgraph $H_{10}$.

Again, these special cases are dealt with lemmas~\ref{lemma:particular_cases} and \ref{lemma:two-point_deletion}. 

\item[(iii)] The configuration with a generalized tadpole on the vertex that already has a tadpole is forbidden, otherwise $G'$ would be disconnected. Hence $G$ contains a type-$I$ tadpole, which we have already excluded. 

\item [(iv)] We delete the dipole in the unique channel $2$ that sends $e_1$ onto $e_3$ (and $e_2$ onto $e_4$). We either obtain a dipole-tadpole or a quartic rung. 

This can create a melon or a double-tadpole only if $G$ contains $H_{10}$. 

\item [(v)] We perform the deletion in the parallel channel. This cannot create a double-tadpole but can create a melon if the graph was in one of the two configurations depicted in figure~\ref{fig:ind_deletion_easy_dipole_d} and \ref{fig:ind_deletion_easy_dipole_e}.

The first one is incompatible with our assumption that the dipole can be deleted in one of the channels $2$ without disconnecting the graph. For the second one, we delete instead the grey dipole in an appropriate channel $2$, for instance the one that sends $e_1$ onto $e_3$ and $e_2$ onto $e_4$. This reduces to the particular case $H_4$, which is covered by lemmas~\ref{lemma:particular_cases} and \ref{lemma:two-point_deletion}. 

\item[(vi)] We instead implement a deletion in an appropriate channel $2$. In more detail, we first try the deletion that sends $e_1$ to $e_3$ (and $e_2$ to $e_4$). The graph is guaranteed to remain connected unless this move maps $e_1'$ to $e_2'$. In that case, we can instead implement a deletion in the channel $2$ that maps $e_1$ to $e_4$, in which case $e_1'$ is mapped to $e_3'$ or $e_4'$. In both situations, the graph remains connected. However, a melon could be created, in which case the graph is the two-point subgraph $H_{11}$ depicted in figure~\ref{fig:particular_cases}.

We then conclude with lemmas~\ref{lemma:particular_cases} and \ref{lemma:two-point_deletion}. 

\item[(vii)] We start by deleting the quartic rung subgraph on the right using lemma~\ref{lemma:melon4pt}. This can create a double-tadpole if $G$ was in one of the configurations depicted in figure~\ref{fig:ind_deletion_easy_dipole_f} and \ref{fig:ind_deletion_easy_dipole_g}. The first one is excluded because it contains a melon. The second one is excluded because the graph disconnects in all three channels $2$ of the dipole we started from. 

This can also create a melon if the graph contains either one of the subgraphs $H_{11}$ and $H_{12}$ of figure~\ref{fig:particular_cases}. We can deal with both of these with lemmas~\ref{lemma:particular_cases} and \ref{lemma:two-point_deletion}. 
\end{itemize}

\end{itemize}

We can now assume that there are no type-$I$ dipoles left. 

\begin{itemize}
\item[\emph{Case E.}] We finally assume that $G$ contains a type-$II$ tadpole or dipole.

Consider a type-$II$ tadpole or dipole $S\subset G$, and assume that the root edge is not contained in $S$. By definition, we know that $S$ is included in a two-point subgraph with one of the structures depicted in figure~\ref{fig:not_easy_2}. Moreover, there is a unique such subgraph that \emph{does not contain the root edge}. We call it the \emph{canonical two-point subgraph} associated to $S$, and denote it by $C_S$.  
\begin{figure}[H]
\centering
\captionsetup[subfigure]{labelformat=empty}
\subfloat[]{\includegraphics[scale=1]{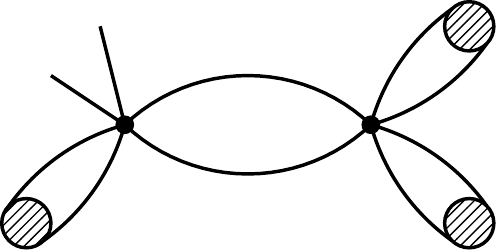}}
\hspace{1cm}
\subfloat[]{\includegraphics[scale=1]{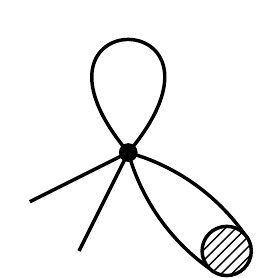}}
\hspace{1cm}
\subfloat[]{\includegraphics[scale=1]{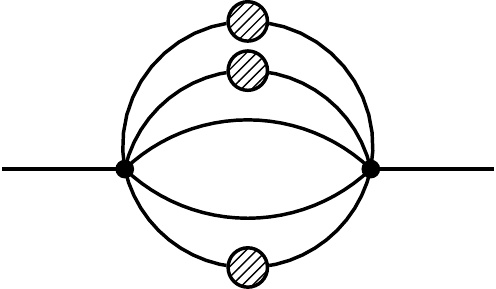}}
\caption{Canonical two-point subgraphs associated to type-$II$ dipoles and tadpoles. Given a type-$II$ tadpole or dipole $S$, $C_S$ is uniquely determined by the requirement that the root edge is not contained in it.}
\label{fig:not_easy_2}
\end{figure}
We then claim that the family of subgraphs $C_S$ forms an \emph{inclusion forest}. That is, given two type-$II$ tadpoles or dipoles $S_1$ and $S_2$, one of the following conditions holds: 
\begin{itemize}
\item $C_{S_1} \subset C_{S_2}$ or $C_{S_2} \subset C_{S_1}$;
\item $C_{S_1}$ and $C_{S_2}$ are (vertex and edge) disjoint. 
\end{itemize}
As a result, provided that this set of subgraphs is non-empty, there exists a dipole or tadpole $S_0$ such that $C_{S_0}$ is minimal for the inclusion. Moreover, $C_{S_0}$ necessarily contains a proper two-point function (i.e. one of the two-point subgraphs represented by blobs in figure~\ref{fig:not_easy_2}, otherwise $G$ would contain a double-tadpole or a melon), which we call $H$. Given that $H\subset C_{S_0}$ and $C_{S_0}$ is minimal for the inclusion among all $C_S$ subgraphs, it follows that $H$ cannot contain any type-$II$ tadpole or dipole. Since we have already assumed the absence of type-$I$ tadpoles or dipoles in $G$, $H$ cannot contain any short face. Consequently, we can use lemmas~\ref{lemma:2PR} and \ref{lemma:two-point_deletion} to construct a suitable graph $G'$ such that $\omega(G)\geq \omega(G')$.  

We are left with one last case to examine: when there is no other type-$II$ tadpole or dipole than one containing the root edge. But in this case $G$ contains at most one short face, so we can immediately conclude that $\omega(G) >0$. 
\end{itemize}

This concludes the proof.
\end{proof}

\section{Melonic dominance}
\label{sec:LO}

Proposition~\ref{prop:positive_degree} and section~\ref{sec:subtraction} immediately imply the existence of the large-$N$ expansion. We now set to prove that leading order graphs are melonic. We start with the following simple observation. 
\begin{lemma}\label{lemma:full-2pt}
Let $\mathcal{G}$ be a (non-amputated) two-point Feynman map. The associated amplitude $\mathcal{A}(\mathcal{G})_{\pmb a,\pmb b}$ can be written as:
\begin{equation*}
\mathcal{A}(\mathcal{G})_{\pmb a,\pmb b}=\lambda^{V(\mathcal{G})}f_{\mathcal{G}}(N)\pmb P_{\pmb a,\pmb b} \, ,
\end{equation*}
where $f_{\mathcal{G}}(N)$ is uniformly bounded.
\end{lemma}
\begin{proof}
The irreducibility of the tensor representation, together with Schur's lemma, immediately imply that the amplitude is proportional to the projector $\pmb P$. Furthermore, consistency with the existence of the large-$N$ expansion requires that $f_{\mathcal{G}}(N)$ is uniformly bounded.
\end{proof}

The next two lemmas demonstrate that many of the stranded configurations which we could not exclude to be of vanishing degree in the previous section, in fact cannot contribute to the leading order. This results from the same type of cancellations we already relied on in section~\ref{sec:subtraction}. But now that the existence of the large-$N$ expansion has been established, we can be more systematic.
\begin{lemma}
Let $\mathcal{G}$ be a (connected and vacuum) Feynman map. If $\mathcal{G}$ has a generalized double-tadpole then it is subleading, that is:
\begin{equation}
\vert A(\cG) \vert \leq K N^4 \,,
\end{equation}
for some constant $K>0$.
\label{lemma:gen_double_tadpole}
\end{lemma}
\begin{proof}
\begin{figure}[htbp]
\centering
\includegraphics[scale=1]{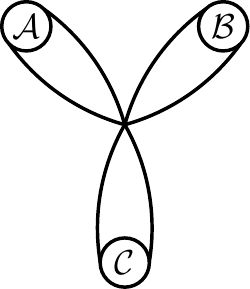}
\caption{Vacuum graph with a generalized double-tadpole}
\label{fig:gen_double_tadpole}
\end{figure}
Up to embedding information (which does not affect large-$N$ scalings), $\mathcal{G}$ must have the configuration depicted in figure~\ref{fig:gen_double_tadpole}, where $\mathcal{A}$, $\mathcal{B}$ and $\mathcal{C}$ are two-point Feynman maps. By lemma~\ref{lemma:full-2pt}, there exists three uniformly bounded functions $f_{\mathcal{A}}$, $f_{\mathcal{B}}$ and $f_{\mathcal{C}}$ such that:
\begin{equation}
\mathcal{A}(\mathcal{G})=\lambda^{V(\mathcal{A})+V(\mathcal{B})+V(\mathcal{C})}f_{\mathcal{A}}(N) f_{\mathcal{B}}(N) f_{\mathcal{C}}(N) \mathcal{A}(\mathcal{G}')\,,
\end{equation}
where $\cG'$ is the map obtained by replacing $\mathcal{A}$, $\mathcal{B}$ and $\mathcal{C}$ with bare propagators $\pmb P$.  $\cG'$ is nothing but a double-tadpole graph, so from the computations of section~\ref{sec:subtraction}, we also know that $\mathcal{A}(\cG')\sim f_1^{\pmb P}(N)\pmb P_{\pmb a,\pmb a} = \mathcal{O}(N^4)$.
\end{proof}

\begin{lemma}\label{lemma:CS}
Let $\cG$ be a (connected and vacuum) Feynman map. If $\cG$ contains a generalized tadpole or a type-$I$ dipole, then it is subleading.
\end{lemma}
\begin{proof}
Let us first assume that $\cG$ contains a generalized tadpole. From lemmas~\ref{lemma:full-2pt} and \ref{lemma:gen_double_tadpole}, it is sufficient to deal with the situation of a single-tadpole, as depicted in figure~\ref{fig:tadpole_LO}, where $\mathcal{S}$ is a \emph{connected} four-point map. We can then apply the Cauchy-Schwarz inequality to find:
\begin{equation}
\mathcal{A}(\cG)^2 = \mathcal{A}\left(\vcenter{\hbox{\includegraphics[scale=.5]{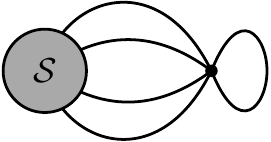}}} \right)^2 \leq \mathcal{A}\left(\vcenter{\hbox{\includegraphics[scale=.5]{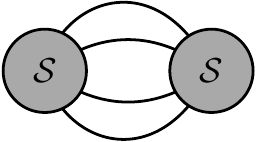}}}\right) \mathcal{A}\left(\vcenter{\hbox{\includegraphics[scale=.5]{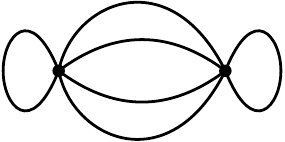}}} \right)\,.
\end{equation}
The first term on the right is the amplitude of a connected map, and is therefore in $\mathcal{O}(N^5)$. Furthermore, from lemma~\ref{lemma:particular_cases} (subcase $H_0$), the degree of any stranded configuration contributing to the second term is at least $1$. Hence this term is in $\mathcal{O}(N^4)$. As a result, $\mathcal{A}(\cG)$ is at most in $\mathcal{O}(N^{9/2})$, which implies it is subleading. 
\begin{figure}[htbp]
\centering
\includegraphics[scale=1]{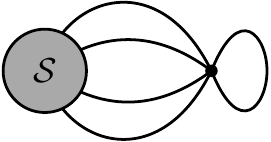}
\caption{$\cG$ contains a tadpole, and the submap $\mathcal{S}$ is assumed to be connected.}
\label{fig:tadpole_LO}
\end{figure}

Let us now assume that $\cG$ contains a type-$I$ dipole. Without loss of generality, and up to embedding, we can assume that we are in one of the situations represented in figure~\ref{fig:dipole_LO}, where the submaps $\mathcal{S}_i$ are all connected.\footnote{In figure~\ref{fig:dipole_LOc}, we have used lemma~\ref{lemma:full-2pt} to suppress one potentially non-trivial two-point subgraph.} In the first case (figure~\ref{fig:dipole_LOa}), we can again invoke the Cauchy-Schwarz inequality, which implies:
\begin{equation}
\mathcal{A}(\cG)^2 = \mathcal{A}\left(\vcenter{\hbox{\includegraphics[scale=.5]{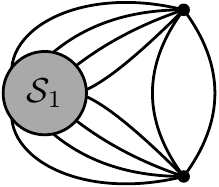}}} \right)^2 \leq \mathcal{A}\left(\vcenter{\hbox{\includegraphics[scale=.5]{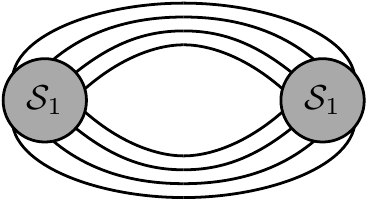}}}\right) \mathcal{A}\left(\vcenter{\hbox{\includegraphics[scale=.5]{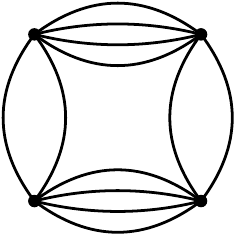}}} \right)\,.
\end{equation}
The first term on the right is the amplitude of a connected map, and the second term is subleading by lemmas~\ref{lemma:melon4pt} and \ref{lemma:particular_cases} (subcase $H_0$). As before, we conclude that $\cG$ is subleading.

The second case (figure~\ref{fig:dipole_LOb}) can be dealt with by successive applications of the Cauchy-Schwarz inequality:
\begin{equation}
\mathcal{A}(\cG)^2 = \mathcal{A}\left(\vcenter{\hbox{\includegraphics[scale=.5]{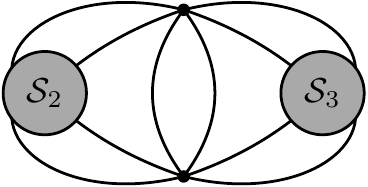}}} \right)^2 \leq \mathcal{A}\left(\vcenter{\hbox{\includegraphics[scale=.5]{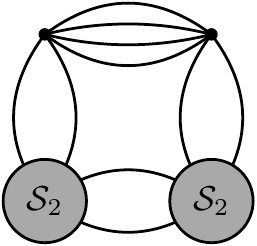}}}\right) \mathcal{A}\left(\vcenter{\hbox{\includegraphics[scale=.5]{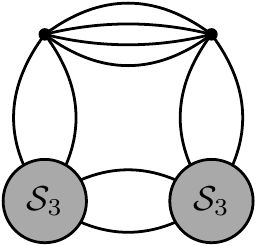}}} \right)\,,
\end{equation}
and
\begin{equation}
\mathcal{A}\left(\vcenter{\hbox{\includegraphics[scale=.5]{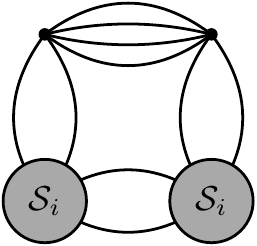}}} \right)^2 \leq \mathcal{A}\left(\vcenter{\hbox{\includegraphics[scale=.5]{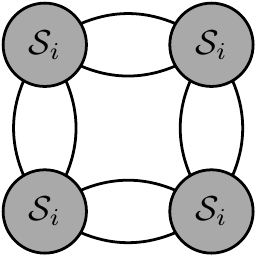}}}\right) \mathcal{A}\left(\vcenter{\hbox{\includegraphics[scale=.5]{dipole_LO_aaa.pdf}}} \right)\,.
\end{equation}
The fact that all the graphs in these relations are connected, while one of them is subleading, allows to conclude that $\cG$ is itself subleading. 

We proceed in a similar way for the last case (figure~\ref{fig:dipole_LOc}) and obtain:
\begin{equation}
\mathcal{A}(\cG)^2 = \mathcal{A}\left(\vcenter{\hbox{\includegraphics[scale=.5]{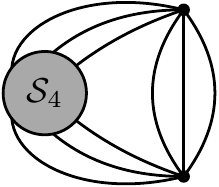}}} \right)^2 \leq \mathcal{A}\left(\vcenter{\hbox{\includegraphics[scale=.5]{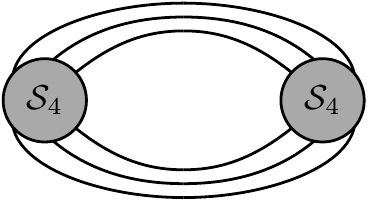}}}\right) \mathcal{A}\left(\vcenter{\hbox{\includegraphics[scale=.5]{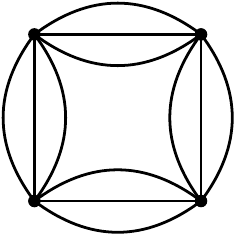}}} \right)\,.
\end{equation}
It is then sufficient to show that the second term on the right is subleading. We can in fact prove that any of the stranded configurations of this map has strictly positive degree.  Indeed, this map has no tadpole and twelve dipoles, thus $F_1=0$ and $F_2\leq 12$. Moreover, any other face has length at least four: $F_3=0$. Using the bounds of appendix~\ref{ap:bounds} with $\mathcal{I}=15\times 4$ and $k=3$, we have $F \leq \lfloor \frac{60+24}{4}\rfloor=21$. Therefore $\omega \geq 5 + 5 \times 4 - 21 = 4$, which concludes the proof. 
\begin{figure}[htbp]
\centering
\subfloat[\label{fig:dipole_LOa}]{\includegraphics[scale=1]{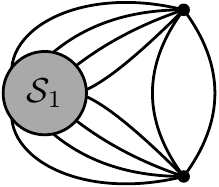}}
\hspace{1cm}
\subfloat[\label{fig:dipole_LOb}]{\includegraphics[scale=1]{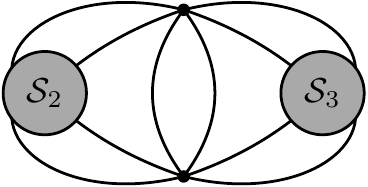}}
\hspace{1cm}
\subfloat[\label{fig:dipole_LOc}]{\includegraphics[scale=1]{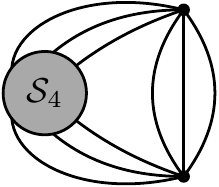}}
\caption{Three configurations of a type-$I$ dipole submap.}
\label{fig:dipole_LO}
\end{figure}
\end{proof}

\begin{proposition}\label{propo:LO}
Let $\cG$ be a (connected and vacuum) Feynman map. $\cG$ is leading order if and only if it is melonic. 
\end{proposition}
\begin{proof}
From section \ref{sec:subtraction}, we already know that melonic graphs are leading order. To prove the converse, let us consider a leading order Feynman map $\cG$. We can start by recursively removing all melon two-point functions from $\cG$, to obtain a leading-order map $\cG'$ with no melon. By definition, $\cG$ is melonic if and only if $\cG'$ is the ring map. Let us assume it is not. Then $\cG'$ must be able to support short faces, otherwise it could not be leading order. Given lemma~\ref{lemma:CS}, the only possibility left is that $\cG'$ contains type-$II$ dipoles whose canonically associated two-point functions are generalized melons (as in the right panel of figure~\ref{fig:not_easy_2}). Considering a minimal such submap for the inclusion, which we call $\mathcal{S}$, leads to a contradiction. Indeed, at least one of the two-point functions in the generalized melon $\mathcal{S}$ must be non-empty, otherwise $\cG'$ would contain a melon submap. By minimality, this two-point function cannot contain any dipole or tadpole, therefore it is necessarily subleading. By lemma~\ref{lemma:full-2pt}, $\cG'$ itself must be subleading, which yields the desired contradiction. Consequently, $\cG'$ is the ring map and $\cG$ is melonic, as claimed.    
\end{proof}

\section{Conclusions}

In this chapter we considered a real $O(N)$ tensor transforming in one of the seven inequivalent irreducible representations of rank $5$ perturbed by a sextic interaction with the combinatorics of a $5$-simplex. The main results of the chapter are the existence of a large-$N$ expansion, and the proof that the leading order graphs are melons. Let us summarize the main steps of the proof. 

\paragraph{Perturbative expansion.} We first expanded the free energy into Feynman amplitudes which are labeled by rooted connected combinatorial maps. We then went to a more detailed representation in terms of stranded graphs, translating the specific tensorial structures of the propagators. The perturbative expansion can be written in terms of those stranded graphs and their amplitudes scale as $N^{-\omega(G)}$, with $\omega$ the \emph{degree} of the graph (see \eqref{eq:degree}), which is an integer quantity. However, the naive conjecture that it is bounded from below is not true in general: stranded graphs with arbitrarily negative degrees do exist. We therefore needed to rely on a subtler strategy: for any map, we proved that none of its stranded configurations with negative degree (if they exist) actually contribute to the full amplitude.

\paragraph{Subtracting and deleting.} The proof of our main results then proceeded in two steps.
\begin{itemize}
\item We first identified a family of stranded graphs supporting arbitrarily negative degrees. Those problematic graphs happen to be generated by melon and double-tadpole maps. Thanks to the irreducibility of the representation, we could straightforwardly prove that the amplitudes of those maps are in fact well-behaved at large $N$. For convenience, we subtracted them through a partial resummation of the perturbative series, governed by a closed and algebraic Schwinger-Dyson equation. 
\item We then proved that, once the problematic configurations have been subtracted, all the remaining stranded graphs have non-negative degree. This was done by induction on the number of vertices of the graphs, through suitable combinatorial deletions of subgraphs.
\end{itemize} 

\paragraph{Leading-order.} After proving the existence of the large-$N$ expansion, the last step was to show that it is dominated by melon diagrams. At this stage, one might be tempted to prove the following improved statement: any stranded graph with no melon and no double-tadpole has in fact strictly positive degree. Again, this is not so simple, as stranded graphs with no melon or tadpole can have vanishing degrees. However, there are again cancellations, and it turns out that none of those configurations can actually contribute to the full amplitudes of their parent maps. We implicitly accounted for such cancellations by mean of Cauchy-Schwarz inequalities which, once the large-$N$ expansion has been established, can be used to directly bound the full amplitudes of non-melonic Feynman maps (without having to resort to the stranded representation). We concluded that a Feynman map is leading order if and only if it is melonic.
\bigskip
\medskip

We have thus established in this chapter that irreducible tensor models with $5$-simplex interaction admit a melonic large-$N$ expansion. Along the way, we had to estimate the large-$N$ behavior of a number of four- and eight-point functions. From this analysis, it is straightforward to include other effective interactions in our models. Any boundary graph we have explicitly investigated may lead to non-vanishing interaction terms if it does not contain self-loops. This includes, for instance, the boundary graphs represented in figures~\ref{fig:tadpole_config_d}, \ref{fig:tadpole_config_e}, or \ref{fig:dipole_config}. With a bit more effort, one could determine the optimal scaling of all effective $n$-point interactions which contribute at leading order, for (say) $n \leq 6$. This would be a prerequisite for potential applications of our results to large-$N$ QFT, where any such interaction that is also relevant in the renormalization sense would have to be included in the bare action. Even though we chose to work in vanishing dimension for simplicity, our main theorems hold in higher dimensions with minimal changes, namely: the algebraic equation defining $F_{\pmb P}^{(0)}$ in theorem~\ref{theorem_LO} should in general be replaced by a suitable (integro-differential) Schwinger-Dyson equation. 

In a similar spirit, it would be interesting to investigate whether our results can be generalized to fermionic tensor fields transforming under the compact symplectic group $Sp(N)$, by analogy with the rank-$3$ construction of \cite{Carrozza:2018psc}. 

Beyond rank-$5$, a number of generalizations could be explored. First, it would be interesting to study irreducible rank-$4$ models with $4$-simplex interaction. The main difficulty we may expect in this case is that, just like in rank-$3$, problematic configurations requiring a detailed combinatorial analysis will also include triangle submaps.\footnote{This was actually one of the reasons why we decided to focus on rank-$5$ in the present chapter.} Finally, it does not seem completely unrealistic to imagine that the present proof could be generalized to arbitrary rank $r \geq 6$. This could be a worthwhile endeavor in view of potential applications of symmetric random tensors to statistics and applied mathematics\cite{Evnin:2020ddw, Gurau:2020ehg}. However, at a minimum, one would need to find a more systematic way of investigating and bounding particular two-point stranded subgraphs, such as those of lemma~\ref{lemma:particular_cases}. Even if one could succeed in this, this would presumably lead to a very technical proof. From this point of view, it would be highly desirable to develop alternative methods that do not rely so heavily on inductive combinatorial moves.

\begin{subappendices}

\section{Bounds on the number of faces}
\label{ap:bounds}

We wish to bound the number of faces of a generic two-point graph. In order to do so, we will follow a method developed in appendix C of \cite{Benedetti:2017qxl}. We label $x,y$ the external legs of the graph. We call $V$, $E$, $F$ and $F_i$, the number of vertices, edges, faces and faces of length $i$ in the subgraph. We also define $l$ as the sum of the length of the open strands of the subgraph. Then, we can write the sum of the length of the internal faces as:
\begin{equation}
\mathcal{I}=5E-l=15V-5-l \, ,
\end{equation}
because for a $6$-valent $2$-point graph, $6V=2E+2$.

Moreover, as stated in \cite{Benedetti:2017qxl}, for $F=\sum_i F_i$ and $\mathcal{I}=\sum_i iF_i$, we have the following bounds:
\begin{equation}
\forall k\geq 2~, \qquad F\leq \lfloor \frac{\mathcal{I}}{k+1} +\sum_{1\leq i\leq k} \frac{k+1-i}{k+1}F_i \rfloor \, .
\label{eq:comb_bounds}
\end{equation}
For $k=2$, we obtain $F\leq \lfloor(\mathcal{I}+2F_1+F_2)/3 \rfloor$. 

Likewise, we can write:
\begin{equation}
l  = \sum_{i \geq 0} i l_i \,, \quad 5p = \sum_i l_i \,,
\end{equation}
where $l_i$ is the number of external strands of length $i$, and $2p$ is the number of external legs (here, $p=2$). We then have the bounds:
\begin{equation}
\forall k\geq 1, \qquad 5p \leq \lfloor \frac{l}{k+1} +\sum_{0 \leq i\leq k} \frac{k+1-i}{k+1} l_i \rfloor  \, .
\label{eq:comb_bounds_l}
\end{equation}
For $k=1$ and $p=2$, this yields in particular: $l \geq 20 - 2 l_0 - l_1$.

\section{Proof of the lemma for particular two-point graphs}
\label{app:particular}

\begin{proof}

We want to gain a factor $N$ by deleting the two-point graphs of figure~\ref{fig:particular_cases}. Therefore, the lemma will follow if we can prove $d(S_{\partial},B_u)\leq 5V-1-F(S)$ for each subgraph $S$ of figure~\ref{fig:particular_cases} and with $B_u$ the boundary graph of the unbroken propagator. 
In order to bound the number of internal faces, we are going to follow the method of \cite{Benedetti:2017qxl} and use the bounds of appendix \ref{ap:bounds}. 

\textbf{Graph $H_0$.} Here $V=2$ and $\mathcal{I}=25-l$. We must have $d(S_{\partial},B_u)\leq 9-F(S)$. We have two tadpoles and three dipoles thus $F_1 \leq 2$ and $F_2\leq 3$. 

\begin{figure}[H]
\centering
\includegraphics[scale=1]{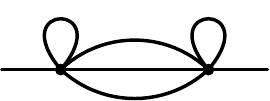}
\caption{The special case $H_0$}
\end{figure}

\begin{itemize}
\item Unbroken case: All external strands traverse and $d(S_{\partial},B_u)=0$. We can thus delete at most $9$ internal faces. We have at most three external strands of length one and two of length two. We thus have $l\geq 7$ and $\mathcal{I}\leq 18$. Then we have $F \leq \lfloor \frac{18+4+3}{3}\rfloor \leq 8$.
\item Broken case: Two external strands loop back. In this case, $d(S_{\partial},B_u)=1$ so we can delete at most $8$ internal faces. We can have all five external strands of length one: $l\geq 5$ and $\mathcal{I}\leq 10$. Here we have to consider faces of length three. We will need equation \eqref{eq:comb_bounds} for $k=3$:
\begin{equation}
F\leq \lfloor\frac{\mathcal{I}+3F_1+2F_2+F_3}{4}\rfloor \, .
\end{equation}
We have $F_3\leq 6$. However, if $F_2=3$, $F_3=0$, if $F_2=2$, $F_3\leq 2$, if $F_2=1$, $F_3\leq 4$ and if $F_2=0$, $F_3\leq 6$. Therefore, $2F_2+F_3\leq 6$ and $F \leq \lfloor \frac{20+6+6}{4}\rfloor \leq 8$.
\item Doubly-broken case: Four external strands loop back. In this case, $d(S_{\partial},B_u)=2$ so we can delete at most $7$ internal faces. We can have three external strands of length one and two of length two: $l\geq 7$ and $\mathcal{I}\leq 18$. 
We thus have $F \leq \lfloor \frac{18+6+6}{4}\rfloor \leq 7$.
\end{itemize}

\textbf{Graph $H_1$.} We now look at the two-point subgraph $H_1$ represented in figure~\ref{fig:h1_label}. It has $3$ vertices: $\mathcal{I}=40-l$. We want $d(S_{\partial},B_u)\leq 14-F(S)$.  

\begin{figure}[htbp]
\centering
\includegraphics[scale=1]{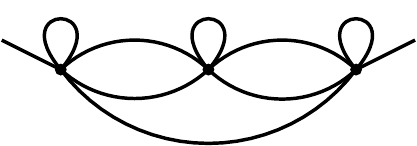}
\caption{The special case $H_1$.}
\label{fig:h1_label}
\end{figure}

\begin{itemize}
\item Unbroken case: In this case we can delete at most $14$ internal faces. The five external strands have at least length one. There can be at most one of length one and two of length two. We thus have $l\geq 11$ and $\mathcal{I} \leq 29$. There are three tadpoles and two dipoles, thus $F_1 \leq 3$ and $F_2 \leq 2$. We then have $F \leq \lfloor \frac{29+6+2}{3}\rfloor \leq 12$. 
\item Broken case:  we can delete at most $13$ internal faces. Now, there can be at most three external strands of length one and two of length two. Thus $l\geq 7$ and $\mathcal{I}\leq 33$. We still have $F_1\leq 3$ and $F_2 \leq 2$. We then have $F \leq \lfloor \frac{33+6+2}{3}\rfloor \leq 13$.
\item Doubly-broken case: we can delete at most $12$ internal faces and we still have $l\geq 7$. We need again to consider the faces of length three.
We can have at most eight faces of length three (four using the tadpoles and four using the strands between the two external points). However, if $F_2=2$ then $F_3\leq 4$, if $F_2=1$ then $F_3\leq 6$ and if $F_2=0$, $F_3\leq 8$. Thus we have $2F_2+F_3\leq 8$ and $F \leq \lfloor \frac{33+9+8}{4}\rfloor \leq 12$.
\end{itemize}

\textbf{Graph $H_2$.} Again $V=3$ and $\mathcal{I}=40-l$. There are two tadpoles and three dipoles so $F_1 \leq 2$ and $F_2 \leq 3$. We again want to prove $d(S_{\partial},B_u)\leq 14-F(S)$.
\begin{figure}[htbp]
\centering
\includegraphics[scale=1]{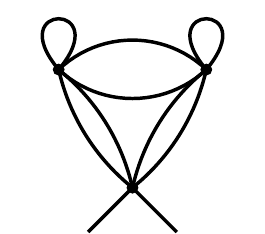}
\caption{The special case $H_2$}
\label{fig:h2_label}
\end{figure}

For this subgraph, in all cases (unbroken, broken and doubly-broken), there is always an external strand of length zero between the two external points as there are connected to the same vertex. There are at most two external strands of length $2$. Thus $l \geq 10$ and $\mathcal{I} \leq 30$. We then have $F\lfloor \frac{30+4+3}{3}\rfloor \leq 12$. This gives us the right bounds for all three cases. 
\\

\textbf{Graph $H_3$.} Here $V=4$ and $\mathcal{I}=55-l$. We now need to prove the following bound: $d(S_{\partial},B_u)\leq 19-F(S)$. Moreover, we have two tadpoles and four dipoles so $F_1 \leq 2$ and $F_2 \leq 4$.

\begin{figure}[htbp]
\centering
\includegraphics[scale=1]{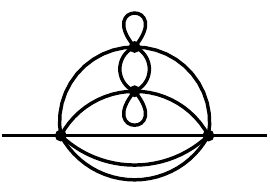}
\caption{The special case $H_3$}
\label{fig:h4_label}
\end{figure}

\begin{itemize}
\item Unbroken case: We have at most three external strands of length one and two of length two. Thus $l \geq 7$ and $\mathcal{I} \leq 48$. Therefore, we have $F \leq \lfloor \frac{48+4+4}{3}\rfloor \leq 18$.
\item Broken case: We have at most three external strands of length one. Then the two remaining external strands must loop back so they have at least length three. In this case $l\geq 9$ and $\mathcal{I}\leq 46$. Therefore, $F \leq \lfloor \frac{46+4+4}{3} \rfloor \leq 18$. 
\item Doubly-broken case: We have at most one external strand of length one as four strands must loop back. We have at most two strands of length two (we only have the two internal corners of the dipole that was not used for the external strand of length one) and two strands of length three.  So, $l \geq 11$ and $\mathcal{I} \leq 44$. Thus we have $F \leq \lfloor \frac{44+4+4}{3} \rfloor \leq 17$.
\end{itemize}

\textbf{Graph $H_4$.} Here, $V=2$ and $\mathcal{I}=25-l$. We need to prove that $d(S_{\partial},B_u)\leq 9-F(S)$. There are one tadpole and six dipoles so $F_1\leq 1$ and $F_2 \leq 6$.

\begin{figure}[htbp]
\centering
\includegraphics[scale=1]{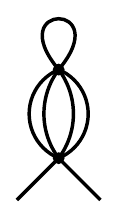}
\caption{The special case $H_4$}
\label{fig:h5_label}
\end{figure}

For all cases (unbroken, broken or doubly-broken), we always have one external strand of length zero as they are connected to the same vertex. The others have at least length two. So $l \geq 8$ and $\mathcal{I}\leq 17$.

\begin{itemize}
\item If $l=8$: we have four external strands of length two. Thus four internal corners of the dipoles are used for the external strands: there are at most two faces of length two. Thus, $F \leq \lfloor \frac{17+2+2}{3} \rfloor \leq 7$.
\item If $l=9$: there is one external strand of length three. The corners of only three dipoles are  now taken by the external strands: $F_2\leq 3$. Thus we have $F \leq \lfloor \frac{16+3+2}{3} \rfloor \leq 7$.
\item If $l\geq 10$, $F_2 \leq l-4$ thus we have $F \leq \lfloor \frac{25-l+2+l-4}{3} =\frac{23}{3}\rfloor \leq 7$.
\end{itemize}

\textbf{Graph $H_5$.} Here $V=3$ and $\mathcal{I}=40-l$. Again, we want to prove the following bound $d(S_{\partial},B_u)\leq 14-F(S)$. There are one tadpole and five dipoles so $F_1 \leq 1$ and $F_2\leq 5$. 

\begin{figure}[htbp]
\centering
\includegraphics[scale=1]{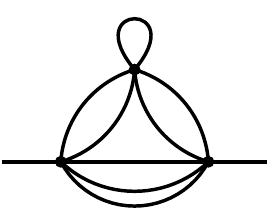}
\caption{The special case $H_5$}
\label{fig:h6_label}
\end{figure}

\begin{itemize}
\item Unbroken case: We can have at most three external strands of length one so $l \geq 7$ and $\mathcal{I}\leq 33$. Thus, $F \leq \lfloor \frac{33+2+5}{3} \rfloor \leq 13$.
\item Broken case: We still have $l\geq 7$ and thus $F \leq 13$. 
\item Doubly-broken case: As only one strand traverses, we can have at most one external strand of length one. Then, $l \geq 9$ and $\mathcal{I}\leq 31$. Thus, $F \leq \lfloor \frac{31+2+5}{3} \rfloor \leq 12$. 
\end{itemize}

\textbf{Graph $H_6$.} Here $V=3$ and $\mathcal{I}=40-l$. We still need to prove that $d(S_{\partial},B_u)\leq 14-F(S)$. There are one tadpole and seven dipoles so $F_1 \leq 1$ and $F_2 \leq 7$. 

\begin{figure}[htbp]
\centering
\includegraphics[scale=1]{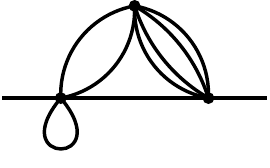}
\caption{The special case $H_6$}
\label{fig:h7_label}
\end{figure}

\begin{itemize}
\item Unbroken case: We have at most one external strand of length one and two of length two (each using one of the edges of the dipole on the left of the graph). Then,  $l\geq 11$ and $\mathcal{I} \leq 29$. Thus, we have $F \leq \lfloor \frac{29+2+7}{3} \rfloor \leq 12$.
\item Broken case: As two strands loop back, we can have one more external strand of length one and three external strands of length two. Then, $l \geq 8$ and $\mathcal{I}\leq 32$. Thus, we have $F \leq \lfloor \frac{32+2+7}{3} \rfloor \leq 13$. 
\item Doubly-broken case: we still have $l\geq 8$. Let us consider the faces of length three. There are at most $3$ faces of length three. However, if $F_2=7$ then $F_3 \leq 2$ and if $F_2\leq 6$, $F_3\leq 3$. Thus $2F_2+F_3\leq 16$ and $F \leq \lfloor \frac{32+3+16}{4} \rfloor \leq 12$.
\end{itemize}

\textbf{Graph $H_7$.} Here $V=3$ and $\mathcal{I}=40-l$. Again, we want to have $d(S_{\partial},B_u)\leq 14-F(S)$. There are no tadpoles and seven dipoles so $F_1=0$ and $F_2 \leq 7$.
For this graph, in all cases (unbroken, broken and doubly-broken), there is always an external strand of length zero and there are at most two external strands of length two. So we then have $l \geq 10$ and $\mathcal{I}\leq 30$. Thus, $F \leq \lfloor \frac{30+0+7}{3} \rfloor \leq 12$ which is okay for unbroken, broken and doubly-broken propagators. 
\\

\begin{figure}[htbp]
\centering
\includegraphics[scale=1]{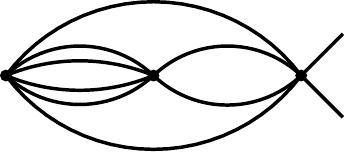}
\caption{The special case $H_7$}
\label{fig:h9_label}
\end{figure}

\textbf{Graph $H_{8}$.} Here $V=3$ and $\mathcal{I}=40-l$. We still need to have $d(S_{\partial},B_u)\leq 14-F(S)$. There are no tadpoles and seven dipoles so $F_1=0$ and $F_2\leq 7$.

\begin{figure}[htbp]
\centering
\includegraphics[scale=1]{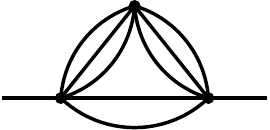}
\caption{The special case $H_{8}$}
\label{fig:h12_label}
\end{figure}

\begin{itemize}
\item Unbroken case: There can be two external strands of length one and three of length two: $l\geq 8$ and $\mathcal{I}\leq 32$. Thus, we have  $F \leq \lfloor \frac{32+0+7}{3} \rfloor \leq 13$.
\item Broken case: Two external strands must loop back but we still have $l\geq 8$ so $F\leq 13$.
\item Doubly-broken case: Four strands must loop back: we can have only one external strand of length one and two of length two. Indeed, calling $x$ the external strand on the left and $a,b,c$ the three edges on the left, if we have an external strand $xabx$ we cannot have a second one using $c$ as both corners $xa$ and $xb$ are already used. Thus, $l\geq 11$ and  $F \leq \lfloor \frac{29+0+7}{3} \rfloor \leq 12$.
\end{itemize}

\textbf{Graph $H_{9}$.} Here $V=3$ and $\mathcal{I}=40-l$ and we again have to prove that $d(S_{\partial},B_u)\leq 14-F(S)$. There are no tadpoles and eight dipoles so $F_2 \leq 8$ and $F_1=0$. Here for all cases (unbroken, broken and doubly-broken), we always have one external strand of length zero. We can also have at most two external strands of length two and two of length three. Then, $l\geq 10$ and $\mathcal{I}\leq 30$. This gives  $F \leq \lfloor \frac{30+0+8}{3} \rfloor \leq 12$.

\begin{figure}[htbp]
\centering
\includegraphics[scale=1]{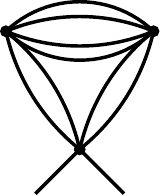}
\caption{The special case $H_{9}$}
\label{fig:h13_label}
\end{figure}

\textbf{Graph $H_{10}$.} Here $V=4$ and $\mathcal{I}=55-l$. We need to prove the following bound $d(S_{\partial},B_u)\leq 19-F(S)$. There are no tadpoles and five dipoles so $F_1=0$ and $F_2\leq 5$. In all cases, there can be at most one external strand of length one and four of length two. Then, $l\geq 9$ and $\mathcal{I}\leq 46$. This gives  $F \leq \lfloor \frac{46+0+5}{3} \rfloor \leq 17$.

\begin{figure}[htbp]
\centering
\includegraphics[scale=1]{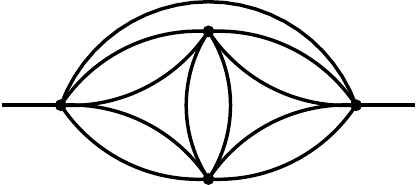}
\caption{The special case $H_{10}$}
\label{fig:h14_label}
\end{figure}

\textbf{Graph $H_{11}.$} Here $V=4$ and $\mathcal{I}=55-l$. In this case we also need to prove that $d(S_{\partial},B_u)\leq 19-F(S)$. There are no tadpoles and nine dipoles so $F_1 =0$ and $F_2 \leq 9$.

\begin{figure}[H]
\centering
\includegraphics[scale=1]{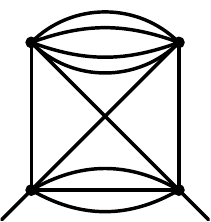}
\caption{The special case $H_{11}$}
\label{fig:hn2_label}
\end{figure}

\begin{itemize}
\item Unbroken case: We can have at most three strands of length one and two of length two. Thus, $l\geq 7$ and $\mathcal{I}\leq 48$. We then have $F \leq \lfloor \frac{48+9}{3} \rfloor \leq 19$.
\item Broken case: We can now have three external strands of length one and two of length three or one of length one and four of length two. Thus we have $l\geq 9$ and $F \leq \lfloor \frac{46+9}{3} \rfloor \leq 18$.
\item Doubly-broken case: We can now have one external strand of length one, two of length two and two of length three. Thus $l\geq 11$ and we have $F \leq \lfloor \frac{44+9}{3} \rfloor \leq 17$.
\end{itemize}

\textbf{Graph $H_{12}$.} Here $V=4$ and $\mathcal{I}=55-l$. In this case we again need to prove that $d(S_{\partial},B_u)\leq 19-F(S)$. There are no tadpoles and twelve dipoles so $F_1 =0$ and $F_2 \leq 12$.

\begin{figure}[htbp]
\centering
\includegraphics[scale=1]{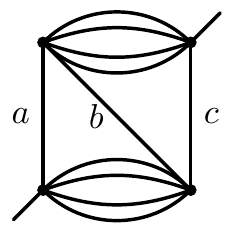}
\caption{The special case $H_{12}$}
\label{fig:hn3_label}
\end{figure}

For all cases (unbroken, broken and doubly-broken), we can have five external strands of length two. Thus, $l\geq 10$. Let us consider faces of length $3$: there can be at most two faces of length three (one using the corner $ab$ and one edge of the bottom quartic rung and one using the corner $bc$ and one edge of the top quartic rung). We thus have: $F \leq \lfloor \frac{45+24+2}{4} \rfloor \leq 17$.

\end{proof}

\end{subappendices}

\chapter*{Conclusion and perspectives}
\label{chap:conclu}
\addcontentsline{toc}{chapter}{Conclusion and perspectives}
\markboth{CONCLUSION AND PERSPECTIVES}{CONCLUSION AND PERSPECTIVES}
In this thesis, we studied the renormalization of tensor field theories as well as the properties of the CFTs they reach in the infrared. 
We started by considering a bosonic quartic model in rank $3$. For the short-range version of the model, the CFT at the fixed point is complex and non-unitary at large $N$. At next-to-leading order, the critical exponents acquire a real part rendering the fixed point stable, the CFT still being non-unitary. In the long-range version of the model the situation is reversed. At large $N$, there exists four lines of fixed points parametrized by the tetrahedral coupling. When choosing the latter to be purely imaginary, one of them is a real, strongly interacting, infrared attractive line of fixed points. Moreover, the values of the OPE coefficients and the spectrum of bilinear operators are consistent with a unitary CFT at large $N$. The complete proof of unitarity would require computing the spectrum of all primary operators. At next-to-leading order, the fixed points and critical exponents acquire an imaginary part: unitarity is broken. Finally, considering the large-$N$ fixed points for the long-range case, we showed that it satisfies the weak version of the $F$-theorem: the free energy on the sphere in dimension $3$ is greater at the UV fixed point than at the IR fixed point. This is also an indicator of possible unitarity at large $N$ as the $F$-theorem was originally proven for unitary CFTs. We can then wonder if the $F$-theorem is always satisfied by long-range melonic CFTs.

We continued with the study of a sextic bosonic model with either bipartite interactions in rank $3$ or non-bipartite interactions in rank $5$. In rank $3$, the situation is similar to the quartic case. For the short-range version of the model, we find two real stable fixed points at large $N$. However, the stability matrix is non-diagonalizable, indicating a logarithmic, and thus a non-unitary, CFT. At next-to-leading order, we still find two stable fixed points and the stability matrix is now diagonalizable. This does not mean the CFT could be unitary as the correction terms in $1/N$ cannot compensate logarithmic contributions from the leading order. For the long-range version of the model, as for the quartic case, we find a line of real stable fixed points with real spectrum of bilinear operators. However, at next-to-leading order, the corrections to the fixed point are non-perturbative. We find no precursor of the large-$N$ fixed point. It would be interesting to further study the properties of the fixed-point CFT at large $N$. In particular, we could compute OPE coefficients or determine the sign of the logarithmic CFT. 
Surprisingly, the results for the rank-$5$ model are very different than the quartic case and the only fixed point is the Gaussian one. 

Finally, in the last chapter, we showed that irreducible tensor models in rank $5$ have a melonic large-$N$ limit. A natural extension of this work would be to consider field theories based on irreducible models. Do they admit stable melonic fixed points? If so, are they different from the fixed points obtained for the uncolored models? What are the properties of the CFTs at these fixed points? Furthermore, the advantage of studying a tensor field theory with irreducible symmetry is to be able to consider maximally single trace interactions (such as the tetrahedron) for fermionic models. This has been done in rank 3 for a one-dimensional SYK-like model with $Sp(N)$ symmetry in \cite{Carrozza:2018psc}. We can then wonder how this generalizes to higher dimensions.

We will now detail some interesting questions for future work. A first generalization would be to extend the class of models exhibiting a melonic CFT by studying models with different symmetry groups or order of interactions. Steps were taken in this direction with the study of cubic interactions in \cite{Benedetti:2020iku} or with supersymmetric models in \cite{Popov:2019nja} and more recently in \cite{Lettera:2020uay}. Another interesting point is about scale versus conformal invariance \cite{Nakayama:2013is}. Fixed points of the renormalization group flow are scale invariant but not necessarily conformal invariant. Conformal invariance was proven for the fixed point of the long-range $O(N)^3$ model in \cite{Benedetti:2020yvb} following the proof of \cite{Paulos:2015jfa} for the long-range Ising model. Such studies were carried out for different types of field theories \cite{Cardy:1985yy,Osborn:1993cr} and we can wonder how this applies to tensor models or more broadly, long-range models. 

While studying the fixed points of the $O(N)^3$ bosonic model, we noticed an interesting phenomenon. The line of fixed points is real for some range of the tetrahedral coupling. However, for some value of this coupling, two fixed points merge and a complex CFT appears. It would be very interesting to study this phenomenon in more detail and see how characteristic of tensor models it is. Indeed, fixed-point merging happens in a broad range of models. A standard example is the Banks-Zaks fixed point where a merge occurs when varying the number of fermions in a $SU(N)$ Yang-Mills theory \cite{Banks:1981nn}. This phenomenon was then especially investigated for conformal theories where it was shown to be linked to mass hierarchy generating and walking behavior \cite{Faedo:2021ksi,Gorbenko:2018dtm}.

In the same spirit, for some values of the tetrahedral coupling in the quartic case or the wheel coupling in the sextic case, complex dimensions of bilinears appear. In the short-range sextic model, these complex dimensions appear for some value of $\epsilon$. This should be compared to the results of \cite{Kim:2019upg} where the authors considered a one-dimensional model with two Majorana fermions and $O(N)^3$ invariance. This model has no pillow or double-trace couplings, but it has several tetrahedral couplings whose relative strength can be adjusted by tuning a parameter. In particular, there exists a critical value of the tuning parameter at which the conformal dimension of an off-diagonal mass bilinear becomes complex, and is $d/2+\im r$, with real $r$. We can give a similar interpretation to this complex dimension and those of our models. Indeed, from the bulk point of view, they can be seen as the consequence of particles which violate the Breitenlohner-Freedman bound $m^2\geq -\frac{d^2}{4}$ \cite{Breitenlohner:1982jf}. Moreover, beyond the critical value of the tuning parameter, the mass bilinear of \cite{Kim:2019upg} acquires a non-zero vacuum expectation value (VEV) which spontaneously breaks conformal invariance, as well as the discrete symmetries of the model. This indicates a second order phase transition between broken and unbroken symmetry phases.  It is likely that in our quartic and sextic bosonic models the mass bilinear also acquires a non-zero VEV. It would be quite involved to check this properly but this would be an interesting endeavor. Steps in this direction were taken in \cite{Benedetti:2021qyk} where it was proven that $d$-dimensional CFTs with operators of scaling dimensions $h=d/2+\im r$ with non-vanishing real $r$ are unstable.

Let us conclude with a broader issue: what is the AdS dual of melonic CFTs? This is a highly non-trivial question as constructing the AdS dual of tensor models is much more challenging than for other large-$N$ theories such as matrix or vector models. The main reason for this is the factorial growth of the number of gauge invariant operators in tensor models. One solution presented in \cite{DeMelloKoch:2019tmo} is to identify subsets of dynamically closed invariants, such as those given by the melonic limit. This led to inconclusive results: other methods thus need to be developed in order to answer this question. A promising one is the adaptation to tensors of the derivation of AdS/CFT for vector models developed in \cite{deMelloKoch:2018ivk} and \cite{Aharony:2020omh}. Another potential path would be to use multi-matrix models, introduced in \cite{Ferrari:2017ryl}, where only two tensor indices take values from $1$ to $N$ while one index takes value from $1$ to $D$. The melonic limit can then be viewed as a new limit (large $N$ and large $D$) of multi-matrix models potentially making the search for an AdS dual easier.

\backmatter
\appendix
\chapter*{Résumé de la thèse}
\addcontentsline{toc}{chapter}{Summary in French}
\markboth{RÉSUMÉ DE LA THÈSE}{RÉSUMÉ DE LA THÈSE}
Dans cette thèse nous nous intéressons à des théories quantiques des champs tensoriels qui présentent un nouveau type de limite infrarouge accessible analytiquement. Les tenseurs aléatoires ont d'abord été introduits en dimension zéro dans un contexte de géométrie aléatoire et gravité quantique \cite{Ambjorn:1990ge,Sasakura:1990fs,RTM}. Ils peuvent être vus comme une généralisation en rang supérieur des modèles de matrices aléatoires qui sont utilisés pour décrire la gravité quantique en dimension $2$. Les modèles de tenseurs aléatoires ont ensuite été étendus en dimension $1$ comme une généralisation du modèle de Sachdev-Ye-Kitaev (SYK) sans désordre (voir section \ref{sec:other_melonic}). Enfin, ils ont été généralisés en dimension $d$ comme des théories quantiques des champs \cite{Giombi:2017dtl,Prakash:2017hwq,Benedetti:2017fmp,Giombi:2018qgp,Benedetti:2018ghn}. Dans ce contexte, ils sont particulièrement intéressants grâce à leur limite à grand $N$. En effet, les modèles de tenseurs présentent une limite à grand $N$ melonique \cite{Bonzom:2011zz,RTM} (voir figure \ref{fig:melon_backgnd} pour une représentation des graphes meloniques). Cette limite melonique est plus riche que la limite à grand $N$ des modèles de vecteurs \cite{Guida:1998bx,Moshe:2003xn} mais plus simple que la limite planaire des modèles de matrices \cite{'tHooft:1973jz,Brezin:1977sv,DiFrancesco:1993nw}. On peut donc espérer obtenir des modèles non-triviaux accessibles analytiquement. En particulier, il a été montré qu'en dimension $d$, les modèles de tenseurs donnent naissance à un nouveau type de théorie conforme des champs (TCC) au point fixe infrarouge que l'on appelle TCC melonique. Cependant, comme toute théorie des champs, les théories tensorielles présentent des divergences. Le but de cette thèse était d'étudier rigoureusement le flux de renormalisation des théories des champs tensoriels ainsi que les propriétés des TCC meloniques au point fixe. 

Dans le chapitre \ref{ch:background}, nous introduisons un certain nombre de notions utiles pour le reste de la thèse. En particulier, nous présentons le groupe de renormalisation Wilsonien, les théories conformes des champs ainsi que les modèles de tenseurs aléatoires.
Le chapitre \ref{chap:3loops} considère un type de théorie des champs qui sera utile pour étudier les modèles de tenseurs: les modèles à longue portée. Ces modèles sont des théories des champs avec une puissance non triviale du Laplacien $0<\zeta<1$. Le paramètre $\zeta$ est choisi positif pour obtenir une limite thermodynamique bien définie et inférieur à $1$ pour préserver la positivité par réflexion. Les modèles à longue portée présentent de nombreuses applications et un grand intérêt d'un point de vue théorique \cite{Campa:2009rev}. Cependant, le paramètre $\zeta$ rend leur étude analytique plus compliquée et ils ont donc été moins étudiés que leurs équivalents à courte portée. Notamment, les calculs ont été arrêtés à deux boucles \cite{Fisher:1972zz,Yamazaki:1977pt} alors que les fonctions bêta sont connues à six boucles pour les modèles à courte portée \cite{Kompaniets:2017yct}. Dans le chapitre \ref{chap:3loops}, nous calculons donc les fonctions bêta d'un modèle multi-scalaire avec interactions quartiques à trois boucles. Nous spécifions ensuite différents groupes de symétries: Ising, symétrie $O(N)$, symétrie cubique et symétrie bi-fondamentale $O(N_1)\times O(N_2)$. Pour chacun de ces modèles, nous calculons les points fixes et exposants critiques et nous les comparons à des résultats numériques lorsque ceux-ci sont disponibles. 

Nous nous intéressons ensuite à un modèle bosonique avec interactions quartiques en rang $3$ et une symétrie $O(N)^3$. Ce modèle a d'abord été introduit en dimension zéro par \cite{Carrozza:2015adg} puis étendu en dimensions supérieures dans \cite{Klebanov:2016xxf}. Il est donc parfois appelé modèle Carrozza-Tanasa-Klebanov-Tarnopolsky (CTKT). Du fait de la symétrie $O(N)^3$, ce modèle comporte trois interactions quartiques différentes : tétraèdre, coussin et double-trace. Ces interactions sont représentées sous forme de graphes colorés en figure \ref{fig:trace_invariants}.
En utilisant une régularisation dimensionnelle avec $d=4-\epsilon$ et $\epsilon\ll 1$, \cite{Giombi:2017dtl} a prouvé que ce modèle admettait un point fixe melonique. Cependant, ce point fixe est complexe et instable et la TCC n'est pas unitaire à cause de la présence d'opérateurs de dimensions complexes. Dans le chapitre \ref{chap:CTKT}, nous utilisons le groupe de renormalisation Wilsonien pour étudier une variation de ce modèle. Nous commençons par considérer un propagateur à longue portée reproduisant le comportement infrarouge du point fixe en choisissant $\zeta=d/4$ pour $d<4$ fixé. Le terme cinétique n'est plus local mais les interactions sont ainsi rendues marginales. Dans ce contexte, il est alors possible de résoudre analytiquement l'équation de Schwinger-Dyson pour obtenir la fonction à deux points dans l'infrarouge. Nous montrons ensuite que ce modèle admet quatre lignes de points fixes non-perturbatifs paramétrisées par le couplage tétraédrique à grand $N$. De plus, une des lignes de points fixes est stable et attractive dans l'infrarouge lorsque l'on choisit un couplage tétraédrique purement imaginaire. 
Nous étudions ensuite les propriétés de la TCC au point fixe. Nous calculons les dimensions des opérateurs bilinéaires ainsi que les coefficients de l'expansion en produit d'opérateurs. Les résultat sont cohérents avec une TCC unitaire au point fixe à grand $N$. 

Dans le chapitre \ref{chap:trif}, nous étudions les corrections en $1/N$ pour ce point fixe. Nous considérons tout d'abord une théorie tri-fondamentale avec symétrie $O(N_1)\times O(N_2)\times O(N_3)$. Nous calculons les points fixes pour différentes limites à grand $N$ et petit $\epsilon$. Dans le cas du modèle à courte portée, $\epsilon$ est la déviation par rapport à la dimension critique tandis que dans le cas à longue portée il s'agit de la déviation par rapport à l'exposant critique du propagateur. Nous concluons qu'en général il n'existe pas de point fixe réel stable avec un couplage tétraédrique non nul. Dans le cas homogène avec symétrie $O(N)^3$, nous trouvons néanmoins des points fixes stables complexes avec un couplage tétraédrique non nul. En particulier, dans le cas à courte portée, à grand $N$ le point fixe est complexe et instable. Cependant, les exposants critiques acquièrent une partie réelle au premier ordre en correction et le point fixe de \cite{Giombi:2017dtl} devient stable. Dans la cas à longue portée la situation est quelque peu inversée. A grand $N$, pour un couplage tétraédrique purement imaginaire, les autres couplages sont réels et le point fixe est stable et unitaire mais au premier ordre en correction, les exposants critiques acquièrent une partie imaginaire. L'unitarité est donc brisée. 

Dans le chapitre \ref{chap:Ftheorem}, nous étudions plus en détails les propriétés du modèle bosonique $O(N)^3$. Plus précisément, nous testons le théorème $F$ pour ce modèle. Selon le théorème $F$, l'énergie libre d'une TCC sur la sphère en dimension $3$ décroit le long du flux du groupe de renormalisation \cite{Klebanov:2011gs}. Ce théorème a été prouvée pour des TCC unitaires, il est donc intéressant de le tester pour notre modèle qui peut être unitaire seulement à grand $N$. Notre conclusion est que ce théorème est effectivement vérifié pour le modèle $O(N)^3$ à longue portée. Nous avons tout d'abord considéré un flux entre deux théories libres généralisées puis testé notre méthode de calcul dans le cas $O(N)$ pour lequel le résultat est connu. En effet, pour le modèle $O(N)^3$, le calcul de l'énergie libre requiert la resommation d'une famille infinie de graphes. Pour cela nous avons utilisé l'expansion en onde partielle qui est une méthode standard en TCC mais qui n'avait jamais été utilisée dans ce contexte. 

Nous pouvons ensuite nous demander comment ces résultats dépendent de l'ordre de l'interaction et du rang des tenseurs. Pour commencer à répondre à cette question, nous considérons dans le chapitre \ref{chap:sextic} des théories des champs tensoriels avec interactions sextiques. Nous considérons deux modèles: un modèle bi-partite en rang $3$ et un modèle non bi-partite en rang $5$. Pour ces deux modèles nous étudions les cas à courte et longue portée. En rang $3$, nous trouvons deux points fixes infrarouges stables pour le modèle à courte portée et une ligne de points fixes infrarouges stables dans le modèle à longue portée. Dans les deux cas, nous trouvons une fenêtre avec un spectre réel des opérateurs bilinéaires. Cependant, dans le cas à courte portée, la matrice de stabilité n'est pas diagonalisable. Cela est caractéristique d'une TCC logarithmique. La TCC au point fixe n'est donc pas unitaire. 
En rang $5$, étonnamment, le seul point fixe est le point fixe Gaussien. Nous étudions ensuite les corrections en $1/N$ pour le modèle en rang $3$. Dans le cas à courte portée, nous trouvons toujours deux points fixes infrarouges stables. De plus, la matrice de stabilité est maintenant diagonalisable. Cependant, cela ne signifie pas que la TCC au point fixe peut être unitaire à $N$ fini. En effet, les corrections en $1/N$ ne peuvent pas compenser les termes logarithmiques de l'ordre dominant. Pour le modèle à longue portée, les corrections au point fixe ne sont pas perturbatives et donc non fiables. Nous ne trouvons pas de précurseur au point fixe à grand $N$. 

Enfin, pour utiliser les tenseurs dans le contexte de la théorie des champs ou de la géométrie aléatoire, la caractéristique clé est l'existence d'une limite melonique à grand $N$. Cependant, cette limite n'est pas connue pour tous les modèles. En particulier, elle manquait pour les modèles avec des représentations ordinaires de $O(N)$ ou $Sp(N)$, telles que antisymétrique ou symétrique à trace nulle. Récemment, il a été prouvé que cette limite existe en rang $3$ pour des modèles dans une représentation irréductible de $O(N)$ ou $Sp(N)$ \cite{Benedetti:2017qxl, Carrozza:2018ewt, Carrozza:2018psc}. Dans la chapitre \ref{chap:rank5}, nous généralisons cette preuve en rang $5$. Cette généralisation repose sur des bornes récursives dérivées d'une analyse combinatoire détaillée des graphes de Feynman impliqués dans l'expansion perturbative du modèle. Nous suivons les méthodes combinatoires développées dans \cite{Benedetti:2017qxl}. Certains aspects sont simplifiés par le rang $5$ (il n'est plus nécessaire de considérer les graphes triangulaires) mais d'autres sont plus compliqués (par exemple, des bornes doivent être calculées sur des graphes à huit points en plus des graphes à deux et quatre points). 
Cette preuve étend la classe de modèles présentant une limite melonique et il serait intéressant de pouvoir la généraliser en rang arbitraire. 

\chapter*{Zusammenfassung}
\addcontentsline{toc}{chapter}{Summary in German}
\markboth{ZUSAMMENFASSUNG}{ZUSAMMENFASSUNG}
In dieser Promotionsarbeit untersuchen wir Tensorfeldtheorien, die ein neuartiges, mit analytischen Methoden zugängliches Infrarotlimit aufweisen. Zufallstensoren wurden erstmals in null Dimensionen als statistische Modelle für stochastische Geometrie und Quantengravitation eingeführt \cite{Ambjorn:1990ge,Sasakura:1990fs,RTM}. Sie können als eine Verallgemeinerung von Matrixmodellen, zu höherem Rang, betrachtet werden, letztere dienen zur Beschreibung von zweidimensionaler Quantengravitation. Zunächst wurden Zufallstensoren zu eindimensionalen Modellen erweitert, als solche sind sie eine Alternative zum Sachdev-Ye-Kitaev-Modell, jedoch ohne quenched Disorder (siehe Abschnitt \ref{sec:other_melonic}). Schließlich wurden sie zu Quantenfeldtheorien in beliebigen Raumzeitdimensionen verallgemeinert\cite{Giombi:2017dtl,Prakash:2017hwq,Benedetti:2017fmp,Giombi:2018qgp,Benedetti:2018ghn}.
In diesem Zusammenhang sind sie insbesondere wegen ihres Large-$N$-Limits interessant. Dieses Limit wird von melonischen Graphen dominiert\cite{Bonzom:2011zz,RTM} (siehe Abbildung \ref{fig:melon_backgnd} für eine Darstellung von melonischen Graphen). Dieses melonische Limit ist vielfältiger als das Large-$N$-Limit von Vektormodellen \cite{Guida:1998bx,Moshe:2003xn}, aber einfacher als das planare Limit von Matrixmodellen \cite{'tHooft:1973jz,Brezin:1977sv,DiFrancesco:1993nw}. Dies lässt darauf hoffen, dass sich nichttriviale, aber analytisch zugängliche Modelle finden lassen. Insbesondere wurde gezeigt, dass Tensormodelle in $d$ Dimensionen, an ihren Infrarotfixpunkten zu einer neuen Art von konformen Feldtheorien führen, diese werden als melonische CFTs bezeichnet. Wie üblich für Feldtheorien zeigen Tensormodelle jedoch Divergenzen. Daher war es das Ziel dieser Arbeit den Renormierungsgruppenfluss von Tensorfeldtheorien, sowie die Eigenschaften von melonischen CFTs am Fixpunkt genau zu untersuchen. 

In Kapitel \ref{ch:background} führen wir für den Rest der Arbeit wichtige Begriffe ein. Insbesondere stellen wir die Wilsonsche Renormierungsgruppe, konforme Feldtheorien und Zufallstensoren vor. Anschließend wird in Kapitel \ref{chap:3loops} eine Art von Feldtheorie betrachtet, die für die Untersuchung von Tensormodellen nützlich sein wird: Langreichweitige Modelle. Diese Modelle sind Feldtheorien mit einer nicht trivialen Potenz des Laplacian $0<\zeta<1$. Der Parameter $\zeta$ muss positiv sein, um eine wohldefiniertes thermodynamisches Limit zu erhalten, und streng kleiner als eins, um die Reflexionspositivität zu erhalten. Langreichweitige Modelle haben eine Vielzahl von Anwendungen und sind aus theoretischer Sicht von großem Interesse \cite{Campa:2009rev}. Die Potenz $\zeta$ erschwert jedoch ihre analytische Untersuchung erheblich. Aus diesem Grund wurden sie weniger untersucht als ihre kurzreichweitigen Gegenstücke. Insbesondere wurden Berechnungen nur bis zu zwei Loops durchgeführt \cite{Fisher:1972zz,Yamazaki:1977pt}, während Betafunktionen für kurzreichweitige Modelle bis zu sechs Loops bekannt sind \cite{Kompaniets:2017yct}. In Kapitel \ref{chap:3loops} berechnen wir die Betafunktionen für ein generisches Multiskalarmodell mit quartischen Wechselwirkungen bis zu drei Loops. Danach spezifizieren wir die Untersuchung für verschiedene Symmetriegruppen: Ising, $O(N)$-Symmetrie, kubische Symmetrie und bifundamentale $O(N_1)\times O(N_2)$-Symmetrie. Für jedes dieser Modelle berechnen wir Fixpunkte und kritische Exponenten, und vergleichen diese mit verfügbaren numerischen Ergebnissen.

Dann betrachten wir ein quartisches bosonisches Rank-$3$-Tensormodell mit $O(N)^3$-Symmetrie. Dieses Modell wurde zuerst von \cite{Carrozza:2015adg} in null Dimensionen eingeführt und dann von \cite{Klebanov:2016xxf} zu höheren Dimensionen erweitert. Es wird daher als CTKT-Modell bezeichnet. Aufgrund der $O(N)^3$-Symmetrie hat dieses Modell drei quartische Wechselwirkungen: Tetraeder, Kissen und Doppel-Spur. Wir stellen diese Wechselwirkungen als kolorierte Graphen in Abbildung \ref{fig:trace_invariants} dar. Durch die Verwendung von dimensionaler Regularisierung mit $d=4-\epsilon$, für kleines $\epsilon$, zeigte \cite{Giombi:2017dtl}, dass dieses Modell einen melonischen Fixpunkt zulässt. Dieser Fixpunkt ist jedoch komplex und aufgrund der komplexen Dimensionen der bilinearen Operatoren instabil. In Kapitel \ref{chap:CTKT} verwenden wir die Wilsonsche Renormierungsgruppe, um eine Variante dieses Modells zu untersuchen. Wir beginnen mit einem langreichweitigen Propagator und reproduzieren die Infrarotskalierung indem wir $\zeta=d/4$ mit $d<4$ fest wählen. Der kinetische Term ist somit nichtlokal, aber die Wechselwirkungen werden marginal. In diesem Zusammenhang ist es möglich, die Schwinger-Dyson-Gleichung analytisch zu lösen, und die Zweipunktfunktion im Infraroten zu erhalten. Wir zeigen nichtperturbativ, aber bei großem $N$, dass dieses Modell vier Linien von Fixpunkten zulässt, die durch die tetraedrische Kopplung parametrisiert sind. Wählt man die tetraedrische Kopplung als rein imaginär, so ist eine der Linien von Fixpunkten stabil und im Infrarot attraktiv. Im zweiten Teil des Kapitels untersuchen wir die Eigenschaften der melonischen CFT am Fixpunkt und berechnen die Dimensionen von bilinearen Operatoren und OPE-Koeffizienten. Die Ergebnisse sind kohärent mit einer unitären CFT am Large-$N$ Fixpunkt.

In Kapitel \ref{chap:trif} untersuchen wir die $1/N$-Korrekturen zu diesem Fixpunkt. Wir betrachten zunächst eine tri-fundamentale Theorie mit $O(N_1)\times O(N_2)\times O(N_3)$-Symmetrie. Wir berechnen Fixpunkte für verschiedene Large-$N$-Limits und kleines $\epsilon$. Im kurzreichweitigem Fall ist $\epsilon$ die Abweichung von der kritische Dimension, während es im langreichweitigen Modell die Abweichung von der kritischen Skalierung des Propagators ist. Wir kommen zu dem Schluss, dass es im Allgemeinen keinen realen Fixpunkt mit nichtverschwindender Tetraeder-Wechselwirkung gibt. Wir finden jedoch stabile komplexe Fixpunkte mit nichtverschwindender Tetraeder-Wechselwirkung im homogenen Fall mit $O(N)^3$-Symmetrie. Insbesondere im kurzreichweitigem Fall ist der Fixpunkt bei großem $N$ komplex und instabil. Die kritischen Exponenten erhalten jedoch einen reellen Teil bei nächsthöherer Ordnung und der Fixpunkt von \cite{Giombi:2017dtl} wird infrarotstabil. Im Langstreckenfall ist die Situation umgekehrt. Bei großem $N$ sind, für eine rein imaginäre tetraedrische Kopplung, die anderen Kopplungen reell und der Fixpunkt ist stabil und unitär, aber bei der nächsthöheren Ordnung erhalten die kritischen Exponenten einen Imaginärteil. Die Unitarität wird also in der nächsthöheren Ordnung gebrochen.

In Kapitel \ref{chap:Ftheorem} untersuchen wir weitere Eigenschaften des bosonischen $O(N)^3$-Modells. Genauer gesagt prüfen wir das $F$-Theorem für dieses Modell. Das $F$-Theorem besagt, dass die freie Energie einer $3$-dimensionalen CFT auf der Sphäre entlang des Flusses der Renormierungsgruppe abnimmt \cite{Klebanov:2011gs}. Dieses Theorem wurde für unitäre CFTs bewiesen. Somit war es interessant es für unser Modell, das nur im strikten large-$N$-Limit unitär ist, zu überprüfen. Wir können das $F$-Theorem für das langreichweitige bosonische $O(N)^3$-Modell  verifizieren. Zunächst betrachteten wir einen Fluss zwischen zwei verallgemeinerten freien Feldtheorien und testeten so unsere Berechnungsmethode für das $O(N)$-Modell, für das das Ergebnis bekannt ist. Für das $O(N)^3$-Modell erforderte die Berechnung der freien Energie die Resummation einer unendlichen Familie von Graphen. Dazu benutzten wir die Expansion in partielle konforme Wellen, eine Standardmethode in CFTs, die aber in diesem Zusammenhang noch nie zum Einsatz kam.

Wir können uns nun fragen, inwiefern diese Ergebnisse von der Ordnung der Wechselwirkungen und vom Rang der Tensoren abhängen. Wir beginnen mit der Beantwortung dieser Fragen in Kapitel \ref{chap:sextic}, indem wir sextische Tensorfeldtheorien betrachten. Wir untersuchen zwei Modelle: ein bipartites Modell im Rang $3$ und ein nicht bipartites Modell im Rang $5$. Für jedes Modell untersuchen wir sowohl den kurz-, als auch den langreichweitigen Fall. Im Rang $3$ finden wir zwei infrarotstabile Fixpunkte im kurzreichweitigem Modell und eine Reihe von infrarotstabilen Fixpunkten für das langreichweitige Modell. In beiden Fällen kann ein Bereich mit einem reellen Spektrum der bilinearen Operatoren gefunden werden. Im Fall des kurzreichweitigen Modells ist die Stabilitätsmatrix jedoch nicht diagonalisierbar. Dies ist charakteristisch für eine logarithmische CFT: die Fixpunkt-CFT ist also nicht unitär. Überraschenderweise ist der einzige Fixpunkt in Rang $5$ der Gaußsche Fixpunkt. Im zweiten Teil des Kapitels untersuchen wir $1/N$-Korrekturen für das Rang-$3$-Modell. Im kurzreichweitigem Fall finden wir weiterhin zwei infrarotstabile Fixpunkte. Außerdem ist die Stabilitätsmatrix jetzt diagonalisierbar. Dies bedeutet jedoch nicht, dass die CFT am Fixpunkt unitär ist. In der Tat können die $1/N$-Korrekturen die logarithmischen Terme der ersten Ordnung nicht kompensieren. Für das langreichweitige Modell sind die $1/N$-Korrekturen des Fixpunkts nichtperturbativ und daher unzuverlässig. Wir finden keinen Vorläufer für den Large-$N$-Fixpunkt. 

Im Kontext von Feldtheorien oder stochastischer Geometrie ist das melonische Large-$N$-Limit die wichtigste Eigenschaft von Tensormodellen. Dieses Limit ist jedoch nicht für alle Tensormodelle etabliert worden. Insbesondere fehlte sie für Tensormodelle mit gewöhnlichen Darstellungen von $O(N)$ oder $Sp(N)$, wie z.B. die antisymmetrische oder symmetrisch spurfreie Darstellung. Diese Lücke wurde kürzlich für Rang $3$ geschlossen. Es wurde bewiesen, dass der melonische Limes für Modelle mit irreduziblen Darstellungen von $O(N)$ oder $Sp(N)$ existiert \cite{Benedetti:2017qxl, Carrozza:2018ewt, Carrozza:2018psc}. Im Kapitel \ref{chap:rank5} verallgemeinern wir diesen Beweis auf Rang $5$. Diese Verallgemeinerung stützt sich auf rekursive Schranken, die mittels einer detaillierten kombinatorischen Analyse der Feynman-Graphen in der Störungsreihe unseres Modells gewonnen werden. Wir nutzen in \cite{Benedetti:2017qxl} entwickelte kombinatorische Methoden. Einige Aspekte vereinfachen sich im Rang $5$ (es ist nicht notwendig, Dreiecksuntergraphen zu untersuchen), aber andere Aspekte sind komplizierter (zum Beispiel müssen Schranken für Acht-Punkt-Graphen zusätzlich zu den Schranken für Zwei- und Vier-Punkt-Graphen abgeleitet werden). Unser Beweis erweitert die Klasse der Modelle, die ein melonisches Large-$N$-Limit aufweisen, und es wäre interessant, ihn auf beliebige Ränge zu verallgemeinern. 

\addcontentsline{toc}{chapter}{Bibliography}
\bibliographystyle{JHEP-3}

\bibliography{Refs-CFT,Refs-QFT,Refs-TMV}


\includepdf[noautoscale]{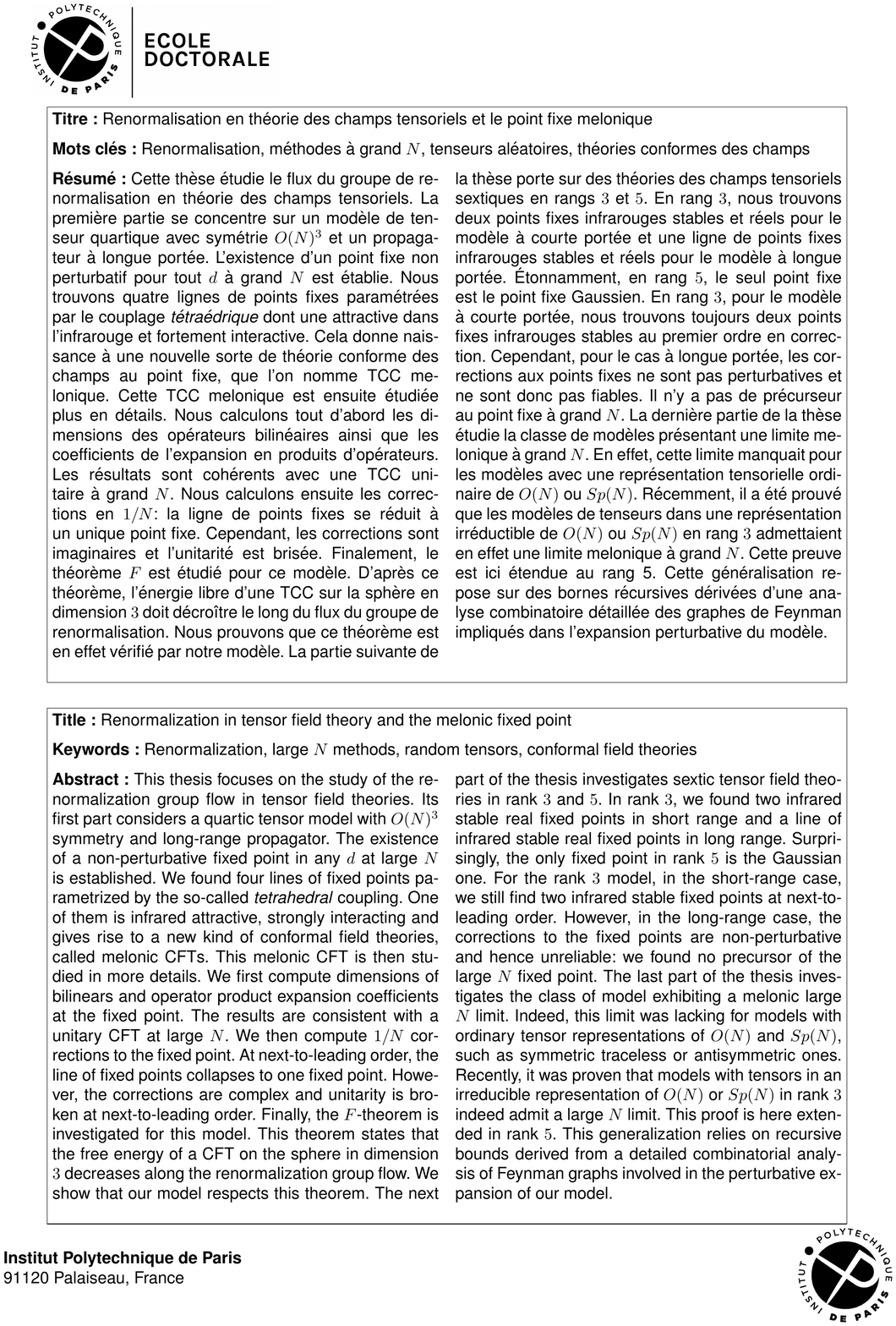}

\end{document}